\documentclass[11pt]{article}
\usepackage{xypic}
\usepackage{mathrsfs}
\usepackage{amssymb}
\usepackage{amsfonts}
\usepackage{amsmath}

\def\ben{\begin{enumerate}}
\def\een{\end{enumerate}}

\def\bit{\begin{itemize}}
\def\eit{\end{itemize}}

\def\0{\leqno}

\input xy
\xyoption{all}
\input{tcilatex}
\begin{document}

\begin{center}
\textbf{THE GENERALIZED LIE ALGEBROIDS\\[0pt]
AND THEIR APPLICATIONS} \bigskip

{CONSTANTIN M. ARCU\c{S} }
\end{center}

\bigskip

\ \ \ \textit{In memory of my uncles}

\ \ \ \textit{Prof. Dr. Gheorghe RADU and Acad. Dr. Doc. Cornelius RADU}

\medskip

\ \ \ \textit{Dedicated to} \textit{Acad. Prof. Dr. Doc. Radu MIRON at his 83%
$^{th}$ anniversary}

\bigskip

\begin{abstract}
In this paper we introduce the notion of generalized Lie algebroid and we
develop a new formalism necessary to obtain a new solution for the
Weistein's Problem [61]. Many applications emphasize the importance and the
utility of this new framework determined by the introduction of generalized
Lie algebroids.

We introduce and develop the exterior differential calculus for generalized
Lie algebroids and, in this general framework, we establish the structure
equations of Maurer-Cartan type. In particular, we obtain a new point of
view over the exterior differential calculus for Lie algebroids.

Using the (generalized) Lie algebroids theory, we build the Lie algebroid
generalized tangent bundle and, using that, we obtain a new method by
determining the (linear) connections for fiber bundles, in general, and for
vector bundles, in particular.

Using the linear connections theory we develop the study of the geometry of
vector bundles. Moreover, using the connections theory, we develop the
geometry of total space of the generalized tangent bundle for a vector
bundle.

We present a geometric description of metrizability for the total space of
the Lie algebroid generalized tangent bundle, where we extend the notions of
generalized Lagrange space, Lagrange space and Finsler space. Using the Lie
algebroid generalized tangent bundle of a generalized Lie algebroid, we
introduce and develop a mechanical systems theory and we present a
Lagrangian formalism for these mechanical systems. In particular, using the
Lie algebroid generalized tangent bundle of a Lie algebroid, we obtain a new
solution for the Weinstein's Problem.

A geometric description of metrizability for the total space of the Lie
algebroid generalized tangent bundle for dual vector bundle is presented. We
extend the notions of generalized Hamilton space, Hamilton space and Cartan
space. Using the Lie algebroid generalized tangent bundle of dual of a
generalized Lie algebroid, we introduce and develop the dual mechanical
systems theory and we present a Hamiltonian formalism for dual mechanical
systems.

Finally, we introduce and develop the concept of (horizontal) Legendre
equivalence between a vector bundle and its dual vector bundle.

We remark that, if the morphisms used are identities morphisms, then we
obtain similar results to the classical results, but which are not classical
results though.\bigskip\newline
\bigskip%
\end{abstract}

\newpage

\newpage

\newpage
\tableofcontents

\newpage

\newpage

\newpage

\newpage

\newpage

\section{Introduction}

The motivation for our researches was the

\begin{quote}
\textbf{Weinstein's Problem:}

\textit{Develop a Lagrangian formalism directly on the given Lie algebroid
similar to Klein's formalism for ordinary Lagrangian Mechanics }$\left[ 25%
\right] $.
\end{quote}

This problem was formulated by A. Weinstein in $[61]$, where the author gave
the theory of Lagrangians on Lie algebroids and obtained the Euler-Lagrange
equations using the dual of a Lie algebroid and the Legendre transformation
defined by a regular Lagrangian.

In $[28]$, P. Liberman showed that such a formalism is not possible if one
consider the tangent bundle of a Lie algebroid as space for developing the
theory. Using the prolongation of a Lie algebroid over a smooth map
introduced by P.J. Higgins and K.~Mackenzie in $\left[ 15\right] $, E.
Martinez solved the \textit{Weinstein's Problem }in $\left[ 61\right]$ (see
also [13],~[27]).

Finding an other space for developing the theory, we discovered the
generalized Lie algebroids which are presented in Subsection 3.3.

Since any Lie algebroid can be regarded as a generalized Lie algebroid, we
proposed to obtain a new solution for the \textit{Weinstein's Problem} using
the new notion of generalized Lie algebroid.

To solve this problem it was necessary to introduce and develop a new
formalism. In order to develop our researches, new and interesting notions
and results appeared, which determined the apparition of new theories which
are naturally integrated in our paper. In particular, using identity
morphisms, we obtain similar results with S. Vacaru [59] (see also [56],
[57], [58]) and L. Popescu (see: [45]-[49]).

So, in Subsection 3.2 we introduce and develop the exterior differential
calculus for generalized Lie algebroids and, using that, we establish the
structure equations of Maurer-Cartan type for generalized Lie algebroids. In
particular, we obtain a new point of view over the exterior differential
calculus for Lie algebroids. We introduced the notion of \emph{interior
differential system }of a generalized Lie algebroid in Paragraph 3.1.3 and
we obtain a theorem of Cartan type. In addition, we introduced the notion of
\emph{exterior differential system} of a generalized Lie algebroid and we
characterized the involutivity of an interior differential system in
Subsection 3.3.

Inspired by the general framework of Yang-Mills theory, presented
synthetically in the following diagram:
\begin{equation*}
\xymatrix{\left( E,\left\langle ,\right\rangle _{E}\right)\ar[d]_\pi & ( TM,
[ , ] _{TM},\ar[d]^{\tau _{M}}&\hspace*{-10mm} ( Id_{TM},Id_{M} ) ,g ) \\ M
\ar[r]^{Id_{M}} & M}
\end{equation*}
where:

\begin{enumerate}
\item $\left( E,\pi ,M\right) $ is a vector bundle,

\item $\left\langle ,\right\rangle _{E}$ is an inner product for the module
of sections $\Gamma \left( E,\pi ,M\right) ,$

\item $\left( \left( Id_{TM},Id_{M}\right) ,\left[ ,\right] _{TM}\right) $
is the usual Lie algebroid structure for the tangent vector bundle $\left(
TM,\tau _{M},M\right) $ and

\item $g\in \Gamma \left( \left( T^{\ast }M,\tau _{M}^{\ast },M\right)
\otimes \left( T^{\ast }M,\tau _{M}^{\ast },M\right) \right) $ such that $%
\left( M,g\right) $ is a Riemannian manifold,
\end{enumerate}

we build the \emph{Lie algebroid generalized tangent bundle} in Subsection
3.3.

Using this in Subsection 3.4, we introduce and develop a (linear)
connections theory for fiber bundles, in general, and for vector bundles, in
particular.

We can define the covariant derivatives with respect to sections of the
generalized Lie algebroid
\begin{equation*}
\left( \left( F,\nu ,N\right) ,\left[ ,\right] _{F,h},\left( \rho ,\eta
\right) \right) .
\end{equation*}

In particular, if we use the generalized Lie algebroid structure
\begin{equation*}
\left( \left[ ,\right] _{TM,Id_{M}},\left( Id_{TM},Id_{M}\right) \right)
\end{equation*}%
for the tangent bundle $\left( TM,\tau _{M},M\right) $ in our theory, then
the linear connections obtained are similar with the classical linear
connections for the vector bundle $\left( E,\pi ,M\right) $, but not
classical linear connections.

It is known that in Yang-Mills theory the set
\begin{equation*}
\begin{array}{c}
Cov_{\left( E,\pi ,M\right) }^{0}%
\end{array}%
\end{equation*}%
of covariant derivatives for the vector bundle $\left( E,\pi ,M\right) $
such that
\begin{equation*}
\begin{array}{c}
X\left( \left\langle u,v\right\rangle _{E}\right) =\left\langle D_{X}\left(
u\right) ,v\right\rangle _{E}+\left\langle u,D_{X}\left( v\right)
\right\rangle _{E} ,%
\end{array}%
\end{equation*}%
for any $X\in \mathcal{X}\left( M\right) $ and $u,v\in \Gamma \left( E,\pi
,M\right) ,$ is very important, because the Yang-Mills theory is a
variational theory which use (cf. $[6]$) the Yang-Mills functional
\begin{equation*}
\begin{array}{rcl}
Cov_{\left( E,\pi ,M\right) }^{0} & ^{\underrightarrow{ \mathcal{YM }}} &
\mathbb{R} \\
D_{X} & \longmapsto & \displaystyle\frac{1}{2}\tint_{M}\left\Vert \mathbb{R}%
^{D_{X}}\right\Vert ^{2}v_{g}%
\end{array}%
\end{equation*}
where $\mathbb{R}^{D_{X}}$ is the curvature.

Using the linear connections theory, we succeed to extend at maximum the set
$Cov_{\left( E,\pi ,M\right) }^{0}$ of Yang-Mills theory, because using all
generalized Lie algebroid structures for the tangent bundle $\left( TM,\tau
_{M},M\right) $, we obtain all possible linear connections for the vector
bundle $\left( E,\pi ,M\right) $.

We emphasize the importance and the utility of linear connections theory for
vector bundles in Chapter IV of our paper, where we present many
applications. In particular, we obtain similar results to the classical
results, but which are not classical results though.

After that we study the geometry of total space of the Lie algebroid
generalized tangent bundle for a vector bundle in Section 5 of our paper,
where we emphasize the importance and the utility of connections theory
presented in Subsection 3.4.

The geometry of Lagrange spaces, introduced and studied in $\left[ 24\right]
$ and $\left[ 35\right] $, was extensively examined in the last two decades
by geometers and physicists from Romania, Japan, Hungary, Canada, Germany,
Italy, Russia and USA. Many international conferences devoted to debate this
subject, proceedings and monographs where published $\left[ 3\right] $, $%
\left[ 4\right] $, $\left[ 41\right] $, $\left[ 42\right] $. A large area of
applicability of this geometry is suggested by the connections to Biology,
Mechanics and Physics and also by its general setting as a generalization of
Finsler and Riemann geometries.

As the (generalized) Lagrange space has been certified as an excellent model
for some important problems in Relativity, Gauge Theory and
Electromagnetism, in Subsections 5.8 and 5.9 we continue and we present a
geometric description of metrizability for the total space of the Lie
algebroid generalized tangent bundle for a vector bundle. We extend the
notions of generalized Lagrange space, Lagrange space and Finsler space and
we define the Einstein equations in this general framework.

Subsection 5.11 is devoted to introduce and study of a new class of
mechanical systems called by us \emph{mechanical} $\left( \rho ,\eta \right)
$\emph{-systems,} \emph{generalized Lagrange mechanical }$\left( \rho ,\eta
\right) $\emph{-systems}, \emph{Lagrange mechanical }$\left( \rho ,\eta
\right) $\emph{-systems and Finsler mechanical }$\left( \rho ,\eta \right) $%
\emph{-systems}.

For these mechanical systems we develop a theory of semisprays and sprays.
We develop a Lagrangian formalism for Lagrange mechanical systems.

We determine and we study the $\left( \rho ,\eta \right) $-semispray
associated to a regular Lagrangian $L$ and external force $F_{e}$ which are
applied on the total space of a generalized Lie algebroid and we derive the
equations of Euler-Lagrange type.

In particular, using the Lie algebroid generalized tangent bundle of a Lie
algebroid, we obtain a new solution for the \textit{Weinstein's Problem }%
different by the Martinez's solution~[61].

Moreover, if the Lie algebroid used is
\begin{equation*}
\left( \left( TM,\tau _{M},M\right) ,\left[ ,\right] _{TM},\left(
Id_{TM},Id_{M}\right) \right) ,
\end{equation*}%
then we obtain similar results to those presented by I.~Bucataru and
R.~Miron in [7].

It is known that in 1918, immediately after the birth of general
re\-la\-ti\-vity, Weyl proposed the first unified theory of gravitation and
electromagnetism, by generalizing the Riemannian space.

We are interested in finding the answer to the following question:

\begin{itemize}
\item \textit{Could we to extend the study of the Riemannian geometry from
the usual Lie algebroid
\begin{equation*}
\begin{array}{c}
\left( \left( TM,\tau _{M},M\right) ,\left[ ,\right] _{TM},\left(
Id_{TM},Id_{M}\right) \right) ,%
\end{array}%
\end{equation*}%
to an arbitrary (generalized) Lie algebroid and can we obtain a general
framework necessary to unify the theory of gravitation with the theory of
electromagnetism?}
\end{itemize}

The future will show how far our theory can be used in this direction.

Our researches continue in Section 6, where we study the geometry of total
space of the Lie algebroid generalized tangent bundle of a dual vector
bundle and so, we emphasize the importance and the utility of the
generalized connections theory presented in paragraph 3.4.1.

We present the adapted $\left( \rho ,\eta \right) $-basis and adapted dual $%
( \rho ,\eta ) $-basis and remarkable endomorphisms of $( \Gamma ( ( \rho
,\eta ) T\overset{\ast }{E}, ( \rho ,\eta ) \tau _{\overset{\ast }{E}},%
\overset{\ast }{E} ) ,+,\cdot ) $ mo\-dule (projectors, almost product
structure, almost tangent structure, almost complex structure, $\left( \rho
,\eta \right) $-tension endomorphism) and we present the $\left( \rho ,\eta
\right) $-torsion and the $\left( \rho ,\eta \right) $-curvature of a $%
\left( \rho ,\eta \right) $-connection $\left( \rho ,\eta \right) \Gamma $.
We introduce and studied distinguished linear $\left( \rho ,\eta \right) $%
-connections and we build the $\left( g,h\right) $-lift of accelerations for
a differentiable curve. Using the distinguished linear $\left( \rho ,\eta
\right) $-connections theory, we introduced and study the $\left( \rho ,\eta
\right) $-torsion, the $\left( \rho ,\eta \right) $-curvature and we present
the formulas of Ricci type and the identities of Cartan and Bianchi type.

The concept of Hamilton space, introduced in $\left[ 36\right] $, $\left[ 40%
\right] $, was intensively studied in $\left[ 19\right] $, $\left[ 20\right]
$, $\left[ 21\right] $, and it has been successful, as a geometric theory of
the Hamiltonian function. The modern formulation of the geometry of Cartan
spaces was given by R. Miron ($\left[ 36\right] $, $\left[ 38\right] $)
although some results where obtained by \'{E}. Cartan $\left[ 9\right] $ and
A. Kawaguchi $\left[ 23\right] .$ Since the fundamental entity in Mechanics
and Physics is the (generalized) Hamilton space, in Subsections 4.8 and 4.9
we continue to present a geometric description of metrizability for the
total space of the Lie algebroid generalized tangent bundle of dual vector
bundle. We extend the notions of generalized Hamilton space, Hamilton space
and Cartan space and we define the Einstein equations in this general
framework.

Subsection 4.11 is devoted to the introduction and the study of a new class
of mechanical systems, called by us \emph{dual} \emph{mechanical} $\left(
\rho ,\eta \right) $\emph{-systems,} \emph{generalized Hamilton mechanical }$%
\left( \rho ,\eta \right) $\emph{-systems,} \emph{Hamilton mechanical }$%
\left( \rho ,\eta \right) $\emph{-systems and Cartan mechanical }$\left(
\rho ,\eta \right) $\emph{-systems}. For dual mechanical systems we develop
a theory of semisprays and sprays. For Hamilton mechanical systems we
develop a Hamiltonian formalism. We determine and study the $\left( \rho
,\eta \right) $-semispray associated to a regular Hamiltonian $H$ and
external force $\overset{\ast }{F}_{e}$, which are applied on the total
space of the dual of a generalized Lie algebroid and we derive the equations
of Hamilton-Jacobi type. One remarks that, if the morphisms used are
identities, then similar results can be obtained by classical results, but
not classical ones.

The classical Legendre's duality makes possible a natural connection between
Lagrange and Hamilton spaces. It reveals new concepts and geometrical
objects of Hamilton spaces that are dual to those which are similar in
Lagrange spaces. The geometrical theory of Hamilton (Cartan) spaces was
investigated from the Legendre duality point of view in the papers $\left[ 36%
\right] $, $\left[ 38\right] $, $\left[ 20\right] $, $\left[ 21\right] $.

In our paper, we propose a new point of view over the Legendre duality. We
introduce and develop the notion of \emph{(horizontal) Legendre }$\left(
\rho ,\eta ,h\right) $-\emph{equivalence }between an arbitrary vector bundle
and its dual. For this new theory it was necessary to build the $\left( \rho
,\eta \right) $\emph{-tangent application} of the Legendre bundle morphism
associated to a Lagrangian or a Hamiltonian.

We consider that this new theory can be used in the develop of the Poisson
Geometry and Symplectic Geometry.

\section{Preliminaries}

In general, if $\mathcal{C}$ is a category, then we denoted by $\left\vert
\mathcal{C}\right\vert $ the class of objects and we denoted by $%
\overrightarrow{\mathcal{C}}$ the class of arrows (morphisms). For any $A,B{%
\in} \left\vert \mathcal{C}\right\vert $, we denote by $\mathcal{C}\left(
A,B\right) $ the morphisms set of $A$ source and $B$ target.

Let $\mathbf{Vect}$, $\mathbf{Liealg},~\mathbf{Mod}$, $\mathbf{Man}$, $%
\mathbf{B}$\textbf{\ }and $\mathbf{B}^{\mathbf{v}}$ be the category of real
vector spaces, Lie algebras, modules, manifolds, fiber bundles and vector
bundles respectively.

\subsection{The category of Lie algebroids}

We assume that $N\in \left\vert \mathbf{Man}\right\vert $ and let $\left[ ,%
\right] _{TN}$ be the usual Lie bracket such that
\begin{equation*}
\begin{array}{c}
\left( \Gamma \left( TN,\tau _{N},N\right) ,+,\cdot ,\left[ ,\right]
_{TN}\right) \in \left\vert \mathbf{LieAlg}\right\vert .%
\end{array}%
\end{equation*}

\noindent\textbf{Definition 2.1.1 }If $\left( F,\nu ,N\right) \in \left\vert
\mathbf{B}^{\mathbf{v}}\right\vert $ such that there exists
\begin{equation*}
\begin{array}{c}
\left( \rho ,Id_{N}\right) \in \mathbf{B}^{\mathbf{v}}\left( \left( F,\nu
,N\right) ,\left( TN,\tau _{N},N\right) \right)%
\end{array}%
\end{equation*}%
and an operation
\begin{equation*}
\begin{array}{ccc}
\Gamma \left( F,\nu ,N\right) \times \Gamma \left( F,\nu ,N\right) & ^{%
\underrightarrow{\, \left[ ,\right] _{F}\, }} & \Gamma \left( F,\nu ,N\right)
\\
\left( u,v\right) & \longmapsto & \left[ u,v\right] _{F}%
\end{array}%
\end{equation*}%
with the following properties:

\begin{itemize}
\item[$LA_{1}$.] the equality holds good
\begin{equation*}
\begin{array}{c}
\left[ u,f\cdot v\right] _{F}=f\left[ u,v\right] _{F}+\Gamma \left( \rho
,Id_{N}\right) \left( u\right) f\cdot v,%
\end{array}%
\end{equation*}%
for all $u,v\in \Gamma \left( F,\nu ,N\right) $ and $f\in \mathcal{F}\left(
N\right) ,$

\item[$LA_{2}$.] the $4$-tuple
\begin{equation*}
\left( \Gamma \left( F,\nu ,N\right) ,+,\cdot ,\left[ ,\right] _{F}\right)
\end{equation*}%
is a Lie $\mathcal{F}\left( N\right) $-algebra$,$

\item[$LA_{3}$.] the $\mathbf{Mod}$-morphism $\Gamma \left( \rho
,Id_{N}\right) $ is a $\mathbf{LieAlg}$-morphism of
\begin{equation*}
\left( \Gamma \left( F,\nu ,N\right) ,+,\cdot ,\left[ ,\right] _{F}\right)
\end{equation*}%
source and
\begin{equation*}
\left( \Gamma \left( TN,\tau _{N},N\right) ,+,\cdot ,\left[ ,\right]
_{TN}\right)
\end{equation*}%
target,
\end{itemize}

then we will say that \emph{the triple }%
\begin{equation*}
\begin{array}{c}
\left( \left( F,\nu ,N\right) ,\left[ ,\right] _{F},\left( \rho
,Id_{N}\right) \right)%
\end{array}%
\leqno(2.1.1)
\end{equation*}%
\emph{is a Lie algebroid. }

The couple
\begin{equation*}
\left( \left[ ,\right] _{F},\left( \rho ,Id_{N}\right) \right)
\end{equation*}
is called \emph{Lie algebroid structure.}

\bigskip\noindent\textbf{Definition 2.1.2 }We define the morphisms set of
\begin{equation*}
\left( \left( F,\nu ,N\right) ,\left[ ,\right] _{F},\left( \rho
,Id_{N}\right) \right)
\end{equation*}%
source and
\begin{equation*}
\left( \left( F^{\prime },\nu ^{\prime },N^{\prime }\right) ,\left[ ,\right]
_{F^{\prime }},\left( \rho ^{\prime },Id_{N^{\prime }}\right) \right)
\end{equation*}%
target as being the set
\begin{equation*}
\begin{array}{c}
\left\{ \left( \varphi ,\varphi _{0}\right) \in \mathbf{B}^{\mathbf{v}%
}\left( \left( F,\nu ,N\right) ,\left( F^{\prime },\nu ^{\prime },N^{\prime
}\right) \right) \right\}%
\end{array}%
\end{equation*}%
such that the $\mathbf{Mod}$-morphism $\Gamma \left( \varphi ,\varphi
_{0}\right) $ is a $\mathbf{LieAlg}$-morphism of
\begin{equation*}
\left( \Gamma \left( F,\nu ,N\right) ,+,\cdot ,\left[ ,\right] _{F}\right)
\end{equation*}%
source and
\begin{equation*}
\left( \Gamma \left( F^{\prime },\nu ^{\prime },N^{\prime }\right) ,+,\cdot
, \left[ ,\right] _{F^{\prime }}\right)
\end{equation*}%
target.

\bigskip\noindent\textbf{Remark 2.1.1 }Note that we can discuss about \emph{%
the category of Lie algebroids.} This category is denoted by $\mathbf{LA}.$

If
\begin{equation*}
\left( \left( F,\nu ,N\right) ,\left[ ,\right] _{F},\left( \rho
,Id_{N}\right) \right)
\end{equation*}%
is a Lie algebroid, then we assume that $\left( F,\nu ,N\right) $ is a
vector bundle with type fibre the real vector space $\left( \mathbb{R}%
^{p},+,\cdot \right) $ and structure group a Lie subgroup of $\left( \mathbf{%
GL}\left( p,\mathbb{R}\right) ,\cdot \right) .$

We take $(\varkappa ^{\tilde{\imath}},z^{\alpha })$ as canonical local
coordinates on $(F,\nu ,N),$ where $\tilde{\imath}{\in }\overline{1,n}$, $%
\alpha \in \overline{1,p}.$

Consider
\begin{equation*}
\left( \varkappa ^{\tilde{\imath}},z^{\alpha }\right) \longrightarrow \left(
\varkappa ^{\tilde{\imath}%
%TCIMACRO{\U{b4}}%
%BeginExpansion
{\acute{}}%
%EndExpansion
},z^{\alpha
%TCIMACRO{\U{b4}}%
%BeginExpansion
{\acute{}}%
%EndExpansion
}\right)
\end{equation*}%
a change of coordinates on $\left( F,\nu ,N\right) $. Then the coordinates $%
z^{\alpha }$ change to $z^{\alpha
%TCIMACRO{\U{b4}}%
%BeginExpansion
{\acute{}}%
%EndExpansion
}$ by the rule:
\begin{equation*}
\begin{array}{c}
z^{\alpha
%TCIMACRO{\U{b4}}%
%BeginExpansion
{\acute{}}%
%EndExpansion
}=\Lambda _{\alpha }^{\alpha
%TCIMACRO{\U{b4}}%
%BeginExpansion
{\acute{}}%
%EndExpansion
}z^{\alpha }.%
\end{array}%
\leqno(2.1.2)
\end{equation*}%
The coefficients $\rho _{\alpha }^{\tilde{\imath}}$ change to $\rho _{\alpha
%TCIMACRO{\U{b4}}%
%BeginExpansion
{\acute{}}%
%EndExpansion
}^{\tilde{\imath}%
%TCIMACRO{\U{b4}}%
%BeginExpansion
{\acute{}}%
%EndExpansion
}$ by the rule:
\begin{equation*}
\begin{array}{c}
\rho _{\alpha
%TCIMACRO{\U{b4}}%
%BeginExpansion
{\acute{}}%
%EndExpansion
}^{\tilde{\imath}%
%TCIMACRO{\U{b4}}%
%BeginExpansion
{\acute{}}%
%EndExpansion
}=\Lambda _{\alpha
%TCIMACRO{\U{b4}}%
%BeginExpansion
{\acute{}}%
%EndExpansion
}^{\alpha }\rho _{\alpha }^{\tilde{\imath}}\frac{\partial \varkappa ^{\tilde{%
\imath}%
%TCIMACRO{\U{b4}}%
%BeginExpansion
{\acute{}}%
%EndExpansion
}}{\partial \varkappa ^{\tilde{\imath}}},%
\end{array}%
\leqno(2.1.3)
\end{equation*}%
where
\begin{equation*}
\left\Vert \Lambda _{\alpha
%TCIMACRO{\U{b4}}%
%BeginExpansion
{\acute{}}%
%EndExpansion
}^{\alpha }\right\Vert =\left\Vert \Lambda _{\alpha }^{\alpha
%TCIMACRO{\U{b4}}%
%BeginExpansion
{\acute{}}%
%EndExpansion
}\right\Vert ^{-1}.
\end{equation*}%
Locally, we obtain
\begin{equation*}
\begin{array}{c}
\left[ t_{\alpha },t_{\beta }\right] _{F}\overset{put}{=}L_{\alpha \beta
}^{\gamma }t_{\gamma }.%
\end{array}%
\leqno(2.1.4)
\end{equation*}%
The real local functions
\begin{equation*}
L_{\alpha \beta }^{\gamma },~\alpha ,\beta ,\gamma \in \overline{1,p}
\end{equation*}%
will be called \emph{structure functions of the Lie algebroid }%
\begin{equation*}
\left( \left( F,\nu ,N\right) ,\left[ ,\right] _{F},\left( \rho
,Id_{N}\right) \right) .
\end{equation*}%
It is easy to prove that
\begin{equation*}
L_{\alpha \beta }^{\gamma }=-L_{\beta \alpha }^{\gamma },~\forall \alpha
,\beta ,\gamma \in \overline{1,p}.
\end{equation*}

\subsection{The pull-back Lie algebroid of a Lie algebroid}

We consider the following diagram:
\begin{equation*}
\begin{array}{ccl}
~\ \ \  &  & \left( F,\left[ ,\right] _{F},\left( \rho ,Id_{N}\right) \right)
\\
&  & ~\downarrow \nu \\
~\ \ \ E & ^{\underrightarrow{~\ \ \ \ \ \ \pi ~\ \ \ \ \ \ }} & ~N%
\end{array}%
\leqno(2.2.1)
\end{equation*}%
where $\left( E,\pi ,M\right) $ is a fiber bundle and $\left( \left( F,\nu
,N\right) ,\left[ ,\right] _{F},\left( \rho ,Id_{N}\right) \right) $ is a
Lie algebroid.

We assume that $\left( E,\pi ,M\right) $ has the type fibre a manifold of
dimension $r$ and structure group a Lie group $\left( \mathbf{G},\cdot
\right) .$

\bigskip\noindent\textbf{Proposition 2.2.1 }\emph{Using the tangent }$%
\mathbf{B}^{\mathbf{v}}$\emph{-morphism }$\left( T\pi ,\pi \right) $ \emph{%
of }$\left( TE,\tau _{E},E\right) $\emph{\ source and }$\left( TN,\tau
_{N},N\right) $\emph{\ target, we obtain that}%
\begin{equation*}
\frac{\partial f\circ \pi }{\partial x^{\tilde{\imath}}}=\frac{\partial f}{%
\partial x^{\tilde{\imath}}}\circ \pi ,~\forall f\in \mathcal{F}\left(
N\right)\leqno(2.2.2)
\end{equation*}%
\emph{and}%
\begin{equation*}
\frac{\partial f\circ \pi }{\partial y^{a}}=0,~\forall f\in \mathcal{F}%
\left( N\right) .\leqno(2.2.3)
\end{equation*}

Let $\mathcal{AF}_{F}$ be a representative of vector fibred $\left(
n+p\right) $-structure for the vector bundle $\left( F,\nu ,N\right) $ and
let $\mathcal{AF}_{E}$ be a representative of fibred $\left( n+r\right) $%
-structure for the fiber bundle $\left( E,\pi ,N\right) $. Let $\left( \pi
^{\ast }F,\pi ^{\ast }\nu ,E\right) $ be the pull-back vector bundle through
$\pi .$

If $\left( U,\xi _{U}\right) \in \mathcal{AF}_{E}$ and $\left(
V,s_{V}\right) \in \mathcal{AF}_{F}$ such that $U\cap V\neq \phi $, then we
define the application%
\begin{equation*}
\begin{array}{ccc}
\pi ^{\ast }\nu ^{-1}(\pi ^{-1}\left( U{\cap }V\right) ) & {}^{%
\underrightarrow{\tilde{s}_{\pi ^{-1}\left( U{\cap }V\right) }}} & \pi
^{-1}\left( U{\cap }V\right) {\times }\mathbb{R}^{p} \\
\left( u,\tilde{Z}\left( u\right) \right) & \longmapsto & \left( \varkappa
,t_{V,\pi \left( u\right) }^{-1}\tilde{Z}\left( u\right) \right).%
\end{array}%
\end{equation*}

\noindent\textbf{Proposition 2.2.2 }\emph{The set}%
\begin{equation*}
\begin{array}{c}
\widetilde{\mathcal{AF}}_{\pi ^{\ast }F}\overset{put}{=}\underset{U\cap
V\neq \phi }{\underset{\left( U,\xi _{U}\right) \in \mathcal{AF}_{E},~\left(
V,s_{V}\right) \in \mathcal{AF}_{F}}{\tbigcup }}\left\{ \left( \pi
^{-1}\left( U{\cap }V\right) ,\tilde{s}_{\pi ^{-1}\left( U{\cap }V\right)
}\right) \right\}%
\end{array}%
\end{equation*}%
\emph{is a vector fibred }$\left( m+r\right) +p$\emph{-atlas for the vector
bundle }$\left( \pi ^{\ast }F,\pi ^{\ast }\nu ,E\right) .$

\emph{If }%
\begin{equation*}
z=z^{\alpha }t_{\alpha }\in \Gamma \left( F,\nu ,N\right) ,
\end{equation*}%
\emph{then, using the vector fibred }$\left( m+r\right) +p$\emph{-structure }%
$\left[ \widetilde{\mathcal{AF}}_{\pi ^{\ast }F}\right] ,$\emph{\ we obtain
the section}%
\begin{equation*}
\tilde{Z}=\left( z^{\alpha }\circ \pi \right) \tilde{T}_{\alpha }\in \Gamma
\left( \pi ^{\ast }F,\pi ^{\ast }\nu ,E\right)
\end{equation*}%
\emph{such that}%
\begin{equation*}
\tilde{Z}\left( u_{x}\right) =z\left( x\right) ,
\end{equation*}%
\emph{for any }$u_{x}\in \pi ^{-1}\left( U{\cap }V\right) .$\bigskip

The set $\left\{ \tilde{T}_{\alpha },~\alpha \in \overline{1,p}\right\} $ is
a base for the module of sections%
\begin{equation*}
\begin{array}{c}
\left( \Gamma \left( \pi ^{\ast }F,\pi ^{\ast }\nu ,E\right) ,+,\cdot
\right) .%
\end{array}%
\end{equation*}%
Let $\Big({\overset{\pi ^{\ast }F}{\rho }},Id_{E}\Big)$ be the $\mathbf{B}^{%
\mathbf{v}}$-morphism of
\begin{equation*}
\left( \pi ^{\ast }F,\pi ^{\ast }\nu ,E\right)
\end{equation*}%
\ source and
\begin{equation*}
\left( TE,\tau _{E},E\right)
\end{equation*}%
\ target, where%
\begin{equation*}
\begin{array}{rcl}
\pi ^{\ast }F & ^{\underrightarrow{\,\overset{\pi ^{\ast }F}{\rho }\,}} & TE
\\
\tilde{Z}^{\alpha }\tilde{T}_{\alpha }\left( u_{x}\right) & \longmapsto &
\left( \tilde{Z}^{\alpha }\cdot \rho _{\alpha }^{\tilde{\imath}}\circ \pi %
\displaystyle\frac{\partial }{\partial x^{\tilde{\imath}}}\right) \left(
u_{x}\right)%
\end{array}%
\leqno(2.2.4)
\end{equation*}%
We consider the operation
\begin{equation*}
\begin{array}{ccc}
\Gamma \left( \pi ^{\ast }F,\pi ^{\ast }\nu ,E\right) \times \Gamma \left(
\pi ^{\ast }F,\pi ^{\ast }\nu ,E\right) & ^{\underrightarrow{~\ \ \left[ ,%
\right] _{\pi ^{\ast }F}~\ \ }} & \Gamma \left( \pi ^{\ast }F,\pi ^{\ast
}\nu ,E\right)%
\end{array}%
\end{equation*}%
defined by%
\begin{equation*}
\begin{array}{ll}
\left[ \tilde{T}_{\alpha },\tilde{T}_{\beta }\right] _{\pi ^{\ast }F} &
=\left( L_{\alpha \beta }^{\gamma }\circ \pi \right) \tilde{T}_{\gamma },%
\vspace*{1mm} \\
\left[ \tilde{T}_{\alpha },f\tilde{T}_{\beta }\right] _{\pi ^{\ast }F} &
=f\left( L_{\alpha \beta }^{\gamma }\circ \pi \right) \tilde{T}_{\gamma
}+\left( \rho _{\alpha }^{\tilde{\imath}}\circ \pi \right) \displaystyle%
\frac{\partial f}{\partial x^{\tilde{\imath}}}\tilde{T}_{\beta },\vspace*{1mm%
} \\
\left[ f\tilde{T}_{\alpha },\tilde{T}_{\beta }\right] _{\pi ^{\ast }F} & =-
\left[ \tilde{T}_{\beta },f\tilde{T}_{\alpha }\right] _{\pi ^{\ast }F},%
\end{array}%
\leqno(2.2.5)
\end{equation*}%
for any $f\in \mathcal{F}\left( E\right) .$\bigskip

\noindent\textbf{Lemma 2.2.1 }\emph{The following equality holds good }%
\begin{equation*}
\left[ \tilde{U},f\tilde{V}\right] _{\pi ^{\ast }F}=f\left[ \tilde{U},\tilde{%
V}\right] _{\pi ^{\ast }F}+\Gamma \left( {\overset{\pi ^{\ast }F}{\rho }}%
,Id_{E}\right) \left( \tilde{U}\right) f\cdot \tilde{V},
\end{equation*}%
\emph{for any }$\tilde{U},\tilde{V}\in \Gamma \left( \pi ^{\ast }F,\pi
^{\ast }\nu ,E\right) $\emph{\ and for any }$f\in \mathcal{F}\left( E\right)
.$\bigskip

\noindent\emph{Proof.} We observe that for any $\alpha ,\beta \in \overline{%
1,p}$, we obtain%
\begin{equation*}
\left[ \tilde{T}_{\alpha },f\tilde{T}_{\beta }\right] _{\pi ^{\ast }F}=f%
\left[ \tilde{T}_{\alpha },\tilde{T}_{\beta }\right] _{\pi ^{\ast }F}+\Gamma
\left( {\overset{h^{\ast }F}{\rho }},Id_{E}\right) \tilde{T}_{\alpha }\left(
f\right) ,~\forall f\in \mathcal{F}\left( E\right) .
\end{equation*}
Using this equality and the definition of the operation $\ \left[ ,\right]
_{\pi ^{\ast }F}$ it results the conclusion of the lemma. \hfill \emph{q.e.d.%
}\bigskip

\noindent\textbf{Lemma 2.2.2 }\emph{The }$\mathcal{F}\left( E\right) $\emph{%
-algebra }%
\begin{equation*}
\left( \Gamma \left( \pi ^{\ast }F,\pi ^{\ast }\nu ,E\right) ,+,\cdot ,\left[
,\right] _{\pi ^{\ast }F}\right)
\end{equation*}%
\emph{is a Lie }$\mathcal{F}\left( E\right) $\emph{-algebra.}\bigskip

\noindent\emph{Proof.} Using the definition of the operation $\ \left[ ,%
\right] _{\pi ^{\ast }F}$ it results that
\begin{equation*}
\ \left[ \tilde{U},\tilde{V}\right] _{\pi ^{\ast }F}=-\left[ \tilde{V},%
\tilde{U}\right] _{\pi ^{\ast }F},
\end{equation*}%
for any $\tilde{U},\tilde{V}\in \Gamma \left( \pi ^{\ast }F,\pi ^{\ast }\nu
,E\right) .$ Therefore, we obtain%
\begin{equation*}
\begin{array}{c}
\left[ \tilde{U},\tilde{U}\right] _{\pi ^{\ast }F}=0,~\forall \tilde{U}\in
\Gamma \left( \pi ^{\ast }F,\pi ^{\ast }\nu ,E\right) .%
\end{array}%
\leqno(1)
\end{equation*}
Since
\begin{equation*}
\left( \Gamma \left( F,\nu ,N\right) ,+,\cdot ,\left[ ,\right] _{F}\right)
\end{equation*}
is a Lie $\mathcal{F}\left( N\right) $-algebra, we obtain the equality:
\begin{equation*}
\begin{array}{c}
\underset{cyclic\left( \alpha ,\beta ,\gamma \right) }{\tsum }\left(
L_{\beta \gamma }^{\varepsilon }L_{\alpha \varepsilon }^{\delta }+\rho
_{\alpha }^{\tilde{\imath}}\frac{\partial L_{\beta \gamma }^{\delta }}{%
\partial \varkappa ^{\tilde{\imath}}}\right) =0.%
\end{array}%
\end{equation*}
Using $\left( 2.2.2\right) $, we obtain the equality:%
\begin{equation*}
\begin{array}{l}
\underset{cyclic\left( \alpha ,\beta ,\gamma \right) }{\tsum }\left( \left(
L_{\beta \gamma }^{\varepsilon }\circ \pi \right) \left( L_{\alpha
\varepsilon }^{\delta }\circ \pi \right) +\rho _{\alpha }^{\tilde{\imath}%
}\circ \pi \frac{\partial \left( L_{\beta \gamma }^{\delta }\circ \pi
\right) }{\partial \varkappa ^{\tilde{\imath}}}\right) \,=0.%
\end{array}%
\end{equation*}%
and multiplying with $\tilde{T}_{\delta }$, we obtain
\begin{equation*}
\begin{array}{l}
\underset{cyclic\left( \alpha ,\beta ,\gamma \right) }{\tsum }\left( \left(
L_{\beta \gamma }^{\varepsilon }\circ \pi \right) \left( L_{\alpha
\varepsilon }^{\delta }\circ \pi \right) \tilde{T}_{\delta }\,+\rho _{\alpha
}^{\tilde{\imath}}\circ \pi \frac{\partial \left( L_{\beta \gamma }^{\delta
}\circ \pi \right) }{\partial \varkappa ^{\tilde{\imath}}}\tilde{T}_{\delta
}\,\right) \,=0.%
\end{array}%
\end{equation*}%
which is equivalent with
\begin{equation*}
\begin{array}{l}
\underset{cyclic\left( \alpha ,\beta ,\gamma \right) }{\tsum }\left( \left(
L_{\beta \gamma }^{\varepsilon }\circ \pi \right) \left[ \tilde{T}_{\alpha },%
\tilde{T}_{\varepsilon }\right] _{\pi ^{\ast }F}+\rho _{\alpha }^{\tilde{%
\imath}}\circ \pi \frac{\partial L_{\beta \gamma }^{\delta }\circ \pi }{%
\partial \varkappa ^{\tilde{\imath}}}\tilde{T}_{\delta }\right) =0.%
\end{array}%
\end{equation*}
Since this equality implies
\begin{equation*}
\begin{array}{c}
\underset{cyclic\left( \alpha ,\beta ,\gamma \right) }{\tsum }\left[ \tilde{T%
}_{\alpha },\!\left( L_{\beta \gamma }^{\varepsilon }{\circ }\pi \right)
\tilde{T}_{\varepsilon }\right] _{\pi ^{\ast }F}=0,%
\end{array}%
\end{equation*}%
it results that the following Jacobi identity is satisfied
\begin{equation*}
\begin{array}{c}
\underset{cyclic\left( \alpha ,\beta ,\gamma \right) }{\tsum }\left[ \tilde{T%
}_{\alpha },\left[ \tilde{T}_{\beta },\tilde{T}_{\gamma }\right] _{\pi
^{\ast }F}\right] _{\pi ^{\ast }F}=0.%
\end{array}%
\end{equation*}
In general, for any $\tilde{U},\tilde{V},\tilde{Z}\in \Gamma \left( \pi
^{\ast }F,\pi ^{\ast }\nu ,E\right) $, we obtain the Jacobi identity:
\begin{equation*}
\left[ \tilde{U},\left[ \tilde{V},\tilde{Z}\right] _{\pi ^{\ast }F}\right]
_{\pi ^{\ast }F}+\left[ \tilde{Z},\left[ \tilde{U},\tilde{V}\right] _{\pi
^{\ast }F}\right] _{\pi ^{\ast }F}\vspace*{1mm}+\left[ \tilde{V},\left[
\tilde{Z},\tilde{U}\right] _{\pi ^{\ast }F}\right] _{\pi ^{\ast }F}=0.\leqno%
(2)
\end{equation*}
Using the affirmations $\left( 1\right) $ and $\left( 2\right) $ it results
the conclusion of the lemma.\hfill \emph{q.e.d.}\bigskip

\noindent\textbf{Lemma 2.2.3 }\emph{The }$\mathbf{Mod}$\emph{-morphism }%
\begin{equation*}
\Gamma \!\!\left( \overset{\pi ^{\ast }F}{\rho }\!\!,Id_{E}\!\right)
\end{equation*}%
\emph{is a }$\mathbf{Liealg}$\emph{-morphism of}%
\begin{equation*}
\left( \Gamma \left( \pi ^{\ast }F,\pi ^{\ast }\nu ,E\right) ,+,\cdot ,\left[
,\right] _{\pi ^{\ast }F}\right)
\end{equation*}%
\emph{source and }%
\begin{equation*}
\left( \Gamma \!(TE,\tau _{E},\!E),+,\cdot ,\left[ ,\right] _{TE}\right)
\end{equation*}%
\emph{target.}\bigskip

\noindent\emph{Proof.} As the $\mathbf{Mod}$-morphism $\Gamma \!\!\left(
\rho \!\!,Id_{N}\!\right) $ is a $\mathbf{Liealg}$-morphism of%
\begin{equation*}
\left( \Gamma \left( F,\nu ,N\right) ,+,\cdot ,\left[ ,\right] _{F}\right)
\end{equation*}%
source and
\begin{equation*}
\left( \Gamma \!(TN,\tau _{N},\!N),+,\cdot ,\left[ ,\right] _{TN}\right)
\end{equation*}%
target, then we obtain%
\begin{equation*}
\begin{array}{c}
L_{\alpha \beta }^{\gamma }\rho _{\gamma }^{\tilde{k}}=\rho _{\alpha }^{%
\tilde{\imath}}\frac{\partial \left( \rho _{\beta }^{\tilde{k}}\right) }{%
\partial \varkappa ^{\tilde{\imath}}}-\rho _{\beta }^{\tilde{j}}\frac{%
\partial \left( \rho _{\alpha }^{\tilde{k}}\right) }{\partial \varkappa ^{%
\tilde{j}}}%
\end{array}%
\end{equation*}
Using relations $\left( 2.2.2\right) $, we obtain:
\begin{equation*}
\begin{array}{c}
\left( L_{\alpha \beta }^{\gamma }\circ \pi \right) \left( \rho _{\gamma }^{%
\tilde{k}}\circ \pi \right) =\rho _{\alpha }^{\tilde{\imath}}\circ \pi \frac{%
\partial \left( \rho _{\beta }^{\tilde{k}}\circ \pi \right) }{\partial
\varkappa ^{\tilde{\imath}}}-\rho _{\beta }^{\tilde{j}}\circ \pi \frac{%
\partial \left( \rho _{\alpha }^{\tilde{k}}\circ \pi \right) }{\partial
\varkappa ^{\tilde{j}}}%
\end{array}%
\end{equation*}
Multiplying with $\frac{\partial }{\partial \varkappa ^{\tilde{k}}}$, we
obtain the equality%
\begin{equation*}
\begin{array}{c}
\left( L_{\alpha \beta }^{\gamma }\circ \pi \right) \left( \rho _{\gamma }^{%
\tilde{k}}\circ \pi \right) \frac{\partial }{\partial \varkappa ^{\tilde{k}}}%
=\rho _{\alpha }^{\tilde{\imath}}\circ \pi \frac{\partial \left( \rho
_{\beta }^{\tilde{k}}\circ \pi \right) }{\partial \varkappa ^{\tilde{\imath}}%
}\frac{\partial }{\partial \varkappa ^{\tilde{k}}}-\rho _{\beta }^{\tilde{j}%
}\circ \pi \frac{\partial \left( \rho _{\alpha }^{\tilde{k}}\circ \pi
\right) }{\partial \varkappa ^{\tilde{j}}}\frac{\partial }{\partial
\varkappa ^{\tilde{k}}}%
\end{array}%
\end{equation*}%
which is equivalent with the equality
\begin{equation*}
\begin{array}{c}
\Gamma \left( \overset{\pi ^{\ast }F}{\rho },Id_{E}\right) \left[ \tilde{T}%
_{\alpha },\tilde{T}_{\beta }\right] _{_{\pi ^{\ast }F}}=\left[ \Gamma
\left( \overset{\pi ^{\ast }F}{\rho },Id_{E}\right) \tilde{T}_{\alpha
},\Gamma \left( \overset{\pi ^{\ast }F}{\rho },Id_{E}\right) \tilde{T}%
_{\beta }\right] _{TN}%
\end{array}%
\end{equation*}%
for any base sections $\tilde{T}_{\alpha },\tilde{T}_{\beta }.$

In general, we obtain the equality
\begin{equation*}
\Gamma \left( \overset{\pi ^{\ast }F}{\rho },Id_{E}\right) \left[ \tilde{U},%
\tilde{V}\right] _{_{\pi ^{\ast }F}}=\left[ \Gamma \left( \overset{\pi
^{\ast }F}{\rho },Id_{E}\right) \tilde{U},\Gamma \left( \overset{\pi ^{\ast
}F}{\rho },Id_{E}\right) \tilde{V}\right] _{TN},
\end{equation*}%
for any $\tilde{U},\tilde{V}\in \Gamma \left( h^{\ast }F,h^{\ast }\nu
,M\right) .$ \hfill \emph{q.e.d.}\bigskip

Using \emph{Lemmas 2.2.1, 2.2.2 }and\emph{\ 2.2.3}, we obtain the following

\bigskip\noindent \textbf{Theorem 2.2.1 }\emph{The couple }%
\begin{equation*}
\begin{array}{c}
\left( \left[ ,\right] _{\pi ^{\ast }F},\left( \overset{\pi ^{\ast }F}{\rho }%
,Id_{E}\right) \right)%
\end{array}%
\end{equation*}%
\emph{\ is a Lie algebroid structure for the vector bundle }$\left( \pi
^{\ast }F,\pi ^{\ast }\nu ,E\right) .$\medskip

This Lie algebroid will be called \emph{the pull-back Lie algebroid of the
Lie algebroid }%
\begin{equation*}
\begin{array}{c}
\left( \left( F,\nu ,N\right) ,\left[ ,\right] _{F},\left( \rho
,Id_{N}\right) \right) .%
\end{array}%
\end{equation*}

\section{Generalized Lie algebroids, exterior differential calculus and
(linear) connections}

\subsection{The category of generalized Lie algebroids}

We assume that $N\in \left\vert \mathbf{Man}\right\vert $ and let $\left[ ,%
\right] _{TN}$ be the usual Lie bracket such that
\begin{equation*}
\begin{array}{c}
\left( \Gamma \left( TN,\tau _{N},N\right) ,+,\cdot ,\left[ ,\right]
_{TN}\right) \in \left\vert \mathbf{LieAlg}\right\vert .%
\end{array}%
\end{equation*}
Let $h\in \mathbf{Man}\left( M,N\right) $ be a surjective application.

\bigskip\noindent\textbf{Definition 3.1.1 }If $\left( F,\nu ,N\right) \in
\left\vert \mathbf{B}^{\mathbf{v}}\right\vert $ such that there exists
\begin{equation*}
\begin{array}{c}
\left( \rho ,\eta \right) \in \mathbf{B}^{\mathbf{v}}\left( \left( F,\nu
,N\right) ,\left( TM,\tau _{M},M\right) \right)%
\end{array}%
\end{equation*}%
and an operation
\begin{equation*}
\begin{array}{ccc}
\Gamma \left( F,\nu ,N\right) \times \Gamma \left( F,\nu ,N\right) & ^{%
\underrightarrow{\left[ ,\right] _{F,h}}} & \Gamma \left( F,\nu ,N\right) \\
\left( u,v\right) & \longmapsto & \left[ u,v\right] _{F,h}%
\end{array}%
\end{equation*}%
with the following properties:\bigskip

\noindent$GLA_{1}$. the equality holds good
\begin{equation*}
\begin{array}{c}
\left[ u,f\cdot v\right] _{F,h}=f\left[ u,v\right] _{F,h}+\Gamma \left(
Th\circ \rho ,h\circ \eta \right) \left( u\right) f\cdot v,%
\end{array}%
\end{equation*}%
\qquad\quad\ \ for all $u,v\in \Gamma \left( F,\nu ,N\right) $ and $f\in
\mathcal{F}\left( N\right) .$

\medskip \noindent$GLA_{2}$. the $4$-tuple
\begin{equation*}
\left( \Gamma \left( F,\nu ,N\right) ,+,\cdot ,\left[ ,\right] _{F,h}\right)
\end{equation*}%
\qquad\quad\ \ is a Lie $\mathcal{F}\left( N\right) $-algebra,

\medskip \noindent$GLA_{3}$. the $\mathbf{Mod}$-morphism $\Gamma \left(
Th\circ \rho ,h\circ \eta \right) $ is a $\mathbf{LieAlg}$-morphism of
\begin{equation*}
\left( \Gamma \left( F,\nu ,N\right) ,+,\cdot ,\left[ ,\right] _{F,h}\right)
\end{equation*}%
\qquad\quad\ \ source and
\begin{equation*}
\left( \Gamma \left( TN,\tau _{N},N\right) ,+,\cdot ,\left[ ,\right]
_{TN}\right)
\end{equation*}%
\qquad\quad\ \ target,

\medskip \noindent then we will say that \emph{the triple }%
\begin{equation*}
\begin{array}{c}
\left( \left( F,\nu ,N\right) ,\left[ ,\right] _{F,h},\left( \rho ,\eta
\right) \right)%
\end{array}%
\leqno(3.1.1)
\end{equation*}%
\emph{is a generalized Lie algebroid. }

The couple
\begin{equation*}
\begin{array}{c}
\left( \left[ ,\right] _{F,h},\left( \rho ,\eta \right) \right)%
\end{array}%
\end{equation*}%
will be called \emph{generalized Lie algebroid structure.}

\bigskip\noindent\textbf{Definition 3.1.2 }We define the morphisms set of
\begin{equation*}
\left( \left( F,\nu ,N\right) ,\left[ ,\right] _{F,h},\left( \rho ,\eta
\right) \right)
\end{equation*}%
source and
\begin{equation*}
\left( \left( F^{\prime },\nu ^{\prime },N^{\prime }\right) ,\left[ ,\right]
_{F^{\prime },h^{\prime }},\left( \rho ^{\prime },\eta ^{\prime }\right)
\right)
\end{equation*}%
target as being the set
\begin{equation*}
\begin{array}{c}
\left\{ \left( \varphi ,\varphi _{0}\right) \in \mathbf{B}^{\mathbf{v}%
}\left( \left( F,\nu ,N\right) ,\left( F^{\prime },\nu ^{\prime },N^{\prime
}\right) \right) \right\}%
\end{array}%
\end{equation*}%
such that the $\mathbf{Mod}$-morphism $\Gamma \left( \varphi ,\varphi
_{0}\right) $ is a $\mathbf{LieAlg}$-morphism of
\begin{equation*}
\left( \Gamma \left( F,\nu ,N\right) ,+,\cdot ,\left[ ,\right] _{F,h}\right)
\end{equation*}%
source and
\begin{equation*}
\left( \Gamma \left( F^{\prime },\nu ^{\prime },N^{\prime }\right) ,+,\cdot
, \left[ ,\right] _{F^{\prime },h^{\prime }}\right)
\end{equation*}%
target.

\bigskip\noindent\textbf{Remark 3.1.1 }Note that we discuss about \emph{the
category of generalized Lie algebroids.} This category will be denoted by $%
\mathbf{GLA}.$ \bigskip

In the following we will build some examples of generalized Lie algebroids.

We assume that $\left( \left( F,\nu ,N\right) ,\left[ ,\right] _{F},\left(
\rho ,Id_{N}\right) \right) $ is a Lie algebroid and let $h\in \mathbf{Man}%
\left( N,N\right) $ be a surjective application.

Let $\mathcal{AF}_{F}$ be a representative of vector fibred $\left(
n+p\right) $-structure for the vector bundle $\left( F,\nu ,N\right) $ and
let $\mathcal{AF}_{TN}$ be a representative of vector fibred $\left(
n+n\right) $-structure for the vector bundle $\left( TN,\tau _{N},N\right) $.

If $\left( U,\xi _{U}\right) \in \mathcal{AF}_{TN}$ and $\left(
V,s_{V}\right) \in \mathcal{AF}_{F}$ such that $U\cap h^{-1}\left( V\right)
\neq \phi $, then we define the application%
\begin{equation*}
\begin{array}{ccc}
\tau _{N}^{-1}(U{\cap }h^{-1}(V))) & {}^{\underrightarrow{\bar{\xi}_{U{\cap }%
h^{-1}(V)}}} & \left( U{\cap }h^{-1}(V)\right) {\times }\mathbb{R}^{n} \\
\left( \varkappa ,u\left( \varkappa \right) \right) & \longmapsto & \left(
\varkappa ,\xi _{U,\varkappa }^{-1}u\left( \varkappa \right) \right).%
\end{array}%
\end{equation*}

\bigskip\noindent \textbf{Proposition 3.1.1 }\emph{The set}%
\begin{equation*}
\begin{array}{c}
\overline{\mathcal{AF}}_{TN}\overset{put}{=}\underset{U\cap h^{-1}\left(
V\right) \neq \phi }{\underset{\left( U,\xi _{U}\right) \in \mathcal{AF}%
_{TN},~\left( V,s_{V}\right) \in \mathcal{AF}_{F}}{\tbigcup }}\left\{ \left(
U\cap h^{-1}\left( V\right) ,\bar{\xi}_{U{\cap }h^{-1}(V)}\right) \right\}%
\end{array}%
\end{equation*}%
\emph{is a vector fibred }$n+n$\emph{-atlas for the vector bundle }$\left(
TN,\tau _{N},N\right) .$

\emph{If }%
\begin{equation*}
\begin{array}{c}
X=X^{\tilde{\imath}}\frac{\partial }{\partial \varkappa ^{\tilde{\imath}}}%
\in \Gamma \left( TN,\tau _{N},N\right)%
\end{array}%
\end{equation*}%
\emph{then, using the vector fibred }$n+n$\emph{-structure }$\left[
\overline{\mathcal{AF}}_{TN}\right] ,$\emph{\ we obtain the section }%
\begin{equation*}
\begin{array}{c}
\bar{X}=\bar{X}^{\tilde{\imath}}\circ h\frac{\partial }{\partial \bar{%
\varkappa}^{\tilde{\imath}}}\in \Gamma \left( TN,\tau _{N},N\right) ,%
\end{array}%
\end{equation*}%
\emph{such that }%
\begin{equation*}
\bar{X}\left( \bar{\varkappa}\right) =X\left( h\left( \bar{\varkappa}\right)
\right) ,
\end{equation*}%
\emph{for any }$\bar{\varkappa}\in U\cap h^{-1}\left( V\right) .$\bigskip

The set $\left\{ \frac{\partial }{\partial \bar{\varkappa}^{\tilde{\imath}}}%
,~\tilde{\imath}\in \overline{1,n}\right\} $ is a base for the $\mathcal{F}%
\left( N\right) $-module $\left( \Gamma \left( TN,\tau _{N},N\right)
,+,\cdot \right) .$

We consider the operation
\begin{equation*}
\begin{array}{ccc}
\Gamma \left( F,\nu ,N\right) \times \Gamma \left( F,\nu ,N\right) & ^{%
\underrightarrow{~\ \ \left[ ,\right] _{F,h}~\ \ }} & \Gamma \left( F,\nu
,N\right)%
\end{array}%
\end{equation*}%
defined by%
\begin{equation*}
\begin{array}{ll}
\left[ t_{\alpha },t_{\beta }\right] _{F,h} & =\left( L_{\alpha \beta
}^{\gamma }\circ h\right) t_{\gamma }, \vspace*{1,5mm} \\
\left[ t_{\alpha },ft_{\beta }\right] _{F,h} & =f\left( L_{\alpha \beta
}^{\gamma }\circ h\right) t_{\gamma }+\rho _{\alpha }^{\tilde{\imath}}\circ h%
\displaystyle\frac{\partial f}{\partial \bar{\varkappa}^{\tilde{\imath}}}%
t_{\beta }, \vspace*{1,5mm} \\
\left[ ft_{\alpha },t_{\beta }\right] _{F,h} & =-\left[ t_{\beta
},ft_{\alpha }\right] _{F,h},~%
\end{array}%
\end{equation*}%
for any $f\in \mathcal{F}\left( N\right) .$

\bigskip \noindent \textbf{Lemma 3.1.1 }\emph{The following equality holds
good }%
\begin{equation*}
\left[ z,fv\right] _{F,h}=f\left[ z,v\right] _{F,h}+\Gamma \left( Th\circ
\rho ,h\right) \left( z\right) f\cdot v,
\end{equation*}%
\emph{for any }$z,v\in \Gamma \left( F,\nu ,N\right) $\emph{\ and} $f\in
\mathcal{F}\left( N\right) .$

\bigskip\noindent\emph{Proof. }We obtain easily that%
\begin{equation*}
\begin{array}{c}
\left[ t_{\alpha },ft_{\beta }\right] _{F,h}=\left[ t_{\alpha },t_{\beta }%
\right] _{F,h}+\Gamma \left( Th\circ \rho ,h\right) \left( t_{\alpha
}\right) f\cdot t_{\beta }%
\end{array}%
\end{equation*}%
for any $\forall f\in \mathcal{F}\left( N\right) .$

Using this equality and the definition of the operation $\ \left[ ,\right]
_{F,h}$ it results the conclusion of the lemma. \hfill \emph{q.e.d.}

\bigskip\noindent\textbf{Lemma 3.1.2 }\emph{The} $\mathcal{F}\left( N\right)
$\emph{-algebra }%
\begin{equation*}
\left( \Gamma \left( F,\nu ,N\right) ,+,\cdot ,\left[ ,\right] _{F,h}\right)
\end{equation*}%
\emph{is a Lie }$\mathcal{F}\left( N\right) $\emph{-algebra.}\bigskip

\noindent\emph{Proof. }Using the definition of the operation $\left[ ,\right]
_{F,h}$ it results that
\begin{equation*}
\left[ u,v\right] _{F,h}=-\left[ v,u\right] _{F,h},
\end{equation*}%
for any $u,v\in \Gamma \left( F,\nu ,N\right) .$ Therefore, we obtain
\begin{equation*}
\begin{array}{c}
\left[ u,u\right] _{F,h}=0,~\forall u\in \Gamma \left( F,\nu ,N\right) .%
\end{array}%
\leqno(1)
\end{equation*}
Since $\left( \Gamma \left( F,\nu ,N\right) ,+,\cdot ,\left[ ,\right]
_{F}\right) $ is a Lie $\mathcal{F}\left( N\right) $-algebra, it results
that
\begin{equation*}
\begin{array}{l}
\underset{cyclic\left( \alpha ,\beta ,\gamma \right) }{\tsum }\left(
L_{\beta \gamma }^{\varepsilon }L_{\alpha \varepsilon }^{\delta }+\rho
_{\alpha }^{\tilde{\imath}}\frac{\partial L_{\beta \gamma }^{\delta }}{%
\partial \varkappa ^{\tilde{\imath}}}\,\right) =0.%
\end{array}%
\end{equation*}
Therefore,%
\begin{equation*}
\begin{array}{l}
\underset{cyclic\left( \alpha ,\beta ,\gamma \right) }{\tsum }\left( \left(
L_{\beta \gamma }^{\varepsilon }\circ h\right) \left( L_{\alpha \varepsilon
}^{\delta }\circ h\right) +\rho _{\alpha }^{\tilde{\imath}}\circ h\frac{%
\partial \left( L_{\beta \gamma }^{\delta }\circ h\right) }{\partial \bar{%
\varkappa}^{\tilde{\imath}}}\,\right) =0.%
\end{array}%
\end{equation*}
Multiplying with $t_{\delta }$, we obtain the equality%
\begin{equation*}
\begin{array}{l}
\underset{cyclic\left( \alpha ,\beta ,\gamma \right) }{\tsum }\left( \left(
L_{\beta \gamma }^{\varepsilon }\circ h\right) \left( L_{\alpha \varepsilon
}^{\delta }\circ h\right) t_{\delta }+\rho _{\alpha }^{\tilde{\imath}}\circ h%
\frac{\partial \left( L_{\beta \gamma }^{\delta }\circ h\right) }{\partial
\bar{\varkappa}^{\tilde{\imath}}}t_{\delta }\,\right) =0%
\end{array}%
\end{equation*}%
which is equivalent with the following equality:%
\begin{equation*}
\begin{array}{l}
\underset{cyclic\left( \alpha ,\beta ,\gamma \right) }{\tsum }\left( \left(
L_{\beta \gamma }^{\varepsilon }\circ h\right) \left[ t_{\alpha
},t_{\varepsilon }\right] _{F,h}+\rho _{\alpha }^{\tilde{\imath}}\circ h%
\frac{\partial \left( L_{\beta \gamma }^{\delta }\circ h\right) }{\partial
\bar{\varkappa}^{\tilde{\imath}}}t_{\delta }\,\right) =0.%
\end{array}%
\end{equation*}
Therefore, we obtain the Jacobi identity
\begin{equation*}
\begin{array}{l}
\underset{cyclic\left( \alpha ,\beta ,\gamma \right) }{\tsum }\left[
t_{\alpha },\left[ t_{\beta },t_{\gamma }\right] _{F,h}\right] _{F,h}=0.%
\end{array}%
\end{equation*}
For any $u,v,w\in \Gamma \left( F,\nu ,N\right) $, we obtain the Jacobi
identity
\begin{equation*}
\begin{array}{l}
\left[ u,\left[ v,w\right] _{F,h}\right] _{F,h}+\left[ v,\left[ w,u\right]
_{F,h}\right] _{F,h}+\left[ w,\left[ u,v\right] _{F,h}\right] _{F,h}=0,%
\end{array}%
\leqno(2)
\end{equation*}
Using $\left( 1\right) $ and $\left( 2\right) $ it results the conclusion of
lemma.\hfill \emph{q.e.d.}\bigskip

\noindent\textbf{Lemma 3.1.3 }\emph{The }$\mathbf{Mod}$\emph{-morphism }$%
\Gamma \left( Th\circ \rho ,h\right) $\emph{\ is a} $\mathbf{Liealg}$\emph{%
-morphism of}%
\begin{equation*}
\left( \Gamma \left( F,\nu ,N\right) ,+,\cdot ,\left[ ,\right] _{F,h}\right)
\end{equation*}%
\emph{source and}
\begin{equation*}
\left( \Gamma \!(TN,\tau _{N},\!N),+,\cdot ,\left[ ,\right] _{TN}\right)
\end{equation*}%
\emph{target.}\bigskip

\noindent\emph{Proof. }As the $\mathbf{Mod}$-morphism $\Gamma \left( \rho
,Id_{N}\right) $ is a $\mathbf{LieAlg}$-morphisms of
\begin{equation*}
\left( \Gamma \left( F,\nu ,N\right) ,+,\cdot ,\left[ ,\right] _{F}\right)
\end{equation*}%
source and
\begin{equation*}
\left( \Gamma \left( TN,\tau _{N},N\right) ,+,\cdot ,\left[ ,\right]
_{TN}\right)
\end{equation*}%
target, then we obtain
\begin{equation*}
\begin{array}{c}
L_{\alpha \beta }^{\gamma }\rho _{\gamma }^{\tilde{k}}=\rho _{\alpha }^{%
\tilde{\imath}}\frac{\partial \rho _{\beta }^{\tilde{k}}}{\partial \varkappa
^{\tilde{\imath}}}-\rho _{\beta }^{\tilde{j}}\frac{\partial \rho _{\alpha }^{%
\tilde{k}}}{\partial \varkappa ^{\tilde{j}}}.%
\end{array}%
\end{equation*}
Therefore, we obtain%
\begin{equation*}
\begin{array}{c}
\left( L_{\alpha \beta }^{\gamma }\circ h\right) \left( \rho _{\gamma }^{%
\tilde{k}}\circ h\right) =\rho _{\alpha }^{\tilde{\imath}}\circ h\frac{%
\partial \rho _{\beta }^{\tilde{k}}\circ h}{\partial \bar{\varkappa}^{\tilde{%
\imath}}}-\rho _{\beta }^{\tilde{j}}\circ h\frac{\partial \rho _{\alpha }^{%
\tilde{k}}\circ h}{\partial \bar{\varkappa}^{\tilde{j}}}.%
\end{array}%
\end{equation*}
Moreover, we obtain%
\begin{equation*}
\begin{array}{c}
\left( L_{\alpha \beta }^{\gamma }\circ h\right) \left( \rho _{\gamma }^{%
\tilde{k}}\circ h\right) \frac{\partial }{\partial \bar{\varkappa}^{\tilde{k}%
}}=\rho _{\alpha }^{\tilde{\imath}}\circ h\frac{\partial \rho _{\beta }^{%
\tilde{k}}\circ h}{\partial \bar{\varkappa}^{\tilde{\imath}}}\frac{\partial
}{\partial \bar{\varkappa}^{\tilde{k}}}-\rho _{\beta }^{\tilde{j}}\circ h%
\frac{\partial \rho _{\alpha }^{\tilde{k}}\circ h}{\partial \bar{\varkappa}^{%
\tilde{j}}}\frac{\partial }{\partial \bar{\varkappa}^{\tilde{k}}}.%
\end{array}%
\end{equation*}
After some calculations, we obtain that
\begin{equation*}
\begin{array}{c}
\Gamma \left( Th\circ \rho ,h\right) \left[ t_{\alpha },t_{\beta }\right]
_{F,h}=\left[ \Gamma \left( Th\circ \rho ,h\right) t_{\alpha },\Gamma \left(
Th\circ \rho ,h\right) t_{\beta }\right] _{TN}.%
\end{array}%
\end{equation*}
We obtain easily that
\begin{equation*}
\begin{array}{c}
\Gamma \left( Th\circ \rho ,h\right) \left[ u,v\right] _{F,h}=\left[ \Gamma
\left( Th\circ \rho ,h\right) u,\Gamma \left( Th\circ \rho ,h\right) v\right]
_{TN},%
\end{array}%
\end{equation*}%
for any $u,v\in \Gamma \left( F,\nu ,N\right) .$ \hfill \emph{q.e.d.}\bigskip

Using \emph{Lemmas 3.1.1, 3.1.2} and \emph{3.1.3} we obtain the following

\bigskip\noindent \textbf{Theorem 3.1.1 }(example of generalized Lie
algebroid) \emph{The couple }%
\begin{equation*}
\begin{array}{c}
\left( \left[ ,\right] _{F,h},\left( \rho ,Id_{N}\right) \right)%
\end{array}%
\end{equation*}%
\emph{\ is a generalized Lie algebroid structure for the vector bundle }$%
\left( F,\nu ,N\right) .$\bigskip

\noindent \textbf{Definition 3.1.3 }The generalized Lie algebroid
\begin{equation*}
\begin{array}{c}
\left( \left( F,\nu ,N\right) ,\left[ ,\right] _{F,h},\left( \rho
,Id_{N}\right) \right)%
\end{array}%
\end{equation*}%
given by the previous theorem, will be called \emph{the generalized Lie
algebroid associated to the Lie algebroid }%
\begin{equation*}
\begin{array}{c}
\left( \left( F,\nu ,N\right) ,\left[ ,\right] _{F},\left( \rho
,Id_{N}\right) \right)%
\end{array}%
\end{equation*}%
\emph{and to the surjective application }%
\begin{equation*}
\begin{array}{c}
h\in \mathbf{Man}\left( N,N\right) .%
\end{array}%
\end{equation*}%
In particular, if $h=Id_{N},$ then the generalized Lie algebroid
\begin{equation*}
\begin{array}{c}
\left( \left( F,\nu ,N\right) ,\left[ ,\right] _{F,Id_{N}},\left( \rho
,Id_{N}\right) \right)%
\end{array}%
\end{equation*}%
will be called \emph{the generalized Lie algebroid associated to the Lie
algebroid}%
\begin{equation*}
\begin{array}{c}
\left( \left( F,\nu ,N\right) ,\left[ ,\right] _{F},\left( \rho
,Id_{N}\right) \right) .%
\end{array}%
\end{equation*}%
Note that any Lie algebroid can be regarded as a generalized Lie algebroid.

\textbf{Theorem 3.1.2 }(example of generalized Lie algebroid) \emph{Let }$%
M\in \left\vert \mathbf{Man}_{m}\right\vert $\emph{\ and }$g,h\in Iso_{%
\mathbf{Man}}\left( M\right) .$

\emph{Let }$\left[ ,\right] _{TM}$\emph{\ be the usual Lie bracket such that
}%
\begin{equation*}
\left( \Gamma \left( TM,\tau _{M},M\right) ,+,\cdot ,\left[ ,\right]
_{TM}\right) \in \left\vert \mathbf{LieAlg}\right\vert .
\end{equation*}

\emph{Using the tangent }$\mathbf{B}^{\mathbf{v}}$\emph{-morphism }$\left(
Tg,g\right) $\emph{\ and the operation }%
\begin{equation*}
\begin{array}{ccc}
\Gamma \left( TM,\tau _{M},M\right) \times \Gamma \left( TM,\tau
_{M},M\right) & ^{\underrightarrow{~\ \ \left[ ,\right] _{TM,h}~\ \ }} &
\Gamma \left( TM,\tau _{M},M\right) \\
\left( u,v\right) & \longmapsto & \ \left[ u,v\right] _{TM,h}%
\end{array}%
\end{equation*}%
\emph{where }%
\begin{equation*}
\left[ u,v\right] _{TM,h}=\Gamma \left( T\left( h\circ g\right) ^{-1},\left(
h\circ g\right) ^{-1}\right) \left( \left[ \Gamma \left( T\left( h\circ
g\right) ,h\circ g\right) u,\Gamma \left( T\left( h\circ g\right) ,h\circ
g\right) v\right] _{TM}\right) ,
\end{equation*}%
\emph{for any }$u,v\in \Gamma \left( TM,\tau _{M},M\right) $\emph{, then we
obtain that }%
\begin{equation*}
\left( \left( TM,\tau _{M},M\right) ,\left[ u,v\right] _{TM,h},\left(
Tg,g\right) \right)
\end{equation*}%
\emph{is a generalized Lie algebroid.}

\emph{Proof:} As the operation $\left[ ,\right] _{TM,h}$ is biadditive, then
we obtain that
\begin{equation*}
\left( \Gamma \left( TM,\tau _{M},M\right) ,+,\cdot ,\left[ ,\right]
_{TM,h}\right) \in \left\vert \mathbf{Alg}\right\vert .
\end{equation*}

Using the definition of the operation $\left[ ,\right] _{TM,h}$ we obtain
that%
\begin{equation*}
\Gamma \left( T\left( h\circ g\right) ,h\circ g\right) \left( \left[ u,v%
\right] _{TM,h}\right) =\left[ \Gamma \left( T\left( h\circ g\right) ,h\circ
g\right) u,\Gamma \left( T\left( h\circ g\right) ,h\circ g\right) v\right]
_{TM}
\end{equation*}%
for any $u,v\in \Gamma \left( TM,\tau _{M},M\right) .$

1) Therefore, $\Gamma \left( T\left( h\circ g\right) ,h\circ g\right) $ is a
$\mathbf{Alg}$-morphism of
\begin{equation*}
\left( \Gamma \left( TM,\tau _{M},M\right) ,+,\cdot ,\left[ ,\right]
_{TM,h}\right)
\end{equation*}%
source and
\begin{equation*}
\left( \Gamma \left( TM,\tau _{M},M\right) ,+,\cdot ,\left[ ,\right]
_{TM}\right)
\end{equation*}%
target.

For any $u,v\in \Gamma \left( TM,\tau _{M},M\right) $ and $f\in \mathcal{F}%
\left( M\right) $ we obtain that
\begin{equation*}
\begin{array}{cl}
\left[ u,fv\right] _{TM,h} & =\Gamma \left( T\left( h\circ g\right)
^{-1},\left( h\circ g\right) ^{-1}\right) \left( \left[ \Gamma \left(
T\left( h\circ g\right) ,h\circ g\right) u,\Gamma \left( T\left( h\circ
g\right) ,h\circ g\right) fv\right] _{TM}\right) \\
& =\Gamma \left( T\left( h\circ g\right) ^{-1},\left( h\circ g\right)
^{-1}\right) \left( f\cdot \left[ \Gamma \left( T\left( h\circ g\right)
,h\circ g\right) u,\Gamma \left( T\left( h\circ g\right) ,h\circ g\right) v%
\right] _{TM}\right) \\
& +\Gamma \left( T\left( h\circ g\right) ^{-1},\left( h\circ g\right)
^{-1}\right) \left( \Gamma \left( T\left( h\circ g\right) ,h\circ g\right)
u\right) \left( f\right) \cdot \Gamma \left( T\left( h\circ g\right) ,h\circ
g\right) v \\
& =f\cdot \Gamma \left( T\left( h\circ g\right) ^{-1},\left( h\circ g\right)
^{-1}\right) \left[ \Gamma \left( T\left( h\circ g\right) ,h\circ g\right)
u,\Gamma \left( T\left( h\circ g\right) ,h\circ g\right) v\right] _{TM} \\
& +\left( \Gamma \left( T\left( h\circ g\right) ,h\circ g\right) u\right)
\left( f\right) \cdot v%
\end{array}%
\end{equation*}

2) Therefore, we obtain that%
\begin{equation*}
\left[ u,fv\right] _{TM,h}=f\cdot \left[ u,v\right] _{TM,h}+\left( \Gamma
\left( T\left( h\circ g\right) ,h\circ g\right) u\right) \left( f\right)
\cdot v
\end{equation*}%
for any $u,v\in \Gamma \left( TM,\tau _{M},M\right) $ and $f\in \mathcal{F}%
\left( M\right) .$

We remark that
\begin{equation*}
\Gamma \left( T\left( h\circ g\right) ^{-1},\left( h\circ g\right)
^{-1}\right) \left( 0\right) =0.
\end{equation*}

As
\begin{equation*}
\left( \Gamma \left( TM,\tau _{M},M\right) ,+,\cdot ,\left[ ,\right]
_{TM}\right) \in \left\vert \mathbf{LieAlg}\right\vert
\end{equation*}%
and
\begin{equation*}
\begin{array}{cl}
^{\left[ u,\left[ v,z\right] _{TM,h}\right] _{TM,h}} & ^{=\Gamma \left(
T\left( h\circ g\right) ^{-1},\left( h\circ g\right) ^{-1}\right) \left[
\Gamma \left( T\left( h\circ g\right) ,h\circ g\right) u,\Gamma \left(
T\left( h\circ g\right) ,h\circ g\right) \left[ v,z\right] _{TM,h}\right]
_{TM}} \\
& ^{=\Gamma \left( T\left( h\circ g\right) ^{-1},\left( h\circ g\right)
^{-1}\right) \left[ \Gamma \left( T\left( h\circ g\right) ,h\circ g\right) u,%
\left[ \Gamma \left( T\left( h\circ g\right) ,h\circ g\right) v,\Gamma
\left( T\left( h\circ g\right) ,h\circ g\right) z\right] _{TM}\right] _{TM}}%
\end{array}%
\end{equation*}%
for any $u,v,z\in \Gamma \left( TM,\tau _{M},M\right) ,$ it results that%
\begin{equation*}
\left[ u,u\right] _{TM,h}=0
\end{equation*}%
for any $u\in \Gamma \left( TM,\tau _{M},M\right) $ and
\begin{equation*}
\left[ u,\left[ v,z\right] _{TM,h}\right] _{TM,h}+\left[ z,\left[ u,v\right]
_{TM,h}\right] _{TM,h}+\left[ v,\left[ z,u\right] _{TM,h}\right] _{TM,h}=0,
\end{equation*}%
for any $u,v,z\in \Gamma \left( TM,\tau _{M},M\right) .$

3) Therefore, we have that
\begin{equation*}
\left( \Gamma \left( TM,\tau _{M},M\right) ,+,\cdot ,\left[ ,\right]
_{TM,h}\right) \in \left\vert \mathbf{LieAlg}\right\vert .
\end{equation*}

\ Using the affirmations 1), 2) and 3) it results the conclusion of the
theorem.

\textbf{Remark 3.1.2 }For any $\mathbf{Man}$-isomorphisms $g$ and $h$ we
obtain new and interesting generalized Lie algebroid structures for the
tangent vector bundle $\left( TM,\tau _{M},M\right) .$

For any base $\left\{ t_{\alpha },~\alpha \in \overline{1,m}\right\} $ of
the module of sections $\left( \Gamma \left( TM,\tau _{M},M\right) ,+,\cdot
\right) $ we obtain the structure functions%
\begin{equation*}
L_{\alpha \beta }^{\gamma }=\left( \theta _{\alpha }^{i}\frac{\partial
\theta _{\beta }^{j}}{\partial x^{i}}-\theta _{\beta }^{i}\frac{\partial
\theta _{\alpha }^{j}}{\partial x^{i}}\right) \tilde{\theta}_{j}^{\gamma
},~\alpha ,\beta ,\gamma \in \overline{1,m}
\end{equation*}%
where
\begin{equation*}
\theta _{\alpha }^{i},~i,\alpha \in \overline{1,m}
\end{equation*}%
are real local functions such that
\begin{equation*}
\Gamma \left( T\left( h\circ g\right) ,h\circ g\right) \left( t_{\alpha
}\right) =\theta _{\alpha }^{i}\frac{\partial }{\partial x^{i}}
\end{equation*}%
and
\begin{equation*}
\tilde{\theta}_{j}^{\gamma },~i,\gamma \in \overline{1,m}
\end{equation*}%
are real local functions such that
\begin{equation*}
\Gamma \left( T\left( h\circ g\right) ^{-1},\left( h\circ g\right)
^{-1}\right) \left( \frac{\partial }{\partial x^{j}}\right) =\tilde{\theta}%
_{j}^{\gamma }t_{\gamma }.
\end{equation*}

In particular, using arbitrary basis for the module of sections and
arbitrary isometries (symmetries, translations, rotations,...) we obtain a
lot of generalized Lie algebroid structures for the tangent vector bundle $%
\left( T\Sigma ,\tau _{\Sigma },\Sigma \right) $ and we can study its
geometry using our theory which is develop in the next.

\subsubsection{Structure functions for generalized Lie algebroids}

Let
\begin{equation*}
\left( \left( F,\nu ,N\right) ,\left[ ,\right] _{F,h},\left( \rho ,\eta
\right) \right)
\end{equation*}%
be a generalized Lie algebroid given by the diagram:%
\begin{equation*}
\begin{array}{ccl}
~\ \ \  &  & \left( F,\left[ ,\right] _{F,h},\left( \rho ,\eta \right)
\right) \\
&  & ~\downarrow \nu \\
~\ \ \ M & ^{\underrightarrow{~\ \ \ \ \ \ h~\ \ \ \ \ \ }} & ~N%
\end{array}%
\leqno(3.1.1.1)
\end{equation*}

We assume that $\left( F,\nu ,N\right) $ is a vector bundle with type fibre
the real vector space $\left( \mathbb{R}^{p},+,\cdot \right) $ and structure
group a Lie subgroup of $\left( \mathbf{GL}\left( p,\mathbb{R}\right) ,\cdot
\right) .$

We take $\left( x^{i},y^{i}\right) $ as canonical local coordinates on $%
\left( TM,\tau _{M},M\right) ,$ where $i\in \overline{1,m}.$

Consider
\begin{equation*}
\left( x^{i},y^{i}\right) \longrightarrow \left( x^{i%
%TCIMACRO{\U{b4}}%
%BeginExpansion
{\acute{}}%
%EndExpansion
}\left( x^{i}\right) ,y^{i%
%TCIMACRO{\U{b4}}%
%BeginExpansion
{\acute{}}%
%EndExpansion
}\left( x^{i},y^{i}\right) \right)
\end{equation*}%
a change of coordinates on $\left( TM,\tau _{M},M\right) $. Then the
coordinates $y^{i}$ change to $y^{i%
%TCIMACRO{\U{b4}}%
%BeginExpansion
{\acute{}}%
%EndExpansion
}$ by the rule:
\begin{equation*}
\begin{array}{c}
y^{i%
%TCIMACRO{\U{b4}}%
%BeginExpansion
{\acute{}}%
%EndExpansion
}=\frac{\partial x^{i%
%TCIMACRO{\U{b4}}%
%BeginExpansion
{\acute{}}%
%EndExpansion
}}{\partial x^{i}}y^{i}.%
\end{array}%
\leqno(3.1.1.2)
\end{equation*}

We take $(\varkappa ^{\tilde{\imath}},z^{\alpha })$ as canonical local
coordinates on $(F,\nu ,N),$ where $\tilde{\imath}{\in }\overline{1,n}$, $%
\alpha \in \overline{1,p}.$

Consider
\begin{equation*}
\left( \varkappa ^{\tilde{\imath}},z^{\alpha }\right) \longrightarrow \left(
\varkappa ^{\tilde{\imath}%
%TCIMACRO{\U{b4}}%
%BeginExpansion
{\acute{}}%
%EndExpansion
},z^{\alpha
%TCIMACRO{\U{b4}}%
%BeginExpansion
{\acute{}}%
%EndExpansion
}\right)
\end{equation*}%
a change of coordinates on $\left( F,\nu ,N\right) $. Then the coordinates $%
z^{\alpha }$ change to $z^{\alpha
%TCIMACRO{\U{b4}}%
%BeginExpansion
{\acute{}}%
%EndExpansion
}$ by the rule:
\begin{equation*}
\begin{array}{c}
z^{\alpha
%TCIMACRO{\U{b4}}%
%BeginExpansion
{\acute{}}%
%EndExpansion
}=\Lambda _{\alpha }^{\alpha
%TCIMACRO{\U{b4}}%
%BeginExpansion
{\acute{}}%
%EndExpansion
}z^{\alpha }.%
\end{array}%
\leqno(3.1.3)
\end{equation*}

We assume that $\left( \theta ,\mu \right) \overset{put}{=}\left( Th\circ
\rho ,h\circ \eta \right) $.

If $z^{\alpha }t_{\alpha }\in \Gamma \left( F,\nu ,N\right) $ is arbitrary,
then
\begin{equation*}
\begin{array}{c}
\Gamma \left( Th\circ \rho ,h\circ \eta \right) \left( z^{\alpha }t_{\alpha
}\right) f\left( h\circ \eta \left( \varkappa \right) \right) = \vspace*{1mm}
\\
\displaystyle=\left( \theta _{\alpha }^{\tilde{\imath}}z^{\alpha }\frac{%
\partial f}{\partial \varkappa ^{\tilde{\imath}}}\right) \left( h\circ \eta
\left( \varkappa \right) \right) =\left( \left( \rho _{\alpha }^{i}\circ
h\right) \left( z^{\alpha }\circ h\right) \frac{\partial f\circ h}{\partial
x^{i}}\right) \left( \eta \left( \varkappa \right) \right) ,%
\end{array}%
\leqno(3.1.1.4)
\end{equation*}%
for any $f\in \mathcal{F}\left( N\right) $ and $\varkappa \in N.$

The coefficients $\rho _{\alpha }^{i}$ respectively $\theta _{\alpha }^{%
\tilde{\imath}}$ change to $\rho _{\alpha
%TCIMACRO{\U{b4}}%
%BeginExpansion
{\acute{}}%
%EndExpansion
}^{i%
%TCIMACRO{\U{b4}}%
%BeginExpansion
{\acute{}}%
%EndExpansion
}$ respectively $\theta _{\alpha
%TCIMACRO{\U{b4}}%
%BeginExpansion
{\acute{}}%
%EndExpansion
}^{\tilde{\imath}%
%TCIMACRO{\U{b4}}%
%BeginExpansion
{\acute{}}%
%EndExpansion
}$ by the rule:
\begin{equation*}
\begin{array}{c}
\rho _{\alpha
%TCIMACRO{\U{b4}}%
%BeginExpansion
{\acute{}}%
%EndExpansion
}^{i%
%TCIMACRO{\U{b4}}%
%BeginExpansion
{\acute{}}%
%EndExpansion
}=\Lambda _{\alpha
%TCIMACRO{\U{b4}}%
%BeginExpansion
{\acute{}}%
%EndExpansion
}^{\alpha }\rho _{\alpha }^{i}\displaystyle\frac{\partial x^{i%
%TCIMACRO{\U{b4}}%
%BeginExpansion
{\acute{}}%
%EndExpansion
}}{\partial x^{i}},%
\end{array}%
\leqno(3.1.1.5)
\end{equation*}%
respectively%
\begin{equation*}
\begin{array}{c}
\theta _{\alpha
%TCIMACRO{\U{b4}}%
%BeginExpansion
{\acute{}}%
%EndExpansion
}^{\tilde{\imath}%
%TCIMACRO{\U{b4}}%
%BeginExpansion
{\acute{}}%
%EndExpansion
}=\Lambda _{\alpha
%TCIMACRO{\U{b4}}%
%BeginExpansion
{\acute{}}%
%EndExpansion
}^{\alpha }\theta _{\alpha }^{\tilde{\imath}}\displaystyle\frac{\partial
\varkappa ^{\tilde{\imath}%
%TCIMACRO{\U{b4}}%
%BeginExpansion
{\acute{}}%
%EndExpansion
}}{\partial \varkappa ^{\tilde{\imath}}},%
\end{array}%
\leqno(3.1.1.6)
\end{equation*}%
where
\begin{equation*}
\left\Vert \Lambda _{\alpha
%TCIMACRO{\U{b4}}%
%BeginExpansion
{\acute{}}%
%EndExpansion
}^{\alpha }\right\Vert =\left\Vert \Lambda _{\alpha }^{\alpha
%TCIMACRO{\U{b4}}%
%BeginExpansion
{\acute{}}%
%EndExpansion
}\right\Vert ^{-1}.
\end{equation*}

Locally, we set
\begin{equation*}
\begin{array}{c}
\left[ t_{\alpha },t_{\beta }\right] _{F}\overset{put}{=}L_{\alpha \beta
}^{\gamma }t_{\gamma }.%
\end{array}%
\leqno(3.1.1.7)
\end{equation*}

We easily obtain that
\begin{equation*}
L_{\alpha \beta }^{\gamma }=-L_{\beta \alpha }^{\gamma },~\forall \alpha
,\beta ,\gamma \in \overline{1,p}.
\end{equation*}
The real local functions
\begin{equation*}
\left\{ L_{\alpha \beta }^{\gamma },~\alpha ,\beta ,\gamma \in \overline{1,p}%
\right\}
\end{equation*}%
will be called the \emph{structure functions of the generalized Lie
algebroid }%
\begin{equation*}
\left( \left( F,\nu ,N\right) ,\left[ ,\right] _{F,h},\left( \rho ,\eta
\right) \right) .
\end{equation*}

\noindent\textbf{Theorem 3.1.1.1 }\emph{The following equalities hold good:}%
\begin{equation*}
\begin{array}{c}
\displaystyle\rho _{\alpha }^{i}\circ h\frac{\partial f\circ h}{\partial
x^{i}}=\left( \theta _{\alpha }^{\tilde{\imath}}\frac{\partial f}{\partial
\varkappa ^{\tilde{\imath}}}\right) \circ h,\forall f\in \mathcal{F}\left(
N\right) .%
\end{array}%
\leqno(3.1.1.8)
\end{equation*}%
\emph{and }%
\begin{equation*}
\begin{array}{c}
\displaystyle\left( L_{\alpha \beta }^{\gamma }\circ h\right) \left( \rho
_{\gamma }^{k}\circ h\right) =\left( \rho _{\alpha }^{i}\circ h\right) \frac{%
\partial \left( \rho _{\beta }^{k}\circ h\right) }{\partial x^{i}}-\left(
\rho _{\beta }^{j}\circ h\right) \frac{\partial \left( \rho _{\alpha
}^{k}\circ h\right) }{\partial x^{j}}.%
\end{array}%
\leqno(3.1.1.9)
\end{equation*}

\noindent\emph{Proof. }Using the relation $\left( 3.1.1.4\right) $, we
obtain the equality $\left( 3.1.1.8\right) .$ Since
\begin{equation*}
\begin{array}{l}
\displaystyle\Gamma \left( Th\circ \rho ,h\circ \eta \right) \left[
t_{\alpha },t_{\beta }\right] _{F}\left( f\right) \vspace*{1mm} \\
\displaystyle=\left[ \Gamma \left( \left( Th,h\right) \circ \left( \rho
,\eta \right) \right) t_{\alpha },\Gamma \left( \left( Th,h\right) \circ
\left( \rho ,\eta \right) \right) t_{\beta }\right] _{F}\left( f\right)
\vspace*{1mm} \\
\displaystyle=\Gamma \left( Th,h\right) \left( \left[ \Gamma \left( \rho
,\eta \right) t_{\alpha },\Gamma \left( \rho ,\eta \right) t_{\beta }\right]
_{TM}\right) \left( f\right) ,~\forall f\in \mathcal{F}\left( N\right) ,%
\end{array}%
\end{equation*}%
it results that%
\begin{equation*}
\begin{array}{c}
\left( L_{\alpha \beta }^{\gamma }\circ h\right) \left( \rho _{\gamma
}^{k}\circ h\right) \frac{\partial f\circ h}{\partial x^{k}}=\left( \left(
\rho _{\alpha }^{i}\circ h\right) \frac{\partial \left( \rho _{\beta
}^{k}\circ h\right) }{\partial x^{i}}-\left( \rho _{\beta }^{j}\circ
h\right) \frac{\partial \left( \rho _{\alpha }^{k}\circ h\right) }{\partial
x^{j}}\right) \frac{\partial f\circ h}{\partial x^{k}},%
\end{array}%
\end{equation*}%
for any $f\in \mathcal{F}\left( N\right) .$\hfill \emph{q.e.d.}

\subsubsection{The pull-back Lie algebroid of a generalized Lie algebroid}

Let
\begin{equation*}
\left( \left( F,\nu ,N\right) ,\left[ ,\right] _{F,h},\left( \rho ,\eta
\right) \right)
\end{equation*}%
be a generalized Lie algebroid given by the diagram $(3.1.1.1).$

Let $\mathcal{AF}_{F}$ be a representative of vector fibred $\left(
n+p\right) $-structure for the vector bundle $\left( F,\nu ,N\right) $ and
let $\mathcal{AF}_{TM}$ be a representative of vector fibred $\left(
m+m\right) $-structure for the vector bundle $\left( TM,\tau _{M},M\right) $.

Let $\left( h^{\ast }F,h^{\ast }\nu ,M\right) $ be the pull-back vector
bundle through $h.$

If $\left( U,\xi _{U}\right) \in \mathcal{AF}_{TM}$ and $\left(
V,s_{V}\right) \in \mathcal{AF}_{F}$ such that $U\cap h^{-1}\left( V\right)
\neq \phi $, then we define the application%
\begin{equation*}
\begin{array}{ccc}
h^{\ast }\nu ^{-1}(U{\cap }h^{-1}(V))) & {}^{\underrightarrow{\bar{s}_{U{%
\cap }h^{-1}(V)}}} & \left( U{\cap }h^{-1}(V)\right) {\times }\mathbb{R}^{p}
\\
\left( \varkappa ,z\left( h\left( \varkappa \right) \right) \right) &
\longmapsto & \left( \varkappa ,t_{V,h\left( \varkappa \right) }^{-1}z\left(
h\left( \varkappa \right) \right) \right).%
\end{array}%
\end{equation*}

\textbf{Proposition 3.1.2.1 }\emph{The set}%
\begin{equation*}
\begin{array}{c}
\overline{\mathcal{AF}}_{F}\overset{put}{=}\underset{U\cap h^{-1}\left(
V\right) \neq \phi }{\underset{\left( U,\xi _{U}\right) \in \mathcal{AF}%
_{TM},~\left( V,s_{V}\right) \in \mathcal{AF}_{F}}{\tbigcup }}\left\{ \left(
U\cap h^{-1}\left( V\right) ,\bar{s}_{U{\cap }h^{-1}(V)}\right) \right\}%
\end{array}%
\end{equation*}%
\emph{is a vector fibred }$m+p$\emph{-atlas for the vector bundle }$\left(
h^{\ast }F,h^{\ast }\nu ,M\right) .$

\emph{If }%
\begin{equation*}
\begin{array}{c}
z=z^{\alpha }t_{\alpha }\in \Gamma \left( F,\nu ,N\right) ,%
\end{array}%
\end{equation*}%
\emph{then, using the vector fibred }$m+p$\emph{-structure }$\left[
\overline{\mathcal{AF}}_{F}\right] ,$\emph{\ we obtain the section}%
\begin{equation*}
\begin{array}{c}
Z=\left( z^{\alpha }\circ h\right) T_{\alpha }\in \Gamma \left( h^{\ast
}F,h^{\ast }\nu ,M\right)%
\end{array}%
\end{equation*}%
\emph{such that}%
\begin{equation*}
\begin{array}{c}
Z\left( x\right) =z\left( h\left( x\right) \right) ,%
\end{array}%
\end{equation*}%
\emph{for any }$x\in U\cap h^{-1}\left( V\right) .$\bigskip

The set $\left\{ T_{\alpha },~\alpha \in \overline{1,p}\right\} $ is a base
for the module of sections%
\begin{equation*}
\begin{array}{c}
\left( \Gamma \left( h^{\ast }F,h^{\ast }\nu ,M\right) ,+,\cdot \right) .%
\end{array}%
\end{equation*}

Let $\Big({\overset{h^{\ast }F}{\rho }},Id_{M}\Big)$ be the $\mathbf{B}^{%
\mathbf{v}}$-morphism of
\begin{equation*}
\left( h^{\ast }F,h^{\ast }\nu ,M\right)
\end{equation*}%
\ source and
\begin{equation*}
\left( TM,\tau _{M},M\right)
\end{equation*}%
\ target, where%
\begin{equation*}
\begin{array}{rcl}
h^{\ast }F & ^{\underrightarrow{\overset{h^{\ast }F}{\rho }}} & TM \\
\displaystyle Z^{\alpha }T_{\alpha }\left( x\right) & \longmapsto & %
\displaystyle\left( Z^{\alpha }\cdot \rho _{\alpha }^{i}\circ h\right) \frac{%
\partial }{\partial x^{i}}\left( x\right)%
\end{array}%
\leqno(3.1.2.1)
\end{equation*}

We consider the operation
\begin{equation*}
\begin{array}{ccc}
\Gamma \left( h^{\ast }F,h^{\ast }\nu ,M\right) \times \Gamma \left( h^{\ast
}F,h^{\ast }\nu ,M\right) & ^{\underrightarrow{~\ \ \left[ ,\right]
_{h^{\ast }F}~\ \ }} & \Gamma \left( h^{\ast }F,h^{\ast }\nu ,M\right)%
\end{array}%
\end{equation*}%
defined by%
\begin{equation*}
\begin{array}{ll}
\left[ T_{\alpha },T_{\beta }\right] _{h^{\ast }F} & =\left( L_{\alpha \beta
}^{\gamma }\circ h\right) T_{\gamma }, \vspace*{1mm} \\
\left[ T_{\alpha },fT_{\beta }\right] _{h^{\ast }F} & \displaystyle =f\left(
L_{\alpha \beta }^{\gamma }\circ h\right) T_{\gamma }+\left( \rho _{\alpha
}^{i}\circ h\right) \frac{\partial f}{\partial x^{i}}T_{\beta }, \vspace*{%
1,5mm} \\
\left[ fT_{\alpha },T_{\beta }\right] _{h^{\ast }F} & =-\left[ T_{\beta
},fT_{\alpha }\right] _{h^{\ast }F},%
\end{array}%
\leqno(3.1.2.2)
\end{equation*}%
for any $f\in \mathcal{F}\left( M\right) .$

\bigskip\noindent\textbf{Lemma 3.1.2.1 }\emph{The following equality holds
good }%
\begin{equation*}
\left[ U,fV\right] _{h^{\ast }F}=f\left[ U,V\right] _{h^{\ast }F}+\Gamma
\left( {\overset{h^{\ast }F}{\rho }},Id_{M}\right) \left( U\right) f\cdot V,
\end{equation*}%
\emph{for any }$U,V\in \Gamma \left( h^{\ast }F,h^{\ast }\nu ,M\right) $%
\emph{\ and for any }$f\in \mathcal{F}\left( M\right) .$

\bigskip\noindent\emph{Proof. }We observe that for any $\alpha ,\beta \in
\overline{1,p}$, we obtain%
\begin{equation*}
\left[ T_{\alpha },fT_{\beta }\right] _{h^{\ast }F}=f\left[ T_{\alpha
},T_{\beta }\right] _{h^{\ast }F}+\Gamma \left( {\overset{h^{\ast }F}{\rho }}%
,Id_{M}\right) \left( T_{\alpha }\right) f\cdot T_{\beta },~\forall f\in
\mathcal{F}\left( N\right) .
\end{equation*}
Using this equality and the definition of the operation $\ \left[ ,\right]
_{h^{\ast }F}$ it results the conclusion of the lemma. \hfill \emph{q.e.d.}%
\bigskip

\noindent\textbf{Lemma 3.1.2.1 }\emph{The} $\mathcal{F}\left( M\right) $%
\emph{-algebra }%
\begin{equation*}
\left( \Gamma \left( h^{\ast }F,h^{\ast }\nu ,M\right) ,+,\cdot ,\left[ ,%
\right] _{h^{\ast }F}\right)
\end{equation*}%
\emph{is a Lie} $\mathcal{F}\left( M\right) $\emph{-algebra.}

\bigskip\noindent\emph{Proof. }Using the definition of the operation $\ %
\left[ ,\right] _{h^{\ast }F}$ it results that
\begin{equation*}
\ \left[ U,V\right] _{h^{\ast }F}=-\left[ V,U\right] _{h^{\ast }F},
\end{equation*}%
for any $U,V\in \Gamma \left( h^{\ast }F,h^{\ast }\nu ,M\right) .$
Therefore, we obtain%
\begin{equation*}
\begin{array}{c}
\left[ U,U\right] _{h^{\ast }F}=0,~\forall U\in \Gamma \left( h^{\ast
}F,h^{\ast }\nu ,M\right) .%
\end{array}%
\leqno(1)
\end{equation*}
Since $\left( \Gamma \left( F,\nu ,M\right) ,+,\cdot ,\left[ ,\right]
_{F}\right) $ is a Lie $\mathcal{F}\left( M\right) $-algebra, we obtain the
equality:
\begin{equation*}
\begin{array}{c}
\displaystyle\underset{cyclic\left( \alpha ,\beta ,\gamma \right) }{\tsum }%
\left( L_{\beta \gamma }^{\varepsilon }L_{\alpha \varepsilon }^{\delta
}+\theta _{\alpha }^{\tilde{\imath}}\frac{\partial L_{\beta \gamma }^{\delta
}}{\partial \varkappa ^{\tilde{\imath}}}\right) =0.%
\end{array}%
\end{equation*}
Using $\left( 3.1.1.9\right) $, we obtain the equality:%
\begin{equation*}
\begin{array}{l}
\displaystyle\underset{cyclic\left( \alpha ,\beta ,\gamma \right) }{\tsum }%
\left( \left( L_{\beta \gamma }^{\varepsilon }\circ h\right) \left(
L_{\alpha \varepsilon }^{\delta }\circ h\right) +\rho _{\alpha }^{i}\circ h%
\frac{\partial \left( L_{\beta \gamma }^{\delta }\circ h\right) }{\partial
x^{i}}\right) \,=0.%
\end{array}%
\end{equation*}%
and multiplying with $T_{\delta }$, we obtain
\begin{equation*}
\begin{array}{l}
\displaystyle\underset{cyclic\left( \alpha ,\beta ,\gamma \right) }{\tsum }%
\left( \left( L_{\beta \gamma }^{\varepsilon }\circ h\right) \left(
L_{\alpha \varepsilon }^{\delta }\circ h\right) T_{\delta }\,+\rho _{\alpha
}^{i}\circ h\frac{\partial \left( L_{\beta \gamma }^{\delta }\circ h\right)
}{\partial x^{i}}T_{\delta }\,\right) \,=0.%
\end{array}%
\end{equation*}%
which is equivalent with
\begin{equation*}
\begin{array}{l}
\displaystyle\underset{cyclic\left( \alpha ,\beta ,\gamma \right) }{\tsum }%
\left( \left( L_{\beta \gamma }^{\varepsilon }\circ h\right) \left[
T_{\alpha },T_{\varepsilon }\right] _{h^{\ast }F}+\rho _{\alpha }^{i}\circ h%
\frac{\partial L_{\beta \gamma }^{\delta }\circ h}{\partial x^{i}}T_{\delta
}\right) =0.%
\end{array}%
\leqno(2)
\end{equation*}%
Since this equality implies
\begin{equation*}
\begin{array}{c}
\underset{cyclic\left( \alpha ,\beta ,\gamma \right) }{\tsum }\left[
T_{\alpha },\!\left( L_{\beta \gamma }^{\varepsilon }{\circ }h\right)
T_{\varepsilon }\right] _{h^{\ast }F}=0,%
\end{array}%
\end{equation*}%
it results that it is satisfied the Jacobi identity
\begin{equation*}
\begin{array}{c}
\underset{cyclic\left( \alpha ,\beta ,\gamma \right) }{\tsum }\left[
T_{\alpha },\left[ T_{\beta },T_{\gamma }\right] _{h^{\ast }F}\right]
_{h^{\ast }F}=0.%
\end{array}%
\end{equation*}
In general, for any $U,V,Z\in \Gamma \left( h^{\ast }F,h^{\ast }\nu
,M\right) $, we obtain the Jacobi identity:
\begin{equation*}
\left[ U,\left[ V,Z\right] _{h^{\ast }F}\right] _{h^{\ast }F}+\left[ Z,\left[
U,V\right] _{h^{\ast }F}\right] _{h^{\ast }F}\vspace*{1mm}+\left[ V,\left[
Z,U\right] _{h^{\ast }F}\right] _{h^{\ast }F}=0.\leqno(2)
\end{equation*}
Using affirmations $\left( 1\right) $ and $\left( 2\right) $, we get the
conclusion of the lemma.\hfill \emph{q.e.d.}\bigskip

\noindent\textbf{Lemma 3.1.2.3 }\emph{The }$\mathbf{Mod}$\emph{-morphism }%
\begin{equation*}
\Gamma \!\!\left( \overset{h^{\ast }F}{\rho }\!\!,Id_{M}\!\right)
\end{equation*}%
\emph{is a }$\mathbf{Liealg}$\emph{-morphism of}%
\begin{equation*}
\left( \Gamma \left( h^{\ast }F,h^{\ast }\nu ,M\right) ,+,\cdot ,\left[ ,%
\right] _{h^{\ast }F}\right)
\end{equation*}%
\emph{source and }%
\begin{equation*}
\left( \Gamma \!(TM,\tau _{M},\!M),+,\cdot ,\left[ ,\right] _{TM}\right)
\end{equation*}%
\emph{target.}\bigskip

\noindent\emph{Proof. }Using relations $\left( 3.1.1.9\right) $, we obtain:
\begin{equation*}
\begin{array}{c}
\displaystyle\left( L_{\alpha \beta }^{\gamma }\circ h\right) \left( \rho
_{\gamma }^{k}\circ h\right) \frac{\partial }{\partial x^{k}}=\left( \rho
_{\alpha }^{i}\circ h\right) \frac{\partial \left( \rho _{\beta }^{k}\circ
h\right) }{\partial x^{i}}\frac{\partial }{\partial x^{k}}-\left( \rho
_{\beta }^{j}\circ h\right) \frac{\partial \left( \rho _{\alpha }^{k}\circ
h\right) }{\partial x^{j}}\frac{\partial }{\partial x^{k}},%
\end{array}%
\end{equation*}
Therefore,
\begin{equation*}
\displaystyle\Gamma \left( \overset{h^{\ast }F}{\rho },Id_{M}\right) \left[
T_{\alpha },T_{\beta }\right] _{_{h^{\ast }F}}=\left[ \Gamma \left( \overset{%
h^{\ast }F}{\rho },Id_{M}\right) T_{\alpha },\Gamma \left( \overset{h^{\ast
}F}{\rho },Id_{M}\right) T_{\beta }\right] _{TM},
\end{equation*}%
for any base sections $T_{\alpha },T_{\beta }.$

In general, we obtain the equality
\begin{equation*}
\displaystyle\Gamma \left( \overset{h^{\ast }F}{\rho },Id_{M}\right) \left[
U,V\right] _{_{h^{\ast }F}}=\left[ \Gamma \left( \overset{h^{\ast }F}{\rho }%
,Id_{M}\right) U,\Gamma \left( \overset{h^{\ast }F}{\rho },Id_{M}\right) U%
\right] _{TM},
\end{equation*}%
for any $U,V\in \Gamma \left( h^{\ast }F,h^{\ast }\nu ,M\right) .$ \hfill
\emph{q.e.d.}

Using Lemmas 3.1.2.1, 3.1.2.2 and 3.1.2.3, we obtain the following

\bigskip\noindent \textbf{Theorem 3.1.2.1 }\emph{The couple }%
\begin{equation*}
\left( \left[ ,\right] _{h^{\ast }F},\left( \overset{h^{\ast }F}{\rho }%
,Id_{M}\right) \right)
\end{equation*}%
\emph{\ is a Lie algebroid structure for the vector bundle }$\left( h^{\ast
}F,h^{\ast }\nu ,M\right) .$\bigskip

This Lie algebroid will be called \emph{the pull-back Lie algebroid of the
generalized Lie algebroid }%
\begin{equation*}
\begin{array}{c}
\left( \left( F,\nu ,N\right) ,\left[ ,\right] _{F,h},\left( \rho ,\eta
\right) \right) .%
\end{array}%
\end{equation*}

\subsubsection{Interior Differential Systems}

\qquad \qquad \qquad %
%
%
%
%
%\end{equation*}

We consider a generalized Lie algebroid $\left( \left( F,\nu ,N\right) ,%
\left[ ,\right] _{F,h},\left( \rho ,\eta \right) \right) $ given by the
diagrams:%
\begin{equation*}
\begin{array}{ccccccc}
&  & F & ^{\underrightarrow{~\ \rho ~\ }} & TM & ^{\underrightarrow{~\ Th~\ }%
} & TN \\
&  & ~\downarrow \nu &  & ~\ \ \ \downarrow \tau _{M} &  & ~\ \ \ \downarrow
\tau _{N} \\
M & ^{\underrightarrow{~\ h~\ }} & N & ^{\underrightarrow{~\ \eta ~\ }} & M
& ^{\underrightarrow{~\ h~\ }} & N%
\end{array}%
\end{equation*}

Let $\left( \left( h^{\ast }F,h^{\ast }\nu ,M\right) ,\left[ ,\right]
_{h^{\ast }F},\left( \overset{h^{\ast }F}{\rho },Id_{M}\right) \right) $ be
the pull-back Lie algebroid.

\textbf{Definition 3.1.3.1 }Any vector subbundle $\left( E,\pi ,M\right) $
of the vector bundle $\left( h^{\ast }F,h^{\ast }\nu ,M\right) $ will be
called \emph{interior differential system (IDS) of the generalized Lie
algebroid }$\left( \left( F,\nu ,N\right) ,\left[ ,\right] _{F,h},\left(
\rho ,\eta \right) \right) .$

In particular, if $h=Id_{N}=\eta $, then any vector subbundle $\left( E,\pi
,N\right) $ of the vector bundle $\left( F,\nu ,N\right) $ will be called
\emph{interior differential system of the Lie algebroid }$\left( \left(
F,\nu ,N\right) ,\left[ ,\right] _{F},\left( \rho ,Id_{N}\right) \right) .$

\textbf{Remark 3.1.3.1 }If $\left( E,\pi ,M\right) $ is an IDS of the
generalized Lie algebroid $\left( \left( F,\nu ,N\right) ,\left[ ,\right]
_{F,h},\left( \rho ,\eta \right) \right) $, then $\left( \Gamma \left( E,\pi
,M\right) ,+,\cdot \right) $ is a $\mathcal{F}\left( M\right) $-submodule of
the $\mathcal{F}\left( M\right) $-module $\left( \Gamma \left( h^{\ast
}F,h^{\ast }\nu ,M\right) ,+,\cdot \right) .$

In addition, if%
\begin{equation*}
\Gamma \left( E^{\perp },\pi ^{\perp },M\right) \overset{put}{=}\left\{
\Omega \in \Gamma \left( \overset{\ast }{h^{\ast }F},\overset{\ast }{h^{\ast
}\nu },M\right) :\Omega \left( S\right) =0,~\forall S\in \Gamma \left( E,\pi
,M\right) \right\} ,
\end{equation*}%
then $\left( \Gamma \left( E^{\perp },\pi ^{\perp },M\right) ,+,\cdot
\right) $ is $\mathcal{F}\left( M\right) $-submodule of the $\mathcal{F}%
\left( M\right) $-module $\left( \Gamma \left( \overset{\ast }{h^{\ast }F},%
\overset{\ast }{h^{\ast }\nu },M\right) ,+,\cdot \right) .$

We obtain a vector subbundle $\left( E^{\perp },\pi ^{\perp },M\right) $ of
the vector bundle $\left( \overset{\ast }{h^{\ast }F},\overset{\ast }{%
h^{\ast }\nu },M\right) $ which will be called \emph{the annihilator vector
subbundle for the IDS }$\left( E,\pi ,M\right) .$

\textbf{Proposition 3.1.3.1 }\emph{Let }$\left( E,\pi ,M\right) $\emph{\ be
an IDS of the generalized Lie algebroid }$\left( \left( F,\nu ,N\right) ,%
\left[ ,\right] _{F,h},\left( \rho ,\eta \right) \right) .$

\emph{If }$dim_{\mathcal{F}\left( M\right) }\Gamma \left( E,\pi ,M\right)
=r\leq p=dim_{\mathcal{F}\left( M\right) }\Gamma \left( h^{\ast }F,h^{\ast
}\nu ,M\right) $\emph{, then }$dim_{\mathcal{F}\left( M\right) }\Gamma
\left( E^{\perp },\pi ^{\perp },M\right) =p-r.$

\emph{Therefore, if }$\Gamma \left( E,\pi ,M\right) =\left\langle
S_{1},...,S_{r}\right\rangle $\emph{, then it exists }$\Theta
^{r+1},...,\Theta ^{p}\in \Gamma \left( \overset{\ast }{h^{\ast }F},\overset{%
\ast }{h^{\ast }\nu },M\right) $\emph{\ linearly independent such that }$%
\Gamma \left( E^{\perp },\pi ^{\perp },M\right) =\left\langle \Theta
^{r+1},...,\Theta ^{p}\right\rangle .$

\emph{Conversely, if }$\Gamma \left( E^{\perp },\pi ^{\perp },M\right)
=\left\langle \Theta ^{r+1},...,\Theta ^{p}\right\rangle $\emph{, then it
exists }$S_{1},...,S_{r}\in \Gamma \left( E,\pi ,M\right) $\emph{\ linearly
independent such that }$\Gamma \left( E,\pi ,M\right) =\left\langle
S_{1},...,S_{r}\right\rangle .$

\textbf{Definition 3.1.3.2 }The IDS $\left( E,\pi ,M\right) $ of the
generalized Lie algebroid $\left( \left( F,\nu ,N\right) ,\left[ ,\right]
_{F,h},\left( \rho ,\eta \right) \right) $ will be called \emph{involutive}
if
\begin{equation*}
\left[ S,T\right] _{h^{\ast }F}\in \Gamma \left( E,\pi ,M\right) ,~\forall
S,T\in \Gamma \left( E,\pi ,M\right) .
\end{equation*}

\textbf{Proposition 3.1.3.2 }\emph{Let }$\left( E,\pi ,M\right) $\emph{\ be
an IDS of the generalized Lie algebroid }$\left( \left( F,\nu ,N\right) ,%
\left[ ,\right] _{F,h},\left( \rho ,\eta \right) \right) .$

\emph{If }$\left\{ S_{1},...,S_{r}\right\} $\emph{\ is a base for the }$%
\mathcal{F}\left( M\right) $\emph{-submodule }$\left( \Gamma \left( E,\pi
,M\right) ,+,\cdot \right) $\emph{\ then }$\left( E,\pi ,M\right) $\emph{\
is involutive if and only if}
\begin{equation*}
\left[ S_{a},S_{b}\right] _{h^{\ast }F}\in \Gamma \left( E,\pi ,M\right)
,~\forall a,b\in \overline{1,r}.
\end{equation*}

\bigskip

\subsection{Exterior differential calculus for generalized Lie algebroids}

We propose an exterior differential calculus in the general framework of
generalized Lie algebroids. As any Lie algebroid can be regarded as a
generalized Lie algebroid, in particular, we obtain a new point of view over
the exterior differential calculus for Lie algebroids.

Let
\begin{equation*}
\left( \left( F,\nu ,N\right) ,\left[ ,\right] _{F,h},\left( \rho ,\eta
\right) \right)
\end{equation*}%
be a generalized Lie algebroid given by the diagram $\left( 3.1.1.1\right) .$

\bigskip\noindent\textbf{Definition 3.2.1 }For any $q\in \mathbb{N}$ we
denote by $\left( \Sigma _{q},\circ \right) $ the permutations group of the
set $\left\{ 1,2,...,q\right\} .$

\bigskip\noindent\textbf{Definition 3.2.2 }We denoted by $\Lambda ^{q}\left(
F,\nu ,N\right) $ the set of $q$-linear applications
\begin{equation*}
\begin{array}{ccc}
\Gamma \left( F,\nu ,N\right) ^{q} & ^{\underrightarrow{\ \ \omega \ \ }} &
\mathcal{F}\left( N\right) \\
\left( z_{1},...,z_{q}\right) & \longmapsto & \omega \left(
z_{1},...,z_{q}\right)%
\end{array}%
\end{equation*}%
such that
\begin{equation*}
\begin{array}{c}
\omega \left( z_{\sigma \left( 1\right) },...,z_{\sigma \left( q\right)
}\right) =sgn\left( \sigma \right) \cdot \omega \left( z_{1},...,z_{q}\right)%
\end{array}%
\end{equation*}%
for any $z_{1},...,z_{q}\in \Gamma \left( F,\nu ,N\right) $ and for any $%
\sigma \in \Sigma _{q}$.

The elements of $\Lambda ^{q}\left( F,\nu ,N\right) $ will be called \emph{%
differential forms of degree }$q$ or \emph{differential }$q$\emph{-forms}$.$

\bigskip \noindent \textbf{Remark 3.2.1 }If $\omega \in \Lambda ^{q}\left(
F,\nu ,N\right) $, then $\omega \left( z_{1},...,z,...,z,...z_{q}\right) =0.$
Therefore, if $\omega \in \Lambda ^{q}\left( F,\nu ,N\right) $, then
\begin{equation*}
\begin{array}{c}
\omega \left( z_{1},...,z_{i},...,z_{j},...z_{q}\right) =-\omega \left(
z_{1},...,z_{j},...,z_{i},...z_{q}\right) .%
\end{array}%
\end{equation*}

\textbf{Theorem 3.2.1 }\emph{If }$q\in N$\emph{, then }$\left( \Lambda
^{q}\left( F,\nu ,N\right) ,+,\cdot \right) $\emph{\ is a} $\mathcal{F}%
\left( N\right) $\emph{-module.}

\bigskip\noindent\textbf{Definition 3.2.3 }If $\omega \in \Lambda ^{q}\left(
F,\nu ,N\right) $ and $\theta \in \Lambda ^{r}\left( F,\nu ,N\right) $, then
the $\left( q+r\right) $-form $\omega \wedge \theta $ defined by%
\begin{equation*}
\begin{array}{ll}
\omega \wedge \theta \left( z_{1},...,z_{q+r}\right) & =\underset{\sigma
\left( q+1\right) <...<\sigma \left( q+r\right) }{\underset{\sigma \left(
1\right) <...<\sigma \left( q\right) }{\tsum }}sgn\left( \sigma \right)
\omega \left( z_{\sigma \left( 1\right) },...,z_{\sigma \left( q\right)
}\right) \theta \left( z_{\sigma \left( q+1\right) },...,z_{\sigma \left(
q+r\right) }\right)\vspace{3mm} \\
& \displaystyle=\frac{1}{q!r!}\underset{\sigma \in \Sigma _{q+r}}{\tsum }%
sgn\left( \sigma \right) \omega \left( z_{\sigma \left( 1\right)
},...,z_{\sigma \left( q\right) }\right) \theta \left( z_{\sigma \left(
q+1\right) },...,z_{\sigma \left( q+r\right) }\right) ,%
\end{array}%
\end{equation*}%
for any $z_{1},...,z_{q+r}\in \Gamma \left( F,\nu ,N\right) ,$ will be
called \emph{the exterior product of the forms }$\omega $ \emph{and}~$\theta
.$

Using the previous definition, we obtain

\bigskip\noindent\textbf{Theorem 3.2.2 }\emph{The following affirmations
hold good:} \medskip

\noindent1. \emph{If }$\omega \in \Lambda ^{q}\left( F,\nu ,N\right) $\emph{%
\ and }$\theta \in \Lambda ^{r}\left( F,\nu ,N\right) $\emph{, then}%
\begin{equation*}
\begin{array}{c}
\omega \wedge \theta =\left( -1\right) ^{q\cdot r}\theta \wedge \omega .%
\end{array}%
\leqno(3.2.1)
\end{equation*}

\noindent2. \emph{For any }$\omega \in \Lambda ^{q}\left( F,\nu ,N\right) $%
\emph{, }$\theta \in \Lambda ^{r}\left( F,\nu ,N\right) $\emph{\ and }$\eta
\in \Lambda ^{s}\left( F,\nu ,N\right) $\emph{\ we obtain}%
\begin{equation*}
\begin{array}{c}
\left( \omega \wedge \theta \right) \wedge \eta =\omega \wedge \left( \theta
\wedge \eta \right) .%
\end{array}%
\leqno(3.2.2)
\end{equation*}

\noindent3. \emph{For any }$\omega ,\theta \in \Lambda ^{q}\left( F,\nu
,N\right) $\emph{\ and }$\eta \in \Lambda ^{s}\left( F,\nu ,N\right) $\emph{%
\ we obtain }%
\begin{equation*}
\begin{array}{c}
\left( \omega +\theta \right) \wedge \eta =\omega \wedge \eta +\theta \wedge
\eta .%
\end{array}%
\leqno(3.2.3)
\end{equation*}

\noindent4. \emph{For any }$\omega \in \Lambda ^{q}\left( F,\nu ,N\right) $%
\emph{\ and }$\theta ,\eta \in \Lambda ^{s}\left( F,\nu ,N\right) $\emph{\
we obtain }%
\begin{equation*}
\begin{array}{c}
\omega \wedge \left( \theta +\eta \right) =\omega \wedge \theta +\omega
\wedge \eta .%
\end{array}%
\leqno(3.2.4)
\end{equation*}

\noindent5. \emph{For any} $f\in \mathcal{F}\left( N\right) $, $\omega \in
\Lambda ^{q}\left( F,\nu ,N\right) $ \emph{and }$\theta \in \Lambda
^{s}\left( F,\nu ,N\right) $\emph{\ we obtain }%
\begin{equation*}
\begin{array}{c}
\left( f\cdot \omega \right) \wedge \theta =f\cdot \left( \omega \wedge
\theta \right) =\omega \wedge \left( f\cdot \theta \right) .%
\end{array}%
\leqno(3.2.5)
\end{equation*}

\noindent \textbf{Theorem 3.2.3 }\emph{If }%
\begin{equation*}
\Lambda \left( F,\nu ,N\right) =\underset{q\geq 0}{\oplus }\Lambda
^{q}\left( F,\nu ,N\right) ,
\end{equation*}%
\emph{then }%
\begin{equation*}
\begin{array}{c}
\left( \Lambda \left( F,\nu ,N\right) ,+,\cdot ,\wedge \right)%
\end{array}%
\end{equation*}%
\emph{is a} $\mathcal{F}\left( N\right) $\emph{-algebra.} This algebra will
be called \emph{the exterior differential algebra of the vector bundle }$%
\left( F,\nu ,N\right) .$

\bigskip\noindent\textbf{Remark 3.2.2 }If $\left\{ t^{\alpha },~\alpha \in
\overline{1,p}\right\} $ is the coframe associated to the frame $\left\{
t_{\alpha },~\alpha \in \overline{1,p}\right\} $ of the vector bundle $%
\left( F,\nu ,N\right) $ in the vector local $\left( n+p\right) $-chart $U$,
then
\begin{equation*}
%
%
\end{equation*}%
for any $q\in \overline{1,p}.$

\bigskip\noindent\textbf{Remark 3.2.3 }If $\left\{ t^{\alpha },~\alpha \in
\overline{1,p}\right\} $ is the coframe associated to the frame $\left\{
t_{\alpha },~\alpha \in \overline{1,p}\right\} $ of the vector bundle $%
\left( F,\nu ,N\right) $ in the vector local $\left( n+p\right) $-chart $U$,
then, for any $q\in \overline{1,p}$ we define $C_{p}^{q}$ exterior
differential forms of the type
\begin{equation*}
\begin{array}{c}
t^{\alpha _{1}}\wedge ...\wedge t^{\alpha _{q}}%
\end{array}%
\end{equation*}%
such that $1\leq \alpha _{1}<...<\alpha _{q}\leq p.$

The set%
\begin{equation*}
\begin{array}{c}
\left\{ t^{\alpha _{1}}\wedge ...\wedge t^{\alpha _{q}},1\leq \alpha
_{1}<...<\alpha _{q}\leq p\right\}%
\end{array}%
\end{equation*}%
is a base for the $\mathcal{F}\left( N\right) $-module
\begin{equation*}
\left( \Lambda ^{q}\left( F,\nu ,N\right) ,+,\cdot \right) .
\end{equation*}
Therefore, if $\omega \in \Lambda ^{q}\left( F,\nu ,N\right) $, then%
\begin{equation*}
\begin{array}{c}
\omega =\omega _{\alpha _{1}...\alpha _{q}}t^{\alpha _{1}}\wedge ...\wedge
t^{\alpha _{q}}.%
\end{array}%
\end{equation*}
In particular, if $\omega $ is an exterior differential $p$-form $\omega ,$
then we can written
\begin{equation*}
\begin{array}{c}
\omega =a\cdot t^{1}\wedge ...\wedge t^{p},%
\end{array}%
\end{equation*}%
where $a\in \mathcal{F}\left( N\right) .$

\bigskip\noindent\textbf{Definition 3.2.4 }If
\begin{equation*}
\begin{array}{c}
\omega =\omega _{\alpha _{1}...\alpha _{q}}t^{\alpha _{1}}\wedge ...\wedge
t^{\alpha _{q}}\in \Lambda ^{q}\left( F,\nu ,N\right)%
\end{array}%
\end{equation*}%
such that
\begin{equation*}
\begin{array}{c}
\omega _{\alpha _{1}...\alpha _{q}}\in C^{r}\left( N\right) ,%
\end{array}%
\end{equation*}%
for any $1\leq \alpha _{1}<...<\alpha _{q}\leq p$, then we will say that
\emph{the }$q$\emph{-form }$\omega $\emph{\ is differentiable of }$C^{r}$%
\emph{-class.}

\bigskip\noindent\textbf{Definition 3.2.5 }For any $z\in \Gamma \left( F,\nu
,N\right) $, the $\mathcal{F}\left( N\right) $-multilinear application
\begin{equation*}
%
%
\end{equation*}%
for any $\omega \in \Lambda ^{q}\mathbf{\ }\left( F,\nu ,N\right) $ and $%
z_{1},...,z_{q}\in \Gamma \left( F,\nu ,N\right) ,$ will be called \emph{the
covariant Lie derivative with respect to the section }$z.$

\bigskip\noindent\textbf{Theorem 3.2.4 }\emph{If }$z\in \Gamma \left( F,\nu
,N\right) ,$ $\omega \in \Lambda ^{q}\left( F,\nu ,N\right) $\emph{\ and }$%
\theta \in \Lambda ^{r}\left( F,\nu ,N\right) $\emph{, then}%
\begin{equation*}
\begin{array}{c}
L_{z}\left( \omega \wedge \theta \right) =L_{z}\omega \wedge \theta +\omega
\wedge L_{z}\theta .%
\end{array}%
\leqno(3.2.6)
\end{equation*}

\noindent\emph{Proof. }Let $z_{1},...,z_{q+r}\in \Gamma \left( F,\nu
,N\right) $ be arbitrary. Since
\begin{equation*}
\begin{array}{l}
L_{z}\left( \omega \wedge \theta \right) \left( z_{1},...,z_{q+r}\right)
=\Gamma \left( Th\circ \rho ,h\circ \eta \right) z\left( \left( \omega
\wedge \theta \right) \left( z_{1},...,z_{q+r}\right) \right) \\
-\overset{q+r}{\underset{i=1}{\tsum }}\left( \omega \wedge \theta \right)
\left( \left( z_{1},...,\left[ z,z_{i}\right] _{F,h},...,z_{q+r}\right)
\right) \\
=\Gamma \left( Th\circ \rho ,h\circ \eta \right) z\left( \underset{\sigma
\left( q+1\right) <...<\sigma \left( q+r\right) }{\underset{\sigma \left(
1\right) <...<\sigma \left( q\right) }{\tsum }}sgn\left( \sigma \right)
\cdot \omega \left( z_{\sigma \left( 1\right) },...,z_{\sigma \left(
q\right) }\right) \right. \\
\qquad\left. \cdot \theta \left( z_{\sigma \left( q+1\right) },...,z_{\sigma
\left( q+r\right) }\right) \right) -\overset{q+r}{\underset{i=1}{\tsum }}%
\left( \omega \wedge \theta \right) \left( \left( z_{1},...,\left[ z,z_{i}%
\right] _{F,h},...,z_{q+r}\right) \right) \\
=\underset{\sigma \left( q+1\right) <...<\sigma \left( q+r\right) }{\underset%
{\sigma \left( 1\right) <...<\sigma \left( q\right) }{\tsum }}sgn\left(
\sigma \right) \cdot \Gamma \left( Th\circ \rho ,h\circ \eta \right) z\left(
\omega \left( z_{\sigma \left( 1\right) },...,z_{\sigma \left( q\right)
}\right) \right) \\
\qquad\cdot \theta \left( z_{\sigma \left( q+1\right) },...,z_{\sigma \left(
q+r\right) }\right) +\underset{\sigma \left( q+1\right) <...<\sigma \left(
q+r\right) }{\underset{\sigma \left( 1\right) <...<\sigma \left( q\right) }{%
\tsum }}sgn\left( \sigma \right) \cdot \omega \left( z_{\sigma \left(
1\right) },...,z_{\sigma \left( q\right) }\right) \\
\qquad\cdot \Gamma \left( Th\circ \rho ,h\circ \eta \right) z\left( \theta
\left( z_{\sigma \left( q+1\right) },...,z_{\sigma \left( q+r\right)
}\right) \right) -\underset{\sigma \left( q+1\right) <...<\sigma \left(
q+r\right) }{\underset{\sigma \left( 1\right) <...<\sigma \left( q\right) }{%
\tsum }}sgn\left( \sigma \right) \\
\end{array}%
\end{equation*}
\begin{equation*}
\begin{array}{l}
-\underset{\sigma \left( q+1\right) <...<\sigma \left( q+r\right) }{\underset%
{\sigma \left( 1\right) <...<\sigma \left( q\right) }{\tsum }}sgn\left(
\sigma \right) \overset{q+r}{\underset{i=q+1}{\tsum }}\omega \left(
z_{\sigma \left( 1\right) },...,z_{\sigma \left( q\right) }\right) \\
\qquad\cdot \theta \left( z_{\sigma \left( q+1\right) },...,\left[
z,z_{\sigma \left( i\right) }\right] _{F,h},...,z_{\sigma \left( q+r\right)
}\right) \\
=\underset{\sigma \left( q+1\right) <...<\sigma \left( q+r\right) }{\underset%
{\sigma \left( 1\right) <...<\sigma \left( q\right) }{\tsum }}sgn\left(
\sigma \right) L_{z}\omega \left( z_{\sigma \left( 1\right) },...,\left[
z,z_{\sigma \left( i\right) }\right] _{F,h},...,z_{\sigma \left( q+r\right)
}\right) \\
\qquad\cdot \theta \left( z_{\sigma \left( q+1\right) },...,z_{\sigma \left(
q+r\right) }\right) +\underset{\sigma \left( q+1\right) <...<\sigma \left(
q+r\right) }{\underset{\sigma \left( 1\right) <...<\sigma \left( q\right) }{%
\tsum }}sgn\left( \sigma \right) \overset{q+r}{\underset{i=q+1}{\tsum }}%
\omega \left( z_{\sigma \left( 1\right) },...,z_{\sigma \left( q\right)
}\right) \\
\qquad\cdot L_{z}\theta \left( z_{\sigma \left( q+1\right) },...,\left[
z,z_{\sigma \left( i\right) }\right] _{F,h},...,z_{\sigma \left( q+r\right)
}\right) \\
=\left( L_{z}\omega \wedge \theta +\omega \wedge L_{z}\theta \right) \left(
z_{1},...,z_{q+r}\right)%
\end{array}%
\end{equation*}
it results the conclusion of the theorem. \hfill \emph{q.e.d.}

\bigskip\noindent\textbf{Definition 3.2.6 }For any $z\in \Gamma \left( F,\nu
,N\right) $, the $\mathcal{F}\left( N\right) $-multilinear application
\begin{equation*}
\begin{array}{rcl}
\Lambda \left( F,\nu ,N\right) & ^{\underrightarrow{ \ \ i_{z} \ \ }} &
\Lambda \left( F,\nu ,N\right) \\
\Lambda ^{q}\left( F,\nu ,N\right) \ni \omega & \longmapsto & i_{z}\omega
\in \Lambda ^{q-1}\left( F,\nu ,N\right),%
\end{array}%
\end{equation*}%
where
\begin{equation*}
\begin{array}{c}
i_{z}\omega \left( z_{2},...,z_{q}\right) =\omega \left(
z,z_{2},...,z_{q}\right) ,%
\end{array}%
\end{equation*}%
for any $z_{2},...,z_{q}\in \Gamma \left( F,\nu ,N\right) $, will be called
the \emph{interior product associated to the section}~$z.$\bigskip

For any $f\in \mathcal{F}\left( N\right) $, we define $\ i_{z}f=0.$

\bigskip\noindent\textbf{Remark 3.2.4 }If $z\in \Gamma \left( F,\nu
,N\right) ,$ $\omega \in $\ $\Lambda ^{p}\left( F,\nu ,N\right) $ and $U$\
is an open subset of $N$\ such that $z_{|U}=0$\ or $\omega _{|U}=0,$\ then $%
\left( i_{z}\omega \right) _{|U}=0.$

\bigskip\noindent\textbf{Theorem 3.2.5 }\emph{If }$z\in \Gamma \left( F,\nu
,N\right) $\emph{, then for any }$\omega \in $\emph{\ }$\Lambda ^{q}\left(
F,\nu ,N\right) $\emph{\ and }$\theta \in $\emph{\ }$\Lambda ^{r}\left(
F,\nu ,N\right) $\emph{\ we obtain}
\begin{equation*}
\begin{array}{c}
i_{z}\left( \omega \wedge \theta \right) =i_{z}\omega \wedge \theta +\left(
-1\right) ^{q}\omega \wedge i_{z}\theta .%
\end{array}%
\leqno(3.2.7)
\end{equation*}

\noindent\emph{Proof. }Let $z_{1},...,z_{q+r}\in \Gamma \left( F,\nu
,N\right) $ be arbitrary. We observe that
\begin{equation*}
\begin{array}{l}
i_{z_{1}}\left( \omega \wedge \theta \right) \left( z_{2},...,z_{q+r}\right)
=\left( \omega \wedge \theta \right) \left( z_{1},z_{2},...,z_{q+r}\right)
\\
=\underset{\sigma \left( q+1\right) <...<\sigma \left( q+r\right) }{\underset%
{\sigma \left( 1\right) <...<\sigma \left( q\right) }{\tsum }}sgn\left(
\sigma \right) \cdot \omega \left( z_{\sigma \left( 1\right) },...,z_{\sigma
\left( q\right) }\right) \cdot \theta \left( z_{\sigma \left( q+1\right)
},...,z_{\sigma \left( q+r\right) }\right) \\
=\underset{\sigma \left( q+1\right) <...<\sigma \left( q+r\right) }{\underset%
{1=\sigma \left( 1\right) <\sigma \left( 2\right) <...<\sigma \left(
q\right) }{\tsum }}sgn\left( \sigma \right) \cdot \omega \left(
z_{1},z_{\sigma \left( 2\right) },...,z_{\sigma \left( q\right) }\right)
\cdot \theta \left( z_{\sigma \left( q+1\right) },...,z_{\sigma \left(
q+r\right) }\right) \\
+\underset{1=\sigma \left( q+1\right) <\sigma \left( q+2\right) <...<\sigma
\left( q+r\right) }{\underset{\sigma \left( 1\right) <...<\sigma \left(
q\right) }{\tsum }}sgn\left( \sigma \right) \cdot \omega \left( z_{\sigma
\left( 1\right) },...,z_{\sigma \left( q\right) }\right) \cdot \theta \left(
z_{1},z_{\sigma \left( q+2\right) },...,z_{\sigma \left( q+r\right) }\right)
\\
=\underset{\sigma \left( q+1\right) <...<\sigma \left( q+r\right) }{\underset%
{\sigma \left( 2\right) <...<\sigma \left( q\right) }{\tsum }}sgn\left(
\sigma \right) \cdot i_{z_{1}}\omega \left( z_{\sigma \left( 2\right)
},...,z_{\sigma \left( q\right) }\right) \cdot \theta \left( z_{\sigma
\left( q+1\right) },...,z_{\sigma \left( q+r\right) }\right) \\
+\underset{\sigma \left( q+2\right) <...<\sigma \left( q+r\right) }{\underset%
{\sigma \left( 1\right) <...<\sigma \left( q\right) }{\tsum }}sgn\left(
\sigma \right) \cdot \omega \left( z_{\sigma \left( 1\right) },...,z_{\sigma
\left( q\right) }\right) \cdot i_{z_{1}}\theta \left( z_{\sigma \left(
q+2\right) },...,z_{\sigma \left( q+r\right) }\right) .%
\end{array}%
\end{equation*}
In the second sum, we have the permutation%
\begin{equation*}
\sigma =\left(
\begin{array}{ccccccc}
1 & ... & q & q+1 & q+2 & ... & q+r \\
\sigma \left( 1\right) & ... & \sigma \left( q\right) & 1 & \sigma \left(
q+2\right) & ... & \sigma \left( q+r\right)%
\end{array}%
\right) .
\end{equation*}
We observe that $\sigma =\tau \circ \tau ^{\prime }$, where%
\begin{equation*}
\tau =\left(
\begin{array}{ccccccc}
1 & 2 & ... & q+1 & q+2 & ... & q+r \\
1 & \sigma \left( 1\right) & ... & \sigma \left( q\right) & \sigma \left(
q+2\right) & ... & \sigma \left( q+r\right)%
\end{array}%
\right)
\end{equation*}%
and
\begin{equation*}
\tau ^{\prime }=\left(
\begin{array}{cccccccc}
1 & 2 & ... & q & q+1 & q+2 & ... & q+r \\
2 & 3 & ... & q+1 & 1 & q+2 & ... & q+r%
\end{array}%
\right) .
\end{equation*}
Since $\tau \left( 2\right) <...<\tau \left( q+1\right) $ and $\tau ^{\prime
}$ has $q$ inversions, it results that
\begin{equation*}
sgn\left( \sigma \right) =\left( -1\right) ^{q}\cdot sgn\left( \tau \right) .
\end{equation*}
Therefore,
\begin{equation*}
\begin{array}{l}
i_{z_{1}}\left( \omega \wedge \theta \right) \left( z_{2},...,z_{q+r}\right)
=\left( i_{z_{1}}\omega \wedge \theta \right) \left( z_{2},...,z_{q+r}\right)
\\
+\left( -1\right) ^{q}\underset{\tau \left( q+2\right) <...<\tau \left(
q+r\right) }{\underset{\tau \left( 2\right) <...<\tau \left( q\right) }{%
\tsum }}sgn\left( \tau \right) \cdot \omega \left( z_{\tau \left( 2\right)
},...,z_{\tau \left( q\right) }\right) \cdot i_{z_{1}}\theta \left( z_{\tau
\left( q+2\right) },...,z_{\tau \left( q+r\right) }\right) \\
=\left( i_{z_{1}}\omega \wedge \theta \right) \left(
z_{2},...,z_{q+r}\right) +\left( -1\right) ^{q}\left( \omega \wedge
i_{z_{1}}\theta \right) \left( z_{2},...,z_{q+r}\right) .%
\end{array}%
\end{equation*}

\hfill \emph{q.e.d.}

\bigskip\noindent\textbf{Theorem 3.2.6 }\emph{For any }$z,v\in \Gamma \left(
F,\nu ,N\right) $\emph{\ we obtain}%
\begin{equation*}
\begin{array}{c}
L_{v}\circ i_{z}-i_{z}\circ L_{v}=i_{\left[ z,v\right] _{F,h}}.%
\end{array}%
\leqno(3.2.8)
\end{equation*}

\noindent\emph{Proof.} Let $\omega \in $\emph{\ }$\Lambda ^{q}\left( F,\nu
,N\right) $ be arbitrary. Since
\begin{equation*}
\begin{array}{l}
i_{z}\left( L_{v}\omega \right) \left( z_{2},...z_{q}\right) =L_{v}\omega
\left( z,z_{2},...z_{q}\right) \\
=\Gamma \left( Th\circ \rho ,h\circ \eta \right) v\left( \omega \left(
z,z_{2},...,z_{q}\right) \right) -\omega \left( \left[ v,z\right]
_{F,h},z_{2},...,z_{q}\right) \\
-\overset{q}{\underset{i=2}{\tsum }}\omega \left( \left( z,z_{2},...,\left[
v,z_{i}\right] _{F,h},...,z_{q}\right) \right) \\
=\Gamma \left( Th\circ \rho ,h\circ \eta \right) v\left( i_{z}\omega \left(
z_{2},...,z_{q}\right) \right) -\overset{q}{\underset{i=2}{\tsum }}%
i_{z}\omega \left( z_{2},...,\left[ v,z_{i}\right] _{F,h},...,z_{q}\right)
\\
-i_{\left[ v,z\right] _{F}}\left( z_{2},...,z_{q}\right) =\left( L_{v}\left(
i_{z}\omega \right) -i_{\left[ v,z\right] _{F,h}}\right) \left(
z_{2},...,z_{q}\right) ,%
\end{array}%
\end{equation*}%
for any $z_{2},...,z_{q}\in \Gamma \left( F,\nu ,N\right) $ it result the
conclusion of the theorem.\hfill \emph{q.e.d.}

\bigskip\noindent\textbf{Definition 3.2.7 }If $f\in \mathcal{F}\left(
N\right) $ and $z\in \Gamma \left( F,\nu ,N\right) ,$ then we define \textbf{%
\ }%
\begin{equation*}
\begin{array}{c}
d^{F}f\left( z\right) =\Gamma \left( Th\circ \rho ,h\circ \eta \right)
\left( z\right) f.%
\end{array}%
%
\end{equation*}%
\emph{for any }$z_{0},z_{1},...,z_{q}\in \Gamma \left( F,\nu ,N\right) ,$
\emph{is unique with the following property:}%
\begin{equation*}
\begin{array}{c}
L_{z}=d^{F}\circ i_{z}+i_{z}\circ d^{F},~\forall z\in \Gamma \left( F,\nu
,N\right) .%
\end{array}%
\leqno(3.2.9)
\end{equation*}%
This $\mathcal{F}\left( N\right) $-multilinear application will be called%
\emph{\ the exterior differentiation ope\-ra\-tor for the exterior
differential algebra of the generalized Lie algebroid }$((F,\nu
,N),[,]_{F,h},(\rho ,\eta )).$\bigskip

\noindent\emph{Proof. }We verify the property $\left( 3.2.9\right) $ Since
\begin{equation*}
\begin{array}{l}
\left( i_{z_{0}}\circ d^{F}\right) \omega \left( z_{1},...,z_{q}\right)
=d\omega \left( z_{0},z_{1},...,z_{q}\right) \\
=\overset{q}{\underset{i=0}{\tsum }}\left( -1\right) ^{i}\Gamma \left(
Th\circ \rho ,h\circ \eta \right) z_{i}\left( \omega \left( z_{0},z_{1},...,%
\hat{z}_{i},...,z_{q}\right) \right) \\
+\underset{0\leq i<j}{\tsum }\left( -1\right) ^{i+j}\omega \left( \left[
z_{i},z_{j}\right] _{F,h},z_{0},z_{1},...,\hat{z}_{i},...,\hat{z}%
_{j},...,z_{q}\right) \\
=\Gamma \left( Th\circ \rho ,h\circ \eta \right) z_{0}\left( \omega \left(
z_{1},...,z_{q}\right) \right) \\
+\overset{q}{\underset{i=1}{\tsum }}\left( -1\right) ^{i}\Gamma \left(
Th\circ \rho ,h\circ \eta \right) z_{i}\left( \omega \left( z_{0},z_{1},...,%
\hat{z}_{i},...,z_{q}\right) \right) \\
+\underset{i=1}{\overset{q}{\tsum }}\left( -1\right) ^{i}\omega \left( \left[
z_{0},z_{i}\right] _{F,h},z_{1},...,\hat{z}_{i},...,z_{q}\right) \\
+\underset{1\leq i<j}{\tsum }\left( -1\right) ^{i+j}\omega \left( \left[
z_{i},z_{j}\right] _{F,h},z_{0},z_{1},...,\hat{z}_{i},...,\hat{z}%
_{j},...,z_{q}\right) \\
=\Gamma \left( Th\circ \rho ,h\circ \eta \right) z_{0}\left( \omega \left(
z_{1},...,z_{q}\right) \right) \\
-\underset{i=1}{\overset{q}{\tsum }}\omega \left( z_{1},...,\left[
z_{0},z_{i}\right] _{F,h},...,z_{q}\right) \\
-\overset{q}{\underset{i=1}{\tsum }}\left( -1\right) ^{i-1}\Gamma \left(
Th\circ \rho ,h\circ \eta \right) z_{i}\left( i_{z_{0}}\omega \left( \left(
z_{1},...,\hat{z}_{i},...,z_{q}\right) \right) \right) \\
-\underset{1\leq i<j}{\tsum }\left( -1\right) ^{i+j-2}i_{z_{0}}\omega \left(
\left( \left[ z_{i},z_{j}\right] _{F,h},z_{1},...,\hat{z}_{i},...,\hat{z}%
_{j},...,z_{q}\right) \right) \\
=\left( L_{z_{0}}-d^{F}\circ i_{z_{0}}\right) \omega \left(
z_{1},...,z_{q}\right) ,%
\end{array}%
\end{equation*}%
for any $z_{0},z_{1},...,z_{q}\in \Gamma \left( F,\nu ,N\right) $ it results
that the property $\left( 3.2.9\right) $ is satisfied.

In the following, we verify the uniqueness of the operator $d^{F}.$

Let $d^{\prime F}$ be an another exterior differentiation operator
satisfying the property $\left( 3.2.9\right) .$

Let $S=\left\{ q\in \mathbb{N}:d^{F}\omega =d^{\prime F}\omega ,~\forall
\omega \in \Lambda ^{q}\left( F,\nu ,N\right) \right\} $ be.

Let $z\in \Gamma \left( F,\nu ,N\right) $ be arbitrary.

We observe that $\left( 3.2.9\right) $ is equivalent with
\begin{equation*}
\begin{array}{c}
i_{z}\circ \left( d^{F}-d^{\prime F}\right) +\left( d^{F}-d^{\prime
F}\right) \circ i_{z}=0.%
\end{array}%
\leqno(1)
\end{equation*}
Since $i_{z}f=0,$ for any $f\in \mathcal{F}\left( N\right) ,$ it results
that
\begin{equation*}
\begin{array}{c}
\left( \left( d^{F}-d^{\prime F}\right) f\right) \left( z\right) =0,~\forall
f\in \mathcal{F}\left( N\right) .%
\end{array}%
\end{equation*}
Therefore, we obtain that
\begin{equation*}
\begin{array}{c}
0\in S.%
\end{array}%
\leqno(2)
\end{equation*}
In the following, we prove that
\begin{equation*}
\begin{array}{c}
q\in S\Longrightarrow q+1\in S%
\end{array}%
\leqno(3)
\end{equation*}

Let $\omega \in \Lambda ^{p+1}\left( F,\nu ,N\right) $ be arbitrary$.$ Since
$i_{z}\omega \in \Lambda ^{q}\left( F,\nu ,N\right) $, using the equality $%
\left( 1\right) $, it results that
\begin{equation*}
i_{z}\circ \left( d^{F}-d^{\prime F}\right) \omega =0.
\end{equation*}
We obtain that, $\left( \left( d^{F}-d^{\prime F}\right) \omega \right)
\left( z_{0},z_{1},...,z_{q}\right) =0,$ for any $z_{1},...,z_{q}\in \Gamma
\left( F,\nu ,N\right) .$ Therefore $d^{F}\omega =d^{\prime F}\omega ,$
namely $q+1\in S.$

Using the \textit{Peano's Axiom }and the affirmations $\left( 2\right) $ and
$\left( 3\right) $ it results that $S=\mathbb{N}.$ Therefore, the uniqueness
is verified.\hfill \emph{q.e.d.}\medskip

Note that if $\omega =\omega _{\alpha _{1}...\alpha _{q}}t^{\alpha
_{1}}\wedge ...\wedge t^{\alpha _{q}}\in \Lambda ^{q}\left( F,\nu ,N\right) $%
, then
\begin{eqnarray*}
d^{F}\omega \left( t_{\alpha _{0}},t_{\alpha _{1}},...,t_{\alpha
_{q}}\right) &=&\overset{q}{\underset{i=0}{\tsum }}\left( -1\right)
^{i}\theta _{\alpha _{i}}^{\tilde{k}}\frac{\partial \omega _{\alpha _{0},...,%
\widehat{\alpha _{i}}...\alpha _{q}}}{\partial \varkappa ^{\tilde{k}}} \\%
[1mm]
&&+\underset{i<j}{\tsum }\left( -1\right) ^{i+j}L_{\alpha _{i}\alpha
_{j}}^{\alpha }\cdot \omega _{\alpha ,\alpha _{0},...,\widehat{\alpha _{i}}%
,...,\widehat{\alpha _{j}},...,\alpha _{q}}.
\end{eqnarray*}
Therefore, we obtain%
\begin{equation*}
\begin{array}{l}
d^{F}\omega =\left( \overset{q}{\underset{i=0}{\tsum }}\left( -1\right)
^{i}\theta _{\alpha _{i}}^{\tilde{k}}\displaystyle\frac{\partial \omega
_{\alpha _{0},...,\widehat{\alpha _{i}}...\alpha _{q}}}{\partial \varkappa ^{%
\tilde{k}}}\right. \\
\left. +\underset{i<j}{\tsum }\left( -1\right) ^{i+j}L_{\alpha _{i}\alpha
_{j}}^{\alpha }\cdot \omega _{\alpha ,\alpha _{0},...,\widehat{\alpha _{i}}%
,...,\widehat{\alpha _{j}},...,\alpha _{q}}\right) t^{\alpha _{0}}\wedge
t^{\alpha _{1}}\wedge ...\wedge t^{\alpha _{q}}.%
\end{array}%
\leqno(3.2.10)
\end{equation*}

\noindent\textbf{Remark 3.2.4 }If $d^{F}$ is the exterior differentiation
operator for the generalized Lie algebroid
\begin{equation*}
\left( \left( F,\nu ,N\right) ,\left[ ,\right] _{F,h},\left( \rho ,\eta
\right) \right) ,
\end{equation*}%
$\omega \in $\emph{\ }$\Lambda ^{q}\left( F,\nu ,N\right) $\ and $U$\ is an
open subset of $N$\ such that $\omega _{|U}=0,$\ then $\left( d^{F}\omega
\right) _{|U}=0.$

\bigskip\noindent\textbf{Theorem 3.2.8} \emph{The exterior differentiation
operator }$d^{F}$\emph{\ given by the previous theorem has the following
properties:}\medskip

\noindent1. \emph{\ For any }$\omega \in $\emph{\ }$\Lambda ^{q}\left( F,\nu
,N\right) $\emph{\ and }$\theta \in $\emph{\ }$\Lambda ^{r}\left( F,\nu
,N\right) $\emph{\ we obtain }%
\begin{equation*}
\begin{array}{c}
d^{F}\left( \omega \wedge \theta \right) =d^{F}\omega \wedge \theta +\left(
-1\right) ^{q}\omega \wedge d^{F}\theta .%
\end{array}%
\leqno(3.2.11)
\end{equation*}

\noindent2.\emph{\ For any }$z\in \Gamma \left( F,\nu ,N\right) $ we obtain
\begin{equation*}
\begin{array}{c}
L_{z}\circ d^{F}=d^{F}\circ L_{z}.%
\end{array}%
\leqno(3.2.12)
\end{equation*}

\noindent3.\emph{\ }$d^{F}\circ d^{F}=0.$

\bigskip\noindent\emph{Proof.}\newline
1. Let $S=\left\{ q\in \mathbb{N}:d^{F}\left( \omega \wedge \theta \right)
=d^{F}\omega \wedge \theta +\left( -1\right) ^{q}\omega \wedge d^{F}\theta
,~\forall \omega \in \Lambda ^{q}\left( F,\nu ,N\right) \right\} $ be. Since
\begin{equation*}
\begin{array}{l}
d^{F}\left( f\wedge \theta \right) \left( z,v\right) =d^{F}\left( f\cdot
\theta \right) \left( z,v\right) \\
=\Gamma \left( Th\circ \rho ,h\circ \eta \right) z\left( f\omega \left(
v\right) \right) -\Gamma \left( Th\circ \rho ,h\circ \eta \right) v\left(
f\omega \left( z\right) \right) -f\omega \left( \left[ z,v\right]
_{F,h}\right) \\
=\Gamma \left( Th\circ \rho ,h\circ \eta \right) z\left( f\right) \cdot
\omega \left( v\right) +f\cdot \Gamma \left( Th\circ \rho ,h\circ \eta
\right) z\left( \omega \left( v\right) \right) \\
-\Gamma \left( Th\circ \rho ,h\circ \eta \right) v\left( f\right) \cdot
\omega \left( z\right) -f\cdot \Gamma \left( Th\circ \rho ,h\circ \eta
\right) v\left( \omega \left( z\right) \right) -f\omega \left( \left[ z,v%
\right] _{F,h}\right) \\
=d^{F}f\left( z\right) \cdot \omega \left( v\right) -d^{F}f\left( v\right)
\cdot \omega \left( z\right) +f\cdot d^{F}\omega \left( z,v\right) \\
=\left( d^{F}f\wedge \omega \right) \left( z,v\right) +\left( -1\right)
^{0}f\cdot d^{F}\omega \left( z,v\right) \\
=\left( d^{F}f\wedge \omega \right) \left( z,v\right) +\left( -1\right)
^{0}\left( f\wedge d^{F}\omega \right) \left( z,v\right) ,~\forall z,v\in
\Gamma \left( F,\nu ,N\right) ,%
\end{array}%
\end{equation*}%
it results that
\begin{equation*}
\begin{array}{c}
0\in S.%
\end{array}%
\leqno(1.1)
\end{equation*}
In the following we prove that
\begin{equation*}
\begin{array}{c}
q\in S\Longrightarrow q+1\in S.%
\end{array}%
\leqno(1.2)
\end{equation*}
Without restricting the generality, we consider that $\theta \in \Lambda
^{r}\left( F,\nu ,N\right) .$ Since
\begin{equation*}
\begin{array}{l}
d^{F}\left( \omega \wedge \theta \right) \left(
z_{0},z_{1},...,z_{q+r}\right) =i_{z_{0}}\circ d^{F}\left( \omega \wedge
\theta \right) \left( z_{1},...,z_{q+r}\right) \\
=L_{z_{0}}\left( \omega \wedge \theta \right) \left(
z_{1},...,z_{q+r}\right) -d^{F}\circ i_{z_{0}}\left( \omega \wedge \theta
\right) \left( z_{1},...,z_{q+r}\right) \\
=\left( L_{z_{0}}\omega \wedge \theta +\omega \wedge L_{z_{0}}\theta \right)
\left( z_{1},...,z_{q+r}\right) \\
-\left[ d^{F}\circ \left( i_{z_{0}}\omega \wedge \theta +\left( -1\right)
^{q}\omega \wedge i_{z_{0}}\theta \right) \right] \left(
z_{1},...,z_{q+r}\right) \\
=\left( L_{z_{0}}\omega \wedge \theta +\omega \wedge L_{z_{0}}\theta -\left(
d^{F}\circ i_{z_{0}}\omega \right) \wedge \theta \right) \left(
z_{1},...,z_{q+r}\right) \\
-\left( \left( -1\right) ^{q-1}i_{z_{0}}\omega \wedge d^{F}\theta +\left(
-1\right) ^{q}d^{F}\omega \wedge i_{z_{0}}\theta \right) \left(
z_{1},...,z_{q+r}\right) \\
-\left( -1\right) ^{2q}\omega \wedge d^{F}\circ i_{z_{0}}\theta \left(
z_{1},...,z_{q+r}\right) \\
=\left( \left( L_{z_{0}}\omega -d^{F}\circ i_{z_{0}}\omega \right) \wedge
\theta \right) \left( z_{1},...,z_{q+r}\right) \\
\end{array}%
\end{equation*}%
for any $z_{0},z_{1},...,z_{q+r}\in \Gamma \left( F,\nu ,N\right) $, it
results $\left( 1.2\right) .$

Using the \textit{Peano's Axiom }and the affirmations $\left( 1.1\right) $
and $\left( 1.2\right) $ it results that $S=\mathbb{N}.$

Therefore, it results the conclusion of affirmation 1.\medskip

2. Let $z\in \Gamma \left( F,\nu ,N\right) $ be arbitrary.

Let $S=\left\{ q\in \mathbb{N}:\left( L_{z}\circ d^{F}\right) \omega =\left(
d^{F}\circ L_{z}\right) \omega ,~\forall \omega \in \Lambda ^{q}\left( F,\nu
,N\right) \right\} $ be.

Let $f\in \mathcal{F}\left( N\right) $ be arbitrary. Since
\begin{equation*}
\begin{array}{l}
\left( d^{F}\circ L_{z}\right) f\left( v\right) =i_{v}\circ \left(
d^{F}\circ L_{z}\right) f=\left( i_{v}\circ d^{F}\right) \circ L_{z}f \\
=\left( L_{v}\circ L_{z}\right) f-\left( \left( d^{F}\circ i_{v}\right)
\circ L_{z}\right) f \\
=\left( L_{v}\circ L_{z}\right) f-L_{\left[ z,v\right] _{F,h}}f+d^{F}\circ
i_{\left[ z,v\right] _{F,h}}f-d^{F}\circ L_{z}\left( i_{v}f\right) \\
=\left( L_{v}\circ L_{z}\right) f-L_{\left[ z,v\right] _{F,h}}f+d^{F}\circ
i_{\left[ z,v\right] _{F,h}}f-0 \\
=\left( L_{v}\circ L_{z}\right) f-L_{\left[ z,v\right] _{F,h}}f+d^{F}\circ
i_{\left[ z,v\right] _{F,h}}f-L_{z}\circ d^{F}\left( i_{v}f\right) \\
=\left( L_{z}\circ i_{v}\right) \left( d^{F}f\right) -L_{\left[ z,v\right]
_{F,h}}f+d^{F}\circ i_{\left[ z,v\right] _{F,h}}f \\
=\left( i_{v}\circ L_{z}\right) \left( d^{F}f\right) +L_{\left[ z,v\right]
_{F,h}}f-L_{\left[ z,v\right] _{F,h}}f \\
=i_{v}\circ \left( L_{z}\circ d^{F}\right) f=\left( L_{z}\circ d^{F}\right)
f\left( v\right) ,~\forall v\in \Gamma \left( F,\nu ,N\right) ,%
\end{array}%
\end{equation*}%
it results that
\begin{equation*}
\begin{array}{c}
0\in S.%
\end{array}%
\leqno(2.1)
\end{equation*}
In the following we prove that
\begin{equation*}
\begin{array}{c}
q\in S\Longrightarrow q+1\in S.%
\end{array}%
\leqno(2.2)
\end{equation*}
Let $\omega \in \Lambda ^{q}\left( F,\nu ,N\right) $ be arbitrary. Since
\begin{equation*}
\begin{array}{l}
\left( d^{F}\circ L_{z}\right) \omega \left( z_{0},z_{1},...,z_{q}\right)
=i_{z_{0}}\circ \left( d^{F}\circ L_{z}\right) \omega \left(
z_{1},...,z_{q}\right) \vspace*{1mm} \\
=\left( i_{z_{0}}\circ d^{F}\right) \circ L_{z}\omega \left(
z_{1},...,z_{q}\right) \vspace*{1mm} \\
=\left[ \left( L_{z_{0}}\circ L_{z}\right) \omega -\left( \left( d^{F}\circ
i_{z_{0}}\right) \circ L_{z}\right) \omega \right] \left(
z_{1},...,z_{q}\right) \vspace*{1mm} \\
=\left[ \left( L_{z_{0}}\circ L_{z}\right) \omega -L_{\left[ z,z_{0}\right]
_{F,h}}\omega \right] \left( z_{1},...,z_{q}\right) \vspace*{1mm} \\
+\left[ d^{F}\circ i_{\left[ z,z_{0}\right] _{F,h}}\omega -d^{F}\circ
L_{z}\left( i_{z_{0}}\omega \right) \right] \left( z_{1},...,z_{q}\right)
\vspace*{1mm} \\
\overset{ip.}{=}\left[ \left( L_{z_{0}}\circ L_{z}\right) \omega -L_{\left[
z,z_{0}\right] _{F,h}}\omega \right] \left( z_{1},...,z_{q}\right) \vspace*{%
1mm} \\
+\left[ d^{F}\circ i_{\left[ z,z_{0}\right] _{F,h}}\omega -L_{z}\circ
d^{F}\left( i_{z_{0}}\omega \right) \right] \left( z_{1},...,z_{q}\right)
\vspace*{1mm} \\
=\left[ \left( L_{z}\circ i_{z_{0}}\right) \left( d^{F}\omega \right) -L_{%
\left[ z,z_{0}\right] _{F,h}}\omega +d^{F}\circ i_{\left[ z,z_{0}\right]
_{F,h}}\omega \right] \left( z_{1},...,z_{q}\right) \vspace*{1mm} \\
=\left[ \left( i_{z_{0}}\circ L_{z}\right) \left( d^{F}\omega \right) +L_{%
\left[ z,z_{0}\right] _{F,h}}\omega -L_{\left[ z,z_{0}\right] _{F,h}}\omega %
\right] \left( z_{1},...,z_{q}\right) \vspace*{1mm} \\
=i_{z_{0}}\circ \left( L_{z}\circ d^{F}\right) \omega \left(
z_{1},...,z_{q}\right) \vspace*{1mm} \\
=\left( L_{z}\circ d^{F}\right) \omega \left( z_{0},z_{1},...,z_{q}\right)
,~\forall z_{0},z_{1},...,z_{q}\in \Gamma \left( F,\nu ,N\right) ,%
\end{array}%
\end{equation*}%
it results $\left( 2.2\right) .$

Using the \textit{Peano's Axiom }and the affirmations $\left( 2.1\right) $
and $\left( 2.2\right) $ it results that $S=\mathbb{N}.$

Therefore, it results the conclusion of affirmation 2.\medskip

3. It is remarked that
\begin{equation*}
\begin{array}{l}
i_{z}\circ \left( d^{F}\circ d^{F}\right) =\left( i_{z}\circ d^{F}\right)
\circ d^{F}=L_{z}\circ d^{F}-\left( d^{F}\circ i_{z}\right) \circ d^{F}
\vspace*{1mm} \\
=L_{z}\circ d^{F}-d^{F}\circ L_{z}+d^{F}\circ \left( d^{F}\circ i_{z}\right)
=\left( d^{F}\circ d^{F}\right) \circ i_{z},%
\end{array}%
\end{equation*}%
for any $z\in \Gamma \left( F,\nu ,N\right) .$

Let $\omega \in \Lambda ^{q}\left( F,\nu ,N\right) $ be arbitrary. Since
\begin{equation*}
\begin{array}{l}
\left( d^{F}\circ d^{F}\right) \omega \left( z_{1},...,z_{q+2}\right)
=i_{z_{q+2}}\circ ...\circ i_{z_{1}}\circ \left( d^{F}\circ d^{F}\right)
\omega =... \vspace*{1mm} \\
=i_{z_{q+2}}\circ \left( d^{F}\circ d^{F}\right) \circ i_{z_{q+1}}\left(
\omega \left( z_{1},...,z_{q}\right) \right) \vspace*{1mm} \\
=i_{z_{q+2}}\circ \left( d^{F}\circ d^{F}\right) \left( 0\right) =0,~\forall
z_{1},...,z_{q+2}\in \Gamma \left( F,\nu ,N\right) ,%
\end{array}%
\end{equation*}%
it results the conclusion of affirmation 3. \hfill \emph{q.e.d.}\bigskip

\noindent \textbf{Theorem 3.2.9 \ } \emph{If }$((F,\nu ,N),[,]_{F,h},(\rho
,\eta ))$ \emph{is a generalized Lie algebroid and} $d^{F}$\break \emph{is
the ex\-te\-rior differentiation operator for the exterior differential} $%
\mathcal{F}(N)$\emph{-algebra}\break $(\Lambda (F,\nu ,N),+,\cdot ,\wedge )$%
, \emph{then we obtain the structure equations of Maurer-Cartan type }%
\begin{equation*}
\begin{array}{c}
d^{F}t^{\alpha }=-\displaystyle\frac{1}{2}L_{\beta \gamma }^{\alpha
}t^{\beta }\wedge t^{\gamma },~\alpha \in \overline{1,p}%
\end{array}%
\leqno(\mathcal{C}_{1})
\end{equation*}%
\emph{and\ }%
\begin{equation*}
\begin{array}{c}
d^{F}\varkappa ^{\tilde{\imath}}=\theta _{\alpha }^{\tilde{\imath}}t^{\alpha
},~\tilde{\imath}\in \overline{1,n},%
\end{array}%
\leqno(\mathcal{C}_{2})
\end{equation*}%
\emph{where }$\left\{ t^{\alpha },\alpha \in \overline{1,p}\right\} ~$\emph{%
is the coframe of the vector bundle }$\left( F,\nu ,N\right) .$\bigskip

\noindent This equations will be called \emph{the structure equations of
Maurer-Cartan type associa\-ted to the generalized Lie algebroid }$\left(
\left( F,\nu ,N\right) ,\left[ ,\right] _{F,h},\left( \rho ,\eta \right)
\right) .$

\bigskip\noindent\emph{Proof.} Let $\alpha \in \overline{1,p}$ be arbitrary.
Since
\begin{equation*}
\begin{array}{c}
d^{F}t^{\alpha }\left( t_{\beta },t_{\gamma }\right) =-L_{\beta \gamma
}^{\alpha },~\forall \beta ,\gamma \in \overline{1,p}%
\end{array}%
\end{equation*}%
it results that
\begin{equation*}
\begin{array}{c}
d^{F}t^{\alpha }=-\underset{\beta <\gamma }{\tsum }L_{\beta \gamma }^{\alpha
}t^{\beta }\wedge t^{\gamma }.%
\end{array}%
\leqno(1)
\end{equation*}
Since $L_{\beta \gamma }^{\alpha }=-L_{\gamma \beta }^{\alpha }$ and $%
t^{\beta }\wedge t^{\gamma }=-t^{\gamma }\wedge t^{\beta }$, for nay $\beta
,\gamma \in \overline{1,p},$ it results that
\begin{equation*}
\begin{array}{c}
\underset{\beta <\gamma }{\tsum }L_{\beta \gamma }^{\alpha }t^{\beta }\wedge
t^{\gamma }=\displaystyle\frac{1}{2}L_{\beta \gamma }^{\alpha }t^{\beta
}\wedge t^{\gamma }%
\end{array}%
\leqno(2)
\end{equation*}%
Using the equalities $\left( 1\right) $ and $\left( 2\right) $ it results
the structure equation $(\mathcal{C}_{1}).$

Let $\tilde{\imath}\in \overline{1,n}$ be arbitrarily. Since
\begin{equation*}
\begin{array}{c}
d^{F}\varkappa ^{\tilde{\imath}}\left( t_{\alpha }\right) =\theta _{\alpha
}^{\tilde{\imath}},~\forall \alpha \in \overline{1,p}%
\end{array}%
\end{equation*}%
it results the structure equation $(\mathcal{C}_{2}).$\hfill \emph{q.e.d.}%
\bigskip

\noindent \textbf{Corollary 3.2.1 }\emph{If }$\left( \left( h^{\ast
}F,h^{\ast }\nu ,M\right) ,\left[ ,\right] _{h^{\ast }F},\left( \overset{%
h^{\ast }F}{\rho },Id_{M}\right) \right) $ \emph{\ is the pull-back Lie
algebroid associated to the generalized Lie algebroid} $%
\begin{array}{c}
\left( \left( F,\nu ,N\right) ,\left[ ,\right] _{F,h},\left( \rho ,\eta
\right) \right)%
\end{array}%
$ \emph{and }$d^{h^{\ast }F}$\emph{\ is the exterior differentiation
operator for the exterior differential }$\mathcal{F}\left( M\right) $\emph{%
-algebra}\break $\left( \Lambda \left( h^{\ast }F,h^{\ast }\nu ,M\right)
,+,\cdot ,\wedge \right) $\emph{, then we obtain the following structure
equations of Maurer-Cartan type \ }%
\begin{equation*}
\begin{array}{c}
d^{h^{\ast }F}T^{\alpha }=-\displaystyle\frac{1}{2}\left( L_{\beta \gamma
}^{\alpha }\circ h\right) T^{\beta }\wedge T^{\gamma },~\alpha \in \overline{%
1,p}%
\end{array}%
\leqno(\mathcal{C}_{1}^{\prime })
\end{equation*}%
\emph{and\ }%
\begin{equation*}
\begin{array}{c}
d^{h^{\ast }F}x^{i}=\left( \rho _{\alpha }^{i}\circ h\right) T^{\alpha
},~i\in \overline{1,m}.%
\end{array}%
\leqno(\mathcal{C}_{2}^{\prime })
\end{equation*}%
This equations will be called \emph{the structure equations of Maurer-Cartan
type associated to the pull-back Lie algebroid }%
\begin{equation*}
\left( \left( h^{\ast }F,h^{\ast }\nu ,M\right) ,\left[ ,\right] _{h^{\ast
}F},\left( \overset{h^{\ast }F}{\rho },Id_{M}\right) \right) .
\end{equation*}

\textbf{Theorem 3.2.10 (}of Cartan type)\textbf{\ }\emph{Let }$\left( E,\pi
,M\right) $\emph{\ be an IDS of the generalized Lie algebroid }$\left(
\left( F,\nu ,N\right) ,\left[ ,\right] _{F,h},\left( \rho ,\eta \right)
\right) .$\emph{\ If }$\left\{ \Theta ^{r+1},...,\Theta ^{p}\right\} $\emph{%
\ is a base for the }$\mathcal{F}\left( M\right) $\emph{-submodule }$\left(
\Gamma \left( E^{\perp },\pi ^{\perp },M\right) ,+,\cdot \right) $\emph{,
then the IDS }$\left( E,\pi ,M\right) $\emph{\ is involutive if and only if
it exists }%
\begin{equation*}
\Omega _{\beta }^{\alpha }\in \Lambda ^{1}\left( h^{\ast }F,h^{\ast }\nu
,M\right) ,~\alpha ,\beta \in \overline{r+1,p}
\end{equation*}%
\emph{such that}
\begin{equation*}
d^{h^{\ast }F}\Theta ^{\alpha }=\Sigma _{\beta \in \overline{r+1,p}}\Omega
_{\beta }^{\alpha }\wedge \Theta ^{\beta }\in \mathcal{I}\left( \Gamma
\left( E^{\perp },\pi ^{\perp },M\right) \right) .
\end{equation*}

\textbf{Proof: }Let $\left\{ S_{1},...,S_{r}\right\} $ be a base for the $%
\mathcal{F}\left( M\right) $-submodule $\left( \Gamma \left( E,\pi ,M\right)
,+,\cdot \right) $

Let $\left\{ S_{r+1},...,S_{p}\right\} \in \Gamma \left( h^{\ast }F,h^{\ast
}\nu ,M\right) $ such that
\begin{equation*}
\left\{ S_{1},...,S_{r},S_{r+1},...,S_{p}\right\}
\end{equation*}%
is a base for the $\mathcal{F}\left( M\right) $-module
\begin{equation*}
\left( \Gamma \left( h^{\ast }F,h^{\ast }\nu ,M\right) ,+,\cdot \right) .
\end{equation*}

Let $\Theta ^{1},...,\Theta ^{r}\in \Gamma \left( \overset{\ast }{h^{\ast }F}%
,\overset{\ast }{h^{\ast }\nu },M\right) $ such that
\begin{equation*}
\left\{ \Theta ^{1},...,\Theta ^{r},\Theta ^{r+1},...,\Theta ^{p}\right\}
\end{equation*}
is a base for the $\mathcal{F}\left( M\right) $-module
\begin{equation*}
\left( \Gamma \left( \overset{\ast }{h^{\ast }F},\overset{\ast }{h^{\ast
}\nu },M\right) ,+,\cdot \right) .
\end{equation*}

For any $a,b\in \overline{1,r}$ and $\alpha ,\beta \in \overline{r+1,p}$, we
have the equalities:%
\begin{equation*}
\begin{array}{ccc}
\Theta ^{a}\left( S_{b}\right) & = & \delta _{b}^{a} \\
\Theta ^{a}\left( S_{\beta }\right) & = & 0 \\
\Theta ^{\alpha }\left( S_{b}\right) & = & 0 \\
\Theta ^{\alpha }\left( S_{\beta }\right) & = & \delta _{\beta }^{\alpha }%
\end{array}%
\end{equation*}

We remark that the set of the $2$-forms%
\begin{equation*}
\left\{ \Theta ^{a}\wedge \Theta ^{b},\Theta ^{a}\wedge \Theta ^{\beta
},\Theta ^{\alpha }\wedge \Theta ^{\beta },~a,b\in \overline{1,r}\wedge
\alpha ,\beta \in \overline{r+1,p}\right\}
\end{equation*}%
is a base for the $\mathcal{F}\left( M\right) $-module
\begin{equation*}
\left( \Lambda ^{2}\left( h^{\ast }F,h^{\ast }\nu ,M\right) ,+,\cdot \right)
.
\end{equation*}

Therefore, we have%
\begin{equation*}
d^{h^{\ast }F}\Theta ^{\alpha }=\Sigma _{b<c}A_{bc}^{\alpha }\Theta
^{b}\wedge \Theta ^{c}+\Sigma _{b,\gamma }B_{b\gamma }^{\alpha }\Theta
^{b}\wedge \Theta ^{\gamma }+\Sigma _{\beta <\gamma }C_{\beta \gamma
}^{\alpha }\Theta ^{\beta }\wedge \Theta ^{\gamma },\leqno\left( 1\right)
\end{equation*}%
where, $A_{bc}^{\alpha },B_{b\gamma }^{\alpha }$ and $C_{\beta \gamma
}^{\alpha },~a,b,c\in \overline{1,r}\wedge \alpha ,\beta ,\gamma \in
\overline{r+1,p}$ are real local functions such that $A_{bc}^{\alpha
}=-A_{cb}^{\alpha }$ and $C_{\beta \gamma }^{\alpha }=-C_{\gamma \beta
}^{\alpha }.$

Using the formula%
\begin{equation*}
d^{h^{\ast }F}\Theta ^{\alpha }\left( S_{b},S_{c}\right) =\Gamma \left(
\overset{h^{\ast }F}{\rho },Id_{M}\right) S_{b}\left( \Theta ^{\alpha
}\left( S_{c}\right) \right) -\Gamma \left( \overset{h^{\ast }F}{\rho }%
,Id_{M}\right) S_{c}\left( \Theta ^{\alpha }\left( S_{b}\right) \right)
-\Theta ^{\alpha }\left( \left[ S_{b},S_{c}\right] _{h^{\ast }F}\right) ,%
\leqno\left( 2\right)
\end{equation*}%
we obtain that
\begin{equation*}
A_{bc}^{\alpha }=-\Theta ^{\alpha }\left( \left[ S_{b},S_{c}\right]
_{h^{\ast }F}\right) ,~\forall \left( b,c\in \overline{1,r}\wedge \alpha \in
\overline{r+1,p}\right) .\leqno\left( 3\right)
\end{equation*}

We admit that $\left( E,\pi ,M\right) $ is an involutive IDS of the
generalized Lie algebroid $\left( \left( F,\nu ,N\right) ,\left[ ,\right]
_{F,h},\left( \rho ,\eta \right) \right) .$

As
\begin{equation*}
\left[ S_{b},S_{c}\right] _{h^{\ast }F}\in \Gamma \left( E,\pi ,M\right)
,~\forall b,c\in \overline{1,r}
\end{equation*}%
it results that
\begin{equation*}
\Theta ^{\alpha }\left( \left[ S_{b},S_{c}\right] _{h^{\ast }F}\right)
=0,~\forall \left( b,c\in \overline{1,r}\wedge \alpha \in \overline{r+1,p}%
\right) .
\end{equation*}

Therefore,
\begin{equation*}
A_{bc}^{\alpha }=0,~\forall \left( b,c\in \overline{1,r}\wedge \alpha \in
\overline{r+1,p}\right)
\end{equation*}%
and we obtain%
\begin{equation*}
\begin{array}{ccl}
d^{h^{\ast }F}\Theta ^{\alpha } & = & \Sigma _{b,\gamma }B_{b\gamma
}^{\alpha }\Theta ^{b}\wedge \Theta ^{\gamma }+\frac{1}{2}C_{\beta \gamma
}^{\alpha }\Theta ^{\beta }\wedge \Theta ^{\gamma } \\
& = & \left( B_{b\gamma }^{\alpha }\Theta ^{b}+\frac{1}{2}C_{\beta \gamma
}^{\alpha }\Theta ^{\beta }\right) \wedge \Theta ^{\gamma }.%
\end{array}%
\end{equation*}

As
\begin{equation*}
\Omega _{\gamma }^{\alpha }\overset{put}{=}B_{b\gamma }^{\alpha }\Theta ^{b}+%
\frac{1}{2}C_{\beta \gamma }^{\alpha }\Theta ^{\beta }\in \Lambda ^{1}\left(
h^{\ast }F,h^{\ast }\nu ,M\right) ,~\forall \alpha ,\beta \in \overline{r+1,p%
}
\end{equation*}%
it results the first implication.

Conversely, we admit that it exists
\begin{equation*}
\Omega _{\beta }^{\alpha }\in \Lambda ^{1}\left( h^{\ast }F,h^{\ast }\nu
,M\right) ,~\alpha ,\beta \in \overline{r+1,p}
\end{equation*}%
such that
\begin{equation*}
d^{h^{\ast }F}\Theta ^{\alpha }=\Sigma _{\beta \in \overline{r+1,p}}\Omega
_{\beta }^{\alpha }\wedge \Theta ^{\beta },~\forall \alpha \in \overline{%
r+1,p}.\leqno\left( 4\right)
\end{equation*}

Using the affirmations $\left( 1\right) ,\left( 2\right) $ and $\left(
4\right) $ we obtain that
\begin{equation*}
A_{bc}^{\alpha }=0,~\forall \left( b,c\in \overline{1,r}\wedge \alpha \in
\overline{r+1,p}\right) .
\end{equation*}

Using the affirmation $\left( 3\right) $, we obtain
\begin{equation*}
\Theta ^{\alpha }\left( \left[ S_{b},S_{c}\right] _{h^{\ast }F}\right)
=0,~\forall \left( b,c\in \overline{1,r}\wedge \alpha \in \overline{r+1,p}%
\right) .
\end{equation*}

Therefore,
\begin{equation*}
\left[ S_{b},S_{c}\right] _{h^{\ast }F}\in \Gamma \left( E,\pi ,M\right)
,~\forall b,c\in \overline{1,r}.
\end{equation*}

Using the \emph{Proposition 3.1.3.2}, we obtain the second
implication.\hfill \emph{q.e.d.}\medskip

\bigskip

Let $\left( \left( F^{\prime },\nu ^{\prime },N^{\prime }\right) ,\left[ ,%
\right] _{F^{\prime },h^{\prime }},\left( \rho ^{\prime },\eta ^{\prime
}\right) \right) $ be an another generalized Lie algebroid.

\bigskip\noindent\textbf{Definition 3.2.8} For any morphism $\left( \varphi
,\varphi _{0}\right) $ of $\left( \left( F,\nu ,N\right) ,\left[ ,\right]
_{F,h},\left( \rho ,\eta \right) \right) $ source and $\left( \left(
F^{\prime },\nu ^{\prime },N^{\prime }\right) ,\left[ ,\right] _{F^{\prime
},h^{\prime }},\left( \rho ^{\prime },\eta ^{\prime }\right) \right) $
target we define the application
\begin{equation*}
\begin{array}{ccc}
\Lambda ^{q}\left( F^{\prime },\nu ^{\prime },N^{\prime }\right) & ^{%
\underrightarrow{ \ \left( \varphi ,\varphi _{0}\right) ^{\ast } \ }} &
\Lambda ^{q}\left( F,\nu ,N\right) \\
\omega ^{\prime } & \longmapsto & \left( \varphi ,\varphi _{0}\right) ^{\ast
}\omega ^{\prime }%
\end{array}%
,
\end{equation*}%
where
\begin{equation*}
\begin{array}{c}
\left( \left( \varphi ,\varphi _{0}\right) ^{\ast }\omega ^{\prime }\right)
\left( z_{1},...,z_{q}\right) =\omega ^{\prime }\left( \Gamma \left( \varphi
,\varphi _{0}\right) \left( z_{1}\right) ,...,\Gamma \left( \varphi ,\varphi
_{0}\right) \left( z_{q}\right) \right) ,%
\end{array}%
\end{equation*}%
for any $z_{1},...,z_{q}\in \Gamma \left( F,\nu ,N\right) .$

\newpage\noindent\textbf{Remark 3.2.5 }It is remarked that the $\mathbf{B}^{%
\mathbf{v}}$-morphism $\left( Th\circ \rho ,h\circ \eta \right) $ is a $%
\mathbf{GLA}$-morphism~of
\begin{equation*}
\left( \left( F,\nu ,N\right) ,\left[ ,\right] _{F,h},\left( \rho ,\eta
\right) \right)
\end{equation*}%
\emph{\ }source and
\begin{equation*}
\left( \left( TN,\tau _{N},N\right) ,\left[ ,\right] _{TN,Id_{N}},\left(
Id_{TN},Id_{N}\right) \right)
\end{equation*}
target.

Moreover, for any $\tilde{\imath}\in \overline{1,n}$, we obtain%
\begin{equation*}
\begin{array}{c}
\left( Th\circ \rho ,h\circ \eta \right) ^{\ast }\left( d\varkappa ^{\tilde{%
\imath}}\right) =d^{F}\varkappa ^{\tilde{\imath}},%
\end{array}%
\end{equation*}%
where $d$ is the exterior differentiation operator associated to the
exterior differential Lie $\mathcal{F}\left( N\right) $-algebra\emph{\ }%
\begin{equation*}
\begin{array}{c}
\left( \Lambda \left( TN,\tau _{N},N\right) ,+,\cdot ,\wedge \right) .%
\end{array}%
\end{equation*}

\noindent \textbf{Theorem 3.2.11 }\emph{If }$\left( \varphi ,\varphi
_{0}\right) $\emph{\ is a morphism of }%
\begin{equation*}
\left( \left( F,\nu ,N\right) ,\left[ ,\right] _{F,h},\left( \rho ,\eta
\right) \right)
\end{equation*}%
\emph{\ source and }%
\begin{equation*}
\left( \left( F^{\prime },\nu ^{\prime },N^{\prime }\right) ,\left[ ,\right]
_{F^{\prime },h^{\prime }},\left( \rho ^{\prime },\eta ^{\prime }\right)
\right)
\end{equation*}%
\emph{target, then the following affirmations are satisfied:}\medskip

\noindent1. \emph{For any }$\omega ^{\prime }\in \Lambda ^{q}\left(
F^{\prime },\nu ^{\prime },N^{\prime }\right) $\emph{\ and }$\theta ^{\prime
}\in \Lambda ^{r}\left( F^{\prime },\nu ^{\prime },N^{\prime }\right) $\emph{%
\ we obtain}%
\begin{equation*}
\begin{array}{c}
\left( \varphi ,\varphi _{0}\right) ^{\ast }\left( \omega ^{\prime }\wedge
\theta ^{\prime }\right) =\left( \varphi ,\varphi _{0}\right) ^{\ast }\omega
^{\prime }\wedge \left( \varphi ,\varphi _{0}\right) ^{\ast }\theta ^{\prime
}.%
\end{array}%
\leqno(3.2.13)
\end{equation*}

\noindent2.\emph{\ For any }$z\in \Gamma \left( F,\nu ,N\right) $\emph{\ and
}$\omega ^{\prime }\in \Lambda ^{q}\left( F^{\prime },\nu ^{\prime
},N^{\prime }\right) $\emph{\ we obtain}%
\begin{equation*}
\begin{array}{c}
i_{z}\left( \left( \varphi ,\varphi _{0}\right) ^{\ast }\omega ^{\prime
}\right) =\left( \varphi ,\varphi _{0}\right) ^{\ast }\left( i_{\varphi
\left( z\right) }\omega ^{\prime }\right) .%
\end{array}%
\leqno(3.2.14)
\end{equation*}

\noindent3. \emph{If }$N=N^{\prime }$ \emph{and }%
\begin{equation*}
\left( Th\circ \rho ,h\circ \eta \right) =\left( Th^{\prime }\circ \rho
^{\prime },h^{\prime }\circ \eta ^{\prime }\right) \circ \left( \varphi
,\varphi _{0}\right) ,
\end{equation*}%
\emph{then we obtain}
\begin{equation*}
\begin{array}{c}
\left( \varphi ,\varphi _{0}\right) ^{\ast }\circ d^{F^{\prime }}=d^{F}\circ
\left( \varphi ,\varphi _{0}\right) ^{\ast }.%
\end{array}%
\leqno(3.2.15)
\end{equation*}

\bigskip\noindent\textit{Proof.}\smallskip

1. Let $\omega ^{\prime }\in \Lambda ^{q}\left( F^{\prime },\nu ^{\prime
},N^{\prime }\right) $ and $\theta ^{\prime }\in \Lambda ^{r}\left(
F^{\prime },\nu ^{\prime },N^{\prime }\right) $ be arbitrary. Since
\begin{equation*}
\begin{array}{l}
\displaystyle\left( \varphi ,\varphi _{0}\right) ^{\ast }\left( \omega
^{\prime }\wedge \theta ^{\prime }\right) \left( z_{1},...,z_{q+r}\right)
=\left( \omega ^{\prime }\wedge \theta ^{\prime }\right) \left( \Gamma
\left( \varphi ,\varphi _{0}\right) z_{1},...,\Gamma \left( \varphi ,\varphi
_{0}\right) z_{q+r}\right) \vspace*{1mm} \\
\qquad\displaystyle=\frac{1}{\left( q+r\right) !}\underset{\sigma \in \Sigma
_{q+r}}{\tsum }sgn\left( \sigma \right) \cdot \omega ^{\prime }\left( \Gamma
\left( \varphi ,\varphi _{0}\right) z_{1},...,\Gamma \left( \varphi ,\varphi
_{0}\right) z_{q}\right) \vspace*{1mm} \\
\hfill\cdot \theta ^{\prime }\left( \Gamma \left( \varphi ,\varphi
_{0}\right) z_{q+1},...,\Gamma \left( \varphi ,\varphi _{0}\right)
z_{q+r}\right) \vspace*{1mm} \\
\qquad\displaystyle=\frac{1}{\left( q+r\right) !}\underset{\sigma \in \Sigma
_{q+r}}{\tsum }sgn\left( \sigma \right) \cdot \left( \varphi ,\varphi
_{0}\right) ^{\ast }\omega ^{\prime }\left( z_{1},...,z_{q}\right) \left(
\varphi ,\varphi _{0}\right) ^{\ast }\theta ^{\prime }\left(
z_{q+1},...,z_{q+r}\right) \vspace*{1mm} \\
\qquad\displaystyle=\left( \left( \varphi ,\varphi _{0}\right) ^{\ast
}\omega ^{\prime }\wedge \left( \varphi ,\varphi _{0}\right) ^{\ast }\theta
^{\prime }\right) \left( z_{1},...,z_{q+r}\right) ,%
\end{array}%
\end{equation*}%
for any $z_{1},...,z_{q+r}\in \Gamma \left( F,\nu ,N\right) $, it results
the conclusion of affirmation 1.\medskip

2. Let $z\in \Gamma \left( F,\nu ,N\right) $\emph{\ }and\emph{\ }$\omega
^{\prime }\in \Lambda ^{q}\left( F^{\prime },\nu ^{\prime },N^{\prime
}\right) $ be arbitrary. Since
\begin{equation*}
\begin{array}{ll}
i_{z}\left( \left( \varphi ,\varphi _{0}\right) ^{\ast }\omega ^{\prime
}\right) \left( z_{2},...,z_{q}\right) & \displaystyle =\omega ^{\prime
}\left( \Gamma \left( \varphi ,\varphi _{0}\right) z,\Gamma \left( \varphi
,\varphi _{0}\right) z_{2},...,\Gamma \left( \varphi ,\varphi _{0}\right)
z_{q}\right) \vspace*{1mm} \\
& \displaystyle=i_{\Gamma \left( \varphi ,\varphi _{0}\right) z}\omega
^{\prime }\left( \Gamma \left( \varphi ,\varphi _{0}\right) z_{2},...,\Gamma
\left( \varphi ,\varphi _{0}\right) z_{q}\right) \vspace*{1mm} \\
& \displaystyle=\left( \varphi ,\varphi _{0}\right) ^{\ast }\left( i_{\Gamma
\left( \varphi ,\varphi _{0}\right) z}\omega ^{\prime }\right) \left(
z_{2},...,z_{q}\right) ,%
\end{array}%
\end{equation*}%
for any $z_{2},...,z_{q}\in \Gamma \left( F,\nu ,N\right) $, it results the
conclusion of affirmation 2$.$\smallskip

3. Let $\omega ^{\prime }\in \Lambda ^{q}\left( F^{\prime },\nu ^{\prime
},N^{\prime }\right) $ and $z_{0},...,z_{q}\in \Gamma \left( F,\nu ,N\right)
$ be arbitrary. Since
\begin{equation*}
\begin{array}{l}
\left( \left( \varphi ,\varphi _{0}\right) ^{\ast }d^{F^{\prime }}\omega
^{\prime }\right) \left( z_{0},...,z_{q}\right) =\left( d^{F^{\prime
}}\omega ^{\prime }\right) \left( \Gamma \left( \varphi ,\varphi _{0}\right)
z_{0},...,\Gamma \left( \varphi ,\varphi _{0}\right) z_{q}\right) \vspace*{%
1mm} \\
=\overset{q}{\underset{i=0}{\tsum }}\left( -1\right) ^{i}\Gamma \left(
Th^{\prime }\circ \rho ^{\prime },h^{\prime }\circ \eta ^{\prime }\right)
\left( \Gamma \left( \varphi ,\varphi _{0}\right) z_{i}\right) \vspace*{1mm}
\\
\hfill\cdot \omega ^{\prime }\left( \left( \Gamma \left( \varphi ,\varphi
_{0}\right) z_{0},\Gamma \left( \varphi ,\varphi _{0}\right) z_{1},...,%
\widehat{\Gamma \left( \varphi ,\varphi _{0}\right) z_{i}},...,\Gamma \left(
\varphi ,\varphi _{0}\right) z_{q}\right) \right) \vspace*{1mm} \\
+\underset{0\leq i<j}{\tsum }\left( -1\right) ^{i+j}\cdot \omega ^{\prime
}\left( \Gamma \left( \varphi ,\varphi _{0}\right) \left[ z_{i},z_{j}\right]
_{F},\Gamma \left( \varphi ,\varphi _{0}\right) z_{0},\Gamma \left( \varphi
,\varphi _{0}\right) z_{1},...,\right. \vspace*{1mm} \\
\hfill\left. \cdot\widehat{\Gamma \left( \varphi ,\varphi _{0}\right) z_{i}}%
,...,\widehat{\Gamma \left( \varphi ,\varphi _{0}\right) z_{j}},...,\Gamma
\left( \varphi ,\varphi _{0}\right) z_{q}\right)%
\end{array}%
\end{equation*}%
and
\begin{equation*}
\begin{array}{l}
d^{F}\left( \left( \varphi ,\varphi _{0}\right) ^{\ast }\omega ^{\prime
}\right) \left( z_{0},...,z_{q}\right) \vspace*{1mm} \\
=\overset{q}{\underset{i=0}{\tsum }}\left( -1\right) ^{i}\Gamma \left(
Th\circ \rho ,h\circ \eta \right) \left( z_{i}\right) \cdot \left( \left(
\varphi ,\varphi _{0}\right) ^{\ast }\omega ^{\prime }\right) \left(
z_{0},...,\widehat{z_{i}},...,z_{q}\right) \vspace*{1mm} \\
+\underset{0\leq i<j}{\tsum }\left( -1\right) ^{i+j}\cdot \left( \left(
\varphi ,\varphi _{0}\right) ^{\ast }\omega ^{\prime }\right) \left( \left[
z_{i},z_{j}\right] _{F,h},z_{0},...,\widehat{z_{i}},...,\widehat{z_{j}}%
,...,z_{q}\right)\vspace*{1mm} \\
=\overset{q}{\underset{i=0}{\tsum }}\left( -1\right) ^{i}\Gamma \left(
Th\circ \rho ,h\circ \eta \right) \left( z_{i}\right) \cdot \omega ^{\prime
}\left( \Gamma \left( \varphi ,\varphi _{0}\right) z_{0},...,\widehat{\Gamma
\left( \varphi ,\varphi _{0}\right) z_{i}},...,\Gamma \left( \varphi
,\varphi _{0}\right) z_{q}\right) \vspace*{1mm} \\
+\underset{0\leq i<j}{\tsum }\left( -1\right) ^{i+j}\cdot \omega ^{\prime
}\left( \Gamma \left( \varphi ,\varphi _{0}\right) \left[ z_{i},z_{j}\right]
_{F,h},\Gamma \left( \varphi ,\varphi _{0}\right) z_{0},\Gamma \left(
\varphi ,\varphi _{0}\right) z_{1},...,\right. \vspace*{1mm} \\
\left. \widehat{\Gamma \left( \varphi ,\varphi _{0}\right) z_{i}},...,%
\widehat{\Gamma \left( \varphi ,\varphi _{0}\right) z_{j}},...,\Gamma \left(
\varphi ,\varphi _{0}\right) z_{q}\right)%
\end{array}%
\end{equation*}%
it results the conclusion of affirmation 3. \hfill \emph{q.e.d.}

\bigskip\noindent\textbf{Definition 3.2.9} For any $q\in \overline{1,n}$ we
define%
\begin{equation*}
\begin{array}{c}
\mathcal{Z}^{q}\left( F,\nu ,N\right) =\left\{ \omega \in \Lambda ^{q}\left(
F,\nu ,N\right) :d\omega =0\right\} ,%
\end{array}%
\end{equation*}%
the set of \emph{closed differential exterior }$q$\emph{-forms} and
\begin{equation*}
\begin{array}{c}
\mathcal{B}^{q}\left( F,\nu ,N\right) =\left\{ \omega \in \Lambda ^{q}\left(
F,\nu ,N\right) :\exists \eta \in \Lambda ^{q-1}\left( F,\nu ,N\right)
~|~d\eta =\omega \right\} ,%
\end{array}%
\end{equation*}%
the set of \emph{exact differential exterior }$q$\emph{-forms}.

\newpage

\subsubsection{Exterior Differential Systems}

\qquad \qquad \qquad %
%
%
%
%
%\end{equation*}

We consider a generalized Lie algebroid $\left( \left( F,\nu ,N\right) ,%
\left[ ,\right] _{F,h},\left( \rho ,\eta \right) \right) $ given by the
diagrams:%
\begin{equation*}
\begin{array}{ccccccc}
&  & F & ^{\underrightarrow{~\ \rho ~\ }} & TM & ^{\underrightarrow{~\ Th~\ }%
} & TN \\
&  & ~\downarrow \nu &  & ~\ \ \ \downarrow \tau _{M} &  & ~\ \ \ \downarrow
\tau _{N} \\
M & ^{\underrightarrow{~\ h~\ }} & N & ^{\underrightarrow{~\ \eta ~\ }} & M
& ^{\underrightarrow{~\ h~\ }} & N%
\end{array}%
\end{equation*}

Let $\left( \left( h^{\ast }F,h^{\ast }\nu ,M\right) ,\left[ ,\right]
_{h^{\ast }F},\left( \overset{h^{\ast }F}{\rho },Id_{M}\right) \right) $ be
the pull-back Lie algebroid.

\textbf{Definition 3.2.1.1 }Any ideal $\left( \mathcal{I},+,\cdot \right) $
of the exterior differential algebra of the pull-back Lie algebroid $\left(
\left( h^{\ast }F,h^{\ast }\nu ,M\right) ,\left[ ,\right] _{h^{\ast
}F},\left( \overset{h^{\ast }F}{\rho },Id_{M}\right) \right) $ closed under
differentiation operator $d^{h^{\ast }F},$ namely $d^{h^{\ast }F}\mathcal{%
I\subseteq I},$ will be called \emph{differential ideal of the generalized
Lie algebroid }$\left( \left( F,\nu ,N\right) ,\left[ ,\right] _{F,h},\left(
\rho ,\eta \right) \right) .$

In particular, if $h=Id_{N}=\eta $, then any ideal $\left( \mathcal{I}%
,+,\cdot \right) $ of the exterior differential algebra of the Lie algebroid
$\left( \left( F,\nu ,N\right) ,\left[ ,\right] _{F},\left( \rho
,Id_{M}\right) \right) $ closed under differentiation operator $d^{F}$ $,$
namely $d^{F}\mathcal{I\subseteq I},$ will be called \emph{differential
ideal of the Lie algebroid }$\left( \left( F,\nu ,N\right) ,\left[ ,\right]
_{F},\left( \rho ,Id_{M}\right) \right) .$

\textbf{Definition 3.2.1.2 }Let $\left( \mathcal{I},+,\cdot \right) $ be a
differential ideal of the generalized Lie algebroid $\left( \left( F,\nu
,N\right) ,\left[ ,\right] _{F,h},\left( \rho ,\eta \right) \right) $ or of
the Lie algebroid $\left( \left( F,\nu ,N\right) ,\left[ ,\right]
_{F},\left( \rho ,Id_{M}\right) \right) $ respectively$.$

If it exists an IDS $\left( E,\pi ,M\right) $ such that for all $k\in
\mathbb{N}^{\ast }$ and $\omega \in \mathcal{I}\cap \Lambda ^{k}\left(
h^{\ast }F,h^{\ast }\nu ,M\right) $ we have $\omega \left(
u_{1},...,u_{k}\right) =0,$ for any $u_{1},...,u_{k}\in \Gamma \left( E,\pi
,M\right) ,$ then we will say that $\left( \mathcal{I},+,\cdot \right) $%
\emph{\ is an exterior differential system (EDS) of the generalized Lie
algebroid }$\left( \left( F,\nu ,N\right) ,\left[ ,\right] _{F,h},\left(
\rho ,\eta \right) \right) .$

In particular, if $h=Id_{N}=\eta $ and it exists an IDS $\left( E,\pi
,M\right) $ such that for all $k\in \mathbb{N}^{\ast }$ and $\omega \in
\mathcal{I}\cap \Lambda ^{k}\left( F,\nu ,M\right) $ we have $\omega \left(
u_{1},...,u_{k}\right) =0,$ for any $u_{1},...,u_{k}\in \Gamma \left( E,\pi
,M\right) ,$ then we will say that $\left( \mathcal{I},+,\cdot \right) $%
\emph{\ is an exterior differential system (EDS) of the Lie algebroid }$%
\left( \left( F,\nu ,N\right) ,\left[ ,\right] _{F},\left( \rho
,Id_{N}\right) \right) .$

\textbf{Theorem 3.2.1.1} \emph{The IDS }$\left( E,\pi ,M\right) $\emph{\ of
the generalized Lie algebroid }$\left( \left( F,\nu ,N\right) ,\left[ ,%
\right] _{F,h},\left( \rho ,\eta \right) \right) $\emph{\ is involutive, if
and only if the ideal generated by the }$\mathcal{F}\left( M\right) $\emph{%
-submodule }$\left( \Gamma \left( E^{\perp },\pi ^{\perp },M\right) ,+,\cdot
\right) $\emph{\ is an EDS of the generalized Lie algebroid }$\left( \left(
F,\nu ,N\right) ,\left[ ,\right] _{F,h},\left( \rho ,\eta \right) \right) .$

\textbf{Proof: }Let $\left( E,\pi ,M\right) $ be an involutive IDS of the
generalized Lie algebroid $\left( \left( F,\nu ,N\right) ,\left[ ,\right]
_{F,h},\left( \rho ,\eta \right) \right) .$

Let $\left\{ \Theta ^{r+1},...,\Theta ^{p}\right\} $ be a base for the $%
\mathcal{F}\left( M\right) $-submodule $\left( \Gamma \left( E^{\perp },\pi
^{\perp },M\right) ,+,\cdot \right) .$

We know that
\begin{equation*}
\mathcal{I}\left( \Gamma \left( E^{\perp },\pi ^{\perp },M\right) \right)
=\cup _{q\in \mathbb{N}}\left\{ \Omega _{\alpha }\wedge \Theta ^{\alpha
},~\left\{ \Omega _{r+1},...,\Omega _{p}\right\} \subset \Lambda ^{q}\left(
h^{\ast }F,h^{\ast }\nu ,M\right) \right\} .
\end{equation*}

Let
\begin{equation*}
S=\left\{ q\in \mathbb{N}:d^{h^{\ast }F}\left( \Omega _{\alpha }\wedge
\Theta ^{\alpha }\right) \in \mathcal{I}\left( \Gamma \left( E^{\perp },\pi
^{\perp },M\right) \right) ,~\forall \left\{ \Omega _{r+1},...,\Omega
_{p}\right\} \subset \Lambda ^{q}\left( h^{\ast }F,h^{\ast }\nu ,M\right)
\right\} .
\end{equation*}

Let $\left\{ \Omega _{r+1},...,\Omega _{p}\right\} \subset \Lambda
^{0}\left( h^{\ast }F,h^{\ast }\nu ,M\right) $ be arbitrary.

Using the \emph{Theorem 3.2.10}, we obtain
\begin{equation*}
\begin{array}{ccl}
d^{h^{\ast }F}\left( \Omega _{\alpha }\wedge \Theta ^{\alpha }\right) & = &
d^{h^{\ast }F}\Omega _{\alpha }\wedge \Theta ^{\alpha }+\left( -1\right)
^{0}\Omega _{\alpha }\wedge d^{h^{\ast }F}\Theta ^{\alpha } \\
& = & \left( d^{h^{\ast }F}\Omega _{\alpha }+\Omega _{\beta }\cdot \Omega
_{\alpha }^{\beta }\right) \wedge \Theta ^{\alpha }.%
\end{array}%
\end{equation*}

As
\begin{equation*}
d^{h^{\ast }F}\Omega _{\beta }+\Omega _{\alpha }\cdot \Omega _{\beta
}^{\alpha }\in \Lambda ^{1}\left( h^{\ast }F,h^{\ast }\nu ,M\right)
\end{equation*}
it results that
\begin{equation*}
d^{h^{\ast }F}\left( \Omega _{\beta }\wedge \Theta ^{\beta }\right) \in
\mathcal{I}\left( \Gamma \left( E^{\perp },\pi ^{\perp },M\right) \right)
\end{equation*}

Therefore,
\begin{equation*}
0\in S.\leqno\left( 1\right)
\end{equation*}

In the following, we prove that
\begin{equation*}
q\in S\Longrightarrow q+1\in S.\leqno\left( 2\right)
\end{equation*}

Let $\left\{ \Omega _{r+1},...,\Omega _{p}\right\} \subset \Lambda
^{q+1}\left( h^{\ast }F,h^{\ast }\nu ,M\right) $ be arbitrary.

Using the \emph{Theorem 3.2.10}, we obtain
\begin{equation*}
\begin{array}{ccl}
d^{h^{\ast }F}\left( \Omega _{\alpha }\wedge \Theta ^{\alpha }\right) & = &
d^{h^{\ast }F}\Omega _{\alpha }\wedge \Theta ^{\alpha }+\left( -1\right)
^{q+1}\Omega _{\beta }\wedge d^{h^{\ast }F}\Theta ^{\beta } \\
& = & \left( d^{h^{\ast }F}\Omega _{\alpha }+\left( -1\right) ^{q+1}\Omega
_{\beta }\wedge \Omega _{\alpha }^{\beta }\right) \wedge \Theta ^{\alpha }.%
\end{array}%
\end{equation*}

As
\begin{equation*}
d^{h^{\ast }F}\Omega _{\alpha }+\left( -1\right) ^{q+1}\Omega _{\beta
}\wedge \Omega _{\alpha }^{\beta }\in \Lambda ^{q+2}\left( h^{\ast
}F,h^{\ast }\nu ,M\right)
\end{equation*}%
it results that
\begin{equation*}
d^{h^{\ast }F}\left( \Omega _{\beta }\wedge \Theta ^{\beta }\right) \in
\mathcal{I}\left( \Gamma \left( E^{\perp },\pi ^{\perp },M\right) \right)
\end{equation*}

Therefore,
\begin{equation*}
q+1\in S.~\
\end{equation*}

Using the \textbf{Peano's Axiom} and the affirmations $\left( 1\right) $ and
$\left( 2\right) ,$ it results that $S=\mathbb{N}.$

Therefore,
\begin{equation*}
d^{h^{\ast }F}\mathcal{I}\left( \Gamma \left( E^{\perp },\pi ^{\perp
},M\right) \right) \subseteq \mathcal{I}\left( \Gamma \left( E^{\perp },\pi
^{\perp },M\right) \right) .
\end{equation*}

Conversely, let $\left( E,\pi ,M\right) $ be an IDS of the generalized Lie
algebroid $\left( \left( F,\nu ,N\right) ,\left[ ,\right] _{F,h},\left( \rho
,\eta \right) \right) $ such that the $\mathcal{F}\left( M\right) $%
-submodule $\left( \mathcal{I}\left( \Gamma \left( E^{\perp },\pi ^{\perp
},M\right) \right) ,+,\cdot \right) $ is an EDS of the generalized Lie
algebroid $\left( \left( F,\nu ,N\right) ,\left[ ,\right] _{F,h},\left( \rho
,\eta \right) \right) .$ We have:
\begin{equation*}
d^{h^{\ast }F}\mathcal{I}\left( \Gamma \left( E^{\perp },\pi ^{\perp
},M\right) \right) \subseteq \mathcal{I}\left( \Gamma \left( E^{\perp },\pi
^{\perp },M\right) \right) .\leqno\left( 3\right)
\end{equation*}

Let $\left\{ \Theta ^{r+1},...,\Theta ^{p}\right\} $ be a base for the $%
\mathcal{F}\left( M\right) $-submodule $\left( \Gamma \left( E^{\perp },\pi
^{\perp },M\right) ,+,\cdot \right) .$

Using the affirmation $\left( 3\right) $, it results that it exists
\begin{equation*}
\Omega _{\beta }^{\alpha }\in \Lambda ^{1}\left( h^{\ast }F,h^{\ast }\nu
,M\right) ,~\alpha ,\beta \in \overline{r+1,p}
\end{equation*}%
such that
\begin{equation*}
d^{h^{\ast }F}\Theta ^{\alpha }=\Sigma _{\beta \in \overline{r+1,p}}\Omega
_{\beta }^{\alpha }\wedge \Theta ^{\beta }\in \mathcal{I}\left( \Gamma
\left( E^{\perp },\pi ^{\perp },M\right) \right) .
\end{equation*}

Using the \emph{Theorem 3.2.10}, it results that $\left( E,\pi ,M\right) $
is an involutive IDS of the generalized Lie algebroid $\left( \left( F,\nu
,N\right) ,\left[ ,\right] _{F,h},\left( \rho ,\eta \right) \right) .$\hfill
\emph{q.e.d.}\medskip

\bigskip

\subsection{The generalized tangent bundle}

We consider the following diagram:
\begin{equation*}
\begin{array}{c}
\xymatrix{E\ar[d]_\pi&\left( F,\left[ ,\right] _{F,h},\left( \rho ,\eta
\right) \right)\ar[d]^\nu\\ M\ar[r]^h&N}%
\end{array}%
\leqno(3.3.1)
\end{equation*}%
where $\left( E,\pi ,M\right) $ is a fiber bundle and $\left( \left( F,\nu
,N\right) ,\left[ ,\right] _{F,h},\left( \rho ,\eta \right) \right) $ is a
generalized Lie algebroid.

We assume that the $r$-dimensional manifold $\mathbf{V}$ is the type fibre
and the Lie group $\left( \mathbf{G},\cdot \right) $ is the structure group
for the fiber bundle $\left( E,\pi ,M\right) .$

We take $\left( x^{i},y^{a}\right) $ as canonical local coordinates on $%
\left( E,\pi ,M\right) ,$ where $i\in \overline{1,m}$ and $a\in \overline{1,r%
}.$

Let
\begin{equation*}
\left( x^{i},y^{a}\right) \longrightarrow \left( x^{i%
%TCIMACRO{\U{b4}}%
%BeginExpansion
{\acute{}}%
%EndExpansion
}\left( x^{i}\right) ,y^{a%
%TCIMACRO{\U{b4}}%
%BeginExpansion
{\acute{}}%
%EndExpansion
}\left( x^{i},y^{a}\right) \right)
\end{equation*}%
be a change of coordinates on $\left( E,\pi ,M\right) $. Then the
coordinates $y^{a}$ change to $y^{a%
%TCIMACRO{\U{b4}}%
%BeginExpansion
{\acute{}}%
%EndExpansion
}$ by the rule:
\begin{equation*}
\begin{array}{c}
y^{a%
%TCIMACRO{\U{b4}}%
%BeginExpansion
{\acute{}}%
%EndExpansion
}=\displaystyle\frac{\partial y^{a%
%TCIMACRO{\U{b4}}%
%BeginExpansion
{\acute{}}%
%EndExpansion
}}{\partial y^{a}}y^{a}.%
\end{array}%
\leqno(3.3.2)
\end{equation*}
In particular, if $\left( E,\pi ,M\right) $ is vector bundle, then the
coordinates $y^{a}$ change to $y^{a%
%TCIMACRO{\U{b4}}%
%BeginExpansion
{\acute{}}%
%EndExpansion
}$ by the rule:
\begin{equation*}
\begin{array}{c}
y^{a%
%TCIMACRO{\U{b4}}%
%BeginExpansion
{\acute{}}%
%EndExpansion
}=M_{a}^{a%
%TCIMACRO{\U{b4}}%
%BeginExpansion
{\acute{}}%
%EndExpansion
}y^{a}.%
\end{array}%
\leqno(3.3.2^{\prime })
\end{equation*}

Let
\begin{equation*}
\begin{array}{c}
\left( h^{\ast }F,h^{\ast }\nu ,M\right) ,\left[ ,\right] _{h^{\ast
}F},\left( \overset{h^{\ast }F}{\rho },Id_{M}\right)%
\end{array}%
\end{equation*}%
be the pull-back Lie algebroid of the generalized Lie algebroid
\begin{equation*}
\begin{array}{c}
\left( \left( F,\nu ,N\right) ,\left[ ,\right] _{F,h},\left( \rho ,\eta
\right) \right) .%
\end{array}%
\end{equation*}

Let
\begin{equation*}
\begin{array}{c}
\left( \pi ^{\ast }\left( h^{\ast }F\right) ,\pi ^{\ast }\left( h^{\ast }\nu
\right) ,E\right) ,\left[ ,\right] _{\pi ^{\ast }\left( h^{\ast }F\right)
},\left( \overset{\pi ^{\ast }\left( h^{\ast }F\right) }{\rho },Id_{E}\right)%
\end{array}%
\end{equation*}%
be the pull-back Lie algebroid of the Lie algebroid
\begin{equation*}
\begin{array}{c}
\left( h^{\ast }F,h^{\ast }\nu ,M\right) ,\left[ ,\right] _{h^{\ast
}F},\left( \overset{h^{\ast }F}{\rho },Id_{M}\right) .%
\end{array}%
\end{equation*}

If
\begin{equation*}
z=z^{\alpha }t_{\alpha }\in \Gamma \left( F,\nu ,N\right) ,
\end{equation*}%
then, using the vector fibred $\left( m+r\right) +p$-structure $\left[
\widetilde{\mathcal{AF}}_{\pi ^{\ast }\left( h^{\ast }F\right) }\right] ,$\
we obtain the section%
\begin{equation*}
\tilde{Z}=\left( z^{\alpha }\circ h\circ \pi \right) \tilde{T}_{\alpha }\in
\Gamma \left( \pi ^{\ast }\left( h^{\ast }F\right) ,\pi ^{\ast }\left(
h^{\ast }\nu \right) ,E\right)
\end{equation*}%
such that%
\begin{equation*}
\tilde{Z}\left( u_{x}\right) =z\left( h\left( x\right) \right) ,
\end{equation*}%
for any $u_{x}\in \pi ^{-1}\left( U{\cap h}^{-1}V\right) .$

The set $\left\{ \tilde{T}_{\alpha },~\alpha \in \overline{1,p}\right\} $ is
a base for the module of sections%
\begin{equation*}
\begin{array}{c}
\left( \Gamma \left( \pi ^{\ast }\left( h^{\ast }F\right) ,\pi ^{\ast
}\left( h^{\ast }\nu \right) ,E\right) ,+,\cdot \right) .%
\end{array}%
\end{equation*}

Since $t_{\alpha
%TCIMACRO{\U{b4}}%
%BeginExpansion
{\acute{}}%
%EndExpansion
}=\Lambda _{\alpha
%TCIMACRO{\U{b4}}%
%BeginExpansion
{\acute{}}%
%EndExpansion
}^{\alpha }t_{\alpha }$, it results that
\begin{equation*}
\begin{array}{c}
\tilde{T}_{\alpha
%TCIMACRO{\U{b4}}%
%BeginExpansion
{\acute{}}%
%EndExpansion
}=\Lambda _{\alpha
%TCIMACRO{\U{b4}}%
%BeginExpansion
{\acute{}}%
%EndExpansion
}^{\alpha }\circ \left( h\circ \pi \right) \tilde{T}_{\alpha }.%
\end{array}%
\leqno(3.3.3)
\end{equation*}%
Therefore,
\begin{equation*}
\begin{array}{c}
\left\Vert \Lambda _{\alpha }^{\alpha
%TCIMACRO{\U{b4}}%
%BeginExpansion
{\acute{}}%
%EndExpansion
}\circ \left( h\circ \pi \right) \right\Vert%
\end{array}%
\leqno(3.3.4)
\end{equation*}%
is the matrix of coordinate transformation on $\left( \pi ^{\ast }\left(
h^{\ast }F\right) ,\pi ^{\ast }\left( h^{\ast }\nu \right) ,E\right) .$

Let $\Big({\overset{\pi ^{\ast }\left( h^{\ast }F\right) }{\rho }},Id_{E}%
\Big)$ be the $\mathbf{B}^{\mathbf{v}}$-morphism of $\left( \pi ^{\ast
}\left( h^{\ast }F\right) ,\pi ^{\ast }\left( h^{\ast }\nu \right) ,E\right)
$\ source and $\left( TE,\tau _{E},E\right) $\ target, where%
\begin{equation*}
\begin{array}{rcl}
\pi ^{\ast }\left( h^{\ast }F\right) & ^{\underrightarrow{\overset{\pi
^{\ast }\left( h^{\ast }F\right) }{\rho }}} & TE\vspace*{1mm} \\
\tilde{Z}^{\alpha }\tilde{T}_{\alpha }\left( u_{x}\right) & \longmapsto &
\tilde{Z}^{\alpha }\cdot \left( \rho _{\alpha }^{i}\circ h\circ \pi \right)
\cdot \displaystyle\frac{\partial }{\partial x^{i}}\left( u_{x}\right) .%
\end{array}%
\leqno(3.3.5)
\end{equation*}%
Using the operation
\begin{equation*}
\begin{array}{ccc}
\Gamma \left( \pi ^{\ast }\left( h^{\ast }F\right) ,\pi ^{\ast }\left(
h^{\ast }\nu \right) ,E\right) ^{2} & ^{\underrightarrow{~\ \ \left[ ,\right]
_{\pi ^{\ast }\left( h^{\ast }F\right) }~\ \ }} & \Gamma \left( \pi ^{\ast
}\left( h^{\ast }F\right) ,\pi ^{\ast }\left( h^{\ast }\nu \right) ,E\right)%
\end{array}%
\end{equation*}%
defined by%
\begin{equation*}
\begin{array}{ll}
\left[ \tilde{T}_{\alpha },\tilde{T}_{\beta }\right] _{\pi ^{\ast }\left(
h^{\ast }F\right) } & =\left( L_{\alpha \beta }^{\gamma }\circ h\circ \pi
\right) \tilde{T}_{\gamma },\vspace*{1mm} \\
\left[ \tilde{T}_{\alpha },f\tilde{T}_{\beta }\right] _{\pi ^{\ast }\left(
h^{\ast }F\right) } & =f\left( L_{\alpha \beta }^{\gamma }\circ h\circ \pi
\right) \tilde{T}_{\gamma }+\left( \rho _{\alpha }^{i}\circ h\circ \pi
\right) \displaystyle\frac{\partial f}{\partial x^{i}}\tilde{T}_{\beta },%
\vspace*{1mm} \\
\left[ f\tilde{T}_{\alpha },\tilde{T}_{\beta }\right] _{\pi ^{\ast }\left(
h^{\ast }F\right) } & =-\left[ \tilde{T}_{\beta },f\tilde{T}_{\alpha }\right]
_{\pi ^{\ast }\left( h^{\ast }F\right) },%
\end{array}%
\leqno(3.3.6)
\end{equation*}%
for any $f\in \mathcal{F}\left( E\right) ,$ we obtain the following

\bigskip \noindent \textbf{Theorem 3.3.1 }\emph{The couple }%
\begin{equation*}
\begin{array}{c}
\left( \left[ ,\right] _{\pi ^{\ast }\left( h^{\ast }F\right) },\left(
\overset{\pi ^{\ast }\left( h^{\ast }F\right) }{\rho },Id_{E}\right) \right)%
\end{array}%
\end{equation*}%
\emph{\ is a Lie algebroid structure for the vector bundle }$\left( \pi
^{\ast }\left( h^{\ast }F\right) ,\pi ^{\ast }\left( h^{\ast }\nu \right)
,E\right) .$\newline

It is known that the tangent bundle $\left( TE,\tau _{E},E\right) $ is a
vector bundle with type fibre the real space $\left( \mathbb{R}%
^{m+r},+,\cdot \right) $ and structure group the Lie group $\mathbf{GL}%
\left( m+r,\mathbb{R}\right) .$

\bigskip\noindent\textbf{Theorem 3.3.2 }\emph{The set }%
\begin{equation*}
\begin{array}{c}
\pi ^{\ast }\left( h^{\ast }F\right) \oplus TE\,=\underset{u\in E}{\dbigcup }%
\pi ^{\ast }\left( h^{\ast }F\right) _{u}\oplus \left( TE\right) _{u}\,%
\end{array}%
\leqno(3.3.7)
\end{equation*}%
\emph{is the total space of a vector bundle with the base }$E$\emph{,
canonical projection denoted }$\overset{\oplus }{\pi },$\emph{\ type fibre
the real space} $\left( \mathbb{R}^{p+\left( m+r\right) },+,\cdot \right) $%
\emph{\ and structure group, a Lie subgroup of }$\left( \mathbf{GL}\left(
p+\left( m+r\right) ,\mathbb{R}\right) ,\cdot \right) .$\bigskip

Let
\begin{equation*}
\begin{array}{c}
\left( \displaystyle\frac{\partial }{\partial x^{i}},\frac{\partial }{%
\partial y^{a}}\right)%
\end{array}%
\end{equation*}
be the natural base for the sections Lie algebra $\left( \Gamma \left(
TE,\tau _{E},E\right) ,+,\cdot ,\left[ ,\right] _{TE}\right) .$

\bigskip \noindent \textbf{Remark 3.3.1 }The sections
\begin{equation*}
\begin{array}{c}
\left( \tilde{T}_{\alpha },\left( \displaystyle\frac{\partial }{\partial
x^{i}},\frac{\partial }{\partial y^{a}}\right) \right)%
\end{array}%
\leqno(3.3.8)
\end{equation*}%
determined the bases for the $\mathcal{F}\left( M\right) $ module $\left(
\Gamma \left( \pi ^{\ast }\left( h^{\ast }F\right) \oplus TE,\overset{\oplus
}{\pi },E\right) ,+,\cdot \right) .$

The matrix of coordinate transformation on
\begin{equation*}
\left( \pi ^{\ast }\left( h^{\ast }F\right) \oplus TE,\overset{\oplus }{\pi }%
,E\right)
\end{equation*}%
at a change of fibred charts is
\begin{equation*}
\left\Vert
\begin{array}{ccc}
\Lambda _{\alpha }^{\alpha
%TCIMACRO{\U{b4}}%
%BeginExpansion
{\acute{}}%
%EndExpansion
}\circ h\circ \pi & 0 & 0\vspace*{2mm} \\
0 & \displaystyle\frac{\partial x^{i%
%TCIMACRO{\U{b4}}%
%BeginExpansion
{\acute{}}%
%EndExpansion
}}{\partial x_{i}}\circ \pi & 0\vspace*{2mm} \\
0 & \displaystyle\frac{\partial y^{a%
%TCIMACRO{\U{b4}}%
%BeginExpansion
{\acute{}}%
%EndExpansion
}}{\partial x^{i}} & \displaystyle\frac{\partial y^{a%
%TCIMACRO{\U{b4}}%
%BeginExpansion
{\acute{}}%
%EndExpansion
}}{\partial y^{a}}%
\end{array}%
\right\Vert .\leqno(3.3.9)
\end{equation*}
In particular, if $\left( E,\pi ,M\right) $ is a vector bundle, then we
consider that the local coordinates on $\left( E,\pi ,M\right) $ is changed
by:
\begin{equation*}
\begin{array}{c}
\left( x^{i},y^{a}\right) \longrightarrow \left( x^{i%
%TCIMACRO{\U{b4}}%
%BeginExpansion
{\acute{}}%
%EndExpansion
}\left( x^{i}\right) ,y^{a%
%TCIMACRO{\U{b4}}%
%BeginExpansion
{\acute{}}%
%EndExpansion
}=M_{a}^{a%
%TCIMACRO{\U{b4}}%
%BeginExpansion
{\acute{}}%
%EndExpansion
}\left( x^{i}\right) y^{a}\right) .%
\end{array}%
\end{equation*}%
Then the matrix of coordinate transformation on
\begin{equation*}
\left( \pi ^{\ast }\left( h^{\ast }F\right) ^{\ast }F\oplus TE,\overset{%
\oplus }{\pi },E\right)
\end{equation*}%
at a change of fibred charts is
\begin{equation*}
\left\Vert
\begin{array}{ccc}
\Lambda _{\alpha }^{\alpha
%TCIMACRO{\U{b4}}%
%BeginExpansion
{\acute{}}%
%EndExpansion
}\circ h\circ \pi & 0 & 0\vspace*{2mm} \\
0 & \displaystyle\frac{\partial x^{i%
%TCIMACRO{\U{b4}}%
%BeginExpansion
{\acute{}}%
%EndExpansion
}}{\partial x_{i}}\circ \pi & 0\vspace*{2mm} \\
0 & \displaystyle\frac{\partial M_{a}^{a%
%TCIMACRO{\U{b4}}%
%BeginExpansion
{\acute{}}%
%EndExpansion
}\circ \pi }{\partial x^{i}}y^{a} & M_{a}^{a%
%TCIMACRO{\U{b4}}%
%BeginExpansion
{\acute{}}%
%EndExpansion
}\circ \pi%
\end{array}%
\right\Vert .\leqno(3.3.10)
\end{equation*}
For any sections%
\begin{equation*}
\begin{array}{c}
\tilde{Z}^{\alpha }\tilde{T}_{\alpha }\in \Gamma \left( \pi ^{\ast }\left(
h^{\ast }F\right) ,\pi ^{\ast }\left( h^{\ast }F\right) ,E\right)%
\end{array}%
\end{equation*}%
and%
\begin{equation*}
\begin{array}{c}
Y^{a}\displaystyle\frac{\partial }{\partial y^{a}}\in \Gamma \left( VTE,\tau
_{E},E\right)%
\end{array}%
\end{equation*}%
we construct the section%
\begin{equation*}
%
%
\end{equation*}
Since we have
\begin{equation*}
\begin{array}{c}
\tilde{Z}^{\alpha }\displaystyle\frac{\partial }{\partial \tilde{z}^{\alpha }%
}+Y^{a}\frac{\partial }{\partial \tilde{y}^{a}}=0 \\
\Updownarrow \\
\tilde{Z}^{\alpha }\tilde{T}_{\alpha }=0~\wedge \tilde{Z}^{\alpha }\left(
\rho _{\alpha }^{i}\circ h\circ \pi \right) \displaystyle\frac{\partial }{%
\partial x^{i}}+Y^{a}\frac{\partial }{\partial y^{a}}=0,%
\end{array}%
\end{equation*}%
it implies $\tilde{Z}^{\alpha }=0,~\alpha \in \overline{1,p}$ and $%
Y^{a}=0,~a\in \overline{1,r}.$

Therefore the sections $\displaystyle\frac{\partial }{\partial \tilde{z}^{1}}%
,...,\frac{\partial }{\partial \tilde{z}^{p}},\frac{\partial }{\partial
\tilde{y}^{1}},...,\frac{\partial }{\partial \tilde{y}^{r}}$ are linearly
independent.\smallskip

We consider the vector subbundle $\left( \left( \rho ,\eta \right) TE,\left(
\rho ,\eta \right) \tau _{E},E\right) $ of the vector bundle\break $\left(
\pi ^{\ast }\left( h^{\ast }F\right) \oplus TE,\overset{\oplus }{\pi }%
,E\right) ,$ for which the $\mathcal{F}\left( E\right) $-module of sections
is the $\mathcal{F}\left( E\right) $-submodule of $\left( \Gamma \left( \pi
^{\ast }\left( h^{\ast }F\right) \oplus TE,\overset{\oplus }{\pi },E\right)
,+,\cdot \right) ,$ generated by the family of sections $\left( \displaystyle%
\frac{\partial }{\partial \tilde{z}^{\alpha }},\frac{\partial }{\partial
\tilde{y}^{a}}\right) .$

The base sections
\begin{equation*}
\begin{array}{c}
\left( \displaystyle\frac{\partial }{\partial \tilde{z}^{\alpha }},\frac{%
\partial }{\partial \tilde{y}^{a}}\right) \overset{put}{=}\left( \tilde{%
\partial}_{\alpha },\overset{\cdot }{\tilde{\partial}}_{a}\right)%
\end{array}%
\leqno(3.3.11)
\end{equation*}%
will be called the \emph{natural }$\left( \rho ,\eta \right) $\emph{-base.}

The matrix of coordinate transformation on $\left( \left( \rho ,\eta \right)
TE,\left( \rho ,\eta \right) \tau _{E},E\right) $ at a change of fibred
charts is
\begin{equation*}
\left\Vert
%
%
\right\Vert .\leqno(3.3.12)
\end{equation*}
In particular, if $\left( E,\pi ,M\right) $ is a vector bundle, then the
matrix of coordinate transformation on $\left( \left( \rho ,\eta \right)
TE,\left( \rho ,\eta \right) \tau _{E},E\right) $ at a change of fibred
charts is
\begin{equation*}
\left\Vert
%
%
\right\Vert .\leqno(3.3.13)
\end{equation*}

Next, we consider the operation
\begin{equation*}
%
%
\leqno(3.3.14)
\end{equation*}%
for any $\left( \tilde{Z}_{1}^{\alpha }\displaystyle\frac{\partial }{%
\partial \tilde{z}^{\alpha }}+Y_{1}^{a}\frac{\partial }{\partial \tilde{y}%
^{a}}\right) $ and $\left( \tilde{Z}_{2}^{\beta }\displaystyle\frac{\partial
}{\partial \tilde{z}^{\beta }}+Y_{2}^{b}\frac{\partial }{\partial \tilde{y}%
^{b}}\right) .$\smallskip

Let $\left( \tilde{\rho},Id_{E}\right) $\ be the $\mathbf{B}^{\mathbf{v}}$%
-morphism of $\left( \left( \rho ,\eta \right) TE,\left( \rho ,\eta \right)
\tau _{E},E\right) $\ source and $\left( TE,\tau _{E},E\right) $\ target,
where
\begin{equation*}
%
\leqno(3.3.15)
\end{equation*}

\noindent \textbf{Lemma 3.3.1 }\emph{The operation }$\left[ ,\right]
_{\left( \rho ,\eta \right) TE}$\emph{\ is a Lie bracket, namely it
satisfies }%
\begin{equation*}
%
%
\leqno(4)
\end{equation*}
In general, after some calculations, we obtain
\begin{equation*}
%
%
\leqno(3.3.17)
\end{equation*}%
\emph{In particular, we obtain: }%
\begin{equation*}
%
%
\leqno(2)
\end{equation*}%
Moreover, we obtain
\begin{equation*}
%
\leqno(6)
\end{equation*}
In general, for any $\tilde{U},\tilde{V}\in \Gamma \left( \left( \rho ,\eta
\right) TE,\left( \rho ,\eta \right) \tau _{E},E\right) ,$ we have:
\begin{equation*}
%
%
\end{equation*}

\hfill\emph{q.e.d.}

\bigskip\noindent\textbf{Lemma 3.3.3 }\emph{We have the Jacobi identity: }%
\begin{equation*}
%
%
\leqno(3.3.19)
\end{equation*}%
\emph{for any }$\tilde{U},\tilde{V},\tilde{Z}\in \Gamma \left( \left( \rho
,\eta \right) TE,\left( \rho ,\eta \right) \tau _{E},E\right) .$

\bigskip\noindent\textit{Proof.} After some calculations, using the sections
of natural $\left( \rho ,\eta \right) $-base, we obtain the following Jacobi
identities:

\begin{itemize}
\item[(1)] $\displaystyle\left[ \frac{\partial }{\partial \tilde{z}^{\alpha }%
},\left[ \frac{\partial }{\partial \tilde{z}^{\beta }},\frac{\partial }{%
\partial \tilde{z}^{\gamma }}\right] _{\left( \rho ,\eta \right) TE}\right]
_{\left( \rho ,\eta \right) TE}+\left[ \frac{\partial }{\partial \tilde{z}%
^{\beta }},\left[ \frac{\partial }{\partial \tilde{z}^{\gamma }},\frac{%
\partial }{\partial \tilde{z}^{\alpha }}\right] _{\left( \rho ,\eta \right)
TE}\right] _{\left( \rho ,\eta \right) TE}\vspace*{1mm}\newline
\hfill +\left[ \frac{\partial }{\partial \tilde{z}^{\gamma }},\left[ \frac{%
\partial }{\partial \tilde{z}^{\alpha }},\frac{\partial }{\partial \tilde{z}%
^{\beta }}\right] _{\left( \rho ,\eta \right) TE}\right] _{\left( \rho ,\eta
\right) TE}=0_{\left( \rho ,\eta \right) TE},$

\item[(2)] $\displaystyle\left[ \frac{\partial }{\partial \tilde{y}^{a}},%
\left[ \frac{\partial }{\partial \tilde{y}^{b}},\frac{\partial }{\partial
\tilde{z}^{\gamma }}\right] _{\left( \rho ,\eta \right) TE}\right] _{\left(
\rho ,\eta \right) TE}+\left[ \frac{\partial }{\partial \tilde{y}^{b}},\left[
\frac{\partial }{\partial \tilde{z}^{\gamma }},\frac{\partial }{\partial
\tilde{y}^{a}}\right] _{\left( \rho ,\eta \right) TE}\right] _{\left( \rho
,\eta \right) TE}\vspace*{1mm}\newline
\hfill +\left[ \frac{\partial }{\partial \tilde{z}^{\gamma }},\left[ \frac{%
\partial }{\partial \tilde{y}^{a}},\frac{\partial }{\partial \tilde{y}^{b}}%
\right] _{\left( \rho ,\eta \right) TE}\right] _{\left( \rho ,\eta \right)
TE}=0_{\left( \rho ,\eta \right) TE},$

\item[(3)] $\displaystyle\left[ \frac{\partial }{\partial \tilde{y}^{a}},%
\left[ \frac{\partial }{\partial \tilde{y}^{b}},\frac{\partial }{\partial
\tilde{y}^{c}}\right] _{\left( \rho ,\eta \right) TE}\right] _{\left( \rho
,\eta \right) TE}+\left[ \frac{\partial }{\partial \tilde{y}^{b}},\left[
\frac{\partial }{\partial \tilde{y}^{c}},\frac{\partial }{\partial \tilde{y}%
^{a}}\right] _{\left( \rho ,\eta \right) TE}\right] _{\left( \rho ,\eta
\right) TE}\vspace*{1mm}\newline
\hfill +\left[ \frac{\partial }{\partial \tilde{y}^{c}},\left[ \frac{%
\partial }{\partial \tilde{y}^{a}},\frac{\partial }{\partial \tilde{y}^{b}}%
\right] _{\left( \rho ,\eta \right) TE}\right] _{\left( \rho ,\eta \right)
TE}=0_{\left( \rho ,\eta \right) TE}.$
\end{itemize}

After some calculations, we obtain the Jacobi identity
\begin{equation*}
\begin{array}{l}
\left[ \tilde{U},\left[ \tilde{V},\tilde{Z}\right] _{\left( \rho ,\eta
\right) TE}\right] _{\left( \rho ,\eta \right) TE}+\left[ \tilde{Z},\left[
\tilde{U},\tilde{V}\right] _{\left( \rho ,\eta \right) TE}\right] _{\left(
\rho ,\eta \right) TE}\vspace*{1mm} \\
\qquad\quad+\left[ \tilde{V},\left[ \tilde{Z},\tilde{U}\right] _{\left( \rho
,\eta \right) TE}\right] _{\left( \rho ,\eta \right) TE}=0_{\left( \rho
,\eta \right) TE},%
\end{array}%
\end{equation*}%
for any $\tilde{U},\tilde{V},\tilde{Z}\in \Gamma \left( \left( \left( \rho
,\eta \right) TE,\left( \rho ,\eta \right) \tau _{E},E\right) \right) $%
\hfill \emph{q.e.d.}

\bigskip\noindent\textbf{Lemma 3.3.4 }\emph{The }$\mathbf{Mod}$\emph{%
-morphism }%
\begin{equation*}
\Gamma \left( \tilde{\rho},Id_{E}\right)
\end{equation*}%
\emph{is a} $\mathbf{Liealg}$\emph{-morphism of}%
\begin{equation*}
\left( \Gamma \left( \left( \rho ,\eta \right) TE,\left( \rho ,\eta \right)
\tau _{E},E\right) ,+,\cdot ,\left[ ,\right] _{\left( \rho ,\eta \right)
TE}\right)
\end{equation*}%
\emph{source and }%
\begin{equation*}
\left( \Gamma \left( TE,\tau _{E},E\right) ,+,\cdot ,\left[ ,\right]
_{TE}\right)
\end{equation*}%
\emph{target.}

\bigskip\noindent\textit{Proof.} Indeed, we have:
\begin{equation*}
%
\leqno(4)
\end{equation*}
In general, for any $\tilde{U},\tilde{Z}\in \Gamma \left( \left( \rho ,\eta
\right) TE,\left( \rho ,\eta \right) \tau _{E},E\right) ,$ we obtain:
\begin{equation*}
\left[ \Gamma \left( \tilde{\rho},Id_{E}\right) (\tilde{U}),\Gamma \left(
\tilde{\rho},Id_{E}\right) (\tilde{Z})\right] _{TE}=\Gamma \left( \tilde{\rho%
},Id_{E}\right) \left( \left[ \tilde{U},\tilde{Z}\right] _{\left( \rho ,\eta
\right) TE}\right) .
\end{equation*}

\hfill \emph{q.e.d.}\bigskip

Using\emph{\ Lemmas 3.3.1, 3.3.2, 3.3.3} and \emph{3.3.4,} we obtain the
following

\bigskip\noindent\textbf{Theorem 3.3.4 }\emph{The couple }%
\begin{equation*}
\left( \left[ ,\right] _{\left( \rho ,\eta \right) TE},\left( \tilde{\rho}%
,Id_{E}\right) \right)
\end{equation*}%
\emph{\ is a Lie algebroid structure for the vector bundle }%
\begin{equation*}
\left( \left( \rho ,\eta \right) TE,\left( \rho ,\eta \right) \tau
_{E},E\right) .
\end{equation*}

\smallskip \noindent \textbf{Remark 3.3.2 }In particular, if $h=Id_{M}$ and $%
\left[ ,\right] _{TM}$ is the usual Lie bracket, it results that the Lie
algebroid
\begin{equation*}
\begin{array}{c}
\left( \left( \left( Id_{TM},Id_{M}\right) TE,\left( Id_{TM},Id_{M}\right)
\tau _{E},E\right) ,\left[ ,\right] _{\left( Id_{TM},Id_{M}\right)
TE},\left( \widetilde{Id_{TM}},Id_{E}\right) \right)%
\end{array}%
\end{equation*}%
is isomorphic with the usual Lie algebroid
\begin{equation*}
\begin{array}{c}
\left( \left( TE,\tau _{E},E\right) ,\left[ ,\right] _{TE},\left(
Id_{TE},Id_{E}\right) \right) .%
\end{array}%
\end{equation*}

This is a reason for which the Lie algebroid
\begin{equation*}
\begin{array}{c}
\left( \left( \left( \rho ,\eta \right) TE,\left( \rho ,\eta \right) \tau
_{E},E\right) ,\left[ ,\right] _{\left( \rho ,\eta \right) TE},\left( \tilde{%
\rho},Id_{E}\right) \right)%
\end{array}%
,
\end{equation*}%
will be called the \emph{Lie algebroid generalized tangent bundle.}

The vector bundle $\left( \left( \rho ,\eta \right) TE,\left( \rho ,\eta
\right) \tau _{E},E\right) $ will be called the \emph{generalized tangent
bundle.}

\subsubsection{The generalized tangent bundle of dual vector bundle}

Let $\left( E,\pi ,M\right) $ be a vector bundle. We build the generalized
tangent bundle of dual vector bundle $\left( \overset{\ast }{E},\overset{%
\ast }{\pi },M\right) $ using the diagram:%
\begin{equation*}
\begin{array}{rcl}
\overset{\ast }{E} &  & \left( F,\left[ ,\right] _{F,h},\left( \rho ,\eta
\right) \right) \\
\overset{\ast }{\pi }\downarrow &  & ~\downarrow \nu \\
M & ^{\underrightarrow{~\ \ \ \ h~\ \ \ \ }} & ~\ N%
\end{array}%
,\leqno(3.3.1.1)
\end{equation*}%
where $\left( \left( F,\nu ,N\right) ,\left[ ,\right] _{F,h},\left( \rho
,\eta \right) \right) $ is a generalized Lie algebroid.

We take $\left( x^{i},p_{a}\right) $ as canonical local coordinates on $%
\left( \overset{\ast }{E},\overset{\ast }{\pi },M\right) ,$ where $i\in
\overline{1,m}$ and $a\in \overline{1,r}.$

Consider
\begin{equation*}
\left( x^{i},p_{a}\right) \longrightarrow \left( x^{i%
%TCIMACRO{\U{b4}}%
%BeginExpansion
{\acute{}}%
%EndExpansion
}\left( x^{i}\right) ,p_{a%
%TCIMACRO{\U{b4}}%
%BeginExpansion
{\acute{}}%
%EndExpansion
}\left( x^{i},p_{a}\right) \right)
\end{equation*}%
a change of coordinates on $\left( \overset{\ast }{E},\overset{\ast }{\pi }%
,M\right) $. Then the coordinates $p_{a}$ change to $p_{a%
%TCIMACRO{\U{b4}}%
%BeginExpansion
{\acute{}}%
%EndExpansion
}$ by the rule:
\begin{equation*}
\begin{array}{c}
p_{a%
%TCIMACRO{\U{b4}}%
%BeginExpansion
{\acute{}}%
%EndExpansion
}=M_{a%
%TCIMACRO{\U{b4}}%
%BeginExpansion
{\acute{}}%
%EndExpansion
}^{a}p_{a}.%
\end{array}%
\leqno(3.3.1.2)
\end{equation*}

Let
\begin{equation*}
\left( \frac{\partial }{\partial x^{i}},\frac{\partial }{\partial p_{a}}%
\right)
\end{equation*}%
be the natural base for the sections Lie algebra $\left( \Gamma \left( T%
\overset{\ast }{E},\tau _{\overset{\ast }{E}},\overset{\ast }{E}\right)
,+,\cdot ,\left[ ,\right] _{T\overset{\ast }{E}}\right) .$

The sections
\begin{equation*}
\left( \tilde{T}_{\alpha },\left( \frac{\partial }{\partial x^{i}},\frac{%
\partial }{\partial p_{a}}\right) \right)\leqno(3.3.1.3)
\end{equation*}%
determine a base for the module $\Gamma \left( \overset{\ast }{\pi }^{\ast
}\left( h^{\ast }F\right) \oplus T\overset{\ast }{E},\overset{\oplus }{%
\overset{\ast }{\pi }},\overset{\ast }{E}\right) .$

The matrix of coordinate transformation on
\begin{equation*}
\left( \overset{\ast }{\pi }^{\ast }\left( h^{\ast }F\right) \oplus T\overset%
{\ast }{E},\overset{\oplus }{\overset{\ast }{\pi }},\overset{\ast }{E}\right)
\end{equation*}%
at a change of fibred charts is
\begin{equation*}
\left\Vert
%
\right\Vert .\leqno(3.3.1.4)
\end{equation*}
For any sections%
\begin{equation*}
\tilde{Z}^{\alpha }\tilde{T}_{\alpha }\in \Gamma \left( \overset{\ast }{\pi }%
^{\ast }\left( h^{\ast }F\right) ,\overset{\ast }{\pi }^{\ast }\left(
h^{\ast }\nu \right) ,\overset{\ast }{E}\right)
\end{equation*}%
and%
\begin{equation*}
Y_{a}\frac{\partial }{\partial p_{a}}\in \Gamma \left( VT\overset{\ast }{E}%
,\tau _{\overset{\ast }{E}},\overset{\ast }{E}\right) ,
\end{equation*}%
we construct the section%
\begin{equation*}
%
%
\end{equation*}
Since we have
\begin{equation*}
%
%
\end{equation*}%
it implies $\tilde{Z}^{\alpha }=0,~\alpha \in \overline{1,p}$ and $%
Y_{a}=0,~a\in \overline{1,r}.$

Therefore, the sections
\begin{equation*}
%
%
\end{equation*}
are linearly independent.

We consider the vector subbundle
\begin{equation*}
%
%
\end{equation*}%
of vector bundle
\begin{equation*}
\left( \overset{\ast }{\pi }^{\ast }\left( h^{\ast }F\right) \oplus T\overset%
{\ast }{E},\overset{\oplus }{\overset{\ast }{\pi }},\overset{\ast }{E}%
\right) ,
\end{equation*}%
for which the $\mathcal{F}\left( \overset{\ast }{E}\right) $-module of
sections is the $\mathcal{F}\left( \overset{\ast }{E}\right) $-submodule of
\begin{equation*}
\left( \Gamma \left( \overset{\ast }{\pi }^{\ast }\left( h^{\ast }F\right)
\oplus T\overset{\ast }{E},\overset{\oplus }{\overset{\ast }{\pi }},\overset{%
\ast }{E}\right) ,+,\cdot \right) ,
\end{equation*}%
generated by the family of sections $\left( \displaystyle\frac{\partial }{%
\partial \tilde{z}^{\alpha }},\displaystyle\frac{\partial }{\partial \tilde{p%
}_{a}}\right) .$

The base sections
\begin{equation*}
\left( \frac{\partial }{\partial \tilde{z}^{\alpha }},\frac{\partial }{%
\partial \tilde{p}_{a}}\right) \overset{put}{=}\left( \tilde{\partial}%
_{\alpha },\overset{\cdot }{\tilde{\partial}}^{a}\right)\leqno(3.3.1.5)
\end{equation*}%
will be called the \emph{natural }$\left( \rho ,\eta \right) $\emph{-base.}

The matrix of coordinate transformation on $\left( \left( \rho ,\eta \right)
T\overset{\ast }{E},\left( \rho ,\eta \right) \tau _{\overset{\ast }{E}},%
\overset{\ast }{E}\right) $ at a change of fibred charts is
\begin{equation*}
\left\Vert
\begin{array}{cc}
\Lambda _{\alpha }^{\alpha
%TCIMACRO{\U{b4}}%
%BeginExpansion
{\acute{}}%
%EndExpansion
}\circ h\circ \overset{\ast }{\pi } & 0 \\
\left( \rho _{\alpha }^{i}\circ h\circ \overset{\ast }{\pi }\right) %
\displaystyle\frac{\partial M_{a%
%TCIMACRO{\U{b4}}%
%BeginExpansion
{\acute{}}%
%EndExpansion
}^{b}\circ \overset{\ast }{\pi }}{\partial x_{i}}p_{b} & M_{a%
%TCIMACRO{\U{b4}}%
%BeginExpansion
{\acute{}}%
%EndExpansion
}^{a}\circ \overset{\ast }{\pi }%
\end{array}%
\right\Vert .\leqno(3.3.1.6)
\end{equation*}

We consider the operation $\left[ ,\right] _{\left( \rho ,\eta \right) T%
\overset{\ast }{E}}$ defined by%
\begin{equation*}
\begin{array}{l}
\displaystyle\left[ \left( \tilde{Z}_{1}^{\alpha }\frac{\partial }{\partial
\tilde{z}^{\alpha }}+Y_{1a}\frac{\partial }{\partial \tilde{p}_{a}}\right)
,\left( \tilde{Z}_{2}^{\beta }\frac{\partial }{\partial \tilde{z}^{\beta }}%
+Y_{2b}\frac{\partial }{\partial \tilde{p}_{b}}\right) \right] _{\left( \rho
,\eta \right) T\overset{\ast }{E}}=\vspace*{1mm} \\
\displaystyle=\left[ \tilde{Z}_{1}^{\alpha }\tilde{T}_{a},\tilde{Z}%
_{2}^{\beta }\tilde{T}_{\beta }\right] _{\overset{\ast }{\pi }^{\ast }\left(
h^{\ast }F\right) }\oplus \left[ \left( \rho _{\alpha }^{i}\circ h\circ \pi
^{\ast }\right) \tilde{Z}_{1}^{\alpha }\frac{\partial }{\partial x^{i}}%
+Y_{1a}\frac{\partial }{\partial p_{a}},\right. \\
\hfill \left. \displaystyle\left( \rho _{\beta }^{i}\circ h\circ \pi ^{\ast
}\right) \tilde{Z}_{2}^{\beta }\frac{\partial }{\partial x^{i}}+Y_{2b}\frac{%
\partial }{\partial p_{b}}\right] _{T\overset{\ast }{E}},%
\end{array}%
\leqno(3.3.1.7)
\end{equation*}%
for any sections $\left( \tilde{Z}_{1}^{\alpha }\displaystyle\frac{\partial
}{\partial \tilde{z}^{\alpha }}+Y_{1a}\displaystyle\frac{\partial }{\partial
\tilde{p}_{a}}\right) $ and $\left( \tilde{Z}_{2}^{\beta }\displaystyle\frac{%
\partial }{\partial \tilde{z}^{\beta }}+Y_{2b}\displaystyle\frac{\partial }{%
\partial \tilde{p}_{b}}\right) .$

Let $\left( \overset{\ast }{\tilde{\rho}},Id_{\overset{\ast }{E}}\right) $\
be the $\mathbf{B}^{\mathbf{v}}$-morphism of $\left( \left( \rho ,\eta
\right) T\overset{\ast }{E},\left( \rho ,\eta \right) \tau _{\overset{\ast }{%
E}},\overset{\ast }{E}\right) $\ source and $\left( T\overset{\ast }{E},\tau
_{\overset{\ast }{E}},\overset{\ast }{E}\right) $\ target, where
\begin{equation*}
\leqno(3.3.1.8)%
\begin{array}{rcl}
\displaystyle\left( \rho ,\eta \right) T\overset{\ast }{E}\!\!\! & \!\!^{%
\underrightarrow{\ \,\tilde{\rho}^{\ast }\ \,}}\!\!\! & \!\!T\overset{\ast }{%
E} \\
\displaystyle\!\left( \tilde{Z}^{\alpha }\frac{\partial }{\partial \tilde{z}%
^{\alpha }}+Y_{a}\displaystyle\frac{\partial }{\partial \tilde{p}_{a}}%
\right) \!(\overset{\ast }{u}_{x})\!\!\!\! & \!\!\longmapsto \!\!\! &
\!\!\left( \!\tilde{Z}^{\alpha }\!\left( \rho _{\alpha }^{i}{\circ }h{\circ }%
\overset{\ast }{\pi }\!\right) \!\displaystyle\frac{\partial }{\partial x^{i}%
}{+}Y_{a}\displaystyle\frac{\partial }{\partial p_{a}}\right) \!(\overset{%
\ast }{u}_{x})\!\!%
\end{array}%
\bigskip
\end{equation*}

The Lie algebroid generalized tangent bundle of the dual vector bundle $%
\left( \overset{\ast }{E},\overset{\ast }{\pi },M\right) $ will be denoted
\begin{equation*}
\begin{array}{c}
\left( \left( \left( \rho ,\eta \right) T\overset{\ast }{E},\left( \rho
,\eta \right) \tau _{\overset{\ast }{E}},\overset{\ast }{E}\right) ,\left[ ,%
\right] _{\left( \rho ,\eta \right) T\overset{\ast }{E}},\left( \overset{%
\ast }{\tilde{\rho}},Id_{\overset{\ast }{E}}\right) \right).%
\end{array}%
\end{equation*}

\subsection{(Linear) $\left( \protect\rho ,\protect\eta \right) $-connections%
}

The theory of (linear) connections constitutes undoubtedly one of most
beautiful and most important chapter of differential geometry, which has
been widely explored in the literature (see [8, 11, 14, 26, 31, 41, 42, 45,
46, 47, 50, 51, 59, 60, 62, 63]).

In the following, we consider the diagram:
\begin{equation*}
\begin{array}{c}
\xymatrix{E\ar[d]_\pi&\left( F,\left[ ,\right] _{F,h},\left( \rho ,\eta
\right) \right)\ar[d]^\nu\\ M\ar[r]^h&N}%
\end{array}%
\end{equation*}%
where $\left( E,\pi ,M\right) \in \left\vert \mathbf{B}\right\vert $ and $%
\left( \left( F,\nu ,N\right) ,\left[ ,\right] _{F,h},\left( \rho ,\eta
\right) \right) $ is a generalized Lie algebroid.

Let
\begin{equation*}
\left( \left( \left( \rho ,\eta \right) TE,\left( \rho ,\eta \right) \tau
_{E},E\right) ,\left[ ,\right] _{\left( \rho ,\eta \right) TE},\left( \tilde{%
\rho},Id_{E}\right) \right)
\end{equation*}%
\ be the Lie algebroid generalized tangent bundle of fiber bundle $\left(
E,\pi ,M\right) $.

We consider the $\mathbf{B}^{\mathbf{v}}$-morphism
\begin{equation*}
\left( \left( \rho ,\eta \right) \pi !,Id_{E}\right)
\end{equation*}%
given by the commutative diagram%
\begin{equation*}
\begin{array}{c}
\xymatrix{\left( \rho ,\eta \right) TE\ar[r]^{( \rho ,\eta ) \pi
!}\ar[d]_{(\rho,\eta)\tau_E}&\pi ^{\ast }\left( h^{\ast
}F\right)\ar[d]^{pr_1} \\ E\ar[r]^{id_{E}}& E}%
\end{array}%
\leqno(3.4.1)
\end{equation*}%
Using the components, this is defined as:%
\begin{equation*}
\left( \rho ,\eta \right) \pi !\left( \left( \tilde{Z}^{\alpha }\frac{%
\partial }{\partial \tilde{z}^{\alpha }}+Y^{a}\frac{\partial }{\partial
\tilde{y}^{a}}\right) \left( u_{x}\right) \right) =\left( \tilde{Z}^{\alpha }%
\tilde{T}_{\alpha }\right) \left( u_{x}\right) ,\leqno(3.4.2)
\end{equation*}%
for any $\displaystyle\left( \tilde{Z}^{\alpha }\frac{\partial }{\partial
\tilde{z}^{\alpha }}+Y^{a}\displaystyle\frac{\partial }{\partial \tilde{y}%
^{a}}\right) \in \Gamma \left( \left( \rho ,\eta \right) TE,\left( \rho
,\eta \right) \tau _{E},E\right) .$\medskip

We define the \emph{tangent }$\left( \rho ,\eta \right) $\emph{-application }%
as being a $\mathbf{B}^{\mathbf{v}}$-morphism
\begin{equation*}
\begin{array}{c}
\left( \left( \rho ,\eta \right) T\pi ,h\circ \pi \right) =\left(
pr_{2},h\circ \pi \right) \circ \left( \left( \rho ,\eta \right) \pi
!,Id_{E}\right)%
\end{array}%
\leqno(3.4.3)
\end{equation*}%
of $\left( \left( \rho ,\eta \right) TE,\left( \rho ,\eta \right) \tau
_{E},E\right) $ source and $\left( F,\nu ,N\right) $ target.

\bigskip\noindent\textbf{Definition~3.4.1} The kernel of the tangent $\left(
\rho ,\eta \right) $-application
\begin{equation*}
\left( \left( \rho ,\eta \right) T\pi ,h\circ \pi \right)
\end{equation*}%
\ is written as%
\begin{equation*}
\left( V\left( \rho ,\eta \right) TE,\left( \rho ,\eta \right) \tau
_{E},E\right)
\end{equation*}%
and will be called \emph{the vertical subbundle}.\bigskip

The set $\left\{ \displaystyle\frac{\partial }{\partial \tilde{y}^{a}},~a\in
\overline{1,r}\right\} $ is a base for the $\mathcal{F}\left( E\right) $%
-module
\begin{equation*}
\left( \Gamma \left( V\left( \rho ,\eta \right) TE,\left( \rho ,\eta \right)
\tau _{E},E\right) ,+,\cdot \right) .
\end{equation*}

\smallskip \noindent\textbf{Proposition 3.4.1} \emph{The short sequence of
vector bundles}%
\begin{equation*}
\begin{array}{c}
\xymatrix{0\ar@{^(->}[r]^i\ar[d]&V(\rho,\eta)TE\ar[d]\ar@{^(->}[r]^i&(\rho,%
\eta)TE\ar[r]^{(\rho,\eta)\pi!}\ar[d]&\pi ^{\ast }\left( h^{\ast
F}\right)\ar[r]\ar[d]&0\ar[d]\\
E\ar[r]^{Id_E}&E\ar[r]^{Id_E}&E\ar[r]^{Id_E}&E\ar[r]^{Id_E}&E}%
\end{array}%
\leqno(3.4.4)
\end{equation*}%
\emph{is exact.}

\bigskip \noindent \textbf{Definition 3.4.2} \textit{A }$\mathbf{Man}$%
-morphism $\left( \rho ,\eta \right) \Gamma $ of $\left( \rho ,\eta \right)
TE$ source and $V\left( \rho ,\eta \right) TE$ target defined by%
\begin{equation*}
\left( \rho ,\eta \right) \Gamma \left( \tilde{Z}^{\alpha }\frac{\partial }{%
\partial \tilde{z}^{\alpha }}+Y^{a}\frac{\partial }{\partial \tilde{y}^{a}}%
\right) \left( u_{x}\right) =\left( Y^{a}+\left( \rho ,\eta \right) \Gamma
_{\alpha }^{a}\tilde{Z}^{\alpha }\right) \frac{\partial }{\partial \tilde{y}%
^{a}}\left( u_{x}\right) ,\leqno(3.4.5)
\end{equation*}%
such that the $\mathbf{B}^{\mathbf{v}}$-morphism $\left( \left( \rho ,\eta
\right) \Gamma ,Id_{E}\right) $ is a split to the left in the previous exact
sequence, will be called $\left( \rho ,\eta \right) $\emph{-connection for
the fiber bundle }$\left( E,\pi ,M\right) $.

The differentiable real local functions $\left( \rho ,\eta \right) \Gamma
_{\alpha }^{a}$ will be called the \emph{components of }$\left( \rho ,\eta
\right) $\emph{-connection }$\left( \rho ,\eta \right) \Gamma .$

The $\left( \rho ,Id_{M}\right) $-connection for the fiber bundle $\left(
E,\pi ,M\right) $ will be called $\rho $\emph{-connection for the fiber
bundle }$\left( E,\pi ,M\right) $ and will be denoted $\rho \Gamma $\emph{.}

The $\left( Id_{TM},Id_{M}\right) $-connection for the fiber bundle $\left(
E,\pi ,M\right) $ will be called \emph{connection for the fiber bundle }$%
\left( E,\pi ,M\right) $ and will be denoted $\Gamma $\emph{.}

\bigskip\noindent\textbf{Definition 3.4.3 }If $\left( \rho ,\eta \right)
\Gamma $ is a $\left( \rho ,\eta \right) $-connection for the fiber bundle $%
\left( E,\pi ,M\right) $, then the kernel of the $\mathbf{B}^{\mathbf{v}}$%
-morphism $\left( \left( \rho ,\eta \right) \Gamma ,Id_{E}\right) $\ is
written as%
\begin{equation*}
\left( H\left( \rho ,\eta \right) TE,\left( \rho ,\eta \right) \tau
_{E},E\right)
\end{equation*}%
and will be called the \emph{horizontal vector subbundle}.

\bigskip \noindent \textbf{Definition 3.4.4} If $\left( E,\pi ,M\right) \in
\left\vert \mathbf{B}\right\vert $, then the $\mathbf{B}$-morphism $\left(
\Pi ,\pi \right) $ defined by the commutative diagram%
\begin{equation*}
\begin{array}{c}
\xymatrix{ {V\left( \rho ,\eta \right)T}{E\ar[r]^{\qquad \Pi}}
\ar[d]_{(\rho,\eta)\tau_E} &E \ar[d]^\pi \\ E\ar[r]^\pi&M}%
\end{array}%
\leqno(3.4.6)
\end{equation*}%
such that the components of the image of vector $Y^{a}\frac{\partial }{%
\partial \tilde{y}^{a}}\left( u_{x}\right) $ are the real numbers $%
Y^{1}\left( u_{x}\right) ,...,Y^{r}\left( u_{x}\right) $ will be called the
\emph{canonical projection }$\mathbf{B}$\emph{-morphism.}\medskip

\textbf{Example 3.4.1 }If $\left( E,\pi ,M\right) \in \left\vert \mathbf{B}^{%
\mathbf{v}}\right\vert $, then the $\mathbf{B}^{\mathbf{v}}$-morphism $%
\left( \Pi ,\pi \right) $ defined by the commutative diagram $\left(
3.4.6\right) $, where $\Pi $\textit{\ }is defined by\textit{\ }%
\begin{equation*}
\Pi \left( Y^{a}\frac{\partial }{\partial \tilde{y}^{a}}\left( u_{x}\right)
\right) =Y^{a}\left( u_{x}\right) s_{a}\left( x\right) ,\leqno(3.4.7)
\end{equation*}%
is canonical projection $\mathbf{B}^{\mathbf{v}}$-morphism$.$

The set $\left\{ s_{a},a\in \overline{1,r}\right\} $ is the base of $%
\mathcal{F}\left( M\right) $-module of sections $\left( \Gamma \left( E,\pi
,M\right) ,+,\cdot \right) .$

\bigskip\noindent\textbf{Theorem 3.4.1 }\emph{If }$\left( \rho ,\eta \right)
\Gamma $\emph{\ is a }$\left( \rho ,\eta \right) $\emph{-connection for the
fiber bundle }$\left( E,\pi ,M\right) ,$\emph{\ then its components satisfy
the law of transformation }%
\begin{equation*}
\left( \rho ,\eta \right) \Gamma _{\gamma
%TCIMACRO{\U{b4}}%
%BeginExpansion
{\acute{}}%
%EndExpansion
}^{a%
%TCIMACRO{\U{b4}}%
%BeginExpansion
{\acute{}}%
%EndExpansion
}=\frac{\partial y^{a%
%TCIMACRO{\U{b4}}%
%BeginExpansion
{\acute{}}%
%EndExpansion
}}{\partial y^{a}}\left[ \rho _{\gamma }^{i}\circ \left( h\circ \pi \right)
\frac{\partial y^{a}}{\partial x^{i}}+\left( \rho ,\eta \right) \Gamma
_{\gamma }^{a}\right] \Lambda _{\gamma
%TCIMACRO{\U{b4}}%
%BeginExpansion
{\acute{}}%
%EndExpansion
}^{\gamma }\circ \left( h\circ \pi \right) .\leqno(3.4.8)
\end{equation*}

\emph{If }$\left( \rho ,\eta \right) \Gamma $\emph{\ is a }$\left( \rho
,\eta \right) $\emph{-connection for the vector bundle }$\left( E,\pi
,M\right) ,$\emph{\ then its components satisfy the law of transformation }%
\begin{equation*}
\left( \rho ,\eta \right) \Gamma _{\gamma
%TCIMACRO{\U{b4}}%
%BeginExpansion
{\acute{}}%
%EndExpansion
}^{a%
%TCIMACRO{\U{b4}}%
%BeginExpansion
{\acute{}}%
%EndExpansion
}{=}M_{a}^{a%
%TCIMACRO{\U{b4}}%
%BeginExpansion
{\acute{}}%
%EndExpansion
}{\circ }\pi \!\!\left[ \rho _{\gamma }^{i}{\circ }\left( h{\circ }\pi
\right) \!\frac{\partial M_{b%
%TCIMACRO{\U{b4}}%
%BeginExpansion
{\acute{}}%
%EndExpansion
}^{a}\circ \pi }{\partial x^{i}}y^{b%
%TCIMACRO{\U{b4}}%
%BeginExpansion
{\acute{}}%
%EndExpansion
}{+}\left( \rho ,\eta \right) \!\Gamma _{\gamma }^{a}\right] \!\!\Lambda
_{\gamma
%TCIMACRO{\U{b4}}%
%BeginExpansion
{\acute{}}%
%EndExpansion
}^{\gamma }{\circ }\left( h{\circ }\pi \right) .\leqno(3.4.8^{\prime })
\end{equation*}

\emph{If }$\rho \Gamma $\emph{\ is a }$\rho $\emph{-connection for the
vector bundle }$\left( E,\pi ,M\right) $ \emph{and }$h=Id_{M},$\emph{\ then
relations }$\left( 3.4.8^{\prime }\right) $\emph{\ become}%
\begin{equation*}
\rho \Gamma _{\gamma
%TCIMACRO{\U{b4}}%
%BeginExpansion
{\acute{}}%
%EndExpansion
}^{a%
%TCIMACRO{\U{b4}}%
%BeginExpansion
{\acute{}}%
%EndExpansion
}=M_{a}^{a%
%TCIMACRO{\U{b4}}%
%BeginExpansion
{\acute{}}%
%EndExpansion
}\circ \pi \left[ \rho _{\gamma }^{i}\circ \pi \frac{\partial M_{b%
%TCIMACRO{\U{b4}}%
%BeginExpansion
{\acute{}}%
%EndExpansion
}^{a}\circ \pi }{\partial x^{i}}y^{b%
%TCIMACRO{\U{b4}}%
%BeginExpansion
{\acute{}}%
%EndExpansion
}+\rho \Gamma _{\gamma }^{a}\right] \Lambda _{\gamma
%TCIMACRO{\U{b4}}%
%BeginExpansion
{\acute{}}%
%EndExpansion
}^{\gamma }\circ \pi .\leqno(3.4.8^{\prime \prime })
\end{equation*}

\emph{In particular, if }$\left( \rho ,\eta \right) =\left(
Id_{TM},Id_{M}\right) $\emph{, then the relations }$\left( 3.4.8^{\prime
\prime }\right) $\emph{\ become}%
\begin{equation*}
\Gamma _{k%
%TCIMACRO{\U{b4}}%
%BeginExpansion
{\acute{}}%
%EndExpansion
}^{i%
%TCIMACRO{\U{b4}}%
%BeginExpansion
{\acute{}}%
%EndExpansion
}=\frac{\partial x^{i%
%TCIMACRO{\U{b4}}%
%BeginExpansion
{\acute{}}%
%EndExpansion
}}{\partial x^{i}}\circ \pi \left[ \frac{\partial }{\partial x^{k}}\left(
\frac{\partial x^{i}}{\partial x^{j%
%TCIMACRO{\U{b4}}%
%BeginExpansion
{\acute{}}%
%EndExpansion
}}\circ \pi \right) y^{j%
%TCIMACRO{\U{b4}}%
%BeginExpansion
{\acute{}}%
%EndExpansion
}+\Gamma _{k}^{i}\right] \frac{\partial x^{k}}{\partial x^{k%
%TCIMACRO{\U{b4}}%
%BeginExpansion
{\acute{}}%
%EndExpansion
}}\circ \pi . \leqno(3.4.8^{\prime \prime \prime })
\end{equation*}

\bigskip\noindent\textit{Proof.} Let $\left( \Pi ,\pi \right) $ be the
canonical projection $\mathbf{B}$-morphism.

Obviously, the components of
\begin{equation*}
\Pi \circ \left( \rho ,\eta \right) \Gamma \left( \tilde{Z}^{\alpha
%TCIMACRO{\U{b4}}%
%BeginExpansion
{\acute{}}%
%EndExpansion
}\frac{\partial }{\partial \tilde{z}^{\alpha
%TCIMACRO{\U{b4}}%
%BeginExpansion
{\acute{}}%
%EndExpansion
}}+Y^{a%
%TCIMACRO{\U{b4}}%
%BeginExpansion
{\acute{}}%
%EndExpansion
}\frac{\partial }{\partial \tilde{y}^{a%
%TCIMACRO{\U{b4}}%
%BeginExpansion
{\acute{}}%
%EndExpansion
}}\right) \left( u_{x}\right)
\end{equation*}%
are the real numbers%
\begin{equation*}
\displaystyle\left( Y^{a%
%TCIMACRO{\U{b4}}%
%BeginExpansion
{\acute{}}%
%EndExpansion
}+\left( \rho ,\eta \right) \Gamma _{\gamma
%TCIMACRO{\U{b4}}%
%BeginExpansion
{\acute{}}%
%EndExpansion
}^{a%
%TCIMACRO{\U{b4}}%
%BeginExpansion
{\acute{}}%
%EndExpansion
}\tilde{Z}^{\gamma
%TCIMACRO{\U{b4}}%
%BeginExpansion
{\acute{}}%
%EndExpansion
}\right) \left( u_{x}\right) .
\end{equation*}%
Since
\begin{equation*}
\begin{array}{l}
\displaystyle\left( \tilde{Z}^{\alpha
%TCIMACRO{\U{b4}}%
%BeginExpansion
{\acute{}}%
%EndExpansion
}\frac{\partial }{\partial \tilde{z}^{\alpha
%TCIMACRO{\U{b4}}%
%BeginExpansion
{\acute{}}%
%EndExpansion
}}+Y^{a%
%TCIMACRO{\U{b4}}%
%BeginExpansion
{\acute{}}%
%EndExpansion
}\frac{\partial }{\partial \tilde{y}^{a%
%TCIMACRO{\U{b4}}%
%BeginExpansion
{\acute{}}%
%EndExpansion
}}\right) \left( u_{x}\right) =\tilde{Z}^{\alpha
%TCIMACRO{\U{b4}}%
%BeginExpansion
{\acute{}}%
%EndExpansion
}\Lambda _{\alpha
%TCIMACRO{\U{b4}}%
%BeginExpansion
{\acute{}}%
%EndExpansion
}^{\alpha }\circ h\circ \pi \frac{\partial }{\partial \tilde{z}^{\alpha }}%
\left( u_{x}\right) \vspace*{1mm} \\
\qquad \displaystyle+\left( \tilde{Z}^{\alpha
%TCIMACRO{\U{b4}}%
%BeginExpansion
{\acute{}}%
%EndExpansion
}\rho _{\alpha
%TCIMACRO{\U{b4}}%
%BeginExpansion
{\acute{}}%
%EndExpansion
}^{i%
%TCIMACRO{\U{b4}}%
%BeginExpansion
{\acute{}}%
%EndExpansion
}\circ h\circ \pi \frac{\partial y^{a}}{\partial x^{i%
%TCIMACRO{\U{b4}}%
%BeginExpansion
{\acute{}}%
%EndExpansion
}}+\frac{\partial y^{a}}{\partial y^{a%
%TCIMACRO{\U{b4}}%
%BeginExpansion
{\acute{}}%
%EndExpansion
}}Y^{a%
%TCIMACRO{\U{b4}}%
%BeginExpansion
{\acute{}}%
%EndExpansion
}\right) \frac{\partial }{\partial \tilde{y}^{a}}\left( u_{x}\right) ,%
\end{array}%
\end{equation*}%
it results that the components of%
\begin{equation*}
\Pi \circ \left( \rho ,\eta \right) \Gamma \left( \tilde{Z}^{\alpha
%TCIMACRO{\U{b4}}%
%BeginExpansion
{\acute{}}%
%EndExpansion
}\frac{\partial }{\partial \tilde{z}^{\alpha
%TCIMACRO{\U{b4}}%
%BeginExpansion
{\acute{}}%
%EndExpansion
}}+Y^{a%
%TCIMACRO{\U{b4}}%
%BeginExpansion
{\acute{}}%
%EndExpansion
}\frac{\partial }{\partial \tilde{y}^{a%
%TCIMACRO{\U{b4}}%
%BeginExpansion
{\acute{}}%
%EndExpansion
}}\right) \left( u_{x}\right)
\end{equation*}%
are the real numbers
\begin{equation*}
\left( \tilde{Z}^{\alpha
%TCIMACRO{\U{b4}}%
%BeginExpansion
{\acute{}}%
%EndExpansion
}\rho _{\alpha
%TCIMACRO{\U{b4}}%
%BeginExpansion
{\acute{}}%
%EndExpansion
}^{i%
%TCIMACRO{\U{b4}}%
%BeginExpansion
{\acute{}}%
%EndExpansion
}\circ h\circ \pi \frac{\partial y^{a}}{\partial x^{i%
%TCIMACRO{\U{b4}}%
%BeginExpansion
{\acute{}}%
%EndExpansion
}}+\frac{\partial y^{a}}{\partial y^{a%
%TCIMACRO{\U{b4}}%
%BeginExpansion
{\acute{}}%
%EndExpansion
}}Y^{a%
%TCIMACRO{\U{b4}}%
%BeginExpansion
{\acute{}}%
%EndExpansion
}+\left( \rho ,\eta \right) \Gamma _{\alpha }^{a}\tilde{Z}^{\alpha
%TCIMACRO{\U{b4}}%
%BeginExpansion
{\acute{}}%
%EndExpansion
}\Lambda _{\alpha
%TCIMACRO{\U{b4}}%
%BeginExpansion
{\acute{}}%
%EndExpansion
}^{\alpha }\circ h\circ \pi \right) \left( u_{x}\right) \frac{\partial y^{a%
%TCIMACRO{\U{b4}}%
%BeginExpansion
{\acute{}}%
%EndExpansion
}}{\partial y^{a}},
\end{equation*}%
where
\begin{equation*}
\left\Vert \frac{\partial y^{a}}{\partial y^{a%
%TCIMACRO{\U{b4}}%
%BeginExpansion
{\acute{}}%
%EndExpansion
}}\right\Vert =\left\Vert \frac{\partial y^{a%
%TCIMACRO{\U{b4}}%
%BeginExpansion
{\acute{}}%
%EndExpansion
}}{\partial y^{a}}\right\Vert ^{-1}.
\end{equation*}%
Therefore, we have:%
\begin{equation*}
\left( \tilde{Z}^{\alpha
%TCIMACRO{\U{b4}}%
%BeginExpansion
{\acute{}}%
%EndExpansion
}\rho _{\alpha
%TCIMACRO{\U{b4}}%
%BeginExpansion
{\acute{}}%
%EndExpansion
}^{i%
%TCIMACRO{\U{b4}}%
%BeginExpansion
{\acute{}}%
%EndExpansion
}\circ h\circ \pi \frac{\partial y^{a}}{\partial x^{i%
%TCIMACRO{\U{b4}}%
%BeginExpansion
{\acute{}}%
%EndExpansion
}}+\frac{\partial y^{a}}{\partial y^{a%
%TCIMACRO{\U{b4}}%
%BeginExpansion
{\acute{}}%
%EndExpansion
}}Y^{a%
%TCIMACRO{\U{b4}}%
%BeginExpansion
{\acute{}}%
%EndExpansion
}+\left( \rho ,\eta \right) \Gamma _{\alpha }^{a}\tilde{Z}^{\alpha
%TCIMACRO{\U{b4}}%
%BeginExpansion
{\acute{}}%
%EndExpansion
}\Lambda _{\alpha
%TCIMACRO{\U{b4}}%
%BeginExpansion
{\acute{}}%
%EndExpansion
}^{\alpha }\circ h\circ \pi \right) \frac{\partial y^{a%
%TCIMACRO{\U{b4}}%
%BeginExpansion
{\acute{}}%
%EndExpansion
}}{\partial y^{a}}=Y^{a%
%TCIMACRO{\U{b4}}%
%BeginExpansion
{\acute{}}%
%EndExpansion
}+\left( \rho ,\eta \right) \Gamma _{\alpha
%TCIMACRO{\U{b4}}%
%BeginExpansion
{\acute{}}%
%EndExpansion
}^{a%
%TCIMACRO{\U{b4}}%
%BeginExpansion
{\acute{}}%
%EndExpansion
}\tilde{Z}^{\alpha
%TCIMACRO{\U{b4}}%
%BeginExpansion
{\acute{}}%
%EndExpansion
}.
\end{equation*}%
After some calculations we obtain:
\begin{equation*}
\left( \rho ,\eta \right) \Gamma _{\alpha
%TCIMACRO{\U{b4}}%
%BeginExpansion
{\acute{}}%
%EndExpansion
}^{a%
%TCIMACRO{\U{b4}}%
%BeginExpansion
{\acute{}}%
%EndExpansion
}=\frac{\partial y^{a%
%TCIMACRO{\U{b4}}%
%BeginExpansion
{\acute{}}%
%EndExpansion
}}{\partial y^{a}}\left( \rho _{\alpha }^{i}\circ \left( h\circ \pi \right)
\frac{\partial y^{a}}{\partial x^{i}}+\left( \rho ,\eta \right) \Gamma
_{\alpha }^{a}\right) \Lambda _{\alpha
%TCIMACRO{\U{b4}}%
%BeginExpansion
{\acute{}}%
%EndExpansion
}^{\alpha }\circ h\circ \pi .\eqno{{q.e.d.}}
\end{equation*}

\smallskip \noindent\textbf{Remark 3.4.1} If we have a set of real local
functions $\left( \rho ,\eta \right) \Gamma _{\gamma }^{a}$ which satisfies
the relations of passing $\left( 3.4.8\right) ,$ then we have a $\left( \rho
,\eta \right) $-connection $\left( \rho ,\eta \right) \Gamma $ for the fiber
bundle $\left( E,\pi ,M\right) .$

\bigskip\noindent\textbf{Example 3.4.1} If $\Gamma $ is a classical
connection for the vector bundle $\left( E,\pi ,M\right) $ on components $%
\Gamma _{k}^{a},$ then the differentiable real local functions
\begin{equation*}
\left( \rho ,\eta \right) \Gamma _{\gamma }^{a}=\left( \rho _{\gamma
}^{k}\circ h\circ \pi \right) \Gamma _{k}^{a}
\end{equation*}%
are the components of a $\left( \rho ,\eta \right) $-connection $\left( \rho
,\eta \right) \Gamma $ for the vector bundle $\left( E,\pi ,M\right) $ which
will be called the $\left( \rho ,\eta \right) $\emph{-connection associated
to the connection }$\Gamma .$

\bigskip \noindent \textbf{Definition 3.4.5 }If $\left( \rho ,\eta \right)
\Gamma $ is a $\left( \rho ,\eta \right) $-connection for the vector bundle $%
\left( E,\pi ,M\right) $, then for any
\begin{equation*}
z=z^{\alpha }t_{\alpha }\in \Gamma \left( F,\nu ,N\right)
\end{equation*}%
the application%
\begin{equation*}
\begin{array}{rcl}
\Gamma \left( E,\pi ,M\right) & ^{\underrightarrow{\left( \rho ,\eta \right)
D_{v}}} & \Gamma \left( E,\pi ,M\right) \\
u=u^{a}s_{a} & \longmapsto & \left( \rho ,\eta \right) D_{z}u%
\end{array}%
\leqno(3.4.9)
\end{equation*}%
where
\begin{equation*}
\left( \rho ,\eta \right) D_{z}u=z^{\alpha }\circ h\left( \rho _{\alpha
}^{i}\circ h\frac{\partial u^{a}}{\partial x^{i}}+\left( \rho ,\eta \right)
\Gamma _{\alpha }^{a}\circ u\right) s_{a}
\end{equation*}%
will be called the \emph{covariant }$\left( \rho ,\eta \right) $\emph{%
-derivative associated to }$\left( \rho ,\eta \right) $\emph{-con\-nec\-tion
}$\left( \rho ,\eta \right) \Gamma $\emph{\ with respect to the section }$z$%
\emph{.}

If $h=Id_{M}$ and $\eta =Id_{M},$ then we obtain the \emph{covariant }$\rho $%
\emph{-derivative associated to }$\rho $\emph{-connection }$\rho \Gamma $%
\emph{\ with respect to the section }$z$\emph{.}

In addition, if $\rho =Id_{TM}$, then we obtain the \emph{covariant
derivative associated to connection }$\Gamma $\emph{\ with respect to the
vector field }$z$.

\bigskip\noindent\textbf{Remark 3.4.2 }If $\left( \rho ,\eta \right) \Gamma $
is a $\left( \rho ,\eta \right) $-connection for the vector bundle $\left(
E,\pi ,M\right) $, then the operator
\begin{equation*}
\begin{array}{rcl}
\Gamma \left( F,\nu ,N\right) \times \Gamma \left( E,\pi ,M\right) & ^{%
\underrightarrow{\left( \rho ,\eta \right) D}} & \Gamma \left( E,\pi
,M\right) \\
\left( z,u\right) & \longmapsto & \left( \rho ,\eta \right) D_{z}u%
\end{array}%
\end{equation*}%
satisfies the following properties:

\begin{itemize}
\item[(i)] $\left( \rho ,\eta \right) D$ is $\mathbb{R}$-bilinear;

\item[(ii)] $\left( \rho ,\eta \right)
D_{f_{1}z_{1}+f_{2}z_{2}}u=f_{1}\left( \rho ,\eta \right)
D_{z_{1}}u+f_{2}\left( \rho ,\eta \right) D_{z_{2}}u;$

\item[(iii)] if $u\in \Gamma \left( E,\pi ,M\right) $ is null on a nonempty
subset of $M,$ then $\left( \rho ,\eta \right) D_{z}u$ is null on the same
nonempty subset, for any $z\in \Gamma \left( F,\nu ,N\right) .$
\end{itemize}

\smallskip \noindent\textbf{Definition 3.4.6 }We will say that the $\left(
\rho ,\eta \right) $\emph{-connection }$\left( \rho ,\eta \right) \Gamma $%
\emph{\ is homogeneous} or \emph{linear }if the local real functions $\left(
\rho ,\eta \right) \Gamma _{\gamma }^{a}$ are homogeneous or linear on the
fibre of the fiber bundle $\left( E,\pi ,M\right) $.

\bigskip\noindent\textbf{Remark 3.4.3} If $\left( \rho ,\eta \right) \Gamma $
is a linear $\left( \rho ,\eta \right) $-connection for the fiber bundle $%
\left( E,\pi ,M\right) $, then for each local vector $\left( m+r\right) $%
-chart $\left( U,s_{U}\right) $ and for each local vector $\left( n+p\right)
$-chart $\left( V,t_{V}\right) $ such that $U\cap h^{-1}\left( V\right) \neq
\phi $, it exists the differentiable real functions $\rho \Gamma _{b\gamma
}^{a}$ defined on $U\cap h^{-1}\left( V\right) $ such that
\begin{equation*}
\begin{array}{c}
\left( \rho ,\eta \right) \Gamma _{\gamma }^{a}\circ u=\left( \rho ,\eta
\right) \Gamma _{b\gamma }^{a}\cdot u^{b},\forall u=u^{b}s_{b}\in \Gamma
\left( E,\pi ,M\right) .%
\end{array}%
\leqno(3.4.10)
\end{equation*}

The differentiable real local functions $\left( \rho ,\eta \right) \Gamma
_{b\alpha }^{a}$ will be called the \emph{Christoffel coefficients of linear
}$\left( \rho ,\eta \right) $\emph{-connection }$\left( \rho ,\eta \right)
\Gamma .$

\bigskip\noindent\textbf{Theorem 3.4.2} \emph{If }$\left( \rho ,\eta \right)
\Gamma $\emph{\ is a linear }$\left( \rho ,\eta \right) $\emph{-connection
for the fiber bundle }$\left( E,\pi ,M\right) ,$\emph{\ then its components
satisfy the law of transformation }%
\begin{equation*}
(\rho ,\eta )\Gamma _{b%
%TCIMACRO{\U{b4}}%
%BeginExpansion
{\acute{}}%
%EndExpansion
\gamma
%TCIMACRO{\U{b4}}%
%BeginExpansion
{\acute{}}%
%EndExpansion
}^{a%
%TCIMACRO{\U{b4}}%
%BeginExpansion
{\acute{}}%
%EndExpansion
}{=}\frac{\partial y^{a%
%TCIMACRO{\U{b4}}%
%BeginExpansion
{\acute{}}%
%EndExpansion
}}{\partial y^{a}}\left[ \rho _{\gamma }^{k}{\circ }h\!\frac{\partial }{%
\partial x^{k}}\!\left( \frac{\partial y^{a}}{\partial y^{b%
%TCIMACRO{\U{b4}}%
%BeginExpansion
{\acute{}}%
%EndExpansion
}}\right) {+}\left( \rho ,\eta \right) \Gamma _{b\gamma }^{a}\frac{\partial
y^{b}}{\partial y^{b%
%TCIMACRO{\U{b4}}%
%BeginExpansion
{\acute{}}%
%EndExpansion
}}\right] \!\Lambda _{\gamma
%TCIMACRO{\U{b4}}%
%BeginExpansion
{\acute{}}%
%EndExpansion
}^{\gamma }{\circ }h. \leqno(3.4.11)
\end{equation*}

\emph{If }$\left( \rho ,\eta \right) \Gamma $\emph{\ is a linear }$\left(
\rho ,\eta \right) $\emph{-connection for the vector bundle }$\left( E,\pi
,M\right) ,$\emph{\ then its components satisfy the law of transformation }%
\begin{equation*}
(\rho ,\eta )\Gamma _{b%
%TCIMACRO{\U{b4}}%
%BeginExpansion
{\acute{}}%
%EndExpansion
\gamma
%TCIMACRO{\U{b4}}%
%BeginExpansion
{\acute{}}%
%EndExpansion
}^{a%
%TCIMACRO{\U{b4}}%
%BeginExpansion
{\acute{}}%
%EndExpansion
}{=}M_{a}^{a%
%TCIMACRO{\U{b4}}%
%BeginExpansion
{\acute{}}%
%EndExpansion
}\!\left[ \rho _{\gamma }^{k}{\circ }h\frac{\partial M_{b%
%TCIMACRO{\U{b4}}%
%BeginExpansion
{\acute{}}%
%EndExpansion
}^{a}}{\partial x^{k}}{+}(\rho ,\eta )\Gamma _{b\gamma }^{a}M_{b%
%TCIMACRO{\U{b4}}%
%BeginExpansion
{\acute{}}%
%EndExpansion
}^{b}\right] \!\!\Lambda _{\gamma
%TCIMACRO{\U{b4}}%
%BeginExpansion
{\acute{}}%
%EndExpansion
}^{\gamma }{\circ }h. \leqno(3.4.11^{\prime })
\end{equation*}

\emph{If }$\rho \Gamma $\emph{\ is a }$\rho $\emph{-connection for the
vector bundle }$\left( E,\pi ,M\right) $\emph{\ and }$h=Id_{M},$\emph{\ then
the relations }$\left( 3.4.11^{\prime }\right) $\emph{\ become}%
\begin{equation*}
\rho \Gamma _{b%
%TCIMACRO{\U{b4}}%
%BeginExpansion
{\acute{}}%
%EndExpansion
\gamma
%TCIMACRO{\U{b4}}%
%BeginExpansion
{\acute{}}%
%EndExpansion
}^{a%
%TCIMACRO{\U{b4}}%
%BeginExpansion
{\acute{}}%
%EndExpansion
}=M_{a}^{a%
%TCIMACRO{\U{b4}}%
%BeginExpansion
{\acute{}}%
%EndExpansion
}\left[ \rho _{\gamma }^{k}\frac{\partial M_{b%
%TCIMACRO{\U{b4}}%
%BeginExpansion
{\acute{}}%
%EndExpansion
}^{a}}{\partial x^{k}}+\rho \Gamma _{b\gamma }^{a}M_{b%
%TCIMACRO{\U{b4}}%
%BeginExpansion
{\acute{}}%
%EndExpansion
}^{b}\right] \Lambda _{\gamma
%TCIMACRO{\U{b4}}%
%BeginExpansion
{\acute{}}%
%EndExpansion
}^{\gamma }. \leqno(3.4.11^{\prime \prime })
\end{equation*}

\emph{In particular, if }$\left( \rho ,\eta \right) =\left(
Id_{TM},Id_{M}\right) $\emph{, then the relations }$\left( 3.4.11^{\prime
\prime }\right) $\emph{\ become}%
\begin{equation*}
\Gamma _{j%
%TCIMACRO{\U{b4}}%
%BeginExpansion
{\acute{}}%
%EndExpansion
k%
%TCIMACRO{\U{b4}}%
%BeginExpansion
{\acute{}}%
%EndExpansion
}^{i%
%TCIMACRO{\U{b4}}%
%BeginExpansion
{\acute{}}%
%EndExpansion
}=\frac{\partial x^{i%
%TCIMACRO{\U{b4}}%
%BeginExpansion
{\acute{}}%
%EndExpansion
}}{\partial x^{i}}\left[ \frac{\partial }{\partial x^{k}}\left( \frac{%
\partial x^{i}}{\partial x^{j%
%TCIMACRO{\U{b4}}%
%BeginExpansion
{\acute{}}%
%EndExpansion
}}\right) +\Gamma _{jk}^{i}\frac{\partial x^{j}}{\partial x^{j%
%TCIMACRO{\U{b4}}%
%BeginExpansion
{\acute{}}%
%EndExpansion
}}\right] \frac{\partial x^{k}}{\partial x^{k%
%TCIMACRO{\U{b4}}%
%BeginExpansion
{\acute{}}%
%EndExpansion
}}. \leqno(3.4.11^{\prime \prime \prime })
\end{equation*}

\smallskip \noindent\textbf{Definition 3.4.7 }We say that the (linear) $%
\left( \rho ,\eta \right) $-connection $\left( \rho ,\eta \right) \Gamma $
for the fiber bundle $\left( E,\pi ,M\right) $ is differentiable of $C^{r}$
class, if its components are differentiable of $C^{r}$ class.

\bigskip \noindent \textbf{Definition 3.4.8 }If $\left( \rho ,\eta \right)
\Gamma $ is a linear $\left( \rho ,\eta \right) $-connection for the vector
bundle $\left( E,\pi ,M\right) $, then for any
\begin{equation*}
z=z^{\alpha }t_{\alpha }\in \Gamma \left( F,\nu ,N\right)
\end{equation*}%
the application%
\begin{equation*}
\begin{array}{rcl}
\Gamma \left( E,\pi ,M\right) \!\! & \!\!^{\underrightarrow{\left( \rho
,\eta \right) D_{z}}}\!\! & \!\!\Gamma \left( E,\pi ,M\right) \\
u{=}u^{a}s_{a}\!\! & \!\!\longmapsto \!\! & \!\!(\rho ,\eta )D_{z}u%
\end{array}%
\leqno(3.4.12)
\end{equation*}%
defined by
\begin{equation*}
\!\!(\rho ,\eta )D_{z}u{=}z^{\alpha }{\circ }h\left( \rho _{\alpha }^{i}{%
\circ }h\frac{\partial u^{a}}{\partial x^{i}}{+}(\rho ,\eta )\Gamma
_{b\alpha }^{a}\cdot u^{b}\right) s_{a},
\end{equation*}%
will be called the \emph{covariant }$\left( \rho ,\eta \right) $\emph{%
-derivative associated to linear }$\left( \rho ,\eta \right) $\emph{%
-connection }$\left( \rho ,\eta \right) \Gamma $\emph{\ with respect to the
section }$z$.

If $h=Id_{M}$ and $\eta =Id_{M},$ then we obtain the \emph{covariant }$\rho $%
\emph{-derivative associated to linear }$\rho $\emph{-connection }$\rho
\Gamma $\emph{\ with respect to the section }$z$\emph{.}

In addition, if $\rho =Id_{TM}$, then we obtain the \emph{covariant
derivative associated to linear connection }$\Gamma $\emph{\ with respect to
the vector field }$z$.

\subsubsection{(Linear) $\left( \protect\rho ,\protect\eta \right) $%
-connections for dual of vector bundles}

Let $\left( E,\pi ,M\right) $ be a vector bundle.

We consider the following diagram:%
\begin{equation*}
\begin{array}{rcl}
\overset{\ast }{E} &  & \left( F,\left[ ,\right] _{F,h},\left( \rho ,\eta
\right) \right) \\
\overset{\ast }{\pi }\downarrow &  & ~\downarrow \nu \\
M & ^{\underrightarrow{~\ \ \ \ h~\ \ \ \ }} & ~\ N%
\end{array}%
,\leqno(3.4.1.1)
\end{equation*}%
where $\left( \left( F,\nu ,N\right) ,\left[ ,\right] _{F,h},\left( \rho
,\eta \right) \right) $ is a generalized Lie algebroid.

Let
\begin{equation*}
\left( \left( \left( \rho ,\eta \right) T\overset{\ast }{E},\left( \rho
,\eta \right) \tau _{\overset{\ast }{E}},\overset{\ast }{E}\right) ,\left[ ,%
\right] _{\left( \rho ,\eta \right) T\overset{\ast }{E}},\left( \overset{%
\ast }{\tilde{\rho}},Id_{\overset{\ast }{E}}\right) \right)
\end{equation*}%
\ be the Lie algebroid generalized tangent bundle of the vector bundle $%
\left( \overset{\ast }{E},\overset{\ast }{\pi },M\right) $.

We consider the $\mathbf{B}^{\mathbf{v}}$-morphism $\left( \left( \rho ,\eta
\right) \overset{\ast }{\pi }!,Id_{\overset{\ast }{E}}\right) $ given by the
commutative diagram%
\begin{equation*}
\begin{array}{c}
\xymatrix{\left( \rho ,\eta \right) T\overset{\ast }{E}\ar[r]^{( \rho ,\eta
) \overset{\ast }{\pi!} }\ar[d]_{(\rho,\eta)\tau_{\overset{\ast }{E}}}&
\overset{\ast }{\pi }^{\ast }\left( h^{\ast }F\right) \ar[d]^{pr_1} \\
\overset{\ast }{E}\ar[r]^{id_{\overset{\ast }{E}}}& \overset{\ast }{E}}%
\end{array}%
\leqno(3.4.1.2)
\end{equation*}%
Using the components, this is defined as:%
\begin{equation*}
\left( \rho ,\eta \right) \overset{\ast }{\pi }!\left( \tilde{Z}^{\alpha }%
\frac{\partial }{\partial \tilde{z}^{\alpha }}+Y_{a}\frac{\partial }{%
\partial \tilde{p}_{a}}\right) \left( \overset{\ast }{u}_{x}\right) =\left(
\tilde{Z}^{\alpha }\tilde{T}_{\alpha }\right) \left( \overset{\ast }{u}%
_{x}\right) ,\leqno(3.4.1.3)
\end{equation*}%
for any $\displaystyle\tilde{Z}^{\alpha }\frac{\partial }{\partial \tilde{z}%
^{\alpha }}+Y_{a}\frac{\partial }{\partial \tilde{p}_{a}}\in \left( \left(
\rho ,\eta \right) T\overset{\ast }{E},\left( \rho ,\eta \right) \tau _{%
\overset{\ast }{E}},\overset{\ast }{E}\right) .$\medskip

We define the \emph{tangent }$\left( \rho ,\eta \right) $\emph{-application }%
as being a $\mathbf{B}^{\mathbf{v}}$-morphism
\begin{equation*}
\begin{array}{c}
\left( \left( \rho ,\eta \right) T\overset{\ast }{\pi },h\circ \overset{\ast
}{\pi }\right) =\left( pr_{2},h\circ \overset{\ast }{\pi }\right) \circ
\left( \left( \rho ,\eta \right) \overset{\ast }{\pi }!,Id_{\overset{\ast }{E%
}}\right)%
\end{array}%
\leqno(3.4.1.4)
\end{equation*}%
of $\left( \left( \rho ,\eta \right) T\overset{\ast }{E},\left( \rho ,\eta
\right) \tau _{\overset{\ast }{E}},\overset{\ast }{E}\right) $ source and $%
\left( F,\nu ,N\right) $ target.

\bigskip\noindent\textbf{Definition~3.4.1.1} The kernel of the tangent $%
\left( \rho ,\eta \right) $-application\break
\begin{equation*}
\left( \left( \rho ,\eta \right) T\overset{\ast }{\pi },h\circ \overset{\ast
}{\pi }\right)
\end{equation*}%
\ is written as%
\begin{equation*}
\left( V\left( \rho ,\eta \right) T\overset{\ast }{E},\left( \rho ,\eta
\right) \tau _{\overset{\ast }{E}},\overset{\ast }{E}\right)
\end{equation*}%
and will be called the \emph{vertical subbundle}.\bigskip

The set $\left\{ \displaystyle\frac{\partial }{\partial \tilde{p}_{a}},~a\in
\overline{1,r}\right\} $ is a base for the $\mathcal{F}\left( \overset{\ast }%
{E}\right) $-module
\begin{equation*}
\left( \Gamma \left( V\left( \rho ,\eta \right) T\overset{\ast }{E},\left(
\rho ,\eta \right) \tau _{\overset{\ast }{E}},\overset{\ast }{E}\right)
,+,\cdot \right) .
\end{equation*}

\smallskip \noindent\textbf{Proposition 3.4.1.1} \emph{The short sequence of
vector bundles }%
\begin{equation*}
\begin{array}{c}
\xymatrix{0\ar@{^(->}[r]^i\ar[d]&V(\rho,\eta)T\overset{\ast
}{E}\ar[d]\ar@{^(->}[r]^i&(\rho,\eta)T\overset{\ast
}{E}\ar[r]^{(\rho,\eta)\overset{\ast }{\pi }!}\ar[d]&\overset{\ast }{\pi
}^{\ast }\left( h^{\ast }F\right)\ar[r]\ar[d]&0\ar[d]\\ \overset{\ast
}{E}\ar[r]^{Id_{\overset{\ast }{E}}}&\overset{\ast
}{E}\ar[r]^{Id_{\overset{\ast }{E}}}&\overset{\ast
}{E}\ar[r]^{Id_{\overset{\ast }{E}}}&\overset{\ast }
{E}\ar[r]^{Id_{\overset{\ast }{E}}}&\overset{\ast }{E}}%
\end{array}%
\end{equation*}%
\emph{is exact.}

\bigskip \noindent \textbf{Definition 3.4.1.2} \textit{A }$\mathbf{Man}$%
-morphism $\left( \rho ,\eta \right) \overset{\ast }{\Gamma }$ of $\left(
\rho ,\eta \right) T\overset{\ast }{E}$ source and $V\left( \rho ,\eta
\right) T\overset{\ast }{E}$ target defined by%
\begin{equation*}
\left( \rho ,\eta \right) \Gamma \left( \tilde{Z}^{\alpha }\frac{\partial }{%
\partial \tilde{z}^{\alpha }}+Y_{b}\frac{\partial }{\partial \tilde{p}_{b}}%
\right) \left( \overset{\ast }{u}_{x}\right) =\left( Y_{b}-\left( \rho ,\eta
\right) \overset{\ast }{\Gamma }_{b\alpha }\tilde{Z}^{\alpha }\right) \frac{%
\partial }{\partial \tilde{p}_{b}}\left( \overset{\ast }{u}_{x}\right) ,%
\leqno(3.4.1.5)
\end{equation*}%
such that the $\mathbf{B}^{\mathbf{v}}$-morphism $\left( \left( \rho ,\eta
\right) \overset{\ast }{\Gamma },Id_{\overset{\ast }{E}}\right) $ is a split
to the left in the previous exact sequence, will be called $\left( \rho
,\eta \right) $\emph{-connection for the dual vector bundle }$\left( \overset%
{\ast }{E},\overset{\ast }{\pi },M\right) $.

The differentiable real local functions $\left( \rho ,\eta \right) \overset{%
\ast }{\Gamma }_{b\alpha }$ will be called the \emph{components of }$\left(
\rho ,\eta \right) $\emph{-connection }$\left( \rho ,\eta \right) \overset{%
\ast }{\Gamma }.$

The $\left( \rho ,Id_{M}\right) $-connection for the dual vector bundle $%
\left( \overset{\ast }{E},\overset{\ast }{\pi },M\right) $ will be called $%
\rho $\emph{-connection for the dual vector bundle }$\left( \overset{\ast }{E%
},\overset{\ast }{\pi },M\right) $ and will be denoted $\rho \overset{\ast }{%
\Gamma }$.

The $\left( Id_{TM},Id_{M}\right) $-connection for the dual vector bundle $%
\left( \overset{\ast }{E},\overset{\ast }{\pi },M\right) $ will be called
\emph{connection for the dual vector bundle }$\left( \overset{\ast }{E},%
\overset{\ast }{\pi },M\right) $ and will be denoted $\overset{\ast }{\Gamma
}$.

Let $\left\{ s^{a},~a\in \overline{1,r}\right\} $ be the dual base of the
base $\left\{ s_{a},a\in \overline{1,r}\right\} .$

The $\mathbf{B}^{\mathbf{v}}$-morphism $\left( \overset{\ast }{\Pi },\overset%
{\ast }{\pi }\right) $ defined by the commutative diagram%
\begin{equation*}
\begin{array}{c}
\xymatrix{V\left( \rho ,\eta \right) T\overset{\ast }{E}\ar[r]^{\qquad
\overset{\ast }{\Pi } }\ar[d]_{(\rho,\eta)\tau_{\overset{\ast
}{E}}}&\overset{\ast }{E} \ar[d]^{\overset{\ast }{\pi }} \\ \overset{\ast
}{E}\ar[r]^{\overset{\ast }{\pi }}&M}%
\end{array}%
,\leqno(3.4.1.6)
\end{equation*}%
where, $\overset{\ast }{\Pi }$\textit{\ }is defined by\textit{\ }%
\begin{equation*}
\overset{\ast }{\Pi }\left( Y_{a}\frac{\partial }{\partial \tilde{p}_{a}}%
\left( \overset{\ast }{u}_{x}\right) \right) =Y_{a}\left( \overset{\ast }{u}%
_{x}\right) s^{a}\left( \overset{\ast }{\pi }\left( \overset{\ast }{u}%
_{x}\right) \right) ,\leqno(3.4.1.7)
\end{equation*}%
is canonical projection $\mathbf{B}^{\mathbf{v}}$-morphism$.$

\bigskip\noindent\textbf{Theorem 3.4.1.1 }\emph{If }$\left( \rho ,\eta
\right) \overset{\ast }{\Gamma }$\emph{\ is a }$\left( \rho ,\eta \right) $%
\emph{-connection for the vector bundle }$\left( \overset{\ast }{E},\overset{%
\ast }{\pi },M\right) ,$\emph{\ then its components satisfy the law of
transformation }%
\begin{equation*}
\left( \rho ,\eta \right) \overset{\ast }{\Gamma }_{b%
%TCIMACRO{\U{b4}}%
%BeginExpansion
{\acute{}}%
%EndExpansion
\gamma
%TCIMACRO{\U{b4}}%
%BeginExpansion
{\acute{}}%
%EndExpansion
}=M_{b%
%TCIMACRO{\U{b4}}%
%BeginExpansion
{\acute{}}%
%EndExpansion
}^{b}{\circ }\overset{\ast }{\pi }\left[ -\rho _{\gamma }^{i}\circ h\circ
\overset{\ast }{\pi }\cdot \frac{\partial M_{b}^{a%
%TCIMACRO{\U{b4}}%
%BeginExpansion
{\acute{}}%
%EndExpansion
}{\circ }\overset{\ast }{\pi }}{\partial x^{i}}p_{a%
%TCIMACRO{\U{b4}}%
%BeginExpansion
{\acute{}}%
%EndExpansion
}+\left( \rho ,\eta \right) \overset{\ast }{\Gamma }_{b\gamma }\right]
\Lambda _{\gamma
%TCIMACRO{\U{b4}}%
%BeginExpansion
{\acute{}}%
%EndExpansion
}^{\gamma }\circ \left( h\circ \overset{\ast }{\pi }\right) . \leqno(3.4.1.8)
\end{equation*}
\emph{In particular, if }$h=Id_{M}$\emph{, then the relations }$\left(
3.4.1.8\right) $\emph{\ become}%
\begin{equation*}
\left( \rho ,\eta \right) \overset{\ast }{\Gamma }_{b%
%TCIMACRO{\U{b4}}%
%BeginExpansion
{\acute{}}%
%EndExpansion
\gamma
%TCIMACRO{\U{b4}}%
%BeginExpansion
{\acute{}}%
%EndExpansion
}=M_{b%
%TCIMACRO{\U{b4}}%
%BeginExpansion
{\acute{}}%
%EndExpansion
}^{b}{\circ }\overset{\ast }{\pi }\!\!\left[ -\rho _{\gamma }^{i}\circ
\overset{\ast }{\pi }\cdot \frac{\partial M_{b}^{a%
%TCIMACRO{\U{b4}}%
%BeginExpansion
{\acute{}}%
%EndExpansion
}{\circ }\overset{\ast }{\pi }}{\partial x^{i}}p_{a%
%TCIMACRO{\U{b4}}%
%BeginExpansion
{\acute{}}%
%EndExpansion
}+\left( \rho ,\eta \right) \overset{\ast }{\Gamma }_{b\gamma }\right]
\Lambda _{\gamma
%TCIMACRO{\U{b4}}%
%BeginExpansion
{\acute{}}%
%EndExpansion
}^{\gamma }\circ \overset{\ast }{\pi }. \leqno(3.4.1.8^{\prime })
\end{equation*}
\emph{In particular, if }$\left( \rho ,\eta \right) =\left(
Id_{TM},Id_{M}\right) $\emph{, then the relations }$\left( 3.4.1.8^{\prime
}\right) $\emph{\ become}%
\begin{equation*}
\overset{\ast }{\Gamma }_{j%
%TCIMACRO{\U{b4}}%
%BeginExpansion
{\acute{}}%
%EndExpansion
k%
%TCIMACRO{\U{b4}}%
%BeginExpansion
{\acute{}}%
%EndExpansion
}=\frac{\partial x^{j}}{\partial x^{j%
%TCIMACRO{\U{b4}}%
%BeginExpansion
{\acute{}}%
%EndExpansion
}}{\circ }\overset{\ast }{\pi }\!\!\left[ -\frac{\partial }{\partial x^{i}}%
\left( \frac{\partial x^{i%
%TCIMACRO{\U{b4}}%
%BeginExpansion
{\acute{}}%
%EndExpansion
}}{\partial x^{j}}{\circ }\overset{\ast }{\pi }\right) p_{i%
%TCIMACRO{\U{b4}}%
%BeginExpansion
{\acute{}}%
%EndExpansion
}+\overset{\ast }{\Gamma }_{jk}\right] \frac{\partial x^{k}}{\partial x^{k%
%TCIMACRO{\U{b4}}%
%BeginExpansion
{\acute{}}%
%EndExpansion
}}\circ \overset{\ast }{\pi }. \leqno(3.4.1.8^{\prime \prime })
\end{equation*}

\smallskip \noindent\textit{Proof.} Let $\left( \overset{\ast }{\Pi },%
\overset{\ast }{\pi }\right) $ be the canonical projection $\mathbf{B}$%
-morphism.

Obviously, the components of
\begin{equation*}
\Pi ^{\ast }\circ \left( \rho ,\eta \right) \overset{\ast }{\Gamma }\left(
\tilde{Z}^{\alpha
%TCIMACRO{\U{b4}}%
%BeginExpansion
{\acute{}}%
%EndExpansion
}\frac{\partial }{\partial \tilde{z}^{\alpha
%TCIMACRO{\U{b4}}%
%BeginExpansion
{\acute{}}%
%EndExpansion
}}+Y_{b%
%TCIMACRO{\U{b4}}%
%BeginExpansion
{\acute{}}%
%EndExpansion
}\frac{\partial }{\partial \tilde{p}_{b%
%TCIMACRO{\U{b4}}%
%BeginExpansion
{\acute{}}%
%EndExpansion
}}\right) \left( \overset{\ast }{u}_{x}\right)
\end{equation*}%
are the real numbers%
\begin{equation*}
\left( Y_{b%
%TCIMACRO{\U{b4}}%
%BeginExpansion
{\acute{}}%
%EndExpansion
}-\left( \rho ,\eta \right) \overset{\ast }{\Gamma }_{b%
%TCIMACRO{\U{b4}}%
%BeginExpansion
{\acute{}}%
%EndExpansion
\gamma
%TCIMACRO{\U{b4}}%
%BeginExpansion
{\acute{}}%
%EndExpansion
}\tilde{Z}^{\gamma
%TCIMACRO{\U{b4}}%
%BeginExpansion
{\acute{}}%
%EndExpansion
}\right) \left( \overset{\ast }{u}_{x}\right) .
\end{equation*}%
Since
\begin{equation*}
\begin{array}{l}
\displaystyle\left( \tilde{Z}^{\alpha
%TCIMACRO{\U{b4}}%
%BeginExpansion
{\acute{}}%
%EndExpansion
}\frac{\partial }{\partial \tilde{z}^{\alpha
%TCIMACRO{\U{b4}}%
%BeginExpansion
{\acute{}}%
%EndExpansion
}}+Y_{b%
%TCIMACRO{\U{b4}}%
%BeginExpansion
{\acute{}}%
%EndExpansion
}\frac{\partial }{\partial \tilde{p}_{b%
%TCIMACRO{\U{b4}}%
%BeginExpansion
{\acute{}}%
%EndExpansion
}}\right) \left( \overset{\ast }{u}_{x}\right) =\tilde{Z}^{\alpha
%TCIMACRO{\U{b4}}%
%BeginExpansion
{\acute{}}%
%EndExpansion
}\Lambda _{\alpha
%TCIMACRO{\U{b4}}%
%BeginExpansion
{\acute{}}%
%EndExpansion
}^{\alpha }\circ h\circ \overset{\ast }{\pi }\frac{\partial }{\partial
\tilde{z}^{\alpha }}\left( \overset{\ast }{u}_{x}\right) \vspace*{1mm} \\
\qquad \displaystyle+\left( \tilde{Z}^{\alpha
%TCIMACRO{\U{b4}}%
%BeginExpansion
{\acute{}}%
%EndExpansion
}\rho _{\alpha
%TCIMACRO{\U{b4}}%
%BeginExpansion
{\acute{}}%
%EndExpansion
}^{i%
%TCIMACRO{\U{b4}}%
%BeginExpansion
{\acute{}}%
%EndExpansion
}\circ h\circ \overset{\ast }{\pi }\frac{\partial M_{b}^{a%
%TCIMACRO{\U{b4}}%
%BeginExpansion
{\acute{}}%
%EndExpansion
}{\circ }\pi }{\partial x^{i%
%TCIMACRO{\U{b4}}%
%BeginExpansion
{\acute{}}%
%EndExpansion
}}p_{a%
%TCIMACRO{\U{b4}}%
%BeginExpansion
{\acute{}}%
%EndExpansion
}+M_{b}^{b%
%TCIMACRO{\U{b4}}%
%BeginExpansion
{\acute{}}%
%EndExpansion
}Y_{b%
%TCIMACRO{\U{b4}}%
%BeginExpansion
{\acute{}}%
%EndExpansion
}\right) \frac{\partial }{\partial \tilde{p}_{b}}\left( \overset{\ast }{u}%
_{x}\right) ,%
\end{array}%
\end{equation*}%
it results that the components of%
\begin{equation*}
\Pi ^{\ast }\circ \left( \rho ,\eta \right) \overset{\ast }{\Gamma }\left(
\tilde{Z}^{\alpha
%TCIMACRO{\U{b4}}%
%BeginExpansion
{\acute{}}%
%EndExpansion
}\frac{\partial }{\partial \tilde{z}^{\alpha
%TCIMACRO{\U{b4}}%
%BeginExpansion
{\acute{}}%
%EndExpansion
}}+Y_{b%
%TCIMACRO{\U{b4}}%
%BeginExpansion
{\acute{}}%
%EndExpansion
}\frac{\partial }{\partial \tilde{p}_{b%
%TCIMACRO{\U{b4}}%
%BeginExpansion
{\acute{}}%
%EndExpansion
}}\right) \left( \overset{\ast }{u}_{x}\right)
\end{equation*}%
are the real numbers
\begin{equation*}
\left( \tilde{Z}^{\alpha
%TCIMACRO{\U{b4}}%
%BeginExpansion
{\acute{}}%
%EndExpansion
}\rho _{\alpha
%TCIMACRO{\U{b4}}%
%BeginExpansion
{\acute{}}%
%EndExpansion
}^{i%
%TCIMACRO{\U{b4}}%
%BeginExpansion
{\acute{}}%
%EndExpansion
}\circ h\circ \overset{\ast }{\pi }\frac{\partial M_{b}^{a%
%TCIMACRO{\U{b4}}%
%BeginExpansion
{\acute{}}%
%EndExpansion
}{\circ }\overset{\ast }{\pi }}{\partial x^{i%
%TCIMACRO{\U{b4}}%
%BeginExpansion
{\acute{}}%
%EndExpansion
}}p_{a%
%TCIMACRO{\U{b4}}%
%BeginExpansion
{\acute{}}%
%EndExpansion
}+M_{b}^{b%
%TCIMACRO{\U{b4}}%
%BeginExpansion
{\acute{}}%
%EndExpansion
}\circ \overset{\ast }{\pi }Y_{b%
%TCIMACRO{\U{b4}}%
%BeginExpansion
{\acute{}}%
%EndExpansion
}-\left( \rho ,\eta \right) \overset{\ast }{\Gamma }_{b\alpha }\tilde{Z}%
^{\alpha
%TCIMACRO{\U{b4}}%
%BeginExpansion
{\acute{}}%
%EndExpansion
}\Lambda _{\alpha
%TCIMACRO{\U{b4}}%
%BeginExpansion
{\acute{}}%
%EndExpansion
}^{\alpha }\circ h\circ \overset{\ast }{\pi }\right) M_{b%
%TCIMACRO{\U{b4}}%
%BeginExpansion
{\acute{}}%
%EndExpansion
}^{b}\circ \overset{\ast }{\pi }\left( \overset{\ast }{u}_{x}\right) ,
\end{equation*}%
where $\left\Vert M_{b%
%TCIMACRO{\U{b4}}%
%BeginExpansion
{\acute{}}%
%EndExpansion
}^{b}\right\Vert =\left\Vert M_{b}^{b%
%TCIMACRO{\U{b4}}%
%BeginExpansion
{\acute{}}%
%EndExpansion
}\right\Vert ^{-1}.$\medskip

Therefore, we have:%
\begin{equation*}
\begin{array}{l}
\displaystyle\left( \tilde{Z}^{\alpha
%TCIMACRO{\U{b4}}%
%BeginExpansion
{\acute{}}%
%EndExpansion
}\rho _{\alpha
%TCIMACRO{\U{b4}}%
%BeginExpansion
{\acute{}}%
%EndExpansion
}^{i%
%TCIMACRO{\U{b4}}%
%BeginExpansion
{\acute{}}%
%EndExpansion
}\circ h\circ \overset{\ast }{\pi }\frac{\partial M_{b}^{a%
%TCIMACRO{\U{b4}}%
%BeginExpansion
{\acute{}}%
%EndExpansion
}{\circ }\overset{\ast }{\pi }}{\partial x^{i%
%TCIMACRO{\U{b4}}%
%BeginExpansion
{\acute{}}%
%EndExpansion
}}p_{a%
%TCIMACRO{\U{b4}}%
%BeginExpansion
{\acute{}}%
%EndExpansion
}+M_{b}^{b%
%TCIMACRO{\U{b4}}%
%BeginExpansion
{\acute{}}%
%EndExpansion
}\circ \overset{\ast }{\pi }Y_{b%
%TCIMACRO{\U{b4}}%
%BeginExpansion
{\acute{}}%
%EndExpansion
}-\left( \rho ,\eta \right) \overset{\ast }{\Gamma }_{b\alpha }\tilde{Z}%
^{\alpha
%TCIMACRO{\U{b4}}%
%BeginExpansion
{\acute{}}%
%EndExpansion
}\Lambda _{\alpha
%TCIMACRO{\U{b4}}%
%BeginExpansion
{\acute{}}%
%EndExpansion
}^{\alpha }\circ h\circ \overset{\ast }{\pi }\right) M_{b%
%TCIMACRO{\U{b4}}%
%BeginExpansion
{\acute{}}%
%EndExpansion
}^{b}\circ \overset{\ast }{\pi }\vspace*{1mm} \\
\qquad\qquad\displaystyle=Y_{b%
%TCIMACRO{\U{b4}}%
%BeginExpansion
{\acute{}}%
%EndExpansion
}-\left( \rho ,\eta \right) \overset{\ast }{\Gamma }_{b%
%TCIMACRO{\U{b4}}%
%BeginExpansion
{\acute{}}%
%EndExpansion
\alpha
%TCIMACRO{\U{b4}}%
%BeginExpansion
{\acute{}}%
%EndExpansion
}\tilde{Z}^{\alpha
%TCIMACRO{\U{b4}}%
%BeginExpansion
{\acute{}}%
%EndExpansion
}.%
\end{array}%
\end{equation*}%
After some calculations we obtain:
\begin{equation*}
\left( \rho ,\eta \right) \overset{\ast }{\Gamma }_{b%
%TCIMACRO{\U{b4}}%
%BeginExpansion
{\acute{}}%
%EndExpansion
\alpha
%TCIMACRO{\U{b4}}%
%BeginExpansion
{\acute{}}%
%EndExpansion
}=M_{b%
%TCIMACRO{\U{b4}}%
%BeginExpansion
{\acute{}}%
%EndExpansion
}^{b}\circ \overset{\ast }{\pi }\left( -\rho _{\alpha }^{i}\circ h\circ
\overset{\ast }{\pi }\cdot \frac{\partial M_{b}^{a%
%TCIMACRO{\U{b4}}%
%BeginExpansion
{\acute{}}%
%EndExpansion
}{\circ }\overset{\ast }{\pi }}{\partial x^{i}}p_{a%
%TCIMACRO{\U{b4}}%
%BeginExpansion
{\acute{}}%
%EndExpansion
}+\left( \rho ,\eta \right) \overset{\ast }{\Gamma }_{b\alpha }\right)
\Lambda _{\alpha
%TCIMACRO{\U{b4}}%
%BeginExpansion
{\acute{}}%
%EndExpansion
}^{\alpha }\circ h\circ \overset{\ast }{\pi }.\eqno{q.e.d.}
\end{equation*}

\smallskip \noindent\textbf{Remark 3.4.1.1} If we have a set of real local
functions $\left( \rho ,\eta \right) \overset{\ast }{\Gamma }_{b\gamma }$
which satisfies the relations of passing $\left( 3.4.1.8\right) ,$ then we
have a $\left( \rho ,\eta \right) $-connection $\left( \rho ,\eta \right)
\overset{\ast }{\Gamma }$ for the fiber bundle $\left( \overset{\ast }{E},%
\overset{\ast }{\pi },M\right) .$

\bigskip\noindent\textbf{Example 3.4.1.1} If $\overset{\ast }{\Gamma }$ is a
classical connection for the vector bundle $\left( \overset{\ast }{E},%
\overset{\ast }{\pi },M\right) $ on components $\overset{\ast }{\Gamma }%
_{bk},$ then the differentiable real local functions
\begin{equation*}
\left( \rho ,\eta \right) \overset{\ast }{\Gamma }_{b\gamma }=\left( \rho
_{\gamma }^{k}\circ h\circ \overset{\ast }{\pi }\right) \overset{\ast }{%
\Gamma }_{bk}
\end{equation*}%
are the components of a $\left( \rho ,\eta \right) $-connection $\left( \rho
,\eta \right) \overset{\ast }{\Gamma }$ for the vector bundle $\left(
\overset{\ast }{E},\overset{\ast }{\pi },M\right) $ which will be called the
$\left( \rho ,\eta \right) $\emph{-connection associated to the connection }$%
\overset{\ast }{\Gamma }.$

\bigskip \noindent \textbf{Definition 3.4.1.3 }If $\left( \rho ,\eta \right)
\overset{\ast }{\Gamma }$ is a $\left( \rho ,\eta \right) $-connection for
the vector bundle $\left( \overset{\ast }{E},\overset{\ast }{\pi },M\right) $%
, then for any
\begin{equation*}
z=z^{\alpha }t_{\alpha }\in \Gamma \left( F,\nu ,N\right)
\end{equation*}%
the application%
\begin{equation*}
\begin{array}{rcl}
\Gamma \left( \overset{\ast }{E},\overset{\ast }{\pi },M\right) & ^{%
\underrightarrow{\left( \rho ,\eta \right) D_{z}}} & \Gamma \left( \overset{%
\ast }{E},\overset{\ast }{\pi },M\right) \\
\overset{\ast }{u}=u_{a}s^{a} & \longmapsto & \left( \rho ,\eta \right) D_{z}%
\overset{\ast }{u}%
\end{array}%
\leqno(3.4.1.9)
\end{equation*}%
defined by
\begin{equation*}
\left( \rho ,\eta \right) D_{z}\overset{\ast }{u}=z^{\alpha }\circ h\left(
\rho _{\alpha }^{i}\circ h\frac{\partial u_{b}}{\partial x^{i}}-\left( \rho
,\eta \right) \overset{\ast }{\Gamma }_{b\alpha }\circ \overset{\ast }{u}%
\right) s^{b},
\end{equation*}%
will be called the \emph{covariant }$\left( \rho ,\eta \right) $\emph{%
-derivative associated to }$\left( \rho ,\eta \right) $\emph{-con\-nec\-tion
}$\left( \rho ,\eta \right) \overset{\ast }{\Gamma }$\emph{\ with respect to
the section }$z$\emph{.}

If $h=Id_{M}$ and $\eta =Id_{M},$ then we obtain the \emph{covariant }$\rho $%
\emph{-derivative associated to }$\rho $\emph{-connection }$\rho \overset{%
\ast }{\Gamma }$\emph{\ with respect to the section }$z$\emph{.}

In addition, if $\rho =Id_{TM}$, then we obtain the \emph{covariant
derivative associated to connection }$\overset{\ast }{\Gamma }$\emph{\ with
respect to the vector field }$z$.

\bigskip\noindent\textbf{Definition 3.4.1.4 }We will say that the $\left(
\rho ,\eta \right) $\emph{-connection }$\left( \rho ,\eta \right) \overset{%
\ast }{\Gamma }$\emph{\ is homogeneous} or \emph{linear }if the local real
functions $\left( \rho ,\eta \right) \overset{\ast }{\Gamma }_{b\gamma }$
are homogeneous or linear on the fibre of vector bundle $\left( \overset{%
\ast }{E},\overset{\ast }{\pi },M\right) $ respectively.

\bigskip \noindent \textbf{Remark 3.4.1.2} If $\left( \rho ,\eta \right)
\overset{\ast }{\Gamma }$ is a linear $\left( \rho ,\eta \right) $%
-connection for the vector bundle $\left( \overset{\ast }{E},\overset{\ast }{%
\pi },M\right) $, then for each local vector $\left( m+r\right) $-chart $%
\left( U,\overset{\ast }{s}_{U}\right) $ and for each local vector $\left(
n+p\right) $-chart $\left( V,t_{V}\right) $ such that $U\cap h^{-1}\left(
V\right) \neq \phi $, there exists the differentiable real functions $\rho
\Gamma _{b\gamma }^{a}$ defined on $U\cap h^{-1}\left( V\right) $ such that
\begin{equation*}
\begin{array}{c}
\left( \rho ,\eta \right) \overset{\ast }{\Gamma }_{b\gamma }\circ \overset{%
\ast }{u}=\left( \rho ,\eta \right) \Gamma _{b\gamma }^{a}\cdot u_{a},%
\end{array}%
\leqno(3.4.1.10)
\end{equation*}%
for any $\overset{\ast }{u}=u_{a}s^{a}\in \Gamma \left( \overset{\ast }{E},%
\overset{\ast }{\pi },M\right) .$

The differentiable real local functions $\left( \rho ,\eta \right) \Gamma
_{b\alpha }^{a}$ will be called the \emph{Christoffel coefficients of linear
}$\left( \rho ,\eta \right) $\emph{-connection }$\left( \rho ,\eta \right)
\Gamma .$

\bigskip\noindent\textbf{Theorem 3.4.1.2} \emph{If }$\left( \rho ,\eta
\right) \Gamma $\emph{\ is a linear }$\left( \rho ,\eta \right) $\emph{%
-connection for the vector bundle }$\left( \overset{\ast }{E},\overset{\ast }%
{\pi },M\right) ,$\emph{\ then its components satisfy the law of
transformation }%
\begin{equation*}
\left( \rho ,\eta \right) \Gamma _{b%
%TCIMACRO{\U{b4}}%
%BeginExpansion
{\acute{}}%
%EndExpansion
\gamma
%TCIMACRO{\U{b4}}%
%BeginExpansion
{\acute{}}%
%EndExpansion
}^{a%
%TCIMACRO{\U{b4}}%
%BeginExpansion
{\acute{}}%
%EndExpansion
}=M_{b%
%TCIMACRO{\U{b4}}%
%BeginExpansion
{\acute{}}%
%EndExpansion
}^{b}\left[ -\rho _{\gamma }^{i}\circ h\frac{\partial M_{b}^{a%
%TCIMACRO{\U{b4}}%
%BeginExpansion
{\acute{}}%
%EndExpansion
}}{\partial x^{i}}+\left( \rho ,\eta \right) \Gamma _{b\gamma }^{a}M_{a}^{a%
%TCIMACRO{\U{b4}}%
%BeginExpansion
{\acute{}}%
%EndExpansion
}\right] \Lambda _{\gamma
%TCIMACRO{\U{b4}}%
%BeginExpansion
{\acute{}}%
%EndExpansion
}^{\gamma }\circ h. \leqno(3.4.1.11)
\end{equation*}
\emph{In particular, if }$\left( \rho ,\eta \right) =\left(
Id_{TM},Id_{M}\right) $\emph{\ and }$h=Id_{M}$\emph{, then the relations }$%
\left( 3.4.1.11\right) $\emph{\ become}%
\begin{equation*}
\Gamma _{j%
%TCIMACRO{\U{b4}}%
%BeginExpansion
{\acute{}}%
%EndExpansion
k%
%TCIMACRO{\U{b4}}%
%BeginExpansion
{\acute{}}%
%EndExpansion
}^{i%
%TCIMACRO{\U{b4}}%
%BeginExpansion
{\acute{}}%
%EndExpansion
}=\frac{\partial x^{j}}{\partial x^{j%
%TCIMACRO{\U{b4}}%
%BeginExpansion
{\acute{}}%
%EndExpansion
}}\left[ -\frac{\partial }{\partial x^{i}}\left( \frac{\partial x^{i%
%TCIMACRO{\U{b4}}%
%BeginExpansion
{\acute{}}%
%EndExpansion
}}{\partial x^{j}}\right) +\Gamma _{jk}^{i}\frac{\partial x^{i%
%TCIMACRO{\U{b4}}%
%BeginExpansion
{\acute{}}%
%EndExpansion
}}{\partial x^{i}}\right] \frac{\partial x^{k}}{\partial x^{k%
%TCIMACRO{\U{b4}}%
%BeginExpansion
{\acute{}}%
%EndExpansion
}}. \leqno(3.4.1.11^{\prime })
\end{equation*}

\smallskip \noindent\textbf{Remark 3.4.1.3} Since
\begin{equation*}
\frac{\partial M_{b}^{a%
%TCIMACRO{\U{b4}}%
%BeginExpansion
{\acute{}}%
%EndExpansion
}}{\partial x^{i}}M_{b%
%TCIMACRO{\U{b4}}%
%BeginExpansion
{\acute{}}%
%EndExpansion
}^{b}+\frac{\partial M_{b%
%TCIMACRO{\U{b4}}%
%BeginExpansion
{\acute{}}%
%EndExpansion
}^{b}}{\partial x^{i}}M_{b}^{a%
%TCIMACRO{\U{b4}}%
%BeginExpansion
{\acute{}}%
%EndExpansion
}=0,
\end{equation*}%
it results that the relations $(3.4.11)$ are equivalent with the relations $%
(3.4.1.11^{\prime }).$

\bigskip \noindent \textbf{Definition 3.4.1.5 }If $\left( \rho ,\eta \right)
\Gamma $ is a linear $\left( \rho ,\eta \right) $-connection for the vector
bundle $\left( \overset{\ast }{E},\overset{\ast }{\pi },M\right) $, then for
any
\begin{equation*}
z=z^{\alpha }t_{\alpha }\in \Gamma \left( F,\nu ,N\right)
\end{equation*}%
the application%
\begin{equation*}
\begin{array}{rcl}
\Gamma \left( \overset{\ast }{E},\overset{\ast }{\pi },M\right) \!\! & \!\!^{%
\underrightarrow{\left( \rho ,\eta \right) D_{z}}}\!\! & \!\!\Gamma \left(
\overset{\ast }{E},\overset{\ast }{\pi },M\right) \\
\overset{\ast }{u}{=}u_{a}s^{a}\!\! & \!\!\longmapsto \!\! & \!\!(\rho ,\eta
)D_{z}\overset{\ast }{u}%
\end{array}%
\leqno(3.4.1.12)
\end{equation*}%
defined by
\begin{equation*}
(\rho ,\eta )D_{z}\overset{\ast }{u}{=}z^{\alpha }{\circ }h\left( \rho
_{\alpha }^{i}{\circ }h\frac{\partial u_{b}}{\partial x^{i}}{-}(\rho ,\eta
)\Gamma _{b\alpha }^{a}\cdot u_{a}\right) s^{b}
\end{equation*}%
will be called the \emph{covariant }$\left( \rho ,\eta \right) $\emph{%
-derivative associated to linear }$\left( \rho ,\eta \right) $\emph{%
-connection }$\left( \rho ,\eta \right) \Gamma $\emph{\ with respect to the
section }$z$.

If $h=Id_{M}$ and $\eta =Id_{M},$ then we obtain the \emph{covariant }$\rho $%
\emph{-derivative associated to linear }$\rho $\emph{-connection }$\rho
\Gamma $\emph{\ with respect to the section }$z$\emph{.}

In addition, if $\rho =Id_{TM}$, then we obtain the \emph{covariant
derivative associated to linear connection }$\Gamma $\emph{\ with respect to
vector field }$z$.

\bigskip

\textbf{Note. }In the next we use the same notation $\left( \rho ,\eta
\right) \Gamma $ for the linear $\left( \rho ,\eta \right) $-connection for
the vector bundle $\left( E,\pi ,M\right) $ or for its dual $\left( \overset{%
\ast }{E},\overset{\ast }{\pi },M\right) $

\bigskip \noindent \textbf{Remark 3.4.1.4 }If $\left( \rho ,\eta \right)
\Gamma $ is a linear $\left( \rho ,\eta \right) $-connection for the vector
bundle $\left( E,\pi ,M\right) $ or for the vector bundle $\left( \overset{%
\ast }{E},\overset{\ast }{\pi },M\right) $ then, the tensor fields algebra
\begin{equation*}
\left( \mathcal{T}\left( E,\pi ,M\right) ,+,\cdot ,\otimes \right)
\end{equation*}%
is endowed with the $\left( \rho ,\eta \right) $-derivative
\begin{equation*}
\begin{array}{rcl}
\Gamma \left( F,\nu ,N\right) \times \mathcal{T}\left( E,\pi ,M\right) & ^{%
\underrightarrow{\left( \rho ,\eta \right) D}} & \mathcal{T}\left( E,\pi
,M\right) \\
\left( z,T\right) & \longmapsto & \left( \rho ,\eta \right) D_{z}T%
\end{array}%
\leqno(3.4.1.13)
\end{equation*}%
defined for a tensor field $T\in \mathcal{T}_{q}^{p}\left( E,\pi ,M\right) $
by the relation:
\begin{equation*}
\begin{array}{l}
\left( \rho ,\eta \right) D_{z}T\left( \overset{\ast }{u}_{1},...,\overset{%
\ast }{u}_{p},u_{1},...,u_{q}\right) =\Gamma \left( \rho ,\eta \right)
\left( z\right) \left( T\left( \overset{\ast }{u}_{1},...,\overset{\ast }{u}%
_{p},u_{1},...,u_{q}\right) \right) \vspace*{1mm} \\
-T\left( \left( \rho ,\eta \right) D_{z}\overset{\ast }{u}_{1},...,\overset{%
\ast }{u}_{p},u_{1},...,u_{q}\right) -...-T\left( \overset{\ast }{u}%
_{1},...,\left( \rho ,\eta \right) D_{z}\overset{\ast }{u}%
_{p},u_{1},...,u_{q}\right) \vspace*{1mm} \\
-T\left( \overset{\ast }{u}_{1},...,\overset{\ast }{u}_{p},\left( \rho ,\eta
\right) D_{z}u_{1},...,u_{q}\right) -...-T\left( \overset{\ast }{u}_{1},...,%
\overset{\ast }{u}_{p},u_{1},...,\left( \rho ,\eta \right) D_{z}u_{q}\right)
.%
\end{array}%
\leqno(3.4.1.14)
\end{equation*}%
Moreover, it satisfies the condition
\begin{equation*}
\begin{array}{c}
\left( \rho ,\eta \right) D_{f_{1}z_{1}+f_{2}z_{2}}T=f_{1}\left( \rho ,\eta
\right) D_{z_{1}}T+f_{2}\left( \rho ,\eta \right) D_{z_{2}}T.%
\end{array}%
\leqno(3.4.1.15)
\end{equation*}%
Consequently, if the tensor algebra $\left( \mathcal{T}\left( E,\pi
,M\right) ,+,\cdot ,\otimes \right) $ is endowed with a $\left( \rho ,\eta
\right) $-derivative $\left( 3.4.1.13\right) $ defined for a tensor field $%
T\in \mathcal{T~}_{q}^{p}\left( E,\pi ,M\right) $ by $\left( 3.4.1.14\right)
$ which satisfies the condition $\left( 3.4.1.15\right) $, then we can
endowed $\left( E,\pi ,M\right) $ with a linear $\left( \rho ,\eta \right) $%
-connection $\left( \rho ,\eta \right) \Gamma $ such that its components are
defined by the equality:
\begin{equation*}
%
\end{equation*}%
The $\left( \rho ,\eta \right) $-derivative $\left( 3.4.1.13\right) $ will
be called the \emph{covariant }$\left( \rho ,\eta \right) $\emph{-derivative.%
}

After some calculations, we obtain:
\begin{equation*}
%
%
\leqno(3.4.1.16)
\end{equation*}

If $\left( \rho ,\eta \right) \Gamma $ is the linear $\left( \rho ,\eta
\right) $\textit{-}connection associated to linear connection $\Gamma ,$
namely $\left( \rho ,\eta \right) \Gamma _{b\alpha }^{a}=\left( \rho
_{\alpha }^{k}\circ h\right) \Gamma _{bk}^{a},$ then
\begin{equation*}
%
%
\leqno(3.4.1.17)
\end{equation*}%
In particular, if $h=Id_{M},$ then we obtain the formula:
\begin{equation*}
%
%
\leqno(3.4.1.18)
\end{equation*}

\section{The geometry of base of the Lie algebroid generalized tangent
bundle for a vector bundle}

In this section, we present new applications of generalized Lie algebroids
in the study of the geometry of vector bundles using the theory of
generalized linear connections.

\subsection{Torsion and curvature. Formulas of Ricci type}

We apply the theory for the diagram:
\begin{equation*}
%
%
,\leqno(4.1.1)
\end{equation*}%
where $\left( E,\pi ,M\right) \in \left\vert \mathbf{B}^{\mathbf{v}%
}\right\vert $ and $\left( \left( F,\nu ,M\right) ,\left[ ,\right]
_{F,h},\left( \rho ,Id_{M}\right) \right) \in \left\vert \mathbf{GLA}%
\right\vert .$

Let $\rho \Gamma $ be a linear $\rho $-connection for the vector bundle $%
\left( E,\pi ,M\right) $ by components $\rho \Gamma _{b\alpha }^{a}.$

Using the components of the linear $\rho $-connection $\rho \Gamma $, then
we obtain a linear $\rho $-connection $\rho \dot{\Gamma}$ for the vector
bundle $\left( E,\pi ,M\right) $ given by the diagram:
\begin{equation*}
\begin{array}{ccl}
~\ \ \ E &  & \left( h^{\ast }F,\left[ ,\right] _{h^{\ast }F},\left( \overset%
{h^{\ast }F}{\rho },Id_{M}\right) \right) \\
\pi \downarrow &  & ~\ \ \downarrow h^{\ast }\nu \\
~\ \ \ M & ^{\underrightarrow{~\ \ \ \ Id_{M}~\ \ }} & ~\ \ \ M%
\end{array}%
.\leqno(4.1.2)
\end{equation*}

If $\left( E,\pi ,M\right) =\left( F,\nu ,N\right) ,$ then, using the
components of the linear $\rho $-connection $\rho \Gamma ,$ we can consider
a linear $\rho $-connection $\rho \ddot{\Gamma}$ for the vector bundle $%
\left( h^{\ast }E,h^{\ast }\pi ,M\right) $ given by the diagram:
\begin{equation*}
\begin{array}{ccl}
~\ \ ~\ \ \ \ h^{\ast }E &  & \left( h^{\ast }E,\left[ ,\right] _{h^{\ast
}E},\left( \overset{h^{\ast }E}{\rho },Id_{M}\right) \right) \\
h^{\ast }\pi \downarrow &  & ~\ \ \downarrow h^{\ast }\pi \\
~\ \ \ \ \ M & ^{\underrightarrow{~\ \ \ \ Id_{M}~\ \ }} & ~\ \ M%
\end{array}%
,\leqno(4.1.3)
\end{equation*}

In the following, we will use the exterior differentiation operators $%
d,~d^{E} $ and $d^{h^{\ast }E}$ res\-pec\-tively for the exterior
differential $\mathcal{F}\left( M\right) $-algebras $\left( \Lambda \left(
TM,\tau _{M},M\right) ,+,\cdot ,\wedge \right) $,\break $\left( \Lambda
\left( E,\pi ,M\right) ,+,\cdot ,\wedge \right) $ and $\left( \left( h^{\ast
}E,h^{\ast }\pi ,M\right) ,+,\cdot ,\wedge \right) $ respectively.

\bigskip\noindent\textbf{Definition 4.1.1 }If $\left( E,\pi ,M\right)
=\left( F,\nu ,N\right) $, then the application
\begin{equation*}
\begin{array}{ccc}
\Gamma \left( h^{\ast }E,h^{\ast }\pi ,M\right) ^{2} & ^{\underrightarrow{\
\left( \rho ,h\right) \mathbb{T}\ }} & \Gamma \left( h^{\ast }E,h^{\ast }\pi
,M\right) \\
\left( U,V\right) & \longrightarrow & \rho \mathbb{T}\left( U,V\right)%
\end{array}%
\leqno(4.1.4)
\end{equation*}%
defined by:
\begin{equation*}
\begin{array}{c}
\left( \rho ,h\right) \mathbb{T}\left( U,V\right) =\rho \ddot{D}_{U}V-\rho
\ddot{D}_{V}U-\left[ U,V\right] _{h^{\ast }E},\,%
\end{array}%
\leqno(4.1.5)
\end{equation*}%
for any $U,V\in \Gamma \left( h^{\ast }E,h^{\ast }\pi ,M\right) ,$ will be
called $\left( \rho ,h\right) $\emph{-torsion associated to linear }$\rho $%
\emph{-connection }$\rho \Gamma .$

\bigskip\noindent\textbf{Remark 4.1.1 }In particular,\ if $h=Id_{M}$, then
we obtain the application
\begin{equation*}
\begin{array}{ccc}
\Gamma \left( E,\pi ,M\right) ^{2} & ^{\underrightarrow{\ \rho \mathbb{T}\ }}
& \Gamma \left( E,\pi ,M\right) \\
\left( u,v\right) & \longrightarrow & \rho \mathbb{T}\left( u,v\right)%
\end{array}%
\leqno(4.1.4^{\prime })
\end{equation*}%
defined by:
\begin{equation*}
\begin{array}{c}
\rho \mathbb{T}\left( u,v\right) =\rho D_{u}v-\rho D_{v}u-\left[ u,v\right]
_{E},\,%
\end{array}%
\leqno(4.1.5^{\prime })
\end{equation*}%
for any $u,v\in \Gamma \left( E,\pi ,M\right) ,$\ which will be called $\rho
$\emph{-torsion associated to linear }$\rho $\emph{-connection }$\rho \Gamma
.$

Moreover, if $\rho =Id_{TM}$, then we obtain the torsion $\mathbb{T}$
associated to linear connection~$\Gamma .$

\bigskip\noindent\textbf{Proposition 4.1.1 }\emph{The }$\left( \rho
,h\right) $\emph{-torsion }$\left( \rho ,h\right) \mathbb{T}$\emph{\
associated to linear }$\rho $\emph{-connection }$\rho \Gamma $\emph{\ is }$%
\mathbb{R}$\emph{-bilinear and antisymmetric.}

\emph{If }%
\begin{equation*}
\left( \rho ,h\right) \mathbb{T}\left( S_{a},S_{b}\right) \overset{put}{=}%
\left( \rho ,h\right) \mathbb{T}_{~ab}^{c}S_{c}
\end{equation*}%
\emph{\ then }%
\begin{equation*}
\begin{array}{c}
\left( \rho ,h\right) \mathbb{T}_{~ab}^{c}=\rho \Gamma _{ab}^{c}-\rho \Gamma
_{ba}^{c}-L_{ab}^{c}\circ h.%
\end{array}%
\leqno(4.1.6)
\end{equation*}
\emph{In particular, if }$h=Id_{M}$\emph{\ and }$\rho \mathbb{T}\left(
s_{a},s_{b}\right) \overset{put}{=}\rho \mathbb{T}_{ab}^{c}s_{c}$\emph{,
then }%
\begin{equation*}
\begin{array}{c}
\rho \mathbb{T}_{~ab}^{c}=\rho \Gamma _{ab}^{c}-\rho \Gamma
_{ba}^{c}-L_{ab}^{c}.%
\end{array}%
\leqno(4.1.6^{\prime })
\end{equation*}
\emph{Moreover, if }$\rho =Id_{TM}$\emph{, then the equality }$\left(
4.1.6^{\prime }\right) $\emph{\ becomes:}
\begin{equation*}
\begin{array}{c}
\mathbb{T}_{~jk}^{i}=\Gamma _{jk}^{i}-\Gamma _{kj}^{i}.%
\end{array}%
\leqno(4.1.6^{\prime \prime })
\end{equation*}

\smallskip \noindent\textbf{Definition 4.1.2 }The application
\begin{equation*}
\begin{array}{ccl}
(\Gamma \left( h^{\ast }F,h^{\ast }\nu ,M\right) ^{2}{\times }\Gamma (E,\pi
,M) & ^{\underrightarrow{\ \left( \rho ,h\right) \mathbb{R}\ }} & \Gamma
(E,\pi ,M) \\
((Z,V),u) & \longrightarrow & \rho \mathbb{R}(Z,V)u%
\end{array}%
\leqno(4.1.7)
\end{equation*}%
defined by
\begin{equation*}
\left( \rho ,h\right) \mathbb{R}\left( Z,V\right) u=\rho \dot{D}_{Z}\left(
\rho \dot{D}_{V}u\right) -\rho \dot{D}_{V}\left( \rho \dot{D}_{Z}u\right)
-\rho \dot{D}_{\left[ Z,V\right] _{h^{\ast }F}}u,\,\leqno(4.1.8)
\end{equation*}%
for any $Z,V\in \Gamma \left( h^{\ast }F,h^{\ast }\nu ,M\right) ,~u\in
\Gamma \left( E,\pi ,M\right) ,$ will be called $\left( \rho ,h\right) $%
\emph{-curvature associated to linear }$\rho $\emph{-connection }$\rho
\Gamma .$

\bigskip\noindent\textbf{Remark 4.1.1} In particular, if $h=Id_{M}$, then we
obtain the application
\begin{equation*}
\begin{array}{ccl}
\Gamma \left( F,\nu ,M\right) ^{2}{\times }\Gamma (E,\pi ,M) & ^{%
\underrightarrow{\ \rho \mathbb{R}\ }} & \Gamma (E,\pi ,M) \\
((z,v),u) & \longrightarrow & \rho \mathbb{R}(z,v)u%
\end{array}%
\leqno(4.1.7^{\prime })
\end{equation*}%
defined by
\begin{equation*}
\rho \mathbb{R}\left( z,v\right) u=\rho D_{z}\left( \rho D_{v}u\right) -\rho
D_{v}\left( \rho D_{z}u\right) -\rho D_{\left[ z,v\right] _{F}}u,\,\leqno%
(4.1.8^{\prime })
\end{equation*}%
for any $z,v\in \Gamma \left( F,\nu ,M\right) ,~u\in \Gamma \left( E,\pi
,M\right) ,$ which will be called $\rho $\emph{-curvature associated to
linear }$\rho $\emph{-connection }$\rho \Gamma .$

Moreover, if $\rho =Id_{TM}$, then we obtain the curvature $\mathbb{R}$
associated to linear connection $\Gamma .$

\bigskip\noindent \textbf{Proposition 4.1.2 }\emph{The }$\left( \rho
,h\right) $\emph{-curvature }$\left( \rho ,h\right) \mathbb{R}$\emph{\
associated to linear }$\rho $\emph{-connection }$\rho \Gamma $\emph{, is }$%
\mathbb{R}$\emph{-linear in each argument and antisymmetric in the first two
arguments.}

\emph{If }%
\begin{equation*}
\left( \rho ,h\right) \mathbb{R}\left( T_{\beta },T_{\alpha }\right) s_{b}%
\overset{put}{=}\left( \rho ,h\right) \mathbb{R}_{b~\alpha \beta }^{a}s_{a},
\end{equation*}%
\emph{then }%
\begin{equation*}
\left( \rho ,h\right) \mathbb{R}_{b~\alpha \beta }^{a}=\rho _{\beta
}^{j}\circ h\frac{\partial \rho \Gamma _{b\alpha }^{a}}{\partial x^{j}}+\rho
\Gamma _{e\beta }^{a}\rho \Gamma _{b\alpha }^{e}-\rho _{\alpha }^{i}\circ h%
\frac{\partial \rho \Gamma _{b\beta }^{a}}{\partial x^{i}} -\rho \Gamma
_{e\alpha }^{a}\rho \Gamma _{b\beta }^{e}+\rho \Gamma _{b\gamma
}^{a}L_{\alpha \beta }^{\gamma }\circ h. \leqno(4.1.9)
\end{equation*}
\emph{In particular, if }$h=Id_{M}$\emph{\ and }$\rho \mathbb{R}\left(
t_{\beta },t_{\alpha }\right) s_{b}\overset{put}{=}\rho \mathbb{R}_{b~\alpha
\beta }^{a}s_{a}$\emph{, then }%
\begin{equation*}
\rho \mathbb{R}_{b~\alpha \beta }^{a}=\rho _{\beta }^{j}\frac{\partial \rho
\Gamma _{b\alpha }^{a}}{\partial x^{j}}+\rho \Gamma _{e\beta }^{a}\rho
\Gamma _{b\alpha }^{e}-\rho _{\alpha }^{i}\frac{\partial \rho \Gamma
_{b\beta }^{a}}{\partial x^{i}} -\rho \Gamma _{e\alpha }^{a}\rho \Gamma
_{b\beta }^{e}+\rho \Gamma _{b\gamma }^{a}L_{\alpha \beta }^{\gamma }. \leqno%
(4.1.9^{\prime })
\end{equation*}
\emph{Moreover, if }$\rho =Id_{TM}$\emph{, then equality }$\left(
4.1.9^{\prime }\right) $\emph{\ becomes:}
\begin{equation*}
\mathbb{R}_{b~hk}^{a}=\frac{\partial \Gamma _{bh}^{a}}{\partial x^{k}}%
+\Gamma _{ek}^{a}\Gamma _{bh}^{e}-\frac{\partial \Gamma _{bk}^{a}}{\partial
x^{h}}-\Gamma _{eh}^{a}\Gamma _{bk}^{e}. \leqno(4.1.9^{\prime \prime })
\end{equation*}

\smallskip \noindent\textbf{Theorem 4.1.1 }\emph{For any } $u^{a}s_{a}\in
\Gamma \left( E,\pi ,M\right) $ \emph{we shall use the notation}
\begin{equation*}
u_{~|\alpha \beta }^{a}=\rho _{\beta }^{j}\circ h\frac{\partial }{\partial
x^{j}}\left( u_{~|\alpha }^{a_{1}}\right) +\rho \Gamma _{b_{\beta
}}^{a_{1}}u_{~|\alpha }^{b}, \leqno(4.1.10)
\end{equation*}%
\emph{and we verify the formulas: }
\begin{equation*}
u_{~|\alpha \beta }^{a_{1}}-u_{~|\beta \alpha }^{a_{1}}=u^{a}\left( \rho
,h\right) \mathbb{R}_{a~\alpha \beta }^{a_{1}}-u_{~|\gamma
}^{a_{1}}L_{\alpha \beta }^{\gamma }\circ h. \leqno(4.1.11)
\end{equation*}
\emph{After some calculations, we obtain}
\begin{equation*}
\left( \rho ,h\right) \mathbb{R}_{a~\alpha \beta }^{a_{1}}=u_{a}\left(
u_{~|\alpha \beta }^{a_{1}}-u_{~|\beta \alpha }^{a_{1}}+u_{~|\gamma
}^{a_{1}}L_{\alpha \beta }^{\gamma }\circ h\right) , \leqno(4.1.12)
\end{equation*}%
\emph{where }$u_{a}s^{a}\in \Gamma \left( \overset{\ast }{E},\overset{\ast }{%
\pi },M\right) $\emph{\ such that }$u_{a}u^{b}=\delta _{a}^{b}.$

\emph{In particular, if }$h=Id_{M}$\emph{, then the relations }$\left(
4.1.12\right) $\emph{\ become}%
\begin{equation*}
\rho \mathbb{R}_{a~\alpha \beta }^{a_{1}}=u_{a}\left( u_{~|\alpha \beta
}^{a_{1}}-u_{~|\beta \alpha }^{a_{1}}+u_{~|\gamma }^{a_{1}}L_{\alpha \beta
}^{\gamma }\right) . \leqno(4.1.12^{\prime })
\end{equation*}
\emph{Moreover, if }$\rho =id_{TM},$ \emph{then the relations }$\left(
4.1.12^{\prime }\right) $\emph{\ become}%
\begin{equation*}
\mathbb{R}_{a~ij}^{a_{1}}=u_{a}\left(
u_{~|ij}^{a_{1}}-u_{~|ji}^{a_{1}}\right) . \leqno\left( 4.1.12^{\prime
\prime }\right)
\end{equation*}

\smallskip \noindent\textit{Proof.} Since
\begin{equation*}
\begin{split}
u_{~|\alpha \beta }^{a_{1}}& =\rho _{\beta }^{j}\circ h\left( \frac{\partial
}{\partial x^{j}}\left( \rho _{\alpha }^{i}\circ h\frac{\partial u^{a_{1}}}{%
\partial x^{i}}+\rho \Gamma _{a\alpha }^{a_{1}}u^{a}\right) \right) \\
& +\rho \Gamma _{b\beta }^{a_{1}}\left( \rho _{\alpha }^{i}\circ h\frac{%
\partial u^{b}}{\partial x^{i}}+\rho \Gamma _{a\alpha }^{b}u^{a}\right) \\
& =\rho _{\beta }^{j}\circ h\frac{\partial \rho _{\alpha }^{i}\circ h}{%
\partial x^{j}}\frac{\partial u^{a_{1}}}{\partial x^{i}}+\rho _{\beta
}^{j}\circ h\rho _{\alpha }^{i}\circ h\frac{\partial }{\partial x^{j}}\left(
\frac{\partial u^{a_{1}}}{\partial x^{i}}\right) \\
& +\rho _{\beta }^{j}\circ h\frac{\partial \rho \Gamma _{a\alpha }^{a_{1}}}{%
\partial x^{j}}u^{a}+\rho _{\beta }^{j}\circ h\rho \Gamma _{a\alpha }^{a_{1}}%
\frac{\partial u^{a}}{\partial x^{j}} \\
& +\rho _{\alpha }^{i}\circ h\rho \Gamma _{b\beta }^{a_{1}}\frac{\partial
u^{b}}{\partial x^{i}}+\rho \Gamma _{b\beta }^{a_{1}}\rho \Gamma _{a\alpha
}^{b}u^{a}
\end{split}%
\end{equation*}%
and%
\begin{equation*}
\begin{split}
u_{~|\beta \alpha }^{a_{1}}& =\rho _{\alpha }^{i}\circ h\left( \frac{%
\partial }{\partial x^{i}}\left( \rho _{\beta }^{j}\circ h\frac{\partial
u^{a_{1}}}{\partial x^{j}}+\rho \Gamma _{a\beta }^{a_{1}}u^{a}\right) \right)
\\
& +\rho \Gamma _{b\alpha }^{a_{1}}\left( \rho _{\beta }^{j}\circ h\frac{%
\partial u^{b}}{\partial x^{j}}+\rho \Gamma _{a\beta }^{b}u^{a}\right) \\
& =\rho _{\alpha }^{i}\circ h\frac{\partial \rho _{\beta }^{j}\circ h}{%
\partial x^{i}}\frac{\partial u^{a_{1}}}{\partial x^{j}}+\rho _{\beta
}^{j}\circ h\rho _{\alpha }^{i}\circ h\frac{\partial }{\partial x^{i}}\left(
\frac{\partial u^{a_{1}}}{\partial x^{j}}\right) \\
& +\rho _{\alpha }^{i}\circ h\frac{\partial \rho \Gamma _{a\beta }^{a_{1}}}{%
\partial x^{i}}u^{a}+\rho _{\alpha }^{i}\circ h\rho \Gamma _{a\beta }^{a_{1}}%
\frac{\partial u^{a}}{\partial x^{i}} \\
& +\rho _{\beta }^{j}\circ h\rho \Gamma _{b\alpha }^{a_{1}}\frac{\partial
u^{b}}{\partial x^{j}}+\rho \Gamma _{b\alpha }^{a_{1}}\rho \Gamma _{a\beta
}^{b}u^{a},
\end{split}%
\end{equation*}%
it results that
\begin{equation*}
\begin{split}
u_{~|\alpha \beta }^{a_{1}}-u_{~|\beta \alpha }^{a_{1}}& =\rho _{\beta
}^{j}\circ h\frac{\partial \rho _{\alpha }^{i}\circ h}{\partial x^{j}}\frac{%
\partial u^{a_{1}}}{\partial x^{i}}-\rho _{\alpha }^{i}\circ h\frac{\partial
\rho _{\beta }^{j}\circ h}{\partial x^{i}}\frac{\partial u^{a_{1}}}{\partial
x^{j}} \\
& +\left( \rho _{\beta }^{j}\circ h\rho _{\alpha }^{i}\circ h\frac{\partial
^{2}u^{a_{1}}}{\partial x^{i}\partial x^{j}}-\rho _{\beta }^{j}\circ h\rho
_{\alpha }^{i}\circ h\frac{\partial ^{2}u^{a_{1}}}{\partial x^{j}\partial
x^{i}}\right) \\
& +\left( \rho _{\beta }^{j}\circ h\frac{\partial \rho \Gamma _{a\alpha
}^{a_{1}}}{\partial x^{j}}u^{a}-\rho _{\alpha }^{i}\circ h\frac{\partial
\rho \Gamma _{a\beta }^{a_{1}}}{\partial x^{i}}u^{a}\right) \\
& +\left( \rho _{\beta }^{j}\circ h\rho \Gamma _{a\alpha }^{a_{1}}\frac{%
\partial u^{a}}{\partial x^{j}}-\rho _{\beta }^{j}\circ h\rho \Gamma
_{b\alpha }^{a_{1}}\frac{\partial u^{b}}{\partial x^{j}}\right) \\
& +\left( \rho _{\alpha }^{i}\circ h\rho \Gamma _{b\beta }^{a_{1}}\frac{%
\partial u^{b}}{\partial x^{i}}-\rho _{\alpha }^{i}\circ h\rho \Gamma
_{a\beta }^{a_{1}}\frac{\partial u^{a}}{\partial x^{i}}\right) \\
& +\rho \Gamma _{b\beta }^{a_{1}}\rho \Gamma _{a\alpha }^{b}u^{a}-\rho
\Gamma _{b\alpha }^{a_{1}}\rho \Gamma _{a\beta }^{b}u^{a}.
\end{split}%
\end{equation*}
After some calculations, we obtain:
\begin{equation*}
\begin{split}
u_{~|\alpha \beta }^{a_{1}}-u_{~|\beta \alpha }^{a_{1}}& =L_{\beta \alpha
}^{\gamma }\circ h\rho _{\gamma }^{k}\circ h\frac{\partial u^{a_{1}}}{%
\partial x^{k}} \\
& +\left( \rho _{\beta }^{j}\circ h\frac{\partial \rho \Gamma _{a\alpha
}^{a_{1}}}{\partial x^{j}}u^{a}-\rho _{\alpha }^{i}\circ h\frac{\partial
\rho \Gamma _{a\beta }^{a_{1}}}{\partial x^{i}}u^{a}\right) \\
& +\rho \Gamma _{b\beta }^{a_{1}}\rho \Gamma _{a\alpha }^{b}u^{a}-\rho
\Gamma _{b\alpha }^{a_{1}}\rho \Gamma _{a\beta }^{b}u^{a}.
\end{split}%
\end{equation*}
Since
\begin{eqnarray*}
u^{a}\left( \rho ,h\right) \mathbb{R}_{a~\alpha \beta }^{a_{1}}
&=&u^{a}\left( \rho _{\beta }^{j}\circ h\frac{\partial \rho \Gamma _{a\alpha
}^{a_{1}}}{\partial x^{j}}+\rho \Gamma _{e\beta }^{a_{1}}\rho \Gamma
_{a\alpha }^{e}-\rho _{\alpha }^{i}\circ h\frac{\partial \rho \Gamma
_{a\beta }^{a_{1}}}{\partial x^{i}}\right. \\
&&\left. -\rho \Gamma _{e\alpha }^{a_{1}}\rho \Gamma _{a\beta }^{e}-\rho
\Gamma _{a\gamma }^{a_{1}}L_{\beta \alpha }^{\gamma }\circ h\right) .
\end{eqnarray*}%
and
\begin{equation*}
u_{~|\gamma }^{a_{1}}L_{\alpha \beta }^{\gamma }\circ h=\left( \rho _{\gamma
}^{k}\circ h\frac{\partial u^{a_{1}}}{\partial x^{k}}+\rho \Gamma _{a\gamma
}^{a_{1}}u^{a}\right) L_{\alpha \beta }^{\gamma }\circ h
\end{equation*}%
it results that
\begin{equation*}
\begin{split}
u^{a}\left( \rho ,h\right) \mathbb{R}_{a~\alpha \beta }^{a_{1}}-u_{~|\gamma
}^{a_{1}}L_{\alpha \beta }^{\gamma }\circ h& =-L_{\alpha \beta }^{\gamma
}\circ h\rho _{\gamma }^{k}\circ h\frac{\partial u^{a_{1}}}{\partial x^{k}}
\\
& +\left( \rho _{\beta }^{j}\circ h\frac{\partial \rho \Gamma _{a\alpha
}^{a_{1}}}{\partial x^{j}}u^{a}-\rho _{\alpha }^{i}\circ h\frac{\partial
\rho \Gamma _{a\beta }^{a_{1}}}{\partial x^{i}}u^{a}\right) \\
& +\rho \Gamma _{b\beta }^{a_{1}}\rho \Gamma _{a\alpha }^{b}u^{a}-\rho
\Gamma _{b\alpha }^{a_{1}}\rho \Gamma _{a\beta }^{b}u^{a}.
\end{split}%
\end{equation*}

\begin{flushright}
\emph{q.e.d.}
\end{flushright}

\bigskip\noindent\textbf{Lemma 4.1.1 }\emph{If }$\left( E,\pi ,M\right)
=\left( F,\nu ,N\right) $\emph{, then, for any }
\begin{equation*}
u^{a}s_{a}\in \Gamma \left( E,\pi ,M\right) ,
\end{equation*}%
\emph{\ we have that } $u_{~|c}^{a},~a,c\in \overline{1,n}$ are the \emph{%
components of a tensor field of }$\left( 1,1\right) $\emph{\ type.}

\bigskip\noindent\textit{Proof.} Let $U$ and $U^{\prime }$ be two vector
local $\left( m+n\right) $-charts such that $U\cap U^{\prime }\neq \phi .$

Since $u^{a^{\prime }}\left( x\right) =M_{a}^{a^{\prime }}\left( x\right)
u^{a}\left( x\right) ,~$for any $x\in U\cap U^{\prime },$ it results that
\begin{equation}
\rho _{c^{\prime }}^{k^{\prime }}\circ h\left( x\right) \frac{\partial
u^{a^{\prime }}\left( x\right) }{\partial x^{k^{\prime }}}=\rho _{c^{\prime
}}^{k^{\prime }}\circ h\left( x\right) \frac{\partial }{\partial
x^{k^{\prime }}}\left( M_{a}^{a^{\prime }}\left( x\right) \right)
u^{a}\left( x\right) +M_{a}^{a^{\prime }}\left( x\right) \rho _{c^{\prime
}}^{k^{\prime }}\circ h\left( x\right) \frac{\partial u^{a}\left( x\right) }{%
\partial x^{k^{\prime }}}.
\end{equation}

Since , for any $x\in U\cap U^{\prime }$, we have
\begin{equation}
\rho \Gamma _{b^{\prime }c^{\prime }}^{a^{\prime }}(x)=M_{a}^{a^{\prime
}}(x)\left( \rho _{c}^{k}\circ h(x)\frac{\partial }{\partial x^{k}}%
(M_{b^{\prime }}^{a}(x))+\rho \Gamma _{bc}^{a}(x)M_{b^{\prime
}}^{b}(x)\right) M_{c^{\prime }}^{c}(x),
\end{equation}%
and
\begin{equation}
0=\frac{\partial }{\partial x^{k^{\prime }}}\left( M_{a}^{a^{\prime }}\left(
x\right) M_{b^{\prime }}^{a}\left( x\right) \right) =\frac{\partial }{%
\partial x^{k^{\prime }}}\left( M_{a}^{a^{\prime }}\left( x\right) \right)
M_{b^{\prime }}^{a}\left( x\right) +M_{a}^{a^{\prime }}\left( x\right) \frac{%
\partial }{\partial x^{k^{\prime }}}\left( M_{b^{\prime }}^{a}\left(
x\right) \right)
\end{equation}%
it results that
\begin{equation}
\begin{split}
\rho \Gamma _{b^{\prime }c^{\prime }}^{a^{\prime }}\left( x\right)
u^{b^{\prime }}\left( x\right) & =-\rho _{c^{\prime }}^{k^{\prime }}\circ
h\left( x\right) \frac{\partial }{\partial x^{k^{\prime }}}\left(
M_{a}^{a^{\prime }}\left( x\right) \right) u^{a}\left( x\right) \\
& +M_{a}^{a^{\prime }}\left( x\right) \rho \Gamma _{bc}^{a}\left( x\right)
u^{b}\left( x\right) M_{c^{\prime }}^{c}\left( x\right) .
\end{split}%
\end{equation}

Summing the equalities $\left( 1\right) $ and $\left( 4\right) $, it results
the conclusion of lemma.\hfill \emph{q.e.d.}

\bigskip\noindent\textbf{Theorem 4.1.2 }\emph{If }$\left( E,\pi ,M\right)
=\left( F,\nu ,N\right) $\emph{, then, for any }
\begin{equation*}
%
\end{equation*}%
\emph{\ we shall use the notation }%
\begin{equation*}
%
%
\leqno\left( 4.1.13\right)
\end{equation*}%
\emph{and we verify the formulas of Ricci type }%
\begin{equation*}
%
%
\leqno\left( 4.1.14\right)
\end{equation*}
\emph{In particular, if }$h=Id_{M}$\emph{, then the relations }$\left(
4.1.14\right) $\emph{\ become}
\begin{equation*}
%
%
\leqno\left( 4.1.14^{\prime }\right)
\end{equation*}
\emph{Moreover, if }$\rho =id_{TM}$\emph{, then the relations }$\left(
4.1.14^{\prime }\right) $\emph{\ become }%
\begin{equation*}
%
%
\leqno\left( 4.1.14^{\prime \prime }\right)
\end{equation*}

\smallskip \noindent\textbf{Theorem 4.1.3 } \emph{For any }$u_{a}s^{a}\in
\Gamma \left( \overset{\ast }{E},\overset{\ast }{\pi },M\right) $ \emph{we
shall use the notation}
\begin{equation*}
%
%
\leqno\left( 4.1.15\right)
\end{equation*}%
\emph{and we verify the formulas: }%
\begin{equation*}
%
%
\leqno\left( 4.1.16\right)
\end{equation*}
\emph{After some calculations, we obtain}%
\begin{equation*}
%
%
\leqno\left( 4.1.17\right)
\end{equation*}%
\emph{where} $u^{a}s_{a}\in \Gamma \left( E,\pi ,M\right) $\emph{\ such that
}$u_{a}u^{b}=\delta _{a}^{b}.$

\emph{In particular, if }$h=Id_{M}$\emph{, then the relations }$\left(
4.1.17\right) $\emph{\ become}%
\begin{equation*}
%
%
\leqno\left( 4.1.17^{\prime }\right)
\end{equation*}
\emph{Moreover, if }$\rho =id_{TM}$ \emph{then the relations }$\left(
4.1.17^{\prime }\right) $\emph{\ become}%
\begin{equation*}
%
%
\leqno(4.1.17^{\prime \prime })
\end{equation*}

\smallskip \noindent \textit{Proof.} Since
\begin{equation*}
\begin{split}
u_{b_{1}\mid \alpha \beta }& =\rho _{\beta }^{j}\circ h\left( \frac{\partial
}{\partial x^{j}}\left( \rho _{\alpha }^{i}\circ h\frac{\partial u_{b_{1}}}{%
\partial x^{i}}-\rho \Gamma _{b_{1}\alpha }^{b}u_{b}\right) \right) \\
& -\rho \Gamma _{b_{1}\beta }^{b}\left( \rho _{\alpha }^{i}\circ h\frac{%
\partial u_{b}}{\partial x^{i}}-\rho \Gamma _{b\alpha }^{a}u_{a}\right) \\
& =\rho _{\beta }^{j}\circ h\frac{\partial \rho _{\alpha }^{i}\circ h}{%
\partial x^{j}}\frac{\partial u_{b_{1}}}{\partial x^{i}}+\rho _{\beta
}^{j}\circ h\rho _{\alpha }^{i}\circ h\frac{\partial }{\partial x^{j}}\left(
\frac{\partial u_{b_{1}}}{\partial x^{i}}\right) \\
& -\rho _{\beta }^{j}\circ h\frac{\partial \rho \Gamma _{b_{1}\alpha }^{b}}{%
\partial x^{j}}u_{b}-\rho _{\beta }^{j}\circ h\rho \Gamma _{b_{1}\alpha }^{b}%
\frac{\partial u_{b}}{\partial x^{j}} \\
& -\rho _{\alpha }^{i}\circ h\rho \Gamma _{b_{1}\beta }^{b}\frac{\partial
u_{b}}{\partial x^{i}}+\rho \Gamma _{b_{1}\beta }^{b}\rho \Gamma _{b\alpha
}^{a}u_{a}
\end{split}%
\end{equation*}%
and%
\begin{equation*}
\begin{split}
u_{b_{1}\mid \beta \alpha }& =\rho _{\alpha }^{i}\circ h\left( \frac{%
\partial }{\partial x^{i}}\left( \rho _{\beta }^{j}\circ h\frac{\partial
u_{b_{1}}}{\partial x^{j}}-\rho \Gamma _{b_{1}\beta }^{b}u_{b}\right) \right)
\\
& -\rho \Gamma _{b_{1}\alpha }^{b}\left( \rho _{\beta }^{j}\circ h\frac{%
\partial u_{b}}{\partial x^{j}}-\rho \Gamma _{b\beta }^{a}u_{a}\right) \\
& =\rho _{\alpha }^{i}\circ h\frac{\partial \rho _{\beta }^{j}\circ h}{%
\partial x^{i}}\frac{\partial u_{b_{1}}}{\partial x^{i}}+\rho _{\beta
}^{j}\circ h\rho _{\alpha }^{i}\circ h\frac{\partial }{\partial x^{i}}\left(
\frac{\partial u_{b_{1}}}{\partial x^{j}}\right) \\
& -\rho _{\alpha }^{i}\circ h\frac{\partial \rho \Gamma _{b_{1}\beta }^{b}}{%
\partial x^{i}}u_{b}-\rho _{\alpha }^{i}\circ h\rho \Gamma _{b_{1}\beta }^{b}%
\frac{\partial u_{b}}{\partial x^{i}} \\
& -\rho _{\beta }^{j}\circ h\rho \Gamma _{b_{1}\alpha }^{b}\frac{\partial
u_{b}}{\partial x^{i}}+\rho \Gamma _{b_{1}\alpha }^{b}\rho \Gamma _{b\beta
}^{a}u_{a}
\end{split}%
\end{equation*}%
it results that%
\begin{equation*}
\begin{split}
\ u_{b_{1}\mid \alpha \beta }-u_{b_{1}\mid \beta \alpha }& =\rho _{\beta
}^{j}\circ h\frac{\partial \rho _{\alpha }^{i}\circ h}{\partial x^{j}}\frac{%
\partial u_{b_{1}}}{\partial x^{i}}-\rho _{\alpha }^{i}\circ h\frac{\partial
\rho _{\beta }^{j}\circ h}{\partial x^{i}}\frac{\partial u_{b_{1}}}{\partial
x^{j}} \\
& +\rho _{\beta }^{j}\circ h\rho _{\alpha }^{i}\circ h\frac{\partial }{%
\partial x^{j}}\left( \frac{\partial u_{b_{1}}}{\partial x^{i}}\right) -\rho
_{\beta }^{j}\circ h\rho _{\alpha }^{i}\circ h\frac{\partial }{\partial x^{i}%
}\left( \frac{\partial u_{b_{1}}}{\partial x^{j}}\right) \\
& +\rho _{\alpha }^{i}\circ h\frac{\partial \rho \Gamma _{b_{1}\beta }^{b}}{%
\partial x^{i}}u_{b}-\rho _{\beta }^{j}\circ h\frac{\partial \rho \Gamma
_{b_{1}\alpha }^{b}}{\partial x^{j}}u_{b} \\
& +\rho _{\beta }^{j}\circ h\rho \Gamma _{b_{1}\alpha }^{b}\frac{\partial
u_{b}}{\partial x^{j}}-\rho _{\beta }^{j}\circ h\rho \Gamma _{b_{1}\alpha
}^{b}\frac{\partial u_{b}}{\partial x^{j}} \\
& +\rho _{\alpha }^{i}\circ h\rho \Gamma _{b_{1}\alpha }^{b}\frac{\partial
u_{b}}{\partial x^{i}}-\rho _{\alpha }^{i}\circ h\rho \Gamma _{b_{1}\alpha
}^{b}\frac{\partial u_{b}}{\partial x^{i}} \\
& +\rho \Gamma _{b_{1}\beta }^{b}\rho \Gamma _{b\alpha }^{a}u_{a}-\rho
\Gamma _{b_{1}\alpha }^{b}\rho \Gamma _{b\beta }^{a}u_{a}.
\end{split}%
\end{equation*}%
After some calculations, we obtain:
\begin{equation*}
\begin{array}{cl}
u_{b_{1}\mid \alpha \beta }-u_{b_{1}\mid \beta \alpha } & =L_{\beta \alpha
}^{\gamma }\circ h\rho _{\gamma }^{k}\circ h\frac{\partial u_{b_{1}}}{%
\partial x^{k}}\vspace*{1mm} \\
& +\left( \rho _{\alpha }^{i}\circ h\frac{\partial \rho \Gamma _{b_{1}\beta
}^{b}}{\partial x^{i}}u_{b}-\rho _{\beta }^{j}\circ h\frac{\partial \rho
\Gamma _{b_{1}\alpha }^{b}}{\partial x^{j}}u_{b}\right) \vspace*{1mm} \\
& +\rho \Gamma _{b_{1}\beta }^{b}\rho \Gamma _{b\alpha }^{a}u_{a}-\rho
\Gamma _{b_{1}\alpha }^{b}\rho \Gamma _{b\beta }^{a}u_{a}.%
\end{array}%
\end{equation*}%
Since
\begin{equation*}
\begin{split}
u_{b}\left( \rho ,h\right) \mathbb{R}_{b_{1}\alpha \beta }^{b}& =u_{b}\left(
\rho _{\beta }^{j}\circ h\frac{\partial \rho \Gamma _{b_{1}\alpha }^{b}}{%
\partial x^{j}}+\rho \Gamma _{e\beta }^{b}\rho \Gamma _{b_{1}\alpha
}^{e}\right. \\
& -\rho _{\alpha }^{i}\circ h\frac{\partial \rho \Gamma _{b_{1}\beta }^{b}}{%
\partial x^{i}}\left. -\rho \Gamma _{e\alpha }^{b}\rho \Gamma _{b_{1}\beta
}^{e}-\rho \Gamma _{b_{1}\gamma }^{b}L_{\beta \alpha }^{\gamma }\circ
h\right)
\end{split}%
\end{equation*}%
and
\begin{equation*}
u_{b_{1}\mid \gamma }L_{\alpha \beta }^{\gamma }\circ h=\left( \rho _{\gamma
}^{k}\circ h\frac{\partial u_{b_{1}}}{\partial x^{k}}-\rho \Gamma
_{b_{1}\gamma }^{b}u_{b}\right) L_{\alpha \beta }^{\gamma }\circ h
\end{equation*}%
it results that %\begin{equation*}
%
%
\end{equation*}%
\emph{\ we have that }$u_{b~|c},~b,c\in \overline{1,n}$ are the \emph{%
components of a tensor field of }$\left( 0,2\right) $\emph{\ type.}

\bigskip\noindent\textit{Proof.} Let $U$ and $U^{\prime }$ be two vector
local $\left( m+n\right) $-charts such that $U\cap U^{\prime }\neq \phi .$

Since $u_{b^{\prime }}\left( x\right) =M_{b^{\prime }}^{b}\left( x\right)
u_{b}\left( x\right) ,~$for any $x\in U\cap U^{\prime },$ it results that
\begin{equation*}
%
%
\leqno(1)
\end{equation*}
Since, for any $x\in U\cap U^{\prime }$, we have
\begin{equation*}
%
%
\leqno(3)
\end{equation*}
it results that
\begin{equation*}
%
%
\leqno(4)
\end{equation*}
Summing the equalities $\left( 1\right) $ and $\left( 4\right) $, it results
the conclusion of lemma.\hfill \emph{q.e.d.}

\bigskip\noindent\textbf{Theorem 4.1.4 }\emph{If }$\left( E,\pi ,M\right)
=\left( F,\nu ,N\right) $\emph{, then, for any }
\begin{equation*}
%
%
\end{equation*}%
\emph{\ we shall use the notation }%
\begin{equation*}
%
%
\leqno(4.1.18)
\end{equation*}%
\emph{and we verify the formulas of Ricci type }%
\begin{equation*}
%
%
\leqno(4.1.19)
\end{equation*}
\emph{In particular, if }$h=Id_{M}$\emph{, then the relations }$\left(
4.1.19\right) $\emph{\ become}%
\begin{equation*}
%
%
\leqno(4.1.19^{\prime })
\end{equation*}
\emph{Moreover, if }$\rho =id_{TM}$ \emph{then the relations }$\left(
4.1.19^{\prime }\right) $\emph{\ become}%
\begin{equation*}
%
%
\leqno(4.1.19^{\prime \prime })
\end{equation*}

\smallskip \noindent\textbf{Theorem 4.1.5 }\emph{For any tensor field }%
\begin{equation*}
%
%
,
\end{equation*}%
\emph{we verify the equality: }%
\begin{equation*}
%
%
\leqno\left( 4.1.20\right)
\end{equation*}
\emph{In particular, if }$h=Id_{M}$\emph{, then the relations }$\left(
4.1.20\right) $\emph{\ become}%
\begin{equation*}
%
%
\leqno( 4.1.20^{\prime })
\end{equation*}

\smallskip \noindent\textbf{Theorem 4.1.6 }\emph{If }$\left( E,\pi ,M\right)
=\left( F,\nu ,N\right) $\emph{, then we obtain the following formulas of
Ricci type:}%
\begin{equation*}
%
%
\leqno(4.1.21)
\end{equation*}
\emph{In particular, if }$h=Id_{M}$\emph{, then the relations }$\left(
4.1.21\right) $\emph{\ become}%
\begin{equation*}
%
%
\leqno(4.1.21^{\prime })
\end{equation*}

We observe that if the structure functions of generalized Lie algebroid
\begin{equation*}
\left( \left( F,\nu ,M\right) ,\left[ ,\right] _{F,h},\left( \rho
,Id_{M}\right) \right) ,
\end{equation*}
the $\left( \rho ,h\right) $-torsion associated to linear $\rho $-connection
$\rho \Gamma $ and the $\left( \rho ,h\right) $-curvature asso\-cia\-ted to
linear $\rho $-connection $\rho \Gamma $ are null, then we have the
equality:
\begin{equation*}
%
%
\leqno(4.1.22)
\end{equation*}%
which generalizes the Schwartz equality.

\subsection{Torsion and curvature forms. Identities of Cartan and Bianchi
type}

We apply the theory of the generalized linear connections for the diagram:%
\begin{equation*}
%
\leqno(4.2.1)
\end{equation*}%
where $\left( E,\pi ,M\right) \in \left\vert \mathbf{B}^{\mathbf{v}%
}\right\vert $ and $\left( \left( F,\nu ,M\right) ,\left[ ,\right]
_{F,h},\left( \rho ,Id_{M}\right) \right) \in \left\vert \mathbf{GLA}%
\right\vert .$

Let $\rho \Gamma $ be a linear $\rho $-connection for the vector bundle $%
\left( E,\pi ,M\right) .$

\bigskip\noindent\textbf{Definition 4.2.1 }For each $a,b\in \overline{1,n}$,
we obtain the scalar $1$-forms%
\begin{equation*}
%
%
\leqno(4.2.2^{\prime })
\end{equation*}%
which will be called the \emph{form of linear }$\rho $\emph{-connection }$%
\rho \dot{\Gamma}$\emph{\ and }$\rho \Gamma $\emph{\ respectively}$.$

\bigskip\noindent\textbf{Definition 4.2.2 }If $\left( E,\pi ,M\right)
=\left( F,\nu ,M\right) $, then the vector valued 2-form%
\begin{equation*}
%
%
\leqno(4.2.3)
\end{equation*}%
will be called the \emph{vector valued form of }$\left( \rho ,h\right) $%
\emph{-torsion }$\left( \rho ,h\right) \mathbb{T}$\emph{.}

In particular, if $h=Id_{M}$, then the vector valued 2-form
\begin{equation*}
%
%
\leqno(4.2.3^{\prime })
\end{equation*}%
will be called the \emph{vector form of }$\rho $\emph{-torsion }$\rho
\mathbb{T}$\emph{.}

Moreover, if $\rho =Id_{TM}$, then the vector valued form $\left(
4.2.3^{\prime }\right) $ becomes:
\begin{equation*}
%
%
\leqno(4.2.3^{\prime \prime })
\end{equation*}

\smallskip \noindent\textbf{Definition 4.2.3 }For each $c\in \overline{1,n}$
we obtain the \emph{scalar }$2$\emph{-form of }$\left( \rho ,h\right) $\emph{%
-torsion }$\left( \rho ,h\right) \mathbb{T}$
\begin{equation*}
%
%
\leqno(4.2.4)
\end{equation*}
In particular, if $h=Id_{M}$, then, for each $c\in \overline{1,n},$ we
obtain the \emph{scalar }$2$\emph{-form of }$\rho $\emph{-torsion }$\rho
\mathbb{T}$
\begin{equation*}
%
%
\leqno(4.2.4^{\prime })
\end{equation*}
Moreover, if $\rho =Id_{TM}$, then the scalar $2$-form $\left( 4.2.4^{\prime
}\right) $ becomes:
\begin{equation*}
%
%
\leqno(4.2.4^{\prime \prime })
\end{equation*}

\smallskip \noindent\textbf{Definition 4.2.3 }The vector mixed form
\begin{equation*}
%
%
\leqno(4.2.5)
\end{equation*}%
will be called the \emph{vector valued form of }$\left( \rho ,h\right) $%
\emph{-curvature }$\left( \rho ,h\right) \mathbb{R}$\emph{.}

In particular, if $h=Id_{M}$, then the vector mixed form
\begin{equation*}
%
%
\leqno(4.2.5^{\prime })
\end{equation*}%
will be called the \emph{vector valued form of }$\rho $\emph{-curvature }$%
\rho \mathbb{R}$\emph{.}

Moreover, if $\rho =Id_{TM}$, then the vector form $\left( 4.2.5^{\prime
}\right) $ becomes:%
\begin{equation*}
%
%
\leqno(4.2.5^{\prime \prime })
\end{equation*}

\smallskip \noindent\textbf{Definition 4.2.4 }For each $a,b\in \overline{1,n}
$ we obtain the \emph{scalar }$2$\emph{-form of }$\left( \rho ,h\right) $%
\emph{-curvature }$\left( \rho ,h\right) \mathbb{R}$
\begin{equation*}
%
%
\leqno(4.2.6)
\end{equation*}
In particular, if $h=Id_{M}$, then, for each $a,b\in \overline{1,n},$ we
obtain the \emph{scalar }$2$\emph{-form of }$\rho $\emph{-curvature }$\rho
\mathbb{R}$
\begin{equation*}
%
%
\leqno(4.2.6^{\prime })
\end{equation*}
Moreover, if $\rho =Id_{TM}$, then the scalar form $\left( 4.2.6^{\prime
}\right) $ becomes:
\begin{equation*}
%
%
\leqno(C_{2})
\end{equation*}%
\emph{hold good. These will be called the first respectively the second
identity of Cartan type.}

\bigskip\noindent\textit{Proof.} To prove the first identity we consider
that $\left( E,\pi ,M\right) =\left( F,\nu ,M\right) .$ Therefore, $\Omega
_{b}^{a}=\rho \Gamma _{bc}^{a}S^{c}.$ Since
\begin{equation*}
%
%
\end{equation*}%
it results the first identity.

To prove the second identity, we consider that $\left( E,\pi ,M\right) \neq
\left( F,\nu ,M\right) .$ Since
\begin{equation*}
%
%
\end{equation*}%
it results the second identity.

\bigskip\noindent\textbf{Corollary 4.2.1 }\emph{In particular, if }$%
h=Id_{M}, $\emph{\ then the identities }$(C_{1})$\emph{\ and }$(C_{2})$
\emph{become}%
\begin{equation*}
%
%
\leqno(C_{2}^{\prime })
\end{equation*}%
\emph{respectively.}

\emph{Moreover, if }$\rho =Id_{TM}$\emph{, then the identities }$\left(
C_{1}^{\prime }\right) $\emph{\ and }$\left( C_{2}^{\prime }\right) \,$\emph{%
become:}%
\begin{equation*}
%
%
\leqno(B_{2})
\end{equation*}%
\emph{hold good.} \emph{We will called these the first respectively the
second identity of Bianchi type.}

\emph{If the }$\left( \rho ,h\right) $\emph{-torsion is null, then the first
identity of Bianchi type becomes: }%
\begin{equation*}
%
%
\leqno(\tilde{B}_{1})
\end{equation*}

\smallskip \noindent\textit{Proof.} \textbf{\ }We consider $\left( E,\pi
,M\right) =\left( F,\nu ,M\right) .$ Using the first identity of Cartan type
and the equality $d^{h^{\ast }F}\circ d^{h^{\ast }F}=0,$ we obtain:
\begin{equation*}
%
%
\end{equation*}
Using the second identity of Cartan type and the previous identity, we
obtain:
\begin{equation*}
d^{h^{\ast }F}\left( \rho ,h\right) \mathbb{T}^{a} =\left( \left( \rho
,h\right) \mathbb{R}_{b}^{a}-\Omega _{c}^{a}\wedge \Omega _{b}^{c}\right)
\wedge S^{b} -\Omega _{c}^{a}\wedge \left( \left( \rho ,h\right) \mathbb{T}%
^{c}-\Omega _{b}^{c}\wedge S^{b}\right) .
\end{equation*}
After some calculations, we obtain the first identity of Bianchi type.

Using the second identity of Cartan type and the equality $d^{h^{\ast
}F}\circ d^{h^{\ast }F}=0,$ we obtain:
\begin{equation*}
%
\end{equation*}
Using the second of Cartan type and the previous identity, we obtain:%
\begin{equation*}
%
%
\end{equation*}
After some calculations, we obtain the second identity of Bianchi
type.\hfill \hfill \emph{q.e.d.}

\bigskip\noindent\textbf{Corollary 4.2.2 }\emph{In particular, if }$%
h=Id_{M}, $\emph{\ then the identities }$(B_{1})$\emph{\ and }$(B_{2})$
\emph{become }%
\begin{equation*}
%
%
\leqno(B_{2}^{\prime })
\end{equation*}%
\emph{\ respectively.}

\emph{Moreover, if }$\rho =Id_{TM}$\emph{, then the identities }$\left(
B_{1}^{\prime }\right) $\emph{\ and }$\left( B_{2}^{\prime }\right) \,$\emph{%
become:}
\begin{equation*}
%
%
\leqno(B_{2}^{\prime \prime })
\end{equation*}%
\emph{\ respectively.}

\bigskip\noindent \textbf{Theorem 4.2.3 }\emph{If }$(E,\pi ,M){=}(F,\nu ,M)$%
\emph{, then the following relations hold good}
\begin{equation*}
%
%
\leqno(\tilde{B}_{1})
\end{equation*}%
\emph{and}
\begin{equation*}
\underset{cyclic\left( u_{1},u_{2},u_{3},u\right) }{\dsum }\!\!\left\{ \rho
\ddot{D}_{U_{1}}\left( \left( \rho ,h\right) \mathbb{R}\left(
U_{2},U_{3}\right) U\right) {-}\left( \rho ,h\right) \mathbb{R}\left( \left(
\rho ,h\right) \mathbb{T}\left( U_{1},U_{2}\right) ,U_{3}\right) U\right\} {=%
}0. \leqno(\tilde{B}_{2})
\end{equation*}%
\emph{respectively. This identities will be called the first respectively
the second identity of Bianchi type.}

\emph{In particular, if }$h=Id_{M},$\emph{\ then the identities }$(\tilde{B}%
_{1})$\emph{\ and }$(\tilde{B}_{2})$\emph{\ become }%
\begin{equation*}
\underset{cyclic\left( u_{1},u_{2},u_{3}\right) }{\dsum }\left\{ \rho
D_{u_{1}}\left( \rho \mathbb{T}\left( u_{2},u_{3}\right) \right) -\rho
\mathbb{R}\left( u_{1},u_{2}\right) u_{3} +\rho \mathbb{T}\left( \rho
\mathbb{\ T}\left( u_{1},u_{2}\right) ,u_{3}\right) \right\} =0, \leqno(%
\tilde{B}_{1}^{\prime })
\end{equation*}
\begin{equation*}
\begin{array}{c}
\underset{cyclic\left( u_{1},u_{2},u_{3},u\right) }{\dsum }\left\{ \rho
D_{u_{1}}\left( \rho \mathbb{R}\left( u_{2},u_{3}\right) u\right) -\rho
\mathbb{R}\left( \rho \mathbb{T}\left( u_{1},u_{2}\right) ,u_{3}\right)
u\right\} =0.%
\end{array}%
\leqno(\tilde{B}_{2}^{\prime })
\end{equation*}%
\emph{which will be called the first respectively the second identity of
Bianchi type.}

\bigskip\noindent \textbf{Remark 4.2.1} On components, the identities of
Bianchi type $(\tilde{B}_{1})$ and $(\tilde{B}_{2})$ become:
\begin{equation*}
\begin{array}{r}
\underset{cyclic\left( a_{1},a_{2},a_{3}\right) }{\sum }\left\{ \left( \rho
,h\right) \mathbb{T}_{~\ a_{2}a_{3}{\mid a_{1}}}^{a}+\left( \rho ,h\right)
\mathbb{T}_{~ga_{3}}^{a}\cdot \left( \rho ,h\right) \mathbb{T}%
_{~a_{1}a_{2}}^{g}\right\}\vspace*{1mm} \\
=\underset{cyclic\left( a_{1},a_{2},a_{3}\right) }{\sum }\left( \rho
,h\right) \mathbb{R}_{a_{3}~a_{1}a_{2}}^{a}%
\end{array}
\leqno(\tilde{B}_{1}^{\prime \prime })
\end{equation*}%
and
\begin{equation*}
\underset{cyclic\left( a_{1},a_{2},a_{3}\right) }{\sum }\left\{ \left( \rho
,h\right) \mathbb{R}_{b~a_{2}a_{3}{\mid a_{1}}}^{a}+\left( \rho ,h\right)
\mathbb{R}_{b~ga_{3}}^{a}\cdot \left( \rho ,h\right) \mathbb{T}%
_{~a_{1}a_{2}}^{g}\right\} =0. \leqno(\tilde{B}_{2}^{\prime \prime })
\end{equation*}
If the $\left( \rho ,h\right) $-torsion is null, then the identities of
Bianchi type become:
\begin{equation*}
\underset{cyclic\left( a_{1}~a_{2},a_{3}\right) }{\sum }\left( \rho
,h\right) \mathbb{R}_{a_{3},a_{1}a_{2}}^{a}=0 \leqno(\tilde{B}_{1}^{\prime
\prime \prime })
\end{equation*}%
and
\begin{equation*}
\underset{cyclic\left( a_{1},a_{2},a_{3}\right) }{\sum }\left( \rho
,h\right) \mathbb{R}_{b~a_{2}a_{3}\mid _{a_{1}}}^{a}=0. \leqno(\tilde{B}%
_{2}^{\prime \prime \prime })
\end{equation*}

\subsection{(Pseudo)metrizable vector bundles}

We will apply our theory for the diagram:
\begin{equation*}
%
,\leqno(4.3.1)
\end{equation*}%
where $\left( E,\pi ,M\right) \in \left\vert \mathbf{B}^{\mathbf{v}%
}\right\vert $ and $\left( \left( F,\nu ,M\right) ,\left[ ,\right]
_{F,h},\left( \rho ,Id_{M}\right) \right) \in \left\vert \mathbf{GLA}%
\right\vert .$

\bigskip\noindent \textbf{Definition 4.3.1} We will say that the \emph{%
vector bundle }$\left( E,\pi ,M\right) $\emph{\ is endowed with a
pseudometrical structure}\textit{\ }if it exists
\begin{equation*}
%
%
\end{equation*}%
such that for each $x\in M,$ the matrix $\left\Vert g_{ab}\left( x\right)
\right\Vert $ is nondegenerate and symmetric.

Moreover, if for each $x\in M$ the matrix $\left\Vert g_{ab}\left( x\right)
\right\Vert $ has constant signature, then we will say that \emph{the vector
bundle }$\left( E,\pi ,M\right) $\emph{\ is endowed with a metrical
structure.}

If
\begin{equation*}
%
%
\end{equation*}%
is a (pseudo) metrical structure, then, for any $a,b\in \overline{1,r}$ and
for any vector local $\left( m+r\right) $-chart $\left( U,s_{U}\right) $ of $%
\left( E,\pi ,M\right) $, we consider the real functions
\begin{equation*}
%
%
\end{equation*}%
for any $\forall x\in U.$

\smallskip \noindent \textbf{Definition 4.3.2} Let $\left( E,\pi ,M\right) $
be a vector bundle endowed with a (pseudo)metrical structure $g$ and with a
linear $\rho $-connection $\rho \Gamma .$

We will say that the \emph{linear }$\rho $\emph{-connection }$\rho \Gamma $%
\emph{\ is compatible with the (pseudo)metrical structure }$g$ if
\begin{equation*}
%
%
\leqno(4.3.2)
\end{equation*}

\smallskip \noindent \textbf{Definition 4.3.3} We will say that the vector
bundle $(E,\pi ,M)$ is $\rho $\emph{-(pseudo)metrizable,} if it exists a
(pseudo)metrical structure
\begin{equation*}
g\in \mathcal{T}_{2}^{0}\left( E,\pi ,M\right)
\end{equation*}%
and a linear $\rho $-connection $\rho \Gamma $ for $\left( E,\pi ,M\right) $
compatible with $g.$ The $id_{TM}$-(pseudo)metri\-zable vector bundles will
be called\textit{(}\textit{pseudo})\textit{metrizable vector bundles.}

In particular, if $\left( TM,\tau _{M},M\right) $ is a (pseudo)metri\-zable
vector bundle, then we will say that $\left( TM,\tau _{M},M\right) $ \textit{%
is a \emph{(pseudo)Riemannian space}, }and the manifold $M$ will be called
the (\textit{pseudo})\textit{Riemannian manifold.}

The linear connection of a (pseudo)Riemannian space will be called (\textit{%
pseudo})\textit{Rieman\-nian linear connection.}

\bigskip \noindent \textbf{Theorem 4.3.1 }\emph{If }$(E,\pi ,M)=(F,\nu ,M)$%
\emph{\ and }$g\in \mathcal{T}_{2}^{0}\left( h^{\ast }E,h^{\ast }\pi
,M\right) $\emph{\ is a} (\textit{pseudo})\textit{metri\-cal structure, then
the local real functions}%
\begin{equation*}
\begin{array}{cl}
\rho \Gamma _{bc}^{a} & =\displaystyle\frac{1}{2}\tilde{g}^{ad}\left( \rho
_{c}^{k}\circ h\frac{\partial g_{bd}}{\partial x^{k}}+\rho _{b}^{j}\circ h%
\frac{\partial g_{dc}}{\partial x^{j}}-\rho _{d}^{h}\circ h\frac{\partial
g_{bc}}{\partial x^{h}}\right. \vspace*{1mm} \\
& \displaystyle\left. +g_{ec}L_{bd}^{e}\circ h+g_{be}L_{dc}^{e}\circ
h-g_{de}L_{bc}^{e}\circ h\right) .%
\end{array}%
\leqno(4.3.3)
\end{equation*}%
\emph{are the components of a linear }$\rho $\emph{-connection }$\rho \Gamma
$\emph{\ for the vector bundle }$\left( h^{\ast }E,h^{\ast }\pi ,M\right) $%
\emph{\ compatible with }$g$\emph{\ such that }$\left( \rho ,h\right)
\mathbb{T}=0$\emph{.}

\emph{Therefore, the vector bundle }$\left( h^{\ast }E,h^{\ast }\pi
,M\right) $\emph{\ becomes} $\rho $-(\textit{pseudo})\textit{metrizable.}

The linear $\rho $-connection $\rho \Gamma $ will be called the \emph{linear
}$\rho $\emph{-connection of Levi-Civita type.}

\bigskip \noindent \textit{Proof.} Since
\begin{equation*}
\begin{array}{cl}
\left( \rho \ddot{D}_{U}g\right) V\otimes Z & =\Gamma \left( \overset{%
h^{\ast }E}{\rho },Id_{M}\right) \left( U\right) \left( \left( g\left(
V\otimes Z\right) \right) -g\left( \left( \rho \ddot{D}_{U}V\right) \otimes
Z\right) \right. \\
& -g\left( V\otimes \left( \rho \ddot{D}_{U}Z\right) \right) ,\vspace*{1mm}%
~\forall U,V,Z\in \Gamma \left( h^{\ast }E,h^{\ast }\pi ,M\right) .%
\end{array}%
\end{equation*}

It results that, for any $U,V,Z\in \Gamma \left( h^{\ast }E,h^{\ast }\pi
,M\right) ,$ we obtain the equalities:
\begin{eqnarray*}
(1) &&\Gamma \left( \overset{h^{\ast }E}{\rho },Id_{M}\right) \left(
U\right) \left( g\left( V\otimes Z\right) \right) =g\left( \left( \rho \ddot{%
D}_{U}V\right) \otimes Z\right) +g\left( V\otimes \left( \rho \ddot{D}%
_{U}Z\right) \right) , \\
(2) &&\Gamma \left( \overset{h^{\ast }E}{\rho },Id_{M}\right) \left(
Z\right) \left( g\left( U\otimes V\right) \right) =g\left( \left( \rho \ddot{%
D}_{Z}U\right) \otimes V\right) +g\left( U\otimes \left( \rho \ddot{D}%
_{Z}V\right) \right) , \\
(3) &&\Gamma \left( \overset{h^{\ast }E}{\rho },Id_{M}\right) \left(
V\right) \left( g\left( Z\otimes U\right) \right) =g\left( \left( \rho \ddot{%
D}_{V}Z\right) \otimes U\right) +g\left( Z\otimes \left( \rho \ddot{D}%
_{V}U\right) \right) .
\end{eqnarray*}%
We observe that $\left( 1\right) +\left( 3\right) -\left( 2\right) $ is
equivalent with the equality:
\begin{equation*}
%
\end{equation*}%
Using the condition $\left( \rho ,h\right) \mathbb{T}=0,$ which is
equivalent with the equality%
\begin{equation*}
\rho \ddot{D}_{U}V-\rho \ddot{D}_{V}U-\left[ U,V\right] _{h^{\ast }E}=0,
\end{equation*}%
we obtain:
\begin{equation*}
%
%
\end{equation*}%
Therefore, we obtain the equality:
\begin{equation*}
%
%
\end{equation*}%
which is equivalent with:
\begin{equation*}
%
%
\end{equation*}%
Finally, we obtain:
\begin{equation*}
%
%
\end{equation*}%
where $\left\Vert \tilde{g}^{dc}\left( x\right) \right\Vert =\left\Vert
g_{cd}\left( x\right) \right\Vert ^{-1},$ for any $x\in M.$ \hfill \emph{%
q.e.d.}

\bigskip\noindent \textbf{Corollary 4.3.1 }\emph{In particular, if }$%
h=Id_{M},$ $(E,\pi ,M)=(F,\nu ,M)$\emph{\ and }$g\in \mathcal{T~}%
_{2}^{0}\left( E,\pi ,M\right) $\emph{\ is a (pseudo)metrical structure,
then the local real functions}%
\begin{equation*}
\rho \Gamma _{bc}^{a} =\frac{1}{2}\tilde{g}^{ad}\left( \rho _{c}^{k}\frac{%
\partial g_{bd}}{\partial x^{k}}+\rho _{b}^{j}\frac{\partial g_{dc}}{%
\partial x^{j}}-\rho _{d}^{h}\frac{\partial g_{bc}}{\partial x^{h}}
+g_{ec}L_{bd}^{e}+g_{be}L_{dc}^{e}-g_{de}L_{bc}^{e}\right) . \leqno%
(4.3.3^{\prime })
\end{equation*}%
\emph{are the components of a linear }$\rho $\emph{-connection }$\rho \Gamma
$\emph{\ for the vector bundle }$\left( E,\pi ,M\right) $\emph{\ compatible
with }$g$\emph{\ such that }$\rho \mathbb{T}=0$\emph{.}

\emph{Therefore, the vector bundle }$\left( E,\pi ,M\right) $\emph{\ becomes}
$\rho $\emph{-(pseudo)metrizable.}

The linear $\rho $-connection $\rho \Gamma $ will be called the \emph{linear
}$\rho $\emph{-connection of Levi-Civita type.}

In particular, if $\rho =Id_{TM}$, we obtain the classical Levi-Civita
linear connection.

\bigskip\noindent\textbf{Theorem 4.3.2. }\emph{If }$(E,\pi ,M)=(F,\nu ,M),$%
\emph{\ }$g\in \mathcal{T~}_{2}^{0}\left( h^{\ast }E,h^{\ast }\pi ,M\right) $%
\emph{\ is a pseudo(metrical) structure and }$\mathbb{T}\in \mathcal{T~}%
_{2}^{1}\left( h^{\ast }E,h^{\ast }\pi ,M\right) $\emph{\ such that its
components are skew symmetric in the lover indices, then the local real
functions }%
\begin{equation*}
\begin{array}{c}
\rho \mathring{\Gamma}_{bc}^{a}=\rho \Gamma _{bc}^{a}+\displaystyle\frac{1}{2%
}\tilde{g}^{ad}\left( g_{de}\mathbb{T}_{bc}^{e}-g_{be}\mathbb{T}%
_{dc}^{e}+g_{ec}\mathbb{T}_{bd}^{e}\right) ,%
\end{array}%
\leqno(4.3.4)
\end{equation*}%
\emph{are the components of a linear} $\rho $\emph{-connection compatible
with the (pseudo) metrical structure }$g,$ \emph{\ where }$\rho \Gamma
_{bc}^{a}$\emph{\ are the components of linear} $\rho $\emph{-connection of
Levi-Civita type. Therefore, the vector bundle }$\left( h^{\ast }E,h^{\ast
}\pi ,M\right) $\emph{\ becomes} $\rho $\emph{-(pseudo)metrizable.}

\emph{In addition, the tensor field }$\mathbb{T}$\emph{\ is the }$\left(
\rho ,h\right) $\emph{-torsion tensor field.}

\bigskip\noindent \textbf{Corollary 4.3.2 }\emph{In particular, if }$%
h=Id_{M},$ $(E,\pi ,M)=(F,\nu ,M),$\emph{\ }$g\in \mathcal{T~}_{2}^{0}\left(
E,\pi ,M\right) $\emph{\ is a pseudo(metrical) structure and }$T\in \mathcal{%
T~}_{2}^{1}\left( E,\pi ,M\right) $\emph{\ such that its components are skew
symmetric in the lover indices, then the local real functions }%
\begin{equation*}
\begin{array}{c}
\rho \mathring{\Gamma}_{bc}^{a}=\rho \Gamma _{bc}^{a}+\displaystyle\frac{1}{2%
}\tilde{g}^{ad}\left( g_{de}\mathbb{T}_{bc}^{e}-g_{be}\mathbb{T}%
_{dc}^{e}+g_{ec}\mathbb{T}_{bd}^{e}\right) ,%
\end{array}%
\leqno(4.3.4^{\prime })
\end{equation*}%
\emph{are the components of a linear} $\rho $\emph{-connection compatible
with the (pseudo)metrical structure }$g,$ \emph{\ where }$\rho \Gamma
_{bc}^{a}$\emph{\ are the components of linear} $\rho $\emph{-connection of
Levi-Civita type. Therefore, the vector bundle }$(E,\pi ,M)$\emph{\ becomes}
$\rho $\emph{-(pseudo)metrizable.}

\emph{In addition, the tensor field }$\mathbb{T}$\emph{\ is the }$\rho $%
\emph{-torsion tensor field.}

\bigskip\noindent\textbf{Theorem 4.3.3 }\emph{If }$g\in \mathcal{T~}%
_{2}^{0}\left( E,\pi ,M\right) $\emph{\ is a pseudo (metrical) structure and
}$\rho \mathring{\Gamma}$\emph{\ is a linear }$\rho $\emph{-connection for
the vector bundle }$\left( E,\pi ,M\right) $\emph{, then the local real
functions }%
\begin{equation*}
\begin{array}{c}
\overset{k}{\rho \Gamma }_{b\alpha }^{a}=\rho \mathring{\Gamma}_{b\alpha
}^{a}+\frac{1}{2}\tilde{g}^{ac}g_{cb\overset{\circ }{\mid }\alpha }%
\end{array}%
\leqno(4.3.5)
\end{equation*}%
\emph{are the components of a linear }$\rho $\emph{-connection compatible
with the (pseudo) metrical structure }$g.$ \emph{Therefore, the vector
bundle }$(E,\pi ,M)$\emph{\ becomes} $\rho $\emph{-(pseudo)metrizable.}

\bigskip\noindent\textbf{Theorem 4.3.4 }\emph{If }$g\in \mathcal{T~}%
_{2}^{0}\left( E,\pi ,M\right) $\emph{\ is a pseudo (metrical) structure, }$%
\rho \mathring{\Gamma}$\emph{\ is a linear }$\rho $\emph{-connection for the
vector bundle }$\left( E,\pi ,M\right) $\emph{\ and }$T=T_{c\alpha
}^{d}s_{d}\otimes s^{c}\otimes t^{\alpha }$\emph{, then the local real
functions}
\begin{equation*}
%
\leqno(4.3.6)
\end{equation*}%
\emph{are the components of a linear }$\rho $\emph{-connection compatible
with (pseudo) metrical structure }$g,$ \emph{where }%
\begin{equation*}
%
%
\leqno(4.3.7)
\end{equation*}%
\emph{is the Obata operator.}

\emph{Therefore, the vector bundle }$(E,\pi ,M)$\emph{\ becomes} $\rho $%
\emph{-(pseudo)metrizable.}

\subsection{Lifts of differentiable curves}

In this section we extend the notion of lift of a curve $c$ at the total
space of a vector bundle using the new notion of \emph{locally invertible }$%
\mathbf{B}^{\mathbf{v}}$\emph{-morphism.}

\subsubsection{The lift of a differentiable curve at the total space of a
vector bundle}

We consider the following diagram:
\begin{equation*}
%
\leqno(4.4.1.1)
\end{equation*}%
where $\left( E,\pi ,M\right) \in \left\vert \mathbf{B}^{\mathbf{v}%
}\right\vert $ and $\left( \left( F,\nu ,M\right) ,\left[ ,\right]
_{F,h},\left( \rho ,\eta \right) \right) \in \left\vert \mathbf{GLA}%
\right\vert .$

We admit that $\left( \rho ,\eta \right) \Gamma $ is a $\left( \rho ,\eta
\right) $-connection for the vector bundle $\left( E,\pi ,M\right) .$

Let
\begin{equation*}
\begin{array}{ccc}
I & ^{\underrightarrow{\ \ c\ \ }} & M%
\end{array}%
\end{equation*}%
be a differentiable curve.

We say that
\begin{equation*}
%
%
\end{equation*}%
is a vector subbundle of the vector bundle $\left( E,\pi ,M\right) .$

\bigskip\noindent\textbf{Definition 4.4.1.1} Let
\begin{equation*}
%
%
\leqno(4.4.1.2)
\end{equation*}%
be a differentiable curve.

If there exists $g\in \mathbf{Man}\left( E,F\right) $ such that the
following conditions are satisfied:

\begin{itemize}
\item[1.] $\left( g,h\right) \in \mathbf{B}^{v}\left( \left( E,\pi ,M\right)
,\left( F,\nu ,N\right) \right) $ and

\item[2.] $\rho \circ g\circ \dot{c}\left( t\right) =\displaystyle\frac{%
d\left( \eta \circ h\circ c\right) ^{i}\left( t\right) }{dt}\frac{\partial }{%
\partial x^{i}}\left( \left( \eta \circ h\circ c\right) \left( t\right)
\right) ,$ for any $t\in I,$ \smallskip then we will say that $\dot{c}$
\emph{\ is the }$\left( g,h\right) $\emph{-lift of the differentiable curve }%
$c.$
\end{itemize}

\smallskip \noindent\textbf{Remark 4.4.1.1 }Condition $2$ is equivalent with
the following affirmation:
\begin{equation*}
\rho _{\alpha }^{i}\left( \eta \circ h\circ c\left( t\right) \right) \cdot
g_{a}^{\alpha }\left( h\circ c\left( t\right) \right) \cdot y^{a}\left(
t\right) =\frac{d\left( \eta \circ h\circ c\right) ^{i}\left( t\right) }{dt}%
,~i\in \overline{1,m}.\leqno\left( 4.4.1.3\right)
\end{equation*}

\smallskip \noindent\textbf{Definition 4.4.1.2} If
\begin{equation*}
%
\end{equation*}%
is a differentiable $\left( g,h\right) $-lift of the differentiable curve $%
c, $ then the section%
\begin{equation*}
%
%
\leqno(4.4.1.4)
\end{equation*}%
will be called the\emph{\ canonical section associated to the couple }$%
\left( c,\dot{c}\right) .$

We will denote by $\left( T^{E}\left( c,\dot{c}\right) ,\tau ,\func{Im}%
\left( \eta \circ h\circ c\right) \right) $ the vector subbundle with
minimal dimension such that
\begin{equation*}
%
%
\leqno(4.4.1.5)
\end{equation*}%
and will denoted by $\left( S^{E}\left( c,\dot{c}\right) ,\sigma ,\func{Im}%
\left( \eta \circ h\circ c\right) \right) $ the vector subbundle such that
\begin{equation*}
T^{E}\left( c,\dot{c}\right) \oplus S^{E}\left( c,\dot{c}\right) =E_{|\func{%
Im}\left( \eta \circ h\circ c\right) }.
\end{equation*}

\smallskip \noindent \textbf{Definition 4.4.1.3 }If $\left( g,h\right) \in
\mathbf{B}^{\mathbf{v}}\left( \left( E,\pi ,M\right) ,\left( F,\nu ,N\right)
\right) $ has the components
\begin{equation*}
%
%
\end{equation*}%
such that for any local vector $\left( n+p\right) $-chart $\left(
V,t_{V}\right) $ of $\left( F,\nu ,N\right) $ there exists the real
functions
\begin{equation*}
%
%
\end{equation*}%
for any $\varkappa \in V,$ then we will say that \emph{the }$\mathbf{B}^{%
\mathbf{v}}$\emph{-morphism }$\left( g,h\right) $\emph{\ is locally
invertible.}

\bigskip\noindent\textbf{Remark 4.4.2.2 }In particular, if $\left(
Id_{TM},Id_{M},Id_{M}\right) =\left( \rho ,\eta ,h\right) $ and the $\mathbf{%
B}^{\mathbf{v}}$ morphism $\left( g,Id_{M}\right) $ is locally invertible,
then we have the differentiable $\left( g,Id_{M}\right) $-lift
\begin{equation*}
%
%
.\leqno(4.4.1.6)
\end{equation*}
Moreover, if $g=Id_{TM}$, then we obtain the usual lift of tangent vectors%
\begin{equation*}
%
%
\leqno(4.4.1.7)
\end{equation*}%
is a differentiable $\left( g,h\right) $-lift of differentiable curve $c,$
such that its components functions $\left( y^{a},~a\in \overline{1,n}\right)
$ are solutions for the differentiable system of equations:%
\begin{equation*}
\frac{du^{a}}{dt}+\left( \rho ,\eta \right) \Gamma _{\alpha }^{a}\circ
u\left( c,\dot{c}\right) \circ \left( \eta \circ h\circ c\right) \cdot
g_{b}^{\alpha }\circ h\circ c\cdot u^{b}=0,\leqno(4.4.1.8)
\end{equation*}%
then we will say that \emph{the }$\left( g,h\right) $\emph{-lift }$\dot{c}$%
\emph{\ is parallel with respect to the }$\left( \rho ,\eta \right) $\emph{%
-connection }$\left( \rho ,\eta \right) \Gamma .$

\bigskip\noindent\textbf{Remark 4.4.1.3 }In particular, if $\left( \rho
,\eta ,h\right) =\left( Id_{TM},Id_{M},Id_{M}\right) $ and the $\mathbf{B}^{%
\mathbf{v}}$ morphism $\left( g,Id_{M}\right) $ is locally invertible, then
the differentiable $\left( g,Id_{TM}\right) $-lift
\begin{equation*}
\begin{array}{ccl}
I & ^{\underrightarrow{\ \ \dot{c}\ \ }} & TM \vspace*{1,5mm} \\
t & \longmapsto & \displaystyle\left( \tilde{g}_{j}^{i}\circ c\cdot \frac{%
dc^{j}}{dt}\right) \frac{\partial }{\partial x^{i}}\left( c\left( t\right)
\right) ,%
\end{array}%
\leqno(4.4.1.9)
\end{equation*}%
is parallel with respect to the connection $\Gamma $ if the component
functions
\begin{equation*}
\left( \tilde{g}_{j}^{i}\circ c\cdot \frac{dc^{j}}{dt},~i\in \overline{1,n}%
\right)
\end{equation*}
are solutions for the differentiable system of equations%
\begin{equation*}
\frac{du^{i}}{dt}+\Gamma _{k}^{i}\circ u\left( c,\dot{c}\right) \circ c\cdot
g_{h}^{k}\circ c\cdot u^{h}=0,\leqno(4.4.1.10)
\end{equation*}%
namely%
\begin{equation*}
\begin{array}{l}
\displaystyle\frac{d}{dt}\left( \tilde{g}_{j}^{i}\left( c\left( t\right)
\right) \cdot \frac{dc^{j}\left( t\right) }{dt}\right) \vspace*{1,5mm} \\
\qquad\displaystyle+\Gamma _{k}^{i}\left( c\left( t\right) ,\left( \tilde{g}%
_{j}^{i}\left( c\left( t\right) \right) \cdot \frac{dc^{j}\left( t\right) }{%
dt}\right) \cdot \frac{\partial }{\partial x^{i}}\left( c\left( t\right)
\right) \right) \cdot \frac{dc^{k}\left( t\right) }{dt}=0.%
\end{array}%
\leqno(4.4.1.10)^{\prime }
\end{equation*}
Moreover, if $g=Id_{TM}$, then the usual lift of tangent vectors $\left(
4.4.1.6\right) ^{\prime }$ is parallel with respect to the connection $%
\Gamma $ if the component functions $\left( \frac{dc^{j}}{dt},~j\in
\overline{1,n}\right) $ are solutions for the differentiable system of
equations
\begin{equation*}
\frac{du^{i}}{dt}+\Gamma _{k}^{i}\circ u\left( c,\dot{c}\right) \circ c\cdot
u^{k}=0,\leqno(4.4.1.10)^{\prime \prime }
\end{equation*}%
namely%
\begin{equation*}
\frac{d}{dt}\left( \frac{dc^{j}\left( t\right) }{dt}\right) +\Gamma
_{k}^{i}\left( c\left( t\right) ,\frac{dc^{j}\left( t\right) }{dt}\cdot
\frac{\partial }{\partial x^{i}}\left( c\left( t\right) \right) \right)
\cdot \frac{dc^{k}\left( t\right) }{dt}=0.\leqno(4.4.1.10)^{\prime \prime
\prime }
\end{equation*}

\subsubsection{The lift of a differentiable curve at the total space of dual
vector bundle}

We consider the following diagram:
\begin{equation*}
\begin{array}{c}
\xymatrix{\overset{\ast }{E}\ar[d]_{\overset{\ast }{\pi }}&\left( F,\left[
,\right] _{F,h},\left( \rho ,\eta \right) \right)\ar[d]^\nu\\ M\ar[r]^h&N}%
\end{array}
\leqno(4.4.2.1)
\end{equation*}%
where $\left( E,\pi ,M\right) \in \left\vert \mathbf{B}^{\mathbf{v}%
}\right\vert $ and $\left( \left( F,\nu ,N\right) ,\left[ ,\right]
_{F,h},\left( \rho ,\eta \right) \right) \in \left\vert \mathbf{GLA}%
\right\vert .$

We admit that $\left( \rho ,\eta \right) \overset{\ast }{\Gamma }$ is a $%
\left( \rho ,\eta \right) $-connection for the vector bundle $\left( \overset%
{\ast }{E},\overset{\ast }{\pi },M\right) .$

Let
\begin{equation*}
\begin{array}{ccc}
I & ^{\underrightarrow{\ \ c\ \ }} & M%
\end{array}%
\end{equation*}%
be a differentiable curve. We say that
\begin{equation*}
\begin{array}{c}
\left( \overset{\ast }{E}_{|\func{Im}\left( \eta \circ h\circ c\right) },%
\overset{\ast }{\pi }_{|\func{Im}\left( \eta \circ h\circ c\right) },\func{Im%
}\left( \eta \circ h\circ c\right) \right)%
\end{array}%
\end{equation*}%
is a vector subbundle of the vector bundle $\left( \overset{\ast }{E},%
\overset{\ast }{\pi },M\right) .$

\bigskip\noindent\textbf{Definition 4.4.2.1} Let
\begin{equation*}
\begin{array}{ccl}
I & ^{\underrightarrow{\ \ \dot{c}\ \ }} & \overset{\ast }{E}_{|\func{Im}%
\left( \eta \circ h\circ c\right) } \vspace*{1,5mm} \\
t & \longmapsto & p_{a}\left( t\right) s^{a}\left( \eta \circ h\circ c\left(
t\right) \right)%
\end{array}%
\leqno(4.4.2.2)
\end{equation*}%
be a differentiable curve.

If there exists $g\in \mathbf{Man}\left( \overset{\ast }{E},F\right) $ such
that the following conditions are satisfied:

\begin{itemize}
\item[1.] $\left( g,h\right) \in \mathbf{B}^{v}\left( \left( \overset{\ast }{%
E},\overset{\ast }{\pi },M\right) ,\left( F,\nu ,N\right) \right) $ and

\item[2.] $\rho \circ g\circ \dot{c}\left( t\right) =\displaystyle\frac{%
d\left( \eta \circ h\circ c\right) ^{i}\left( t\right) }{dt}\frac{\partial }{%
\partial x^{i}}\left( \left( \eta \circ h\circ c\right) \left( t\right)
\right) ,$ for any $t\in I,$\smallskip then we will say that $\dot{c}$\emph{%
\ is the }$\left( g,h\right) $\emph{-lift of the differentiable curve }$c.$
\end{itemize}

\smallskip \noindent\textbf{Remark 4.4.2.1 }Condition 2 is equivalent with
the following affirmation:
\begin{equation*}
\rho _{\alpha }^{i}\left( \eta \circ h\circ c\left( t\right) \right)
g^{\alpha a}\left( h\circ c\left( t\right) \right) p_{a}\left( t\right) =%
\frac{d\left( \eta \circ h\circ c\right) ^{i}\left( t\right) }{dt},~i\in
\overline{1,m}.\leqno(4.4.2.3)
\end{equation*}

\smallskip \noindent\textbf{Definition 4.4.2.2} If
\begin{equation*}
\begin{array}{ccc}
I & ^{\underrightarrow{\ \ \dot{c}\ \ }} & \overset{\ast }{E}_{|\func{Im}%
\left( \eta \circ h\circ c\right) }%
\end{array}%
\end{equation*}%
is a differentiable $\left( g,h\right) $-lift of the differentiable curve $%
c, $ then the section%
\begin{equation*}
\begin{array}{ccc}
\func{Im}\left( \eta \circ h\circ c\right) & ^{\underrightarrow{\overset{%
\ast }{u}\left( c,\dot{c}\right) }} & \overset{\ast }{E}_{|\func{Im}\left(
\eta \circ h\circ c\right) } \vspace*{1,5mm} \\
\eta \circ h\circ c\left( t\right) & \longmapsto & \dot{c}\left( t\right)%
\end{array}%
\leqno(4.4.2.4)
\end{equation*}%
will be called the\emph{\ canonical section associated to the couple } $%
\left( c,\dot{c}\right) .$

We will denote by $\left( T^{\overset{\ast }{E}}\left( c,\dot{c}\right)
,\tau ,\func{Im}\left( \eta \circ h\circ c\right) \right) $ the vector
subbundle with minimal dimension such that
\begin{equation*}
\begin{array}{c}
\overset{\ast }{u}\left( c,\dot{c}\right) \in \Gamma \left( T^{\overset{\ast
}{E}}\left( c,\dot{c}\right) ,\tau ,\func{Im}\left( \eta \circ h\circ
c\right) \right)%
\end{array}%
\leqno(4.4.2.5)
\end{equation*}%
and will denoted by $\left( S^{\overset{\ast }{E}}\left( c,\dot{c}\right)
,\sigma ,\func{Im}\left( \eta \circ h\circ c\right) \right) $ the vector
subbundle such that
\begin{equation*}
T^{\overset{\ast }{E}}\left( c,\dot{c}\right) \oplus S^{\overset{\ast }{E}%
}\left( c,\dot{c}\right) =\overset{\ast }{E}_{|\func{Im}\left( \eta \circ
h\circ c\right) }.
\end{equation*}

\smallskip \noindent\textbf{Definition 4.4.2.3 }If $\left( g,h\right) \in
\mathbf{B}^{\mathbf{v}}\left( \left( \overset{\ast }{E},\overset{\ast }{\pi }%
,M\right) ,\left( F,\nu ,N\right) \right) $ has the components
\begin{equation*}
\begin{array}{c}
g^{\alpha a};a\in \overline{1,r},~\alpha \in \overline{1,p}%
\end{array}%
\end{equation*}%
such that for any vector local $\left( n+p\right) $-chart $\left(
V,t_{V}\right) $ of $\left( F,\nu ,N\right) $ there exists the real
functions
\begin{equation*}
\begin{array}{ccc}
V & ^{\underrightarrow{~\ \ \ \tilde{g}_{a\alpha }~\ \ }} & \mathbb{R}%
\end{array}%
;~a\in \overline{1,r},~\alpha \in \overline{1,p}
\end{equation*}%
such that
\begin{equation*}
\begin{array}{c}
\tilde{g}_{a\alpha }\left( \varkappa \right) \cdot g^{\alpha b}\left(
\varkappa \right) =\delta _{a}^{b},~\forall \varkappa \in V,%
\end{array}%
\end{equation*}%
then we will say that \emph{the }$\mathbf{B}^{\mathbf{v}}$\emph{-morphism }$%
\left( g,h\right) $\emph{\ is locally invertible.}

\bigskip\noindent\textbf{Remark 4.4.2.2 }In particular, if $\left(
Id_{TM},Id_{M},Id_{M}\right) =\left( \rho ,\eta ,h\right) $ and the $\mathbf{%
B}^{\mathbf{v}}$ morphism $\left( g,Id_{M}\right) $ is locally invertible,
then we have the differentiable $\left( g,Id_{M}\right) $-lift
\begin{equation*}
\begin{array}{ccl}
I & ^{\underrightarrow{\ \ \dot{c}\ \ }} & \overset{\ast }{TM} \\
t & \longmapsto & \displaystyle\tilde{g}_{ji}\left( c\left( t\right) \right)
\frac{dc^{j}\left( t\right) }{dt}dx^{i}\left( c\left( t\right) \right)%
\end{array}%
.\leqno(4.4.2.6)
\end{equation*}

\bigskip\noindent\textbf{Definition 4.4.2.4 }If
\begin{equation*}
\begin{array}{ccl}
I & ^{\underrightarrow{\ \ \dot{c}\ \ }} & \overset{\ast }{E}_{|\func{Im}%
\left( \eta \circ h\circ c\right) }%
\end{array}%
\leqno(4.4.2.7)
\end{equation*}%
is a differentiable $\left( g,h\right) $-lift for the curve $c$ such that
its components functions $\left( p_{b},~b\in \overline{1,r}\right) $ are
solutions for the differentiable system of equations:%
\begin{equation*}
\frac{du_{b}}{dt}+\left( \rho ,\eta \right) \overset{\ast }{\Gamma }%
_{b\alpha }\circ \overset{\ast }{u}\left( c,\dot{c}\right) \circ \left( \eta
\circ h\circ c\right) \cdot g^{a\alpha }\circ h\circ c\cdot u_{a}=0,\leqno%
(4.4.2.8)
\end{equation*}%
then we will say that \emph{the }$\left( g,h\right) $\emph{-lift }$\dot{c}$%
\emph{\ is parallel with respect to the }$\left( \rho ,\eta \right) $\emph{%
-connection }$\left( \rho ,\eta \right) \overset{\ast }{\Gamma }.$

\bigskip \noindent \textbf{Remark 4.4.2.3 }In particular, if $\left(
Id_{TM},Id_{M},Id_{M}\right) =\left( \rho ,\eta ,h\right) $ and the $\mathbf{%
B}^{\mathbf{v}}$ morphism $\left( g,Id_{M}\right) $ is locally invertible,
then the differentiable $\left( g,Id_{M}\right) $-lift $(4.4.2.6)$ is
parallel with respect to the connection $\Gamma $ if the component functions
$\left( \tilde{g}_{ji}\circ c\cdot \displaystyle\frac{dc^{i}}{dt},~j\in
\overline{1,m}\right) $ are solutions for the differentiable system of
equations%
\begin{equation*}
\frac{du_{j}}{dt}+\Gamma _{jk}\circ \overset{\ast }{u}\left( c,\dot{c}%
\right) \circ c\cdot g^{kh}\circ c\cdot u_{h}=0,\leqno(4.4.2.9)
\end{equation*}%
namely%
\begin{equation*}
\begin{array}{l}
\displaystyle\frac{d}{dt}\left( \tilde{g}_{ji}\circ c\left( t\right) \cdot
\frac{dc^{i}\left( t\right) }{dt}\right) \vspace*{1mm} \\
\qquad \displaystyle+\Gamma _{jk}\left( c\left( t\right) ,\left( \tilde{g}%
_{ji}\circ c\left( t\right) \cdot \frac{dc^{j}\left( t\right) }{dt}\right)
\cdot dx^{i}\left( c\left( t\right) \right) \right) \cdot \frac{dc^{k}\left(
t\right) }{dt}=0,%
\end{array}%
\leqno(4.4.2.9)^{\prime }
\end{equation*}

\subsection{Parallel transport}

We consider the following diagram:
\begin{equation*}
\begin{array}{ccccl}
&  & E & ^{\underrightarrow{~\ \ \ \ g~\ \ \ \ }} & \left( F,\left[ ,\right]
_{F},\left( \rho ,Id_{M}\right) \right) \\
&  & ~\downarrow \pi &  & ~\downarrow \nu \\
I & ^{\underrightarrow{~\ \ \ \ c~\ \ \ \ }} & M & ^{\underrightarrow{~\ \ \
\ Id_{M}~\ \ \ \ }} & ~M%
\end{array}
\leqno(4.5.1)
\end{equation*}%
where $\left( E,\pi ,M\right) \in \left\vert \mathbf{B}^{\mathbf{v}%
}\right\vert $, $\left( \left( F,\nu ,M\right) ,\left[ ,\right] _{F},\left(
\rho ,Id_{M}\right) \right) \in \left\vert \mathbf{LA}\right\vert ,$ $\left(
g,Id_{M}\right) $ is a $\mathbf{B}^{\mathbf{v}}$-morphism and $c$ is a
differentiable curve$.$

Let $\dot{c}$ be a $\left( g,Id_{M}\right) $-lift of the curve $c$.

We admit that $\rho \Gamma $ is a linear $\rho $-connection for the vector
bundle $\left( E,\pi ,M\right) .$

\bigskip \noindent \textbf{Definition 4.5.1 }We will called \emph{parallel
transport of tensor fields of }$\left( r,s\right) $\emph{\ type along a
curve }$c$ any family
\begin{equation*}
\begin{array}{c}
\mathcal{P}_{c}=\left\{ P_{t_{1},t_{2}}\in Izo\left( \mathcal{T}%
_{q}^{p}\left( E,\pi ,M\right) _{c\left( t_{1}\right) },\mathcal{T}%
_{q}^{p}\left( E,\pi ,M\right) _{c\left( t_{2}\right) }\right)
,\,t_{1},t_{2}\in I\right\}%
\end{array}%
\end{equation*}%
with the following properties:

\begin{itemize}
\item[1.] For any $t_{1},t_{2}\in I$ it exists a unique isomorphism $%
P_{t_{1},t_{2}}\in \mathcal{P}_{c}$ such that\newline
$\left( P_{t_{1},t_{2}}\right) ^{-1}=P_{t_{2},t_{1}}.$

\item[2.] For any $t_{1},t_{2},t_{3}\in I$ \ we have that $%
P_{t_{2},t_{3}}\circ P_{t_{1},t_{2}}=P_{t_{1},t_{3}}$.
\end{itemize}

\smallskip \noindent \textbf{Theorem 4.5.1 }\emph{If }$t_{0},t{\in }I$\emph{%
\ and }$U$\emph{\ is a local vector }$\left( m{+}n\right) $\emph{-chart such
that }$c\left( t_{0}\right) ,c\left( t\right) {\in }U,$ \emph{then it exists
an unique isomorphism }%
\begin{equation*}
P_{t_{0},t}\in Izo\left( \mathcal{T}_{q}^{p}\left( E,\pi ,M\right) _{c\left(
t_{0}\right) },\mathcal{T}_{q}^{p}\left( E,\pi ,M\right) _{c\left( t\right)
}\right)
\end{equation*}%
\emph{such that }$\left( P_{t_{0},t}\right) ^{-1}=P_{t,t_{0}}$\emph{\ which
not depend on the local vector chart used}$.$

\bigskip \noindent \textit{Proof.} Let $T_{c\left( t_{0}\right) }\in
\mathcal{T}_{q}^{p}\left( E,\pi ,M\right) _{c\left( t_{0}\right) }$ be. We
admit that
\begin{equation*}
T_{\pi \circ c\left( t_{0}\right) }=\left(
T_{b_{1},...,b_{q}}^{a_{1},...,a_{p}}s_{a_{1}}\otimes ...\otimes
s_{a_{p}}\otimes s^{b_{1}}\otimes ...\otimes s^{b_{q}}\right) \left( c\left(
t_{0}\right) \right)
\end{equation*}%
and
\begin{equation*}
%
\end{equation*}%
where the matrices
\begin{equation*}
\left\Vert A_{a_{1}}^{\tilde{a}_{1}}\left( t_{0},t\right) \right\Vert
,...\left\Vert A_{a_{p}}^{\tilde{a}_{p}}\left( t_{0},t\right) \right\Vert
,\left\Vert B_{\tilde{b}_{1}}^{b_{1}}\left( t_{0},t\right) \right\Vert
,...,\left\Vert B_{\tilde{b}_{q}}^{b_{q}}\left( t_{0},t\right) \right\Vert
\end{equation*}%
are the matrices used for base transformation. Using the equality%
\begin{equation*}
%
%
\end{equation*}%
and the notation
\begin{equation*}
%
%
\end{equation*}%
we obtain the equality%
\begin{equation*}
%
%
\end{equation*}%
Since the differentiable equations:%
\begin{equation*}
%
%
\end{equation*}%
are equivalent with the following differentiable equations%
\begin{equation*}
%
%
\end{equation*}%
which has unique solutions which not depend on the local vector chart used,
it results the conclusion of \ the theorem.\hfill \hfill \emph{q.e.d.}

\bigskip\noindent\textbf{Corollary 4.5.1} \emph{For any }$p,q\in \mathbb{N}$%
\emph{, it exists a parallel transport }$\mathcal{P}_{c}$\emph{\ between the
tensors of }$\left( p,q\right) $\emph{\ type.}

This parallel transport will be called the \emph{parallel transport along
the curve }$c$\emph{\ associated to linear }$\rho $\emph{-connection }$\rho
\Gamma .$

\bigskip\noindent\textit{Proof.} \textbf{\ }Let $p,q\in \mathbb{N}$ and $%
t_{0},t\in I$ be. Without restricting the generality, we admit that not
exists a vector local $m+r$-chart $U$ which contain the points $c\left(
t_{0}\right) $ and $c\left( t\right) .$

Since $I$ is a conex manifold, it results that it exist a finite numbers of
real numbers $t_{1},t_{2},...,t_{r}=t$ such that for each $j\in \overline{1,r%
},$ the points $c\left( t_{j-1}\right) $ and $c\left( t_{j}\right) $ belongs
to the same vector local $m+r$-chart.

Using the previous theorem, we build the linear isomorphisms $%
P_{t_{0},t_{1}}, $ $\,P_{t_{1},t_{2}},$ $\,...\,,P_{t_{r-1},t}$.

The linear isomorphism $P_{t_{r-1},t}\circ ...\circ P_{t_{1},t_{2}}\circ
P_{t_{0},t_{1}}=P_{t_{0},t}$ not depend on the vector local $m+r$-charts
used.\hfill \hfill \emph{q.e.d.}

\bigskip\noindent\textbf{Remark 4.5.1 }Using the notations of the previous
theorem we obtain:%
\begin{equation*}
\begin{array}{cl}
-\displaystyle\frac{d}{dt}\tilde{T}_{\tilde{b}_{1},...,\tilde{b}_{q}}^{%
\tilde{a}_{1},...,\tilde{a}_{p}}c\left( t\right) & =T_{\tilde{b}_{1},...,%
\tilde{b}_{q}}^{a\tilde{a}_{2},...,\tilde{a}_{p}}c\left( t\right) \rho
\Gamma _{a\alpha }^{\tilde{a}_{1}}c\left( t\right) g_{c}^{\alpha }\left(
x\left( t\right) \right) y^{c}\left( t\right) +... \vspace*{1,5mm} \\
& +T_{\tilde{b}_{1},...,\tilde{b}_{q}}^{\tilde{a}_{1},...,\tilde{a}%
_{p-1}a}c\left( t\right) \rho \Gamma _{a\alpha }^{\tilde{a}_{p}}c\left(
t\right) g_{c}^{\alpha }\left( x\left( t\right) \right) y^{c}\left( t\right)
+ \vspace*{1,5mm} \\
& -T_{b\tilde{b}_{2},...,\tilde{b}_{q}}^{\tilde{a}_{1},...,\tilde{a}%
_{p}}c\left( t\right) \rho \Gamma _{\tilde{b}_{1}\alpha }^{b}c\left(
t\right) g_{c}^{\alpha }\left( x\left( t\right) \right) y^{c}\left( t\right)
-...\vspace*{1,5mm} \\
& -T_{\tilde{b}_{1},...,\tilde{b}_{q-1}b}^{\tilde{a}_{1},...,\tilde{a}%
_{p}}c\left( t\right) \rho \Gamma _{\tilde{b}_{q}\alpha }^{b}c\left(
t\right) g_{c}^{\alpha }\left( x\left( t\right) \right) y^{c}\left( t\right)
.%
\end{array}%
\leqno(4.5.2)
\end{equation*}

\smallskip \noindent \textbf{Theorem 4.5.2 }\emph{If }$\left( E,\pi
,M\right) =\left( F,\nu ,N\right) $\emph{\ and }$\mathcal{P}_{c}$\emph{\ is
the parallel transport along the curve }$c$\emph{\ associated to linear }$%
\rho $\emph{-connection }$\rho \Gamma $\emph{, then, for any }$t\in I$\emph{%
\ we obtain: }%
\begin{equation*}
\underset{h\longrightarrow 0}{\lim }\frac{P_{t+h,t}\left( T_{c\left(
t+h\right) }\right) -T_{c\left( t\right) }}{h}=\left( \rho D_{u\left( c,\dot{%
c}\right) }T\right) c\left( t\right) ,\leqno(4.5.3)
\end{equation*}%
\emph{for any }$T\in \mathcal{T}_{q}^{p}\left( E,\pi ,M\right) .$

\bigskip \noindent \textit{Proof.} \textbf{\ }Let be $T\in \mathcal{T}%
_{q}^{p}\left( E,\pi ,M\right) $. Let $t\in I$ and $h>0$ be such that $\left]
t-h,t+h\right[ \subset I.$

For any $\tilde{a}_{1},...,\tilde{a}_{p},\tilde{b}_{1},...,\tilde{b}_{q}\in
\overline{1,n}$ we build the following application
\begin{equation*}
\begin{array}{rcr}
\left[ t,t+h\right] & ^{\underrightarrow{z_{\tilde{b}_{1},...,\tilde{b}%
_{q}}^{\tilde{a}_{1},...,\tilde{a}_{p}}}\,} & \mathbb{R} \vspace*{1,5mm} \\
\theta & \longmapsto & z_{\tilde{b}_{1},...,\tilde{b}_{q}}^{\tilde{a}%
_{1},...,\tilde{a}_{p}}\left( \theta \right)%
\end{array}%
\end{equation*}%
defined by
\begin{equation*}
z_{\tilde{b}_{1},...,\tilde{b}_{q}}^{\tilde{a}_{1},...,\tilde{a}_{p}}\left(
\theta \right) {=}T_{\tilde{b}_{1},...,\tilde{b}_{q}}^{\tilde{a}_{1},...,%
\tilde{a}_{p}}c\left( t{+}h\right) A_{a_{1}}^{\tilde{a}_{1}}\left( t{+}%
h,\theta \right) \cdot ...\cdot A_{a_{p}}^{\tilde{a}_{p}}\left( t+h,\theta
\right) \cdot B_{\tilde{b}_{1}}^{b_{1}}\left( t{+}h,\theta \right) ...\cdot
B_{\tilde{b}_{q}}^{b_{q}}\left( t+h,\theta \right)
\end{equation*}
Using the main theorem, it exists a unique real number
\begin{equation*}
\xi _{\tilde{b}_{1},...,\tilde{b}_{q}}^{\tilde{a}_{1},...,\tilde{a}_{p}}\in %
\left] t,t+h\right[
\end{equation*}
such that
\begin{equation*}
%
\end{equation*}
Since the component by indices $_{\tilde{b}_{1},...,\tilde{b}_{q}}^{\tilde{a}%
_{1},...,\tilde{a}_{p}}$ of a tensor $P_{t+h,t}\left( T_{c\left( t+h\right)
}\right) $ is $z_{\tilde{b}_{1},...,\tilde{b}_{q}}^{\tilde{a}_{1},...,\tilde{%
a}_{p}}\left( t\right) $, it results that
\begin{equation*}
%
%
\end{equation*}
Using Remark 4.5.1 and the equality
\begin{equation*}
\frac{d}{dt}T_{\tilde{b}_{1},...,\tilde{b}_{q}}^{\tilde{a}_{1},...,\tilde{a}%
_{p}}c\left( t\right) =\frac{dx^{i}}{dt}\frac{\partial T_{\tilde{b}_{1},...,%
\tilde{b}_{q}}^{\tilde{a}_{1},...,\tilde{a}_{p}}c\left( t\right) }{\partial
x^{i}}=g_{c}^{\alpha }\left( x\left( t\right) \right) y^{c}\left( t\right)
\rho _{\alpha }^{i}\frac{\partial T_{\tilde{b}_{1},...,\tilde{b}_{q}}^{%
\tilde{a}_{1},...,\tilde{a}_{p}}c\left( t\right) }{\partial x^{i}},
\end{equation*}%
it results the conclusion of theorem.\hfill \emph{q.e.d.}

\bigskip \noindent \textbf{Definition 4.5.2 }\emph{The tensor field }$T\in
\mathcal{T}_{q}^{p}\left( E,\pi ,M\right) $\emph{\ is parallel along the
curve }$c$\emph{\ }with respect to the linear $\rho $-connection $\rho
\Gamma $ if for any $t_{1},t_{2}\in I$ it results that
\begin{equation*}
%
\leqno(4.5.4)
\end{equation*}

\smallskip \noindent \textbf{Theorem 4.5.3 }The tensor field $T\in \mathcal{T%
}_{q}^{p}\left( E,\pi ,M\right) $\ is parallel along the curve $c$ with
respect to linear $\rho $-connection $\rho \Gamma $ if and only if
\begin{equation*}
%
%
\leqno(4.5.5)
\end{equation*}

\smallskip \noindent \textbf{Corollary 4.5.2} The tensor field $T\in
\mathcal{T}_{q}^{p}\left( E,\pi ,M\right) $\ is parallel along the curve $c$
with respect to linear $\rho $-connection $\rho \Gamma $ if and only if
\begin{equation*}
%
%
\end{equation*}

\subsection{Formulas of Gauss-Weingarten type}

Using the main ideas of the theory of Myller configurations, introduced by
R. Miron in $[37]$ and applied to Finsler spaces by O. Constantinescu in $%
[13]$ and his Ph.D. Thesis, we present the Gauss-Weingarten formulas for
generalized Lie algebroids.

Let
\begin{equation*}
\left( \left( F,\nu ,N\right) ,\left[ ,\right] _{F,h},\left( \rho ,\eta
\right) \right)
\end{equation*}%
be a generalized Lie algebroid given by the diagram:
\begin{equation*}
%
\leqno(4.6.1)
\end{equation*}
The geometry of the couple $\left( M,h\right) $ is the geometry of the
pull-back vector bundle $\left( h^{\ast }F,h^{\ast }\nu ,M\right) $ using
the diagram%
\begin{equation*}
%
\leqno(4.6.2)
\end{equation*}

Let $I ^{\underrightarrow{\ \ c\ \ }} M $ be a differentiable curve and let $%
M^{\prime }=Im\left( \eta \circ h\circ c\right) $ be.

Let $I ^{\underrightarrow{\ \ \dot{c}\ \ }} h^{\ast }F_{|M^{\prime }}$ be
the $\left( Id_{h^{\ast }F},Id_{M}\right) $-lift of the curve $c.$

Let $\left\{ T_{\alpha },\alpha \in \overline{1,p}\right\} ,$ $\left\{
s_{a},a\in \overline{1,q}\right\} $ and $\left\{ \chi _{_{i}},i\in \overline{%
1,s}\right\} $ be the base for
\begin{equation*}
\mbox{$\Gamma \left( h^{\ast }F_{|M^{\prime }},h^{\ast
}\nu _{|M^{\prime }},M^{\prime }\right) ,$ $\Gamma \left( T^{h^{\ast
}F}\left( c,\dot{c}\right) ,\tau ,M^{\prime }\right) $ and $\Gamma \left(
S^{h^{\ast }F}\left( c,\dot{c}\right) ,\sigma ,M^{\prime }\right) $,}
\end{equation*}
respectively.

The dimension of type fibre of the vector bundle $\left( h^{\ast
}F_{|M^{\prime }},h^{\ast }\nu _{|M^{\prime }},M^{\prime }\right) $ is $%
q+s=p.$

Consequently, for any $a\in \overline{1,q}$ we have%
\begin{equation*}
%
\leqno(4.6.3)
\end{equation*}%
and for any $i\in \overline{1,s}$ we have%
\begin{equation*}
%
%
\leqno(4.6.4)
\end{equation*}

Let $g=g_{\alpha \beta }t^{\alpha }\otimes t^{\beta }\in \mathcal{T~}%
_{2}^{0}\left( F,\nu ,N\right) $ be a (pseudo)metrical structure.

\bigskip\noindent\textbf{Remark 4.6.1 }The following affirmations are
satisfied:\medskip

\noindent1. The section $\overset{h^{\ast }F}{g}=\overset{h^{\ast }F}{g}%
_{\alpha \beta }T^{\alpha }\otimes T^{\beta }$ defined by
\begin{equation*}
%
%
\leqno(4.6.7)
\end{equation*}%
is a (pseudo)metrical structure.\bigskip

\noindent\textbf{Remark 4.6.2 }Using the diagram $\left( 4.6.2\right) $ we
can construct the linear $\rho $-connection of Levi Civita type $\overset{%
h^{\ast }F}{\rho }\Gamma $ of components $\overset{h^{\ast }F}{\rho }\Gamma
_{\beta \gamma }^{\alpha }.$

We have the covariant $\rho $-derivative defined by%
\begin{equation*}
%
%
\leqno(4.6.8)
\end{equation*}%
where $\left( \overset{h^{\ast }F}{\rho }\right) _{\gamma }^{k}$ are the
components of the map $\overset{h^{\ast }F}{\rho }$.

\bigskip \noindent \textbf{Definition 4.6.1 }If we can defined%
\begin{equation*}
%
%
\leqno(4.6.12)
\end{equation*}%
and we can consider the bilinear applications
\begin{equation*}
%
%
\end{equation*}%
which satisfy the following relations%
\begin{equation*}
%
%
\leqno(4.6.15)
\end{equation*}%
then we will say that the\emph{\ relations }$(4.6.13),$\emph{\ }$(4.6.14)$%
\emph{\ and }$(4.6.15)$\emph{\ are formulas of Gauss-Weingarten type
associated to differentiable curve }$c$\emph{, metrical structure }$g$\emph{%
\ and bilinear applications }$H$\emph{\ and }$A$\emph{.}

The bilinear application $H$ will be called\emph{\ the second fundamental
form of differentiable curve }$c$.

\bigskip \noindent \textbf{Remark 4.6.2} Using the base sections, then the
formulas of Gauss-Weingarten type become:%
%\[
%\left( 5.1.5.17\right) ^{\prime }~~%
\begin{equation*}
\Lambda _{c}^{\gamma }\left( \left( \overset{h^{\ast }F}{\rho }\right)
_{\gamma }^{k}\frac{\partial \Lambda _{b}^{\alpha }}{\partial x^{k}}+\overset%
{h^{\ast }F}{\rho }\Gamma _{\beta \gamma }^{\alpha }\Lambda _{b}^{\beta
}\right) =\overset{h^{\ast }F}{\rho }\Gamma _{bc}^{a}\Lambda _{a}^{\alpha
}+H_{bc}^{i}\Lambda _{i}^{\alpha },\vspace*{-3mm}\leqno(4.6.13^{\prime })
\end{equation*}%
\begin{equation*}
\Lambda _{c}^{\gamma }\left( \left( \overset{h^{\ast }F}{\rho }\right)
_{\gamma }^{h}\frac{\partial \Lambda _{j}^{\alpha }}{\partial x^{h}}+\overset%
{h^{\ast }F}{\rho }\Gamma _{\beta \gamma }^{\alpha }\Lambda _{j}^{\beta
}\right) =-A_{jc}^{a}\Lambda _{a}^{\alpha }+\overset{h^{\ast }F}{\rho }%
\Gamma _{jc}^{i}\Lambda _{i}^{\alpha }\vspace*{-3mm}\leqno(4.6.14^{\prime })
\end{equation*}%
\begin{equation*}
\overset{S^{h^{\ast }F}}{g}_{ij}H_{bc}^{i}=\overset{T^{h^{\ast }F}}{g}%
_{ab}A_{jc}^{a}.\leqno(4.6.15^{\prime })
\end{equation*}

\section{The geometry of total space of the Lie algebroid generalized
tangent bundle for a vector bundle}

\subsection{Adapted $\left( \protect\rho ,\protect\eta \right) $-basis and
adapted dual $\left( \protect\rho ,\protect\eta \right) $-basis}

In the following, we consider the following diagram:
\begin{equation*}
\begin{array}{c}
\xymatrix{{E}\ar[d]_{{\pi }}&\left( F,\left[ ,\right] _{F,h},\left( \rho
,\eta \right) \right)\ar[d]^\nu\\ M\ar[r]^h&N}%
\end{array}%
\end{equation*}%
where $\left( E,\pi ,M\right) \in \left\vert \mathbf{B}^{\mathbf{v}%
}\right\vert $ and $\left( \left( F,\nu ,N\right) ,\left[ ,\right]
_{F,h},\left( \rho ,\eta \right) \right) $ is a generalized Lie algebroid.

Let $\left( \rho ,\eta \right) \Gamma $ be a $\left( \rho ,\eta \right) $%
-connection for the vector bundle $\left( E,\pi ,M\right) .$

If we put the problem of finding a base for the $\mathcal{F}\left( E\right) $%
-module
\begin{equation*}
\left( \Gamma \left( H\left( \rho ,\eta \right) TE,\left( \rho ,\eta \right)
\tau _{E},E\right) ,+,\cdot \right)
\end{equation*}%
of the type\textbf{\ }
\begin{equation*}
\frac{\delta }{\delta \tilde{z}^{\alpha }}=\tilde{Z}_{\alpha }^{\beta }\frac{%
\partial }{\partial \tilde{z}^{\alpha }}+Y_{\alpha }^{a}\frac{\partial }{%
\partial \tilde{y}^{a}},\alpha \in \overline{1,r}
\end{equation*}%
which satisfies the following conditions:
\begin{equation*}
\begin{array}{rcl}
\displaystyle\Gamma \left( \left( \rho ,\eta \right) \pi !,Id_{E}\right)
\left( \frac{\delta }{\delta \tilde{z}^{\alpha }}\right) & = & \tilde{T}%
_{\alpha }\vspace*{2mm}, \\
\displaystyle\Gamma \left( \left( \rho ,\eta \right) \Gamma ,Id_{E}\right)
\left( \frac{\delta }{\delta \tilde{z}^{\alpha }}\right) & = & 0,%
\end{array}%
\leqno(5.1.1)
\end{equation*}%
then we obtain the sections
\begin{equation*}
\frac{\delta }{\delta \tilde{z}^{\alpha }}=\frac{\partial }{\partial \tilde{z%
}^{\alpha }}-\left( \rho ,\eta \right) \Gamma _{\alpha }^{a}\frac{\partial }{%
\partial \tilde{y}^{a}}.\leqno(5.1.2)
\end{equation*}%
We observe that their law of change is a tensorial law under a change of
vector fiber charts.

\bigskip\noindent \textbf{Definition 5.1.1} The base
\begin{equation*}
\left( \frac{\delta }{\delta \tilde{z}^{\alpha }},\frac{\partial }{\partial
\tilde{y}^{a}}\right) \overset{put}{=}\left( \tilde{\delta}_{\alpha },%
\overset{\cdot }{\tilde{\partial}}_{a}\right)
\end{equation*}%
will be called the \emph{adapted }$\left( \rho ,\eta \right) $\emph{-base.}

The following equality holds good%
\begin{equation*}
%
\leqno(5.1.3)
\end{equation*}%
where $\left( \partial _{i},\dot{\partial}_{a}\right) $ is the natural base
for the $\mathcal{F}\left( E\right) $-module $\left( \Gamma \left( TE,\tau
_{E},E\right) ,+,\cdot \right) .$

Moreover, if $\rho \Gamma $ is the $\rho $-connection associated to the
connection $\Gamma $, then we obtain
\begin{equation*}
%
%
\leqno(5.1.4)
\end{equation*}%
where $\left( \delta _{i},\dot{\partial}_{a}\right) $ is the adapted base
for the $\mathcal{F}\left( E\right) $-module $\left( \Gamma \left( TE,\tau
_{E},E\right) ,+,\cdot \right) .$

\bigskip \noindent \textbf{Theorem 5.1.1 }\emph{The following equality holds
good\ }%
\begin{equation*}
%
%
\leqno(5.1.6)
\end{equation*}%
\emph{Moreover, we have: }%
\begin{equation*}
%
%
\leqno(5.1.8)
\end{equation*}

Let $\left( d\tilde{z}^{\alpha },d\tilde{y}^{b}\right) $ be the natural dual
$\left( \rho ,\eta \right) $-base.

If we consider the problem of finding a base for the $\mathcal{F}\left(
E\right) $-module
\begin{equation*}
\left( \Gamma \left( \left( V\left( \rho ,\eta \right) TE\right) ^{\ast
},\left( \left( \rho ,\eta \right) \tau _{E}\right) ^{\ast },E\right)
,+,\cdot \right)
\end{equation*}%
of the type
\begin{equation*}
%
%
\end{equation*}%
which satisfies the following conditions:
\begin{equation*}
%
%
\leqno(5.1.9)
\end{equation*}%
then we obtain the sections
\begin{equation*}
%
%
\leqno(5.1.10)
\end{equation*}
We observe that their changing rule is tensorial under a change of vector
fiber charts.

\bigskip\noindent \textbf{Definition 5.1.2} The base $\left( d\tilde{z}%
^{\alpha },\delta \tilde{y}^{a}\right) $ will be called the \emph{adapted
dual }$\left( \rho ,\eta \right) $\emph{-base.}

\subsection{Remarkable $\mathbf{Mod}$-endomorphisms}

In the following we consider the diagram:
\begin{equation*}
%
%
\end{equation*}%
where $\left( E,\pi ,M\right) \in \left\vert \mathbf{B}^{\mathbf{v}%
}\right\vert $ and $\left( \left( F,\nu ,N\right) ,\left[ ,\right]
_{F,h},\left( \rho ,\eta \right) \right) $ is a generalized Lie algebroid.

\bigskip\noindent\textbf{Definition 5.2.1 }For any $\mathbf{Mod}$%
-endomorphism $e$ of
\begin{equation*}
\left( \Gamma \left( \left( \rho ,\eta \right) TE,\left( \rho ,\eta \right)
\tau _{E},E\right) ,+,\cdot \right)
\end{equation*}%
we define the application of Nijenhuis \ type
\begin{equation*}
\!\!\Gamma \!((\rho ,\!\eta )TE,\!(\rho ,\eta )\tau _{E},E)^{2~\
\underrightarrow{~\ \ N_{e}~\ \ }}~\ \Gamma \!((\rho ,\eta )TE,\!(\rho ,\eta
)\tau _{E},\!E)\vspace*{1mm}
\end{equation*}%
defined by
\begin{equation*}
\begin{array}{c}
N_{e}\left( X,Y\right) =\left[ eX,eY\right] _{\rho TE}+e^{2}\left[ X,Y\right]
_{\rho TE}-e\left[ eX,Y\right] _{\rho TE}-e\left[ X,eY\right] _{\rho TE},%
\end{array}%
\end{equation*}%
for any $X,Y\in \Gamma \!((\rho ,\eta )TE,\!(\rho ,\eta )\tau _{E},\!E)%
\vspace*{1mm}.$

\smallskip \noindent\textbf{Remark 5.2.1 }The vertical and the horizontal
vector subbundles are interior differential systems for the Lie algebroid
generalized tangent bundle
\begin{equation*}
\begin{array}{c}
\left( \!((\rho ,\eta )TE,\!(\rho ,\eta )\tau _{E},\!E),\left[ ,\right]
_{\left( \rho ,\eta \right) TE},\left( \tilde{\rho},Id_{E}\right) \right) .%
\end{array}%
\end{equation*}
These interior differential systems will be called \emph{vertical }and \emph{%
horizontal interior dif\-fe\-ren\-tial systems.}

\subsubsection{\noindent Projectors}

\textbf{Definition 5.2.1.1 }Any $\mathbf{Mod}$-endomorphism $e$ of
\begin{equation*}
\Gamma \left( (\rho ,\eta )TE,\break (\rho ,\eta \ )\tau _{E},E\right)
\end{equation*}%
with the property
\begin{equation*}
\begin{array}{l}
e^{2}=e%
\end{array}%
\leqno(5.2.1.1)
\end{equation*}%
will be called \emph{projector.}

\bigskip\noindent \textbf{Example 5.2.1.1 }The $\mathbf{Mod}$-endomorphism
\begin{equation*}
\begin{array}{rcl}
\Gamma \left( \left( \rho ,\eta \right) TE,\left( \rho ,\eta \right) \tau
_{E},E\right) & ^{\underrightarrow{\ \ \mathcal{V}\ \ }} & \Gamma \left(
\left( \rho ,\eta \right) TE,\left( \rho ,\eta \right) \tau _{E},E\right) \\
\tilde{Z}^{\alpha }\tilde{\delta}_{\alpha }+Y^{a}\overset{\cdot }{\tilde{%
\partial}}_{a} & \longmapsto & Y^{a}\overset{\cdot }{\tilde{\partial}}_{a}%
\end{array}%
\end{equation*}%
is a projector which will be called \textit{the vertical projector.}

\bigskip\noindent \textbf{Remark 5.2.1.1 } We have $\mathcal{V}\left( \tilde{%
\delta}_{\alpha }\right) =0$ and $\mathcal{V}\left( \overset{\cdot }{\tilde{%
\partial}}_{a}\right) =\overset{\cdot }{\tilde{\partial}}_{a}.$ Therefore,
it follows\vspace*{-2mm}
\begin{equation*}
\mathcal{V}\left( \tilde{\partial}_{\alpha }\right) =\left( \rho ,\eta
\right) \Gamma _{\alpha }^{a}\overset{\cdot }{\tilde{\partial}}_{a}.
\end{equation*}

\smallskip\noindent \textbf{Theorem 5.2.1.1 }\emph{A }$(\rho ,\eta )$\emph{%
-connection for the vector bundle} $(E,\pi ,M)$ \emph{is characterized by
the existence of a} $\mathbf{Mod}$\emph{-endomorphism} $\mathcal{V}$ \emph{of%
}
\begin{equation*}
\left( \Gamma \left( \left( \rho ,\eta \right) TE,\left( \rho ,\eta \right)
\tau _{E},E\right) ,+,\cdot \right)
\end{equation*}%
\emph{with the properties:}
\begin{equation*}
\begin{array}{c}
\mathcal{V}\left( \Gamma \left( \left( \rho ,\eta \right) TE,\left( \rho
,\eta \right) \tau _{E},E\right) \right) \subset \Gamma \left( V\left( \rho
,\eta \right) TE,\left( \rho ,\eta \right) \tau _{E},E\right) \vspace*{1,5mm}
\\
\mathcal{V}\left( X\right) =X\Longleftrightarrow X\in \Gamma \left( V\left(
\rho ,\eta \right) TE,\left( \rho ,\eta \right) \tau _{E},E\right)%
\end{array}%
\leqno(5.2.1.2)
\end{equation*}

\smallskip\noindent \textbf{Example 5.2.1.2 }The $\mathbf{Mod}$-endomorphism
\begin{equation*}
\begin{array}{rcl}
\Gamma \left( \left( \rho ,\eta \right) TE,\left( \rho ,\eta \right) \tau
_{E},E\right) & ^{\underrightarrow{\ \ \mathcal{H}\ \ }} & \Gamma \left(
\left( \rho ,\eta \right) TE,\left( \rho ,\eta \right) \tau _{E},E\right) \\
\tilde{Z}^{\alpha }\tilde{\delta}_{\alpha }+Y^{a}\overset{\cdot }{\tilde{%
\partial}}_{a} & \longmapsto & \tilde{Z}^{\alpha }\tilde{\delta}_{\alpha }%
\end{array}%
\end{equation*}%
is a projector which will be called the \emph{horizontal projector.}

\bigskip\noindent \textbf{Remark 5.2.1.2} We have $\mathcal{H}\left( \tilde{%
\delta}_{\alpha }\right) {=}\tilde{\delta}_{\alpha }$ and $\mathcal{H}\big(%
\overset{\cdot }{\tilde{\partial}}_{a}\big){=}0.$ Therefore, we obtain $%
\mathcal{H}\left( \tilde{\partial}_{\alpha }\right) {=}\tilde{\delta}%
_{\alpha }.$

\bigskip\noindent\textbf{Theorem 5.2.1.2 }\emph{A }$\left( \rho ,\eta
\right) $\emph{-connection for the vector bundle} $\left( E,\pi ,M\right)$
\emph{\ is characterized by the existence of a }$\mathbf{Mod}$\emph{%
-endomorphism }$\mathcal{H}$\emph{\ of}
\begin{equation*}
\left( \Gamma \left( \left( \rho ,\eta \right) TE,\left( \rho ,\eta \right)
\tau _{E},E\right) ,+,\cdot \right)
\end{equation*}%
\emph{with the properties:}
\begin{equation*}
\begin{array}{c}
\mathcal{H}\left( \Gamma \left( \left( \rho ,\eta \right) TE,\left( \rho
,\eta \right) \tau _{E},E\right) \right) \subset \Gamma \left( H\left( \rho
,\eta \right) TE,\left( \rho ,\eta \right) \tau _{E},E\right) \vspace*{1,5mm}
\\
\mathcal{H}\left( X\right) =X\Longleftrightarrow X\in \Gamma \left( H\left(
\rho ,\eta \right) TE,\left( \rho ,\eta \right) \tau _{E},E\right) .%
\end{array}%
\leqno(5.2.1.3)
\end{equation*}

\smallskip\noindent \textbf{Corollary 5.2.1.1} \emph{A }$\left( \rho ,\eta
\right) $\emph{-connection for the vector bundle} $\left( E,\pi ,M\right) $
\emph{\ is characterized by the existence of a }$\mathbf{Mod}$\emph{%
-endomorphism }$\mathcal{H}$\emph{\ of}
\begin{equation*}
\left( \Gamma \left( \left( \rho ,\eta \right) TE,\left( \rho ,\eta \right)
\tau _{E},E\right) ,+,\cdot \right)
\end{equation*}%
\emph{\ with the properties:}
\begin{equation*}
\begin{array}{c}
\mathcal{H}^{2}=\mathcal{H} \vspace*{1,5mm} \\
Ker\left( \mathcal{H}\right) =\left( \Gamma \left( V\left( \rho ,\eta
\right) TE,\left( \rho ,\eta \right) \tau _{E},E\right) ,+,\cdot \right) .%
\end{array}%
\leqno(5.2.1.4)
\end{equation*}

\smallskip \noindent\textbf{Remark 5.2.1.3 }For any
\begin{equation*}
X\in \Gamma \left( \left( \rho ,\eta \right) TE,\left( \rho ,\eta \right)
\tau _{E},E\right)
\end{equation*}%
we obtain the following unique decomposition
\begin{equation*}
X=\mathcal{H}X+\mathcal{V}X.
\end{equation*}

\smallskip \noindent \textbf{Proposition 5.2.1.1} \emph{After some
calculations we obtain }%
\begin{equation*}
\begin{array}{c}
N_{\mathcal{V}}\left( X,Y\right) =\mathcal{V}\left[ \mathcal{H}X,\mathcal{H}Y%
\right] _{\left( \rho ,\eta \right) TE}=N_{\mathcal{H}}\left( X,Y\right) ,%
\end{array}%
\leqno(5.2.1.5)
\end{equation*}%
\textit{for any }$X,Y\in \Gamma \left( \left( \rho ,\eta \right) TE,\left(
\rho ,\eta \right) \tau _{E},E\right) .$

\bigskip \noindent \textbf{Corollary 5.2.1.2} \emph{The horizontal interior
differential system }%
\begin{equation*}
\left( H\left( \rho ,\eta \right) TE,\left( \rho ,\eta \right) \tau
_{E},E\right)
\end{equation*}%
\emph{is involutive if and only if }$N_{\mathcal{V}}=0$\emph{\ or }$N_{%
\mathcal{H}}=0.$

\subsubsection{\noindent The almost product structure}

\noindent \textbf{Definition 5.2.2.1 }Any $\mathbf{Mod}$-endomorphism $e$ of
\begin{equation*}
\left( \Gamma \left( (\rho ,\eta )TE,\break (\rho ,\eta \ )\tau
_{E},E\right) ,+,\cdot \right)
\end{equation*}%
with the property%
\begin{equation*}
\begin{array}{c}
e^{2}=Id%
\end{array}%
\leqno(5.2.2.1)
\end{equation*}%
will be called the \emph{almost product structure}.

\bigskip\noindent\textbf{Example 5.2.2.1 }The $\mathbf{Mod}$-endomorphism
\begin{equation*}
\begin{array}{rcl}
\Gamma \left( \left( \rho ,\eta \right) TE,\left( \rho ,\eta \right) \tau
_{E},E\right) & ^{\underrightarrow{\ \ \mathcal{P}\ \ }} & \Gamma \left(
\left( \rho ,\eta \right) TE,\left( \rho ,\eta \right) \tau _{E},E\right) \\
\tilde{Z}^{\alpha }\tilde{\delta}_{\alpha }+Y^{a}\overset{\cdot }{\tilde{%
\partial}}_{a} & \longmapsto & \tilde{Z}^{\alpha }\tilde{\delta}_{\alpha
}-Y^{a}\overset{\cdot }{\tilde{\partial}}_{a}%
\end{array}%
\end{equation*}%
is an almost product structure.

\bigskip\noindent\textbf{Remark 5.2.2.1 }The previous almost product
structure has the properties:
\begin{equation*}
\begin{array}{l}
\mathcal{P}=2\mathcal{H}-Id; \\
\mathcal{P}=Id-2\mathcal{V}; \\
\mathcal{P}=\mathcal{H}-\mathcal{V}.%
\end{array}%
\leqno(5.2.2.2)
\end{equation*}

\smallskip\noindent \textbf{Remark 5.2.2.2 } We obtain that $\mathcal{P}%
\left( \tilde{\delta}_{\alpha }\right) =\tilde{\delta}_{\alpha }$ and $%
\mathcal{P}\left( \overset{\cdot }{\tilde{\partial}}_{a}\right) =-\overset{%
\cdot }{\tilde{\partial}}_{a}.$ Therefore, it follows \vspace*{-2mm}
\begin{equation*}
\mathcal{P}\left( \tilde{\partial}_{\alpha }\right) =\tilde{\delta}_{\alpha
}-\rho \Gamma _{\alpha }^{a}\overset{\cdot }{\tilde{\partial}}_{a}.
\end{equation*}

\smallskip\noindent \textbf{Theorem 5.2.2.1 }\emph{A }$\left( \rho ,\eta
\right) $\emph{-connection for the vector bundle }$\left( E,\pi ,M\right) $%
\emph{\ is characterized by the existence of a }$\mathbf{Mod}$\emph{%
-endomorphism }$\mathcal{P}$\emph{\ of }%
\begin{equation*}
\left( \Gamma \left( \left( \rho ,\eta \right) TE,\left( \rho ,\eta \right)
\tau _{E},E\right) ,+,\cdot \right)
\end{equation*}%
\emph{with the following property:}
\begin{equation*}
\begin{array}{c}
\mathcal{P}\left( X\right) =-X\Longleftrightarrow X\in \Gamma \left( V\left(
\rho ,\eta \right) TE,\left( \rho ,\eta \right) \tau _{E},E\right) .%
\end{array}%
\leqno(5.2.2.3)
\end{equation*}

\smallskip\noindent \textbf{Proposition 5.2.2.1} \emph{After some
calculations, we obtain }%
\begin{equation*}
N_{\mathcal{P}}\left( X,Y\right) =4\mathcal{V}\left[ \mathcal{H}X,\mathcal{H}%
Y\right] ,
\end{equation*}%
\emph{for any }$X,Y\in \Gamma \left( \left( \rho ,\eta \right) TE,\left(
\rho ,\eta \right) \tau _{E},E\right) .$

\bigskip\noindent\textbf{Corollary 5.2.2.1} \emph{The horizontal interior
differential system }%
\begin{equation*}
\left( H\left( \rho ,\eta \right) TE,\left( \rho ,\eta \right) \tau
_{E},E\right)
\end{equation*}%
\emph{is involutive if and only if} $N_{\mathcal{P}}=0.$

\subsubsection{\noindent The almost tangent structure}

\noindent\textbf{Definition 5.2.3.1 }Any $\mathbf{Mod}$-endomorphism $e$ of
\begin{equation*}
\left( \Gamma \!((\rho ,\eta )TE,\break (\rho ,\eta )\tau _{E},E),+,\cdot
\right)
\end{equation*}%
with the property
\begin{equation*}
\begin{array}{c}
e^{2}=0%
\end{array}%
\leqno(5.2.3.1)
\end{equation*}%
will be called the \emph{almost tangent structure.}

\bigskip \noindent \textbf{Example 5.2.3.1 }If $\left( E,\pi ,M\right)
=\left( F,\nu ,N\right) $, $g\in \mathbf{Man}\left( E,E\right) $ such that $%
\left( g,h\right) $ is a $\mathbf{B}^{\mathbf{v}}$\textit{-}morphism locally
invertible, then the $\mathbf{Mod}$-endomorphism
\begin{equation*}
\begin{array}{rcl}
\Gamma \left( \left( \rho ,\eta \right) TE,\left( \rho ,\eta \right) \tau
_{E},E\right) & ^{\underrightarrow{\mathcal{J}_{\left( g,h\right) }}} &
\Gamma \left( \left( \rho ,\eta \right) TE,\left( \rho ,\eta \right) \tau
_{E},E\right) \\
\tilde{Z}^{a}\tilde{\delta}_{a}+\tilde{Y}^{b}\overset{\cdot }{\tilde{\partial%
}}_{b} & \longmapsto & \left( \tilde{g}_{a}^{b}\circ h\circ \pi \right)
\tilde{Z}^{a}\overset{\cdot }{\tilde{\partial}}_{b}%
\end{array}%
\end{equation*}%
is an almost tangent structure which will be called the \emph{almost tangent
structure associated to }$\mathbf{B}^{\mathbf{v}}$\emph{-morphism }$\left(
g,h\right) $. (See: \textbf{Definition 4.4.1.3)}

\bigskip\noindent\textbf{Remark 5.2.3.1 }We obtain that
\begin{equation*}
\mbox{$\mathcal{J}_{\left( g,h\right)
}\left( \tilde{\delta}_{a}\right) =\mathcal{J}_{\left( g,h\right) }\left(
\tilde{\partial}_{a}\right) =\left( \tilde{g}_{a}^{b}\circ h\circ \pi
\right) \overset{\cdot }{\tilde{\partial}}_{b}$ and $\mathcal{J}_{\left(
g,h\right) }\left( \overset{\cdot }{\tilde{\partial}}_{b}\right) =0.$}
\end{equation*}

\smallskip \noindent \textbf{Remark 5.2.3.2} The previous almost tangent
structure has the following properties:
\begin{equation*}
%
\leqno(5.2.3.2)
\end{equation*}

\subsubsection{\noindent The almost complex structure}

Let us consider in the case $\left( E,\pi ,M\right) =\left( F,\nu ,N\right) $%
.

\bigskip\noindent\textbf{Definition 5.2.4.1 }Any $\mathbf{Mod}$-endomorphism
$e$ of
\begin{equation*}
\left( \Gamma (\left( \rho ,\eta \right) TE,\break (\rho ,\eta )\tau
_{E},E),+,\cdot \right)
\end{equation*}%
with the property
\begin{equation*}
%
%
\leqno(5.3.4.1)
\end{equation*}%
will be called the \emph{almost complex structure}.

\bigskip \noindent \textbf{Example 5.2.4.1} If $\left( g,h\right) $ is a $%
\mathbf{B}^{\mathbf{v}}$\textit{-}morphism of $\left( E,\pi ,M\right) $
source and target locally invertible, then the $\mathbf{Mod}$-endomorphism
\begin{equation*}
%
%
\end{equation*}%
is an almost complex structure.

\bigskip \noindent \textbf{Remark 5.2.4.1 }We have
\begin{equation*}
%
%
\end{equation*}%
Therefore, we obtain:
\begin{equation*}
%
%
\end{equation*}

\smallskip \noindent\textbf{Remark 5.2.4.2 }The previous almost complex
structure has the following properties:
\begin{equation*}
%
%
\leqno(5.2.4.2)
\end{equation*}

\subsubsection{\noindent The $\left( \protect\rho ,\protect\eta \right) $%
-tension endomorphism}

Since
\begin{equation*}
\frac{\partial \left( \rho ,\eta \right) \Gamma _{\alpha
%TCIMACRO{\U{b4}}%
%BeginExpansion
{\acute{}}%
%EndExpansion
}^{a%
%TCIMACRO{\U{b4}}%
%BeginExpansion
{\acute{}}%
%EndExpansion
}}{\partial y^{b%
%TCIMACRO{\U{b4}}%
%BeginExpansion
{\acute{}}%
%EndExpansion
}}=M_{a}^{a%
%TCIMACRO{\U{b4}}%
%BeginExpansion
{\acute{}}%
%EndExpansion
}\left( \rho _{\alpha }^{i}\frac{\partial M_{b%
%TCIMACRO{\U{b4}}%
%BeginExpansion
{\acute{}}%
%EndExpansion
}^{a}}{\partial x^{i}}+\frac{\partial \left( \rho ,\eta \right) \Gamma
_{c}^{a}}{\partial y^{b}}M_{b%
%TCIMACRO{\U{b4}}%
%BeginExpansion
{\acute{}}%
%EndExpansion
}^{b}\right) \Lambda _{\alpha
%TCIMACRO{\U{b4}}%
%BeginExpansion
{\acute{}}%
%EndExpansion
}^{\alpha },
\end{equation*}%
it results that
\begin{equation*}
\left( \rho ,\eta \right) \Gamma _{\alpha
%TCIMACRO{\U{b4}}%
%BeginExpansion
{\acute{}}%
%EndExpansion
}^{a%
%TCIMACRO{\U{b4}}%
%BeginExpansion
{\acute{}}%
%EndExpansion
}-y^{b%
%TCIMACRO{\U{b4}}%
%BeginExpansion
{\acute{}}%
%EndExpansion
}\frac{\partial \left( \rho ,\eta \right) \Gamma _{\alpha
%TCIMACRO{\U{b4}}%
%BeginExpansion
{\acute{}}%
%EndExpansion
}^{a%
%TCIMACRO{\U{b4}}%
%BeginExpansion
{\acute{}}%
%EndExpansion
}}{\partial y^{b%
%TCIMACRO{\U{b4}}%
%BeginExpansion
{\acute{}}%
%EndExpansion
}}=M_{a}^{a%
%TCIMACRO{\U{b4}}%
%BeginExpansion
{\acute{}}%
%EndExpansion
}\left( \left( \rho ,\eta \right) \Gamma _{\alpha }^{a}-y^{b}\frac{\partial
\left( \rho ,\eta \right) \Gamma _{\alpha }^{a}}{\partial y^{b}}\right)
\Lambda _{\alpha
%TCIMACRO{\U{b4}}%
%BeginExpansion
{\acute{}}%
%EndExpansion
}^{\alpha },
\end{equation*}
Therefore, we can introduce the following

\bigskip\noindent \textbf{Definition 5.2.5.1 }The\textbf{\ }$\mathbf{Mod}$%
-endomorphism
\begin{equation*}
\Gamma \left( \left( \rho ,\eta \right) TE,\left( \rho ,\eta \right) \tau
_{E},E\right) \,\ \,^{\underrightarrow{\left( \rho ,\eta \right) \mathbb{H}}%
}\,\,\ \Gamma \left( \left( \rho ,\eta \right) TE,\left( \rho ,\eta \right)
\tau _{E},E\right)
\end{equation*}%
defined by
\begin{equation*}
%
\leqno(5.2.5\vspace*{1mm}.1)
\end{equation*}%
will be called the $\left( \rho ,\eta \right) $\emph{-tension of }$\left(
\rho ,\eta \right) $\emph{-connection }$\left( \rho ,\eta \right) \Gamma .$

In particular, if $h=Id_{M}$ and $\left( \rho ,\eta \right) {=}\left(
Id_{TM},Id_{M}\right) $, then we obtain the \emph{tension of connection}~$%
\Gamma .$

\bigskip\noindent\textbf{Proposition 5.2.5.1} \emph{We obtain the following
equalities}%
\begin{equation*}
%
%
\end{equation*}

\subsection{The $\left( \protect\rho ,\protect\eta ,h\right) $-torsion and
the $\left( \protect\rho ,\protect\eta ,h\right) $-curvature of a $\left(
\protect\rho ,\protect\eta \right) $-connection}

We consider the following diagram:
\begin{equation*}
%
%
\end{equation*}%
where $\left( E,\pi ,M\right) \in \left\vert \mathbf{B}^{\mathbf{v}%
}\right\vert $ and $\left( \left( F,\nu ,N\right) ,\left[ ,\right]
_{F,h},\left( \rho ,\eta \right) \right) $ is a generalized Lie algebroid.

\bigskip \noindent \textbf{Definition 5.3.1} If $\left( E,\pi ,M\right)
=\left( F,\nu ,N\right) $, then the $\mathcal{F}\left( E\right) $-bilinear
application
\begin{equation*}
%
%
\!\!\leqno(5.3.1)
\end{equation*}%
will be called the $\left( \rho ,\eta ,h\right) $\emph{-torsion associated
to }$\left( \rho ,\eta \right) $\emph{-connection }$\left( \rho ,\eta
\right) \Gamma .$

In particular, if $h=Id_{M}$, then we obtain the $\left( \rho ,\eta \right) $%
\emph{-torsion associated to }$\left( \rho ,\eta \right) $\emph{-connection }%
$\left( \rho ,\eta \right) \Gamma .$

Moreover, if $\left( \rho ,\eta \right) =\left( Id_{TM},Id_{M}\right) $,
then we obtain the \emph{torsion associated to connection }$\Gamma .$

\bigskip\noindent\textbf{Remark 5.3.1 }If $\left( \rho ,\eta ,h\right)
\mathbb{T}$ is the $\left( \rho ,\eta ,h\right) $-torsion associated to $%
\left( \rho ,\eta \right) $-connection $\left( \rho ,\eta \right) \Gamma $,
then
\begin{equation*}
\begin{array}{c}
\left( \rho ,\eta ,h\right) \mathbb{T}(X,Y)=-\left( \rho ,\eta ,h\right)
\mathbb{T}(Y,X),\vspace*{1mm}%
\end{array}%
\leqno(5.3.2)
\end{equation*}%
for any $X,Y\in \Gamma \left( \left( \rho ,\eta \right) TE,\left( \rho ,\eta
\right) \tau _{E},E\right) .$

\bigskip\noindent\textbf{Definition 5.3.2 }If we consider the notation
\begin{equation*}
\left( \rho ,\eta ,h\right) \mathbb{T}_{~bc}^{a}\overset{put}{=}\frac{%
\partial \left( \rho ,\eta \right) \Gamma _{c}^{a}}{\partial y^{b}}-\frac{%
\partial \left( \rho ,\eta \right) \Gamma _{b}^{a}}{\partial y^{c}}%
-L_{bc}^{a}\circ h\circ \pi\leqno(5.3.3)
\end{equation*}%
then the tensor field
\begin{equation*}
\left( \rho ,\eta ,h\right) \mathbb{T}_{~bc}^{a}\frac{\delta }{\delta \tilde{%
z}^{a}}\otimes d\tilde{z}^{b}\otimes d\tilde{z}^{c}\leqno(5.3.4)
\end{equation*}%
will be called the $\left( \rho ,\eta ,h\right) $\emph{-torsion tensor field
associated to }$\left( \rho ,\eta \right) $\emph{-connection }$\left( \rho
,\eta \right) \Gamma .$

\bigskip\noindent\textbf{Proposition 5.3.1 }\emph{We obtain }%
\begin{equation*}
\mathcal{J}_{\left( Id_{E},Id_{M}\right) }\circ \left( \rho ,\eta \right)
\mathbb{T}=0
\end{equation*}%
\emph{and }%
\begin{eqnarray*}
\left( \rho ,\eta ,h\right) \mathbb{T}\left( \mathcal{J}_{\left(
Id_{E},Id_{M}\right) }X,Y\right) &=&\left( \rho ,\eta \right) \mathbb{T}%
\left( \mathcal{J}_{\left( Id_{E},Id_{M}\right) }X,\mathcal{J}_{\left(
Id_{E},Id_{M}\right) }Y\right) \vspace*{1,5mm} \\
&=&\left( \rho ,\eta \right) \mathbb{T}\left( X,\mathcal{J}_{\left(
Id_{E},Id_{M}\right) }Y\right) ,
\end{eqnarray*}%
\emph{for any} $X,Y\in \Gamma \left( \left( \rho ,\eta \right) TE,\left(
\rho ,\eta \right) \tau _{E},E\right) .$

\bigskip\noindent\textbf{Theorem 5.3.1} \emph{Using the }$\left( \rho ,\eta
\right) $\emph{-tension tensor field }%
\begin{equation*}
\left( \rho ,\eta \right) \mathbb{H}_{b}^{a}\frac{\partial }{\partial \tilde{%
y}^{a}}\otimes d\tilde{z}^{b}=\left( \left( \rho ,\eta \right) \Gamma
_{b}^{a}-y^{c}\frac{\partial \left( \rho ,\eta \right) \Gamma _{b}^{a}}{%
\partial y^{c}}\right) \frac{\partial }{\partial \tilde{y}^{a}}\otimes d%
\tilde{z}^{b},\leqno(5.3.5)
\end{equation*}%
\emph{and the }$\left( \rho ,\eta ,h\right) $\emph{-deflection of the }$%
\left( \rho ,\eta \right) $\emph{-connection }$\left( \rho ,\eta \right)
\Gamma $\emph{\ }%
\begin{equation*}
\left( \rho ,\eta ,h\right) \mathbb{D}_{b}^{a}=-\left( \rho ,\eta \right)
\Gamma _{b}^{a}+y^{c}\frac{\partial \left( \rho ,\eta \right) \Gamma _{c}^{a}%
}{\partial y^{b}}-y^{c}L_{bc}^{a}\circ h\circ \pi ,\leqno(5.3.6)
\end{equation*}%
\emph{we obtain that }$(\rho ,\eta ,h)\mathbb{D}_{b}^{a}{=}0$ \emph{if and
only if} $(\rho ,\eta )\mathbb{H}_{b}^{a}{=}0$ \emph{and} $(\rho ,\eta ,h)%
\mathbb{T}_{~bc}^{a}{=}0.$

\bigskip\noindent\bigskip\noindent\textit{Proof.} If $\left( \rho ,\eta
,h\right) \mathbb{D}_{b}^{a}\mathbb{=}0,$ then deriving with respect to $%
y^{c},$ we obtain: \vspace*{-5mm}
\begin{equation*}
-\displaystyle\frac{\partial \left( \rho ,\eta \right) \Gamma _{b}^{a}}{%
\partial y^{c}}+\frac{\partial \left( \rho ,\eta \right) \Gamma _{c}^{a}}{%
\partial y^{b}}-L_{bc}^{a}\circ h\circ \pi =0\Longleftrightarrow \left( \rho
,\eta ,h\right) \mathbb{T}_{~bc}^{a}=0.
\end{equation*}%
The equality $\left( \rho ,\eta ,h\right) \mathbb{D}_{b}^{a}\mathbb{=}0$
implies:
\begin{equation*}
\left( \rho ,\eta \right) \Gamma _{b}^{a}=y^{c}\frac{\partial \left( \rho
,\eta \right) \Gamma _{c}^{a}}{\partial y^{b}}-y^{c}L_{bc}^{a}\circ h\circ
\pi .\leqno(1)
\end{equation*}%
Since
\begin{equation*}
\begin{array}{rl}
\displaystyle\left( \rho ,\eta \right) \mathbb{H}_{b}^{a} & \displaystyle%
=\left( \rho ,\eta \right) \Gamma _{b}^{a}-y^{c}\frac{\partial \left( \rho
,\eta \right) \Gamma _{b}^{a}}{\partial y^{c}}=\vspace*{1mm} \\
& \displaystyle=y^{c}\frac{\partial \left( \rho ,\eta \right) \Gamma _{c}^{a}%
}{\partial y^{b}}-y^{c}L_{bc}^{a}\circ h\circ \pi -y^{c}\frac{\partial
\left( \rho ,\eta \right) \Gamma _{b}^{a}}{\partial y^{c}}=y^{c}\left( \rho
,\eta ,h\right) \mathbb{T}_{~bc}^{a}%
\end{array}%
\end{equation*}%
it results the equality $\left( \rho ,\eta \right) \mathbb{H}_{b}^{a}\mathbb{%
=}0.$

Conservely, if $\left( \rho ,\eta ,h\right) \mathbb{T}_{~bc}^{a}\mathbb{=}0,$
then, multiplying with $y^{c},$ we obtain:
\begin{equation*}
\frac{\partial \left( \rho ,\eta \right) \Gamma _{c}^{a}}{\partial y^{b}}%
y^{c}-\frac{\partial \left( \rho ,\eta \right) \Gamma _{b}^{a}}{\partial
y^{c}}y^{c}-y^{c}L_{bc}^{a}\circ h\circ \pi =0.\leqno(2)
\end{equation*}%
The equality $\left( \rho ,\eta \right) \mathbb{H}_{b}^{a}\mathbb{=}0$ is
equivalent with:
\begin{equation*}
\left( \rho ,\eta \right) \Gamma _{b}^{a}=y^{c}\frac{\partial \left( \rho
,\eta \right) \Gamma _{b}^{a}}{\partial y^{c}}.\leqno(3)
\end{equation*}%
Using $\left( 2\right) $ and $\left( 3\right) $, it results the equality $%
\left( \rho ,\eta ,h\right) \mathbb{D}_{b}^{a}=0.$ \hfill \emph{q.e.d.}

\bigskip\noindent\textbf{Definition 5.3.3 }The $\mathcal{F}\left( E\right) $%
-bilinear application
\begin{equation*}
\Gamma \left( \left( \rho ,\eta \right) TE,\left( \rho ,\eta \right) \tau
_{E},E\right) ^{2}\,\ \,^{\underrightarrow{\left( \rho ,\eta ,h\right)
\mathbb{R}}}\,\,\ \Gamma \left( \left( \rho ,\eta \right) TE,\left( \rho
,\eta \right) \tau _{E},E\right)
\end{equation*}%
defined by
\begin{equation*}
\begin{array}{l}
\left( \rho ,\eta ,h\right) \mathbb{R}\left( \tilde{\delta}_{\alpha },\tilde{%
\delta}_{\beta }\right) =\left( \rho ,\eta ,h\right) \mathbb{R}_{\,\ \
\alpha \beta }^{a}\overset{\cdot }{\tilde{\partial}}_{a}; \vspace*{1,5mm} \\
\left( \rho ,\eta ,h\right) \mathbb{R}\left( \tilde{\delta}_{\alpha },%
\overset{\cdot }{\tilde{\partial}}_{b}\right) =0=\left( \rho ,\eta ,h\right)
\mathbb{R}\left( \overset{\cdot }{\tilde{\partial}}_{b},\tilde{\delta}%
_{\alpha }\right) ;\vspace*{1,5mm} \\
\left( \rho ,\eta ,h\right) \mathbb{R}\left( \overset{\cdot }{\tilde{\partial%
}}_{a},\overset{\cdot }{\tilde{\partial}}_{b}\right) =0;%
\end{array}%
\leqno(5.3.7)
\end{equation*}%
will be called the $\left( \rho ,\eta ,h\right) $\emph{-curvature associated
to }$\left( \rho ,\eta \right) $\emph{-connection }$\left( \rho ,\eta
\right) \Gamma .$

In particular, if $h=Id_{M}$, then we obtain the $\left( \rho ,\eta \right) $%
\emph{-curvature associated to }$\left( \rho ,\eta \right) $\emph{%
-connection }$\left( \rho ,\eta \right) \Gamma .$

Moreover, if $\left( \rho ,\eta \right) =\left( Id_{TM},Id_{M}\right) $,
then we obtain the \emph{curvature associated to connection }$\Gamma .$

\bigskip\noindent\textbf{Remark 5.3.2 }If $\left( \rho ,\eta ,h\right)
\mathbb{R}$ is the $\left( \rho ,\eta ,h\right) $-curvature associated to $%
\left( \rho ,\eta \right) $-connection $\left( \rho ,\eta \right) \Gamma $,
then
\begin{equation*}
\begin{array}{c}
\left( \rho ,\eta ,h\right) \mathbb{R}\left( X,Y\right) =-\left( \rho ,\eta
,h\right) \mathbb{R}\left( Y,X\right) ,%
\end{array}%
\vspace*{1mm}\leqno(5.3.8)
\end{equation*}%
for any $X,Y\in \Gamma \left( \left( \rho ,\eta \right) TE,\left( \rho ,\eta
\right) \tau _{E},E\right) .$

\bigskip\noindent\textbf{Definition 5.3.4 }The tensor field
\begin{equation*}
\left( \rho ,\eta ,h\right) \mathbb{R}_{\,\ \ \alpha \beta }^{a}\frac{%
\partial }{\partial \tilde{y}^{a}}\otimes d\tilde{z}^{\alpha }\otimes d%
\tilde{z}^{\beta }\leqno(5.3.9)
\end{equation*}%
will be called the $\left( \rho ,\eta ,h\right) $\emph{-curvature tensor
field associated to the }$\left( \rho ,\eta \right) $\emph{-connection }$%
\left( \rho ,\eta \right) \Gamma .$

Using equality $\left( 5.1.5\right) $, we obtain

\bigskip\noindent\textbf{Remark 5.3.3} The horizontal interior differential
system $\left( H\left( \rho ,\eta \right) TE,\left( \rho ,\eta \right) \tau
_{E},E\right) $ is involutive if and only if the $\left( \rho ,\eta
,h\right) $-curvature tensor field associated to the $\left( \rho ,\eta
\right) $-connection $\left( \rho ,\eta \right) \Gamma $ is null.

\bigskip\noindent\textbf{Theorem 5.3.2} \emph{If }$\mathcal{F}$\emph{\ is
the almost complex structure presented in Example $5.2.4.1$, then }$\left(
\rho ,\eta ,h\right) \mathbb{T}=0$\emph{\ and }$\left( \rho ,\eta ,h\right)
\mathbb{R}=0$\emph{\ if and only if }$N_{\mathcal{F}}=0.$

\bigskip\noindent\textit{Proof.} \textbf{\ }After some calculations, we
obtain the relations:
\begin{equation*}
\begin{array}{l}
N_{\mathcal{F}}\left( \tilde{\delta}_{b},\tilde{\delta}_{c}\right) =\left(
\rho ,\eta ,h\right) \mathbb{T}_{~bc}^{a}\tilde{\delta}_{a}-\left( \rho
,\eta ,h\right) \mathbb{R}_{~\ \ bc}^{a}\overset{\cdot }{\tilde{\partial}}%
_{a},\vspace*{1mm} \\
N_{\mathcal{F}}\left( \tilde{\delta}_{b},\overset{\cdot }{\tilde{\partial}}%
_{c}\right) =\left( \rho ,\eta ,h\right) \mathbb{R}_{~\ \ bc}^{a}\tilde{%
\delta}_{a}-\left( \rho ,\eta ,h\right) \mathbb{T}_{~bc}^{a}\overset{\cdot }{%
\tilde{\partial}}_{a},\vspace*{1mm} \\
N_{\mathcal{F}}\left( \overset{\cdot }{\tilde{\partial}}_{b},\overset{\cdot }%
{\tilde{\partial}}_{c}\right) =-\left( \rho ,\eta ,h\right) \mathbb{T}%
_{~bc}^{a}\tilde{\delta}_{a}-\left( \rho ,\eta ,h\right) \mathbb{R}_{~\ \
bc}^{a}\overset{\cdot }{\tilde{\partial}}_{a}.%
\end{array}%
\end{equation*}

Obviously, $\left( \rho ,\eta ,h\right) \mathbb{T=}0$ and $\left( \rho ,\eta
,h\right) \mathbb{R=}0$ imply $N_{\mathcal{F}}=0.$

Conservely, if $N_{\mathcal{F}}=0,$ then we have the equalities:
\begin{equation*}
\begin{array}{r}
\left( \rho ,\eta ,h\right) \mathbb{T}_{~bc}^{a}\tilde{\delta}_{a}-\left(
\rho ,\eta ,h\right) \mathbb{R}_{~\ \ bc}^{a}\overset{\cdot }{\tilde{\partial%
}}_{a}=0, \\
\left( \rho ,\eta ,h\right) \mathbb{R}_{~\ bc}^{a}\tilde{\delta}_{a}-\left(
\rho ,\eta ,h\right) \mathbb{T}_{~bc}^{a}\overset{\cdot }{\tilde{\partial}}%
_{a}=0, \\
-\left( \rho ,\eta ,h\right) \mathbb{T}_{~bc}^{a}\tilde{\delta}_{a}-\left(
\rho ,\eta ,h\right) \mathbb{R}_{~\ bc}^{a}\overset{\cdot }{\tilde{\partial}}%
_{a}=0.%
\end{array}%
\end{equation*}

\hfill \emph{q.e.d.}

\subsection{Tensor $d$-fields. Distinguished linear $\left( \protect\rho ,%
\protect\eta \right) $-connections}

We consider the following diagram:
\begin{equation*}
\begin{array}{c}
\xymatrix{E\ar[d]_\pi&\left( F,\left[ ,\right] _{F,h},\left( \rho ,\eta
\right) \right)\ar[d]^\nu\\ M\ar[r]^h&N}%
\end{array}%
\end{equation*}%
where $\left( E,\pi ,M\right) \in \left\vert \mathbf{B}^{\mathbf{v}%
}\right\vert $ and $\left( \left( F,\nu ,N\right) ,\left[ ,\right]
_{F,h},\left( \rho ,\eta \right) \right) $ is a generalized Lie algebroid.

Let
\begin{equation*}
\left( \mathcal{T}~_{q,s}^{p,r}\left( \left( \rho ,\eta \right) TE,\left(
\rho ,\eta \right) \tau _{E},E\right) ,+,\cdot \right)
\end{equation*}%
be the $\mathcal{F}\left( E\right) $-module of tensor fields by $\left(
_{q,s}^{p,r}\right) $-type from the generalized tangent bundle
\begin{equation*}
\left( H\left( \rho ,\eta \right) TE\oplus V\left( \rho ,\eta \right)
TE,\left( \rho ,\eta \right) \tau _{E},E\right) .
\end{equation*}

An arbitrarily tensor field $T$ is written as
\begin{equation*}
\begin{array}{c}
T=T_{\beta _{1}...\beta _{q}b_{1}...b_{s}}^{\alpha _{1}...\alpha
_{p}a_{1}...a_{r}}\tilde{\delta}_{\alpha _{1}}\otimes ...\otimes \tilde{%
\delta}_{\alpha _{p}}\otimes d\tilde{z}^{\beta _{1}}\otimes ...\otimes d%
\tilde{z}^{\beta _{q}}\otimes \\
\overset{\cdot }{\tilde{\partial}}_{a_{1}}\otimes ...\otimes \overset{\cdot }%
{\tilde{\partial}}_{a_{r}}\otimes \delta \tilde{y}^{b_{1}}\otimes ...\otimes
\delta \tilde{y}^{b_{s}}.%
\end{array}%
\end{equation*}

Let
\begin{equation*}
\left( \mathcal{T}~\left( \left( \rho ,\eta \right) TE,\left( \rho ,\eta
\right) \tau _{E},E\right) ,+,\cdot ,\otimes \right)
\end{equation*}%
be the tensor fields algebra of generalized tangent bundle $\left( \left(
\rho ,\eta \right) TE,\left( \rho ,\eta \right) \tau _{E},E\right) $.

If $T_{1} \in \mathcal{T}_{q_{1},s_{1}}^{p_{1},r_{1}}(( \rho ,\eta ) TE,(
\rho ,\eta ) \tau _{E},E) $ and $T_{2}{\in} \mathcal{T}%
_{q_{2},s_{2}}^{p_{2},r_{2}}( ( \rho ,\eta ) TE,( \rho ,\eta ) \tau _{E},E) $%
, then the components of product tensor field $T_{1}\otimes T_{2}$ are the
products of local components of $T_{1}$ and $T_{2}.$

Therefore, we obtain $T_{1}\otimes T_{2}\in \mathcal{T}%
_{q_{1}+q_{2},s_{1}+s_{2}}^{p_{1}+p_{2},r_{1}+r_{2}}\left( \left( \rho ,\eta
\right) TE,\left( \rho ,\eta \right) \tau _{E},E\right) .$\smallskip

Let $\mathcal{DT}( ( \rho ,\eta ) TE,( \rho ,\eta ) \tau _{E},E) $ be the
family of tensor fields
\begin{equation*}
T\in \mathcal{T}( ( \rho ,\eta ) TE,( \rho ,\eta ) \tau _{E},E)
\end{equation*}
for which there exists
\begin{equation*}
\mbox{$T_{1}\in \mathcal{T}_{q,0}^{p,0}(
( \rho ,\eta ) TE,( \rho ,\eta ) \tau _{E},E) $
and $T_{2}\in \mathcal{T}_{0,s}^{0,r}( ( \rho ,\eta )
TE,( \rho ,\eta ) \tau _{E},E) $}
\end{equation*}
such that $T=T_{1}+T_{2}.$

The $\mathcal{F}\left( E\right) $-module $\left( \mathcal{DT}\left( \left(
\rho ,\eta \right) TE,\left( \rho ,\eta \right) \tau _{E},E\right) ,+,\cdot
\right) $ will be called the \emph{module of distinguished tensor fields} or
the \emph{module of tensor }$d$-\emph{fields.}

\bigskip\noindent\textbf{Remark 5.4.1 }The elements of
\begin{equation*}
\Gamma \left( \left( \rho ,\eta \right) TE,\left( \rho ,\eta \right) \tau
_{E},E\right)
\end{equation*}%
respectively
\begin{equation*}
\Gamma (((\rho ,\eta )TE)^{\ast },\break ((\rho ,\eta )\tau _{E})^{\ast },E)
\end{equation*}%
are tensor $d$-fields.

\bigskip\noindent \textbf{Definition 5.4.1 }Let $\left( E,\pi ,M\right) $ be
a vector bundle endowed with a $\left( \rho ,\eta \right) $-connection $%
\left( \rho ,\eta \right) \Gamma $ and let
\begin{equation*}
\begin{array}{l}
\left( X,T\right) ^{\ \underrightarrow{\left( \rho ,\eta \right) D}\,}%
\vspace*{1mm}\left( \rho ,\eta \right) D_{X}T%
\end{array}%
\leqno(5.4.1)
\end{equation*}%
be a covariant $\left( \rho ,\eta \right) $-derivative for the tensor
algebra of generalized tangent bundle
\begin{equation*}
\left( \left( \rho ,\eta \right) TE,\left( \rho ,\eta \right) \tau
_{E},E\right)
\end{equation*}
which preserves the horizontal and vertical IDS by parallelism.

The real local functions
\begin{equation*}
\left( \left( \rho ,\eta \right) H_{\beta \gamma }^{\alpha },\left( \rho
,\eta \right) H_{b\gamma }^{a},\left( \rho ,\eta \right) V_{\beta c}^{\alpha
},\left( \rho ,\eta \right) V_{bc}^{a}\right)
\end{equation*}%
defined by the following equalities:
\begin{equation*}
\begin{array}{ll}
\left( \rho ,\eta \right) D_{\tilde{\delta}_{\gamma }}\tilde{\delta}_{\beta
}=\left( \rho ,\eta \right) H_{\beta \gamma }^{\alpha }\tilde{\delta}%
_{\alpha }, & \left( \rho ,\eta \right) D_{\tilde{\delta}_{\gamma }}\overset{%
\cdot }{\tilde{\partial}}_{b}=\left( \rho ,\eta \right) H_{b\gamma }^{a}%
\overset{\cdot }{\tilde{\partial}}_{a} \\
\left( \rho ,\eta \right) D_{\overset{\cdot }{\tilde{\partial}}_{c}}\tilde{%
\delta}_{\beta }=\left( \rho ,\eta \right) V_{\beta c}^{\alpha }\tilde{\delta%
}_{\alpha }, & \left( \rho ,\eta \right) D_{\overset{\cdot }{\tilde{\partial}%
}_{c}}\overset{\cdot }{\tilde{\partial}}_{b}=\left( \rho ,\eta \right)
V_{bc}^{a}\overset{\cdot }{\tilde{\partial}}_{a}%
\end{array}%
\leqno(5.4.2)
\end{equation*}%
are the components of a linear $\left( \rho ,\eta \right) $-connection
\begin{equation*}
\left( \left( \rho ,\eta \right) H,\left( \rho ,\eta \right) V\right)
\end{equation*}%
for the generalized tangent bundle $\left( \left( \rho ,\eta \right)
TE,\left( \rho ,\eta \right) \tau _{E},E\right) $ which will be called the
\emph{distinguished linear }$\left( \rho ,\eta \right) $\emph{-connection.}

If $h=Id_{M},$ then the distinguished linear $\left( Id_{TM},Id_{M}\right) $%
-connection will be called the \emph{distinguished linear connection.}

The components of a distinguished linear connection $\left( H,V\right) $
will be denoted
\begin{equation*}
\left( H_{jk}^{i},H_{bk}^{a},V_{jc}^{i},V_{bc}^{a}\right) .
\end{equation*}

\smallskip \noindent\textbf{Theorem 5.4.1 }\emph{If }$((\rho ,\eta )H,(\rho
,\eta )V)$ \emph{is a distinguished linear} $(\rho ,\eta )$- \emph{%
connection for the generalized tangent bundle }$\left( \left( \rho ,\eta
\right) TE,\left( \rho ,\eta \right) \tau _{E},E\right) $\emph{, then its
components satisfy the change relations: }

\begin{equation*}
%
\leqno(5.4.3)
\end{equation*}%
\emph{The components of a distinguished linear connection $\left( H,V\right)
$ verify the change relations:}
\begin{equation*}
%
%
\leqno(5.4.3^{\prime })
\end{equation*}

\smallskip\noindent \textbf{Example 5.4.1} If $\left( E,\pi ,M\right) $ is a
vector bundle endowed with the $\left( \rho ,\eta \right) $-connection $%
\left( \rho ,\eta \right) \Gamma $, then the local real functions%
\begin{equation*}
\left( \frac{\partial \left( \rho ,\eta \right) \Gamma _{\gamma }^{a}}{%
\partial y^{b}},\frac{\partial \left( \rho ,\eta \right) \Gamma _{\gamma
}^{a}}{\partial y^{b}},0,0\right)
\end{equation*}%
are the components of a distinguished linear $\left( \rho ,\eta \right) $%
\textit{-}connection for $\left( \left( \rho ,\eta \right) TE,\left( \rho
,\eta \right) \tau _{E},E\right) ,$ which will by called the \emph{Berwald
linear }$\left( \rho ,\eta \right) $\emph{-connection.}

The Berwald linear $(Id_{TM},Id_{M})$-connection will be called the \emph{%
Ber\-wald linear connection.}

\bigskip \noindent \textbf{Theorem 5.4.2} \emph{If the generalized tangent
bundle} $\!(\!(\rho ,\!\eta )T\!E,\!(\rho ,\!\eta )\tau _{E},\!E\!)$ \emph{%
is endowed with a distinguished linear} $\!(\rho ,\!\eta )$\emph{-connection}
$((\rho ,\eta )H,(\rho ,\eta )V)$, \emph{then, for any}
\begin{equation*}
X=\tilde{Z}^{\alpha }\tilde{\delta}_{\alpha }+Y^{a}\overset{\cdot }{\tilde{%
\partial}}_{a}\in \Gamma (\!(\rho ,\eta )TE,\!(\rho ,\!\eta )\tau _{E},\!E)
\end{equation*}%
\emph{and for any}
\begin{equation*}
T\in \mathcal{T~}_{qs}^{pr}\!(\!(\rho ,\eta )TE,\!(\rho ,\eta )\tau
_{E},\!E),
\end{equation*}%
\emph{we obtain the formula:}
\begin{equation*}
%
\end{equation*}

\smallskip \noindent\textbf{Definition 5.4.2 }We assume that $\left( E,\pi
,M\right) =\left( F,\nu ,N\right) .$

If $\left( \rho ,\eta \right) \Gamma $ is a $\left( \rho ,\eta \right) $%
-connection for the vector bundle $\left( E,\pi ,M\right) $ and
\begin{equation*}
\left( \left( \rho ,\eta \right) H_{bc}^{a},\left( \rho ,\eta \right) \tilde{%
H}_{bc}^{a},\left( \rho ,\eta \right) V_{bc}^{a},\left( \rho ,\eta \right)
\tilde{V}_{bc}^{a}\right)
\end{equation*}%
are the components of a distinguished linear $\left( \rho ,\eta \right) $%
\textit{-}connection for the generalized tangent bundle $\left( \left( \rho
,\eta \right) TE,\left( \rho ,\eta \right) \tau _{E},E\right) $ such that
\begin{equation*}
\left( \rho ,\eta \right) H_{bc}^{a}=\left( \rho ,\eta \right) \tilde{H}%
_{bc}^{a}\mbox{ and }\left( \rho ,\eta \right) V_{bc}^{a}=\left( \rho ,\eta
\right) \tilde{V}_{bc}^{a},
\end{equation*}%
then we will say that \emph{the generalized tangent bundle }$\!(\! ( \rho
,\!\eta ) TE, ( \rho ,\!\eta ) \tau _{E},\!E ) $ \emph{is endowed with a
normal distinguished linear }$\left( \rho ,\eta \right) $\emph{-connection
on components }$\left( \left( \rho ,\eta \right) H_{bc}^{a},\left( \rho
,\eta \right) V_{bc}^{a}\right) $.

The components of a normal distinguished linear $\left(
Id_{TM},Id_{M}\right) $-con\-nec\-tion $\left( H,V\right) $ will be denoted $%
\left( H_{jk}^{i},V_{jk}^{i}\right) $.

\subsection{The lift of accelerations for a differentiable curve}

We consider the following diagram:
\begin{equation*}
\begin{array}{c}
\xymatrix{E\ar[d]_\pi&(F,[,]_{F,h},(\rho,\eta))\ar[d]^\nu\\ M\ar[r]^h&N}%
\end{array}
\leqno(5.5.1)
\end{equation*}%
where $\left( E,\pi ,M\right) \in \left\vert \mathbf{B}^{\mathbf{v}%
}\right\vert $ and $\left( \left( F,\nu ,N\right) ,\left[ ,\right]
_{F,h},\left( \rho ,\eta \right) \right) \in \left\vert \mathbf{GLA}%
\right\vert .$

Let $\left( \rho ,\eta \right) \Gamma $ be a $\left( \rho ,\eta \right) $%
-connection for the vector bundle $\left( E,\pi ,M\right) .$

We admit that $\left( \left( \rho ,\eta \right) H,\left( \rho ,\eta \right)
V\right) $ is a distinguished linear $\left( \rho ,\eta \right) $-connection
for the vector bundle $\left( \left( \rho ,\eta \right) TE,\left( \rho ,\eta
\right) \tau _{E},E\right) .$

Let $g\in \mathbf{Man}\left( E,F\right) $ be such that $\left( g,h\right) $
is a $\mathbf{B}^{\mathbf{v}}$-morphism of $\left( E,\pi ,M\right) $ source
and $\left( F,\nu ,N\right) $ target.

Let%
\begin{equation*}
%
\leqno(5.5.2)
\end{equation*}%
be the $\left( g,h\right) $-lift of differentiable curve $%
%
%
$

\bigskip\noindent\textbf{Definition 5.5.1 }The differentiable curve
\begin{equation*}
%
%
\leqno(5.5.3)
\end{equation*}%
will be called the \emph{differentiable }$\left( g,h\right) $\emph{-lift of
accelerations of the differentiable curve }$c.$

The section
\begin{equation*}
%
%
\leqno(5.5.4)
\end{equation*}%
will be called the \emph{canonical section associated to the triple }$\left(
c,\dot{c},\ddot{c}\right) .$

\bigskip\noindent\textbf{Remark 5.5.1 }For any $t\in I$, we obtain:
%\begin{equation*}
%\left( 5.2.3.14\right) ~\ \
\begin{equation*}
%
%
\leqno(5.5.5)
\end{equation*}

We observe easily that $u\left( c,\dot{c},\ddot{c}\right) \left( \dot{c}%
\left( t\right) \right) \in H\left( \rho ,\eta \right) TE_{|\func{Im}\left(
\dot{c}\right) }$ if and only if the components functions $\left(
y^{a},~a\in \overline{1,n}\right) $ are solutions for the differentiable
equations %\begin{equation*}
%\left( 5.2.3.15\right) ~\ \
\begin{equation*}
\frac{du^{a}}{dt}+\left( \rho ,\eta \right) \Gamma _{\alpha }^{a}\circ
u\left( c,\dot{c}\right) \circ \eta \circ h\circ c\cdot \left( g_{b}^{\alpha
}\circ h\circ c\right) \cdot u^{b}=0,~a\in \overline{1,r}.\leqno(5.5.6)
\end{equation*}

\smallskip \noindent\textbf{Remark 5.5.2 }In particular, if $\left( \rho
,\eta ,h\right) =\left( Id_{TM},Id_{M},Id_{M}\right) $ and $\left(
g,Id_{M}\right) $ is locally invertible, then, using the $\left(
g,Id_{M}\right) $-lift
\begin{equation*}
\begin{array}{rcl}
I & ^{\underrightarrow{\ \ \dot{c}\ \ }} & TM \vspace*{1,5mm} \\
t & \longmapsto & \displaystyle\tilde{g}_{j}^{i}\frac{dc^{j}\left( t\right)
}{dt}\frac{\partial }{\partial x^{i}}\left( c\left( t\right) \right) ,%
\end{array}%
\leqno(5.5.7)
\end{equation*}
the differentiable $\left( g,Id_{M}\right) $-lift of accelerations for the
differentiable curve $c$ is
\begin{equation*}
\begin{array}{rcl}
I & ^{\underrightarrow{\ \ \ddot{c}\ \ }} & \left( Id_{TM},Id_{M}\right)
TTM_{|\func{Im}\left( \dot{c}\right) } \vspace*{1,5mm} \\
t & \longmapsto & \displaystyle\frac{dc^{i}\left( t\right) }{dt}\frac{%
\partial }{\partial \tilde{z}^{i}}\left( \dot{c}\left( t\right) \right) +%
\tilde{g}_{j}^{i}\left( c\left( t\right) \right) \frac{dc^{j}\left( t\right)
}{dt}\frac{\partial }{\partial \tilde{y}^{i}}\left( \dot{c}\left( t\right)
\right).%
\end{array}%
\leqno(5.5.8)
\end{equation*}

\smallskip \noindent\textbf{Definition 5.5.2 }If the component functions
\begin{equation*}
\left( \left( g_{a}^{\alpha }\circ h\circ c\right) y^{a},~a\in \overline{1,r}%
\right)
\end{equation*}%
are solutions for the differentiable system of equations%
%\begin{equation*}
%\left( 5.2.3.18\right) ~\ \
\begin{equation*}
\frac{dz^{\alpha }}{dt}+\left( \rho ,\eta \right) H_{\beta \gamma }^{\alpha
}\circ u\left( c,\dot{c}\right) \circ \eta \circ h\circ c\cdot z^{\beta
}\cdot z^{\gamma }=0,~\alpha \in \overline{1,p},\leqno(5.5.9)
\end{equation*}%
then the differentiable curve $\dot{c}$ will be called \emph{horizontal
parallel with respect to the distinguished linear }$\left( \rho ,\eta
\right) $\emph{-connection }$\left( \left( \rho ,\eta \right) H,\left( \rho
,\eta \right) V\right) .$

If the component functions $\left( y^{a},~a\in \overline{1,n}\right) $ are
solutions for the differentiable system of equations%
\begin{equation*}
\frac{du^{a}}{dt}+\left( \rho ,\eta \right) V_{bc}^{a}\circ u\left( c,\dot{c}%
\right) \circ \eta \circ h\circ c\cdot u^{b}\cdot u^{c}=0,~a\in \overline{1,r%
}, \leqno(5.5.10)
\end{equation*}%
then the differentiable curve $\dot{c}$ will be called \emph{vertical
parallel with respect to the distinguished linear }$\left( \rho ,\eta
\right) $\emph{-connection }$\left( \left( \rho ,\eta \right) H,\left( \rho
,\eta \right) V\right) .$

\bigskip\noindent\textbf{Remark 5.5.3 }In particular, if $\left( \rho ,\eta
,h\right) =\left( Id_{TM},Id_{M},Id_{M}\right) $ and $\left( g,Id_{M}\right)
$ is locally invertible, then the $\left( g,Id_{M}\right) $-lift of tangent
vectors $\left( 5.5.7\right) $ is horizontal parallel with res\-pect to the
distinguished linear connection $\left( H,V\right) $ if the component
functions\break $\left( \displaystyle\frac{dc^{i}}{dt},~i\in \overline{1,m}%
\right) $ are solutions for the differentiable system of equations%
\begin{equation*}
\frac{dz^{i}}{dt}+H_{jk}^{i}\circ u\left( c,\dot{c}\right) \circ c\cdot
z^{j}\cdot z^{k}=0,~i\in \overline{1,m}.\leqno(5.5.12)
\end{equation*}
Moreover, the $\left( g,Id_{M}\right) $-lift of tangent vectors $\left(
5.5.7\right) $ is vertical parallel with respect to the distinguished linear
connection $\left( H,V\right) $ if the component functions
\begin{equation*}
\left( \tilde{g}_{j}^{i}\circ c\cdot \displaystyle\frac{dc^{j}\left(
t\right) }{dt},~i\in \overline{1,m}\right)
\end{equation*}
are solutions for the differentiable system of equations%
\begin{equation*}
\frac{du^{i}}{dt}+V_{jk}^{i}\circ u\left( c,\dot{c}\right) \circ c\cdot
u^{j}\cdot u^{k}=0,~i\in \overline{1,m}.\leqno(5.5.13)
\end{equation*}

\subsection{The $\left( \protect\rho ,\protect\eta ,h\right) $-torsion and
the $\left( \protect\rho ,\protect\eta ,h\right) $-curvature of a
distinguished linear $\left( \protect\rho ,\protect\eta \right) $-connection}

We consider the following diagram:
\begin{equation*}
\begin{array}{c}
\xymatrix{E\ar[d]_\pi&(F,[,]_{F,h},(\rho,\eta))\ar[d]^\nu\\ M\ar[r]^h&N}%
\end{array}%
\end{equation*}%
where $\left( E,\pi ,M\right) \in \left\vert \mathbf{B}^{\mathbf{v}%
}\right\vert $ and $\left( \left( F,\nu ,M\right) ,\left[ ,\right]
_{F.h},\left( \rho ,\eta \right) \right) \in \left\vert \mathbf{GLA}%
\right\vert .$ Let $\left( \rho ,\eta \right) \Gamma $ be a $\left( \rho
,\eta \right) $-connection for the vector bundle $\left( E,\pi ,M\right) $
and let $\left( \left( \rho ,\eta \right) H,\left( \rho ,\eta \right)
V\right) $ be a distinguished linear $\left( \rho ,\eta \right) $-connection
for the generalized tangent bundle $\left( \left( \rho ,\eta \right)
TE,\left( \rho ,\eta \right) \tau _{E},E\right) .$

\bigskip\noindent\textbf{Definition 5.6.1 }The application
\begin{equation*}
\begin{array}{rcl}
\Gamma \left( \left( \rho ,\eta \right) TE,\left( \rho ,\eta \right) \tau
_{E},E\right) ^{2} & ^{\underrightarrow{\left( \rho ,\eta ,h\right) \mathbb{T%
}}} & \Gamma \left( \left( \rho ,\eta \right) TE,\left( \rho ,\eta \right)
\tau _{E},E\right) \vspace*{2mm} \\
\left( X,Y\right) & \longmapsto & \left( \rho ,\eta \right) \mathbb{T}\left(
X,Y\right)%
\end{array}%
\end{equation*}%
defined by
\begin{equation*}
\begin{array}{c}
\left( \rho ,\eta ,h\right) \mathbb{T}\left( X,Y\right) =\left( \rho ,\eta
\right) D_{X}Y-\left( \rho ,\eta \right) D_{Y}X-\left[ X,Y\right] _{\left(
\rho ,\eta \right) TE},%
\end{array}%
\leqno(5.6.1)
\end{equation*}%
for any $X,Y\in \Gamma \left( \left( \rho ,\eta \right) TE,\left( \rho ,\eta
\right) \tau _{E},E\right) ,$ will be called the $\left( \rho ,\eta
,h\right) $\textit{-torsion associated to distinguished linear }$\left( \rho
,\eta \right) $\textit{-connection }$\left( \left( \rho ,\eta \right)
H,\left( \rho ,\eta \right) V\right) .$

The applications
\begin{equation*}
\mathcal{H}\left( \rho ,\eta ,h\right) \mathbb{T}\left( \mathcal{H}\left(
\cdot \right) ,\mathcal{H}\left( \cdot \right) \right) ,\,\mathcal{V}\left(
\rho ,\eta ,h\right) \mathbb{T}\left( \mathcal{H}\left( \cdot \right) ,%
\mathcal{H}\left( \cdot \right) \right) ,...,\mathcal{V}\left( \rho ,\eta
,h\right) \mathbb{T}\left( \mathcal{V}\left( \cdot \right) ,\mathcal{V}%
\left( \cdot \right) \right)
\end{equation*}%
are called $\mathcal{H}\left( \mathcal{HH}\right) ,\,\mathcal{V}\left(
\mathcal{HH}\right) ,...,\mathcal{V}\left( \mathcal{VV}\right) $ $\left(
\rho ,\eta ,h\right) $\textit{-torsions associated to distinguished linear }$%
\left( \rho ,\eta \right) $\textit{-connection }$\left( \left( \rho ,\eta
\right) H,\left( \rho ,\eta \right) V\right) .$

\bigskip\noindent\textbf{Proposition 5.6.1 }\emph{The }$\left( \rho ,\eta
,h\right) $\emph{-torsion }$\left( \rho ,\eta ,h\right) \mathbb{T}$\emph{\
associated to distinguished linear }$\left( \rho ,\eta \right) $\emph{%
-connection }$\left( \left( \rho ,\eta \right) H,\left( \rho ,\eta \right)
V\right) $\emph{, is }$\mathbb{R}$\emph{-bilinear and antisymmetric in the
lower indices.}\medskip

{Using the notations: }%
\begin{equation*}
%
\leqno(5.6.2)
\end{equation*}
{we can easily prove the following}

\bigskip\noindent\textbf{Theorem 5.6.1 }\emph{The }$\left( \rho ,\eta
,h\right) $\emph{-torsion }$\left( \rho ,\eta ,h\right) \mathbb{T}$\emph{\
associated to the distinguished linear }$\left( \rho ,\eta \right) $\emph{%
-connection }$\left( \left( \rho ,\eta \right) H,\left( \rho ,\eta \right)
V\right) $\emph{, is characterized by the\ tensor fields with local
components: }%
\begin{equation*}
%
%
\leqno(5.6.3)
\end{equation*}
\emph{In particular, when }$\left( \rho ,\eta ,h\right) =\left(
Id_{TM},Id_{M},Id_{M}\right) ,$\emph{\ we regain the local components of
torsion associated to distinguished linear connection }$\left( H,V\right) $%
\emph{: }%
\begin{equation*}
%
%
\leqno(18.4)
\end{equation*}%
for any $X,Y,Z\in \Gamma \left( \left( \rho ,\eta \right) TE,\left( \rho
,\eta \right) \tau _{E},E\right) ,$ will be called the $\left( \rho ,\eta
,h\right) $\emph{-curvature asso\-cia\-ted to distinguished linear }$\left(
\rho ,\eta \right) $\emph{-connection }$\left( \left( \rho ,\eta \right)
H,\left( \rho ,\eta \right) V\right) .$

\bigskip\noindent \textbf{Proposition 5.6.2 }\emph{The }$\left( \rho ,\eta
,h\right) $\emph{-curvature }$\left( \rho ,\eta ,h\right) \mathbb{R}$\emph{\
associated to distinguished linear }$\left( \rho ,\eta \right) $\emph{%
-connection }$\left( \left( \rho ,\eta \right) H,\left( \rho ,\eta \right)
V\right) $\emph{, is }$\mathbb{R}$\emph{-linear in each argument and
antisymmetric in the first two arguments.}\medskip

Using the notations:
\begin{equation*}
%
\leqno(5.6.5)
\end{equation*}%
we can easily prove the following

\bigskip\noindent\textbf{Theorem 5.6.2 }\emph{The }$\left( \rho ,\eta
,h\right) $\emph{-curvature }$\left( \rho ,\eta ,h\right) \mathbb{R}$\emph{\
associated to distinguished linear }$\left( \rho ,\eta \right) $\emph{%
-connection }$\left( \left( \rho ,\eta \right) H,\left( \rho ,\eta \right)
V\right) $\emph{, is characterized by the\ tensor fields with local
components:}\bigskip

\noindent$(5.6.6)\ \ \ \ \left\{%
%
%
\right. $\bigskip

\noindent\emph{In particular, when }$\left( \rho ,\eta ,h\right) =\left(
Id_{TM},Id_{M},Id_{M}\right) ,$\emph{\ we see the local components of the
curvature associated to distinguished linear connection }$\left( H,V\right) $
\emph{in the followings:}
\begin{equation*}
%
%
\hspace*{6mm} \leqno(5.6.8^{\prime })
\end{equation*}

\smallskip \noindent \textbf{Definition 5.6.3 }The tensor field
%\begin{equation*}
%\left( 5.2.4.9\right) ~\ \
\begin{equation*}
%
%
,\leqno(5.6.10)
\end{equation*}%
will be called \emph{the Ricci tensor field associated to distinguished
linear }$\left( \rho ,\eta \right) $\emph{-connection }$\left( \left( \rho
,\eta \right) H,\left( \rho ,\eta \right) V\right) .$

This tensor field will be used to write the Einstein equations in Subsection
5.10.

\subsection{Formulas of Ricci type. Identities of Cartan and Bianchi type}

We consider the following diagram:
\begin{equation*}
%
%
\end{equation*}%
where $\left( E,\pi ,M\right) \in \left\vert \mathbf{B}^{\mathbf{v}%
}\right\vert $ and $\left( \left( F,\nu ,M\right) ,\left[ ,\right]
_{F.h},\left( \rho ,\eta \right) \right) \in \left\vert \mathbf{GLA}%
\right\vert .$ Let $\left( \rho ,\eta \right) \Gamma $ be a $\left( \rho
,\eta \right) $-connection for the vector bundle $\left( E,\pi ,M\right) $
and let $\left( \left( \rho ,\eta \right) H,\left( \rho ,\eta \right)
V\right) $ be a distinguished linear $\left( \rho ,\eta \right) $-connection
for the generalized tangent bundle $\left( \left( \rho ,\eta \right)
TE,\left( \rho ,\eta \right) \tau _{E},E\right) .$

\bigskip \noindent \textbf{Theorem 5.7.1} \emph{Using the definition of }$%
\left( \rho ,\eta ,h\right) $\emph{-curvature associated to the
distinguished linear }$\left( \rho ,\eta \right) $\emph{-connection }$\left(
\left( \rho ,\eta \right) H,\left( \rho ,\eta \right) V\right) $\emph{, it
results the following formulas:}%
\begin{equation*}
\left\{
%
\right. \leqno(\mathcal{R}_{2})
\end{equation*}

Using the previous theorem, the horizontal and vertical sections of adapted
base and an arbitrary section
\begin{equation*}
Z^{\alpha }\frac{\partial }{\partial \tilde{z}^{\alpha }}+Y^{a}\frac{%
\partial }{\partial \tilde{y}^{a}}\in \Gamma \left( \left( \rho ,\eta
\right) TE,\left( \rho ,\eta \right) \tau _{E},E\right),
\end{equation*}%
we propose the following

\bigskip \noindent \textbf{Theorem 5.7.2 }\emph{We obtain the following
formulas of Ricci type: }%
\begin{equation*}
\left\{
%
%
\right. \leqno(\mathcal{R}_{2})
\end{equation*}%
\emph{In particular, if }$\left( \rho ,\eta ,h\right) =\left(
Id_{TM},Id_{M},id_{M}\right) $\emph{\ and the Lie bracket }$\left[ ,\right]
_{TM}$\emph{\ is the usual Lie bracket, then the formulas of Ricci type }$%
\left( \mathcal{R}_{1}\right) $\emph{\ and }$\left( \mathcal{R}_{2}\right) $%
\emph{\ become:}%
\begin{equation*}
\left( \mathcal{R}_{1}\right) ^{\prime }~~\left\{
%
%
\right.
\end{equation*}%
Using the $1$-forms associated to distinguished linear $\left( \rho ,\eta
\right) $-connection $\left( \left( \rho ,\eta \right) H,\left( \rho ,\eta
\right) V\right) $%
%\begin{equation*}
%\left( 5.2.4.11\right) ~\ \ \left\{
\begin{equation*}
%
%
\right. \leqno(5.7.2)
\end{equation*}%
and the curvature $2$-forms%
\begin{equation*}
\left\{
%
%
\right. \leqno(5.7.3)
\end{equation*}%
we obtain the following

\bigskip\noindent\textbf{Theorem 5.7.3} \emph{We obtain the following
identities of Cartan type:}%
%\begin{equation*}
%\left( C_{1}\right) ~\ \ \left\{
\begin{equation*}
%
\leqno(\mathcal{C}_{2})
\end{equation*}
In particular, if $\left( \rho ,\eta ,h\right) =\left(
Id_{TM},Id_{M},Id_{M}\right) $ and the Lie bracket $\left[ ,\right] _{TM}$
is the usual Lie bracket, then the identities of Cartan type $\left(
\mathcal{C}_{1}\right) $ and $\left( \mathcal{C}_{2}\right) $ become:
\begin{equation*}
%
\leqno(\mathcal{C}_{2})^{\prime }
\end{equation*}

\smallskip \noindent\textbf{Remark 5.7.1 } For any $X,Y,Z\in \Gamma \left(
\left( \rho ,\eta \right) TE,\left( \rho ,\eta \right) \tau _{E},E\right) $,
the following identities
\begin{equation*}
%
%
\leqno(5.7.6)
\end{equation*}%
hold. Using the formulas of Bianchi type from Theorem 4.2.3 and Remark
5.7.1, we obtain the following

\bigskip\noindent\textbf{Theorem 5.7.4} \emph{The identities of Bianchi
type: }%
\begin{equation*}
\left\{
%
%
\right.\leqno\left( \mathcal{B}_{2}\right)
\end{equation*}%
\emph{hold good for any }$X,Y,Z\in \Gamma \left( \left( \rho ,\eta \right)
TE,\left( \rho ,\eta \right) \tau _{E},E\right) .$

\bigskip\noindent\textbf{Corollary 5.7.1 }\emph{Using the following sections
}$\left( \delta _{\theta },\delta _{\gamma },\delta _{\beta }\right) $\emph{%
, the identities }$\left( \mathcal{B}_{1}\right) $\emph{\ become:}%
\begin{equation*}
\left\{
%
%
\right.\leqno (\mathcal{B}_{1})^{\prime }
\end{equation*}%
\emph{and using the following sections }$\left( \delta _{\lambda },\delta
_{\theta },\delta _{\gamma },\delta _{\beta }\right) $\emph{, the identities
}$\left( \mathcal{B}_{2}\right) $\emph{\ become: }%
\begin{equation*}
\left\{
%
%
\right.\leqno (\mathcal{B}_{2})^{\prime }
\end{equation*}

Using another base of sections, we shall obtain new identities of Bianchi
type necessary in the applications.

\subsection{The $\left( \protect\rho ,\protect\eta \right) $%
-(pseudo)metrizability}

We consider the following diagram:
\begin{equation*}
%
%
\end{equation*}%
where $\left( E,\pi ,M\right) \in \left\vert \mathbf{B}^{\mathbf{v}%
}\right\vert $ and $\left( \left( F,\nu ,M\right) ,\left[ ,\right]
_{F,h},\left( \rho ,\eta \right) \right) \in \left\vert \mathbf{GLA}%
\right\vert .$ Let $\left( \rho ,\eta \right) \Gamma $ be a $\left( \rho
,\eta \right) $-connection for the vector bundle $\left( E,\pi ,M\right) $
and let $\left( \left( \rho ,\eta \right) H,\left( \rho ,\eta \right)
V\right) $ be a distinguished linear $\left( \rho ,\eta \right) $-connection
for the generalized tangent bundle $\left( \left( \rho ,\eta \right)
TE,\left( \rho ,\eta \right) \tau _{E},E\right) .$

\bigskip \noindent \textbf{Definition 5.8.1} A tensor $d$-field
\begin{equation*}
G=g_{\alpha \beta }d\tilde{z}^{\alpha }\otimes d\tilde{z}^{\beta
}+g_{ab}\delta \tilde{y}^{a}\otimes \delta \tilde{y}^{b}\in \mathcal{DT}%
_{22}^{00}\left( \left( \rho ,\eta \right) TE,\left( \rho ,\eta \right) \tau
_{E},E\right)
\end{equation*}%
will be called \emph{pseudometrical structure }if its components are
symmetric and the matrices $\left\Vert g_{\alpha \beta }\left( u_{x}\right)
\right\Vert $and $\left\Vert g_{ab}\left( u_{x}\right) \right\Vert $ are
nondegenerate, for any point $u_{x}\in E.$

Moreover, if the matrices $\left\Vert g_{\alpha \beta }\left( u_{x}\right)
\right\Vert $ and $\left\Vert g_{ab}\left( u_{x}\right) \right\Vert $ has
constant signature, then the tensor $d$-field $G$ will be called \emph{%
metrical structure}\textit{.}

Let
\begin{equation*}
G=g_{\alpha \beta }d\tilde{z}^{\alpha }\otimes d\tilde{z}^{\beta
}+g_{ab}\delta \tilde{y}^{a}\otimes \delta \tilde{y}^{b}
\end{equation*}%
be a (pseudo)metrical structure. If $\alpha ,\beta \in \overline{1,p}$ and $%
a,b\in \overline{1,r},$ then for any vector local $\left( m+r\right) $-chart
$\left( U,\overset{\ast }{s}_{U}\right) $ of $\left( \overset{\ast }{E},%
\overset{\ast }{\pi },M\right) $, we consider the real functions
\begin{equation*}
\begin{array}{ccc}
\pi ^{-1}\left( U\right) & ^{\underrightarrow{~\ \ \tilde{g}^{\beta \alpha
}~\ \ }} & \mathbb{R}%
\end{array}%
\end{equation*}%
and
\begin{equation*}
\begin{array}{ccc}
\pi ^{-1}\left( U\right) & ^{\underrightarrow{~\ \ \tilde{g}^{ba}~\ \ }} &
\mathbb{R}%
\end{array}%
\end{equation*}%
such that%
\begin{equation*}
\begin{array}{c}
\left\Vert \tilde{g}^{\beta \alpha }\left( u_{x}\right) \right\Vert
=\left\Vert g_{\alpha \beta }\left( u_{x}\right) \right\Vert ^{-1}%
\end{array}%
\end{equation*}%
and
\begin{equation*}
\begin{array}{c}
\left\Vert \tilde{g}^{ba}\left( u_{x}\right) \right\Vert =\left\Vert
g_{ab}\left( u_{x}\right) \right\Vert ^{-1},%
\end{array}%
\end{equation*}%
for any $u_{x}\in \pi ^{-1}\left( U\right) \backslash \left\{ 0_{x}\right\} $%
.

\bigskip \noindent \textbf{Definition 5.8.2} We will say that the \emph{%
(pseudo)metrical structure \ }%
\begin{equation*}
G=g_{\alpha \beta }d\tilde{z}^{\alpha }\otimes d\tilde{z}^{\beta
}+g_{ab}\delta \tilde{y}^{a}\otimes \delta \tilde{y}^{b}
\end{equation*}%
\emph{is Riemannian (pseudo)metrical structure}\textit{\ }if around each
point $x\in M$ it exists a local vector $m+r$-chart $\left( U,s_{U}\right) $
and a local $m$-chart $\left( U,\xi _{U}\right) $ such that $g_{\alpha \beta
}\circ s_{U}^{-1}\circ \left( \xi _{U}^{-1}\times Id_{\mathbb{R}^{m}}\right)
\left( x,y\right) $ and $g_{ab}\circ s_{U}^{-1}\circ \left( \xi
_{U}^{-1}\times Id_{\mathbb{R}^{m}}\right) \left( x,y\right) $ depends only
on $x$, for any $u_{x}\in \pi ^{-1}\left( U\right) .$

If only the condition is verified:

\textquotedblright $g_{\alpha \beta }\circ s_{U}^{-1}\circ \left( \xi
_{U}^{-1}\times Id_{\mathbb{R}^{m}}\right) \left( x,y\right) $\textit{\
depends only on }$x$\textit{, for any }$u_{x}\in \pi ^{-1}\left( U\right) $%
\textit{" res\-pec\-ti\-vely \textquotedblright }$g_{ab}\circ
s_{U}^{-1}\circ \left( \xi _{U}^{-1}\times Id_{\mathbb{R}^{m}}\right) \left(
x,y\right) $\textit{\ depends only on }$x$\textit{, for any }$u_{x}\in \pi
^{-1}\left( U\right) $", then we will say that the \textit{(pseudo)metrical
structure }$G$ \textit{is a }\emph{Riemannian }$\mathcal{H}$\emph{%
-(pseudo)metrical structure}\textit{\ }respectively a \emph{Riemannian }$%
\mathcal{V}$\emph{-(pseudo)metrical structure.}

\noindent

\textbf{Definition 5.8.3} We will say that the \emph{(pseudo)metrical
structure }%
\begin{equation*}
G=g_{\alpha \beta }d\tilde{z}^{\alpha }\otimes d\tilde{z}^{\beta
}+g_{ab}\delta \tilde{y}^{a}\otimes \delta \tilde{y}^{b}
\end{equation*}%
\emph{is a locally Minkowski structure}\textit{\ }if around each point $x\in
M$ there exists a local vector $m+r$-chart $\left( U,s_{U}\right) $ and a
local $m$-chart $\left( U,\xi _{U}\right) $ such that $g_{\alpha \beta
}\circ s_{U}^{-1}\circ \left( \xi _{U}^{-1}\times Id_{\mathbb{R}^{m}}\right)
\left( x,y\right) $ and $g_{ab}\circ s_{U}^{-1}\circ \left( \xi
_{U}^{-1}\times Id_{\mathbb{R}^{m}}\right) \left( x,y\right) $ depends only
on~$y$\textit{, for any }$u_{x}\in \pi ^{-1}\left( U\right) .$

If only the condition is verified:

\textquotedblright $g_{\alpha \beta }\circ s_{U}^{-1}\circ \left( \xi
_{U}^{-1}\times Id_{\mathbb{R}^{m}}\right) \left( x,y\right) $\textit{\
depends only on }$y$\textit{, for any }$u_{x}\in \pi ^{-1}\left( U\right) $"
\textit{respectively \textquotedblright }$g_{ab}\circ s_{U}^{-1}\circ \left(
\xi _{U}^{-1}\times Id_{\mathbb{R}^{m}}\right) \left( x,y\right) $ \textit{%
depends only on }$y$\textit{, for any }$u_{x}\in \pi ^{-1}\left( U\right) $%
\textit{", }\noindent then we will say that \emph{the (pseudo)metrical
structure }$G$\emph{\ is a (pseudo)metrical structure }$\mathcal{H}$\emph{%
-locally Minkowski }and \emph{\ }$\mathcal{V}$\emph{-locally Minkowski,
respectively.}

\textbf{Definition 5.8.4} The generalized tangent bundle $\left( (\rho ,\eta
)TE,\break (\rho ,\eta )\tau _{E},E\right) $ will be called $(\rho ,\eta )$%
\emph{-(pseudo)metrizable} if there exists a (pseudo)metrical structure%
\begin{equation*}
G=g_{\alpha \beta }d\tilde{z}^{\alpha }\otimes d\tilde{z}^{\beta
}+g_{ab}\delta \tilde{y}^{a}\otimes \delta \tilde{y}^{b}
\end{equation*}%
and a distinguished linear $\left( \rho ,\eta \right) $-connection%
\begin{equation*}
\left( \left( \rho ,\eta \right) H,\left( \rho ,\eta \right) V\right)
\end{equation*}%
such that\emph{\ }%
\begin{equation*}
\begin{array}{c}
\left( \rho ,\eta \right) D_{\tilde{X}}G=0,~\forall \tilde{X}\in \Gamma
\left( \left( \rho ,\eta \right) TE,\left( \rho ,\eta \right) \tau
_{E},E\right) .%
\end{array}%
\leqno(5.8.1)
\end{equation*}

Condition $\left( 5.8.1\right) $ is equivalent with the following
equalities:
\begin{equation*}
\begin{array}{c}
g_{\alpha \beta \mid \gamma }=0,\,g_{ab\mid \gamma }=0,\,\,g_{\alpha \beta
}\mid _{c}=0\,,\,\,g_{ab}\mid _{c}=0.%
\end{array}%
\leqno(5.8.2)
\end{equation*}

If $g_{\alpha \beta \mid \gamma }{=}0$ and $\,g_{ab\mid \gamma }{=}0$, then
we will say that \emph{the vector bundle} $((\rho ,\eta )TE,(\rho ,\eta
)\tau _{E},E) $ \emph{is }$\mathcal{H}$\emph{-}$(\rho ,\eta )$\emph{%
-(pseudo)metrizable.}

If $g_{\alpha \beta }|_{c}{=}0$ and $\,g_{ab}|_{c}{=}0$, then we will say
that \emph{the vector bundle} $((\rho ,\eta )TE,(\rho ,\eta )\tau _{E},E) $
\emph{is }$\mathcal{V}$\emph{-}$(\rho ,\eta )$\emph{-(pseudo)\-metrizable.}

\bigskip \noindent \textbf{Theorem 5.8.1} \emph{If} $\left( \left( \rho
,\eta \right) \overset{0}{H},\left( \rho ,\eta \right) \overset{0}{V}\right)
$ \emph{is a distinguished linear }$\left( \rho ,\eta \right) $\emph{%
-connection for the generalized tangent bundle }$\left( \left( \rho ,\eta
\right) TE,\left( \rho ,\eta \right) \tau _{E},E\right) $ \emph{\ and }$%
G=g_{\alpha \beta }d\tilde{z}^{\alpha }\otimes d\tilde{z}^{\beta
}+g_{ab}\delta \tilde{y}^{a}\otimes \delta \tilde{y}^{b}$ \emph{is a
(pseudo)metrical structure, then the following real local functions:}%
\begin{equation*}
\begin{array}{ll}
\left( \rho ,\eta \right) H_{\beta \gamma }^{\alpha }\!\! & =\displaystyle%
\frac{1}{2}\tilde{g}^{\alpha \varepsilon }\left( \Gamma \left( \tilde{\rho}%
,Id_{E}\right) \left( \tilde{\delta}_{\gamma }\right) g_{\varepsilon \beta
}\right. \vspace*{1mm} \\
& +\Gamma \left( \tilde{\rho},Id_{E}\right) \left( \tilde{\delta}_{\beta
}\right) g_{\varepsilon \gamma }-\Gamma \left( \tilde{\rho},Id_{E}\right)
\left( \tilde{\delta}_{\varepsilon }\right) g_{\beta \gamma }\vspace*{1mm}
\\
& \left. +g_{\theta \varepsilon }L_{\gamma \beta }^{\theta }\circ h\circ \pi
-g_{\beta \theta }L_{\gamma \varepsilon }^{\theta }\circ h\circ \pi
-g_{\theta \gamma }L_{\beta \varepsilon }^{\theta }\circ h\circ \pi \right) ,%
\vspace*{2mm} \\
\left( \rho ,\eta \right) H_{b\gamma }^{a}\!\! & =\left( \rho ,\eta \right)
\overset{0}{H}_{b\gamma }^{a}+\displaystyle\frac{1}{2}\tilde{g}^{ac}g_{bc%
\overset{0}{\mid }\gamma },\vspace*{2mm} \\
\left( \rho ,\eta \right) V_{\beta c}^{\alpha }\!\! & =\left( \rho ,\eta
\right) \overset{0}{V}_{\beta c}^{\alpha }+\displaystyle\frac{1}{2}\tilde{g}%
^{\alpha \varepsilon }g_{\beta \varepsilon \overset{0}{\mid }c},\vspace*{2mm}
\\
\left( \rho ,\eta \right) V_{bc}^{a}\!\! & =\displaystyle\frac{1}{2}\tilde{g}%
^{ae}\left( \Gamma \left( \tilde{\rho},Id_{E}\right) \left( \overset{\cdot }{%
\tilde{\partial}}_{c}\right) g_{eb}\right. \vspace*{1mm} \\
& \left. +\Gamma \left( \tilde{\rho},Id_{E}\right) \left( \overset{\cdot }{%
\tilde{\partial}}_{b}\right) g_{ec}-\Gamma \left( \tilde{\rho},Id_{E}\right)
\left( \overset{\cdot }{\tilde{\partial}}_{e}\right) g_{bc}\right)%
\end{array}%
\leqno(5.8.3)
\end{equation*}%
\emph{are components of a distinguished linear }$\left( \rho ,\eta \right) $%
\emph{-connection such that the generalized tangent bundle }$\left( \left(
\rho ,\eta \right) TE,\left( \rho ,\eta \right) \tau _{E},E\right) $ \emph{%
becomes }$\left( \rho ,\eta \right) $\emph{-(pseudo)metrizable.}

\bigskip\noindent\textbf{Corollary 5.8.1 }\emph{If the distinguished linear }%
$\left( \rho ,\eta \right) $\emph{-connection} $\left( \left( \rho ,\eta
\right) \overset{0}{H},\left( \rho ,\eta \right) \overset{0}{V}\right) $
\emph{coincides with the Berwald linear }$\left( \rho ,\eta \right) $\emph{%
-connection, then the local real functions: }
\begin{equation*}
%
\hspace*{-6mm}\leqno(5.8.4)
\end{equation*}%
\emph{are the components of a distinguished linear }$\left( \rho ,\eta
\right) $\emph{-connection such that the generalized tangent bundle }$\left(
\left( \rho ,\eta \right) TE,\left( \rho ,\eta \right) \tau _{E},E\right) $
\emph{becomes }$\left( \rho ,\eta \right) $\emph{-(pseudo)metrizable.}

\emph{Moreover, if the (pseudo)metrical structure }$G$\emph{\ is }$\mathcal{H%
}$\emph{- and }$\mathcal{V}$\emph{-Rieman\-nian, then the local real
functions:}\smallskip \noindent
\begin{equation*}
%
%
\leqno(5.8.5)
\end{equation*}%
\emph{are the components of a distinguished linear }$\left( \rho ,\eta
\right) $\emph{-connection such that the generalized tangent bundle }$\left(
\left( \rho ,\eta \right) TE,\left( \rho ,\eta \right) \tau _{E},E\right) $
\emph{becomes }$\left( \rho ,\eta \right) $\emph{-(pseudo)metrizable.}

\bigskip\noindent\textbf{Theorem 5.8.2} \emph{Let }$\left( \rho ,\eta
\right) \Gamma $\emph{\ be a }$\left( \rho ,\eta \right) $\emph{-connection
for the vector bundle }$\left( E,\pi ,M\right) .$ \emph{Let }%
\begin{equation*}
\left( \left( \rho ,\eta \right) \overset{0}{H},\left( \rho ,\eta \right)
\overset{0}{V}\right)
\end{equation*}%
\emph{be a distinguished linear }$\left( \rho ,\eta \right) $\emph{%
-connection for }$\left( \left( \rho ,\eta \right) TE,\left( \rho ,\eta
\right) \tau _{E},E\right) $\emph{\ and let }%
\begin{equation*}
\begin{array}{c}
G=g_{\alpha \beta }d\tilde{z}^{\alpha }\otimes d\tilde{z}^{\beta
}+g_{ab}\delta \tilde{y}^{a}\otimes \delta \tilde{y}^{b}%
\end{array}%
\end{equation*}%
\emph{be a (pseudo)metrical structure.}

\emph{Let }
\begin{equation*}
\begin{array}{ll}
O_{\beta \gamma }^{\alpha \varepsilon }=\displaystyle\frac{1}{2}\left(
\delta _{\beta }^{\alpha }\delta _{\gamma }^{\varepsilon }-g_{\beta \gamma }%
\tilde{g}^{\alpha \varepsilon }\right) , & O_{\beta \gamma }^{\ast \alpha
\varepsilon }=\displaystyle\frac{1}{2}\left( \delta _{\beta }^{\alpha
}\delta _{\gamma }^{\varepsilon }+g_{\beta \gamma }\tilde{g}^{\alpha
\varepsilon }\right) ,\vspace*{2mm} \\
O_{bc}^{ae}=\displaystyle\frac{1}{2}\left( \delta _{b}^{a}\delta
_{c}^{e}-g_{bc}\tilde{g}^{ae}\right) , & O_{bc}^{\ast ae}=\displaystyle\frac{%
1}{2}\left( \delta _{b}^{a}\delta _{c}^{e}+g_{bc}\tilde{g}^{ae}\right) ,%
\end{array}%
\leqno(5.8.6)
\end{equation*}%
\emph{be the Obata operators}.

\emph{If the real local functions }$X_{\beta \gamma }^{\alpha },X_{\beta
c}^{\alpha },Y_{b\gamma }^{a},Y_{bc}^{a}$ \emph{are components of tensor
fields,} \emph{then the local real functions are given in the following: }%
\vspace*{-3mm}
\begin{equation*}
\begin{array}{ll}
\left( \rho ,\eta \right) H_{\beta \gamma }^{\alpha }=\left( \rho ,\eta
\right) \overset{c}{H_{\beta \gamma }^{\alpha }}+O_{\gamma \eta }^{\alpha
\varepsilon }X_{\varepsilon \beta }^{\eta },\vspace*{1mm} &  \\
\left( \rho ,\eta \right) H_{b\gamma }^{a}=\left( \rho ,\eta \right) \overset%
{c}{H_{b\gamma }^{a}}+O_{bd}^{ae}Y_{e\gamma }^{d},\vspace*{1mm} &  \\
\left( \rho ,\eta \right) V_{\beta c}^{\alpha }=\left( \rho ,\eta \right)
\overset{c}{V_{\beta c}^{\alpha }}+O_{\beta \eta }^{\ast \alpha \varepsilon
}X_{\varepsilon c}^{\eta },\vspace*{1mm} &  \\
\left( \rho ,\eta \right) V_{bc}^{a}=\left( \rho ,\eta \right) \overset{c}{%
V_{bc}^{a}}+O_{bd}^{\ast ae}Y_{ec}^{d}, &
\end{array}%
\leqno(5.8.7)
\end{equation*}%
\emph{are the components of a distinguished linear }$\left( \rho ,\eta
\right) $\emph{-connection such that the generalized tangent bundle }$\left(
\left( \rho ,\eta \right) TE,\left( \rho ,\eta \right) \tau _{E},E\right) $
\emph{becomes }$\left( \rho ,\eta \right) $\emph{-(pseudo)metrizable.}

\bigskip\noindent\textbf{Theorem 5.8.3 }\emph{Let }$\left( \rho ,\eta
\right) \Gamma $\emph{\ be a }$\left( \rho ,\eta \right) $\emph{-connection
for the vector bundle }$\left( E,\pi ,M\right) .$

\emph{If }%
\begin{equation*}
\left( \left( \rho ,\eta \right) \overset{0}{H},\left( \rho ,\eta \right)
\overset{0}{V}\right)
\end{equation*}%
\emph{is a distinguished linear }$\left( \rho ,\eta \right) $\emph{%
-connection for the generalized tangent bundle }%
\begin{equation*}
\left( \left( \rho ,\eta \right) TE,\left( \rho ,\eta \right) \tau
_{E},E\right)
\end{equation*}%
\emph{\ and }%
\begin{equation*}
G=g_{\alpha \beta }d\tilde{z}^{\alpha }\otimes d\tilde{z}^{\beta
}+g_{ab}\delta \tilde{y}^{a}\otimes \delta \tilde{y}^{b}
\end{equation*}%
\emph{is a (pseudo)metrical structure, then the real local functions: }%
\begin{equation*}
\begin{array}{l}
\left( \rho ,\eta \right) H_{\beta \gamma }^{\alpha }=\left( \rho ,\eta
\right) \overset{0}{H_{\beta \gamma }^{\alpha }}+\displaystyle\frac{1}{2}%
\tilde{g}^{\alpha \varepsilon }g_{\varepsilon \beta \overset{0}{\mid }\gamma
},\vspace*{2mm} \\
\left( \rho ,\eta \right) H_{b\gamma }^{a}=\left( \rho ,\eta \right) \overset%
{0}{H_{b\gamma }^{a}}+\displaystyle\frac{1}{2}\tilde{g}^{ae}g_{eb\overset{0}{%
\mid }\gamma },\vspace*{2mm} \\
\left( \rho ,\eta \right) V_{\beta c}^{\alpha }=\left( \rho ,\eta \right)
\overset{0}{V_{\beta c}^{\alpha }}+\displaystyle\frac{1}{2}\tilde{g}^{\alpha
\varepsilon }g_{\varepsilon \beta }\overset{0}{\mid }_{c},\vspace*{2mm} \\
\left( \rho ,\eta \right) V_{bc}^{a}=\left( \rho ,\eta \right) \overset{0}{%
V_{bc}^{a}}+\displaystyle\frac{1}{2}\tilde{g}^{ae}g_{eb}\overset{0}{\mid }%
_{c}%
\end{array}%
\leqno(5.8.8)
\end{equation*}%
\emph{are the components of a distinguished linear }$\left( \rho ,\eta
\right) $\emph{-connection such that the generalized tangent bundle }$\left(
\left( \rho ,\eta \right) TE,\left( \rho ,\eta \right) \tau _{E},E\right) $
\emph{becomes }$\left( \rho ,\eta \right) $\emph{-(pseudo)metrizable.}

\subsection{Generalized Lagrange $\left( \protect\rho ,\protect\eta \right) $%
-spaces, Lagrange $\left( \protect\rho ,\protect\eta \right) $-spaces\newline
and Finsler $\left( \protect\rho ,\protect\eta \right) $-spaces}

We consider the following diagram:
\begin{equation*}
\begin{array}{c}
\xymatrix{E\ar[d]_\pi&(F,[,]_{F,h},(\rho,\eta))\ar[d]^\nu\\ M\ar[r]^h&N}%
\end{array}%
\end{equation*}%
such that $\left( E,\pi ,M\right) =\left( F,\nu ,N\right) $ and the
generalized tangent bundle
\begin{equation*}
\left( \left( \rho ,\eta \right) TE,\left( \rho ,\eta \right) \tau
_{E},E\right)
\end{equation*}%
is $\left( \rho ,\eta \right) $-(pseudo)metrizable.

\bigskip\noindent\textbf{Definition 5.9.1 }A smooth \emph{Lagrange
fundamental function} on the vector bundle\break $\left( E,\pi ,M\right) $
is a mapping
\begin{equation*}
E~\ ^{\underrightarrow{\ \ L\ \ }}~\ \mathbb{R}
\end{equation*}%
which satisfies the following conditions:\medskip

1. $L\circ u\in C^{\infty }\left( M\right) $, for any $u\in \Gamma \left(
E,\pi ,M\right) \setminus \left\{ 0\right\} $;\smallskip

2. $L\circ 0\in C^{0}\left( M\right) $, where $0$ means the null section of $%
\left( E,\pi ,M\right) .$\medskip

Let $L$ be a Lagrangian defined on the total space of the vector bundle $%
\left( E,\pi ,M\right) .$

If $\left( U,s_{U}\right) $ is a local vector $\left( m+r\right) $-chart for
$\left( E,\pi ,M\right) $, then we obtain the following real functions
defined on $\pi ^{-1}\left( U\right) $:%
\begin{equation*}
\begin{array}{cc}
L_{i}\overset{put}{=}\displaystyle\frac{\partial L}{\partial x^{i}}\overset{%
put}{=}\frac{\partial }{\partial x^{i}}\left( L\right) & L_{ib}\overset{put}{%
=}\displaystyle\frac{\partial ^{2}L}{\partial x^{i}\partial y^{b}}\vspace*{%
2mm}\overset{put}{=}\frac{\partial }{\partial x^{i}}\left( \frac{\partial }{%
\partial y^{b}}\left( L\right) \right) \\
L_{a}\overset{put}{=}\displaystyle\frac{\partial L}{\partial y^{a}}\overset{%
put}{=}\frac{\partial }{\partial y^{a}}\left( L\right) & L_{ab}\overset{put}{%
=}\displaystyle\frac{\partial ^{2}L}{\partial y^{a}\partial y^{b}}\overset{%
put}{=}\frac{\partial }{\partial y^{a}}\left( \frac{\partial }{\partial y^{b}%
}\left( L\right) \right) .%
\end{array}%
\leqno(5.9.1)
\end{equation*}

\smallskip \noindent \textbf{Definition 5.9.2 }If for any local vector $m+r$%
-chart $\left( U,s_{U}\right) $ of $\left( E,\pi ,M\right) ,$ we have:
\begin{equation*}
\begin{array}{c}
rank\left\Vert L_{ab}\left( u_{x}\right) \right\Vert =r,%
\end{array}%
\leqno(5.9.2)
\end{equation*}%
for any $u_{x}\in \pi ^{-1}\left( U\right) \backslash \left\{ 0_{x}\right\} $%
, then we will say that \emph{the Lagrangian }$L$\emph{\ is regular.}

\bigskip \noindent \textbf{Proposition 5.9.1} \emph{If the Lagrangian }$L$%
\emph{\ is regular, then for any local vector }$m+r$\emph{-chart }$\left(
U,s_{U}\right) $\emph{\ of }$\left( E,\pi ,M\right) ,$\emph{\ we obtain the
real functions }$\tilde{L}^{ab}$\emph{\ locally defined by}%
\begin{equation*}
\begin{array}{ccc}
\pi ^{-1}\left( U\right) & ^{\underrightarrow{\ \ \tilde{L}^{ab}\ \ }} &
\mathbb{R} \\
u_{x} & \longmapsto & \tilde{L}^{ab}\left( u_{x}\right)%
\end{array}%
,\leqno(5.9.3)
\end{equation*}%
\emph{where }$\left\Vert \tilde{L}^{ab}\left( u_{x}\right) \right\Vert
=\left\Vert L_{ab}\left( u_{x}\right) \right\Vert ^{-1}$\emph{, for any }$%
u_{x}\in \pi ^{-1}\left( U\right) \backslash \left\{ 0_{x}\right\} .$

\bigskip\noindent\textbf{Definition 5.9.3 }A smooth \emph{Finsler
fundamental function} on the vector bundle $\left( E,\pi ,M\right) $ is a
mapping
\begin{equation*}
E~\ ^{\underrightarrow{\ \ F\ \ }}~\ \mathbb{R}_{+}
\end{equation*}%
which satisfies the following conditions:\medskip

1. $F\circ u\in C^{\infty }\left( M\right) $, for any $u\in \Gamma \left(
E,\pi ,M\right) \setminus \left\{ 0\right\} $;\smallskip

2. $F\circ 0\in C^{0}\left( M\right) $, where $0$ means the null section of $%
\left( E,\pi ,M\right) $;\smallskip

3. $F$ is positively $1$-homogenous on the fibres of vector bundle $\left(
E,\pi ,M\right) ;$\smallskip

4. For any local vector $m+r$-chart $\left( U,s_{U}\right) $ of $\left(
E,\pi ,M\right) ,$ the hessian:%
\begin{equation*}
\left\Vert F_{~ab}^{2}\left( u_{x}\right) \right\Vert \leqno(5.9.4)
\end{equation*}%
is positively define for any $u_{x}\in \pi ^{-1}\left( U\right) \backslash
\left\{ 0_{x}\right\} $.

\bigskip\noindent \textbf{Definition 5.9.4 }If the (pseudo)metrical
structure $G$\ is determined by a (pseudo)metrical structure
\begin{equation*}
\begin{array}{c}
g\in \mathcal{T}~_{2}^{0}\left( V\left( \rho ,\eta \right) TE,\left( \rho
,\eta \right) ,{\tau _{E}},E\right) ,%
\end{array}%
\end{equation*}%
then the $\left( \rho ,\eta \right) $-(pseudo)metrizable vector bundle
\begin{equation*}
\begin{array}{c}
\left( \left( \rho ,\eta \right) TE,\left( \rho ,\eta \right) \tau
_{E},E\right)%
\end{array}%
\end{equation*}%
will be called the \emph{generalized Lagrange }$\left( \rho ,\eta \right) $%
\emph{-space.}

In particular, if the (pseudo)metrical structure $g$\ is determined with the
help of a Lagrange fundamental function\ or Finsler fundamental function,
then the $\left( \rho ,\eta \right) $-(pseudo)metrizable vector bundle
\begin{equation*}
\begin{array}{c}
\left( \left( \rho ,\eta \right) TE,\left( \rho ,\eta \right) \tau
_{E},E\right)%
\end{array}%
\end{equation*}
will be called the \emph{Lagrange }$\left( \rho ,\eta \right) $\emph{-space }%
or the \emph{Finsler }$\left( \rho ,\eta \right) $\emph{-space, respectively.%
}

The generalized Lagrange $\left( Id_{TM},Id_{M}\right) $-space, the Lagrange
$\left( Id_{TM},Id_{M}\right) $-space, and the Finsler $\left(
Id_{TM},Id_{M}\right) $-space will be called the \emph{generalized Lagrange
space, Lagrange space}, \emph{Finsler space}.

\bigskip \noindent \textbf{Definition 5.9.5 }The normal distinguished linear
$\left( \rho ,\eta \right) $-connections of a Lagrange or Finsler $\left(
\rho ,\eta \right) $-space will be called \emph{Lagrange }or\emph{\ Finsler
linear }$\left( \rho ,\eta \right) $\emph{-connections.}

The Lagrange and Finsler linear $\left( Id_{TM},Id_{M}\right) $-connections
will be called \emph{Lagrange }and\emph{\ Finsler linear connections},
respectively.

\bigskip \noindent \textbf{Theorem 5.9.1 }\emph{If the (pseudo)metrical
structure }$G$\emph{\ is determined by a (pseudo)metrical structure }%
\begin{equation*}
\begin{array}{c}
g\in \mathcal{T}~_{2}^{0}\left( V\left( \rho ,\eta \right) TE,\left( \rho
,\eta \right) ,{\tau _{E}},E\right) ,%
\end{array}%
\end{equation*}%
\emph{then, the real local functions:}
\begin{equation*}
\begin{array}{ll}
\left( \rho ,\eta \right) H_{bc}^{a}\!\!\! & =\displaystyle\frac{1}{2}\tilde{%
g}^{ae}\left( \Gamma \left( \tilde{\rho},Id_{E}\right) \left( \delta
_{b}\right) g_{ec}+\Gamma \left( \tilde{\rho},Id_{E}\right) \left( \delta
_{c}\right) g_{be}\right. -\Gamma \left( \tilde{\rho},Id_{E}\right) \left(
\delta _{e}\right) g_{bc}\vspace*{1mm} \\
& -\,g_{cd}L_{be}^{d}{\circ }\left( h{\circ }\pi \right) \left.
+g_{bd}L_{ec}^{d}{\circ }\left( h{\circ }\pi \right) -g_{ed}L_{bc}^{d}{\circ
}\left( h{\circ }\pi \right) \right) ,\vspace*{2mm} \\
\left( \rho ,\eta \right) V_{bc}^{a}\!\!\! & =\displaystyle\frac{1}{2}\tilde{%
g}^{ae}\left( \Gamma \left( \tilde{\rho},Id_{E}\right) \left( \overset{\cdot
}{\tilde{\partial}}_{c}\right) g_{eb}\right. \\
& \left. +\Gamma \left( \tilde{\rho},Id_{E}\right) \left( \overset{\cdot }{%
\tilde{\partial}}_{b}\right) g_{ec}-\Gamma \left( \tilde{\rho},Id_{E}\right)
\left( \overset{\cdot }{\tilde{\partial}}_{e}\right) g_{bc}\right)%
\end{array}%
\leqno(5.9.5)
\end{equation*}%
\medskip

\noindent \emph{are the components of a normal distinguished linear }$\left(
\rho ,\eta \right) $\emph{-connection with }$\left( \rho ,\eta \right) $%
\emph{-}$\mathcal{H}\left( \mathcal{HH}\right) $\emph{\ and }$\left( \rho
,\eta \right) $\emph{-}$\mathcal{V}\left( \mathcal{VV}\right) $\emph{\
torsions free such that the generalized tangent bundle}\break $\left( \left(
\rho ,\eta \right) TE,\left( \rho ,\eta \right) \tau _{E},E\right) $\emph{\
derives generalized Lagrange }$\left( \rho ,\eta \right) $\emph{-space.}%
\medskip

This normal distinguished linear $( \rho ,\eta ) $-connection will be called
\emph{generalized linear} $( \rho ,\eta ) $\emph{-connection of Levi-Civita
type.}

If the \textit{(pseudo)metrical structure }$g$ is determined with the help
of a Lagrange and Finsler fundamental function, then the Lagrange and
Finsler linear $\left( \rho ,\eta \right) $-connections will be called \emph{%
canonical Lagrange }and \emph{\ Finsler linear }$\left( \rho ,\eta \right) $%
\emph{-connection, respectively.}

The canonical Lagrange and Finsler linear $\left( Id_{TM},Id_{M}\right) $%
-con\-nec\-tion will be called\emph{\ }the \emph{canonical Lagrange }and%
\emph{\ Finsler linear connection }respectively.

\bigskip\noindent\textbf{Theorem 5.9.2 }\emph{Let\ }$\left( \left( \rho
,\eta \right) H,\left( \rho ,\eta \right) V\right) $\emph{\ be the normal
dis\-tin\-guished linear }$\left( \rho ,\eta \right) $\emph{-connec\-tion
presented in the previous theorem.}

\emph{If }%
\begin{equation*}
\mathbb{T}_{bc}^{a}\tilde{\delta}_{a}\otimes d\tilde{z}^{b}\otimes d\tilde{z}%
^{c}\in \mathcal{T}_{20}^{10}\left( \left( \rho ,\eta \right) TE,\left( \rho
,\eta \right) \tau _{E},E\right)
\end{equation*}%
\emph{and }%
\begin{equation*}
\mathbb{S}_{bc}^{a}\overset{\cdot }{\tilde{\partial}}_{a}\otimes \delta
\tilde{y}^{b}\otimes \delta \tilde{y}^{c}\in \mathcal{T}_{02}^{01}\left(
\left( \rho ,\eta \right) TE,\left( \rho ,\eta \right) \tau _{E},E\right)
\end{equation*}%
\emph{such that they satisfy the conditions:}%
\begin{equation*}
\mathbb{T}_{bc}^{a}=-\mathbb{T}_{cb}^{a},~\mathbb{S}_{bc}^{a}=-\mathbb{S}%
_{cb}^{a},~\forall b,c\in \overline{1,n},
\end{equation*}%
\emph{then the following real local functions:\ }%
\begin{equation*}
\begin{array}{l}
\left( \rho ,\eta \right) \tilde{H}_{bc}^{a}=\left( \rho ,\eta \right)
H_{bc}^{a}+\displaystyle\frac{1}{2}\tilde{g}^{ae}\left( g_{ed}\mathbb{T}%
_{bc}^{d}-g_{bd}\mathbb{T}_{ec}^{d}+g_{cd}\mathbb{T}_{be}^{d}\right) ,%
\vspace*{2mm} \\
\left( \rho ,\eta \right) \tilde{V}_{bc}^{a}=\left( \rho ,\eta \right)
V_{bc}^{a}+\displaystyle\frac{1}{2}\tilde{g}^{ae}\left( g_{ed}\mathbb{S}%
_{bc}^{d}-g_{bd}\mathbb{S}_{ec}^{d}+g_{cd}\mathbb{S}_{be}^{d}\right)%
\end{array}%
\leqno(5.9.6)
\end{equation*}%
\emph{are the components of a normal distinguished linear }$\left( \rho
,\eta \right) $\emph{-connection with }$\left( \rho ,\eta \right) $\emph{-}$%
\mathcal{H}\left( \mathcal{HH}\right) $\emph{\ and }$\left( \rho ,\eta
\right) $\emph{-}$\mathcal{V}\left( \mathcal{VV}\right) $\emph{\ torsions a
priori given such that the generalized tangent bundle}\break $\left( \left(
\rho ,\eta \right) TE,\left( \rho ,\eta \right) \tau _{E},E\right) $\emph{\
derives generalized Lagrange }$\left( \rho ,\eta \right) $\emph{-space.}

\emph{Moreover, we obtain: }%
\begin{equation*}
\begin{array}{l}
\mathbb{T}_{bc}^{a}=\left( \rho ,\eta \right) H_{bc}^{a}-\left( \rho ,\eta
\right) H_{cb}^{a}-L_{bc}^{a}\circ h\circ \pi ,\vspace*{2mm} \\
\mathbb{S}_{bc}^{a}=\left( \rho ,\eta \right) V_{bc}^{a}-\left( \rho ,\eta
\right) V_{cb}^{a}.%
\end{array}%
\leqno(5.9.87
\end{equation*}

\subsection{Einstein equations}

We shall consider a metric structure
\begin{equation*}
\begin{array}{c}
G=g_{\alpha \beta }d\tilde{z}^{\alpha }\otimes d\tilde{z}^{\beta
}+g_{ab}\delta \tilde{y}^{a}\otimes \delta \tilde{y}^{b}%
\end{array}%
\end{equation*}
and a distinguished linear $\left( \rho ,\eta \right) $-connection $\left(
\left( \rho ,\eta \right) H,\left( \rho ,\eta \right) V\right) $ compatible
with the structure metric $G$ having $\mathcal{H}\left( \mathcal{HH}\right) $
and $\mathcal{V}\left( \mathcal{VV}\right) $-torsions prescribed.

\bigskip\noindent\textbf{Definition 5.10.1 }If $\left( \rho ,\eta ,h\right)
\mathbb{R}_{~\alpha ~\beta }$ and $\left( \rho ,\eta ,h\right) \mathbb{S}%
_{~a~b}$ are the components of tensor Ricci associated to distinguished
linear $\left( \rho ,\eta \right) $-connection
\begin{equation*}
\begin{array}{c}
\left( \left( \rho ,\eta \right) H,\left( \rho ,\eta \right) V\right) ,%
\end{array}%
\end{equation*}%
then the scalar
\begin{equation*}
\begin{array}{c}
\left( \rho ,\eta ,h\right) \mathbb{R=}\left( \rho ,\eta ,h\right) \mathbb{R}%
_{~\alpha ~\beta }\tilde{g}^{\alpha \beta }+\left( \rho ,\eta ,h\right)
\mathbb{S}_{~a~b}\tilde{g}^{ab}%
\end{array}%
\leqno(5.10.1)
\end{equation*}%
will be called the \emph{scalar of curvature of distinguished linear }$(
\rho ,\eta ) $\emph{-connection}\break $( ( \rho ,\eta ) H, ( \rho ,\eta ) V
) .$

\bigskip \noindent \textbf{Definition 5.10.2 }The tensor field\textbf{\ }%
\begin{equation*}
\begin{array}{cc}
\left( \rho ,\eta ,h\right) \mathbb{T} & =\left( \rho ,\eta ,h\right)
\mathbb{T}_{~\alpha ~\beta }d\tilde{z}^{\alpha }\otimes d\tilde{z}^{\beta
}+\left( \rho ,\eta ,h\right) \mathbb{T}_{~\alpha ~b}d\tilde{z}^{\alpha
}\otimes \delta \tilde{y}^{b}\vspace*{2mm} \\
& +\left( \rho ,\eta ,h\right) \mathbb{T}_{~a~\beta }\delta \tilde{y}%
^{a}\otimes d\tilde{z}^{\alpha }+\left( \rho ,\eta ,h\right) \mathbb{T}%
_{~a~b}\delta \tilde{y}^{a}\otimes \delta \tilde{y}^{b}%
\end{array}%
\leqno(5.10.2)
\end{equation*}%
such that its components verify the following conditions:
\begin{equation*}
\begin{array}{rl}
\varkappa \left( \rho ,\eta ,h\right) \mathbb{T}_{~\alpha ~\beta } & =\left(
\rho ,\eta ,h\right) \mathbb{R}_{~\alpha ~\beta }-\displaystyle\frac{1}{2}%
\left( \rho ,\eta ,h\right) \mathbb{R\cdot }g_{\alpha \beta },\vspace*{1mm}
\\
-\varkappa \left( \rho ,\eta ,h\right) \mathbb{T}_{~\alpha ~b} & =\left(
\rho ,\eta ,h\right) \mathbb{P}_{~\alpha ~b},\vspace*{2mm} \\
\varkappa \left( \rho ,\eta ,h\right) \mathbb{T}_{~a~\beta } & =\left( \rho
,\eta ,h\right) \mathbb{P}_{~a~\beta },\vspace*{1mm} \\
\varkappa \left( \rho ,\eta ,h\right) \mathbb{T}_{~a~b} & =\left( \rho ,\eta
,h\right) \mathbb{S}_{~a~b}-\displaystyle\frac{1}{2}\left( \rho ,\eta
,h\right) \mathbb{R\cdot }g_{ab},%
\end{array}%
\leqno(5.10.3)
\end{equation*}%
where $\varkappa $ is a constant, will be called \emph{the energy-momentum
tensor field associated to distinguished linear }$\left( \rho ,\eta \right) $%
\emph{-connection }$\left( \left( \rho ,\eta \right) H,\left( \rho ,\eta
\right) V\right) $\emph{\ and metrical structure }$G.$

The equations $\left( 5.10.3\right) $ will be called \emph{the Einstein
equations associated to distinguished linear }$\left( \rho ,\eta \right) $%
\emph{-connection }$\left( \left( \rho ,\eta \right) H,\left( \rho ,\eta
\right) V\right) $\emph{\ and metrical structure }$G$\emph{.}

Formally, the Einstein equations will be written%
\begin{equation*}
\mathbf{Ric}\left( \left( \rho ,\eta \right) H,\left( \rho ,\eta \right)
V\right) -\frac{1}{2}\left( \rho ,\eta ,h\right) \mathbb{R\cdot }G=\varkappa
\cdot \left( \rho ,\eta ,h\right) \mathbb{T}. \leqno(5.10.3^{\prime })
\end{equation*}

\subsection{Mechanical systems}

Using the diagram:
\begin{equation*}
\begin{array}{c}
\xymatrix{E\ar[d]_\pi&\left( E,\left[ ,\right] _{E,h},\left( \rho ,\eta
\right) \right)\ar[d]^\pi \\ M\ar[r]^h&M}%
\end{array}
\leqno(5.11.1)
\end{equation*}%
where $\left( \left( E,\pi ,M\right) ,\left[ ,\right] _{E,h},\left( \rho
,\eta \right) \right) $ is a generalized Lie algebroid, we build the
generalized tangent bundle
\begin{equation*}
\begin{array}{c}
(((\rho ,\eta )TE,(\rho ,\eta )\tau _{E},E),[,]_{(\rho ,\eta )TE},(\tilde{%
\rho},Id_{E})).%
\end{array}%
\end{equation*}

\smallskip \noindent \textbf{Definition 5.11.1 }A triple
\begin{equation*}
\begin{array}{c}
\left( \left( E,\pi ,M\right) ,F_{e},\left( \rho ,\eta \right) \Gamma
\right) ,%
\end{array}%
\leqno(5.11.2)
\end{equation*}%
where
\begin{equation*}
F_{e}=F^{a}\frac{\partial }{\partial \tilde{y}^{a}}\in \Gamma \left( V\left(
\rho ,\eta \right) TE,\left( \rho ,\eta \right) \tau _{E},E\right) \leqno%
(5.11.3)
\end{equation*}%
is an external force and $\left( \rho ,\eta \right) \Gamma $ is a $\left(
\rho ,\eta \right) $-connection, will be called the \emph{mechanical }$%
\left( \rho ,\eta \right) $\emph{-system.}

A mechanical $\left( \rho ,\eta \right) $-system
\begin{equation*}
\begin{array}{c}
\left( \left( E,\pi ,M\right) ,F_{e},\left( \rho ,\eta \right) \Gamma \right)%
\end{array}%
\end{equation*}%
endowed with a (pseudo)metrical structure $G$ determined with the help of a
(pseudo)metrical structure
\begin{equation*}
\begin{array}{c}
g=g_{ab}d\tilde{y}^{a}\otimes d\tilde{y}^{b}\in \mathcal{T~}_{02}^{00}\left(
\left( \rho ,\eta \right) TE,\left( \rho ,\eta \right) \tau _{E},E\right)%
\end{array}%
\vspace*{1mm}
\end{equation*}%
will be denoted
\begin{equation*}
\begin{array}{c}
\left( \left( E,\pi ,M\right) ,F_{e},\left( \rho ,\eta \right) \Gamma
,G\right) .%
\end{array}%
\leqno(5.11.4)
\end{equation*}%
and will be called \emph{generalized Lagrange mechanical }$\left( \rho ,\eta
\right) $\emph{-system.}

Any mechanical $\left( Id_{TM},Id_{M}\right) $-system and any generalized
Lagrange mechanical\break $\left( Id_{TM},Id_{M}\right) $-system will be
called \emph{mechanical system} and \emph{generalized Lagrange mechanical
system, }respectively.

\bigskip \noindent \textbf{Definition 5.11.2 }If $L$ (respectively $F$) is a
smooth Lagrange (respectively Finsler function), then we put the triples
\begin{equation*}
\left( \left( E,\pi ,M\right) ,F_{e},L\right)
\mbox{\ \ \ (
respectively
$
\left( E,F_{e},F\right))
$}
\end{equation*}%
where $F_{e}=F^{a}\displaystyle\frac{\partial }{\partial \tilde{y}^{a}}\in
\Gamma \left( V\left( \rho ,\eta \right) TE,\left( \rho ,\eta \right) \tau
_{E},E\right) $ is an external force. These are called \emph{Lagrange
mechanical }$\left( \rho ,\eta \right) $\emph{-system }(\emph{Finsler
mechanical }$\left( \rho ,\eta \right) $\emph{-system, respectively}).

Any Lagrange mechanical $\left( Id_{TM},Id_{M}\right) $-system and any
Finsler mechanical\break $\left( Id_{TM},Id_{M}\right) $-system will be
called \emph{Lagrange mechanical system }and \emph{Finsler mechanical system}%
, respectively.

\subsubsection{\noindent $(\protect\rho ,\protect\eta )$-semisprays and $(%
\protect\rho ,\protect\eta )$-sprays for mechanical $(\protect\rho ,\protect%
\eta )$-systems}

Let $(\left( E,\pi ,M\right) ,F_{e},(\rho ,\eta )\Gamma )$ be an arbitrary
mechanical $\left( \rho ,\eta \right) $-system.

\bigskip\noindent\textbf{Definition 5.11.1.1 }The\textit{\ }vertical section%
\textit{\ }%
\begin{equation*}
\begin{array}{c}
\mathbb{C}\mathbf{=}y^{a}\overset{\cdot }{\tilde{\partial}}_{a},%
\end{array}%
\leqno(5.11.1.1)
\end{equation*}%
will be called the\textit{\ }\emph{Liouville section.}

\newpage\noindent \textbf{Definition 5.11.1.2 }The section $S\in \Gamma
\left( \left( \rho ,\eta \right) TE,\left( \rho ,\eta \right) \tau
_{E},E\right) $\ will be called $\left( \rho ,\eta \right) $\emph{-semispray}%
\ if there exists an almost tangent structure $e$ such that
\begin{equation*}
\begin{array}{c}
e\left( S\right) =\mathbb{C}.%
\end{array}%
\leqno(5.11.1.2)
\end{equation*}

Let $g\in \mathbf{Man}\left( E,E\right) $ be such that $\left( g,h\right) $
is a locally invertible $\mathbf{B}^{\mathbf{v}}$-morphism of $\left( E,\pi
,M\right) $\ source and $\left( E,\pi ,M\right) $\ target.

\bigskip\noindent\textbf{Theorem 5.11.1.1 }\emph{The section }%
\begin{equation*}
S=\left( g_{b}^{a}\circ h\circ \pi \right) y^{b}\frac{\partial }{\partial
\tilde{z}^{a}}-2\left( G^{a}-\frac{1}{4}F^{a}\right) \frac{\partial }{%
\partial \tilde{y}^{a}} \leqno(5.11.1.3)
\end{equation*}%
\emph{is a }$\left( \rho ,\eta \right) $\emph{-semispray such that the real
local functions }$G^{a},\ a\in \overline{1,n},$\emph{\ satisfy the following
conditions }%
\begin{equation*}
\begin{array}{cl}
\left( \rho ,\eta \right) \Gamma _{c}^{a} & =\tilde{g}_{c}^{e}\circ h\circ
\pi \displaystyle\frac{\partial \left( G^{a}-\frac{1}{4}F^{a}\right) }{%
\partial y^{e}} \vspace*{1,2mm} \\
& -\displaystyle\frac{1}{2}\left( g_{e}^{d}\circ h\circ \pi \cdot
y^{e}\right) L_{dc}^{b}\circ h\circ \pi \cdot \tilde{g}_{b}^{a}\circ h\circ
\pi ,~a,b\in \overline{1,r}.%
\end{array}%
\leqno(5.11.1.4)
\end{equation*}
\emph{In addition, we remark that the local real functions }%
\begin{equation*}
\begin{array}{cl}
\left( \rho ,\eta \right) \mathring{\Gamma}_{c}^{a} & \overset{put}{=}\tilde{%
g}_{c}^{e}{\circ h\circ }\pi \frac{\partial G^{a}}{\partial y^{e}} \\
& -\frac{1}{2}\left( g_{e}^{d}\circ h\circ \pi \cdot y^{e}\right)
L_{dc}^{b}\circ h\circ \pi \cdot \tilde{g}_{b}^{a}\circ h\circ \pi ,~a,b{\in
}\overline{1,r}%
\end{array}%
\leqno(5.11.1.5)
\end{equation*}%
\emph{are the components of a }$\left( \rho ,\eta \right) $\emph{-connection
}$\left( \rho ,\eta \right) \mathring{\Gamma}$\emph{\ for the vector bundle }%
$\left( E,\pi ,M\right) .$

The $\left( \rho ,\eta \right) $-semispray $S$\ will be called \emph{the\
canonical }$\left( \rho ,\eta \right) $\emph{-semispray associated to
mechanical }$\left( \rho ,\eta \right) $\emph{-system }$\left( \left( E,\pi
,M\right) ,F_{e},\left( \rho ,\eta \right) \Gamma \right) $\emph{\ and from
locally invertible }$\mathbf{B}^{\mathbf{v}}$\emph{-morphism }$\left(
g,h\right) .$

\bigskip\noindent\textit{Proof.} We consider the $\mathbf{Mod}$-endomorphism%
\begin{equation*}
%
\end{equation*}

Let $X=\tilde{Z}^{a}\tilde{\partial}_{a}+Y^{a}\overset{\cdot }{\tilde{%
\partial}}_{a}$ be an arbitrary section. Since
\begin{equation*}
%
%
\end{equation*}%
it results that
\begin{equation*}
%
%
\end{equation*}%
it results that%
\begin{equation*}
%
%
\leqno\left( P_{2}\right)
\end{equation*}
We remark that
\begin{equation*}
%
%
\end{equation*}
After some calculations, it results that $\mathbb{P}$ is an almost product
structure.

Using the equality
\begin{equation*}
\mathbb{P}=Id-2\left( \rho ,\eta \right) \Gamma ,
\end{equation*}%
we obtain that%
\begin{equation*}
%
%
\end{equation*}%
it results that relations $\left( 5.11.1.4\right) $ are satisfied. In
addition, since
\begin{equation*}
\left( \rho ,\eta \right) \mathring{\Gamma}_{c}^{a}=\left( \rho ,\eta
\right) \Gamma _{c}^{a}+\frac{1}{4}\tilde{g}_{c}^{d}\circ h\circ \pi \frac{%
\partial F^{a}}{\partial y^{d}}
\end{equation*}
and
\begin{equation*}
%
%
\end{equation*}%
it results the conclusion of the theorem. \hfill \emph{q.e.d.}

\bigskip\noindent\textbf{Remarks}

\begin{itemize}
\item[1.] If $\left( \rho ,\eta \right) =\left( Id_{TM},Id_{M}\right) $, $%
\left( g,h\right) =\left( Id_{E},Id_{M}\right) $,\ and $F_{e}\neq 0$, then
we obtain\ the canonical semispray associated to connection $\Gamma $ which
is not the same canonical semispray presented by I. Bucataru and R.~Miron
in~[7].

\item[2.] If $\left( \rho ,\eta \right) =\left( Id_{TM},Id_{M}\right) $, $%
\left( g,h\right) =\left( Id_{E},Id_{M}\right) $, and $F_{e}=0$, then we
obtain the canonical semispray associated to connection $\Gamma $ which is
not the classical canonical semispray associated to connection $\Gamma $.
\end{itemize}

Using \emph{Theorem 5.11.1.1}, we obtain the following:

\bigskip\noindent\medskip \textbf{Theorem 5.11.1.2 }\emph{The following
properties hold good:}\medskip

$1^{\circ }$\ \emph{Since } $\overset{\circ }{\tilde{\delta}}_{c}=\tilde{%
\partial}_{c}-\left( \rho ,\eta \right) \mathring{\Gamma}_{c}^{a}\overset{%
\cdot }{\tilde{\partial}}_{a},~c\in \overline{1,r}, $ \emph{it results that }%
\begin{equation*}
\overset{\circ }{\tilde{\delta}}_{c}=\tilde{\delta}_{c}-\frac{1}{4}\tilde{g}%
_{c}^{b}\circ h\circ \pi \cdot \frac{\partial F^{a}}{\partial y^{b}}\overset{%
\cdot }{\tilde{\partial}}_{a},~c\in \overline{1,r}.\leqno(5.11.1.6)
\end{equation*}

$2^{\circ }\ $\emph{Since} $\mathring{\delta}\tilde{y}^{a}=\left( \rho ,\eta
\right) \mathring{\Gamma}_{c}^{a}d\tilde{z}^{c}+d\tilde{y}^{a}, $ \emph{it
results that \ \ }%
\begin{equation*}
\mathring{\delta}\tilde{y}^{a}=\delta \tilde{y}^{a}+\frac{1}{4}\tilde{g}%
_{c}^{b}\circ h\circ \pi \frac{\partial F^{a}}{\partial y^{b}}d\tilde{z}%
^{c},~a\in \overline{1,r}.\leqno(5.11.1.7)
\end{equation*}

\smallskip \noindent\textbf{Theorem 5.11.1.3 }\emph{The real local functions}%
\begin{equation*}
\left( \frac{\partial \left( \rho ,\eta \right) \Gamma _{c}^{a}}{\partial
y^{b}},\frac{\partial \left( \rho ,\eta \right) \Gamma _{c}^{a}}{\partial
y^{b}},0,~0\right) ,~a,b,c\in \overline{1,r}, \leqno(5.11.1.8)
\end{equation*}%
\emph{and}
\begin{equation*}
\left( \frac{\partial \left( \rho ,\eta \right) \mathring{\Gamma}_{c}^{a}}{%
\partial y^{b}},\frac{\partial \left( \rho ,\eta \right) \mathring{\Gamma}%
_{c}^{a}}{\partial y^{b}},0,~0\right) ,~a,b,c\in \overline{1,r}, \leqno%
(5.11.1.8^{\prime })
\end{equation*}%
\emph{respectively, are the coefficients to a normal Berwald linear }$\left(
\rho ,\eta \right) $\emph{-connection for the generalized tangent bundle }$%
\left( \left( \rho ,\eta \right) TE,\left( \rho ,\eta \right) \tau
_{E},E\right) $.

\bigskip \noindent \textbf{Theorem 5.11.1.4 }\emph{The tensor of
integrability of the }$\left( \rho ,\eta \right) $\emph{-connection }$\left(
\rho ,\eta \right) \mathring{\Gamma}$\emph{\ is as follows:}%
\begin{equation*}
%
\leqno(5.11.1.9)
\end{equation*}%
\emph{where }$_{|c}$\emph{\ is the }$h$\emph{-covariant derivation with
respect to the normal Berwald linear }$\left( \rho ,\eta \right) $\emph{%
-connection }$(5.11.1.8)$\emph{.}

\bigskip\noindent\textit{Proof.} Since
\begin{equation*}
%
%
\end{equation*}%
it results the conclusion of the theorem.\hfill \emph{q.e.d.}

\bigskip\noindent\textbf{Theorem 5.11.1.5 }\emph{Let }%
\begin{equation*}
\mathbb{T}_{bc}^{a}\delta _{a}\otimes d\tilde{z}^{b}\otimes d\tilde{z}%
^{c}\in \mathcal{T}_{20}^{10}\left( \left( \rho ,\eta \right) TE,\left( \rho
,\eta \right) \tau _{E},E\right)
\end{equation*}%
\emph{and }%
\begin{equation*}
\mathbb{S}_{bc}^{a}\overset{\cdot }{\tilde{\partial}}_{a}\otimes \delta
\tilde{y}^{b}\otimes \delta \tilde{y}^{c}\in \mathcal{T}_{02}^{01}\left(
\left( \rho ,\eta \right) TE,\left( \rho ,\eta \right) \tau _{E},E\right)
\end{equation*}%
\emph{such that they verify the following conditions:}%
\begin{equation*}
\mathbb{T}_{bc}^{a}=-\mathbb{T}_{cb}^{a},~\mathbb{S}_{bc}^{a}=-\mathbb{S}%
_{cb}^{a},~\forall b,c\in \overline{1,r}.
\end{equation*}

\emph{If }$\left( \left( \rho ,\eta \right) \widetilde{H},\left( \rho ,\eta
\right) \widetilde{V}\right) $\emph{\ is the distinguished linear }$\left(
\rho ,\eta \right) $\emph{-connection presented in the Theorem 5.9.2, then
the local real functions:\ }%
\begin{equation*}
\begin{array}{ll}
\left( \rho ,\eta \right) \mathring{H}_{bc}^{a}\!\! & =\displaystyle\left(
\rho ,\eta \right) \widetilde{H}_{bc}^{a}+\frac{1}{8}\tilde{g}^{ae}\left( -%
\tilde{g}_{c}^{f}\circ h\circ \pi \frac{\partial F^{d}}{\partial y^{f}}\frac{%
\partial g_{be}}{\partial y^{d}}\right. \vspace*{1mm} \\
& \displaystyle\left. +\ \tilde{g}_{e}^{f}\circ h\circ \pi \frac{\partial
F^{d}}{\partial y^{f}}\frac{\partial g_{bc}}{\partial y^{d}}-\tilde{g}%
_{b}^{f}\circ h\circ \pi \frac{\partial F^{d}}{\partial y^{f}}\frac{\partial
g_{ec}}{\partial y^{d}}\right) ,\vspace*{2mm} \\
\left( \rho ,\eta \right) \mathring{V}_{bc}^{a}\!\! & =\left( \rho ,\eta
\right) \widetilde{V}_{bc}^{a}%
\end{array}%
\leqno(5.11.1.10)
\end{equation*}%
\emph{are the components of a normal distinguished linear }$\left( \rho
,\eta \right) $\emph{-connection with }$\left( \rho ,\eta \right) $\emph{-}$%
\mathcal{H}\left( \mathcal{HH}\right) $\emph{\ and }$\left( \rho ,\eta
\right) $\emph{-}$\mathcal{V}\left( \mathcal{VV}\right) $\emph{\ torsions a
priori given such that the generalized tangent bundle}\break $\left( \left(
\rho ,\eta \right) TE,\left( \rho ,\eta \right) \tau _{E},E\right) $\emph{\
derives generalized Lagrange }$\left( \rho ,\eta \right) $\emph{-space.}

\emph{In addition, we have: }%
\begin{equation*}
\begin{array}{c}
\left( \rho ,\eta ,h\right) \mathbb{\mathring{T}}_{bc}^{a}=\mathbb{T}%
_{bc}^{a}\vspace*{1mm} \\
\left( \rho ,\eta ,h\right) \mathbb{\mathring{S}}_{bc}^{a}=\mathbb{S}%
_{bc}^{a}.%
\end{array}%
\leqno(5.11.1.11)
\end{equation*}

\smallskip \noindent \textbf{Proposition 5.11.1.1 }\emph{If }$S$\emph{\ is
the canonical }$\left( \rho ,\eta \right) $\emph{-semispray asso\-cia\-ted
to the mechanical }$\left( \rho ,\eta \right) $\emph{-system }$\left( \left(
E,\pi ,M\right) ,F_{e},\left( \rho ,\eta \right) \Gamma \right) $\emph{\ and
from }$\mathbf{B}^{\mathbf{v}}$\emph{-mor\-phism }$\left( g,h\right) $\emph{%
,\ then }%
\begin{equation*}
2G^{a%
%TCIMACRO{\U{b4}}%
%BeginExpansion
{\acute{}}%
%EndExpansion
}=2G^{a}M_{a}^{a%
%TCIMACRO{\U{b4}}%
%BeginExpansion
{\acute{}}%
%EndExpansion
}\circ h\circ \pi -\left( g_{b}^{a}\circ h\circ \pi \right) y^{b}\left( \rho
_{a}^{i}\circ h\circ \pi \right) \frac{\partial y^{a%
%TCIMACRO{\U{b4}}%
%BeginExpansion
{\acute{}}%
%EndExpansion
}}{\partial x^{i}}.\leqno(5.11.1.12)
\end{equation*}

\smallskip \noindent\textit{Proof.} Since the Jacobian matrix of coordinates
transformation is
\begin{equation*}
\left\Vert
%
\right) ,%
\end{array}%
\end{equation*}%
the conclusion results immediately.

In the following, we consider a differentiable curve $I\ ^{\underrightarrow{c%
}}~M$ and its $\left( g,h\right) $-lift $I\ ^{\underrightarrow{\dot{c}}}~E.$%
\hfill \emph{q.e.d.}

\bigskip \noindent \textbf{Definition 5.11.1.3 }The curve $\dot{c}$ is a
integral curve of the $\left( \rho ,\eta \right) $-semispray $S$ of the
mechanical $\left( \rho ,\eta \right) $-system $\left( \left( E,\pi
,M\right) ,F_{e},\left( \rho ,\eta \right) \Gamma \right) $, if it is
verifies the following equality:\textit{\ }%
\begin{equation*}
\frac{d\dot{c}\left( t\right) }{dt}=\Gamma \left( \tilde{\rho},Id_{E}\right)
S\left( \dot{c}\left( t\right) \right) .\leqno(5.11.1.13)
\end{equation*}

\smallskip \noindent \textbf{Theorem 5.11.1.6 }\emph{All }$\left( g,h\right)
$\emph{-lifts solutions of the equations:\ }%
\begin{equation*}
\frac{dy^{a}\left( t\right) }{dt}+2G^{a}\!\circ u\left( c,\dot{c}\right)
\left( x\left( t\right) \right) {=}\frac{1}{2}F^{a}\!\circ u\left( c,\dot{c}%
\right) \left( x\left( t\right) \right) \!,\,a{\in }\overline{1,\!r},\leqno%
(5.11.1.14)
\end{equation*}%
\emph{where }$x\left( t\right) =\left( \eta \circ h\circ c\right) \left(
t\right) ,$ \emph{are } \emph{integral curves of the canonical }$\left( \rho
,\eta \right) $\emph{-semispray asso\-cia\-ted to mechanical }$\left( \rho
,\eta \right) $\emph{-system }$\left( \left( E,\pi ,M\right) ,F_{e},\left(
\rho ,\eta \right) \Gamma \right) $\emph{\ and from locally invertible }$%
\mathbf{B}^{\mathbf{v}}$\emph{-mor\-phism }$\left( g,h\right) .$

\bigskip\noindent\textit{Proof.} Since the equality
\begin{equation*}
\frac{d\dot{c}\left( t\right) }{dt}=\Gamma \left( \tilde{\rho},Id_{E}\right)
S\left( \dot{c}\left( t\right) \right)
\end{equation*}%
is equivalent to
\begin{equation*}
\begin{array}{l}
\displaystyle\frac{d}{dt}((\eta \circ h\circ c)^{i}(t),y^{a}(t)) \vspace*{2mm%
} \\
\qquad=\displaystyle\left( \rho _{a}^{i}\circ \eta \circ h\circ
c(t)g_{b}^{a}\circ h\circ c(t)y^{b}(t),-2\left( G^{a}-\frac{1}{4}%
F^{a}\right) ((\eta \circ h\circ c)^{i}(t),y^{a}(t))\right) ,%
\end{array}%
\end{equation*}%
it results
\begin{equation*}
\begin{array}{l}
\displaystyle\frac{dy^{a}\left( t\right) }{dt}+2G^{a}\left( x^{i}\left(
t\right) ,y^{a}\left( t\right) \right) =\frac{1}{2}F^{a}\left( x^{i}\left(
t\right) ,y^{a}\left( t\right) \right) ,\ \ a\in \overline{1,n},\vspace*{2mm}
\\
\displaystyle\frac{dx^{i}\left( t\right) }{dt}=\rho _{a}^{i}\circ \eta \circ
h\circ c\left( t\right) g_{b}^{a}\circ h\circ c\left( t\right) y^{b}\left(
t\right) ,%
\end{array}%
\end{equation*}%
where $x^{i}\left( t\right) =\left( \eta \circ h\circ c\right) ^{i}\left(
t\right) $. \hfill \emph{q.e.d.}

\bigskip\noindent\textbf{Definition 5.11.1.4 }If $S$\ is a $\left( \rho
,\eta \right) $-semispray, then the vector field
\begin{equation*}
\begin{array}{l}
\left[ \mathbb{C},S\right] _{\left( \rho ,\eta \right) TE}-S%
\end{array}%
\leqno(5.11.1.15)
\end{equation*}%
will be called the \emph{derivation of }$\left( \rho ,\eta \right) $\emph{%
-semispray }$S.$

The $\left( \rho ,\eta \right) $-semispray $S$ will be called $\left( \rho
,\eta \right) $\emph{-spray} if the following conditions are
verified:\medskip

1. $S\circ 0\in C^{1},$\ where $0$\ is the null section;\smallskip

2. Its derivation is the null vector field.\medskip

The $\left( \rho ,\eta \right) $-semispray $S$\ will be called \emph{%
quadratic }$\left( \rho ,\eta \right) $\emph{-spray }if there are verified
the following conditions:\medskip

1. $S\circ 0\in C^{2},$\ where $0$\ is the null section;\smallskip

2. Its derivation is the null vector field.\medskip

In particular, \ if $\ \left( \rho ,\eta \right) =\left(
id_{TM},Id_{M}\right) $ and $\left( g,h\right) =\left( Id_{E},Id_{M}\right)
, $ \ then \ we \ obtain \ the \ \emph{spray} \ and the \emph{quadratic
spray }which is similar with the classical spray and quadratic spray\textit{.%
}

\bigskip \noindent \textbf{Theorem 5.11.1.7 }\emph{If }$S$\emph{\ is the
canonical }$\left( \rho ,\eta \right) $\emph{-spray associated to mechanical
}$\left( \rho ,\eta \right) $\emph{-system }$\left( \left( E,\pi ,M\right)
,F_{e},\left( \rho ,\eta \right) \Gamma \right) $\emph{\ and from locally
invertible }$\mathbf{B}^{\mathbf{v}}$\emph{-morphism }$\left( g,h\right) $%
\emph{, then}%
\begin{equation*}
\begin{array}{l}
2\left( G^{a}-\displaystyle\frac{1}{4}F^{a}\right) \displaystyle=\left( \rho
,\eta \right) \Gamma _{c}^{a}\left( g_{f}^{c}\circ h\circ \pi \cdot
y^{f}\right) \vspace*{2mm} \\
\qquad \displaystyle+\frac{1}{2}\left( g_{e}^{d}\circ h\circ \pi \cdot
y^{e}\right) L_{dc}^{b}\circ h\circ \pi \tilde{g}_{b}^{a}\circ h\circ \pi
\left( g_{f}^{c}\circ h\circ \pi \cdot y^{f}\right) ,~a\in \overline{1,r}.%
\end{array}%
\leqno(5.11.1.16)
\end{equation*}%
\emph{Then, we obtain the spray }%
\begin{equation*}
\begin{array}{ll}
S & \displaystyle=\left( g_{b}^{a}\circ h\circ \pi \right) y^{b}\frac{%
\partial }{\partial \tilde{z}^{a}}+\left( \rho ,\eta \right) \Gamma
_{c}^{a}\left( g_{f}^{c}\circ h\circ \pi \cdot y^{f}\right) \frac{\partial }{%
\partial \tilde{y}^{a}}\vspace*{2mm} \\
& \displaystyle+\frac{1}{2}\left( g_{e}^{d}\circ h\circ \pi \cdot
y^{e}\right) L_{dc}^{b}\circ h\circ \pi \cdot \tilde{g}_{b}^{a}\circ h\circ
\pi \left( g_{f}^{c}\circ h\circ \pi \cdot y^{f}\right) \frac{\partial }{%
\partial \tilde{y}^{a}}.%
\end{array}%
\leqno(5.11.1.17)
\end{equation*}

\emph{This }$\left( \rho ,\eta \right) $\emph{-spray will be called the
canonical }$\left( \rho ,\eta \right) $\emph{-spray associated to mechanical
system }$\left( \left( E,\pi ,M\right) ,F_{e},\left( \rho ,\eta \right)
\Gamma \right) $\emph{\ and from locally invertible }$\mathbf{B}^{\mathbf{v}%
} $\emph{-morphism }$(g,h).$

\emph{In particular, if }$\left( \rho ,\eta \right) =\left(
id_{TM},Id_{M}\right) $\emph{\ and }$\left( g,h\right) =\left(
Id_{E},Id_{M}\right) ,$\emph{\ then we get the canonical spray associated to
connection }$\Gamma $\emph{\ which is similar with the classical canonical
spray associated to connection }$\Gamma $.

\bigskip\noindent\textit{Proof.} Since
\begin{equation*}
\left[ \mathbb{C},S\right] _{\left( \rho ,\eta \right) TE} =\left[ y^{a}%
\overset{\cdot }{\tilde{\partial}}_{a},\left( g_{e}^{b}\circ h\circ \pi
\cdot y^{e}\right) \tilde{\partial}_{b}\right] _{\left( \rho ,\eta \right)
TE} -2\left[ y^{a}\overset{\cdot }{\tilde{\partial}}_{a},\left( G^{b}-\frac{1%
}{4}F^{b}\right) \overset{\cdot }{\tilde{\partial}}_{b}\right] _{\left( \rho
,\eta \right) TE},
\end{equation*}

\begin{equation*}
\!\!%
%
\end{equation*}%
it results that
\begin{equation*}
%
%
\leqno\left( S_{1}\right)
\end{equation*}%
Using equality $(5.11.1.4)$, it results that
\begin{equation*}
%
%
\leqno\left( S_{2}\right)
\end{equation*}%
Using equalities $\left( S_{1}\right) $ and $\left( S_{2}\right) $, it
results the conclusion of the theorem.\hfill \emph{q.e.d.}

\bigskip\noindent\textbf{Remark 5.11.1.2. }If $\left( \rho ,\eta \right)
=\left( id_{TM},Id_{M}\right) $ and $\left( g,h\right) =\left(
Id_{E},Id_{M}\right) , $ then we get the canonical spray associated to
connection $\Gamma $.

\bigskip \noindent \textbf{Theorem 5.11.1.8 }\emph{\ All }$\left( g,h\right)
$\emph{-lifts solutions of the following system of equations:\ }%
\begin{equation*}
%
%
\leqno(5.11.1.17)
\end{equation*}%
\emph{are the integral curves of canonical }$\left( \rho ,\eta \right) $%
\emph{-spray associated to mechanical }$\left( \rho ,\eta \right) $\emph{%
-system }$\left( \left( E,\pi ,M\right) ,F_{e},\left( \rho ,\eta \right)
\Gamma \right) $\emph{\ and from locally invertible }$\mathbf{B}^{\mathbf{v}%
} $\emph{-morphism\ }$\left( g,h\right) .$

\subsubsection{The Lagrangian formalism for Lagrange mechanical $\left(
\protect\rho ,\protect\eta \right) $-systems}

Let $\left( \left( E,\pi ,M\right) ,F_{e},L\right) $ be an arbitrarily
Lagrange mechanical $\left( \rho ,\eta \right) $-system.

The \emph{natural dual }$\left( \rho ,\eta \right) $\emph{-base }$\left( d%
\tilde{z}^{\alpha },d\tilde{y}^{a}\right) $ of natural $\left( \rho ,\eta
\right) $-base $\left( \displaystyle\frac{\partial }{\partial \tilde{z}%
^{\alpha }},\displaystyle\frac{\partial }{\partial \tilde{y}^{a}}\right) $
is determined by the equations
\begin{equation*}
\begin{array}{c}
\left\{
\begin{array}{cc}
\displaystyle\left\langle d\tilde{z}^{\alpha },\frac{\partial }{\partial
\tilde{z}^{\beta }}\right\rangle =\delta _{\beta }^{\alpha }, & \displaystyle%
\left\langle d\tilde{z}^{\alpha },\frac{\partial }{\partial \tilde{y}^{a}}%
\right\rangle =0, \vspace*{2mm} \\
\displaystyle\left\langle d\tilde{y}^{a},\frac{\partial }{\partial \tilde{z}%
^{\beta }}\right\rangle =0, & \displaystyle \left\langle d\tilde{y}^{a},%
\frac{\partial }{\partial \tilde{y}^{b}}\right\rangle =\delta _{b}^{a}.%
\end{array}%
\right.%
\end{array}%
\end{equation*}
It is very important to remark that the $1$-forms $d\tilde{z}^{\alpha
},~\alpha \in \overline{1,p}$ and $d\tilde{y}^{a},~a\in \overline{1,n}$ are
not the differentials of coordinates functions as in the classical case, but
we will use the same notations. In this case
\begin{equation*}
\left( d\tilde{z}^{\alpha }\right) \neq d^{\left( \rho ,\eta \right)
TE}\left( \tilde{z}^{\alpha }\right) =0,
\end{equation*}%
where $d^{\left( \rho ,\eta \right) TE}$ is the exterior differentiation
operator associated to exterior differential $\mathcal{F}\left( E\right) $%
-algebra
\begin{equation*}
\left( \Lambda \left( \left( \rho ,\eta \right) TE,\left( \rho ,\eta \right)
\tau _{E},E\right) ,+,\cdot ,\wedge \right) .
\end{equation*}

Let $L$ be a regular Lagrangian and let $\left( g,h\right) $\ be a locally
invertible $\mathbf{B}^{\mathbf{v}}$-morphism of $\left( E,\pi ,M\right) $
source and $\left( E,\pi ,M\right) $ target.

\bigskip\noindent\textbf{Definition 5.11.2.1 } The $1$-form
\begin{equation*}
\begin{array}{c}
\theta _{L}=\left( \tilde{g}_{a}^{e}\circ h\circ \pi \cdot L_{e}\right) d%
\tilde{z}^{a}%
\end{array}%
\leqno(5.11.2.1)
\end{equation*}%
will be called the $1$\emph{-form of Poincar\'{e}-Cartan type associated to
the Lagrangian }$L$ \emph{and to the locally invertible }$\mathbf{B}^{%
\mathbf{v}}$\emph{-morphism }$\left( g,h\right) $.\medskip

We obtain easily:
\begin{equation*}
\theta _{L}\left( \frac{\partial }{\partial \tilde{z}^{b}}\right) =\tilde{g}%
_{b}^{e}\circ h\circ \pi \cdot L_{e},\,\,\ \theta _{L}\left( \frac{\partial
}{\partial \tilde{y}^{b}}\right) =0. \leqno(5.11.2.2)
\end{equation*}

\smallskip\noindent \textbf{Definition 5.11.2.2 } The $2$-form
\begin{equation*}
\omega _{L}=d^{\left( \rho ,\eta \right) TE}\theta _{L}
\end{equation*}%
will be called the $2$\emph{-form of Poincar\'{e}-Cartan type associated to
the Lagrangian }$L$\emph{\ and to the locally invertible }$\mathbf{B}^{%
\mathbf{v}}$\emph{-morphism }$\left( g,h\right) $.\medskip

By the definition of $d^{\left( \rho ,\eta \right) TE},$ we obtain:
\begin{equation*}
%
\right. \hspace*{-4mm}\leqno(5.11.2.4)
\end{equation*}

\medskip\noindent \textbf{Definition 5.11.2.3 } The real function
\begin{equation*}
%
%
\leqno(5.11.2.5)
\end{equation*}%
will be called the \emph{energy of regular Lagrangian }$L.$

\bigskip \noindent \textbf{Theorem 5.11.2.1 }\emph{The equation }%
\begin{equation*}
\left( 5.11.2.6\right) \,\
%
%
\end{equation*}%
\emph{has an unique solution }$S_{L}\left( g,h\right) $\emph{\ of the type: }%
\begin{equation*}
\left( g_{e}^{a}\circ h\circ \pi \right) y^{e}\frac{\partial }{\partial
\tilde{z}^{a}}-2\left( G^{a}-\frac{1}{4}F^{a}\right) \frac{\partial }{%
\partial \tilde{y}^{a}},\leqno(5.11.2.7)
\end{equation*}%
\emph{where }%
\begin{equation*}
2G^{a}=g_{e}^{a}\circ h\circ \pi \cdot \tilde{L}^{eb}\cdot E_{b}\left(
L,g,h\right) +\frac{1}{2}F^{a}\leqno(5.11.2.8)
\end{equation*}%
\emph{and}%
\begin{equation*}
%
%
\hspace*{-4mm}\leqno(5.11.2.9)
\end{equation*}%
$S_{L}\left( g,h\right) $\textit{\ }will be called \emph{the\ canonical }$%
\left( \rho ,\eta \right) $\emph{-semispray associated to Lagrange
mechanical }$\left( \rho ,\eta \right) $\emph{-system }$\left( \left( E,\pi
,M\right) ,F_{e},L\right) $\emph{\ and from locally invertible }$\mathbf{B}^{%
\mathbf{v}}$\emph{-morphism }$(g,h).$

\noindent \textit{Proof.} We obtain that
\begin{equation*}
i_{S}\left( \omega _{L}\right) =-d^{\left( \rho ,\eta \right) TE}\left(
\mathcal{E}_{L}\right)
\end{equation*}
if and only if
\begin{equation*}
\omega _{L}\left( S,X\right) =-\Gamma \left( \tilde{\rho},Id_{E}\right)
\left( X\right) \left( \mathcal{E}_{L}\right) ,
\end{equation*}%
for any $X\in \Gamma \left( \left( \rho ,\eta \right) TE,\left( \rho ,\eta
\right) \tau _{E},E\right) .$

Particularly, we obtain:%
\begin{equation*}
\omega _{L}\left( S,\frac{\partial }{\partial \tilde{z}^{b}}\right) =-\Gamma
\left( \tilde{\rho},Id_{E}\right) \left( \frac{\partial }{\partial \tilde{z}%
^{b}}\right) \left( \mathcal{E}_{L}\right) .
\end{equation*}%
If we expand this equality by using $\left( 5.11.2.2\right) $ and $\left(
5.11.2.4\right) $, we obtain%
\begin{equation*}
\begin{array}{l}
\displaystyle g_{e}^{a}\circ h\circ \pi \cdot y^{e}\cdot \left[ \rho _{a}^{i}%
{\circ }h{\circ }\pi \cdot \frac{\partial \left( \tilde{g}_{b}^{e}\circ
h\circ \pi \cdot L_{e}\right) }{\partial x^{i}}-\rho _{b}^{i}{\circ }h{\circ
}\pi \cdot \frac{\partial \left( \tilde{g}_{a}^{e}\circ h\circ \pi \cdot
L_{e}\right) }{\partial x^{i}}\right. \vspace*{2mm} \\
\qquad \displaystyle\left. -\ L_{ab}^{d}{\circ }h{\circ }\pi \cdot \left(
\tilde{g}_{d}^{e}\circ h\circ \pi \cdot L_{e}\right) \right] +2\left( G^{a}-%
\frac{1}{4}F^{a}\right) \left( \tilde{g}_{a}^{e}\circ h\circ \pi \right)
\cdot L_{eb}\vspace*{2mm} \\
\qquad \displaystyle=-\rho _{b}^{i}{\circ }h{\circ }\pi \cdot \left(
g_{e}^{a}\circ h\circ \pi \cdot y^{e}\right) \cdot \frac{\partial \left(
\tilde{g}_{a}^{e}\circ h\circ \pi \cdot L_{e}\right) }{\partial x^{i}}+\rho
_{b}^{i}{\circ }h{\circ }\pi \cdot L_{i}.%
\end{array}%
\end{equation*}%
After some calculations, we obtain
\begin{equation*}
2\left( G^{a}-\frac{1}{4}F^{a}\right) =g_{e}^{a}\circ h\circ \pi \cdot
\tilde{L}^{eb}\cdot E_{b}\left( L,g,h\right) ,
\end{equation*}%
where
\begin{equation*}
\begin{array}{ll}
E_{b}\left( L,g,h\right) & \displaystyle=\rho _{b}^{i}{\circ }h{\circ }\pi
\cdot L_{i}-g_{e}^{a}\circ h\circ \pi \cdot y^{e}\cdot \rho _{a}^{i}{\circ }h%
{\circ }\pi \cdot \frac{\partial \left( \tilde{g}_{b}^{e}\circ h\circ \pi
\cdot L_{e}\right) }{\partial x^{i}}+\vspace*{2mm} \\
& \displaystyle+\ g_{e}^{a}\circ h\circ \pi \cdot y^{e}\cdot L_{ab}^{d}{%
\circ }h{\circ }\pi \cdot \left( \tilde{g}_{d}^{e}\circ h\circ \pi \cdot
L_{e}\right) .%
\end{array}%
\end{equation*}

\hfill\emph{q.e.d.}

\bigskip\noindent\textbf{Remarks }

\begin{itemize}
\item[1.] If $F_{e}=0$ and $\eta =Id_{M},$\ then $S_{L}\left(
Id_{E},Id_{M}\right) \overset{put}{=}S_{L}$\ is the canonical $\rho $%
-semispray associated to regular Lagrangian $L$\ which is similar with the
semispray presented in $\left[ 27\right] $ by M. de Leon, J. Marrero and E.
Martinez.

\item[2.] If $F_{e}\neq 0$ and\ $\left( \rho ,\eta \right) =\left(
Id_{TM},Id_{M}\right) $, then $S_{L}\left( Id_{E},Id_{M}\right) \overset{put}%
{=}S_{L}$\ will be called \emph{the canonical semispray}\ which is not the
same canonical semispray presented by I.~Bucataru and R. Miron in $\left[ 7%
\right] $.

\item[3.] If $F_{e}=0$\ and $\left( \rho ,\eta \right) =\left(
Id_{TM},Id_{M}\right) $, then $S_{L}\left( Id_{M},Id_{E}\right) \overset{put}%
{=}S_{L}$\ will be called \emph{the canonical semispray}\ which is not the
same canonical semispray presented by R.~Miron and M. Anastasiei in~[41].
\end{itemize}

\smallskip \noindent \textbf{Theorem 5.11.2.2 }\emph{The\ real local
functions }%
\begin{equation*}
\begin{array}{cl}
\left( \rho ,\eta \right) \Gamma _{c}^{a} & \displaystyle=\frac{1}{2}\tilde{g%
}_{c}^{e}\circ h\circ \pi \frac{\partial \left( g_{e}^{a}\circ h\circ \pi
\cdot L^{eb}\cdot E_{b}\left( L,g,h\right) \right) }{\partial y^{e}}\vspace*{%
2mm} \\
& \displaystyle-\ \frac{1}{2}\left( g_{e}^{d}\circ h\circ \pi \cdot
y^{e}\right) L_{dc}^{b}\circ h\circ \pi \cdot \tilde{g}_{b}^{a}\circ h\circ
\pi ,~a,c\in \overline{1,r}.%
\end{array}%
\leqno(5.11.2.10)
\end{equation*}%
\emph{are the components of a }$\left( \rho ,\eta \right) $\emph{-connection
}$\left( \rho ,\eta \right) \Gamma $\emph{\ for the vector bundle }$\left(
E,\pi ,M\right) $\emph{\ which will be called the }$\left( \rho ,\eta
\right) $\emph{-connection associated to Lagrange mechanical }$\left( \rho
,\eta \right) $\emph{-system}\break $\left( \left( E,\pi ,M\right)
,F_{e},L\right) $\emph{\ and from }$\mathbf{B}^{\mathbf{v}}$\emph{-morphism}
$(g,h).$

\bigskip \noindent \textbf{Corollary 5.11.2.1 }\emph{The real local functions%
}%
\begin{equation*}
\begin{array}{cl}
(\rho ,\eta )\mathring{\Gamma}^{a}{}_{c}\!\!\! & \displaystyle=\left( \tilde{%
g}_{c}^{b}\circ h\circ \pi \right) \frac{\partial G^{a}}{\partial y^{b}}%
\vspace*{1.2mm} \\
& \displaystyle-\ \frac{1}{2}\left( g_{e}^{d}\circ h\circ \pi \cdot
y^{e}\right) L_{dc}^{b}\circ h\circ \pi \cdot \tilde{g}_{b}^{a}\circ h\circ
\pi ,~a,c\in \overline{1,r}%
\end{array}%
\leqno(5.11.2.11)
\end{equation*}%
\emph{are the components of a }$\left( \rho ,\eta \right) $\emph{-connection
}$\left( \rho ,\eta \right) \mathring{\Gamma}$\emph{\ for the vector bundle }%
$\left( E,\pi ,M\right) .$

\emph{In addition, we have}%
\begin{equation*}
\left( \rho ,\eta \right) \mathring{\Gamma}_{\,c}^{a}=\left( \rho ,\eta
\right) \Gamma _{\,c}^{a}+\frac{1}{4}(\tilde{g}_{c}^{b}\circ h\circ \pi
)\cdot \frac{\partial F^{a}}{\partial y^{b}},~\forall a,c\in \overline{1,r}. %
\leqno(5.11.2.12)
\end{equation*}

\smallskip \noindent \textbf{Theorem 5.11.2.3 }\emph{The parallel }$\left(
g,h\right) $\emph{-lifts with respect to }$\left( \rho ,\eta \right) $\emph{%
-connection }$\left( \rho ,\eta \right) \Gamma $ \emph{are the\ integral
curves of the canonical }$\left( \rho ,\eta \right) $\emph{-semispray
associated to\ mechanical }$\left( \rho ,\eta \right) $\emph{-system }$%
\left( \left( E,\pi ,M\right) ,F_{e},L\right) $ \emph{and from locally
invertible }$\mathbf{B}^{\mathbf{v}}$\emph{-morphism }$\left( g,h\right) .$

\bigskip \noindent \textbf{Definition 5.11.2.4 }The equations
\begin{equation*}
\begin{array}{c}
\,\dfrac{dy^{a}\left( t\right) }{dt}+\left( g_{e}^{a}\circ h\circ \pi \cdot
\tilde{L}^{eb}\cdot E_{b}\left( L,g,h\right) \right) \circ u\left( c,\dot{c}%
\right) \left( x\left( t\right) \right) =0,%
\end{array}%
\leqno(5.11.2.13)
\end{equation*}%
where $x\left( t\right) =\eta \circ h\circ c\left( t\right) $, will be
called the \emph{equations of Euler-Lagrange type associated to Lagrange
mechanical }$\left( \rho ,\eta \right) $\emph{-system }$\left( \left( E,\pi
,M\right) ,F_{e},L\right) $\emph{\ and from locally invertible }$\mathbf{B}^{%
\mathbf{v}}$\emph{-morphism }$\left( g,h\right) .$

The equations
\begin{equation*}
\begin{array}{c}
\dfrac{dy^{a}\left( t\right) }{dt}+\left( \tilde{L}^{ab}\cdot E_{b}\left(
L,Id_{E},Id_{M}\right) \right) \circ u\left( c,\dot{c}\right) \left( x\left(
t\right) \right) =0,%
\end{array}%
\leqno(5.11.2.13^{\prime })
\end{equation*}%
where $x\left( t\right) =h\circ \eta \circ c\left( t\right) $, will be
called the \emph{equations of Euler-Lagrange type associated to Lagrange
mechanical} $\left( \rho ,\eta \right) $\emph{-system }$\left( \left( E,\pi
,M\right) ,F_{e},L\right) $.

\bigskip \noindent \textbf{Remark 5.11.2.1 }The\ integral curves of the
canonical $\left( \rho ,\eta \right) $-semispray associated to mechanical $%
\left( \rho ,\eta \right) $-system $\left( \left( E,\pi ,M\right)
,F_{e},L\right) $\ and from locally invertible\emph{\ }$\mathbf{B}^{\mathbf{v%
}}$-morphism $\left( g,h\right) $\ are the $\left( g,h\right) $-lifts
solutions for the equations of Euler-Lagrange type $\left( 5.11.2.13\right) $%
.

It is known that, in classical sense, a geodesic with respect to a Finsler
metric
\begin{equation*}
TM~~^{\underrightarrow{\ F\ }}~~\mathbb{R}\mathbf{_{+}}
\end{equation*}%
is a curve $c$ on the manifold $M$ such that the components of its tangent
lift
\begin{equation*}
\begin{array}{c}
\displaystyle\frac{dc^{i}}{dt}\cdot \frac{\partial }{\partial x^{i}}%
\end{array}%
\end{equation*}%
are solutions for the Euler-Lagrange equations%
\begin{equation*}
\begin{array}{c}
\frac{d}{dt}\left( \frac{\partial F^{2}}{\partial y^{i}}\right) -\frac{%
\partial F^{2}}{\partial x^{i}}=0,~i\in \overline{1,m}%
\end{array}%
.\leqno(5.11.2.14)
\end{equation*}

If
\begin{equation*}
\left( \left( TM,\tau _{M},M\right) ,\widetilde{\left[ ,\right] }%
_{TM,h},\left( \rho ,\eta \right) \right)
\end{equation*}%
is a generalized Lie algebroid different by the generalized Lie algebroid
\begin{equation*}
\left( \left( TM,\tau _{M},M\right) ,\left[ ,\right] _{TM,Id_{M}},\left(
Id_{TM},Id_{M}\right) \right) ,
\end{equation*}%
then, using the classical method by work, we can not determine the geodesics
on the manifold $M$ such that the components of their lifts (different by
the tangent lift) are solutions for the Euler-Lagrange equations $%
(5.11.2.14) $.

Using our theory, we obtain the following

\bigskip\noindent\textbf{Theorem 5.11.2.4 }\emph{If }$F$\emph{\ is a Finsler
fundamental function, then the\ geodesics on the manifold }$M$\emph{\ are
the curves such that the components of their }$\left( g,h\right) $\emph{%
-lifts are\ solutions for the equations of Euler-Lagrange type }$\left(
5.11.2.13\right).$\bigskip

Therefore, it is natural to propose to extend the study of the Finsler
geometry from the usual Lie algebroid
\begin{equation*}
\begin{array}{c}
\left( \left( TM,\tau _{M},M\right) ,\left[ ,\right] _{TM},\left(
Id_{TM},Id_{M}\right) \right) ,%
\end{array}%
\end{equation*}%
to an arbitrary (generalized) Lie algebroid%
\begin{equation*}
\begin{array}{c}
\left( \left( E,\pi ,M\right) ,\left[ ,\right] _{E,h},\left( \rho ,\eta
\right) \right) .%
\end{array}%
\end{equation*}

\section{The geometry of total space of the generalized tangent bundle for
dual vector bundle}

\subsection{Adapted $\left( \protect\rho ,\protect\eta \right) $-basis and
adapted dual $\left( \protect\rho ,\protect\eta \right) $-basis}

In the following we consider the following diagram:
\begin{equation*}
\begin{array}{c}
\xymatrix{\overset{\ast }{E}\ar[d]_{\overset{\ast }{\pi }}&\left( F,\left[
,\right] _{F,h},\left( \rho ,\eta \right) \right)\ar[d]^\nu\\ M\ar[r]^h&N}%
\end{array}%
\end{equation*}%
where $\left( E,\pi ,M\right) \in \left\vert \mathbf{B}^{\mathbf{v}%
}\right\vert $ and $\left( \left( F,\nu ,N\right) ,\left[ ,\right]
_{F,h},\left( \rho ,\eta \right) \right) $ is a generalized Lie algebroid.

Let $\left( \rho ,\eta \right) \overset{\ast }{\Gamma }$ be a $\left( \rho
,\eta \right) $-connection for the vector bundle $\left( \overset{\ast }{E},%
\overset{\ast }{\pi },M\right) .$

If we put the problem of finding a base for the $\mathcal{F}\left( \overset{%
\ast }{E}\right) $-module
\begin{equation*}
\left( \Gamma \left( H\left( \rho ,\eta \right) T\overset{\ast }{E},\left(
\rho ,\eta \right) \tau _{\overset{\ast }{E}},\overset{\ast }{E}\right)
,+,\cdot \right)
\end{equation*}%
of the type\textbf{\ }
\begin{equation*}
\frac{\overset{\ast }{\delta }}{\delta \tilde{z}^{\alpha }}=\tilde{Z}%
_{\alpha }^{\beta }\frac{\overset{\ast }{\partial }}{\partial \tilde{z}%
^{\beta }}+Y_{b\alpha }\frac{\partial }{\partial \tilde{p}_{b}},\alpha \in
\overline{1,p}
\end{equation*}%
which satisfies the following conditions:
\begin{equation*}
%
\leqno(6.1.1)
\end{equation*}%
then we obtain the sections
\begin{equation*}
\displaystyle%
%
%
.\leqno(6.1.2)
\end{equation*}

We observe that their law of change is a tensorial law under a change of
vector fiber charts.

\medskip \textbf{Definition 6.1.1} The base
\begin{equation*}
\displaystyle%
%
%
\end{equation*}%
will be called the \emph{adapted }$\left( \rho ,\eta \right) $\emph{-base.}

The following equality holds good%
\begin{equation*}
%
%
\leqno(6.1.3)
\end{equation*}%
where $\left( \overset{\ast }{\partial }_{i},\dot{\partial}^{a}\right) $ is
the natural base for the $\mathcal{F}\left( \overset{\ast }{E}\right) $%
-module $\left( \Gamma \left( T\overset{\ast }{E},\tau _{\overset{\ast }{E}},%
\overset{\ast }{E}\right) ,+,\cdot \right) .$

Moreover, if $\left( \rho ,\eta \right) \overset{\ast }{\Gamma }$ is the $%
\left( \rho ,\eta \right) $-connection associated to connection $\overset{%
\ast }{\Gamma }$, then we obtain
\begin{equation*}
%
%
\leqno(6.1.4)
\end{equation*}%
where $\left( \overset{\ast }{\delta }_{i},\dot{\partial}_{a}\right) $ is
the adapted base for the $\mathcal{F}\left( \overset{\ast }{E}\right) $%
-module $\left( \Gamma \left( T\overset{\ast }{E},\tau _{\overset{\ast }{E}},%
\overset{\ast }{E}\right) ,+,\cdot \right) .$

\medskip \textbf{Theorem 6.1.1 }\emph{The following equality holds good\ }%
\begin{equation*}
%
%
\leqno(6.1.6)
\end{equation*}

\emph{Moreover, we have: }%
\begin{equation*}
%
%
\leqno(6.1.8)
\end{equation*}

Let $\left( d\tilde{z}^{\alpha },d\tilde{p}_{a}\right) $ be the natural dual
$\left( \rho ,\eta \right) $-base.

If we consider the problem of finding a base for the $\mathcal{F}\left(
\overset{\ast }{E}\right) $-module
\begin{equation*}
\left( \Gamma \left( \left( V\left( \rho ,\eta \right) T\overset{\ast }{E}%
\right) ^{\ast },\left( \left( \rho ,\eta \right) \tau _{\overset{\ast }{E}%
}\right) ^{\ast },\overset{\ast }{E}\right) ,+,\cdot \right)
\end{equation*}%
of the type
\begin{equation*}
%
%
\end{equation*}%
which satisfies the following conditions:
\begin{equation*}
%
%
\leqno(6.1.9)
\end{equation*}%
then we obtain the sections
\begin{equation*}
%
%
\leqno(6.1.10)
\end{equation*}

We observe that their changing rule is tensorial under a change of vector
fiber charts.

\medskip\noindent \textbf{Definition 6.1.2} The base $\left( d\tilde{z}%
^{\alpha },\delta \tilde{p}_{a}\right) $ will be called the \emph{adapted
dual }$\left( \rho ,\eta \right) $\emph{-base.}

\subsection{Remarkable endomorphisms}

Now, let us consider the following diagram:
\begin{equation*}
%
%
\end{equation*}%
where $\left( E,\pi ,M\right) \in \left\vert \mathbf{B}^{\mathbf{v}%
}\right\vert $ and $\left( \left( F,\nu ,N\right) ,\left[ ,\right]
_{F,h},\left( \rho ,\eta \right) \right) $ is a generalized Lie algebroid.

\bigskip\noindent\textbf{Definition 6.2.1 }For any $\mathbf{Mod}$%
-endomorphism $e$ of
\begin{equation*}
\left( \Gamma \left( \left( \rho ,\eta \right) T\overset{\ast }{E},\left(
\rho ,\eta \right) \tau _{\overset{\ast }{E}},\overset{\ast }{E}\right)
,+,\cdot \right)
\end{equation*}%
we define the application of Nijenhuis \ type
\begin{equation*}
\!\!\Gamma \left( \left( \rho ,\eta \right) T\overset{\ast }{E},\left( \rho
,\eta \right) \tau _{\overset{\ast }{E}},\overset{\ast }{E}\right) ^{2~\
\underrightarrow{~\ \ N_{e}~\ \ }}~\ \Gamma \!\left( \left( \rho ,\eta
\right) T\overset{\ast }{E},\left( \rho ,\eta \right) \tau _{\overset{\ast }{%
E}},\overset{\ast }{E}\right)
\end{equation*}%
defined by
\begin{equation*}
%
\end{equation*}%
for any $X,Y\in \Gamma \!\left( \left( \rho ,\eta \right) T\overset{\ast }{E}%
,\left( \rho ,\eta \right) \tau _{\overset{\ast }{E}},\overset{\ast }{E}%
\right) .$

\smallskip \noindent \textbf{Remark 6.2.1 }The vertical and the horizontal
vector subbundles are interior differential systems for the Lie algebroid
generalized tangent bundle
\begin{equation*}
%
%
\end{equation*}

These interior differential systems will be called \emph{vertical }and \emph{%
horizontal interior differential systems.}

\subsubsection{Projectors}

\medskip \textbf{Definition 6.2.1.1 }Any $\mathbf{Mod}$-endomorphism $e$ of
\begin{equation*}
\Gamma \left( \left( \rho ,\eta \right) T\overset{\ast }{E},\left( \rho
,\eta \right) \tau _{\overset{\ast }{E}},\overset{\ast }{E}\right)
\end{equation*}%
with the property
\begin{equation*}
%
%
\leqno(6.2.1.1)
\end{equation*}%
will be called a \emph{projector.}

\bigskip\noindent \textbf{Example 6.2.1.1 }The $\mathbf{Mod}$-endomorphism
\begin{equation*}
%
%
\end{equation*}%
is a projector which will be called the \emph{the vertical projector.}

\bigskip\noindent\textbf{Remark 6.2.1.1 } We have $\overset{\ast }{\mathcal{V%
}}\left( \overset{\ast }{\tilde{\delta}}_{\alpha }\right) =0$ and $\overset{%
\ast }{\mathcal{V}}\left( \overset{\cdot }{\tilde{\partial}}^{a}\right) =%
\overset{\cdot }{\tilde{\partial}}^{a}.$ Therefore, it follows\vspace*{-2mm}
\begin{equation*}
\overset{\ast }{\mathcal{V}}\left( \overset{\ast }{\tilde{\partial}}_{\alpha
}\right) =-\left( \rho ,\eta \right) \overset{\ast }{\Gamma }_{b\alpha }%
\overset{\cdot }{\tilde{\partial}}^{b}.
\end{equation*}

\smallskip \noindent\textbf{Theorem 6.2.1.1 }\emph{A }$(\rho ,\eta )$\emph{%
-connection for the vector bundle} $\left( \overset{\ast }{E},\overset{\ast }%
{\pi },M\right) $ \emph{is characterized by the existence of a} $\mathbf{Mod}
$\emph{-endomorphism} $\overset{\ast }{\mathcal{V}}$ \emph{of}
\begin{equation*}
\left( \Gamma \left( \left( \rho ,\eta \right) T\overset{\ast }{E},\left(
\rho ,\eta \right) \tau _{\overset{\ast }{E}},\overset{\ast }{E}\right)
,+,\cdot \right)
\end{equation*}%
\emph{\ with the properties:}
\begin{equation*}
\begin{array}{c}
\overset{\ast }{\mathcal{V}}\left( \Gamma \left( \left( \rho ,\eta \right) T%
\overset{\ast }{E},\left( \rho ,\eta \right) \tau _{\overset{\ast }{E}},%
\overset{\ast }{E}\right) \right) \subset \Gamma \left( \left( V\left( \rho
,\eta \right) T\overset{\ast }{E},\left( \rho ,\eta \right) \tau _{\overset{%
\ast }{E}},\overset{\ast }{E}\right) \right) \vspace*{1,5mm} \\
\overset{\ast }{\mathcal{V}}\left( X\right) =X~\Longleftrightarrow ~X\in
\Gamma \left( \left( V\left( \rho ,\eta \right) T\overset{\ast }{E},\left(
\rho ,\eta \right) \tau _{\overset{\ast }{E}},\overset{\ast }{E}\right)
\right)%
\end{array}%
\leqno(6.2.1.2)
\end{equation*}

\smallskip\noindent \textbf{Example 6.2.1.2 }The $\mathbf{Mod}$-endomorphism
\begin{equation*}
\begin{array}{rcl}
\Gamma \left( \left( \rho ,\eta \right) T\overset{\ast }{E},\left( \rho
,\eta \right) \tau _{\overset{\ast }{E}},\overset{\ast }{E}\right) & ^{%
\underrightarrow{\ \ \overset{\ast }{\mathcal{H}}\ \ }} & \Gamma \left(
\left( \rho ,\eta \right) T\overset{\ast }{E},\left( \rho ,\eta \right) \tau
_{\overset{\ast }{E}},\overset{\ast }{E}\right) \vspace*{1,5mm} \\
\tilde{Z}^{\alpha }\overset{\ast }{\tilde{\delta}}_{\alpha }+Y_{a}\overset{%
\cdot }{\tilde{\partial}}^{a} & \longmapsto & \tilde{Z}^{\alpha }\overset{%
\ast }{\tilde{\delta}}_{\alpha }%
\end{array}%
\end{equation*}%
is a projector which will be called the \emph{horizontal projector.}

\noindent\textbf{Remark 6.2.1.2} We have $\ \overset{\ast }{\mathcal{H}}%
\left( \overset{\ast }{\tilde{\delta}}_{\alpha }\right) {=}\overset{\ast }{%
\tilde{\delta}}_{\alpha }$ and $\ \overset{\ast }{\mathcal{H}}\big(\overset{%
\cdot }{\tilde{\partial}}^{a}\big){=}0.$ Therefore, we obtain $\ \overset{%
\ast }{\mathcal{H}}\left( \overset{\ast }{\tilde{\partial}}_{\alpha }\right)
{=}\overset{\ast }{\tilde{\delta}}_{\alpha }.$

\bigskip\noindent \textbf{Theorem 6.2.1.2 }\emph{A }$\left( \rho ,\eta
\right) $\emph{-connection for the vector bundle} $\left( \overset{\ast }{E},%
\overset{\ast }{\pi },M\right) $ \emph{is characterized by the existence of
a }$\mathbf{Mod}$\emph{-endomorphism}$\ \overset{\ast }{\mathcal{H}}$\emph{\
of}
\begin{equation*}
\left( \Gamma \left( \left( \rho ,\eta \right) T\overset{\ast }{E},\left(
\rho ,\eta \right) \tau _{\overset{\ast }{E}},\overset{\ast }{E}\right)
,+,\cdot \right)
\end{equation*}%
\emph{\ with the properties:}
\begin{equation*}
\begin{array}{c}
\ \Gamma \left( \left( \rho ,\eta \right) T\overset{\ast }{E},\left( \rho
,\eta \right) \tau _{\overset{\ast }{E}},\overset{\ast }{E}\right) \subset
\Gamma \left( H\left( \rho ,\eta \right) T\overset{\ast }{E},\left( \rho
,\eta \right) \tau _{\overset{\ast }{E}},\overset{\ast }{E}\right) \vspace*{%
1,5mm} \\
\ \overset{\ast }{\mathcal{H}}\left( X\right) =X\Longleftrightarrow X\in
\Gamma \left( H\left( \rho ,\eta \right) T\overset{\ast }{E},\left( \rho
,\eta \right) \tau _{\overset{\ast }{E}},\overset{\ast }{E}\right) .%
\end{array}%
\leqno(6.2.1.3)
\end{equation*}

\smallskip\noindent \textbf{Corollary 6.2.1.1} \emph{A }$\left( \rho ,\eta
\right) $\emph{-connection for the vector bundle} $\left( \overset{\ast }{E},%
\overset{\ast }{\pi },M\right) $ \emph{is characterized by the existence of
a }$\mathbf{Mod}$\emph{-endomorphism }$\ \overset{\ast }{\mathcal{H}}$\emph{%
\ of}
\begin{equation*}
\left( \Gamma \left( \left( \rho ,\eta \right) T\overset{\ast }{E},\left(
\rho ,\eta \right) \tau _{\overset{\ast }{E}},\overset{\ast }{E}\right)
,+,\cdot \right)
\end{equation*}%
\emph{\ with the properties:}
\begin{equation*}
\begin{array}{c}
\ \overset{\ast }{\mathcal{H}}^{2}=\ \overset{\ast }{\mathcal{H}} \\
Ker\left( \overset{\ast }{\mathcal{H}}\right) =\left( \Gamma \left( V\left(
\rho ,\eta \right) T\overset{\ast }{E},\left( \rho ,\eta \right) \tau _{%
\overset{\ast }{E}},\overset{\ast }{E}\right) ,+,\cdot \right) .%
\end{array}%
\leqno(6.2.1.4)
\end{equation*}

\smallskip\noindent \textbf{Remark 6.2.1.3 }For any
\begin{equation*}
X\in \Gamma \left( \left( \rho ,\eta \right) T\overset{\ast }{E},\left( \rho
,\eta \right) \tau _{\overset{\ast }{E}},\overset{\ast }{E}\right)
\end{equation*}%
we obtain the following unique decomposition
\begin{equation*}
X=\ \overset{\ast }{\mathcal{H}}X+\ \overset{\ast }{\mathcal{V}}X.
\end{equation*}

\smallskip \noindent \textbf{Proposition 6.2.1.1} \emph{After some
calculations we obtain }%
\begin{equation*}
\begin{array}{c}
N_{\overset{\ast }{\mathcal{V}}}\left( X,Y\right) =\overset{\ast }{\mathcal{V%
}}\left[ \ \overset{\ast }{\mathcal{H}}X,\ \overset{\ast }{\mathcal{H}}Y%
\right] _{\left( \rho ,\eta \right) T\overset{\ast }{E}}=N_{\ \overset{\ast }%
{\mathcal{H}}}\left( X,Y\right) ,%
\end{array}%
\leqno(6.2.1.5)
\end{equation*}%
for any $X,Y\in \Gamma \left( \left( \rho ,\eta \right) T\overset{\ast }{E}%
,\left( \rho ,\eta \right) \tau _{\overset{\ast }{E}},\overset{\ast }{E}%
\right) .$

\bigskip\noindent\textbf{Corollary 6.2.1.2} \emph{The horizontal interior
differential system }%
\begin{equation*}
\left( H\left( \rho ,\eta \right) T\overset{\ast }{E},\left( \rho ,\eta
\right) \tau _{\overset{\ast }{E}},\overset{\ast }{E}\right)
\end{equation*}%
\emph{is involutive if and only if }$N_{\overset{\ast }{\mathcal{V}}}=0$%
\emph{\ or }$N_{\overset{\ast }{\mathcal{H}}}=0.$

\subsubsection{The almost product structure}

\medskip \textbf{Definition 6.2.2.1 }Any $\mathbf{Mod}$-endomorphism $e$ of
\begin{equation*}
\left( \Gamma \left( \left( \rho ,\eta \right) T\overset{\ast }{E},\left(
\rho ,\eta \right) \tau _{\overset{\ast }{E}},\overset{\ast }{E}\right)
,+,\cdot \right)
\end{equation*}%
with the property%
\begin{equation*}
\begin{array}{c}
e^{2}=Id%
\end{array}%
\leqno(6.2.2.1)
\end{equation*}%
will be called the \emph{almost product structure}\textit{.}

\bigskip\noindent \textbf{Example 6.2.2.1 }The $\mathbf{Mod}$-endomorphism
\begin{equation*}
\begin{array}{rcl}
\Gamma \left( \left( \rho ,\eta \right) T\overset{\ast }{E},\left( \rho
,\eta \right) \tau _{\overset{\ast }{E}},\overset{\ast }{E}\right) & ^{%
\underrightarrow{\ \ \overset{\ast }{\mathcal{P}}\ \ }} & \Gamma \left(
\left( \rho ,\eta \right) T\overset{\ast }{E},\left( \rho ,\eta \right) \tau
_{\overset{\ast }{E}},\overset{\ast }{E}\right) \vspace*{1,5mm} \\
\tilde{Z}^{\alpha }\overset{\ast }{\tilde{\delta}}_{\alpha }+Y_{a}\overset{%
\cdot }{\tilde{\partial}}^{a} & \longmapsto & \tilde{Z}^{\alpha }\overset{%
\ast }{\tilde{\delta}}_{\alpha }-Y_{a}\overset{\cdot }{\tilde{\partial}}^{a}%
\end{array}%
\end{equation*}%
is an almost product structure\textbf{.}

\bigskip\noindent\textbf{Remark 6.2.2.1 }The previous almost product
structure has the properties:
\begin{equation*}
\begin{array}{l}
\overset{\ast }{\mathcal{P}}=2\overset{\ast }{\mathcal{H}}-Id; \\
\overset{\ast }{\mathcal{P}}=Id-2\overset{\ast }{\mathcal{V}}; \\
\overset{\ast }{\mathcal{P}}=\overset{\ast }{\mathcal{H}}-\overset{\ast }{%
\mathcal{V}}.%
\end{array}%
\leqno(6.2.2.2)
\end{equation*}

\smallskip\noindent \textbf{Remark 6.2.2.2 } We obtain that $\overset{\ast }{%
\mathcal{P}}\left( \overset{\ast }{\tilde{\delta}}_{\alpha }\right) =\overset%
{\ast }{\tilde{\delta}}_{\alpha }$ and $\overset{\ast }{\mathcal{P}}\left(
\overset{\cdot }{\tilde{\partial}}^{a}\right) =-\overset{\cdot }{\tilde{%
\partial}}^{a}.$ Therefore, it follows \vspace*{-2mm}
\begin{equation*}
\overset{\ast }{\mathcal{P}}\left( \overset{\ast }{\tilde{\partial}}_{\alpha
}\right) =\overset{\ast }{\tilde{\delta}}_{\alpha }-\rho \overset{\ast }{%
\Gamma }_{b\alpha }\overset{\cdot }{\tilde{\partial}}^{b}.
\end{equation*}

\smallskip\noindent \textbf{Theorem 6.2.2.1 }\emph{A }$\left( \rho ,\eta
\right) $\emph{-connection for the vector bundle }$\left( \overset{\ast }{E},%
\overset{\ast }{\pi },M\right) $ \emph{is characterized by the existence of
a }$\mathbf{Mod}$\emph{-endomorphism }$\overset{\ast }{\mathcal{P}}$\emph{\
of }%
\begin{equation*}
\left( \Gamma \left( \left( \rho ,\eta \right) T\overset{\ast }{E},\left(
\rho ,\eta \right) \tau _{\overset{\ast }{E}},\overset{\ast }{E}\right)
,+,\cdot \right)
\end{equation*}%
\emph{\ with the following property:}
\begin{equation*}
\begin{array}{c}
\overset{\ast }{\mathcal{P}}\left( X\right) =-X\Longleftrightarrow X\in
\Gamma \left( V\left( \rho ,\eta \right) T\overset{\ast }{E},\left( \rho
,\eta \right) \tau _{\overset{\ast }{E}},\overset{\ast }{E}\right) .%
\end{array}%
\leqno(6.2.2.3)
\end{equation*}

\smallskip\noindent \textbf{Proposition 6.2.2.1} \emph{After some
calculations, we obtain }%
\begin{equation*}
N_{\overset{\ast }{\mathcal{P}}}\left( X,Y\right) =4\overset{\ast }{\mathcal{%
V}}\left[ \overset{\ast }{\mathcal{H}}X,\overset{\ast }{\mathcal{H}}Y\right]
,
\end{equation*}%
\emph{for any }$X,Y\in \Gamma \left( \left( \rho ,\eta \right) T\overset{%
\ast }{E},\left( \rho ,\eta \right) \tau _{\overset{\ast }{E}},\overset{\ast
}{E}\right) .$

\medskip\noindent \textbf{Corollary 6.2.2.1} \emph{The horizontal interior
differential system }%
\begin{equation*}
\left( H\left( \rho ,\eta \right) T\overset{\ast }{E},\left( \rho ,\eta
\right) \tau _{\overset{\ast }{E}},\overset{\ast }{E}\right)
\end{equation*}%
\emph{is involutive if and only if} $N_{\overset{\ast }{\mathcal{P}}}=0.$

\subsubsection{\noindent The almost tangent structure}

\medskip \textbf{Definition 6.2.3.1 }Any $\mathbf{Mod}$-endomorphism $e$ of
\begin{equation*}
\left( \Gamma \!(\left( \rho ,\eta \right) T\overset{\ast }{E},\left( \rho
,\eta \right) \tau _{\overset{\ast }{E}},\overset{\ast }{E}),+,\cdot \right)
\end{equation*}%
with the property
\begin{equation*}
%
\leqno(6.2.3.1)
\end{equation*}%
will be called the \emph{almost tangent structure.}

\bigskip\noindent\textbf{Example 6.2.3.1 }If $\left( E,\pi ,M\right) =\left(
F,\nu ,N\right) $ and $g\in \mathbf{Man}\left( \overset{\ast }{E},E\right) $
such that $\left( g,h\right) $ is a $\mathbf{B}^{\mathbf{v}}$\textit{-}%
morphism locally invertible, then the $\mathbf{Mod}$-endomorphism
\begin{equation*}
%
%
\end{equation*}%
is an almost tangent structure which will be called the \emph{almost tangent
structure asso\-cia\-ted to }$\mathbf{B}^{\mathbf{v}}$\emph{-morphism }$%
\left( g,h\right) $. (See: \textbf{Definition 4.4.2.3})

\bigskip\noindent\textbf{Remark 6.2.3.1 }We obtain that
\begin{equation*}
%
%
\end{equation*}

\smallskip \noindent\textbf{Remark 6.2.3.2} The previous almost tangent
structure has the following properties:
\begin{equation*}
%
%
\leqno(6.2.3.2)
\end{equation*}

\subsubsection{The almost complex structure}

Let us consider in the case $\left( E,\pi ,M\right) =\left( F,\nu ,N\right) $%
.

bk\noindent\textbf{Definition 6.2.4.1 }Any $\mathbf{Mod}$-endomorphism $e$
of
\begin{equation*}
\left( \Gamma (\left( \rho ,\eta \right) T\overset{\ast }{E},\left( \rho
,\eta \right) \tau _{\overset{\ast }{E}},\overset{\ast }{E}),+,\cdot \right)
\end{equation*}%
with the property
\begin{equation*}
%
%
\leqno(6.2.4.1)
\end{equation*}%
will be called the \emph{almost complex structure}.

\bigskip\noindent \textbf{Example 6.2.4.1} If $\left( g,h\right) $ is a $%
\mathbf{B}^{\mathbf{v}}$\textit{-}morphism of $\left( \overset{\ast }{E},%
\overset{\ast }{\pi },M\right) $ source and $\left( E,\pi ,M\right) $ target
locally invertible, then the $\mathbf{Mod}$-endomorphism
\begin{equation*}
%
%
\end{equation*}%
is an almost complex structure.

\bigskip\noindent\textbf{Remark 6.2.4.1 }We have
\begin{equation*}
%
%
\end{equation*}%
Therefore, we obtain:
\begin{equation*}
%
%
\end{equation*}

\smallskip\noindent \textbf{Remark 6.2.4.2 }The previous almost complex
structure has the following properties:
\begin{equation*}
%
%
\leqno(6.2.4.2)
\end{equation*}

\subsubsection{\noindent The $\left( \protect\rho ,\protect\eta \right) $%
-tension endomorphism}

Since
\begin{equation*}
\frac{\partial \left( \rho ,\eta \right) \overset{\ast }{\Gamma }_{b%
%TCIMACRO{\U{b4}}%
%BeginExpansion
{\acute{}}%
%EndExpansion
\alpha
%TCIMACRO{\U{b4}}%
%BeginExpansion
{\acute{}}%
%EndExpansion
}}{\partial p_{a%
%TCIMACRO{\U{b4}}%
%BeginExpansion
{\acute{}}%
%EndExpansion
}}=M_{b%
%TCIMACRO{\U{b4}}%
%BeginExpansion
{\acute{}}%
%EndExpansion
}^{b}\circ \overset{\ast }{\pi }\left( -\rho _{\alpha }^{i}\circ h\frac{%
\partial M_{b}^{a%
%TCIMACRO{\U{b4}}%
%BeginExpansion
{\acute{}}%
%EndExpansion
}\circ \overset{\ast }{\pi }}{\partial x^{i}}+\frac{\partial \left( \rho
,\eta \right) \overset{\ast }{\Gamma }_{bc}}{\partial p_{a}}M_{a}^{a%
%TCIMACRO{\U{b4}}%
%BeginExpansion
{\acute{}}%
%EndExpansion
}\circ \overset{\ast }{\pi }\right) \Lambda _{\alpha
%TCIMACRO{\U{b4}}%
%BeginExpansion
{\acute{}}%
%EndExpansion
}^{\alpha }\circ h,
\end{equation*}%
it results that
\begin{equation*}
\left( \rho ,\eta \right) \overset{\ast }{\Gamma }_{b%
%TCIMACRO{\U{b4}}%
%BeginExpansion
{\acute{}}%
%EndExpansion
\alpha
%TCIMACRO{\U{b4}}%
%BeginExpansion
{\acute{}}%
%EndExpansion
}-p_{a%
%TCIMACRO{\U{b4}}%
%BeginExpansion
{\acute{}}%
%EndExpansion
}\frac{\partial \left( \rho ,\eta \right) \overset{\ast }{\Gamma }_{b%
%TCIMACRO{\U{b4}}%
%BeginExpansion
{\acute{}}%
%EndExpansion
\alpha
%TCIMACRO{\U{b4}}%
%BeginExpansion
{\acute{}}%
%EndExpansion
}}{\partial p_{a%
%TCIMACRO{\U{b4}}%
%BeginExpansion
{\acute{}}%
%EndExpansion
}}=M_{b%
%TCIMACRO{\U{b4}}%
%BeginExpansion
{\acute{}}%
%EndExpansion
}^{b}\circ \overset{\ast }{\pi }\left( \left( \rho ,\eta \right) \overset{%
\ast }{\Gamma }_{b\alpha }-p_{a}\frac{\partial \left( \rho ,\eta \right)
\overset{\ast }{\Gamma }_{b\alpha }}{\partial p_{a}}\right) \Lambda _{\alpha
%TCIMACRO{\U{b4}}%
%BeginExpansion
{\acute{}}%
%EndExpansion
}^{\alpha }\circ h\circ \overset{\ast }{\pi },
\end{equation*}
Therefore, we can introduce the following

\bigskip\noindent \textbf{Definition 6.2.5.1 }The $\mathbf{Mod}$%
-endomorphism
\begin{equation*}
%
\leqno(6.2.5\vspace*{1mm}.1)
\end{equation*}%
will be called the $\left( \rho ,\eta \right) $\emph{-tension of }$\left(
\rho ,\eta \right) $\emph{-connection }$\left( \rho ,\eta \right) \overset{%
\ast }{\Gamma }.$

In particular, if $\left( \rho ,\eta ,h\right) {=}\left(
Id_{TM},Id_{M},Id_{M}\right) $, then we obtain the \emph{tension of
connection}~$\overset{\ast }{\Gamma }.$

\bigskip\noindent \textbf{Proposition 6.2.5.1} \emph{We obtain the following
equalities}%
\begin{equation*}
%
%
\end{equation*}

\subsection{The $\left( \protect\rho ,\protect\eta ,h\right) $-torsion and
the $\left( \protect\rho ,\protect\eta ,h\right) $-curvature of a $\left(
\protect\rho ,\protect\eta \right) $-connection}

We consider the following diagram:
\begin{equation*}
%
%
\end{equation*}%
where $\left( E,\pi ,M\right) \in \left\vert \mathbf{B}^{\mathbf{v}%
}\right\vert $ and $\left( \left( F,\nu ,N\right) ,\left[ ,\right]
_{F,h},\left( \rho ,\eta \right) \right) $ is a generalized Lie algebroid.

\medskip \noindent\textbf{Definition 6.3.1} If $\left( E,\pi ,M\right)
=\left( F,\nu ,N\right) $, then the $\mathcal{F}\left( \overset{\ast }{E}%
\right) $-bilinear application
\begin{equation*}
%
\leqno(6.3.1)
\end{equation*}%
will be called the $\left( \rho ,\eta ,h\right) $\emph{-torsion associated
to }$\left( \rho ,\eta \right) $\emph{-connection }$\left( \rho ,\eta
\right) \Gamma .$

In particular, if $h=Id_{M},$ then we obtain the $\left( \rho ,\eta \right) $%
\emph{-torsion associated to }$\left( \rho ,\eta \right) $\emph{-connection }%
$\left( \rho ,\eta \right) \overset{\ast }{\Gamma }.$

Moreover, if $\left( \rho ,\eta \right) =\left( Id_{TM},Id_{M}\right) $,
then we obtain the \emph{torsion associated to connection }$\overset{\ast }{%
\Gamma }.$

\bigskip\noindent \textbf{Remark 6.3.1 }If $\left( \rho ,\eta ,h\right)
\overset{\ast }{\mathbb{T}}$ is the $\left( \rho ,\eta ,h\right) $-torsion
associated to $\left( \rho ,\eta \right) $-connection $\left( \rho ,\eta
\right) \overset{\ast }{\Gamma }$, then
\begin{equation*}
\begin{array}{c}
\left( \rho ,\eta ,h\right) \overset{\ast }{\mathbb{T}}(X,Y)=-\left( \rho
,\eta ,h\right) \overset{\ast }{\mathbb{T}}(Y,X),\vspace*{1mm}%
\end{array}%
\leqno(6.3.2)
\end{equation*}%
for any $X,Y\in \Gamma \left( \left( \rho ,\eta \right) T\overset{\ast }{E}%
,\left( \rho ,\eta \right) \tau _{\overset{\ast }{E}},\overset{\ast }{E}%
\right) .$

\bigskip\noindent \textbf{Definition 6.3.2 }If we consider the notation
\begin{equation*}
\left( \rho ,\eta ,h\right) \overset{\ast }{\mathbb{T}}_{~bc}^{a}\overset{put%
}{=}\frac{\partial \left( \rho ,\eta \right) \overset{\ast }{\Gamma }_{bc}}{%
\partial p_{a}}-\frac{\partial \left( \rho ,\eta \right) \overset{\ast }{%
\Gamma }_{cb}}{\partial p_{a}}-L_{bc}^{a}\circ h\circ \overset{\ast }{\pi } %
\leqno(6.3.3)
\end{equation*}%
then the tensor field
\begin{equation*}
\begin{array}{l}
\left( \rho ,\eta ,h\right) \overset{\ast }{\mathbb{T}}_{~bc}^{a}\overset{%
\ast }{\tilde{\delta}}_{a}\otimes d\tilde{z}^{b}\otimes d\tilde{z}^{c}%
\end{array}%
\leqno(6.3.4)
\end{equation*}%
will be called the $\left( \rho ,\eta ,h\right) $\emph{-torsion tensor field
associated to }$\left( \rho ,\eta \right) $\emph{-connection }$\left( \rho
,\eta \right) \overset{\ast }{\Gamma }.$

\bigskip\noindent \textbf{Proposition 6.3.1 }\emph{We obtain }%
\begin{equation*}
\overset{\ast }{\mathcal{J}}_{\left( Id_{\overset{\ast }{E}},Id_{M}\right)
}\circ \left( \rho ,\eta \right) \overset{\ast }{\mathbb{T}}=0
\end{equation*}%
\emph{and }%
\begin{eqnarray*}
\left( \rho ,\eta \right) \overset{\ast }{\mathbb{T}}\left( \overset{\ast }{%
\mathcal{J}}_{\left( Id_{\overset{\ast }{E}},Id_{M}\right) }X,Y\right)
&=&\left( \rho ,\eta \right) \overset{\ast }{\mathbb{T}}\left( \overset{\ast
}{\mathcal{J}}_{\left( Id_{\overset{\ast }{E}},Id_{M}\right) }X,\overset{%
\ast }{\mathcal{J}}_{\left( Id_{\overset{\ast }{E}},Id_{M}\right) }Y\right)
\\
&=&\left( \rho ,\eta \right) \overset{\ast }{\mathbb{T}}\left( X,\overset{%
\ast }{\mathcal{J}}_{\left( Id_{\overset{\ast }{E}},Id_{M}\right) }Y\right) ,
\end{eqnarray*}%
\emph{for any }$X,Y\in \Gamma \left( \left( \rho ,\eta \right) T\overset{%
\ast }{E},\left( \rho ,\eta \right) \tau _{\overset{\ast }{E}},\overset{\ast
}{E}\right) .$

\bigskip\noindent \textbf{Theorem 6.3.1} \emph{Using the }$\left( \rho ,\eta
\right) $\emph{-tension tensor field }%
\begin{equation*}
\left( \rho ,\eta \right) \overset{\ast }{\mathbb{H}}_{ba}\overset{\cdot }{%
\tilde{\partial}}^{b}\otimes d\tilde{z}^{a}=\left( \left( \rho ,\eta \right)
\overset{\ast }{\Gamma }_{ba}-p_{c}\frac{\partial \left( \rho ,\eta \right)
\overset{\ast }{\Gamma }_{ba}}{\partial p_{c}}\right) \overset{\cdot }{%
\tilde{\partial}}^{b}\otimes d\tilde{z}^{a}, \leqno(6.3.5)
\end{equation*}%
\emph{and the }$\left( \rho ,\eta ,h\right) $\emph{-deflection of the }$%
\left( \rho ,\eta \right) $\emph{-connection }$\left( \rho ,\eta \right)
\overset{\ast }{\Gamma }$\emph{\ }%
\begin{equation*}
\left( \rho ,\eta ,h\right) \overset{\ast }{\mathbb{D}}_{bc}=-\left( \rho
,\eta \right) \overset{\ast }{\Gamma }_{bc}+p_{a}\frac{\partial \left( \rho
,\eta \right) \overset{\ast }{\Gamma }_{cb}}{\partial p_{a}}+p_{a}\cdot
L_{bc}^{a}\circ h\circ \overset{\ast }{\pi }, \leqno(6.3.6)
\end{equation*}%
\emph{we obtain that }$(\rho ,\eta ,h)\overset{\ast }{\mathbb{D}}_{bc}{=}0$
\emph{if and only if} $(\rho ,\eta )\overset{\ast }{\mathbb{H}}_{bc}{=}0$
\emph{and} $(\rho ,\eta ,h)\overset{\ast }{\mathbb{T}}_{bc}^{a}{=}0.$

\bigskip\noindent\textit{Proof.} If $\left( \rho ,\eta ,h\right) \overset{%
\ast }{\mathbb{D}}_{bc}\mathbb{=}0,$ then deriving with respect to $p_{a},$
we obtain:
\begin{equation*}
-\frac{\partial \left( \rho ,\eta \right) \overset{\ast }{\Gamma }_{bc}}{%
\partial p_{a}}+\frac{\partial \left( \rho ,\eta \right) \overset{\ast }{%
\Gamma }_{cb}}{\partial p_{a}}+L_{bc}^{a}\circ h\circ \overset{\ast }{\pi }%
=0\Longleftrightarrow \left( \rho ,\eta ,h\right) \overset{\ast }{\mathbb{T}}%
_{bc}^{a}=0.
\end{equation*}
The equality $\left( \rho ,\eta ,h\right) \overset{\ast }{\mathbb{D}}_{bc}%
\mathbb{=}0$ implies:
\begin{equation*}
\left( \rho ,\eta \right) \overset{\ast }{\Gamma }_{bc}=p_{a}\frac{\partial
\left( \rho ,\eta \right) \overset{\ast }{\Gamma }_{cb}}{\partial p_{a}}%
+p_{a}L_{bc}^{a}\circ h\circ \overset{\ast }{\pi }. \leqno(1)
\end{equation*}
Since
\begin{equation*}
\begin{array}{ll}
\displaystyle\left( \rho ,\eta \right) \overset{\ast }{\mathbb{H}}_{bc} & %
\displaystyle=\left( \rho ,\eta \right) \overset{\ast }{\Gamma }_{bc}-p_{a}%
\frac{\partial \left( \rho ,\eta \right) \overset{\ast }{\Gamma }_{bc}}{%
\partial p_{a}} \vspace*{1mm} \\
& \displaystyle=p_{a}\frac{\partial \left( \rho ,\eta \right) \overset{\ast }%
{\Gamma }_{cb}}{\partial p_{a}}-p_{a}\frac{\partial \left( \rho ,\eta
\right) \overset{\ast }{\Gamma }_{bc}}{\partial p_{a}}+p_{a}L_{bc}^{a}\circ
h\circ \overset{\ast }{\pi }=p_{a}\left( \rho ,\eta ,h\right) \overset{\ast }%
{\mathbb{T}}_{bc}^{a}%
\end{array}%
\end{equation*}%
it results the equality $\left( \rho ,\eta \right) \overset{\ast }{\mathbb{H}%
}_{bc}\mathbb{=}0.$\medskip

Conservely, if $\left( \rho ,\eta ,h\right) \overset{\ast }{\mathbb{T}}%
_{bc}^{a}\mathbb{=}0,$ then, multiplying with $p_{a},$ we obtain:
\begin{equation*}
p_{a}\frac{\partial \left( \rho ,\eta \right) \overset{\ast }{\Gamma }_{cb}}{%
\partial p_{a}}-p_{a}\frac{\partial \left( \rho ,\eta \right) \overset{\ast }%
{\Gamma }_{bc}}{\partial p_{a}}+p_{a}L_{bc}^{a}\circ h\circ \overset{\ast }{%
\pi }=0.\leqno(2)
\end{equation*}

The equality $\left( \rho ,\eta \right) \overset{\ast }{\mathbb{H}}_{bc}%
\mathbb{=}0$ is equivalent with:
\begin{equation*}
\left( \rho ,\eta \right) \overset{\ast }{\Gamma }_{bc}=p_{a}\frac{\partial
\left( \rho ,\eta \right) \overset{\ast }{\Gamma }_{bc}}{\partial p_{a}}. %
\leqno(3)
\end{equation*}

Using $\left( 2\right) $ and $\left( 3\right) $, it results the equality $%
\left( \rho ,\eta ,h\right) \overset{\ast }{\mathbb{D}}_{bc}=0.$ \hfill
\emph{q.e.d.}

\bigskip \noindent\textbf{Definition 6.3.3 }The $\mathcal{F}\left( \overset{%
\ast }{E}\right) $-bilinear application
\begin{equation*}
\begin{array}{r}
\Gamma \left( \left( \rho ,\eta \right) T\overset{\ast }{E},\left( \rho
,\eta \right) \tau _{\overset{\ast }{E}},\overset{\ast }{E}\right) ^{2}\,\
\,^{\underrightarrow{\left( \rho ,\eta ,h\right) \overset{\ast }{\mathbb{R}}}%
}\,\,\ \Gamma \left( \left( \rho ,\eta \right) T\overset{\ast }{E},\left(
\rho ,\eta \right) \tau _{\overset{\ast }{E}},\overset{\ast }{E}\right)%
\end{array}%
\end{equation*}%
defined by
\begin{equation*}
\begin{array}{l}
\left( \rho ,\eta ,h\right) \overset{\ast }{\mathbb{R}}\left( \overset{\ast }%
{\tilde{\delta}}_{\alpha },\overset{\ast }{\tilde{\delta}}_{\beta }\right)
=\left( \rho ,\eta ,h\right) \overset{\ast }{\mathbb{R}}_{\,b~\alpha \beta }%
\overset{\cdot }{\tilde{\partial}}^{b}; \vspace*{1,5mm} \\
\left( \rho ,\eta ,h\right) \overset{\ast }{\mathbb{R}}\left( \overset{\ast }%
{\tilde{\delta}}_{\alpha },\overset{\cdot }{\tilde{\partial}}^{b}\right)
=0=\left( \rho ,\eta ,h\right) \overset{\ast }{\mathbb{R}}\left( \overset{%
\cdot }{\tilde{\partial}}^{b},\overset{\ast }{\tilde{\delta}}_{\alpha
}\right) ; \vspace*{1,5mm} \\
\left( \rho ,\eta ,h\right) \overset{\ast }{\mathbb{R}}\left( \overset{\cdot
}{\tilde{\partial}}^{a},\overset{\cdot }{\tilde{\partial}}^{b}\right) =0;%
\end{array}%
\leqno(6.3.7)
\end{equation*}%
will be called the $\left( \rho ,\eta ,h\right) $\emph{-curvature associated
to }$\left( \rho ,\eta \right) $\emph{-connection }$\left( \rho ,\eta
\right) \overset{\ast }{\Gamma }.$

In particular, if $\left( \rho ,\eta ,h\right) =\left(
Id_{TM},Id_{M},Id_{M}\right) $, then we obtain the \emph{curvature
asso\-cia\-ted to connection }$\overset{\ast }{\Gamma }.$

\bigskip\noindent \textbf{Remark 6.3.2 }If $\left( \rho ,\eta ,h\right)
\overset{\ast }{\mathbb{R}}$ is the $\left( \rho ,\eta ,h\right) $-curvature
associated to $\left( \rho ,\eta \right) $-connection $\left( \rho ,\eta
\right) \overset{\ast }{\Gamma }$, then
\begin{equation*}
\begin{array}{c}
\left( \rho ,\eta ,h\right) \overset{\ast }{\mathbb{R}}\left( X,Y\right)
=-\left( \rho ,\eta ,h\right) \overset{\ast }{\mathbb{R}}\left( Y,X\right) ,%
\end{array}%
\vspace*{1mm}\leqno(6.3.8)
\end{equation*}%
for any $X,Y\in \Gamma \left( \left( \rho ,\eta \right) T\overset{\ast }{E}%
,\left( \rho ,\eta \right) \tau _{\overset{\ast }{E}},\overset{\ast }{E}%
\right) .$

\bigskip\noindent \textbf{Definition 4.3.4 }The tensor field
\begin{equation*}
\begin{array}{l}
\left( \rho ,\eta ,h\right) \overset{\ast }{\mathbb{R}}_{\,b\ \alpha \beta }%
\overset{\cdot }{\tilde{\partial}}^{b}\otimes d\tilde{z}^{\alpha }\otimes d%
\tilde{z}^{\beta }%
\end{array}%
\leqno(4.3.9)
\end{equation*}%
will be called the $\left( \rho ,\eta ,h\right) $\emph{-curvature tensor
field associated to the }$\left( \rho ,\eta \right) $\emph{-connection }$%
\left( \rho ,\eta \right) \overset{\ast }{\Gamma }.$

Using equality $\left( 4.1.5\right) $ we obtain

\bigskip\noindent \textbf{Remark 6.3.3} The horizontal interior differential
system $\left( H\left( \rho ,\eta \right) T\overset{\ast }{E},\left( \rho
,\eta \right) \tau _{\overset{\ast }{E}},\overset{\ast }{E}\right) $ is
involutive if and only if the $\left( \rho ,\eta ,h\right) $-curvature
tensor field associated to the $\left( \rho ,\eta \right) $-connection $%
\left( \rho ,\eta \right) \overset{\ast }{\Gamma }$ is null.

\subsection{\protect\smallskip Tensor $d$-fields. Distinguished linear $%
\left( \protect\rho ,\protect\eta \right) $-connections}

We consider the following diagram:
\begin{equation*}
\begin{array}{c}
\xymatrix{\overset{\ast }{E}\ar[d]_{\overset{\ast }{\pi }}&\left( F,\left[
,\right] _{F,h},\left( \rho ,\eta \right) \right)\ar[d]^\nu\\ M\ar[r]^h&N}%
\end{array}%
\end{equation*}%
where $\left( E,\pi ,M\right) \in \left\vert \mathbf{B}^{\mathbf{v}%
}\right\vert $ and $\left( \left( F,\nu ,N\right) ,\left[ ,\right]
_{F,h},\left( \rho ,\eta \right) \right) $ is a generalized Lie algebroid.

Let
\begin{equation*}
\left( \mathcal{T}~_{q,s}^{p,r}\left( \left( \rho ,\eta \right) T\overset{%
\ast }{E},\left( \rho ,\eta \right) \tau _{\overset{\ast }{E}},\overset{\ast
}{E}\right) ,+,\cdot \right)
\end{equation*}%
be the $\mathcal{F}\left( \overset{\ast }{E}\right) $-module of tensor
fields by $\left( _{q,s}^{p,r}\right) $-type from the generalized tangent
bundle
\begin{equation*}
\left( H\left( \rho ,\eta \right) T\overset{\ast }{E}\oplus V\left( \rho
,\eta \right) T\overset{\ast }{E},\left( \rho ,\eta \right) \tau _{\overset{%
\ast }{E}},\overset{\ast }{E}\right) .
\end{equation*}

An arbitrarily tensor field $T$ is written by the form:
\begin{equation*}
T=T_{\beta _{1}...\beta _{q}b_{1}...b_{s}}^{\alpha _{1}...\alpha
_{p}a_{1}...a_{r}}\overset{\ast }{\tilde{\delta}}_{\alpha _{1}}\otimes
...\otimes \overset{\ast }{\tilde{\delta}}_{\alpha _{p}}\otimes d\tilde{z}%
^{\beta _{1}}\otimes ...\otimes d\tilde{z}^{\beta _{q}} \otimes \overset{%
\cdot }{\tilde{\partial}}^{b_{1}}\otimes ...\otimes \overset{\cdot }{\tilde{%
\partial}}^{b_{s}}\otimes \delta \tilde{p}_{a_{1}}\otimes ...\otimes \delta
\tilde{p}_{a_{r}}.
\end{equation*}

Let
\begin{equation*}
\left( ~\mathcal{T}\left( \left( \rho ,\eta \right) T\overset{\ast }{E}%
,\left( \rho ,\eta \right) \tau _{\overset{\ast }{E}},\overset{\ast }{E}%
\right) ,+,\cdot ,\otimes \right)
\end{equation*}%
be the tensor fields algebra of generalized tangent bundle $\left( \left(
\rho ,\eta \right) T\overset{\ast }{E},\left( \rho ,\eta \right) \tau _{%
\overset{\ast }{E}},\overset{\ast }{E}\right) $.

If $T_{1}{\in }\mathcal{T}_{q_{1},s_{1}}^{p_{1},r_{1}}\left( \left( \rho
,\eta \right) T\overset{\ast }{E},\left( \rho ,\eta \right) \tau _{\overset{%
\ast }{E}},\overset{\ast }{E}\right) $ and $T_{2}{\in }\mathcal{T}%
_{q_{2},s_{2}}^{p_{2},r_{2}}\left( \left( \rho ,\eta \right) T\overset{\ast }%
{E},\left( \rho ,\eta \right) \tau _{\overset{\ast }{E}},\overset{\ast }{E}%
\right) $, then the components of product tensor field $T_{1}\otimes T_{2}$
are the products of local components of $T_{1}$ and $T_{2}.$

Therefore, we obtain $T_{1}\otimes T_{2}\in \mathcal{T}%
_{q_{1}+q_{2},s_{1}+s_{2}}^{p_{1}+p_{2},r_{1}+r_{2}}\left( \left( \rho ,\eta
\right) T\overset{\ast }{E},\left( \rho ,\eta \right) \tau _{\overset{\ast }{%
E}},\overset{\ast }{E}\right) .$

Let $\mathcal{DT}\left( \left( \rho ,\eta \right) T\overset{\ast }{E},\left(
\rho ,\eta \right) \tau _{\overset{\ast }{E}},\overset{\ast }{E}\right) $ be
the family of tensor fields
\begin{equation*}
T\in \mathcal{T}\left( \left( \rho ,\eta \right) T\overset{\ast }{E},\left(
\rho ,\eta \right) \tau _{\overset{\ast }{E}},\overset{\ast }{E}\right)
\end{equation*}%
for which there exists%
\begin{equation*}
T_{1}{\in }\mathcal{T}_{q,0}^{p,0}\left( \left( \rho ,\eta \right) T\overset{%
\ast }{E},\left( \rho ,\eta \right) \tau _{\overset{\ast }{E}},\overset{\ast
}{E}\right)
\end{equation*}%
and%
\begin{equation*}
T_{2}{\in }\mathcal{T}_{0,s}^{0,r}\left( \left( \rho ,\eta \right) T\overset{%
\ast }{E},\left( \rho ,\eta \right) \tau _{\overset{\ast }{E}},\overset{\ast
}{E}\right)
\end{equation*}
such that $T=T_{1}+T_{2}.$

The $\mathcal{F}\left( \overset{\ast }{E}\right) $-module $\left( \mathcal{DT%
}\left( \left( \rho ,\eta \right) T\overset{\ast }{E},\left( \rho ,\eta
\right) \tau _{\overset{\ast }{E}},\overset{\ast }{E}\right) ,+,\cdot
\right) $ will be called the \emph{module of distinguished tensor fields} or
the \emph{module of tensor }$d$-\emph{fields.}

\bigskip\noindent\textbf{Remark 6.4.1 }The elements of
\begin{equation*}
\Gamma \left( \left( \rho ,\eta \right) T\overset{\ast }{E},\left( \rho
,\eta \right) \tau _{\overset{\ast }{E}},\overset{\ast }{E}\right)
\end{equation*}%
respectively
\begin{equation*}
\Gamma \left( \left( \left( \rho ,\eta \right) T\overset{\ast }{E}\right)
^{\ast },\break ((\rho ,\eta )\tau _{\overset{\ast }{E}})^{\ast },\overset{%
\ast }{E}\right)
\end{equation*}%
are tensor $d$-fields.

\bigskip\noindent \textbf{Definition 6.4.1 }Let $\left( \rho ,\eta \right)
\overset{\ast }{\Gamma }$ be a $\left( \rho ,\eta \right) $-connection $\ $%
for the vector bundle $\left( \overset{\ast }{E},\overset{\ast }{\pi }%
,M\right) $ and let
\begin{equation*}
\begin{array}{l}
\left( X,T\right) ^{\underrightarrow{\left( \rho ,\eta \right) \overset{\ast
}{D}}\,}\vspace*{1mm}\left( \rho ,\eta \right) \overset{\ast }{D}_{X}T%
\end{array}%
\leqno(6.4.1)
\end{equation*}%
be a covariant $\left( \rho ,\eta \right) $-derivative for the tensor
algebra of generalized tangent bundle
\begin{equation*}
\left( \left( \rho ,\eta \right) T\overset{\ast }{E},\left( \rho ,\eta
\right) \tau _{\overset{\ast }{E}},\overset{\ast }{E}\right)
\end{equation*}%
which \ preserves \ the \ horizontal and vertical distributions by
parallelism.

If $\left( U,\overset{\ast }{s}_{U}\right) $ is a vector local $\left(
m+r\right) $-chart for $\left( \overset{\ast }{E},\overset{\ast }{\pi }%
,M\right) ,$ then the real local functions
\begin{equation*}
\left( \left( \rho ,\eta \right) \overset{\ast }{H}_{\beta \gamma }^{\alpha
},\left( \rho ,\eta \right) \overset{\ast }{H}_{b\gamma }^{a},\left( \rho
,\eta \right) \overset{\ast }{V}_{\beta }^{\alpha c},\left( \rho ,\eta
\right) \overset{\ast }{V}_{a}^{bc}\right)
\end{equation*}%
defined on $\overset{\ast }{\pi }^{-1}\left( U\right) $ and determined by
the following equalities:
\begin{equation*}
\begin{array}{ll}
\left( \rho ,\eta \right) \overset{\ast }{D}_{\overset{\ast }{\tilde{\delta}}%
_{\gamma }}\overset{\ast }{\tilde{\delta}}_{\beta }=\left( \rho ,\eta
\right) \overset{\ast }{H}_{\beta \gamma }^{\alpha }\overset{\ast }{\tilde{%
\delta}}_{\alpha }, & \left( \rho ,\eta \right) \overset{\ast }{D}_{\overset{%
\ast }{\tilde{\delta}}_{\gamma }}\overset{\cdot }{\tilde{\partial}}%
^{a}=\left( \rho ,\eta \right) \overset{\ast }{H}_{b\gamma }^{a}\overset{%
\cdot }{\tilde{\partial}}^{b} \\
\left( \rho ,\eta \right) \overset{\ast }{D}_{\overset{\cdot }{\tilde{%
\partial}}^{c}}\overset{\ast }{\tilde{\delta}}_{\beta }=\left( \rho ,\eta
\right) \overset{\ast }{V}_{\beta }^{\alpha c}\overset{\ast }{\tilde{\delta}}%
_{\alpha }, & \left( \rho ,\eta \right) \overset{\ast }{D}_{\overset{\cdot }{%
\tilde{\partial}}^{c}}\overset{\cdot }{\tilde{\partial}}^{b}=\left( \rho
,\eta \right) \overset{\ast }{V}_{a}^{bc}\overset{\cdot }{\tilde{\partial}}%
^{a}%
\end{array}%
\leqno(6.4.2)
\end{equation*}%
are the components of a linear $\left( \rho ,\eta \right) $-connection
\begin{equation*}
\left( \left( \rho ,\eta \right) \overset{\ast }{H},\left( \rho ,\eta
\right) \overset{\ast }{V}\right)
\end{equation*}%
for the generalized tangent bundle $\left( \left( \rho ,\eta \right) T%
\overset{\ast }{E},\left( \rho ,\eta \right) \tau _{\overset{\ast }{E}},%
\overset{\ast }{E}\right) $ which will be called the \emph{distinguished
linear }$\left( \rho ,\eta \right) $\emph{-connection.}

If $h=Id_{M},$ then the distinguished linear $\left( Id_{TM},Id_{M}\right) $%
-connection will be called the \emph{\ distinguished linear connection.}

The components of a distinguished linear connection $\left( \overset{\ast }{H%
},\overset{\ast }{V}\right) $ will be denoted
\begin{equation*}
\left( \overset{\ast }{H}_{jk}^{i},\overset{\ast }{H}_{bk}^{a},\overset{\ast
}{V}_{j}^{ic},\overset{\ast }{V}_{a}^{bc}\right) .
\end{equation*}

\smallskip \noindent\textbf{Theorem 6.4.1 }\emph{If }$(\left( \rho ,\eta
\right) \overset{\ast }{H},\left( \rho ,\eta \right) \overset{\ast }{V})$
\emph{is a distinguished linear} $(\rho ,\eta )$\emph{-connection for the
generalized tangent bundle }$\left( \left( \rho ,\eta \right) T\overset{\ast
}{E},\left( \rho ,\eta \right) \tau _{\overset{\ast }{E}},\overset{\ast }{E}%
\right) $\emph{, then its components satisfy the change relations: }%
\begin{equation*}
%
\leqno(6.4.3)
\end{equation*}

\emph{The components of a distinguished linear connection }$\left( \overset{%
\ast }{H},\overset{\ast }{V}\right) $\emph{\ verify the change relations: }%
\begin{equation*}
%
%
\leqno(6.4.3)^{\prime }
\end{equation*}

\smallskip\noindent \textbf{Example 6.4.1} If $\left( \overset{\ast }{E},%
\overset{\ast }{\pi },M\right) $ is endowed with the $\left( \rho ,\eta
\right) $-connection $\left( \rho ,\eta \right) \overset{\ast }{\Gamma }$,
then the local real functions%
\begin{equation*}
\left( \frac{\partial \left( \rho ,\eta \right) \overset{\ast }{\Gamma }%
_{b\gamma }}{\partial p_{a}},\frac{\partial \left( \rho ,\eta \right)
\overset{\ast }{\Gamma }_{b\gamma }}{\partial p_{a}},0,0\right)
\end{equation*}%
are the components of a distinguished linear $\left( \rho ,\eta \right) $%
\textit{-}connection for the generalized tangent bundle
\begin{equation*}
\left( \left( \rho ,\eta \right) T\overset{\ast }{E},\left( \rho ,\eta
\right) \tau _{\overset{\ast }{E}},\overset{\ast }{E}\right) ,
\end{equation*}%
which will by called the \emph{Berwald linear }$\left( \rho ,\eta \right) $%
\emph{-connection.}

The Berwald linear $(Id_{TM},Id_{M})$-connection will be called the \emph{%
Ber\-wald linear connection.}

\smallskip \noindent \textbf{Theorem 6.4.2} \emph{If the generalized tangent
bundle} $\!\left( \left( \rho ,\eta \right) T\overset{\ast }{E},\left( \rho
,\eta \right) \tau _{\overset{\ast }{E}},\overset{\ast }{E}\right) $ \emph{%
is endowed with a distinguished linear} $\!(\rho ,\!\eta )$\emph{-connection}
$((\rho ,\eta )\overset{\ast }{H},(\rho ,\eta )\overset{\ast }{V})$, \emph{%
then, for any}
\begin{equation*}
X=\tilde{Z}^{\gamma }\overset{\ast }{\tilde{\delta}}_{\gamma }+Y_{a}\overset{%
\cdot }{\tilde{\partial}}^{a}\in \Gamma \left( \left( \rho ,\eta \right) T%
\overset{\ast }{E},\left( \rho ,\eta \right) \tau _{\overset{\ast }{E}},%
\overset{\ast }{E}\right)
\end{equation*}%
\emph{and for any}
\begin{equation*}
T\in \mathcal{T~}_{qs}^{pr}\!\left( \left( \rho ,\eta \right) T\overset{\ast
}{E},\left( \rho ,\eta \right) \tau _{\overset{\ast }{E}},\overset{\ast }{E}%
\right) ,
\end{equation*}%
\emph{we obtain the formula:}
\begin{equation*}
%
\end{equation*}

\smallskip\noindent \textbf{Definition 6.4.2 }We assume that $\left( E,\pi
,M\right) =\left( F,\nu ,N\right) .$

If $\left( \rho ,\eta \right) \overset{\ast }{\Gamma }$ is a $\left( \rho
,\eta \right) $-connection for the vector bundle $\left( \overset{\ast }{E},%
\overset{\ast }{\pi },M\right) $ and
\begin{equation*}
\left( \left( \rho ,\eta \right) \overset{\ast }{H}_{bc}^{a},\left( \rho
,\eta \right) \overset{\ast }{\tilde{H}}_{bc}^{a},\left( \rho ,\eta \right)
\overset{\ast }{V}_{b}^{ac},\left( \rho ,\eta \right) \overset{\ast }{\tilde{%
V}}_{b}^{ac}\right)
\end{equation*}%
are the components of a distinguished linear $\left( \rho ,\eta \right) $%
\textit{-}connection for the generalized tangent bundle $\left( \left( \rho
,\eta \right) T\overset{\ast }{E},\left( \rho ,\eta \right) \tau _{\overset{%
\ast }{E}},\overset{\ast }{E}\right) $ such that
\begin{equation*}
\left( \rho ,\eta \right) \overset{\ast }{H}_{bc}^{a}=\left( \rho ,\eta
\right) \overset{\ast }{\tilde{H}}_{bc}^{a}\mbox{ and }\left( \rho ,\eta
\right) \overset{\ast }{V}_{b}^{ac}=\left( \rho ,\eta \right) \overset{\ast }%
{\tilde{V}}_{b}^{ac},
\end{equation*}%
then we will say that \emph{the generalized tangent bundle }$\!\left( \left(
\rho ,\eta \right) T\overset{\ast }{E},\left( \rho ,\eta \right) \tau _{%
\overset{\ast }{E}},\overset{\ast }{E}\right) $ \emph{is endowed with a
normal distinguished linear }$\left( \rho ,\eta \right) $\emph{-connection
on components }%
\begin{equation*}
\left( \left( \rho ,\eta \right) \overset{\ast }{H}_{bc}^{a},\left( \rho
,\eta \right) \overset{\ast }{V}_{b}^{ac}\right) .
\end{equation*}
The components of a normal distinguished linear $\left(
Id_{TM},Id_{M}\right) $-con\-nec\-tion $\left( \overset{\ast }{H},\overset{%
\ast }{V}\right) $ will be denoted $\left( \overset{\ast }{H}_{jk}^{i},%
\overset{\ast }{V}_{jk}^{i}\right) $.

\subsection{The lift of accelerations for a differentiable curve}

We consider the following diagram:
\begin{equation*}
\begin{array}{c}
\xymatrix{\overset{\ast }{E}\ar[d]_{\overset{\ast }{\pi }}&\left( F,\left[
,\right] _{F,h},\left( \rho ,\eta \right) \right)\ar[d]^\nu\\ M\ar[r]^h&N}%
\end{array}%
\end{equation*}%
where $\left( E,\pi ,M\right) \in \left\vert \mathbf{B}^{\mathbf{v}%
}\right\vert $ and $\left( \left( F,\nu ,N\right) ,\left[ ,\right]
_{F,h},\left( \rho ,\eta \right) \right) \in \left\vert \mathbf{GLA}%
\right\vert .$

Let $\left( \rho ,\eta \right) \overset{\ast }{\Gamma }$ be a $\left( \rho
,\eta \right) $-connection for the vector bundle $\left( \overset{\ast }{E},%
\overset{\ast }{\pi },M\right) .$

We admit that $\left( \left( \rho ,\eta \right) \overset{\ast }{H},\left(
\rho ,\eta \right) \overset{\ast }{V}\right) $ is a distinguished linear $%
\left( \rho ,\eta \right) $-connection for the vector bundle $\left( \left(
\rho ,\eta \right) T\overset{\ast }{E},\left( \rho ,\eta \right) \tau _{%
\overset{\ast }{E}},\overset{\ast }{E}\right) .$

Let $g\in \mathbf{Man}\left( \overset{\ast }{E},E\right) $ be such that $%
\left( g,h\right) $ is a $\mathbf{B}^{\mathbf{v}}$-morphism of $\left(
\overset{\ast }{E},\overset{\ast }{\pi },M\right) $ source and $\left( E,\pi
,M\right) $ target.

Let%
\begin{equation*}
%
\leqno(6.5.1)
\end{equation*}%
be the $\left( g,h\right) $-lift of differentiable curve $I ^{%
\underrightarrow{\ \ c\ \ }} M.$

\bigskip\noindent\textbf{Definition 6.5.1 }The differentiable curve
\begin{equation*}
%
%
\leqno(6.5.2)
\end{equation*}%
will be called the \emph{differentiable }$\left( g,h\right) $\emph{-lift of
accelerations of the differentiable curve }$c.$

The section
\begin{equation*}
%
%
\leqno(6.5.3)
\end{equation*}%
will be called the \emph{canonical section associated to the triple }$\left(
c,\dot{c},\ddot{c}\right) .$

\bigskip\noindent\textbf{Remark 6.5.1 }For any $t\in I$, we obtain:
\begin{equation*}
%
%
\leqno(6.5.4)
\end{equation*}

We observe easily that $\overset{\ast }{u}\left( c,\dot{c},\ddot{c}\right)
\left( \dot{c}\left( t\right) \right) \in H\left( \rho ,\eta \right) T%
\overset{\ast }{E}_{|\func{Im}\left( \dot{c}\right) }$if and only if the
components functions $\left( p_{a},~a\in \overline{1,r}\right) $ are
solutions for the differentiable equations %\begin{equation*}
%\left( 5.2.3.15\right) ~\ \
\begin{equation*}
\frac{du_{b}}{dt}+\left( \rho ,\eta \right) \overset{\ast }{\Gamma }%
_{b\alpha }\circ \overset{\ast }{u}\left( c,\dot{c}\right) \circ \eta \circ
h\circ c\cdot g^{\alpha a}\circ h\circ c\cdot u_{a},~a\in \overline{1,r}. %
\leqno(6.5.5)
\end{equation*}

\smallskip \noindent\textbf{Remark 6.5.2 }In particular, if $\left( \rho
,\eta ,h\right) =\left( Id_{TM},Id_{M},Id_{M}\right) $ , then, using the
differentiable $\left( g,Id_{M}\right) $-lift
\begin{equation*}
\begin{array}{rcl}
I & ^{\underrightarrow{\ \ \dot{c}\ \ }} & \overset{\ast }{TM} \vspace*{1,5mm%
} \\
t & \longmapsto & \tilde{g}_{ji}\left( c\left( t\right) \right) \displaystyle%
\frac{dc^{j}\left( t\right) }{dt}\cdot dx^{i}\left( c\left( t\right) \right)
,%
\end{array}%
\leqno(6.5.6)
\end{equation*}
we obtain the $\left( g,Id_{M}\right) $-lift of accelerations of the
differentiable curve $c$ as being
\begin{equation*}
\begin{array}{rcl}
I & ^{\underrightarrow{\ \ \ddot{c}\ \ }} & \left( Id_{TM},Id_{M}\right) T%
\overset{\ast }{E}_{|\func{Im}\left( \dot{c}\right) } \vspace*{1,5mm} \\
t & \longmapsto & \displaystyle\frac{dc^{i}\left( t\right) }{dt}\cdot \frac{%
\partial }{\partial \tilde{z}^{i}}\left( \dot{c}\left( t\right) \right) +%
\tilde{g}_{ji}\left( c\left( t\right) \right) \frac{dc^{j}\left( t\right) }{%
dt}\cdot \frac{\partial }{\partial \tilde{p}_{i}}\left( \dot{c}\left(
t\right) \right)%
\end{array}%
\leqno(6.5.7)
\end{equation*}

\smallskip \noindent\textbf{Definition 6.5.2 }If the component functions
\begin{equation*}
\left( \left( g^{\alpha a}\circ h\circ c\right) \cdot p_{a},~a\in \overline{%
1,r}\right)
\end{equation*}%
are solutions for the differentiable system of equations
\begin{equation*}
\frac{dz^{\alpha }}{dt}+\left( \rho ,\eta \right) \overset{\ast }{H}_{\beta
\gamma }^{\alpha }\circ \overset{\ast }{u}\left( c,\dot{c}\right) \circ \eta
\circ h\circ c\cdot z^{\beta }\cdot z^{\gamma }=0,~\alpha \in \overline{1,p}%
, \leqno(6.5.8)
\end{equation*}%
then the differentiable curve $\dot{c}$ will be called \emph{horizontal
parallel with respect to the distinguished linear }$\left( \rho ,\eta
\right) $\emph{-connection }$\left( \left( \rho ,\eta \right) \overset{\ast }%
{H},\left( \rho ,\eta \right) \overset{\ast }{V}\right) .$

If the component functions $\left( p_{a},~a\in \overline{1,r}\right) $ are
solutions for the differentiable system of equations%
\begin{equation*}
\frac{du_{b}}{dt}-\left( \rho ,\eta \right) \overset{\ast }{V}_{b}^{ac}\circ
\overset{\ast }{u}\left( c,\dot{c}\right) \circ \eta \circ h\circ c\cdot
u_{a}\cdot u_{c}=0,~b\in \overline{1,r}. \leqno(6.5.9)
\end{equation*}%
then the differentiable curve $\dot{c}$ will be called \emph{vertical
parallel with respect to the distinguished linear }$\left( \rho ,\eta
\right) $\emph{-connection }$\left( \left( \rho ,\eta \right) \overset{\ast }%
{H},\left( \rho ,\eta \right) \overset{\ast }{V}\right) .$

\bigskip \noindent \textbf{Remark 6.5.3 }In particular, if $\left( \rho
,\eta ,h\right) =\left( Id_{TM},Id_{M},Id_{M}\right) $ , then the
differentiable $\left( g,Id_{M}\right) $-lift $\left( 6.5.6\right) $ is
horizontal parallel with respect to the distinguished linear connection $%
\left( \overset{\ast }{H},\overset{\ast }{V}\right) $ if the component
functions $\left( \displaystyle\frac{dc^{j}\left( t\right) }{dt},~i\in
\overline{1,m}\right) $ are solutions for the differentiable system of
equations%
\begin{equation*}
\frac{dz^{i}(t)}{dt}+\overset{\ast }{H}_{jk}^{i}\circ \overset{\ast }{u}%
\left( c,\dot{c}\right) \circ c\cdot z^{j}\cdot z^{k}=0,~i\in \overline{1,m}.%
\leqno(6.5.11)
\end{equation*}

Moreover, the differentiable $\left( g,Id_{M}\right) $-lift $\left(
4.5.6\right) $ is vertical parallel with respect\break to the distinguished
linear connection $\left( \overset{\ast }{H},\overset{\ast }{V}\right) $ if
the component functions\break $\left( \tilde{g}_{ji}\circ c\cdot %
\displaystyle\frac{dc^{j}}{dt},~i\in \overline{1,m}\right) $ are solutions
for the differentiable system of equations%
\begin{equation*}
\frac{du_{j}}{dt}+\overset{\ast }{V}_{j}^{ik}\circ \overset{\ast }{u}\left(
c,\dot{c}\right) \circ c\cdot u_{i}\cdot u_{k}=0,~j\in \overline{1,m}. \leqno%
(6.5.12)
\end{equation*}

\subsection{The $\left( \protect\rho ,\protect\eta ,h\right) $-torsion and
the $\left( \protect\rho ,\protect\eta ,h\right) $-curvature of a
distinguished linear $\left( \protect\rho ,\protect\eta \right) $-connection}

We consider the following diagram:
\begin{equation*}
\begin{array}{c}
\xymatrix{\overset{\ast }{E}\ar[d]_{\overset{\ast }{\pi }}&\left( F,\left[
,\right] _{F,h},\left( \rho ,\eta \right) \right)\ar[d]^\nu\\ M\ar[r]^h&N}%
\end{array}%
\end{equation*}%
where $\left( E,\pi ,M\right) \in \left\vert \mathbf{B}^{\mathbf{v}%
}\right\vert $ and $\left( \left( F,\nu ,M\right) ,\left[ ,\right]
_{F,h},\left( \rho ,\eta \right) \right) \in \left\vert \mathbf{GLA}%
\right\vert .$

Let $\left( \rho ,\eta \right) \overset{\ast }{\Gamma }$ be a $\left( \rho
,\eta \right) $-connection for the vector bundle $\left( \overset{\ast }{E},%
\overset{\ast }{\pi },M\right) $ and let\vspace*{-5mm}
\begin{equation*}
\left( \left( \rho ,\eta \right) \overset{\ast }{H},\left( \rho ,\eta
\right) \overset{\ast }{V}\right)
\end{equation*}
be a distinguished linear $\left( \rho ,\eta \right) $-connection for the
generalized tangent bundle
\begin{equation*}
\left( \left( \rho ,\eta \right) T\overset{\ast }{E},\left( \rho ,\eta
\right) \tau _{\overset{\ast }{E}},\overset{\ast }{E}\right) .
\end{equation*}

\smallskip \noindent \textbf{Definition 6.6.1 }The application
\begin{equation*}
\begin{array}{rcl}
\Gamma \left( \left( \rho ,\eta \right) T\overset{\ast }{E},\left( \rho
,\eta \right) \tau _{\overset{\ast }{E}},\overset{\ast }{E}\right) ^{2} & ^{%
\underrightarrow{\left( \rho ,\eta ,h\right) \mathbb{T}}} & \Gamma \left(
\left( \rho ,\eta \right) T\overset{\ast }{E},\left( \rho ,\eta \right) \tau
_{\overset{\ast }{E}},\overset{\ast }{E}\right) \vspace*{2mm} \\
\left( X,Y\right) & \longmapsto & \left( \rho ,\eta ,h\right) \overset{\ast }%
{\mathbb{T}}\left( X,Y\right)%
\end{array}%
\end{equation*}%
defined by
\begin{equation*}
\begin{array}{c}
\left( \rho ,\eta ,h\right) \overset{\ast }{\mathbb{T}}\left( X,Y\right)
=\left( \rho ,\eta \right) \overset{\ast }{D}_{X}Y-\left( \rho ,\eta \right)
\overset{\ast }{D}_{Y}X-\left[ X,Y\right] _{\left( \rho ,\eta \right) T%
\overset{\ast }{E}},%
\end{array}%
\leqno(6.6.1)
\end{equation*}%
for any $X,Y\in \Gamma \left( \left( \rho ,\eta \right) T\overset{\ast }{E}%
,\left( \rho ,\eta \right) \tau _{\overset{\ast }{E}},\overset{\ast }{E}%
\right) ,$ will be called the $\left( \rho ,\eta ,h\right) $\textit{-torsion
associated to distinguished linear }$\left( \rho ,\eta \right) $\textit{%
-connection }$\left( \left( \rho ,\eta \right) \overset{\ast }{H},\left(
\rho ,\eta \right) \overset{\ast }{V}\right) .$

The applications
\begin{equation*}
\mbox{$\overset{\ast }{\mathcal{H}}\left( \rho ,\eta ,h\right)
\overset{\ast }{\mathbb{T}}\left( \overset{\ast }{\mathcal{H}}\left( \cdot
\right) ,\overset{\ast }{\mathcal{H}}\left( \cdot \right) \right) ,\,\overset{\ast }{\mathcal{H}}\left( \rho ,\eta ,h\right) \overset{\ast }{\mathbb{T}}\left( \overset{\ast }{\mathcal{H}}\left( \cdot \right) ,\overset{\ast }{\mathcal{H}}\left( \cdot \right) \right) ,...,  \overset{\ast }{\mathcal{V}}\left( \rho ,\eta ,h\right) \overset{\ast }{\mathbb{T}}\left(
\overset{\ast }{\mathcal{V}}\left( \cdot \right) ,\overset{\ast }{\mathcal{V}}\left( \cdot \right) \right) $}
\end{equation*}%
are called $\overset{\ast }{\mathcal{H}}\left( \overset{\ast }{\mathcal{H}}%
\overset{\ast }{\mathcal{H}}\right) ,\,\overset{\ast }{\mathcal{V}}\left(
\overset{\ast }{\mathcal{H}}\overset{\ast }{\mathcal{H}}\right) ,...,\overset%
{\ast }{\mathcal{V}}\left( \overset{\ast }{\mathcal{V}}\overset{\ast }{%
\mathcal{V}}\right) $ $\left( \rho ,\eta ,h\right) $\textit{-torsions
associated to distinguished linear }$\left( \rho ,\eta \right) $\textit{%
-connection }$\left( \left( \rho ,\eta \right) \overset{\ast }{H},\left(
\rho ,\eta \right) \overset{\ast }{V}\right) .$

\medskip \noindent \textbf{Proposition 6.6.1 }\emph{The }$\left( \rho ,\eta
,h\right) $\emph{-torsion }$\left( \rho ,\eta ,h\right) \overset{\ast }{%
\mathbb{T}}$\emph{\ associated to distinguished linear }$\left( \rho ,\eta
\right) $\emph{-connection }$\left( \left( \rho ,\eta \right) \overset{\ast }%
{H},\left( \rho ,\eta \right) \overset{\ast }{V}\right) $\emph{, is }$%
\mathbb{R} $\emph{-bilinear and antisymmetric in the lower indices.}\medskip

Using the notations:
\begin{equation*}
%
\leqno(6.6.2)
\end{equation*}%
we can easily prove the following

\bigskip\noindent\textbf{Theorem 6.6.1 }\emph{The }$\left( \rho ,\eta
,h\right) $\emph{-torsion }$\left( \rho ,\eta ,h\right) \overset{\ast }{%
\mathbb{T}}$\emph{\ associated to the distinguished linear }$\left( \rho
,\eta \right) $\emph{-connection }$\left( \left( \rho ,\eta \right) \overset{%
\ast }{H},\left( \rho ,\eta \right) \overset{\ast }{V}\right) $\emph{, is
characterized by the\ tensor fields with local components: }%
\begin{equation*}
%
%
\leqno(6.6.3)
\end{equation*}

\emph{In particular, when }$\left( \rho ,\eta ,h\right) =\left(
Id_{TM},Id_{M},Id_{M}\right) ,$\emph{\ we regain the local components of
torsion associated to distinguished linear connection }$\left( \overset{\ast
}{H},\overset{\ast }{V}\right) $\emph{: }%
\begin{equation*}
%
%
\leqno(6.6.4)
\end{equation*}%
for any $\tilde{X},\tilde{Y},\tilde{Z}\in \Gamma \left( \left( \rho ,\eta
\right) T\overset{\ast }{E},\left( \rho ,\eta \right) \tau _{\overset{\ast }{%
E}},\overset{\ast }{E}\right) ,$ will be called the $\left( \rho ,\eta
,h\right) $\emph{-curvature associated to distinguished linear }$\left( \rho
,\eta \right) $\emph{-connection }$\left( \left( \rho ,\eta \right) \overset{%
\ast }{H},\left( \rho ,\eta \right) \overset{\ast }{V}\right) .$

\bigskip\noindent \textbf{Proposition 6.6.2 }\emph{The }$\left( \rho ,\eta
,h\right) $\emph{-curvature }$\left( \rho ,\eta ,h\right) \overset{\ast }{%
\mathbb{R}}$\emph{\ associated to distinguished linear }$\left( \rho ,\eta
\right) $\emph{-connection }$\left( \left( \rho ,\eta \right) \overset{\ast }%
{H},\left( \rho ,\eta \right) \overset{\ast }{V}\right) $\emph{, is }$%
\mathbb{R} $\emph{-linear in each argument and antisymmetric in the first
two arguments.}\medskip

Using the notations:
\begin{equation*}
%
\leqno(6.6.5)
\end{equation*}%
we can easily prove the following

\medskip \noindent \textbf{Theorem 6.6.2 }\emph{The }$\left( \rho ,\eta
,h\right) $\emph{-curvature }$\left( \rho ,\eta ,h\right) \overset{\ast }{%
\mathbb{R}}$\emph{\ associated to distinguished linear }$\left( \rho ,\eta
\right) $\emph{-connection }$\left( \left( \rho ,\eta \right) \overset{\ast }%
{H},\left( \rho ,\eta \right) \overset{\ast }{V}\right) $\emph{, is
characterized by the\ tensor fields with local components: }
\begin{equation*}
\left\{%
%
%
\right.\leqno(6.6.8)
\end{equation*}

\emph{In particular, when }$\left( \rho ,\eta ,h\right) =\left(
Id_{TM},Id_{M},Id_{M}\right) ,$\emph{\ we see the local components of the
curvature associated to distinguished linear connection }$\left( \overset{%
\ast }{H},\overset{\ast }{V}\right) $ \emph{in the followings:}
\begin{equation*}
%
%
\leqno(6.6.8)^{\prime }
\end{equation*}

\smallskip \noindent \textbf{Definition 6.6.3 }The tensor field
%\begin{equation*}
%\left( 5.2.4.9\right) ~\ \
\begin{equation*}
%
%
,\leqno(6.6.10)
\end{equation*}%
will be called \emph{the Ricci tensor field associated to distinguished
linear }$\left( \rho ,\eta \right) $\emph{-connection }$\left( \left( \rho
,\eta \right) \overset{\ast }{H},\left( \rho ,\eta \right) \overset{\ast }{V}%
\right) .$

This tensor field will be used for writing the Einstein equations in
Subsection 6.10.

\subsection{Formulas of Ricci type. Identities of Cartan and Bianchi type}

We consider the following diagram:
\begin{equation*}
%
%
\end{equation*}%
where $\left( E,\pi ,M\right) \in \left\vert \mathbf{B}^{\mathbf{v}%
}\right\vert $ and $\left( \left( F,\nu ,M\right) ,\left[ ,\right]
_{F.h},\left( \rho ,\eta \right) \right) \in \left\vert \mathbf{GLA}%
\right\vert .$

Let $\left( \rho ,\eta \right) \overset{\ast }{\Gamma }$ be a $\left( \rho
,\eta \right) $-connection for the vector bundle $\left( \overset{\ast }{E},%
\overset{\ast }{\pi },M\right) $ and let
\begin{equation*}
\left( \left( \rho ,\eta \right) \overset{\ast }{H},\left( \rho ,\eta
\right) \overset{\ast }{V}\right)
\end{equation*}
be a distinguished linear $\left( \rho ,\eta \right) $-connection for the
generalized tangent bundle
\begin{equation*}
\left( \left( \rho ,\eta \right) T\overset{\ast }{E},\left( \rho ,\eta
\right) \tau _{\overset{\ast }{E}},\overset{\ast }{E}\right) .
\end{equation*}

\smallskip \noindent \textbf{Theorem 6.7.1} Using the definition of $\left(
\rho ,\eta ,h\right) $-curvature associated to the distinguished linear $%
\left( \rho ,\eta \right) $-connection $\left( \left( \rho ,\eta \right)
\overset{\ast }{H},\left( \rho ,\eta \right) \overset{\ast }{V}\right) $, it
results the following formulas:%
\begin{equation*}
\left\{
%
\end{array}%
\right. \leqno(\mathcal{R}_{2})
\end{equation*}

Using the previous theorem, the horizontal and vertical sections of adapted
base and an arbitrary section
\begin{equation*}
\tilde{Z}^{\alpha }\frac{\partial }{\partial \tilde{z}^{\alpha }}+Y^{a}\frac{%
\partial }{\partial \tilde{y}^{a}}\in \Gamma \left( \left( \rho ,\eta
\right) T\overset{\ast }{E},\left( \rho ,\eta \right) \tau _{\overset{\ast }{%
E}},\overset{\ast }{E}\right)
\end{equation*}%
we can propose the following

\bigskip \noindent \textbf{Theorem 6.7.2 }\emph{We obtain the following
formulas of Ricci type: }%
\begin{equation*}
\left\{
%
\right. \leqno(\mathcal{R}_{2})
\end{equation*}

\emph{In particular, if }$\left( \rho ,\eta ,h\right) =\left(
Id_{TM},Id_{M},id_{M}\right) $\emph{\ and the Lie bracket }$\left[ ,\right]
_{TM}$\emph{\ is the usual Lie bracket, then the formulas of Ricci type }$%
\left( \mathcal{R}_{1}\right) $\emph{\ and }$\left( \mathcal{R}_{2}\right) $%
\emph{\ become:}%
\begin{equation*}
\left( \mathcal{R}_{1}\right) ^{\prime }~~\left\{
%
%
\right.
\end{equation*}%
Using the $1$-forms associated to distinguished linear $\left( \rho ,\eta
\right) $-connection $\left( \left( \rho ,\eta \right) \overset{\ast }{H}%
,\left( \rho ,\eta \right) \overset{\ast }{V}\right) $
\begin{equation*}
%
%
\right. \leqno(6.7.2)
\end{equation*}%
and the curvature $2$-forms%
\begin{equation*}
\left\{
%
%
\right. \leqno(6.7.3)
\end{equation*}%
we obtain the following

\bigskip\noindent\textbf{Theorem 4.7.3} \emph{Identities of Cartan type hold
good:}%
\begin{equation*}
%
\leqno(\mathcal{C}_{2})
\end{equation*}

In particular, if $\left( \rho ,\eta ,h\right) =\left(
Id_{TM},Id_{M},Id_{M}\right) $ and the Lie bracket $\left[ ,\right] _{TM}$
is the usual Lie bracket, then the identities of Cartan type $\left(
\mathcal{C}_{1}\right) $ and $\left( \mathcal{C}_{2}\right) $ become:
\begin{equation*}
%
\leqno(\mathcal{C}_{2})^{\prime }
\end{equation*}

\smallskip \noindent\textbf{Remark 6.7.1 } For any
\begin{equation*}
X,Y,Z\in \Gamma \left( \left( \rho ,\eta \right) T\overset{\ast }{E},\left(
\rho ,\eta \right) \tau _{\overset{\ast }{E}},\overset{\ast }{E}\right)
\end{equation*}%
the following identities hold good:
\begin{equation*}
%
\leqno(6.7.6)
\end{equation*}

Using the formulas of Bianchi type presented in Subsection 4.2 of our paper
and the Remark 6.7.1 we obtain the following

\bigskip\noindent\textbf{Theorem 6.7.4} \emph{The identities of Bianchi
type: }
\begin{equation*}
\left\{
%
%
\right.\leqno\left( \mathcal{B}_{2}\right)
\end{equation*}%
\emph{hold good for any }$X,Y,Z\in \Gamma \left( \left( \rho ,\eta \right) T%
\overset{\ast }{E},\left( \rho ,\eta \right) \tau _{\overset{\ast }{E}},%
\overset{\ast }{E}\right) .$

\bigskip\noindent\textbf{Corollary 6.7.1 }\emph{Using the following sections
}$\left( \delta _{\theta },\delta _{\gamma },\delta _{\beta }\right) $\emph{%
, the identities }$\left( \mathcal{B}_{1}\right) $\emph{\ become:}%
\begin{equation*}
\left\{
%
%
\right.\leqno(\mathcal{B}_{1})^{\prime }
\end{equation*}%
\emph{and using the following sections }$\left( \delta _{\lambda },\delta
_{\theta },\delta _{\gamma },\delta _{\beta }\right) $\emph{, the identities
}$\left( \mathcal{B}_{2}\right) $\emph{\ become: }%
\begin{equation*}
\left\{
%
%
\right.\leqno(\mathcal{B}_{2})^{\prime }
\end{equation*}
Using another base of sections, we shall obtain new identities of Bianchi
type necessary in the applications.

\subsection{The $\left( \protect\rho ,\protect\eta \right) $%
-(pseudo)metrizability}

We consider the following diagram:
\begin{equation*}
%
%
\end{equation*}%
where $\left( E,\pi ,M\right) \in \left\vert \mathbf{B}^{\mathbf{v}%
}\right\vert $ and $\left( \left( F,\nu ,M\right) ,\left[ ,\right]
_{F.h},\left( \rho ,\eta \right) \right) \in \left\vert \mathbf{GLA}%
\right\vert .$ Let $\left( \rho ,\eta \right) \overset{\ast }{\Gamma }$ be a
$\left( \rho ,\eta \right) $-connection for the vector bundle $\left(
\overset{\ast }{E},\overset{\ast }{\pi },M\right) $ and let $\left( \left(
\rho ,\eta \right) \overset{\ast }{H},\left( \rho ,\eta \right) \overset{%
\ast }{V}\right) $ be a distinguished linear $\left( \rho ,\eta \right) $%
-connection for the generalized tangent bundle
\begin{equation*}
\left( \left( \rho ,\eta \right) T\overset{\ast }{E},\left( \rho ,\eta
\right) \tau _{\overset{\ast }{E}},\overset{\ast }{E}\right) .
\end{equation*}

\smallskip \noindent \textbf{Definition 6.8.1} A tensor $d$-field
\begin{equation*}
\overset{\ast }{G}=g_{\alpha \beta }d\tilde{z}^{\alpha }\otimes d\tilde{z}%
^{\beta }+g^{ab}\delta \tilde{p}_{a}\otimes \delta \tilde{p}_{b}\in \mathcal{%
DT}_{20}^{02}\left( \left( \rho ,\eta \right) T\overset{\ast }{E},\left(
\rho ,\eta \right) \tau _{\overset{\ast }{E}},\overset{\ast }{E}\right)
\end{equation*}%
will be called a \emph{pseudometrical structure }if its components are
symmetric and the matrices $\left\Vert g_{\alpha \beta }\left( \overset{\ast
}{u}_{x}\right) \right\Vert $and $\left\Vert g^{ab}\left( \overset{\ast }{u}%
_{x}\right) \right\Vert $ are nondegenerate, for any point $\overset{\ast }{u%
}_{x}\in \overset{\ast }{E}.$

Moreover, if the matrices $\left\Vert g_{\alpha \beta }\left( \overset{\ast }%
{u}_{x}\right) \right\Vert $ and $\left\Vert g^{ab}\left( \overset{\ast }{u}%
_{x}\right) \right\Vert $ has constant signature, then the tensor $d$-field $%
\overset{\ast }{G}$ will be called \emph{metrical structure}\textit{.}

Let
\begin{equation*}
\overset{\ast }{G}=g_{\alpha \beta }d\tilde{z}^{\alpha }\otimes d\tilde{z}%
^{\beta }+g^{ab}\delta \tilde{p}_{a}\otimes \delta \tilde{p}_{b}
\end{equation*}%
be a (pseudo)metrical structure. If $\alpha ,\beta \in \overline{1,p}$ and $%
a,b\in \overline{1,r},$ then for any vector local $\left( m+r\right) $-chart
$\left( U,\overset{\ast }{s}_{U}\right) $ of $\left( \overset{\ast }{E},%
\overset{\ast }{\pi },M\right) $, we consider the real functions
\begin{equation*}
\begin{array}{ccc}
\overset{\ast }{\pi }^{-1}\left( U\right) & ^{\underrightarrow{~\ \ \tilde{g}%
^{\beta \alpha }~\ \ }} & \mathbb{R}%
\end{array}%
\end{equation*}%
and
\begin{equation*}
\begin{array}{ccc}
\overset{\ast }{\pi }^{-1}\left( U\right) & ^{\underrightarrow{~\ \ \tilde{g}%
_{ba}~\ \ }} & \mathbb{R}%
\end{array}%
\end{equation*}%
such that%
\begin{equation*}
\begin{array}{c}
\tilde{g}^{\beta \alpha }\left( \overset{\ast }{u}_{x}\right) \cdot
g_{\alpha \gamma }\left( \overset{\ast }{u}_{x}\right) =\delta _{\gamma
}^{\beta }%
\end{array}%
\end{equation*}%
and
\begin{equation*}
\begin{array}{c}
\tilde{g}_{ba}\left( \overset{\ast }{u}_{x}\right) \cdot g^{ac}\left(
\overset{\ast }{u}_{x}\right) =\delta _{b}^{c},~%
\end{array}%
\end{equation*}%
for any $\overset{\ast }{u}_{x}\in \overset{\ast }{\pi }^{-1}\left( U\right)
\backslash \left\{ 0_{x}\right\} $.

\bigskip \noindent \textbf{Definition 6.8.2} We will say that the \emph{%
(pseudo)metrical structure \ }%
\begin{equation*}
\begin{array}{c}
\overset{\ast }{G}=g_{\alpha \beta }d\tilde{z}^{\alpha }\otimes d\tilde{z}%
^{\beta }+g^{ab}\delta \tilde{p}_{a}\otimes \delta \tilde{p}_{b}%
\end{array}%
\end{equation*}%
\emph{is Riemannian (pseudo)metrical structure}\textit{\ }if around each
point $x\in M$ it exists a local vector $m+r$-chart $\left( U,\overset{\ast }%
{s}_{U}\right) $ and a local $m$-chart $\left( U,\xi _{U}\right) $ such that
$g_{\alpha \beta }\circ s_{U}^{-1}\circ \left( \xi _{U}^{-1}\times Id_{%
\mathbb{R}^{m}}\right) \left( x,p\right) $ and $g^{ab}\circ s_{U}^{-1}\circ
\left( \xi _{U}^{-1}\times Id_{\mathbb{R}^{m}}\right) \left( x,p\right) $
depends only on $x$, for any $\overset{\ast }{u}_{x}\in \overset{\ast }{\pi }%
^{-1}\left( U\right) .$

If only the condition is verified: \textit{\textquotedblright }$g_{\alpha
\beta }\circ s_{U}^{-1}\circ \left( \xi _{U}^{-1}\times Id_{\mathbb{R}%
^{m}}\right) \left( x,p\right) $\textit{\ depends only on }$x$\textit{, for
any }$\overset{\ast }{u}_{x}\in \overset{\ast }{\pi }^{-1}\left( U\right) $%
\textit{\textquotedblright } (res\-pec\-ti\-vely \textit{\textquotedblright }%
$g^{ab}\circ s_{U}^{-1}\circ \left( \xi _{U}^{-1}\times Id_{\mathbb{R}%
^{m}}\right) \left( x,p\right) $\textit{\ depend only on }$x$\textit{, for
any }$\overset{\ast }{u}_{x}\in \overset{\ast }{\pi }^{-1}\left( U\right) $%
\textit{\textquotedblright }), then we will say that the \textit{%
(pseudo)metrical structure }$G$ \textit{is a }\emph{Riemannian }$\overset{%
\ast }{\mathcal{H}}$\emph{-(pseudo)metrical structure}\textit{\ }%
respectively a \emph{Riemannian }$\overset{\ast }{\mathcal{V}}$\emph{%
-(pseudo)metrical structure.}

\bigskip \noindent \textbf{Definition 6.8.3} We will say that the \emph{%
(pseudo)metrical structure }%
\begin{equation*}
\begin{array}{c}
\overset{\ast }{G}=g_{\alpha \beta }d\tilde{z}^{\alpha }\otimes d\tilde{z}%
^{\beta }+g^{ab}\delta \tilde{p}_{a}\otimes \delta \tilde{p}_{b}%
\end{array}%
\end{equation*}%
\emph{is locally Minkowski structure}\textit{\ }if around each point $x\in M$
it exists a local vector $m+r$-chart $\left( U,\overset{\ast }{s}_{U}\right)
$ and a local $m$-chart $\left( U,\xi _{U}\right) $ such that $g_{\alpha
\beta }\circ s_{U}^{-1}\circ \left( \xi _{U}^{-1}\times Id_{\mathbb{R}%
^{m}}\right) (x,p)$ and $g^{ab}\circ s_{U}^{-1}\circ \left( \xi
_{U}^{-1}\times Id_{\mathbb{R}^{m}}\right) (x,p)$ depends only on~$p$, for
any $\overset{\ast }{u}_{x}\in \overset{\ast }{\pi }^{-1}\left( U\right) .$

If only the condition is verified: \textquotedblright $g_{\alpha \beta
}\circ s_{U}^{-1}\circ \left( \xi _{U}^{-1}\times Id_{\mathbb{R}^{m}}\right)
\left( x,p\right) $ depend only on $p$, for any $\overset{\ast }{u}_{x}\in
\overset{\ast }{\pi }^{-1}\left( U\right) $\textquotedblright\ respectively
\textquotedblright $g^{ab}\circ s_{U}^{-1}\circ \left( \xi _{U}^{-1}\times
Id_{\mathbb{R}^{m}}\right) \left( x,p\right) $ depends only on $p$, for any $%
\overset{\ast }{u}_{x}\in \overset{\ast }{\pi }^{-1}\left( U\right) $%
\textquotedblright , then we will say that \emph{the (pseudo)metrical
structure }$\overset{\ast }{G}$\emph{\ is a (pseudo)metrical structure }$%
\overset{\ast }{\mathcal{H}}$\emph{-locally Minkowski }and \emph{\ }$\overset%
{\ast }{\mathcal{V}}$\emph{-locally Minkowski, respectively.}

\medskip\noindent \textbf{Definition 6.8.4} The generalized tangent bundle $%
\left( (\rho ,\eta )T\overset{\ast }{E},\break (\rho ,\eta )\tau _{\overset{%
\ast }{E}},\overset{\ast }{E}\right) $ will be called $(\rho ,\eta )$\emph{%
-(pseudo)metrizable} if it exists a (pseudo)metrical structure
\begin{equation*}
\overset{\ast }{G}=g_{\alpha \beta }d\tilde{z}^{\alpha }\otimes d\tilde{z}%
^{\beta }+g^{ab}\delta \tilde{p}_{a}\otimes \delta \tilde{p}_{b}
\end{equation*}
and a distinguished linear $\left( \rho ,\eta \right) $-connection%
\begin{equation*}
\left( \left( \rho ,\eta \right) \overset{\ast }{H},\left( \rho ,\eta
\right) \overset{\ast }{V}\right)
\end{equation*}
such that
\begin{equation*}
\begin{array}{c}
\left( \rho ,\eta \right) D_{X}G=0,~\forall X\in \Gamma \left( \left( \rho
,\eta \right) T\overset{\ast }{E},\left( \rho ,\eta \right) \tau _{\overset{
\ast }{E}},\overset{\ast }{E}\right) .%
\end{array}
\leqno(6.8.1)
\end{equation*}

The condition $\left( 6.8.1\right) $ is equivalent with the following
equalities:
\begin{equation*}
\begin{array}{c}
g_{\alpha \beta \mid \gamma }=0,\,g_{~\ \ \mid \gamma }^{ab}=0,\,\,g_{\alpha
\beta }|^{c}=0\,,\,\,g^{ab}|^{c}=0.%
\end{array}%
\leqno(6.8.2)
\end{equation*}

If $g_{\alpha \beta \mid \gamma }{=}0$ and $\,g_{~\ \ \mid \gamma }^{ab}{=}0$%
, then we will say that \emph{the vector bundle}
\begin{equation*}
\left( (\rho ,\eta )T\overset{\ast }{E},(\rho ,\eta )\tau _{\overset{\ast }{E%
}},\overset{\ast }{E}\right)
\end{equation*}%
\emph{is }$\overset{\ast }{\mathcal{H}}$\emph{-}$(\rho ,\eta )$\emph{%
-(pseudo)metrizable.}

If $g_{\alpha \beta }|^{c}=0$ and $\,g^{ab}|^{c}=0$, then we will say that
\emph{the vector bundle}
\begin{equation*}
\left( (\rho ,\eta )T\overset{\ast }{E},(\rho ,\eta )\tau _{\overset{\ast }{E%
}},\overset{\ast }{E}\right)
\end{equation*}%
\emph{is }$\overset{\ast }{\mathcal{V}}$\emph{-}$(\rho ,\eta )$\emph{%
-(pseudo)\-metrizable.}

\medskip\noindent \textbf{Theorem 6.8.1} \emph{If}
\begin{equation*}
\left( \left( \rho ,\eta \right) \overset{\ast }{\mathring{H}},\left( \rho
,\eta \right) \overset{\ast }{\mathring{V}}\right)
\end{equation*}%
\emph{is a distinguished linear }$\left( \rho ,\eta \right) $\emph{%
-connection for the generalized tangent bundle }%
\begin{equation*}
\left( (\rho ,\eta )T\overset{\ast }{E},(\rho ,\eta )\tau _{\overset{\ast }{E%
}},\overset{\ast }{E}\right)
\end{equation*}%
\emph{\ and }%
\begin{equation*}
\overset{\ast }{G}=g_{\alpha \beta }d\tilde{z}^{\alpha }\otimes d\tilde{z}%
^{\beta }+g^{ab}\delta \tilde{p}_{a}\otimes \delta \tilde{p}_{b}
\end{equation*}
\emph{is a (pseudo)metrical structure, then the real local functions:}
\begin{equation*}
%
\leqno(6.8.3)
\end{equation*}
\emph{are components of a distinguished linear }$\left( \rho ,\eta \right) $%
\emph{-connection such that the generalized tangent bundle }
\begin{equation*}
\left( (\rho ,\eta )T\overset{\ast }{E},(\rho ,\eta )\tau _{\overset{\ast }{E%
}},\overset{\ast }{E}\right)
\end{equation*}
\emph{becomes }$\left( \rho ,\eta \right) $\emph{-(pseudo)metrizable.}

\medskip\noindent \textbf{Corollary 6.8.1 }\emph{If the distinguished linear
}$\left( \rho ,\eta \right) $\emph{-connection}
\begin{equation*}
\left( \left( \rho ,\eta \right) \overset{\ast }{\mathring{H}},\left( \rho
,\eta \right) \overset{\ast }{\mathring{V}}\right)
\end{equation*}%
\emph{coincide with the Berwald linear }$\left( \rho ,\eta \right) $\emph{%
-connection, then the local real functions: }
\begin{equation*}
\begin{array}{ll}
\left( \rho ,\eta \right) \overset{c}{\overset{\ast }{H}}_{\beta \gamma
}^{\alpha }\!\!\! & =\displaystyle\frac{1}{2}\tilde{g}^{\alpha \varepsilon
}\left( \Gamma \left( \overset{\ast }{\tilde{\rho}},Id_{\overset{\ast }{E}%
}\right) \left( \overset{\ast }{\tilde{\delta}}_{\gamma }\right)
g_{\varepsilon \beta }\right. \vspace*{1mm} \\
& +\Gamma \left( \overset{\ast }{\tilde{\rho}},Id_{\overset{\ast }{E}%
}\right) \left( \overset{\ast }{\tilde{\delta}}_{\beta }\right)
g_{\varepsilon \gamma }-\Gamma \left( \overset{\ast }{\tilde{\rho}},Id_{%
\overset{\ast }{E}}\right) \left( \overset{\ast }{\tilde{\delta}}%
_{\varepsilon }\right) g_{\beta \gamma } \\
& \left. +g_{\theta \varepsilon }L_{\gamma \beta }^{\theta }\circ h\circ
\overset{\ast }{\pi }\vspace*{1mm}-g_{\beta \theta }L_{\gamma \varepsilon
}^{\theta }\circ h\circ \overset{\ast }{\pi }-g_{\theta \gamma }L_{\beta
\varepsilon }^{\theta }\circ h\circ \overset{\ast }{\pi }\right) ,\vspace*{%
2mm} \\
\left( \rho ,\eta \right) \overset{c}{\overset{\ast }{H}}_{b\gamma
}^{a}\!\!\! & =\displaystyle\frac{\partial \left( \rho ,\eta \right) \overset%
{\ast }{\Gamma }_{b\gamma }}{\partial p_{a}}+\frac{1}{2}\tilde{g}_{bc}g_{~\
\ \overset{0}{\mid }\gamma }^{ac},\vspace*{2mm} \\
\left( \rho ,\eta \right) \overset{c}{\overset{\ast }{V}}_{\beta }^{\alpha
c}\!\! & =\displaystyle\frac{1}{2}g_{\beta \varepsilon }\frac{\partial
\tilde{g}^{\alpha \varepsilon }}{\partial p_{c}},\vspace*{2mm} \\
\left( \rho ,\eta \right) \overset{c}{\overset{\ast }{V}}_{a}^{bc}\!\!\! & =%
\displaystyle \frac{1}{2}\tilde{g}_{ae}\left( \frac{\partial g^{eb}}{%
\partial p_{c}}+\frac{\partial g^{ec}}{\partial p_{b}}-\frac{\partial g^{bc}%
}{\partial p_{e}}\right)%
\end{array}%
\hspace*{-6mm}\leqno(6.8.4)
\end{equation*}%
\emph{are the components of a distinguished linear }$\left( \rho ,\eta
\right) $\emph{-connection such that the generalized tangent bundle }%
\begin{equation*}
\left( (\rho ,\eta )T\overset{\ast }{E},(\rho ,\eta )\tau _{\overset{\ast }{E%
}},\overset{\ast }{E}\right)
\end{equation*}%
\emph{becomes }$\left( \rho ,\eta \right) $\emph{-(pseudo)metrizable.}

\emph{Moreover, if the (pseudo)metrical structure }$\overset{\ast }{G}$\emph{%
\ is }$\mathcal{H}$\emph{- and }$\mathcal{V}$\emph{-Rieman\-nian, then the
local real functions:}\smallskip \noindent
\begin{equation*}
\begin{array}{l}
(\rho ,\eta )\overset{c}{\overset{\ast }{H}}_{\beta \gamma }^{\alpha }{=}%
\displaystyle\frac{1}{2}g^{\alpha \varepsilon }\left( \rho _{\gamma }^{k}{%
\circ }h{\circ }\pi \frac{\partial g_{\varepsilon \beta }}{\partial x^{k}}%
+\rho _{\beta }^{j}{\circ }h{\circ }\pi \frac{\partial g_{\varepsilon \gamma
}}{\partial x^{j}}-\rho _{\varepsilon }^{e}{\circ }h{\circ }\pi \frac{%
\partial g_{\beta \gamma }}{\partial x^{e}}+\right. \\
\hfill \left. +g_{\theta \varepsilon }L_{\gamma \beta }^{\theta }{\circ }h{%
\circ }\pi -g_{\beta \theta }L_{\gamma \varepsilon }^{\theta }{\circ }h{%
\circ }\pi -g_{\theta \gamma }L_{\beta \varepsilon }^{\theta }{\circ }h{%
\circ }\pi \right) ,\vspace*{1mm} \\
\left( \rho ,\eta \right) \overset{c}{\overset{\ast }{H}}_{b\gamma }^{a}{=}%
\displaystyle\frac{\partial \left( \rho ,\eta \right) \Gamma _{\gamma }^{a}}{%
\partial y^{b}}+\frac{1}{2}g^{ac}\left( \rho _{\gamma }^{i}{\circ }h{\circ }%
\pi \frac{\partial g_{bc}}{\partial x^{i}}-\frac{\partial \rho \Gamma
_{\gamma }^{e}}{\partial y^{b}}g_{ec}-\frac{\partial \rho \Gamma _{\gamma
}^{e}}{\partial y^{c}}g_{eb}\right) ,\vspace*{2mm} \\
\left( \rho ,\eta \right) \overset{c}{\overset{\ast }{V}}_{\beta c}^{\alpha
}=0,\ \left( \rho ,\eta \right) \overset{c}{\overset{\ast }{V}}_{bc}^{a}=0%
\end{array}%
\leqno(4.8.5)
\end{equation*}
\emph{are the components of a distinguished linear }$\left( \rho ,\eta
\right) $\emph{-connection such that the generalized tangent bundle }%
\begin{equation*}
\left( (\rho ,\eta )T\overset{\ast }{E},(\rho ,\eta )\tau _{\overset{\ast }{E%
}},\overset{\ast }{E}\right)
\end{equation*}
\emph{becomes }$\left( \rho ,\eta \right) $\emph{-(pseudo)metrizable.}

\medskip\noindent \textbf{Theorem 6.8.2} \emph{Let }$\left( \rho ,\eta
\right) \overset{\ast }{\Gamma }$\emph{\ be a }$\left( \rho ,\eta \right) $%
\emph{-connection for the vector bundle }$\left( \overset{\ast }{E},\overset{%
\ast }{\pi },M\right) .$ \emph{Let }%
\begin{equation*}
\left( \left( \rho ,\eta \right) \overset{\ast }{\mathring{H}},\left( \rho
,\eta \right) \overset{\ast }{\mathring{V}}\right)
\end{equation*}%
\emph{be a distinguished linear }$\left( \rho ,\eta \right) $\emph{%
-connection for }
\begin{equation*}
\left( \left( \rho ,\eta \right) T\overset{\ast }{E},\left( \rho ,\eta
\right) \tau _{\overset{\ast }{E}},\overset{\ast }{E}\right)
\end{equation*}
\emph{and let }
\begin{equation*}

\leqno(6.8.6)
\end{equation*}
\emph{be the Obata operators}

\emph{If the real local functions }$X_{\beta \gamma }^{\alpha },X_{\beta
}^{\alpha c},Y_{b\gamma }^{a},Y_{b}^{ac}$ \emph{are components of tensor
fields,} \emph{then the local real functions are given in the following: }%
\vspace*{-3mm}
\begin{equation*}
%
\leqno(6.8.7)
\end{equation*}
\emph{are the components of a distinguished linear }$\left( \rho ,\eta
\right) $\emph{-connection such that the generalized tangent bundle }
\begin{equation*}
\left( \left( \rho ,\eta \right) T\overset{\ast }{E},\left( \rho ,\eta
\right) \tau _{\overset{\ast }{E}},\overset{\ast }{E}\right)
\end{equation*}
\emph{becomes }$\left( \rho ,\eta \right) $\emph{-(pseudo)metrizable.}

\medskip \noindent\textbf{Theorem 6.8.3 }\emph{Let }$\left( \rho ,\eta
\right) \overset{\ast }{\Gamma }$\emph{\ be a }$\left( \rho ,\eta \right) $%
\emph{-connection for the vector bundle }$\left( \overset{\ast }{E},\overset{%
\ast }{\pi },M\right) .$ \emph{If }
\begin{equation*}
\left( \left( \rho ,\eta \right) \overset{\ast }{\mathring{H}},\left( \rho
,\eta \right) \overset{\ast }{\mathring{V}}\right)
\end{equation*}
\emph{is a distinguished linear }$\left( \rho ,\eta \right) $\emph{%
-connection for the generalized tangent bundle }
\begin{equation*}
\left( \left( \rho ,\eta \right) T\overset{\ast }{E},\left( \rho ,\eta
\right) \tau _{\overset{\ast }{E}},\overset{\ast }{E}\right)
\end{equation*}
\emph{and }
\begin{equation*}
\overset{\ast }{G}=g_{\alpha \beta }d\tilde{z}^{\alpha }\otimes d\tilde{z}
^{\beta }+g^{ab}\delta \tilde{p}_{a}\otimes \delta \tilde{p}_{b}
\end{equation*}
\emph{is a (pseudo)metrical structure, then the real local functions: }
\begin{equation*}
\begin{array}{l}
\left( \rho ,\eta \right) \overset{\ast }{H}_{\beta \gamma }^{\alpha
}=\left( \rho ,\eta \right) \overset{\ast }{\mathring{H}}_{\beta \gamma
}^{\alpha }+\frac{1}{2}g_{\beta \varepsilon }\tilde{g}_{~\ \ \ \overset{0}{
\mid }\gamma }^{\varepsilon \alpha },\vspace*{0mm} \\
\left( \rho ,\eta \right) \overset{\ast }{H}_{b\gamma }^{a}=\left( \rho
,\eta \right) \overset{\ast }{\mathring{H}}_{b\gamma }^{a}+\frac{1}{2}\tilde{
g}_{be}g_{~\ \ \ \overset{0}{\mid }\gamma }^{ea},\vspace*{1mm} \\
\left( \rho ,\eta \right) \overset{\ast }{V}_{\beta }^{\alpha c}=\left( \rho
,\eta \right) \overset{\ast }{\mathring{V}}_{\beta }^{\alpha c}+\frac{1}{2}
g_{\beta \varepsilon }\tilde{g}^{\varepsilon \alpha }\overset{0}{\mid }^{c},
\vspace*{1mm} \\
\left( \rho ,\eta \right) \overset{\ast }{V}_{b}^{ac}=\left( \rho ,\eta
\right) \overset{\ast }{\mathring{V}}_{b}^{ac}+\frac{1}{2}\tilde{g}
_{be}g^{ea}\overset{0}{\mid }^{c}%
\end{array}
\leqno(6.8.8)
\end{equation*}
\emph{are the components of a distinguished linear }$\left( \rho ,\eta
\right) $\emph{-connection such that the generalized tangent bundle }%
\begin{equation*}
\left( \left( \rho ,\eta \right) T\overset{\ast }{E},\left( \rho ,\eta
\right) \tau _{\overset{\ast }{E}},\overset{\ast }{E}\right)
\end{equation*}%
\emph{becomes }$\left( \rho ,\eta \right) $\emph{-(pseudo)metrizable.}

\subsection[Generalized Hamilton $\left( \protect\rho ,\protect\eta \right) $%
-spaces, Hamilton $\left( \protect\rho ,\protect\eta \right) $-spaces and
Cartan $\left( \protect\rho ,\protect\eta \right) $-spaces]{Generalized
Hamilton $\left( \rho ,\eta \right) $-spaces, Hamilton $\left( \rho ,\eta
\right) $-spaces\newline
and Cartan $\left( \rho ,\eta \right) $-spaces}

We consider the following diagram:
\begin{equation*}
\begin{array}{c}
\xymatrix{\overset{\ast }{E}\ar[d]_{\overset{\ast }{\pi }}&\left( F,\left[
,\right] _{F,h},\left( \rho ,\eta \right) \right)\ar[d]^\nu\\ M\ar[r]^h&N}%
\end{array}%
\end{equation*}%
such that $\left( E,\pi ,M\right) =\left( F,\nu ,N\right) $ and the
generalized tangent bundle
\begin{equation*}
\left( \left( \rho ,\eta \right) T\overset{\ast }{E},\left( \rho ,\eta
\right) \tau _{\overset{\ast }{E}},\overset{\ast }{E}\right)
\end{equation*}%
is $\left( \rho ,\eta \right) $-(pseudo)metrizable.

\bigskip\noindent\textbf{Definition 6.9.1 }A smooth \emph{Hamilton
fundamental function} on the dual vector bundle $\left( \overset{\ast }{E},%
\overset{\ast }{\pi },M\right) $ is a mapping
\begin{equation*}
\overset{\ast }{E}~\ ^{\underrightarrow{\ \ H\ \ }}~\ \mathbb{R}
\end{equation*}%
which satisfies the following conditions:\medskip

1. $H\circ \overset{\ast }{u}\in C^{\infty }\left( M\right) $, for any $%
\overset{\ast }{u}\in \Gamma \left( \overset{\ast }{E},\overset{\ast }{\pi }%
,M\right) \setminus \left\{ 0\right\} $;\smallskip

2. $H\circ 0\in C^{0}\left( M\right) $, where $0$ means the null section of $%
\left( \overset{\ast }{E},\overset{\ast }{\pi },M\right) .$\medskip

Let $H$ be a differentiable Hamiltonian defined on the total space of the
vector bundle $\left( \overset{\ast }{E},\overset{\ast }{\pi },M\right) .$

If $\left( U,\overset{\ast }{s}_{U}\right) $ is a local vector $\left(
m+r\right) $-chart for $\left( \overset{\ast }{E},\overset{\ast }{\pi }%
,M\right) $, then we obtain the following real functions defined on $\overset%
{\ast }{\pi }^{-1}\left( U\right) $:%
\begin{equation*}
\begin{array}{cc}
H_{i}\overset{put}{=}\displaystyle\frac{\partial H}{\partial x^{i}}\overset{%
put}{=}\frac{\partial }{\partial x^{i}}\left( H\right) & H_{i}^{b}\overset{%
put}{=}\displaystyle\frac{\partial ^{2}H}{\partial x^{i}\partial p_{b}}%
\vspace*{2mm}\overset{put}{=}\frac{\partial }{\partial x^{i}}\left( \frac{%
\partial }{\partial p_{b}}\left( H\right) \right) \\
H^{a}\overset{put}{=}\displaystyle\frac{\partial H}{\partial p_{a}}\overset{%
put}{=}\frac{\partial }{\partial p_{a}}\left( H\right) & H^{ab}\overset{put}{%
=}\displaystyle\frac{\partial ^{2}H}{\partial p_{a}\partial p_{b}}\overset{%
put}{=}\frac{\partial }{\partial p_{a}}\left( \frac{\partial }{\partial p_{b}%
}\left( H\right) \right)%
\end{array}%
.\leqno(6.9.1)
\end{equation*}

\bigskip \noindent \textbf{Definition 6.9.2 }If for any local vector $m+r$%
-chart $\left( U,\overset{\ast }{s}_{U}\right) $ of $\left( \overset{\ast }{E%
},\overset{\ast }{\pi },M\right) ,$ we have:
\begin{equation*}
\begin{array}{c}
rank\left\Vert H^{ab}\left( \overset{\ast }{u}_{x}\right) \right\Vert =r,%
\end{array}%
\leqno(6.9.2)
\end{equation*}%
for any $\overset{\ast }{u}_{x}\in \overset{\ast }{\pi }^{-1}\left( U\right)
\backslash \left\{ 0_{x}\right\} $, then we say that \emph{the Hamiltonian }$%
H$\emph{\ is regular.}

\bigskip \noindent \textbf{Proposition 6.9.1} If the Hamiltonian $H$ is
regular, then for any local vector $m+r$-chart $\left( U,\overset{\ast }{s}%
_{U}\right) $ of $\left( \overset{\ast }{E},\overset{\ast }{\pi },M\right) ,$
we obtain the real functions $\tilde{H}_{ba}$ locally defined by%
\begin{equation*}
\begin{array}{ccc}
\overset{\ast }{\pi }^{-1}\left( U\right) & ^{\underrightarrow{\ \tilde{H}%
_{ba}\ }} & \mathbb{R}\vspace*{1mm} \\
\overset{\ast }{u}_{x} & \longmapsto & H_{ba}\left( \overset{\ast }{u}%
_{x}\right)%
\end{array}%
\leqno(6.9.3)
\end{equation*}%
where $\left\Vert \tilde{H}_{ba}\left( \overset{\ast }{u}_{x}\right)
\right\Vert =\left\Vert H^{ab}\left( \overset{\ast }{u}_{x}\right)
\right\Vert ^{-1}$, for any $\overset{\ast }{u}_{x}\in \overset{\ast }{\pi }%
^{-1}\left( U\right) .$

\bigskip\noindent\textbf{Definition 6.9.3 }A smooth \emph{Cartan fundamental
function} on the vector bundle $\left( \overset{\ast }{E},\overset{\ast }{%
\pi },M\right) $ is a mapping
\begin{equation*}
\overset{\ast }{E}~\ ^{\underrightarrow{\ \ K\ \ }}~\ \mathbb{R}_{+}
\end{equation*}%
which satisfies the following conditions:\medskip

1. $K\circ \overset{\ast }{u}\in C^{\infty }\left( M\right) $, for any $%
\overset{\ast }{u}\in \Gamma \left( \overset{\ast }{E},\overset{\ast }{\pi }%
,M\right) \setminus \left\{ 0\right\} $;\smallskip

2. $K\circ 0\in C^{0}\left( M\right) $, where $0$ means the null section of $%
\left( \overset{\ast }{E},\overset{\ast }{\pi },M\right) $;

3. $K$ is positively $1$-homogenous on the fibres of vector bundle $\left(
\overset{\ast }{E},\overset{\ast }{\pi },M\right) ;$

4. For any local vector $m+r$-chart $\left( U,\overset{\ast }{s}_{U}\right) $
of $\left( \overset{\ast }{E},\overset{\ast }{\pi },M\right) ,$ the hessian:%
\begin{equation*}
\left\Vert K^{2~ab}\left( \overset{\ast }{u}_{x}\right) \right\Vert \leqno%
(6.9.4)
\end{equation*}%
is positively define for any $\overset{\ast }{u}_{x}\in \overset{\ast }{\pi }%
^{-1}\left( U\right) \backslash \left\{ 0_{x}\right\} $.

\medskip\noindent \textbf{Definition 6.9.4 }If the (pseudo)metrical
structure $\overset{\ast }{G}$\ is determined by a (pseudo)metrical
structure
\begin{equation*}
\begin{array}{c}
g\in \mathcal{T}~_{0}^{2}\left( V\left( \rho ,\eta \right) T\overset{\ast }{E%
},\left( \rho ,\eta \right) {\tau _{\overset{\ast }{E}}},\overset{\ast }{E}%
\right) ,%
\end{array}%
\end{equation*}%
then the $\left( \rho ,\eta \right) $-(pseudo)metrizable vector bundle
\begin{equation*}
\begin{array}{c}
\left( \left( \rho ,\eta \right) T\overset{\ast }{E},\left( \rho ,\eta
\right) {\tau _{\overset{\ast }{E}}},\overset{\ast }{E}\right)%
\end{array}%
\end{equation*}%
will be called the \emph{generalized Hamilton }$\left( \rho ,\eta \right) $%
\emph{-space.}

In particular, if the (pseudo)metrical structure $g$\ is determined with the
help of a Hamilton fundamental function\ or Cartan fundamental function,
then the $\left( \rho ,\eta \right) $-(pseudo)metrizable vector bundle
\begin{equation*}
\begin{array}{c}
\left( \left( \rho ,\eta \right) T\overset{\ast }{E},\left( \rho ,\eta
\right) {\tau _{\overset{\ast }{E}}},\overset{\ast }{E}\right)%
\end{array}%
\end{equation*}%
will be called the \emph{Hamilton }$\left( \rho ,\eta \right) $\emph{-space }%
or the \emph{Cartan }$\left( \rho ,\eta \right) $\emph{-space, respectively.}

The generalized Hamilton $\left( Id_{TM},Id_{M}\right) $-space, the Hamilton
$\left( Id_{TM},Id_{M}\right) $-space, and the Cartan $\left(
Id_{TM},Id_{M}\right) $-space will be called the \emph{generalized Hamilton
space, Hamilton space}, \emph{Cartan space}.

\medskip\noindent \textbf{Definition 6.9.5 }The normal distinguished linear $%
\left( \rho ,\eta \right) $-connections of a Hamilton or Cartan $\left( \rho
,\eta \right) $-space will be called \emph{Hamilton }and\emph{\ Cartan
linear }$\left( \rho ,\eta \right) $\emph{-connections.}

The Hamilton and Cartan linear $\left( Id_{TM},Id_{M}\right) $-connections
will be called \emph{Hamilton }and\emph{\ Cartan linear connections},
respectively.

\medskip \noindent \textbf{Theorem 6.9.1 }\emph{If the (pseudo)metrical
structure }$\overset{\ast }{G}$\emph{\ is determined by a (pseudo)metrical
structure }%
\begin{equation*}
\begin{array}{c}
g\in \mathcal{T}~_{0}^{2}\left( V\left( \rho ,\eta \right) T\overset{\ast }{E%
},\left( \rho ,\eta \right) {\tau _{\overset{\ast }{E}}},\overset{\ast }{E}%
\right) ,%
\end{array}%
\end{equation*}%
\emph{then the real local functions:}
\begin{equation*}
\begin{array}{ll}
\left( \rho ,\eta \right) \overset{\ast }{H}_{bc}^{a}\!\!\! & =\displaystyle%
\frac{1}{2}g^{ae}\left( \Gamma \left( \overset{\ast }{\tilde{\rho}},Id_{%
\overset{\ast }{E}}\right) \left( \overset{\ast }{\tilde{\delta}}_{b}\right)
\tilde{g}_{ec}+\Gamma \left( \overset{\ast }{\tilde{\rho}},Id_{\overset{\ast
}{E}}\right) \left( \overset{\ast }{\tilde{\delta}}_{c}\right) \tilde{g}%
_{be}\right. \vspace*{1mm} \\
& -\Gamma \left( \overset{\ast }{\tilde{\rho}},Id_{\overset{\ast }{E}%
}\right) \left( \overset{\ast }{\tilde{\delta}}_{e}\right) \tilde{g}_{bc}-%
\tilde{g}_{cd}\cdot L_{be}^{d}{\circ }h{\circ }\overset{\ast }{\pi }\vspace*{%
1mm} \\
& \left. +\tilde{g}_{bd}\cdot L_{ec}^{d}{\circ }h{\circ }\overset{\ast }{\pi
}-\tilde{g}_{ed}\cdot L_{bc}^{d}{\circ }h{\circ }\overset{\ast }{\pi }%
\right) ,\vspace*{2mm} \\
\left( \rho ,\eta \right) \overset{\ast }{V}_{a}^{bc}\!\!\! & =\displaystyle%
\frac{1}{2}\tilde{g}_{ae}\left( \Gamma \left( \overset{\ast }{\tilde{\rho}}%
,Id_{\overset{\ast }{E}}\right) \left( \overset{\cdot }{\tilde{\partial}}%
^{c}\right) g^{eb}\right. \\
& \left. +\Gamma \left( \overset{\ast }{\tilde{\rho}},Id_{\overset{\ast }{E}%
}\right) \left( \overset{\cdot }{\tilde{\partial}}^{b}\right) g^{ec}-\Gamma
\left( \overset{\ast }{\tilde{\rho}},Id_{\overset{\ast }{E}}\right) \left(
\overset{\cdot }{\tilde{\partial}}^{e}\right) g^{bc}\right)%
\end{array}%
\leqno(6.9.5)
\end{equation*}%
\emph{are the components of a normal distinguished linear }$\left( \rho
,\eta \right) $\emph{-connection with}\break $\left( \rho ,\eta \right) $%
\emph{-}$\overset{\ast }{\mathcal{H}}\left( \overset{\ast }{\mathcal{H}}%
\overset{\ast }{\mathcal{H}}\!\right) $\emph{\ and }$\left( \rho ,\eta
\!\right) $\emph{-}$\overset{\ast }{\mathcal{V}}\left( \!\overset{\ast }{%
\mathcal{V}}\overset{\ast }{\mathcal{V}}\!\right) $\emph{\ torsions free
such that the generalized tangent bundle}
\begin{equation*}
\left( \left( \rho ,\eta \right) T\overset{\ast }{E},\left( \rho ,\eta
\right) {\tau _{\overset{\ast }{E}}},\overset{\ast }{E}\right)
\end{equation*}%
\emph{\ derives generalized Hamilton }$\left( \rho ,\eta \right) $\emph{%
-space.}\medskip

This normal distinguished linear $( \rho ,\eta ) $-connection will be called
\emph{generalized linear} $( \rho ,\eta ) $\emph{-connection of Levi-Civita
type.}

If the \textit{(pseudo)metrical structure }$g$ is determined with the help
of a Hamilton or Cartan fundamental function, then the Hamilton and the
Cartan linear $\left( \rho ,\eta \right) $-connections will be called \emph{%
canonical Hamilton }and \emph{\ Cartan linear }$\left( \rho ,\eta \right) $%
\emph{-connection, respectively.}

The canonical Hamilton and Cartan linear $\left( Id_{TM},Id_{M}\right) $%
-con\-nec\-tion will be called\emph{\ }the \emph{canonical Hamilton }and%
\emph{\ Cartan linear connection }respectively.

\bigskip\noindent\textbf{Theorem 6.9.2 }\emph{Let\ }$\left( \left( \rho
,\eta \right) \overset{\ast }{H},\left( \rho ,\eta \right) \overset{\ast }{V}%
\right) $\emph{\ be the normal dis\-tin\-guished linear }$\left( \rho ,\eta
\right) $\emph{-connection presented in the previous theorem.}

\emph{If }
\begin{equation*}
\overset{\ast }{\mathbb{T}}_{bc}^{a}\overset{\ast }{\tilde{\delta}}%
_{a}\otimes d\tilde{z}^{b}\otimes d\tilde{z}^{c}\in \mathcal{T}%
_{20}^{10}\left( \left( \rho ,\eta \right) T\overset{\ast }{E},\left( \rho
,\eta \right) {\tau _{\overset{\ast }{E}}},\overset{\ast }{E}\right)
\end{equation*}%
\emph{and }%
\begin{equation*}
\overset{\ast }{\mathbb{S}}_{a}^{bc}\overset{\cdot }{\tilde{\partial}}%
^{a}\otimes \delta \tilde{p}_{b}\otimes \delta \tilde{p}_{c}\in \mathcal{T}%
_{01}^{02}\left( \left( \rho ,\eta \right) T\overset{\ast }{E},\left( \rho
,\eta \right) {\tau _{\overset{\ast }{E}}},\overset{\ast }{E}\right)
\end{equation*}%
\emph{such that they satisfy the following conditions:}%
\begin{equation*}
\overset{\ast }{\mathbb{T}}_{bc}^{a}=-\overset{\ast }{\mathbb{T}}_{cb}^{a},~%
\overset{\ast }{\mathbb{S}}_{a}^{bc}=-\overset{\ast }{\mathbb{S}}%
_{a}^{bc},~\forall b,c\in \overline{1,r},
\end{equation*}%
\emph{then the real local functions:\ }%
\begin{equation*}
\begin{array}{l}
\left( \rho ,\eta \right) \overset{\ast }{\tilde{H}}_{bc}^{a}=\left( \rho
,\eta \right) \overset{\ast }{H}_{bc}^{a}+\displaystyle\frac{1}{2}%
g^{ae}\cdot \left( \tilde{g}_{ed}\overset{\ast }{\mathbb{T}}_{bc}^{d}-\tilde{%
g}_{bd}\overset{\ast }{\mathbb{T}}_{ec}^{d}+\tilde{g}_{cd}\overset{\ast }{%
\mathbb{T}}_{be}^{d}\right) ,\vspace*{1mm} \\
\left( \rho ,\eta \right) \overset{\ast }{\tilde{V}}_{a}^{bc}=\left( \rho
,\eta \right) \overset{\ast }{V}_{a}^{bc}+\displaystyle\frac{1}{2}\tilde{g}%
_{ae}\cdot \left( g^{ed}\overset{\ast }{\mathbb{S}}_{d}^{bc}-g^{bd}\overset{%
\ast }{\mathbb{S}}_{d}^{ec}+g^{cd}\overset{\ast }{\mathbb{S}}_{d}^{be}\right)%
\end{array}%
\leqno(6.9.6)
\end{equation*}%
\emph{are the components of a normal distinguished linear }$\left( \rho
,\eta \right) $\emph{-connection with}\break $\left( \rho ,\eta \right) $%
\emph{-}$\overset{\ast }{\mathcal{H}}\left( \overset{\ast }{\mathcal{H}}%
\overset{\ast }{\mathcal{H}}\right) $\emph{\ and }$\left( \rho ,\eta \right)
$\emph{-}$\overset{\ast }{\mathcal{V}}\left( \overset{\ast }{\mathcal{V}}%
\overset{\ast }{\mathcal{V}}\right) $\emph{torsions a priori given such that
the generalized tangent bundle}
\begin{equation*}
\left( \left( \rho ,\eta \right) T\overset{\ast }{E},\left( \rho ,\eta
\right) {\tau _{\overset{\ast }{E}}},\overset{\ast }{E}\right)
\end{equation*}%
\emph{\ derives generalized Hamilton }$\left( \rho ,\eta \right) $\emph{%
-space.}

\emph{Moreover, we obtain: }%
\begin{equation*}
\begin{array}{l}
\overset{\ast }{\mathbb{T}}_{bc}^{a}=\left( \rho ,\eta \right) \overset{\ast
}{H}_{bc}^{a}-\left( \rho ,\eta \right) \overset{\ast }{H}%
_{cb}^{a}-L_{bc}^{a}\circ h\circ \overset{\ast }{\pi },\vspace*{2mm} \\
\overset{\ast }{\mathbb{S}}_{a}^{bc}=\left( \rho ,\eta \right) \overset{\ast
}{V}_{a}^{bc}-\left( \rho ,\eta \right) \overset{\ast }{V}_{a}^{cb}.%
\end{array}%
\leqno(6.9.7)
\end{equation*}

\subsection{Einstein equations}

We shall consider a metric structure
\begin{equation*}
\begin{array}{c}
\overset{\ast }{G}=g_{\alpha \beta }d\tilde{z}^{\alpha }\otimes d\tilde{z}%
^{\beta }+g^{ab}\delta \tilde{p}_{a}\otimes \delta \tilde{p}_{b}%
\end{array}%
\end{equation*}%
and a distinguished linear $\left( \rho ,\eta \right) $-connection $\left(
\left( \rho ,\eta \right) \overset{\ast }{H},\left( \rho ,\eta \right)
\overset{\ast }{V}\right) $ compatible with the structure metric $\overset{%
\ast }{G}$ having $\overset{\ast }{\mathcal{H}}\left( \overset{\ast }{%
\mathcal{H}}\overset{\ast }{\mathcal{H}}\right) $ and $\overset{\ast }{%
\mathcal{V}}\left( \overset{\ast }{\mathcal{V}}\overset{\ast }{\mathcal{V}}%
\right) $-torsions prescribed.

\bigskip\noindent\textbf{Definition 6.10.1 }If $\left( \rho ,\eta ,h\right)
\overset{\ast }{\mathbb{R}}_{~\alpha ~\beta }$ and $\left( \rho ,\eta
,h\right) \overset{\ast }{\mathbb{S}}^{~a~b}$ are the components of tensor
Ricci associated to distinguished linear $\left( \rho ,\eta \right) $%
-connection
\begin{equation*}
%
\end{equation*}%
then the scalar
\begin{equation*}
%
%
\leqno(6.10.1)
\end{equation*}%
will be called the \emph{scalar of curvature of distinguished linear }$%
\left( \rho ,\eta \right) $\emph{-connection}\break$\left( \left( \rho ,\eta
\right) H,\left( \rho ,\eta \right) V\right) .$

\bigskip \noindent \textbf{Definition 6.10.2 }The tensor field
\begin{equation*}
%
%
\leqno(6.10.2)
\end{equation*}%
such that its components verify the following conditions:
\begin{equation*}
%
%
\leqno(6.10.3)
\end{equation*}%
where $\varkappa $ is a constant, will be called \emph{the energy-momentum
tensor field associated to distinguished linear }$\left( \rho ,\eta \right) $%
\emph{-connection }$\left( \left( \rho ,\eta \right) \overset{\ast }{H}%
,\left( \rho ,\eta \right) \overset{\ast }{V}\right) $\emph{\ and metrical
structure }$\overset{\ast }{G}.$

The equations $\left( 4.10.3\right) $ will be called \emph{the Einstein
equations associated to distinguished linear }$\left( \rho ,\eta \right) $%
\emph{-connection }$\left( \left( \rho ,\eta \right) \overset{\ast }{H}%
,\left( \rho ,\eta \right) \overset{\ast }{V}\right) $\emph{\ and metrical
structure }$\overset{\ast }{G}$.

Formally, the Einstein equations will be written%
\begin{equation*}
%
\leqno(6.10.3)^{\prime }
\end{equation*}

\subsection{Dual mechanical systems}

Using the diagram:
\begin{equation*}
%
\leqno(6.11.1)
\end{equation*}%
where $\left( \left( E,\pi ,M\right) ,\left[ ,\right] _{E,h},\left( \rho
,\eta \right) \right) $ is a generalized Lie algebroid, we build the
generalized tangent bundle
\begin{equation*}
%
%
\leqno(6.11.3)
\end{equation*}%
is an external force and $\left( \rho ,\eta \right) \overset{\ast }{\Gamma }$
is a $\left( \rho ,\eta \right) $-connection, will be called \emph{dual
mechanical }$\left( \rho ,\eta \right) $\emph{-system.}

A dual mechanical $\left( \rho ,\eta \right) $-system
\begin{equation*}
%
%
\end{equation*}%
endowed with a (pseudo)metrical structure $\overset{\ast }{G}$ determined
with the help of a (pseudo)metrical structure
\begin{equation*}
%
%
\vspace*{1mm}
\end{equation*}%
will be denoted
\begin{equation*}
%
%
\leqno(6.11.4)
\end{equation*}%
and will be called \emph{generalized Hamilton mechanical }$\left( \rho ,\eta
\right) $\emph{-system.}

Any dual mechanical $\left( Id_{TM},Id_{M}\right) $-system and any
generalized Hamilton mechanical $\left( Id_{TM},Id_{M}\right) $-system will
be called \emph{mechanical system} and \emph{generalized Hamilton mechanical
system, }respectively.

\textbf{Definition 6.11.2 }If $H$ respectively $K$ is a smooth Hamilton
respectively Cartan function, then we put the triple
\begin{equation*}
%
%
\end{equation*}%
is an external force. These are called \emph{Hamilton mechanical }$\left(
\rho ,\eta \right) $\emph{-system }and \emph{Cartan mechanical }$\left( \rho
,\eta \right) $\emph{-system }respectively.

Any Hamilton mechanical $\left( Id_{TM},Id_{M}\right) $-system and any
Cartan mechanical\break $\ \left( Id_{TM},Id_{M}\right) $-system will be
called \emph{Hamilton mechanical system }and \emph{Cartan mechanical system,
respectively.}

\subsubsection{\noindent $(\protect\rho ,\protect\eta )$-semisprays and $(%
\protect\rho ,\protect\eta )$-sprays for dual mechanical $(\protect\rho ,%
\protect\eta )$-systems}

Let $\left( \left( \overset{\ast }{E},\overset{\ast }{\pi },M\right) ,%
\overset{\ast }{F}_{e},\left( \rho ,\eta \right) \overset{\ast }{\Gamma }%
\right) $ be an arbitrary dual mechanical $\left( \rho ,\eta \right) $%
-system.

\bigskip\noindent\textbf{Definition 6.11.1.1 }The\textit{\ }vertical section%
\textit{\ }%
\begin{equation*}
%
\leqno(6.11.1.1)
\end{equation*}%
will be called\textit{\ the }\emph{Liouville section.}

\bigskip\noindent \textbf{Definition 6.11.1.2 }The section $\overset{\ast }{S%
}\in \Gamma \left( \left( \rho ,\eta \right) T\overset{\ast }{E},\left( \rho
,\eta \right) \tau _{\overset{\ast }{E}},\overset{\ast }{E}\right) $\ will
be called $\left( \rho ,\eta \right) $\emph{-semispray}\ if there exists an
almost tangent structure $e$ such that
\begin{equation*}
%
%
\leqno(6.11.1.3)
\end{equation*}%
\emph{is a }$\left( \rho ,\eta \right) $\emph{-semispray such that the real
local functions }$G_{a},~a\in \overline{1,n},$\emph{\ satisfy the following
conditions }
\begin{equation*}
%
%
\leqno(6.11.1.4)
\end{equation*}

\emph{In addition, we remark that the local real functions }%
\begin{equation*}
%
%
\leqno(6.11.1.5)
\end{equation*}%
\emph{are the components of a }$\left( \rho ,\eta \right) $\emph{-connection
}$\left( \rho ,\eta \right) \overset{\ast }{\mathring{\Gamma}}$\emph{\ for
the vector bundle }$\left( \overset{\ast }{E},\overset{\ast }{\pi },M\right)
.$

The $\left( \rho ,\eta \right) $-semispray $\overset{\ast }{S}$\ will be
called \emph{the\ canonical }$\left( \rho ,\eta \right) $\emph{-semispray
associated to mechanical }$\left( \rho ,\eta \right) $\emph{-system }$\left(
\left( \overset{\ast }{E},\overset{\ast }{\pi },M\right) ,\overset{\ast }{F}%
_{e},\left( \rho ,\eta \right) \overset{\ast }{\Gamma }\right) $\emph{\ and
from locally invertible }$\mathbf{B}^{\mathbf{v}}$\emph{-morphism }$\left(
g,h\right) .$

\bigskip\noindent\textit{Proof.} We consider the $\mathbf{Mod}$-endomorphism%
\begin{equation*}
%
\end{equation*}

Let $X=\tilde{Z}^{a}\overset{\ast }{\tilde{\partial}}_{a}+Y_{a}\overset{%
\cdot }{\tilde{\partial}}^{a}$ be an arbitrary section.

Since
\begin{equation*}
%
%
\end{equation*}%
it results that
\begin{equation*}
%
%
\end{equation*}%
it results that%
\begin{equation*}
%
%
\leqno\left( P_{2}\right)
\end{equation*}

We remark that
\begin{equation*}
%
%
\end{equation*}

Using the equalities $\left( P_{1}\right) $ and $\left( P_{2}\right) $ we
obtain:
\begin{equation*}
%
%
\end{equation*}

After some calculations, it results that $\mathbb{P}$ is an almost product
structure.

Using the equality
\begin{equation*}
\mathbb{P}=Id-2\left( \rho ,\eta \right) \overset{\ast }{\Gamma },
\end{equation*}%
we obtain that%
\begin{equation*}
%
%
\end{equation*}%
it results that relations $\left( 4.11.1.4\right) $ are satisfied. In
addition, since
\begin{equation*}
%
%
\end{equation*}%
it results the conclusion of the theorem. \hfill \emph{q.e.d.}

\textbf{Theorem 6.11.1.2 }\emph{The following properties hold:}\smallskip

$1^{\circ }\ $\emph{\ Since }%
\begin{equation*}
\overset{\ast }{\overset{\circ }{\tilde{\delta}}}_{c}=\overset{\ast }{\tilde{%
\partial}}_{c}+\left( \rho ,\eta \right) \overset{\ast }{\mathring{\Gamma}}%
_{bc}\overset{\cdot }{\tilde{\partial}}^{b},~c\in \overline{1,r},
\end{equation*}%
\emph{it results that }%
\begin{equation*}
%
%
\leqno(6.11.1.6)
\end{equation*}

$2^{\circ }\ $\emph{\ Since}%
\begin{equation*}
\mathring{\delta}\tilde{p}_{b}=-\left( \rho ,\eta \right) \overset{\ast }{%
\mathring{\Gamma}}_{bc}d\tilde{z}^{c}+d\tilde{p}_{b},
\end{equation*}%
\emph{it results that \ \ }%
\begin{equation*}
%
%
\leqno(6.11.1.7)
\end{equation*}

\smallskip \textbf{Theorem 6.11.1.3 }\emph{The real local functions}%
\begin{equation*}
%
%
\leqno(6.11.1.8)^{\prime }
\end{equation*}%
\emph{respectively are the coefficients to a normal Berwald linear }$\left(
\rho ,\eta \right) $\emph{-connection for the generalized tangent bundle }$%
\left( \left( \rho ,\eta \right) T\overset{\ast }{E},\left( \rho ,\eta
\right) \tau _{\overset{\ast }{E}},\overset{\ast }{E}\right) $.

\medskip \textbf{Theorem 6.11.1.4 }\emph{The tensor of integrability of the }%
$\left( \rho ,\eta \right) $\emph{-connection }$\left( \rho ,\eta \right)
\overset{\ast }{\mathring{\Gamma}}$\emph{\ is as follows:}%
\begin{equation*}
%
\leqno(6.11.1.9)
\end{equation*}%
\emph{where }$_{|c}$\emph{\ is the }$h$\emph{-covariant derivation with
respect to the normal Berwald linear }$\rho $\emph{-connection }$\left(
6.11.1.8\right) $.

\medskip \bigskip\noindent\textit{Proof.} Since
\begin{equation*}
%
%
\end{equation*}%
it results the conclusion of the theorem.\hfill \emph{q.e.d.}

\medskip \textbf{Theorem 6.11.1.5 }\emph{Let }%
\begin{equation*}
\overset{\ast }{\mathbb{T}}_{bc}^{a}\delta _{a}\otimes d\tilde{z}^{b}\otimes
d\tilde{z}^{c}\in \mathcal{T}_{20}^{10}\left( \left( \rho ,\eta \right) T%
\overset{\ast }{E},\left( \rho ,\eta \right) \tau _{\overset{\ast }{E}},%
\overset{\ast }{E}\right)
\end{equation*}%
\emph{and }%
\begin{equation*}
\overset{\ast }{\mathbb{S}}_{a}^{bc}\overset{\cdot }{\tilde{\partial}}%
_{a}\otimes \delta \tilde{y}^{b}\otimes \delta \tilde{y}^{c}\in \mathcal{T}%
_{01}^{02}\left( \left( \rho ,\eta \right) T\overset{\ast }{E},\left( \rho
,\eta \right) \tau _{\overset{\ast }{E}},\overset{\ast }{E}\right)
\end{equation*}%
\emph{such that they verify the following conditions:}%
\begin{equation*}
\overset{\ast }{\mathbb{T}}_{bc}^{a}=-\overset{\ast }{\mathbb{T}}_{cb}^{a},~%
\overset{\ast }{\mathbb{S}}_{a}^{bc}=-\overset{\ast }{\mathbb{S}}%
_{a}^{bc},~\forall b,c\in \overline{1,r}.
\end{equation*}

\emph{If }$\left( \left( \rho ,\eta \right) \overset{\ast }{\tilde{H}}%
,\left( \rho ,\eta \right) \overset{\ast }{\tilde{V}}\right) $\emph{\ is the
distinguished linear }$\left( \rho ,\eta \right) $\emph{-connection
presented in the Theorem 6.9.2, then the local real functions:\ }%
\begin{equation*}
\begin{array}{ll}
\left( \rho ,\eta \right) \overset{\ast }{\widetilde{\mathring{H}}}%
_{bc}^{a}\!\! & =\left( \rho ,\eta \right) \overset{\ast }{\tilde{H}}%
_{bc}^{a}+\frac{1}{8}g^{ae}\left( -\tilde{g}_{fc}\circ h\circ \overset{\ast }%
{\pi }\frac{\partial F_{d}}{\partial p_{f}}\frac{\partial \tilde{g}_{be}}{%
\partial p_{d}}\right. \vspace*{1mm} \\
& \left. +\ \tilde{g}_{fe}\circ h\circ \overset{\ast }{\pi }\frac{\partial
F_{d}}{\partial p_{f}}\frac{\partial \tilde{g}_{bc}}{\partial p_{d}}-\tilde{g%
}_{fb}\circ h\circ \overset{\ast }{\pi }\frac{\partial F_{d}}{\partial p_{f}}%
\frac{\partial \tilde{g}_{ec}}{\partial p_{d}}\right) ,\vspace*{2mm} \\
\left( \rho ,\eta \right) \overset{\ast }{\widetilde{\mathring{V}}}%
_{bc}^{a}\!\! & =\left( \rho ,\eta \right) \overset{\ast }{\tilde{V}}%
_{bc}^{a}%
\end{array}%
\leqno(6.11.1.10)
\end{equation*}%
\emph{are the components of a normal distinguished linear }$\left( \rho
,\eta \right) $\emph{-connection with }$\left( \rho ,\eta \right) $\emph{-}$%
\overset{\ast }{\mathcal{H}}\left( \overset{\ast }{\mathcal{H}}\overset{\ast
}{\mathcal{H}}\right) $\emph{\ and }$\left( \rho ,\eta \right) $\emph{-}$%
\overset{\ast }{\mathcal{V}}\left( \overset{\ast }{\mathcal{V}}\overset{\ast
}{\mathcal{V}}\right) $\emph{\ torsions a priori given such that the
generalized tangent bundle }$\left( \left( \rho ,\eta \right) T\overset{\ast
}{E},\left( \rho ,\eta \right) \tau _{\overset{\ast }{E}},\overset{\ast }{E}%
\right) $\emph{\ derives generalized Hamilton }$\left( \rho ,\eta \right) $%
\emph{-space.}

\emph{In addition, we have: }%
\begin{equation*}
\begin{array}{c}
\left( \rho ,\eta ,h\right) \overset{\ast }{\widetilde{\mathbb{\mathring{T}}}%
}_{bc}^{a}=\overset{\ast }{\mathbb{T}}_{bc}^{a}\vspace*{1mm} \\
\left( \rho ,\eta ,h\right) \overset{\ast }{\widetilde{\mathbb{\mathring{S}}}%
}_{a}^{bc}=\overset{\ast }{\mathbb{S}}_{a}^{bc}.%
\end{array}%
\leqno(6.11.1.11)
\end{equation*}

\emph{The local functions }$\tilde{g}_{fc},~\tilde{g}_{fe},~\tilde{g}_{fb}$%
\emph{\ are the local functions associated to the locally invertible }$%
\mathbf{B}^{\mathbf{v}}$\emph{-morphism }$\left( g,h\right) .$

\smallskip \textbf{Proposition 6.11.1.1 }\emph{If }$\overset{\ast }{S}$\emph{%
\ is the canonical }$\left( \rho ,\eta \right) $\emph{-semispray
asso\-cia\-ted to the mechanical }$\left( \rho ,\eta \right) $\emph{-system }%
$\left( \left( \overset{\ast }{E},\overset{\ast }{\pi },M\right) ,\overset{%
\ast }{F}_{e},\left( \rho ,\eta \right) \overset{\ast }{\Gamma }\right) $%
\emph{\ and from locally invertible }$\mathbf{B}^{\mathbf{v}}$\emph{%
-mor\-phism }$\left( g,h\right) $\emph{,\ then }%
\begin{equation*}
%
\leqno(6.11.1.12)
\end{equation*}

\bigskip\noindent\textit{Proof.} Since the Jacobian matrix of coordinates
transformation is
\begin{equation*}
\left\Vert
%
%
\right) ,%
\end{array}%
\end{equation*}%
the conclusion results immediately. \hfill \emph{q.e.d.}

In the following we consider a differentiable curve $I~\ ^{\underrightarrow{c%
}}~\ M$ and its $\left( g,h\right) $-lift $\dot{c}.$

\medskip \textbf{Definition 6.11.1.3 }The curve $\dot{c}$ is an integral
curve of the $\left( \rho ,\eta \right) $-semispray $\overset{\ast }{S}$ of
the dual mechanical $\left( \rho ,\eta \right) $-system $\left( \left(
\overset{\ast }{E},\overset{\ast }{\pi },M\right) ,\overset{\ast }{F}%
_{e},\left( \rho ,\eta \right) \overset{\ast }{\Gamma }\right) $, if it is
verify the following equality:\textit{\ }%
\begin{equation*}
\begin{array}{l}
\frac{d\dot{c}\left( t\right) }{dt}=\Gamma \left( \overset{\ast }{\tilde{\rho%
}},Id_{\overset{\ast }{E}}\right) \overset{\ast }{S}\left( \dot{c}\left(
t\right) \right) .%
\end{array}%
\leqno(6.11.1.13)
\end{equation*}

\textbf{Theorem 6.11.1.6 }\emph{The integral curves of the canonical }$%
\left( \rho ,\eta \right) $\emph{-semispray asso\-cia\-ted to the mechanical
}$\left( \rho ,\eta \right) $\emph{-system }$\left( \left( \overset{\ast }{E}%
,\overset{\ast }{\pi },M\right) ,\overset{\ast }{F}_{e},\left( \rho ,\eta
\right) \overset{\ast }{\Gamma }\right) $\emph{\ and from locally invertible
}$\mathbf{B}^{\mathbf{v}}$\emph{-mor\-phism }$\left( g,h\right) $\emph{, are
the }$\left( g,h\right) $\emph{-lifts solutions of the equations:\ }%
\begin{equation*}
\begin{array}{l}
\frac{dp_{b}\left( t\right) }{dt}+2G_{b}\!\circ \overset{\ast }{u}\left( c,%
\dot{c}\right) \left( x\left( t\right) \right) {=}\frac{1}{2}F_{b}\!\circ
\overset{\ast }{u}\left( c,\dot{c}\right) \left( x\left( t\right) \right)
\!,\,b{\in }\overline{1,\!r},%
\end{array}%
\leqno(6.11.1.14)
\end{equation*}%
\emph{where }$x\left( t\right) =\left( \eta \circ h\circ c\right) \left(
t\right) .$\medskip

\bigskip\noindent\textit{Proof.} Since the equality
\begin{equation*}
\begin{array}{c}
\frac{d\dot{c}\left( t\right) }{dt}=\Gamma \left( \overset{\ast }{\tilde{\rho%
}},Id_{\overset{\ast }{E}}\right) \overset{\ast }{S}\left( \dot{c}\left(
t\right) \right)%
\end{array}%
\end{equation*}%
is equivalent with
\begin{equation*}
\begin{array}{c}
\frac{d}{dt}((\eta \circ h\circ c)^{i}(t),p_{b}(t))=\vspace*{1mm} \\
=\left( \rho _{a}^{i}\circ \eta \circ h\circ c(t)g^{ae}\circ h\circ
c(t)p_{e}(t),-2\left( G_{b}-\frac{1}{4}F_{b}\right) ((\eta \circ h\circ
c)(t),p\left( t\right) )\right) ,%
\end{array}%
\end{equation*}%
it results
\begin{equation*}
\begin{array}{c}
\frac{dp_{b}\left( t\right) }{dt}+2G_{b}\!\left( x\left( t\right) ,p\left(
t\right) \right) {=}\frac{1}{2}F_{b}\!\left( x\left( t\right) ,p\left(
t\right) \right) \!,\,b{\in }\overline{1,\!r},\vspace*{1mm} \\
\frac{dx^{i}\left( t\right) }{dt}=\rho _{a}^{i}\circ \eta \circ h\circ
c\left( t\right) g^{ae}\circ h\circ c\left( t\right) p_{e}\left( t\right) ,%
\end{array}%
\end{equation*}%
where $x^{i}\left( t\right) =\left( \eta \circ h\circ c\right) ^{i}\left(
t\right) $. \hfill \emph{q.e.d.}\medskip

\textbf{Definition 6.11.1.4 }If $\overset{\ast }{S}$\ is a $\left( \rho
,\eta \right) $-semispray, then the vector field
\begin{equation*}
\begin{array}{l}
\left[ \overset{\ast }{\mathbb{C}},\overset{\ast }{S}\right] _{\left( \rho
,\eta \right) T\overset{\ast }{E}}-\overset{\ast }{S}%
\end{array}%
\leqno(6.11.1.15)
\end{equation*}%
will be called the \emph{derivation of }$\left( \rho ,\eta \right) $\emph{%
-semispray }$\overset{\ast }{S}.$

The $\left( \rho ,\eta \right) $-semispray $\overset{\ast }{S}$\ will be
called $\left( \rho ,\eta \right) $\emph{-spray} if there are verified the
following conditions:\medskip

1. $\overset{\ast }{S}\circ 0\in C^{1},$\ where $0$\ is the null
section;\smallskip

2. Its derivation is the null vector field.\medskip

The $\left( \rho ,\eta \right) $-semispray $\overset{\ast }{S}$\ will be
called \emph{quadratic }$\left( \rho ,\eta \right) $\emph{-spray }if there
are verified the following conditions:\medskip

1. $\overset{\ast }{S}\circ 0\in C^{2},$\ where $0$\ is the null
section;\smallskip

2. Its derivation is the null vector field.\medskip

In particular, \ if $\ \left( \rho ,\eta \right) =\left(
id_{TM},Id_{M}\right) $ and $\left( g,h\right) =\left( Id_{E},Id_{M}\right)
, $ \ then \ we \ obtain \ the \ \emph{spray} \ and the \emph{quadratic
spray }which is similar with the classical spray and quadratic spray.

\medskip \textbf{Theorem 6.11.1.7 }\emph{If }$\overset{\ast }{S}$\emph{\ is
the canonical }$\left( \rho ,\eta \right) $\emph{-semispray associated to
mechanical }$\left( \rho ,\eta \right) $\emph{-system }$\left( \left(
\overset{\ast }{E},\overset{\ast }{\pi },M\right) ,\overset{\ast }{F}%
_{e},\left( \rho ,\eta \right) \overset{\ast }{\Gamma }\right) $\emph{\ and
from locally invertible }$\mathbf{B}^{\mathbf{v}}$\emph{-morphism }$\left(
g,h\right) $\emph{, then}%
\begin{equation*}
\begin{array}{cl}
2\left( G_{b}-\frac{1}{4}F_{b}\right) & =\left( \rho ,\eta \right) \overset{%
\ast }{\Gamma }_{bc}\left( g^{cf}\circ h\circ \overset{\ast }{\pi }\cdot
p_{f}\right) \\
&
\begin{array}{l}
+\frac{1}{2}\left( g^{de}\circ h\circ \overset{\ast }{\pi }\cdot
p_{e}\right) L_{dc}^{a}\circ h\circ \overset{\ast }{\pi } \\
\cdot \tilde{g}_{ba}\circ h\circ \overset{\ast }{\pi }\left( g^{cf}\circ
h\circ \overset{\ast }{\pi }\cdot p_{f}\right) ,~b\in \overline{1,r}.%
\end{array}%
\end{array}%
\leqno(6.11.1.16)
\end{equation*}

\emph{Then we obtain the spray }
\begin{equation*}
\begin{array}{l}
\overset{\ast }{S}=\left( g^{ae}\circ h\circ \overset{\ast }{\pi }\right)
p_{e}\frac{\partial }{\partial \tilde{z}^{a}}+\left( \rho ,\eta \right)
\overset{\ast }{\Gamma }_{bc}\left( g^{cf}\circ h\circ \overset{\ast }{\pi }%
\cdot p_{f}\right) \frac{\partial }{\partial \tilde{p}_{b}} \\
+\frac{1}{2}\left( g^{de}\circ h\circ \overset{\ast }{\pi }\cdot
p_{e}\right) L_{dc}^{a}\circ h\circ \overset{\ast }{\pi }\cdot \tilde{g}%
_{ba}\circ h\circ \overset{\ast }{\pi }\left( g^{cf}\circ h\circ \overset{%
\ast }{\pi }\cdot p_{f}\right) \frac{\partial }{\partial \tilde{p}_{b}}.%
\end{array}%
\leqno(6.11.1.17)
\end{equation*}

\emph{This }$\left( \rho ,\eta \right) $\emph{-spray will be called the
canonical }$\left( \rho ,\eta \right) $\emph{-spray associated to mechanical
system }$\left( \left( \overset{\ast }{E},\overset{\ast }{\pi },M\right) ,%
\overset{\ast }{F}_{e},\left( \rho ,\eta \right) \overset{\ast }{\Gamma }%
\right) $\emph{\ and from locally invertible }$\mathbf{B}^{\mathbf{v}}$\emph{%
-morphism }$(g,h).$

\emph{In particular, if }$\left( \rho ,\eta ,g,h\right) =\left(
Id_{TM},Id_{M},Id_{M},Id_{M}\right) ,$\emph{\ then we get the canonical
spray associated to connection }$\overset{\ast }{\Gamma }$\emph{\ which is
similar with the classical canonical spray associated to connection }$%
\overset{\ast }{\Gamma }$.

\bigskip\noindent\textit{Proof.} Since
\begin{equation*}
%
\end{equation*}%
it results that
\begin{equation*}
%
%
\leqno\left( S_{1}\right)
\end{equation*}%
Using the equality $(6.11.1.4)$ it results that
\begin{equation*}
%
%
\leqno\left( S_{2}\right)
\end{equation*}%
Using the equalities $\left( S_{1}\right) $ and $\left( S_{2}\right) $ it
results the conclusion of the theorem.\hfill \emph{q.e.d.}\medskip

\textbf{Remark 6.11.1.2. }If $\left( \rho ,\eta ,h\right) =\left(
id_{TM},Id_{M},Id_{M}\right) ,$ then we get the canonical spray associated
to connection $\overset{\ast }{\Gamma }$.

\textbf{Theorem 6.11.1.8 }\emph{The integral curves of canonical }$\left(
\rho ,\eta \right) $\emph{-spray associated to mechanical }$\left( \rho
,\eta \right) $\emph{-system }$\left( \left( \overset{\ast }{E},\overset{%
\ast }{\pi },M\right) ,\overset{\ast }{F}_{e},\left( \rho ,\eta \right)
\overset{\ast }{\Gamma }\right) $\emph{\ and from locally invertible }$%
\mathbf{B}^{\mathbf{v}}$\emph{-morphism\ }$\left( g,h\right) $\emph{\ are
the }$\left( g,h\right) $\emph{-lifts solutions of the following system of
equations:\ }%
\begin{equation*}
\begin{array}{l}
\frac{dp_{b}}{dt}-\left( \rho ,\eta \right) \overset{\ast }{\Gamma }%
_{bc}\circ \overset{\ast }{u}\left( c,\dot{c}\right) \circ \left( \eta \circ
h\circ c\right) \cdot \left( g^{cf}\circ h\circ \overset{\ast }{\pi }\cdot
p_{f}\right) \\
+\frac{1}{2}\left( g^{de}\circ h\circ \overset{\ast }{\pi }\cdot
p_{e}\right) \cdot L_{dc}^{a}\circ h\circ \overset{\ast }{\pi }\cdot \tilde{g%
}_{ba}\circ h\circ \overset{\ast }{\pi }\cdot \left( g^{cf}\circ h\circ
\overset{\ast }{\pi }\cdot p_{f}\right) =0,%
\end{array}%
\leqno(6.11.1.17)
\end{equation*}%
\emph{where }$x\left( t\right) =\eta \circ h\circ c\left( t\right) .$

\subsubsection{The Hamiltonian formalism for Hamilton mechanical $\left(
\protect\rho ,\protect\eta \right) $-systems}

Let $\left( \left( \overset{\ast }{E},\overset{\ast }{\pi },M\right) ,%
\overset{\ast }{F}_{e},H\right) $ be an arbitrary Hamilton mechanical $%
\left( \rho ,\eta \right) $-system.

The \emph{natural dual }$\left( \rho ,\eta \right) $\emph{-base }$\left( d%
\tilde{z}^{\alpha },d\tilde{p}_{a}\right) $ of natural $\left( \rho ,\eta
\right) $-base $\left( \frac{\partial }{\partial \tilde{z}^{\alpha }},\frac{%
\partial }{\partial \tilde{p}_{a}}\right) $ is determined by the equations
\begin{equation*}
\begin{array}{c}
\left\{
\begin{array}{cc}
\left\langle d\tilde{z}^{\alpha },\frac{\partial }{\partial \tilde{z}^{\beta
}}\right\rangle =\delta _{\beta }^{\alpha }, & \left\langle d\tilde{z}%
^{\alpha },\frac{\partial }{\partial \tilde{p}_{b}}\right\rangle =0, \\
\left\langle d\tilde{p}_{a},\frac{\partial }{\partial \tilde{z}^{\beta }}%
\right\rangle =0, & \left\langle d\tilde{p}_{a},\frac{\partial }{\partial
\tilde{p}_{b}}\right\rangle =\delta _{a}^{b}.%
\end{array}%
\right.%
\end{array}%
\end{equation*}

It is very important to remark that the $1$-forms $d\tilde{z}^{\alpha
},~\alpha \in \overline{1,p}$ and $d\tilde{p}_{a},~a\in \overline{1,r}$ are
not the differentials of coordinates functions as in the classical case, but
we will use the same notations.

In this case
\begin{equation*}
\left( d\tilde{z}^{\alpha }\right) \neq d^{\left( \rho ,\eta \right) T%
\overset{\ast }{E}}\left( \tilde{z}^{\alpha }\right) =0,
\end{equation*}%
where $d^{\left( \rho ,\eta \right) T\overset{\ast }{E}}$ is the exterior
differentiation operator associated to exterior differential $\mathcal{F}%
\left( \overset{\ast }{E}\right) $-algebra
\begin{equation*}
\left( \Lambda \left( \left( \rho ,\eta \right) T\overset{\ast }{E},\left(
\rho ,\eta \right) \tau _{\overset{\ast }{E}},\overset{\ast }{E}\right)
,+,\cdot ,\wedge \right) .
\end{equation*}

Let $H$ be a regular Hamiltonian and let $\left( g,h\right) $\ be a $\mathbf{%
B}^{\mathbf{v}}$-morphism locally invertible of $\left( \overset{\ast }{E},%
\overset{\ast }{\pi },M\right) $ source and $\left( E,\pi ,M\right) $ target.

\textbf{Definition 6.11.2.1 } The $1$-form
\begin{equation*}
%
\leqno(6.11.2.1)
\end{equation*}%
will be called the $1$\emph{-form of Poincar\'{e}-Cartan type associated to
the regular Hamiltonian }$H$ \emph{and to the locally invertible }$\mathbf{B}%
^{\mathbf{v}}$\emph{-morphism }$\left( g,h\right) $.\medskip

We obtain easily:
\begin{equation*}
%
%
\leqno(6.11.2.2)
\end{equation*}

\smallskip \noindent\textbf{Definition 6.11.2.2 } The $2$-form
\begin{equation*}
\omega _{H}=d^{\left( \rho ,\eta \right) T\overset{\ast }{E}}\theta _{L}
\end{equation*}%
will be called the $2$\emph{-form of Poincar\'{e}-Cartan type associated to
the regular Hamiltonian }$H$ \emph{and to the locally invertible }$\mathbf{B}%
^{\mathbf{v}}$\emph{-morphism }$\left( g,h\right) $.\medskip

By the definition of $d^{\left( \rho ,\eta \right) T\overset{\ast }{E}},$ we
obtain:
\begin{equation*}
%
%
\right. \hspace*{-4mm}\leqno(6.11.2.4)
\end{equation*}

\medskip \textbf{Definition 6.11.2.3 } The real function
\begin{equation*}
%
%
\leqno(6.11.2.5)
\end{equation*}%
will be called the \emph{energy of regular Hamiltonian }$H.$

\medskip \textbf{Theorem 6.11.2.1 }\emph{The equation }%
\begin{equation*}
\left( 6.11.2.6\right) \,\
%
%
\end{equation*}%
\emph{has an unique solution }$\overset{\ast }{S}_{H}\left( g,h\right) $%
\emph{\ of the type: }%
\begin{equation*}
%
%
\hspace*{-4mm}\leqno(6.11.2.9)
\end{equation*}

$\overset{\ast }{S}_{H}\left( g,h\right) $\textit{\ }will be called \emph{%
the\ canonical }$\left( \rho ,\eta \right) $\emph{-semispray associated to
Hamilton mechanical }$\left( \rho ,\eta \right) $\emph{-system }$\left(
\left( \overset{\ast }{E},\overset{\ast }{\pi },M\right) ,\overset{\ast }{F}%
_{e},H\right) $\emph{\ and from locally invertible }$\mathbf{B}^{\mathbf{v}}$%
\emph{-morphism }$(g,h).$

\noindent \textit{Proof.} We obtain that%
\begin{equation*}
i_{\overset{\ast }{S}}\left( \omega _{H}\right) =-d^{\left( \rho ,\eta
\right) T\overset{\ast }{E}}\left( \mathcal{E}_{H}\right)
\end{equation*}%
if and only if%
\begin{equation*}
\omega _{H}\left( \overset{\ast }{S},X\right) =-\Gamma \left( \overset{\ast }%
{\tilde{\rho}},Id_{\overset{\ast }{E}}\right) \left( X\right) \left(
\mathcal{E}_{H}\right) ,\vspace*{1mm}
\end{equation*}%
for any $X\in \Gamma \left( \left( \rho ,\eta \right) T\overset{\ast }{E}%
,\left( \rho ,\eta \right) \tau _{\overset{\ast }{E}},\overset{\ast }{E}%
\right) .$

Particularly, we obtain:
\begin{equation*}
%
\end{equation*}

If we expand this equality using $\left( 6.11.2.2\right) $ and $\left(
6.11.2.4\right) $, we obtain%
\begin{equation*}
%
%
\end{equation*}%
After some calculations, we obtain
\begin{equation*}
2\left( G_{a}-\frac{1}{4}F_{a}\right) =\left( g^{be}\circ h\circ \pi \cdot
\tilde{H}_{ea}\right) \cdot E_{b}\left( H,g,h\right) ,
\end{equation*}%
where
\begin{equation*}
%
%
\end{equation*}

\hfill\emph{q.e.d.}

\textbf{Theorem 6.11.2.2 }\emph{The\ real local functions }
\begin{equation*}
%
%
\leqno(6.11.2.10)
\end{equation*}%
\emph{are the components of a }$\left( \rho ,\eta \right) $\emph{-connection
}$\left( \rho ,\eta \right) \overset{\ast }{\Gamma }$\emph{\ for the vector
bundle }$\left( \overset{\ast }{E},\overset{\ast }{\pi },M\right) $\emph{\
which will be called the }$\left( \rho ,\eta \right) $\emph{-connection
associated to Hamilton mechanical }$\left( \rho ,\eta \right) $\emph{-system
}$\left( \left( \overset{\ast }{E},\overset{\ast }{\pi },M\right) ,\overset{%
\ast }{F}_{e},H\right) $\emph{\ and from locally invertible }$\mathbf{B}^{%
\mathbf{v}}$\emph{-morphism} $(g,h).$

\medskip \textbf{Corollary 6.11.2.1 }\emph{The real local functions}%
\begin{equation*}
\begin{array}{cl}
(\rho ,\eta )\overset{\ast }{\mathring{\Gamma}}_{bc}\!\!\! & =\left( \tilde{g%
}_{ec}\circ h\circ \pi \right) \frac{\partial G_{b}}{\partial p_{e}}\vspace*{%
1.2mm} \\
& -\frac{1}{2}\left( g^{de}\circ h\circ \overset{\ast }{\pi }\cdot
p_{e}\right) L_{dc}^{a}\circ h\circ \overset{\ast }{\pi }\cdot \tilde{g}%
_{ab}\circ h\circ \overset{\ast }{\pi },~b,c\in \overline{1,r}%
\end{array}%
\leqno(6.11.2.11)
\end{equation*}%
\emph{are the components of a }$\left( \rho ,\eta \right) $\emph{-connection
}$\left( \rho ,\eta \right) \overset{\ast }{\mathring{\Gamma}}$\emph{\ for
the vector bundle }$\left( \overset{\ast }{E},\overset{\ast }{\pi },M\right)
.$

\emph{In addition, we have}
\begin{equation*}
\begin{array}{c}
\left( \rho ,\eta \right) \overset{\ast }{\mathring{\Gamma}}_{\,bc}=\left(
\rho ,\eta \right) \overset{\ast }{\Gamma }_{bc}+\frac{1}{4}(\tilde{g}%
_{ec}\circ h\circ \pi )\cdot \frac{\partial F_{b}}{\partial p_{e}},~\forall
a,c\in \overline{1,r}.%
\end{array}%
\leqno(6.11.2.12)
\end{equation*}

\textbf{Theorem 6.11.2.3 }\emph{The\ integral curves of the canonical }$%
\left( \rho ,\eta \right) $\emph{-semispray associated to }$\left( \left(
\overset{\ast }{E},\overset{\ast }{\pi },M\right) ,\overset{\ast }{F}%
_{e},H\right) $\emph{\ mechanical }$\left( \rho ,\eta \right) $\emph{-system
and from locally invertible }$\mathbf{B}^{\mathbf{v}}$\emph{-morphism }$%
\left( g,h\right) $\emph{\ are the autoparallel lifts with respect to }$%
\left( \rho ,\eta \right) $\emph{-connection }$\left( \rho ,\eta \right)
\overset{\ast }{\Gamma }.$

\medskip \textbf{Definition 6.11.2.4 }The equations
\begin{equation*}
\begin{array}{c}
\,\dfrac{dp_{b}\left( t\right) }{dt}+\left( g^{ae}\circ h\circ \overset{\ast
}{\pi }\cdot \tilde{H}_{eb}\cdot E_{a}\left( H,g,h\right) \right) \circ
\overset{\ast }{u}\left( c,\dot{c}\right) \circ \left( \eta \circ h\circ
c\left( t\right) \right) =0,%
\end{array}%
\leqno(6.11.2.13)
\end{equation*}%
will be called the \emph{equations of Hamilton-Jacobi type associated to
Hamilton mechanical }$\left( \rho ,\eta \right) $\emph{-system }$\left(
\left( \overset{\ast }{E},\overset{\ast }{\pi },M\right) ,\overset{\ast }{F}%
_{e},H\right) $\emph{\ and from locally invertible }$\mathbf{B}^{\mathbf{v}}$%
\emph{-morphism }$\left( g,h\right) .$

\medskip \textbf{Remark 6.11.2.1 }The\ integral curves of the canonical $%
\left( \rho ,\eta \right) $-semispray associated to dual mechanical $\left(
\rho ,\eta \right) $-system $\left( \left( \overset{\ast }{E},\overset{\ast }%
{\pi },M\right) ,\overset{\ast }{F}_{e},H\right) $\ and from locally
invertible\emph{\ }$\mathbf{B}^{\mathbf{v}}$-morphism $\left( g,h\right) $\
are the $\left( g,h\right) $-lifts \ solutions for \ the \ equations of
Hamilton-Jacobi type $\left( 6.11.2.13\right) $.

\section{The (horizontal) Legendre $\left( \protect\rho ,\protect\eta %
,h\right) $-equivalence}

Let $\left( E,\pi ,M\right) $ be a vector bundle.

We take $\left( x^{i},y^{a}\right) $ as canonical local coordinates on $%
\left( E,\pi ,M\right) ,$ where $i\in \overline{1,m}$ and $a\in \overline{1,r%
}.$

Consider
\begin{equation*}
\left( x^{i},y^{a}\right) \longrightarrow \left( x^{i%
%TCIMACRO{\U{b4}}%
%BeginExpansion
{\acute{}}%
%EndExpansion
}\left( x^{i}\right) ,y^{a%
%TCIMACRO{\U{b4}}%
%BeginExpansion
{\acute{}}%
%EndExpansion
}\left( x^{i},y^{a}\right) \right)
\end{equation*}%
a change of coordinates on $\left( E,\pi ,M\right) $. Then the coordinates $%
y^{a}$ change to $y^{a%
%TCIMACRO{\U{b4}}%
%BeginExpansion
{\acute{}}%
%EndExpansion
}$ by the rule:
\begin{equation*}
\begin{array}{c}
y^{a%
%TCIMACRO{\U{b4}}%
%BeginExpansion
{\acute{}}%
%EndExpansion
}=M_{a}^{a%
%TCIMACRO{\U{b4}}%
%BeginExpansion
{\acute{}}%
%EndExpansion
}y^{a}.%
\end{array}%
\leqno(7.1)
\end{equation*}

Let $\left( \overset{\ast }{E},\overset{\ast }{\pi },M\right) $ be the dual
vector bundle of $\left( E,\pi ,M\right) $.

We take $\left( x^{i},p_{a}\right) $ as canonical local coordinates on $%
\left( \overset{\ast }{E},\overset{\ast }{\pi },M\right) ,$ where $i\in
\overline{1,m}$ and $a\in \overline{1,r}.$

Consider
\begin{equation*}
\left( x^{i},p_{a}\right) \longrightarrow \left( x^{i%
%TCIMACRO{\U{b4}}%
%BeginExpansion
{\acute{}}%
%EndExpansion
}\left( x^{i}\right) ,p_{a%
%TCIMACRO{\U{b4}}%
%BeginExpansion
{\acute{}}%
%EndExpansion
}\left( x^{i},p_{a}\right) \right)
\end{equation*}%
a change of coordinates on $\left( \overset{\ast }{E},\overset{\ast }{\pi }%
,M\right) $. Then the coordinates $p_{a}$ change to $p_{a%
%TCIMACRO{\U{b4}}%
%BeginExpansion
{\acute{}}%
%EndExpansion
}$ by the rule:
\begin{equation*}
%
\leqno(7.1^{\prime })
\end{equation*}

If $\left( U,s_{U}\right) $ and $\left( U,\overset{\ast }{s}_{U}\right) $
are vector local $\left( m+r\right) $-charts then
\begin{equation*}
%
%
\end{equation*}

Let $L$ be a differentiable Lagrangian defined on the total space of the
vector bundle $\left( E,\pi ,M\right) .$

If $\left( U,s_{U}\right) $ is a vector local $\left( m+r\right) $-chart for
$\left( E,\pi ,M\right) $, then we obtain the following real functions
defined on $\pi ^{-1}\left( U\right) $:%
\begin{equation*}
%
%
.\leqno(7.3)
\end{equation*}

We build the fiber bundle morphism%
\begin{equation*}
%
%
,
\end{equation*}%
where $\ \varphi _{L}$ is locally defined
\begin{equation*}
%
%
,\leqno(7.4)
\end{equation*}%
for any vector local $\left( m+r\right) $-chart $\left( U,s_{U}\right) $ of $%
\left( E,\pi ,M\right) $ and for any vector local $\left( m+r\right) $-chart
$\left( U,\overset{\ast }{s}_{U}\right) $ of $\left( \overset{\ast }{E},%
\overset{\ast }{\pi },M\right) .$

Using the differentiable Lagrangian $L,$ we build the differentiable
Hamiltonian $H,$ locally defined by
\begin{equation*}
%
%
,\leqno(7.2^{\prime })
\end{equation*}%
for any vector local $\left( m+r\right) $-chart $\left( U,\overset{\ast }{s}%
_{U}\right) $ of $\left( \overset{\ast }{E},\overset{\ast }{\pi },M\right) ,$
where $\left( y^{a},~a\in \overline{1,r}\right) $ are the components
solutions of the differentiable equations%
\begin{equation*}
%
%
\end{equation*}

If $\left( U,\overset{\ast }{s}_{U}\right) $ is a vector local $\left(
m+r\right) $-chart for $\left( \overset{\ast }{E},\overset{\ast }{\pi }%
,M\right) $, then we obtain the following real functions defined on $\overset%
{\ast }{\pi }^{-1}\left( U\right) $:%
\begin{equation*}
%
%
.\leqno(7.3^{\prime })
\end{equation*}

Using this Hamiltonian, we build the fiber bundle morphism%
\begin{equation*}
%
%
,
\end{equation*}%
where $\ \varphi _{H}$ is locally defined%
\begin{equation*}
%
%
,\leqno(7.4^{\prime })
\end{equation*}%
for any vector local $\left( m+r\right) $-chart $\left( U,s_{U}\right) $ of $%
\left( E,\pi ,M\right) $ and for any vector local $\left( m+r\right) $-chart
$\left( U,\overset{\ast }{s}_{U}\right) $ of $\left( \overset{\ast }{E},%
\overset{\ast }{\pi },M\right) .$

Using the $\mathbf{B}$-morphism $\left( \varphi _{L},Id_{M}\right) $, we
build the $\mathbf{B}^{\mathbf{v}}$-morphism $\left( \left( \rho ,\eta
\right) T\varphi _{L},\varphi _{L}\right) $ given by the diagram%
\begin{equation*}
%
%
\leqno(7.6)
\end{equation*}%
for any $\tilde{Z}^{\alpha }\tilde{\partial}_{\alpha }+Y^{a}\overset{\cdot }{%
\tilde{\partial}}_{a}\in \Gamma \left( \left( \rho ,\eta \right) TE,\left(
\rho ,\eta \right) \tau _{E},E\right) .$

The $\mathbf{B}^{\mathbf{v}}$-morphism $\left( \left( \rho ,\eta \right)
T\varphi _{L},\varphi _{L}\right) $ will be called the $\left( \rho ,\eta
\right) $\emph{-tangent application of the Legendre bundle morphism
associated to the Lagrangian }$L$.

Using the $\mathbf{B}$-morphism $\left( \varphi _{H},Id_{M}\right) $, we
build the $\mathbf{B}^{\mathbf{v}}$-morphism $\left( \left( \rho ,\eta
\right) T\varphi _{H},\varphi _{H}\right) $ given by the diagram
\begin{equation*}
\begin{array}{rcl}
\left( \rho ,\eta \right) T\overset{\ast }{E} & ^{\underrightarrow{~\ \left(
\rho ,\eta \right) T\varphi _{H}~\ }} & \left( \rho ,\eta \right) TE \\
\left( \rho ,\eta \right) \tau _{\overset{\ast }{E}}\downarrow &  &
\downarrow \left( \rho ,\eta \right) \tau _{E} \\
E^{\ast } & ^{\underrightarrow{~\ \ \ \ \varphi _{H~\ \ \ \ }}} & E,%
\end{array}
\leqno(7.5^{\prime })
\end{equation*}
such that
\begin{equation*}
\begin{array}{cl}
\Gamma \left( \left( \rho ,\eta \right) T\varphi _{L},\varphi _{L}\right)
\left( \tilde{Z}^{\alpha }\overset{\ast }{\tilde{\partial}}_{\alpha }\right)
& =\left( \tilde{Z}^{\alpha }\circ \varphi _{L}\right) \tilde{\partial}%
_{\alpha }+\left[ \left( \rho _{\alpha }^{i}{\circ }h{\circ }\overset{\ast }{%
\pi }\right) \tilde{Z}^{\alpha }H_{i}^{b}\right] \circ \varphi _{L}\overset{%
\cdot }{\tilde{\partial}}_{b}, \\
\Gamma \left( \left( \rho ,\eta \right) T\varphi _{L},\varphi _{L}\right)
\left( Y_{a}\overset{\cdot }{\tilde{\partial}}^{a}\right) & =\left(
Y_{a}H^{ab}\right) \circ \varphi _{L}\overset{\cdot }{\tilde{\partial}}_{b},%
\end{array}
\leqno(7.6^{\prime })
\end{equation*}
for any $\tilde{Z}^{\alpha }\overset{\ast }{\tilde{\partial}}_{\alpha }+Y_{a}%
\overset{\cdot }{\tilde{\partial}}^{a}\in \Gamma \left( \left( \rho ,\eta
\right) T\overset{\ast }{E},\left( \rho ,\eta \right) \tau _{\overset{\ast }{%
E}},\overset{\ast }{E}\right) .$

The $\mathbf{B}^{\mathbf{v}}$-morphism $\left( \left( \rho ,\eta \right)
T\varphi _{H},\varphi _{H}\right) $ will be called the $\left( \rho ,\eta
\right) $\emph{-tangent application of the Legendre bundle morphism
associated to the Hamiltonian }$H$.

Let
\begin{equation*}
%
\leqno(7.7)
\end{equation*}%
be the natural base for sections Lie algebra $\left( \Gamma \left( TE,\tau
_{E},E\right) ,+,\cdot ,\left[ ,\right] _{TE}\right) .$

Let
\begin{equation*}
%
%
\leqno(7.7^{\prime })
\end{equation*}%
be the natural base for sections Lie algebra $\left( \Gamma \left( T\overset{%
\ast }{E},\tau _{\overset{\ast }{E}},\overset{\ast }{E}\right) ,+,\cdot ,%
\left[ ,\right] _{T\overset{\ast }{E}}\right) .$

Using the diagram:%
\begin{equation*}
%
%
,\leqno(7.8)
\end{equation*}%
where $\left( \left( F,\nu ,N\right) ,\left[ ,\right] _{F,h},\left( \rho
,\eta \right) \right) $ is a generalized Lie algebroid, we build the
generalized tangent bundle
\begin{equation*}
%
%
.\leqno(7.9)
\end{equation*}

The natural $\left( \rho ,\eta \right) $-base of sections is denoted
\begin{equation*}
%
%
\leqno(7.10)
\end{equation*}

Using the diagram:
\begin{equation*}
%
%
,\leqno(7.8^{\prime })
\end{equation*}%
where $\left( \left( F,\nu ,N\right) ,\left[ ,\right] _{F,h},\left( \rho
,\eta \right) \right) $ is a generalized Lie algebroid, we build the
generalized tangent bundle%
\begin{equation*}
%
%
\leqno(7.9^{\prime })
\end{equation*}

The natural $\left( \rho ,\eta \right) $-base of sections is denoted%
\begin{equation*}
%
%
\leqno(7.10^{\prime })
\end{equation*}

\subsection{The duality between mechanical systems}

Let $(\left( E,\pi ,M\right) ,F_{e},(\rho ,\eta )\Gamma )$ be a mechanical $%
\left( \rho ,\eta \right) $-system.

Let $g\in \mathbf{Man}\left( E,E\right) $ such that $\left( g,h\right) $ is
a $\mathbf{B}^{\mathbf{v}}$\textit{-}morphism locally invertible of $\left(
E,\pi ,M\right) $ source and $\left( E,\pi ,M\right) $ target, on components
$g_{b}^{a}$.

The $\mathbf{Mod}$-endomorphism
\begin{equation*}
%
%
\leqno(7.1.1)
\end{equation*}%
is the almost tangent structure associated to\emph{\ }$\mathbf{B}^{\mathbf{v}%
}$-morphism $\left( g,h\right) $.

The\textit{\ }vertical section
\begin{equation*}
%
%
\leqno(7.1.2)
\end{equation*}%
is the Liouville section.

Let $\left( \left( \overset{\ast }{E},\overset{\ast }{\pi },M\right) ,%
\overset{\ast }{F}_{e},(\rho ,\eta )\overset{\ast }{\Gamma }\right) $ be a
dual mechanical $\left( \rho ,\eta \right) $-system.

Let $g\in \mathbf{Man}\left( \overset{\ast }{E},E\right) $ be such that $%
\left( g,h\right) $ is a $\mathbf{B}^{\mathbf{v}}$\textit{-}morphism locally
invertible of $\left( \overset{\ast }{E},\overset{\ast }{\pi },M\right) $
source and $\left( E,\pi ,M\right) $ target, on components $g^{ab}$.

The $\mathbf{Mod}$-endomorphism
\begin{equation*}
%
\leqno(7.1.1)^{\prime }
\end{equation*}%
is the almost tangent structure associated to $\mathbf{B}^{\mathbf{v}}$%
-morphism $\left( g,h\right) $.

The vertical section
\begin{equation*}
%
%
\leqno(7.1.2)^{\prime }
\end{equation*}
is the Liouville section.

Let
\begin{equation*}
%
%
\leqno(7.1.3)
\end{equation*}%
be the $\left( \rho ,\eta \right) $-semispray associated to mechanical $%
\left( \rho ,\eta \right) $-system $\left( \left( E,\pi ,M\right)
,F_{e},\left( \rho ,\eta \right) \Gamma \right) $\ and from locally
invertible $\mathbf{B}^{\mathbf{v}}$-morphism $\left( g,h\right) $ and let
\begin{equation*}
%
%
\leqno(7.1.3)^{\prime }
\end{equation*}%
be the $\left( \rho ,\eta \right) $-semispray associated to the mechanical $%
\left( \rho ,\eta \right) $-system $\left( \left( \overset{\ast }{E},\overset%
{\ast }{\pi },M\right) ,F_{e},\left( \rho ,\eta \right) \overset{\ast }{%
\Gamma }\right) $\ and from locally invertible $\mathbf{B}^{\mathbf{v}}$%
-morphism $\left( g,h\right) .$

\medskip \textbf{Theorem 7.1.1 }\emph{\ If }%
\begin{equation*}
%
%
\leqno(7.1.4)
\end{equation*}%
\emph{then we obtain:}%
\begin{equation*}
%
%
\leqno(7.1.4)^{\prime }
\end{equation*}%
\emph{then we obtain:}%
\begin{equation*}
%
%
\leqno(7.1.6)^{\prime }
\end{equation*}

\subsection{The duality between Lie algebroids structures}

The generalized tangent bundle
\begin{equation*}
%
%
\end{equation*}%
can be endowed with a Lie algebroid structure
\begin{equation*}
\left( \left[ ,\right] _{\left( \rho ,\eta \right) TE},\left( \tilde{\rho}%
,Id_{E}\right) \right) .
\end{equation*}

The Lie bracket $\left[ ,\right] _{\left( \rho ,\eta \right) TE}$ is defined
by%
\begin{equation*}
%
%
\leqno(7.2.1)
\end{equation*}%
for any sections $\left( \tilde{Z}_{1}^{\alpha }\frac{\partial }{\partial
\tilde{z}^{\alpha }}+Y_{1}^{a}\frac{\partial }{\partial \tilde{y}^{a}}%
\right) $ and $\left( \tilde{Z}_{2}^{\beta }\frac{\partial }{\partial \tilde{%
z}^{\beta }}+Y_{2}^{b}\frac{\partial }{\partial \tilde{y}^{b}}\right) .$

The anchor map $\left( \tilde{\rho},Id_{E}\right) $\ is a $\mathbf{B}^{%
\mathbf{v}}$-morphism of $\left( \left( \rho ,\eta \right) TE,\left( \rho
,\eta \right) \tau _{E},E\right) $\ source and $\left( TE,\tau _{E},E\right)
$\ target, where
\begin{equation*}
\left( 5.2.2\right) ~\
%
\bigskip
\end{equation*}

The generalized tangent bundle%
\begin{equation*}
%
%
\end{equation*}%
can be endowed with a Lie algebroid structure%
\begin{equation*}
\left( \left[ ,\right] _{\left( \rho ,\eta \right) T\overset{\ast }{E}%
},\left( \overset{\ast }{\tilde{\rho}},Id_{\overset{\ast }{E}}\right)
\right) .
\end{equation*}

The Lie bracket $\left[ ,\right] _{\left( \rho ,\eta \right) T\overset{\ast }%
{E}}$ is defined by%
\begin{equation*}
%
%
\leqno(7.2.1)^{\prime }
\end{equation*}%
for any sections $\left( \tilde{Z}_{1}^{\alpha }\frac{\overset{\ast }{%
\partial }}{\partial \tilde{z}^{\alpha }}+Y_{1a}\frac{\partial }{\partial
\tilde{p}_{a}}\right) $ and $\left( \tilde{Z}_{2}^{\beta }\frac{\overset{%
\ast }{\partial }}{\partial \tilde{z}^{\beta }}+Y_{2b}\frac{\partial }{%
\partial \tilde{p}_{b}}\right) .$

The anchor map $\left( \overset{\ast }{\tilde{\rho}},Id_{\overset{\ast }{E}%
}\right) $\ is a $\mathbf{B}^{\mathbf{v}}$-morphism of $\left( \left( \rho
,\eta \right) T\overset{\ast }{E},\left( \rho ,\eta \right) \tau _{\overset{%
\ast }{E}},\overset{\ast }{E}\right) $\ source and $\left( T\overset{\ast }{E%
},\tau _{\overset{\ast }{E}},\overset{\ast }{E}\right) $\ target, where
\begin{equation*}
(5.2.2)^{\prime }%
%
\bigskip
\end{equation*}

\textbf{Theorem 7.2.1 }\emph{If the} $\mathbf{B}^{\mathbf{v}}$\emph{%
-morphism }$\left( \left( \rho ,\eta \right) T\varphi _{L},\varphi
_{L}\right) $ \emph{is morphism of Lie algebroids, then we obtain:}
\begin{equation*}
%
%
\leqno(7.2.6)
\end{equation*}

\bigskip\noindent\textit{Proof.} Developing the following equalities
\begin{equation*}
%
%
\end{equation*}%
it results the conclusion of the theorem. \hfill \emph{q.e.d.}

\textbf{Corollary 7.2.1 }\emph{In particular, if }$\left( \rho ,\eta
,h\right) =\left( Id_{TM},Id_{M},Id_{M}\right) $\emph{, then we obtain:}
\begin{equation*}
%
%
\leqno(7.2.6)^{\prime }
\end{equation*}

\textbf{Theorem 7.2.2 }\emph{Dual, if the the }$\mathbf{B}^{\mathbf{v}}$%
\emph{-morphism} $\left( \left( \rho ,\eta \right) T\varphi _{H},\varphi
_{H}\right) $ \emph{is morphism of Lie algebroids, then we obtain:}
\begin{equation*}
%
%
\leqno(7.2.10)
\end{equation*}

\bigskip\noindent\textit{Proof.} Developing the following equalities
\begin{equation*}
%
%
\end{equation*}%
it results the conclusion of the theorem. \hfill \emph{q.e.d.}\bigskip

\textbf{Corollary 7.2.2 }\emph{In particular, if }$\left( \rho ,\eta
,h\right) =\left( Id_{TM},Id_{M},Id_{M}\right) $\emph{, then we obtain:}%
\begin{equation*}
%
%
\leqno(7.2.10)^{\prime }
\end{equation*}

\textbf{Definition 7.2.1 }If $\left( \left( \rho ,\eta \right) T\varphi
_{L},\varphi _{L}\right) $ and $\left( \left( \rho ,\eta \right) T\varphi
_{H},\varphi _{H}\right) $ are Lie algebroids morphisms, then we will say
that $\left( E,\pi ,M\right) $\emph{\ and }$\left( \overset{\ast }{E},%
\overset{\ast }{\pi },M\right) $\emph{\ are Legendre }$\left( \rho ,\eta
,h\right) $\emph{-equivalent.}

We will write%
\begin{equation*}
%
%
\end{equation*}%
\emph{then, using the equalities }$\left( 7.2.3\right) $\emph{\ and }$%
(7.2.7) $\emph{\ it results that for any vcetor local }$\left( m+r\right) $%
\emph{-chart }$\left( U,s_{U}\right) $\emph{\ of }$\left( E,\pi ,M\right) $%
\emph{\ and for any vector local }$\left( m+r\right) $\emph{-chart }$\left(
U,\overset{\ast }{s}_{U}\right) $\emph{\ of }$\left( \overset{\ast }{E},%
\overset{\ast }{\pi },M\right) $\emph{\ we obtain:}
\begin{equation*}
%
%
\leqno(7.2.12)
\end{equation*}

\emph{Therefore, locally, }$\varphi _{L}$\emph{\ is diffeomorphism and }$%
\varphi _{L}^{-1}=\varphi _{H}.$

\subsection{The duality betwen adapted $\left( \protect\rho ,\protect\eta %
\right) $-basis}

If $\left( \rho ,\eta \right) \Gamma $ is a $\left( \rho ,\eta \right) $%
-connection for the vector bundle $\left( E,\pi ,M\right) ,$ then the
adapted $\left( \rho ,\eta \right) $-base of sections is
\begin{equation*}
%
%
\leqno(7.3.1)
\end{equation*}

If $\left( \rho ,\eta \right) \overset{\ast }{\Gamma }$ is a $\left( \rho
,\eta \right) $-connection for the vector bundle $\left( \overset{\ast }{E},%
\overset{\ast }{\pi },M\right) ,$ then the adapted $\left( \rho ,\eta
\right) $-base of sections is
\begin{equation*}
%
%
\leqno(7.3.2)
\end{equation*}

\bigskip\noindent\textit{Proof.} After some calculations, it results the
conclusion of the theorem.\hfill \emph{q.e.d.}\bigskip

\textbf{Corollary 7.3.1 }\emph{In particular, if} $\left( \rho ,\eta
,h\right) =\left( Id_{TM},Id_{M},Id_{M}\right) $\emph{, then the equality}%
\begin{equation*}
%
%
\end{equation*}%
\emph{implies the equality}
\begin{equation*}
%
%
\leqno(7.3.3)
\end{equation*}

\bigskip\noindent\textit{Proof.} After some calculations, it results the
conclusion of the theorem.\hfill \emph{q.e.d.}\bigskip

\textbf{Corollary 7.3.2 }\emph{In particular, if} $\left( \rho ,\eta
,h\right) =\left( Id_{TM},Id_{M},Id_{M}\right) $\emph{, then the equality}%
\begin{equation*}
%
%
\end{equation*}%
\emph{implies the equality}%
\begin{equation*}
%
%
\leqno(7.3.4)^{\prime }
\end{equation*}%
then we will say that $\left( E,\pi ,M\right) $\emph{\ and }$\left( \overset{%
\ast }{E},\overset{\ast }{\pi },M\right) $\emph{\ are horizontal Legendre }$%
\left( \rho ,\eta ,h\right) $\emph{-equivalent.}

We will write%
\begin{equation*}
%
%
\end{equation*}

The dual natural $\left( \rho ,\eta \right) $-base of the natural $\left(
\rho ,\eta \right) $-base $\left( \tilde{\partial}_{\alpha },\overset{\cdot }%
{\tilde{\partial}}_{a}\right) $ is denoted $\left( d\tilde{z}^{\alpha },d%
\tilde{y}^{a}\right) $ and the dual adapted $\left( \rho ,\eta \right) $%
-base of the adapted $\left( \rho ,\eta \right) $-base $\left( \tilde{\delta}%
_{\alpha },\overset{\cdot }{\tilde{\partial}}_{a}\right) $ is denoted
\begin{equation*}
%
%
\leqno(7.3.5)
\end{equation*}

The dual natural $\left( \rho ,\eta \right) $-base of the natural $\left(
\rho ,\eta \right) $-base $\left( \overset{\ast }{\tilde{\partial}}_{\alpha
},\overset{\cdot }{\tilde{\partial}}^{a}\right) $ is denoted $\left( d\tilde{%
z}^{\alpha },d\tilde{p}_{a}\right) $ and the dual adapted $\left( \rho ,\eta
\right) $-base of the adapted $\left( \rho ,\eta \right) $-base $\left(
\overset{\ast }{\tilde{\delta}}_{\alpha },\overset{\cdot }{\tilde{\partial}}%
_{a}\right) $ is denoted
\begin{equation*}
%
%
\leqno(7.3.5)^{\prime }
\end{equation*}

\textbf{Theorem 7.3.3 }\emph{The equality }$(7.3.4)$\emph{\ is equivalent
with the equality:}%
\begin{equation*}
%
%
\leqno(7.3.6)
\end{equation*}%
\emph{and the equality }$(7.3.4)^{\prime }$\emph{\ is equivalent with the
equality:}%
\begin{equation*}
%
%
\end{equation*}%
\emph{then we obtain:}%
\begin{equation*}
%
%
\leqno(7.3.7)^{\prime }
\end{equation*}

If the Lagrangian $L$ is regular, then we will define the real local
functions $\tilde{L}^{ab}$ such that
\begin{equation*}
%
%
\end{equation*}

If the Hamiltonian $H$ is regular, then we will define the real local
functions $\tilde{H}_{ab}$ such that
\begin{equation*}
%
%
\end{equation*}

\textbf{Remark 7.3.1 }If the Lagrangian $L$ is regular and
\begin{equation*}
%
%
\end{equation*}%
then, using the equalities $(7.3.7)$\ and $(7.3.7)^{\prime }$, we obtain:
\begin{equation*}
%
%
\leqno(7.3.9)
\end{equation*}

Therefore, the Hamiltonian $H$ is regular and
\begin{equation*}
%
%
\leqno(7.3.10)
\end{equation*}

It is known that the following equalities hold good\textbf{\ }
\begin{equation*}
%
%
\end{equation*}%
\emph{then, we obtain:}%
\begin{equation*}
%
%
\end{equation*}%
\emph{then we obtain}%
\begin{equation*}
%
%
\leqno(7.3.13)^{\prime }
\end{equation*}

\bigskip\noindent\textit{Proof.} Developing the following equalities%
\begin{equation*}
%
%
\end{equation*}%
it results the conclusion of the theorem.\hfill \emph{q.e.d.}

\subsection{The duality between distinguished linear $\left( \protect\rho ,%
\protect\eta \right) $-connections}

Let $\left( \rho ,\eta \right) \Gamma $ be a $\left( \rho ,\eta \right) $%
-connection for the vector bundle $\left( E,\pi ,M\right) $ and let
\begin{equation*}
%
%
\leqno(7.4.1)
\end{equation*}%
be a covariant $\left( \rho ,\eta \right) $-derivative for the tensor
algebra of generalized tangent bundle
\begin{equation*}
\left( \left( \rho ,\eta \right) TE,\left( \rho ,\eta \right) \tau
_{E},E\right)
\end{equation*}%
which \ preserves \ the \ horizontal and vertical IDS by parallelism.

If $\left( U,s_{U}\right) $ is a vector local $\left( m+r\right) $-chart for
$\left( E,\pi ,M\right) ,$ then the real local functions
\begin{equation*}
\left( \left( \rho ,\eta \right) H_{\beta \gamma }^{\alpha },\left( \rho
,\eta \right) H_{b\gamma }^{a},\left( \rho ,\eta \right) V_{\beta c}^{\alpha
},\left( \rho ,\eta \right) V_{bc}^{a}\right)
\end{equation*}%
defined on $\pi ^{-1}\left( U\right) $ and determined by the following
equalities:
\begin{equation*}
\begin{array}{ll}
\left( \rho ,\eta \right) D_{\tilde{\delta}_{\gamma }}\tilde{\delta}_{\beta
}=\left( \rho ,\eta \right) H_{\beta \gamma }^{\alpha }\tilde{\delta}%
_{\alpha }, & \left( \rho ,\eta \right) D_{\tilde{\delta}_{\gamma }}\overset{%
\cdot }{\tilde{\partial}}_{b}=\left( \rho ,\eta \right) H_{b\gamma }^{a}%
\overset{\cdot }{\tilde{\partial}}_{a} \\
\left( \rho ,\eta \right) D_{\overset{\cdot }{\tilde{\partial}}_{c}}\tilde{%
\delta}_{\beta }=\left( \rho ,\eta \right) V_{\beta _{c}}^{\alpha }\tilde{%
\delta}_{\alpha }, & \left( \rho ,\eta \right) D_{\overset{\cdot }{\tilde{%
\partial}}_{c}}\overset{\cdot }{\tilde{\partial}}_{b}=\left( \rho ,\eta
\right) V_{bc}^{a}\overset{\cdot }{\tilde{\partial}}_{a}%
\end{array}%
\leqno(7.4.2)
\end{equation*}%
are the components of a distinguished linear $\left( \rho ,\eta \right) $%
-connection $\left( \left( \rho ,\eta \right) H,\left( \rho ,\eta \right)
V\right) .$

Let $\left( \rho ,\eta \right) \overset{\ast }{\Gamma }$ be a $\left( \rho
,\eta \right) $-connection for the vector bundle $\left( \overset{\ast }{E},%
\overset{\ast }{\pi },M\right) $ and let
\begin{equation*}
\begin{array}{l}
\left( X,T\right) ^{\underrightarrow{\left( \rho ,\eta \right) \overset{\ast
}{D}}\,}\vspace*{1mm}\left( \rho ,\eta \right) \overset{\ast }{D}_{X}T%
\end{array}%
\leqno(7.4.1)^{\prime }
\end{equation*}%
be a covariant $\left( \rho ,\eta \right) $-derivative for the tensor
algebra of generalized tangent bundle
\begin{equation*}
\left( \left( \rho ,\eta \right) T\overset{\ast }{E},\left( \rho ,\eta
\right) \tau _{\overset{\ast }{E}},\overset{\ast }{E}\right)
\end{equation*}%
which \ preserves \ the \ horizontal and vertical IDS by parallelism.

If $\left( U,\overset{\ast }{s}_{U}\right) $ is a vector local $\left(
m+r\right) $-chart for $\left( \overset{\ast }{E},\overset{\ast }{\pi }%
,M\right) ,$ then the real local functions
\begin{equation*}
\left( \left( \rho ,\eta \right) \overset{\ast }{H}_{\beta \gamma }^{\alpha
},\left( \rho ,\eta \right) \overset{\ast }{H}_{b\gamma }^{a},\left( \rho
,\eta \right) \overset{\ast }{V}_{\beta }^{\alpha c},\left( \rho ,\eta
\right) \overset{\ast }{V}_{b}^{ac}\right)
\end{equation*}%
defined on $\overset{\ast }{\pi }^{-1}\left( U\right) $ and determined by
the following equalities:
\begin{equation*}
%
\leqno(7.4.2)^{\prime }
\end{equation*}%
are the components of a distinguished linear $\left( \rho ,\eta \right) $%
-connection
\begin{equation*}
\left( \left( \rho ,\eta \right) \overset{\ast }{H},\left( \rho ,\eta
\right) \overset{\ast }{V}\right) .
\end{equation*}

\textbf{Theorem 7.4.1 }\emph{If}
\begin{equation*}
%
%
\end{equation*}%
for any $X,Y\in \Gamma \left( \left( \rho ,\eta \right) TE,\left( \rho ,\eta
\right) \tau _{E},E\right) $, then we obtain:%
\begin{equation*}
%
%
\end{equation*}%
for any $X,Y\in \Gamma \left( \left( \rho ,\eta \right) T\overset{\ast }{E}%
,\left( \rho ,\eta \right) \tau _{\overset{\ast }{E}},\overset{\ast }{E}%
\right) $, then we obtain:%
\begin{equation*}
%
%
\leqno(7.4.6)^{\prime }
\end{equation*}

\section*{Acknowledgment}

\addcontentsline{toc}{section}{Acknowledgment}

I am very grateful to Professor Mihai ANASTASIEI from A.I. Cuza University,
Romania, the supervisor of my Doctoral Thesis.

I would like to thank Matsumae International Foundation for the research
grant at Tokai University, during April-September 2008.

I also would like to thank Professors Hideo SHIMADA and Sorin Vasile SABAU
from Tokai University-Japan, for useful discussions and their suggestions
which improved the final manuscript. My gratitude to Professor Joszef
Szilasi from Debrecen University, Hungary, for his continuous support,
encouragement and numerous valuable suggestions.

To Mrs. Elena MOCANU, who helped me to improve the final Latex form of the
text.

I want to express my gratitude to all who helped me in accomplishing this
project.

Last but not least, I want to thank my family for supporting me all the time
in this effort.

\bigskip

\bigskip

\hfill%
\begin{tabular}{c}
SECONDARY SCHOOL \textquotedblleft CORNELIUS RADU\textquotedblright, \\
RADINESTI VILLAGE, 217196, GORJ COUNTY, ROMANIA \\
e-mail: c\_arcus@yahoo.com, c\_arcus@radinesti.ro%
\end{tabular}

\end{document}